%% file: thesis.tex
\include{settings}	

\documentclass[12pt,openany]{book}

\usepackage[paper = a4paper,         
			twoside,                 
			bindingoffset = 2mm,     
			hmargin = 25mm,          
			vmargin = 25mm,          
			dvipdfm]
			{geometry}

\usepackage[T1]{fontenc}	
\usepackage{mathptmx}       
\usepackage[scaled]{helvet} 
\usepackage{sectsty}        

\usepackage{graphicx}       
\usepackage{amsmath,nccmath}
\usepackage{mathtools}
\usepackage{siunitx}
\usepackage{amsfonts}
\usepackage{amssymb}
\usepackage[mathcal]{euscript} 
\usepackage{booktabs}       
\usepackage{setspace}       
\usepackage{forloop}		
\usepackage[round]{natbib}  
\usepackage{hypernat}       
\usepackage{float}
\usepackage{natbib}  
\usepackage[table]{xcolor}
\usepackage[final]{pdfpages}
\usepackage{epigraph}			
\usepackage{textcomp}
\usepackage{sidecap} 
\usepackage{enumitem} 

\usepackage{titlesec} 
\titleformat{\title}[display]
  {\normalfont\rmfamily\huge\bfseries\color{black}}
  {\chaptertitlename\ \thechapter}{20pt}{\Huge}
\titleformat{\chapter}[display]
  {\normalfont\rmfamily\huge\bfseries\color{black}}
  {\chaptertitlename\ \thechapter}{20pt}{\Huge}
\titleformat{\section}
  {\normalfont\rmfamily\Large\bfseries\color{black}}
  {\thesection}{1em}{}
\titleformat{\subsection}
  {\normalfont\rmfamily\Large\bfseries\color{black}}
  {\thesubsection}{1em}{}

\usepackage[numbib]{tocbibind}
\usepackage[tight]{shorttoc}    
\usepackage[english]{minitoc}   
\usepackage[
			backref,         
			colorlinks,      
			citecolor=black,  
			linkcolor=black,  
			urlcolor=blue,   
			]{hyperref}

\usepackage{fancyhdr}
\fancypagestyle{plain}{
	\fancyhead{} 
}
\usepackage[font={small,it,rm}]{caption}
\usepackage{atveryend}
\usepackage{pdfpages}

\include{macros}

\title{{\fontsize{25}{24}\sffamily\textbf{Title}}
	\author{
	\large\textbf{Your Name}
	    \vspace{2em}\\
	    \large\textbf{Supervisor Name}
	    \vspace{4em}\\
		\includegraphics[width=74mm]{figures/uglo}\vspace{4em}\\
		A Preliminary Master's Thesis\\
		Semester Assignment for INFO300 \\
		Department of Information Science and Media Studies \\
		University of Bergen}
	\huge\date{\today}
}

\begin{document}

\dominitoc
\doparttoc


\ifDownscaledFinalDoc
	\fontsize{\TextSize}{\BaseLineSkip}
	\selectfont
\fi

\ifDraft
	\doublespacing
\fi

\renewcommand{\familydefault}{\sfdefault} 

\includepdf[pages={1}]{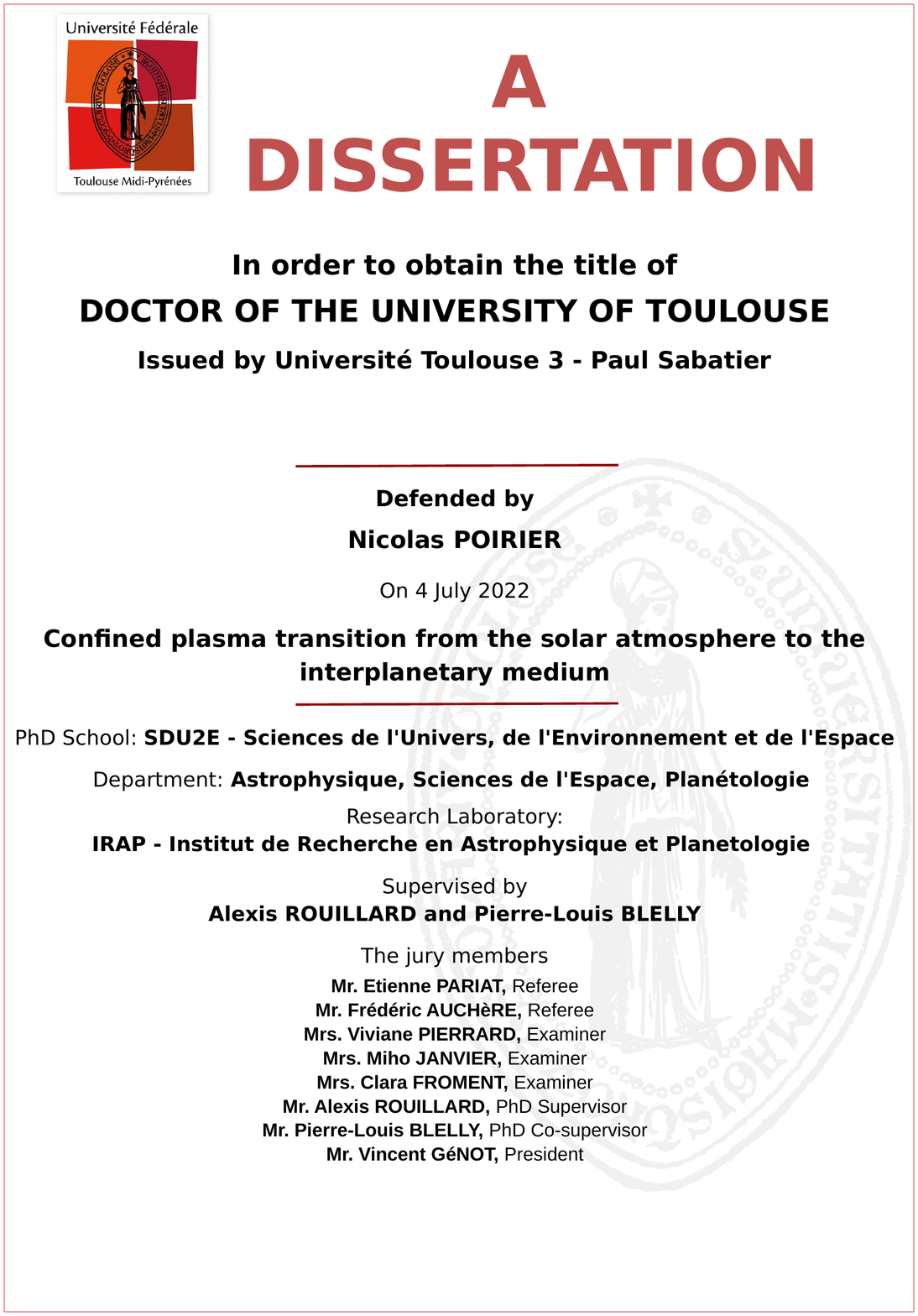}
\includepdf[pages={1}]{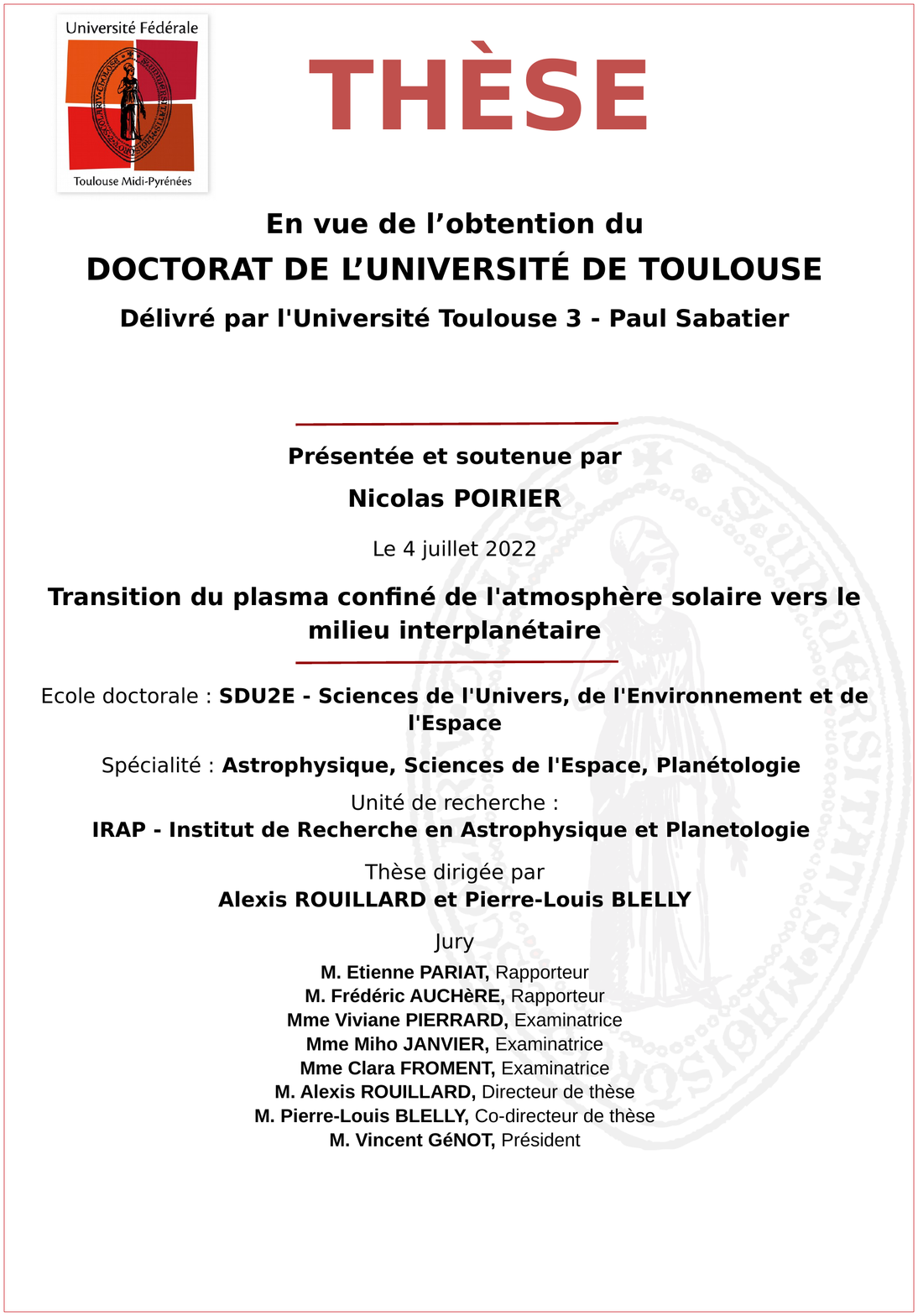}

\frontmatter
\normalfont\rmfamily

\include{chapters/02acknowledgements}
\include{chapters/Abstract}

\shorttoc{Table of contents}{1}

\mainmatter

%
%
\include{chapters/Introduction}      
\include{chapters/Introduction_FR}  
\include{chapters/Instruments_tools}
\include{chapters/Stationnary}
\include{chapters/Dynamics}
\include{chapters/ISAM_v2}

\include{chapters/ISAM_results}
\include{chapters/Conclusion_perspectives}

%
%
\appendix
\include{appendices/ISAM}
\include{appendices/List_papers}
\include{appendices/Communications}

\backmatter

\include{chapters/TableofContents}
\include{chapters/ListofAbreviations}
\include{chapters/ListofFigures}

%
%

%
%

%
%
\include{Journal_abreviations}

\bibliographystyle{AASJournal}

\settocbibname{References}
\bibliography{thesis}

\end{document}

%% file: settings.tex
\newcommand{\TextSize}{13}
\newcommand{\BaseLineSkip}{15}
\newif\ifDownscaledFinalDoc
	\DownscaledFinalDoctrue		

\newif\ifDraft
	\Draftfalse		

%% file: macros.tex
\allowdisplaybreaks

\renewcommand{\equiv}[0]{\ensuremath{:=}}

\newcounter{ct}

\newcommand{\includepaperpages}[2]
{
	\forloop{ct}{0}{\value{ct} < #2}
	{
		\begin{figure}[ht]
			\includegraphics[width=.99\textwidth]{#1\the\value{ct}.ps}
		\end{figure}
		\clearpage
	}
}

%% file: chapters/02acknowledgements.tex
\chapter{Acknowledgements}

Preparing for a Ph.D. is like a box of chocolates, you never know what you're gonna get\footnote{from the original quote of Robert Zemeckis in the Forrest Gump movie}. This is how I would summarize these three years of PhD that went by so fast, a journey punctuated with many obstacles but also happy moments. Anyway, I may never thank enough all the people who have accompanied me from near or far during this adventure. \\

I would like to start by thanking all the jury members: Vincent Génot for having accepted the presidency, Vivianne Pierrard, Miho Janvier and Clara Froment for their participation as examiners, and finally Etienne Pariat and Frédéric Auchère for their dedication as reviewers. I thank you all for your attention in reviewing this manuscript, as well as for your very constructive remarks and questions.

I will never be too grateful to my two Ph.D. supervisors, Alexis Rouillard and Pierre-Louis Blelly, for their guidance and advice during these three years. It has been a pleasure to work with you both from a scientific and human point of view. I must particularly acknowledge the devotion of Alexis Rouillard to his folks even during hard times. Thank you for trusting me over four years, starting from my Master project followed by this thesis. \\

I dedicate this Ph.D. thesis to my first mentor Shahab Fatemi who transmitted me his passion for science, and who allowed me to take my first steps in astrophysics during a 3-month internship (summer 2017). A very enriching first experience at the Swedish Institute of Space Physics (Institutet för rymdfysik, IRF), to which I must commend its members who gave me a warm welcome during these three months in Kiruna (Sweden), including for instance Shahab Fatemi, Kei Masunaga, Audrey Schillings, Moa Persson, George Nicolaou, Mats Holmström and Stas Barabash. \\

These three years of Ph.D. have been rich in meetings, collaborations and discussions (scientific or not) that have driven me to where I am today. I sincerely thank the \textit{WISPR} instrument team including e.g. Russell A. Howard, Angelos Vourlidas and Nour-Edine Raouafi for giving me the opportunity to exploit the very first of these magnificent observations made by WISPR, but also the team from the international working group ISSI for all the fruitful exchanges that we had and the collaborations that resulted, the Modelling and Data Analysis Working Group (MADAWG) in which I could participate to the operations of the \textit{Solar Orbiter} mission, and finally all the people with whom I could interact during conferences and workshops.

A Ph.D. and participation in conferences made possible thanks to the financial support from the European Research Council (ERC) throughout the \textit{SLOW\_SOURCE} project (DLV-819189), a project initiated and directed by Alexis Rouillard. I must also mention all the institutions that have funded the production/development of the data sets and tools that allowed me to carry out this work. Among which the \href{https://cnes.fr/fr}{Centre National des Études Spatiales (CNES)} which supports the \href{http://cdpp.eu/}{Centre de Données de la Physique des Plasmas (CDPP)}, the \href{https://idoc.ias.u-psud.fr/MEDOC}{Multi Experiment Data \& Operation Center (MEDOC)}, and the space weather pole of Toulouse (\href{http://storms-service.irap.omp.eu/}{Solar-Terrestrial Observations and Modelling Service, STORMS}). This includes funding for tools such as \href{http://amda.cdpp.eu/}{AMDA}, \href{http://clweb.irap.omp.eu/}{ClWEB}, the \href{http://propagationtool.cdpp.eu}{propagation tool} and the \href{http://connect-tool.irap.omp.eu/}{Magnetic Connectivity Tool (MCT)}. Also ESA, NASA and their partner institutions for operating the \textit{PSP}, \textit{STEREO}, \textit{SoHO}, \textit{ACE} and \textit{WIND} missions and for providing their observations. This work also exploited the \href{https://ui.adsabs.harvard.edu/}{Astrophysics Data System (ADS)} operated by the Smithsonian Astrophysical Observatory (SAO) and funded by NASA. 

This project has been carried out at the \href{https://www.irap.omp.eu/}{Institute of Research in Astrophysics and Planetology (IRAP)} with a technical and administrative team always in support, and within which I would like to thank in particular Dorine Roma, Josette Garcia, Sandrine Chupin, Alexandre Baudrimont, Philippe Louarn, Nicole Briat, Sélim Benguesmia and Jean-François Botte. \\

I have met many people at IRAP as colleagues but above all friends, who have been like a family and with whom I have shared such wonderful moments. I cannot mention everyone but you will recognize yourselves: Corentin Louis, Kévin Dalmasse, Michael Lavarra, Athanasios Kouloumvakos, Mikel Indurain, Issaad Kacem, Matthieu Alexandre, Naïs Fargette, Victor Réville, Rui Pinto, Léa Griton, Anaïs Amato, Emeline Valette, Sae Aizawa, Rungployphan Kieokaew, Sid Fadanelli, Nathanael Jourdanne. I would also like to mention the new Ph.D students who joined the team recently as well as the many Master students who have been part of the team.

I also sincerely thank my friends from other horizons who have helped me in keeping motivation and smiling even during the most difficult moments of this Ph.D., especially at the height of the Covid-19 pandemic. \\

Finally, I would like to thank my family: my two sisters and parents, cousins, aunts and uncles, and grandparents who have supported me during all these years of study.

\vspace*{\fill}
\begin {flushright}
Nicolas Poirier\\
Toulouse, 04/07/2022
\end{flushright}
\clearpage


\chapter{Remerciements}

Préparer un doctorat c'est comme une boite de chocolat, on ne sait jamais sur quoi on va tomber\footnote{en empruntant la citation originelle de Robert Zemeckis dans le rôle de Forrest Gump.}. C'est de cette manière que je résumerais ces trois années de thèse qui sont passées si vite, un périple ponctué par de nombreux obstacles mais aussi de moments forts heureux. Quoi qu'il en soit, je ne saurais jamais assez remercier l'ensemble des personnes m'ayant accompagnées de près ou de loin durant cette aventure. \\

Je commencerais par remercier l'ensemble des membres de mon jury: Vincent Génot pour en avoir accepté la présidence, Vivianne Pierrard, Miho Janvier et Clara Froment pour leur participation en tant qu'examinatrices, et enfin Etienne Pariat et Frédéric Auchère pour leur implication en tant que rapporteurs. Je vous remercie pour votre attention dans l'examen de ce manuscrit, ainsi que pour vos remarques et questions très constructives.

Je ne serais jamais trop reconnaissant envers mes deux directeurs de thèse, Alexis Rouillard et Pierre-Louis Blelly pour m'avoir guidé et conseillé durant ces trois années. Ce fût un plaisir de travailler avec vous tant d'un point de vue scientifique qu'humain. Je me dois particulièrement de saluer le dévouement d'Alexis Rouillard envers son équipe, et ceux même dans les moments difficiles. Merci de m'avoir accordé ta confiance pendant près de quatre ans, depuis mon projet de Master jusqu'à cette thèse. \\

Je dédie cette thèse à mon premier mentor Shahab Fatemi qui m'a transmis sa passion pour la science, et qui m'a permis de faire mes premiers pas dans l'astrophysique lors d'un stage de trois mois (été 2017). Une première expérience très enrichissante au sein de la Swedish Institute of Space Physics (Institutet för rymdfysik, IRF), à laquelle je dois saluer l'accueil chaleureux que ses membres m'ont réservé pendant ces trois mois à Kiruna (Suède), pour ne citer que Shahab Fatemi, Kei Masunaga, Audrey Schillings, Moa Persson, George Nicolaou, Mats Holmström et Stas Barabash. \\

Ces trois années de thèse ont été riches en rencontres, collaborations et discussions (scientifiques ou non) qui m'ont permis d'en arriver là aujourd'hui. Je remercie pour cela sincèrement l'équipe de l'instrument \textit{WISPR} dont e.g. Russell A. Howard, Angelos Vourlidas et Nour-Edine Raouafi pour m'avoir offert l'opportunité d'exploiter les toutes premières de ces superbes observations réalisées par \textit{WISPR}, mais aussi l'équipe du groupe de travail international ISSI pour tous les échanges fructueux qu'on a pu avoir et les collaborations qui en ont résulté, le Modelling and Data Analysis Working Group (MADAWG) au sein duquel j'ai pu participer aux opérations de la mission \textit{Solar Orbiter}, et enfin l'ensemble des personnes avec lesquelles j'ai pu interagir pendant des conférences et workshops.

Une thèse et des participations aux conférences rendues possibles grâce au conseil européen pour la recherche qui les a financées dans le cadre d'un projet nommé \textit{SLOW\_SOURCE} (DLV-819189), un projet initié et dirigé par Alexis Rouillard. Je me dois également de mentionner l'ensemble des institutions ayant financé la production/développement des jeux de données et outils qui m'ont permis de mener à bien ces travaux. Dont le \href{https://cnes.fr/fr}{Centre National des Études Spatiales (CNES)} qui finance le \href{http://cdpp.eu/}{Centre de Données de la Physique des Plasmas (CDPP)}, le \href{https://idoc.ias.u-psud.fr/MEDOC}{Multi Experiment Data \& Operation Center (MEDOC)}, et le pôle de météo de l'espace de Toulouse (\href{http://storms-service.irap.omp.eu/}{Solar-Terrestrial Observations and Modelling Service, STORMS}). Cela inclut le financement des outils tels que \href{http://amda.cdpp.eu/}{AMDA}, \href{http://clweb.irap.omp.eu/}{ClWEB}, le \href{http://propagationtool.cdpp.eu}{propagation tool} et le \href{http://connect-tool.irap.omp.eu/}{Magnetic Connectivity Tool (MCT)}. L'ESA, la NASA et leurs institutions partenaires pour la mise en œuvre des missions \textit{PSP}, \textit{STEREO}, \textit{SoHO}, \textit{ACE}, \textit{WIND} et la mise à disposition de leurs données d'observations. Ces travaux ont aussi exploité le \href{https://ui.adsabs.harvard.edu/}{Astrophysics Data System (ADS)} opéré par le Smithsonian Astrophysical Observatory (SAO) et financé par la NASA. 

Cette thèse a été réalisée à l'\href{https://www.irap.omp.eu/}{Institut de Recherche en Astrophysique et Planétologie (IRAP)} à Toulouse (France) auprès d'une équipe technique et administrative toujours en soutient, et au sein de laquelle je souhaiterais remercier notamment Dorine Roma, Josette Garcia, Sandrine Chupin, Alexandre Baudrimont, Philippe Louarn, Nicole Briat, Sélim Benguesmia et Jean-François Botte. \\

Des rencontres réalisées aussi à l'IRAP avec des collègues mais aussi et surtout ami(e)s, vous qui avez été comme une famille et avec qui j'ai pu partager de merveilleux moments. Je ne pourrais pas tous vous citer mais vous vous reconnaîtrez: Corentin Louis, Kévin Dalmasse, Michael Lavarra, Athanasios Kouloumvakos, Mikel Indurain, Issaad Kacem, Matthieu Alexandre, Naïs Fargette, Victor Réville, Rui Pinto, Léa Griton, Anaïs Amato, Emeline Valette, Sae Aizawa, Rungployphan Kieokaew, Sid Fadanelli, Nathanael Jourdanne. J'aimerais aussi mentionner les nouveaux doctorant(e)s ayant intégré(e)s l'équipe récemment ainsi que les stagiaires en Master qui en ont fait partie.

Je remercie aussi très sincèrement mes amis d'autres horizons qui m'ont permis de garder la motivation et le sourire même pendant les moments les plus difficiles de cette thèse, surtout au plus fort de la pandémie de Covid-19. \\

Enfin je concluerais en remerciant cette fois-ci ma famille: mes deux s\oe urs et parents, cousins et cousines, oncles et tantes, et grands-parents qui m'ont soutenu durant toutes ces années d'études.

\vspace*{\fill}
\begin {flushright}
Nicolas Poirier\\
Toulouse, 04/07/2022
\end{flushright}
\clearpage


%% file: chapters/Abstract.tex
\chapter{Abstract}
The last 60 years of space exploration have shown that the interplanetary medium is continually perturbed by a myriad of different solar winds and storms that transport solar material across the whole heliosphere. If there is a consensus on the source of the fast solar wind that is known to originate in coronal holes, the question is still largely debated on the origin of the slow solar wind (SSW). The abundance of heavy ions measured in situ provides a precious diagnostic of potential source regions, because the composition is established very low in the solar atmosphere at the interface between the dense chromosphere and the tenuous corona, and remains invariant during transport in the collisionless solar wind. The similar composition measured in situ in the SSW and spectroscopically in coronal loops suggests that a significant fraction of the SSW originates as plasma material that was initially trapped along corona loops and subsequently released in the solar wind. The recent observations from the \textit{Parker Solar Probe} (\textit{PSP}) mission also provide new insights on the nascent solar wind. A great challenge remains to explain both the composition and bulk properties of the SSW in a self-consistent manner. For this purpose we exploit and develop models with various degrees of complexity. This context constitutes the backbone of this thesis which is structured in four major steps: we begin by presenting a new technique that exploits white-light (WL) observations of the SSW taken from multiple vantage points to constrain global models of the solar atmosphere. We then exploit the first images taken by the \textit{Wide-Field Imager for Solar PRobe} (\textit{WISPR}) from inside the solar corona to test our global models at smaller scales because \textit{WISPR} offers an unprecedented close-up view of the fine structure of streamers and of the nascent SSW. This work provides further evidence for the transient release of plasma trapped in coronal loops into the solar wind, that we interpret by exploiting high-resolution magneto-hydro-dynamics (MHD) simulations. Finally we develop and exploit a new multi-specie model of coronal loops called the Irap Solar Atmosphere Model (ISAM) to provide an in-depth analysis of the plasma transport mechanisms at play between the chromosphere and the corona. ISAM solves for the coupled transport of the main constituents of the solar wind with minor ions through a comprehensive treatment of collisions as well as partial ionization and radiative cooling/heating mechanisms near the top of the chromosphere. We use this model to study the different mechanisms that can preferentially extract ions according to their first-ionization-potential (FIP), from the chromosphere to the corona. In this process we compare the relative roles of frictional and thermal diffusive effects in enriching coronal loops with low-FIP elements that could be subsequently expelled in the SSW through the mechanisms discussed in the first part of this thesis.

\clearpage

\chapter{Résumé}
Les 60 dernières années d'exploration spatiale ont montré que le milieu interplanétaire est continuellement perturbé par une myriade de vents et de tempêtes solaires qui transportent de la matière solaire dans toute l'héliosphère. S'il existe un consensus sur la source du vent solaire rapide dont on sait qu'il provient des trous coronaux, la question de l'origine du vent solaire lent (SSW) est encore largement débattue. L'abondance des ions lourds mesurée in situ fournit un précieux diagnostic des potentielles régions sources, car la composition est établie très bas dans l'atmosphère solaire à l'interface entre la dense chromosphère et la couronne ténue, et reste invariable pendant le transport dans le vent solaire sans collisions. Les compositions similaires mesurées à la fois in situ dans le SSW et par spectroscopie dans les boucles coronales suggèrent qu'une fraction significative du SSW proviendrait de plasmas qui seraient initialement piégés le long des boucles coronales et ensuite libérés dans le vent solaire. Les observations récentes de la sonde \textit{Parker Solar Probe} (\textit{PSP}) fournissent également de nouvelles informations sur le vent solaire naissant. Il reste un grand défi cependant d'expliquer à la fois la composition et les propriétés macroscopiques du SSW d'une manière autoconsistante. À cette fin, nous exploitons et développons des modèles de différents degrés de complexité. Ce contexte constitue le fil conducteur de cette thèse qui est structurée en quatre grands axes: nous commençons par présenter une nouvelle technique qui exploite les observations en lumière blanche (WL) du SSW prises depuis de multiples points d'observation afin de mieux contraindre les modèles globaux de l'atmosphère solaire. Nous exploitons ensuite les premières images en lumière blanche du télescope \textit{Wide-Field Imager for Solar PRobe} (\textit{WISPR}) à bord de \textit{PSP}, prises depuis l'intérieur de la couronne solaire dans le but de tester nos modèles globaux à des échelles encore plus petites, car \textit{WISPR} offre une vue rapprochée inédite de la structure fine des streamers et du SSW encore naissant. Ce travail fournit des preuves supplémentaires d'une éjection intermittente de plasmas initiallement piégés dans les boucles coronales dans le vent solaire, que nous interprétons en exploitant des simulations magnéto-hydro-dynamiques (MHD) à haute résolution. Enfin, nous développons et exploitons un nouveau modèle multi-espèces de boucles coronales appelé ISAM (pour Irap Solar Atmosphere Model) pour fournir une analyse approfondie des mécanismes de transport du plasma en action entre la chromosphère et la couronne. ISAM résout le transport couplé des principaux constituants du vent solaire avec les ions mineurs par un traitement complet des collisions ainsi que des mécanismes d'ionisation partielle et de refroidissement/chauffage radiatif qui sont importants dans la partie haute de la chromosphère. Nous utilisons ce modèle pour étudier les différents mécanismes qui peuvent extraire préférentiellement les ions de la chromosphère vers la couronne, en fonction de leur potentiel de première ionisation (FIP). Dans ce processus, nous comparons les rôles relatifs des effets frictionnels et thermiques diffusifs dans l'enrichissement des boucles coronales avec des éléments à faible FIP qui pourraient ensuite être expulsés dans le SSW par les mécanismes discutés dans la première partie de cette thèse.

%% file: chapters/Introduction.tex
\chapter{Introduction}
\label{cha:intro}

\minitoc

The occurrence of the aurora was suspected, as early as the second half of the nineteenth century, to be induced by occasional streams of solar particles impinging on the Earth's atmosphere. Near the middle of the twentieth century the German astronomer Ludwig Biermann \citep{Biermann1951} concluded from observations of comet tails that solar particles interacting with comets were likely part of a continuous (and not sporadic) flow of particles expelled from the solar atmosphere. Sydney Chapman, a British astronomer and geophysicist then argued that the solar atmosphere could extend well beyond the corona probably all the way to at least the Earth's orbit \citep{Chapman1957}. It is only in 1958 that \citet{Parker1958} showed theoretically that the Sun produces a supersonic flow of charged particles that he called the solar wind. This was rapidly validated by the first space missions (\textit{Luna-1,2,3}: see e.g. \citet{Gringauz1961}, \textit{Venera-1}: \citet{Gringauz1964}, and \textit{Mariner-2}: \citet{Neugebauer1962}) and laid the foundations for half a decade of solar physics research. \\

\begin{figure*}[]
\centering
\includegraphics[width=0.85\textwidth]{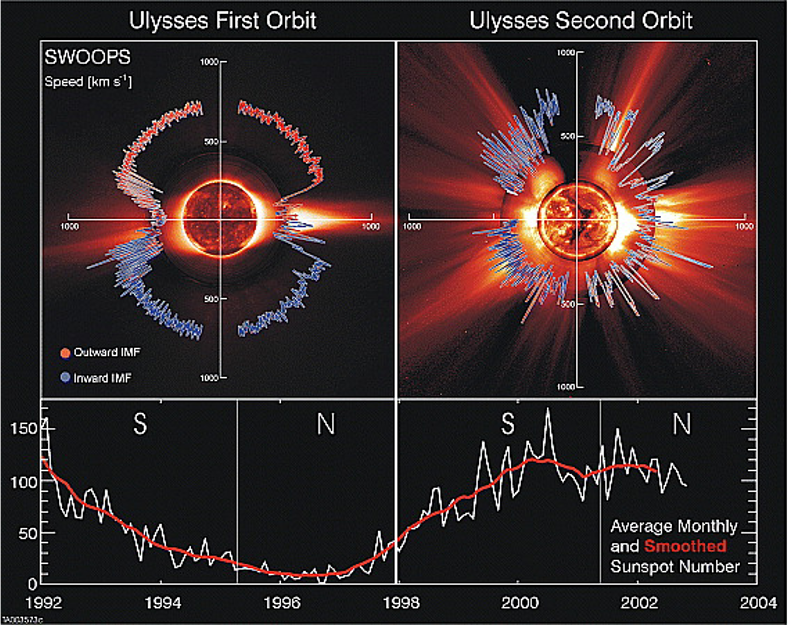}
\caption{ The solar wind speed as sampled by \textit{Ulysses} during its first (left panel) and second (right panel) passage. The large-scale configuration of the solar corona is illustrated with composite images assembled from: solar disk images taken by the \textit{SoHO} \textit{Extreme ultraviolet Imager Telescope} at $195$\si{\angstrom}, Mauna Loa K-coronameter images ($7000$-$9500$\si{\angstrom}) of the inner corona, and WL images from the \textit{SoHO} \textit{LASCO-C2} coronagraph. The bottom panel shows the evolution of the sunspot number that is a good indicator of the level of solar activity. Figure adapted from \citet[][Figure 1]{McComas2003}.
\label{fig:Ulysses}}
\end{figure*}

\begin{figure*}[]
\centering
\includegraphics[width=0.95\textwidth]{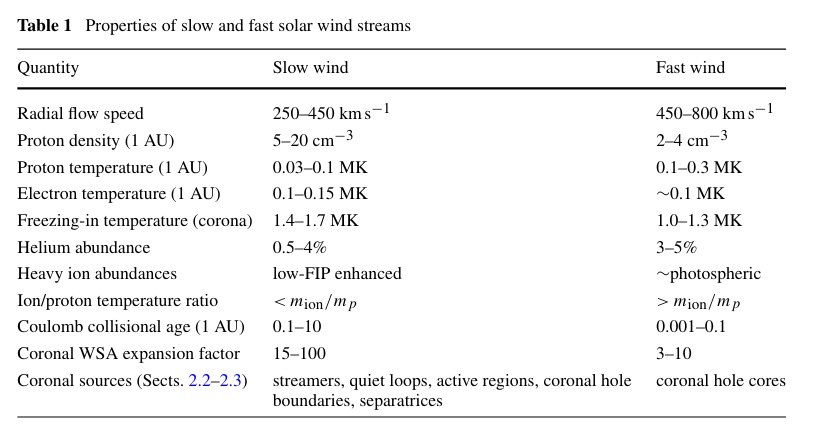}
\caption{Properties of the slow and fast solar wind. Figure taken from \citet[][Table 1]{Cranmer2017}.
\label{tab:Cranmer2017_tab1}}
\end{figure*}

In the late twentieth century, the \textit{Ulysses} joint ESA-NASA mission was the first to explore the solar wind out of the solar ecliptic plane (of about $80^\circ$) collecting unprecedented in situ measurements of the solar corona and heliosphere at high latitudes. The first orbit of \textit{Ulysses} around solar minimum unveiled what appeared to be a bimodal nature of the solar wind with a slow wind present mainly at low latitudes and a fast solar wind predominant in the polar regions (see the left panel of Figure \ref{fig:Ulysses}) \citep{McComas2003}. Indeed the solar wind has been classified in two main regimes, the slow or fast solar wind according to its bulk velocity is lower or greater than $450\ \rm{km/s}$. A substantial review of the main differences between the slow and fast solar winds has been carried out by \citet{Cranmer2017} for which a summary of their properties is given in Table \ref{tab:Cranmer2017_tab1}. A recent study by \citet{Sanchez-Diaz2016} even identified a very slow solar wind regime with typical bulk velocities slower than $300\ \rm{km/s}$.

\section{General considerations on the enigmatic origin of the slow wind}

If there is now a consensus on the source region of the fast solar wind, the origin of the slow wind remains highly debated. The \textit{Parker Solar Probe} \citep[PSP: ][]{Fox2016} and \textit{Solar Orbiter} \citep[SolO: ][]{Muller2013,Muller2020} missions launched in August 2018 and February 2020 respectively, have been specifically designed to address this question. By measuring the solar wind in situ and imaging the solar atmosphere closer to the Sun than ever before, these missions are providing a wealth of new information on the nascent solar winds and their possible origins. \\

The solar wind is primarily composed of ionized Hydrogen (protons) at $\approx 95\%$ and of doubly ionized Helium (alpha particles) at $\approx 4\%$ where the $\approx 1\%$ left includes a myriad of minor heavier ions \citep{Rouillard2021}. In addition to the bimodal speed distribution of the solar wind illustrated by \textit{Ulysses} measurements, additional in situ data from the \textit{Solar Wind Ion Composition Spectrometer} \citep[SWICS: ][]{Gloeckler1992} on \textit{Ulysses} and on the \textit{Advanced Composition Explorer} (\textit{ACE}) has shown a significant variability in the abundances of alpha particles \citep[see e.g.][]{Kasper2007,McGregor2011} and minor ions in the solar wind \citep{Geiss1995,Steiger1996}
, as well as in the charge states of minor ions \citep[see e.g.][]{Neugebauer2002,Liewer2004,Stakhiv2015,Stakhiv2016}, that is further discussed in section \ref{subsubsec:intro_SSW_composition}. Since the composition of the solar wind does not change between the solar corona and the point of in situ measurements in the heliosphere, the different compositions of the fast and slow winds have been related to different source locations. In particular the ionic abundances measured in situ in the slow and fast wind have been associated with those measured by spectroscopy in coronal loops of active regions \citep[see e.g.][]{Ko2002,Brooks2011,Doschek2019} and coronal holes respectively \citep{Feldman1998b}. \\

Two theories stand out for the formation of the slow solar wind. The premises of the coronal and heliospheric white-light imagery with the \textit{Solar and Heliospheric Observatory} \citep[SoHO: ][]{Domingo1995} depicted a highly variable slow wind \citep{Sheeley1997} which was further supported later on by observations from the \textit{Solar-TErrestrial RElations Observatory} \citep[STEREO: ][]{Kaiser2008} \citep[see e.g.][]{Rouillard2009,Rouillard2011a,Plotnikov2016,Sanchez-Diaz2017a,DeForest2018} and recently from \textit{PSP} \citep{Rouillard2020a}, and consequently led up to a myriad of dynamic theories where the slow wind may be sporadically supplied with plasma that is initially confined along coronal loops. This theory explains quite naturally the ionic abundance of the slow wind. In parallel, a quasi-stationary theory has been proposed that unifies the slow and fast wind as the different manifestations of plasma continuously accelerated along open magnetic field lines. Although this quasi-stationary theory struggles to reconcile composition measurements of the slow wind in situ with spectroscopic observations of loops, it has the advantage of providing quantitative theoretical predictions for the bulk properties of the fast and slow solar wind. Hence the quasi-stationary and dynamic theories have their own advantages and caveats, which will be introduced in section \ref{sec:intro_stationnary} and \ref{sec:intro_dynamic} and further discussed throughout this thesis. \\

A similar enrichment in certain minor ions, those that have a low first ionization potential (FIP) like iron and magnesium, measured in situ in the slow wind, has also been observed in the coronal loops that constitute the so-called "closed corona". The physical processes that enrich the corona with low-FIP elements, the so-called "FIP effect", is still debated (see section \ref{subsec:intro_FIP}). In addition to looking at the processes that may expel this enriched plasma into the slow wind (referred to as the \emph{expulsion} process), we must also address how these minor ions, which are much heavier than the main proton constituents, manage to escape the deep solar atmosphere where they are formed to enter the corona (referred to as the \emph{extraction} process). In the context of the possible origins of the slow solar wind, can modeling of this extraction process (or as we shall see fractionation of heavy ions by their first ionisation potential) provide an additional test to separate between a quasi-stationary and dynamic theories of the slow wind? \\

The quasi-stationary theory is elegant in the sense that it can reproduce many bulk properties of both the slow and fast wind, of which the bimodal variation of the solar wind speed but also the mean temperature and density. For that purpose, the MULTI-VP model \citep{PintoRouillard2017} that is introduced in section \ref{subsec:MULTI-VP} is exploited in section \ref{sec:dynamics_insitu} to analyse the structure of the slow wind and streamers from a quasi-stationnary perspective. A close-up analysis of recent remote-sensing observations by \textit{Parker Solar Probe} supports the quasi-stationary behavior of the slow wind at even smaller scales that is discussed in chapter \ref{cha:stationnary}. In section \ref{sec:dynamics_griton2020}, we show that the quasi-stationnary theory can be extended to explain some of the intermittent features observed in white-light remote-sensing observations of the slow wind. Finally we shall see that this theory may be a key to explain the measured abundances of the slow wind in heavy ions, by considering quasi-stationary processes that involve diffusion/collision processes that are introduced in section \ref{subsec:intro_FIP} and analysed in detail in chapter \ref{cha:ISAM_results}.\\

Only dynamic theories of the slow wind implying the phenomenon of magnetic reconnection can allow plasma transfer from coronal loops to the slow wind. Therefore they naturally appear more fitted to explain why the slow wind exhibits abundances that are similar to the ones observed in coronal loops. Yet, it remains highly difficult to pinpoint precisely the source regions of the slow wind because magnetic reconnection with coronal loops can occur at various places in the solar atmosphere. As we shall see in section \ref{sec:intro_dynamic}, most of the slow solar wind originates above streamers and large coronal loops close to the so-called heliospheric plasma sheet (HPS), however other sources that exhibit typical slow wind compositions have also been identified far from the HPS \citep{Zurbuchen2007}. The intermittent nature of the slow wind that is observed in white-light imagery can be explained in part by magnetic reconnection at the tip of streamers, which is further investigated in section \ref{sec:dynamics_tearing} with a magneto-hydrodynamic model called WindPredict-AW \citep{Reville2020a}. A common thought is that the longer a plasma is kept trapped along coronal loops, the more it is susceptible to undergo fractionation processes, such as diffusion effects, gravitational stratification and wave-particle interactions, that may enrich the solar atmosphere in low-FIP elements. For instance an increase of the coronal abundance of minor ions having a low FIP has been observed during the aging of active regions from \textit{Skylab} data \citep{Widing2001}. Those results have been mitigated using recent observations from the \textit{Hinode} mission that suggest that the magnetic field plays a major role in the abundance evolution in active regions where reconnection of the pre-existing field with magnetic flux emerging from the photosphere can mix up coronal and photospheric abundances \citep{Baker2015}. A number of studies have addressed the processes that may separate minor ions according to their FIP which are introduced in section \ref{subsec:intro_FIP}, but to this date there is no consensus on the relative roles of the different invoked processes. To move towards a systematic assessment of the proposed mechanisms, I introduce in chapter \ref{cha:ISAM} a model of the solar atmosphere that I specifically tailored to investigate the fractionation processes of minor ions in coronal loops. First applications of this model are presented in chapter \ref{cha:ISAM_results} which may shed new light on the physical processes that control the composition of minor ions in coronal loops, and consequently on the source regions of the slow wind. \\

Before presenting the dynamic and quasi-stationary theories of the slow wind in greater depth in section \ref{sec:intro_stationnary} and \ref{sec:intro_dynamic} respectively, I present in section \ref{sec:intro_general} some general known properties of the solar atmosphere, the slow solar wind and its candidate source regions. I start by introducing the overall structure of the solar atmosphere in section \ref{subsec:intro_high_atmosphere}. The main characteristics of the slow wind are then discussed in section \ref{subsec:intro_SSW} where both the bulk properties and composition aspects are considered. Then I address the dynamics of the low solar atmosphere in section \ref{sec:intro_low_atmosphere} where I introduce the physical ingredients that form a basis to better understand the plasma and energy transfers from the chromosphere to the solar corona.

\section{Observations of the slow wind and its candidate sources}
\label{sec:intro_general}

\subsection{Structure of the solar atmosphere}
\label{subsec:intro_high_atmosphere}

\subsubsection{The heliospheric plasma and current sheets}
\label{subsubsec:intro_HCS_HPS}

\begin{figure*}[]
\centering
\includegraphics[width=0.95\textwidth]{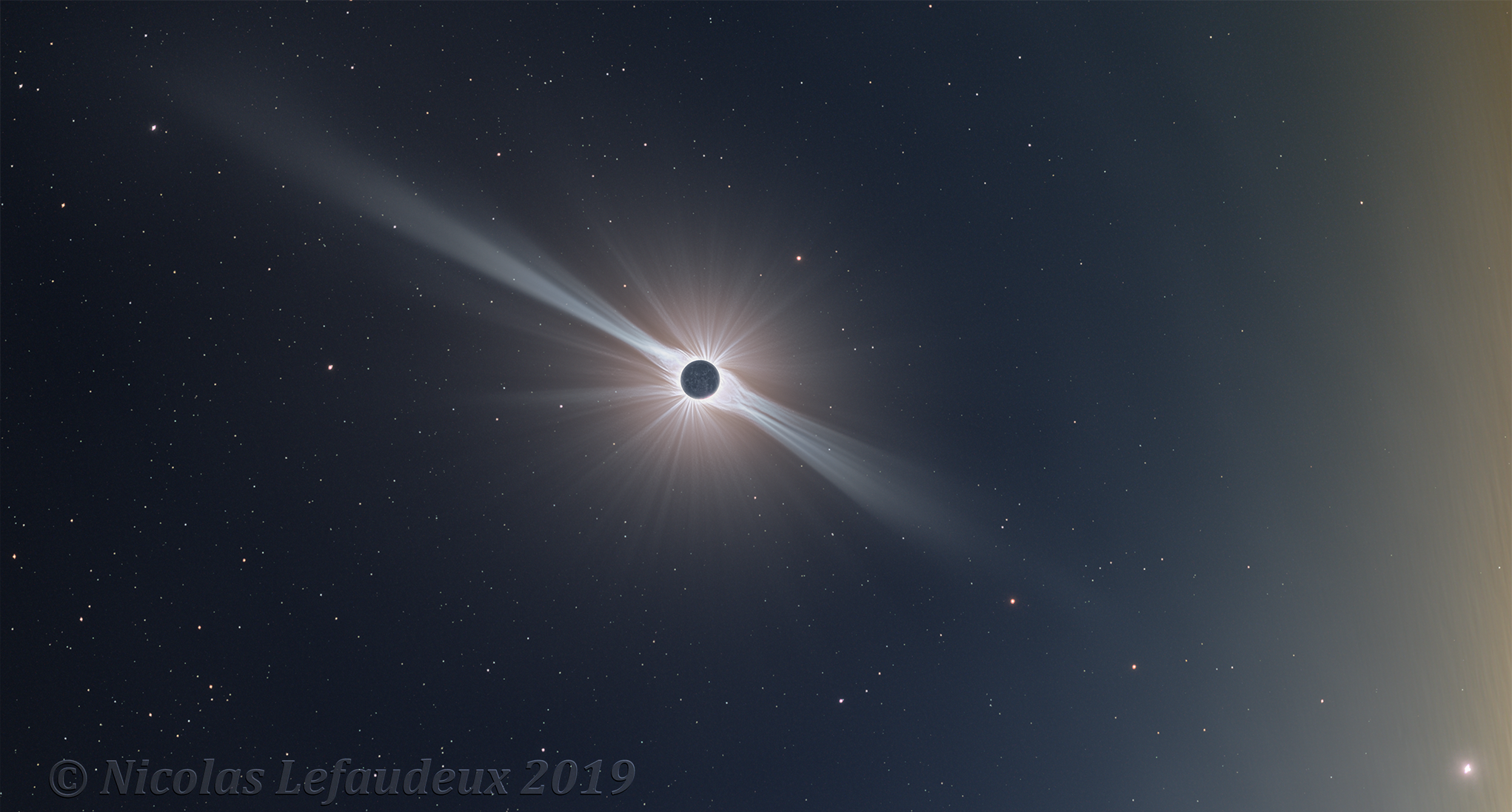}
\caption{Processed photography of the total solar eclipse of 2019. Credits: Nicolas Lefaudeux - \url{https://hdr-astrophotography.com}.
\label{fig:eclipse}}
\end{figure*}

The slow wind is not only slower but tends to be denser, with a plasma density near one astronomical unit (AU) around $5-20\ \rm{cm^{-3}}$, compared to $2-4\ \rm{cm^{-3}}$ measured in the fast wind. This means that overall more electrons are present in the slow wind making it typically well observed in white-light images. This is because electrons are the main scatterers of the photospheric light through the process known as Thomson scattering. The background coronagraphic image shown in the left hand-side panel of Figure \ref{fig:Ulysses} illustrates this relation by showing that the brightest features in the images occur near the solar equator where the slow wind was primarily observed by \textit{Ulysses}. These bright features are called streamers and their stalks extend far out in the corona as unveiled in the enhanced solar eclipse image taken on July 2nd 2019 by Nicolas Lefaudeux shown in Figure \ref{fig:eclipse}. This images shows how far the dense and bright slow wind structures the heliosphere. As we shall discuss in detail throughout this thesis, the heliosperic plasma sheet (HPS) is lodged inside this dense regions of the solar atmosphere. \\

The \textit{Interplanetary Monitoring Platform} (\textit{IMP}) NASA missions unveiled the existence of a sector structure that divides the heliosphere into two magnetic hemispheres, as a structured but continuous belt surrounding the Sun \citep{Ness1964,Wilcox1965}. These observations were confirmed by subsequent in situ measurements taken by \textit{Pioneer-11}  \citep{Smith1978} and interpreted by magneto-hydrodynamic theories \citep{Schulz1973}, this structure was called the heliospheric current sheet (HCS). Thorough analyses of HCS crossings by the \textit{IMP} heliospheric probes have shown a persistent correlation between magnetic field reversals measured in situ and an enhanced plasma density \citep{Wilcox1967}. It was then suggested that the source of the HCS measured in the heliosphere could be the stalks of bright streamers observed in coronagraphs \citep{Howard1974,Hundhausen1977}. \\

The joint ESA-NASA \textit{International Sun-Earth Explorer} series of three spacecraft performed subsequent higher time resolution samplings of the heliospheric plasma. They unveiled that the magnetic field reversals take place inside a thin and dense plasma layer called the heliospheric plasma sheet (HPS) \citep{Winterhalter1994}, and hence suggesting that both the HPS and HCS likely originate in the brightest part of streamers observed in white-light \citep[see also ][]{Guhathakurta1996}. In this thesis we further support this statement by interpreting the novel white-light observations taken from inside the solar corona by the \textit{Wide-Field Imager for Parker Solar Probe} (\textit{WISPR}). 

\subsubsection{The quiet and active Sun}

Sunspots, observed as dark regions on the solar surface because of their cooler local temperature, have been observed for more than two millennia. The number of sunspots on the solar surface varies significantly with the level of solar activity following a well-known 11-year solar cycle. During high solar activity the number of sunspots can reach values over 100, with a peak value that varies from one cycle to another. An example of the sunspots evolution during solar cycles 22 and 23 is shown in the bottom panel of Figure \ref{fig:Ulysses}.

\begin{figure*}[]
\centering
\includegraphics[width=0.5\textwidth]{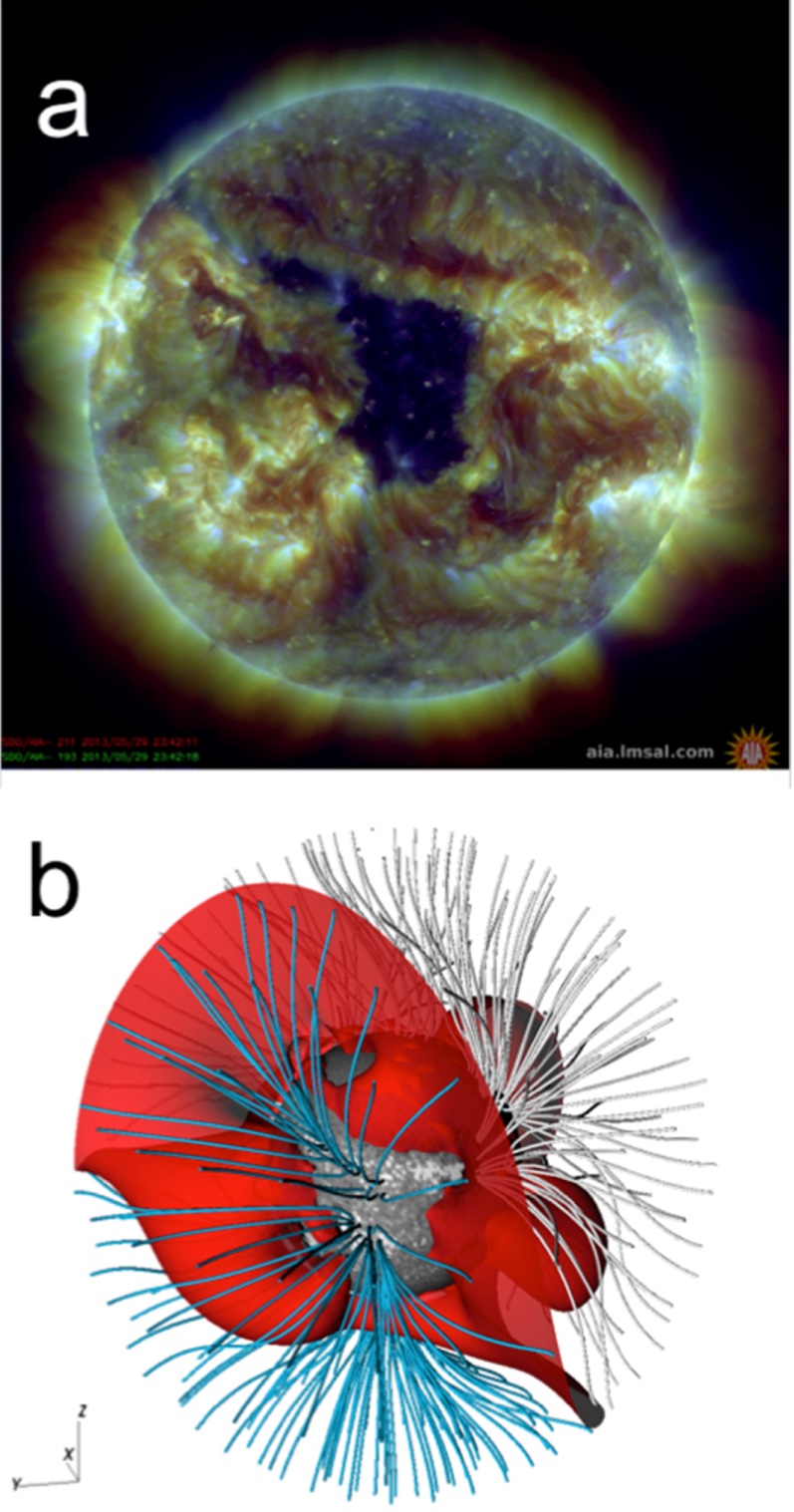}
\caption{ Illustration of a highly structured corona during period of high solar activity. Top panel: combined $193$\si{\angstrom} and $211$\si{\angstrom} EUV image taken by \textit{SDO-AIA} on 2013 May 29. Bottom panel: a potential field source surface (PFSS, see section \ref{subsec:PFSS}) reconstruction of the coronal magnetic field for Carrington rotation 2137, where open field lines colored in blue and grey denote opposite polarities, and where the HCS is plotted as a red sheet. Figure extracted from \citet[][Figure 1]{Sanchez-Diaz2017a}.
\label{fig:Sanchez2017a_fig1}}
\end{figure*}

During periods of low solar activity with only a few sunspots, the HPS and HCS remain located near the equatorial plane. The second passage of \textit{Ulysses} near solar maximum revealed a much more complex configuration of the solar corona as shown in the right-hand side panel of Figure \ref{fig:Ulysses}. This picture reflects a highly structured solar corona in white-light images as well as a heliosphere consisting of slow and fast solar winds present at all latitudes \citep{McComas2003}. In such cases the HCS and also HPS are significantly warped and have been compared to the dress of a ballerina rather than a flat carpet. An example of such high-activity configuration is illustrated in Figure \ref{fig:Sanchez2017a_fig1} for a rare case where a very large coronal hole is present at low latitudes that significantly deflects the HCS towards the polar regions. \\

The structure of the coronal magnetic field is continuously reconfigured at a global scale along the solar cycle, that is coupled with the emergence of active regions and coronal holes at low latitudes. 

Coronal holes are cooler regions of the solar corona best observed in extreme ultraviolet (EUV) images as darker (dimmer) regions of the atmosphere \citep{Waldmeier1981}. Their lower temperatures are induced by the effects of the solar wind, escaping from these regions, dragging matter and energy out of the hole \citep{Aschwanden2014}.

Active regions are regions of the solar atmosphere that are dominated by coronal loops that confine the plasma and where the photospheric magnetic field can reach $\approx 100\ \rm{G}$. The hot coronal temperatures as well as the high density achieved in these loops make them easily detectable in extreme ultraviolet imagers as very bright structures. \\

If coronal holes are primarily concentrated in the polar regions around solar minimum, isolated yet smaller coronal holes can be found wandering at equatorial latitudes during periods of high solar activity. Inversely, more active regions can be found at higher latitudes along with the rising solar activity, which otherwise remain concentrated near the equator within $\approx 20^\circ$ of solar latitude. Both coronal holes and active regions contribute significantly in shaping the coronal magnetic field and hence in a global extent the HCS and HPS as well.

\subsubsection{Streamers and pseudo-streamers}
\label{subsubsec:intro_streamers}

\begin{figure*}[]
\centering
\includegraphics[width=0.95\textwidth]{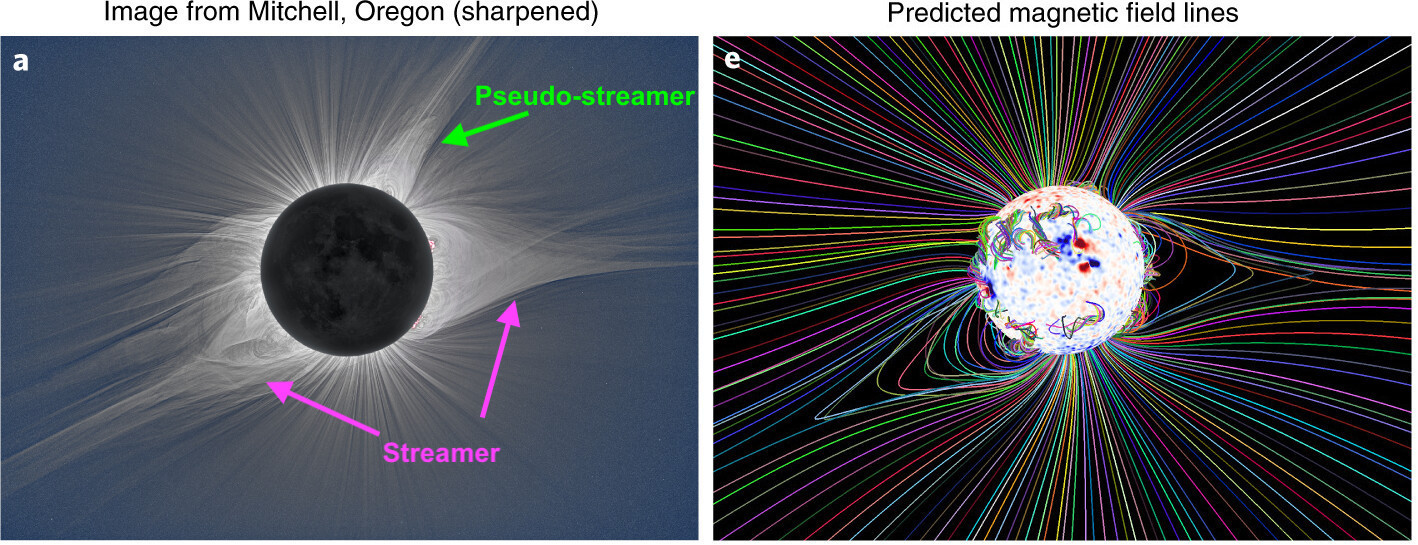}
\caption{Comparison between the observed corona (left) captured during the 21 August 2017 total solar eclipse, and an associated prediction of the magnetic field structure (right). Adapted from Figure 1 in \citet{Mikic2018}. Credit for image a: \copyright 2017 Miloslav Druckmüller, Peter Aniol, Shadia Habbal. \label{fig:Mikic2018_fig1}}
\end{figure*}

As already illustrated, excellent conditions to observe the solar corona are certainly met during total solar eclipses when the Moon hides the solar disk and unveils the faint coronal emissions. Equipped with modern cameras and telescopes, amateur astronomers can make highly resolved images of the solar atmosphere. A processed photography of the total solar eclipse of August 21st 2017 is shown in Figure \ref{fig:Mikic2018_fig1} and reveals the structure of the solar corona with a high level of detail \citep[see also][]{November1996,Druckmuller2014}. A three-dimensional (3-D) modelling of the corona's magnetic field is shown as a comparison in the right-hand side panel, that was produced by the Predictive Science team using advanced magneto-hydrodynamic modeling techniques \citep{Mikic2018}. The bright helmet streamers denoted by the pink arrows in the left-hand side panel enclose regions dominated by closed magnetic fields whereas the streamer stalks are primarily made up of magnetic fields that are connected to the interplanetary magnetic field. The boundary between these two parts is called the cusp where the upper most coronal loops are often significantly stretched by the expansion of the solar wind. \\

\begin{figure*}[]
\centering
\includegraphics[width=0.85\textwidth]{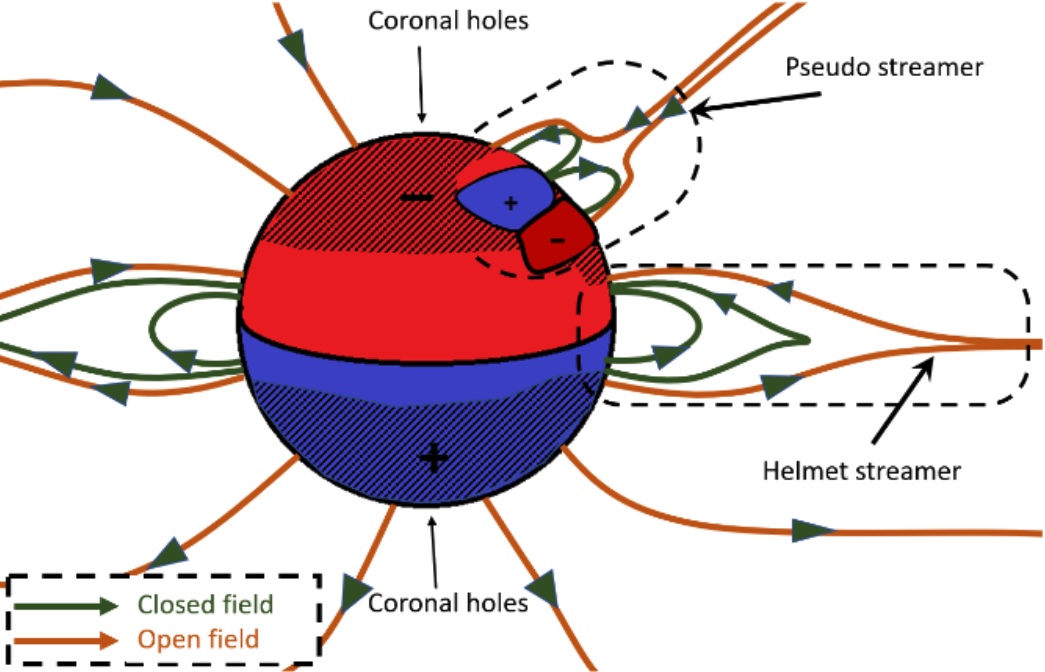}
\caption{ Cartoon of the magnetic field structure of the solar corona. Figure taken from \citet{Pellegrin2021}, with prior permission by
the author.
\label{fig:Pellegrin2021}}
\end{figure*}

Bipolar streamers are defined as bright coronal structures made of loops that connect opposite polarities as shown in Figure \ref{fig:Pellegrin2021}, their extension in the corona is formed by open magnetic fields of opposite polarities that are involved in the formation of the HCS and HPS. The northern streamer seen in Figure \ref{fig:Mikic2018_fig1}, although smaller, resembles a bipolar streamer but is different in essence because it forms in otherwise unipolar regions where both sides of the streamer have the same polarity as schematized in Figure \ref{fig:Pellegrin2021}. This particular type of streamers is called a pseudo-streamer and its extension in the corona and heliosphere does not host a HCS. Throughout this thesis we will simply use the term "streamer" to refer to regular bipolar streamers associated with the HCS and HPS. Although pseudo-streamers are different from bipolar streamers, they clearly produce a solar wind that is denser than observed in the fast wind making them stand out in eclipse images (Figure \ref{fig:Mikic2018_fig1}). 

\begin{figure*}[]
\centering
\includegraphics[width=0.8\textwidth]{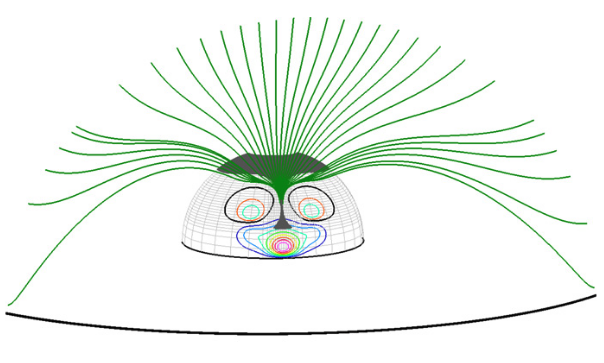}
\caption{Example of a narrow open-field corridor (green lines) forming along an extension of a polar coronal hole (shaded grey area). Contours of the magnetic field amplitude are color plotted at the photosphere. Figure taken from \citet[][Figure 4]{Antiochos2011}.
\label{fig:Antiochos2011_fig4}}
\end{figure*}

In practice pseudo-streamers can form at any place where an isolated bipole emerges in an otherwise unipolar region of the photosphere as shown in Figure \ref{fig:Pellegrin2021}. The formation of isolated coronal holes or even extensions of polar corona holes also triggers the formation of large-scale pseudo-streamers. An example is schematised in Figure \ref{fig:Antiochos2011_fig4} where open fields from a polar coronal hole, depicted by the grey shaded areas at the photosphere \citep[see also][]{Antiochos2011}, extend to the low latitudes. The open-field (green) lines that are associated to the narrow extension of the coronal hole, connect to the heliosphere at a noticeable angular distance away from the main HCS, which is here leaned on the equatorial plane and traced as a black solid line. This specific idealised configuration requires the presence of two dipoles on each side of the polar extension \citep{Antiochos2011}, which create two additional polarity inversion lines that are plotted as black solid lines at the photosphere.\\

\begin{figure*}[]
\centering
\includegraphics[width=0.95\textwidth]{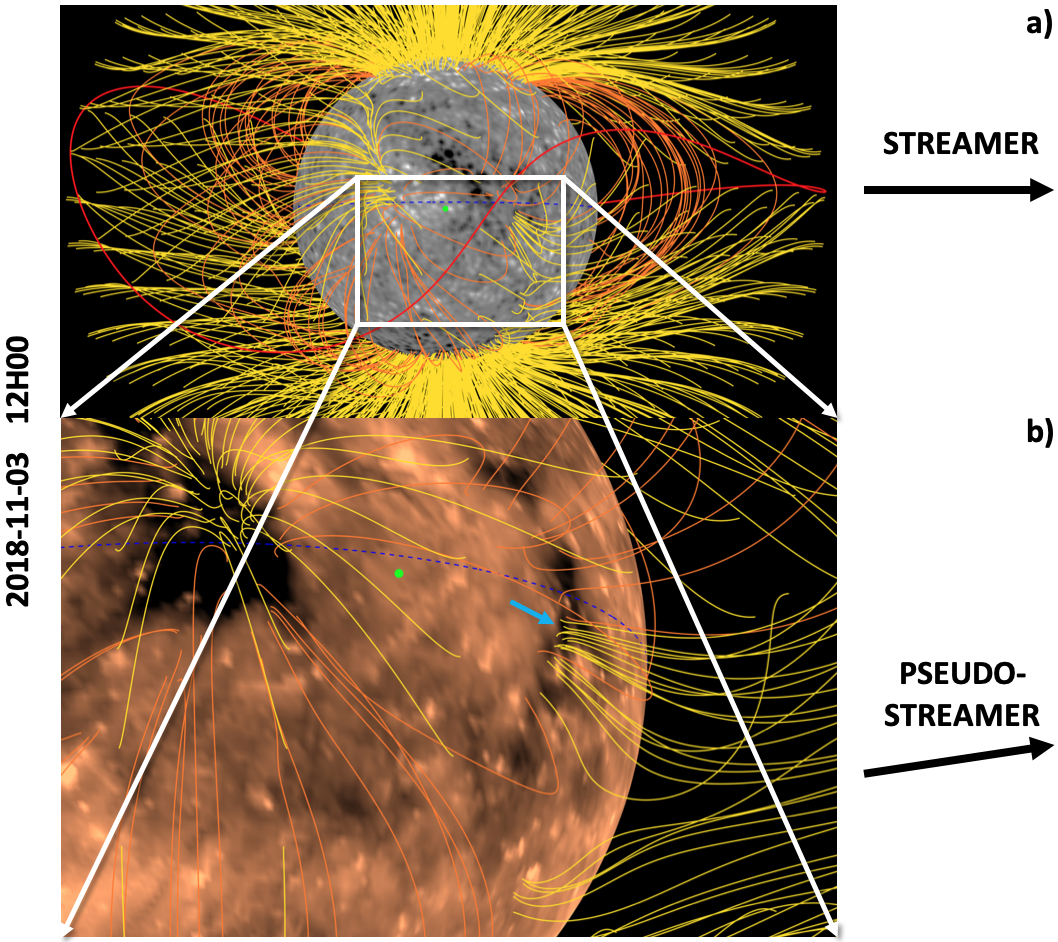}
\caption{ Panel (a): a 3-D reconstruction of the coronal magnetic field on 2018 November 5 using a PFSS extrapolation method (see section \ref{subsec:PFSS}). The magnetic map used as an input to the PFSS extrapolation is displayed in a grey scale at the solar surface. The open and closed magnetic field lines are depicted in yellow and orange, respectively. The polarity inversion line (i.e. the baseline of the HCS) is plotted as a red line at a height of $2.1\ R_\odot$. Panel (b): a zoomed-in view of a pseudo-streamer anchored in an isolated and small equatorial coronal hole (blue arrow). Extreme ultraviolet emissions in the $193$\si{\angstrom} wavelength are shown at the surface.
\label{fig:Poirier2020_fig7}}
\end{figure*}

Well-developed isolated coronal holes and extensions of polar coronal holes are in general well detected in extreme ultraviolet images of the solar disk, an example is shown in Figure \ref{fig:Poirier2020_fig7}. The extension of the polar coronal hole is often, but not always, the precursor of an isolated yet smaller coronal hole developing at low latitudes. The associated 3-D configuration of the open (yellow) field lines is shown in the bottom panel of Figure \ref{fig:Poirier2020_fig7} produced for the \citet{Poirier2020} paper. In this study, the pseudo-streamer that arises from this small equatorial coronal hole was responsible for the appearance of additional bright rays in \textit{WISPR} white-light images, that is further discussed in section \ref{sec:stationnary_poirier2020}. Another study from \citet{Griton2021} and for which I contributed to, identified a small equatorial coronal hole as one of the source regions of the slow solar wind sampled by \textit{PSP} during its second passage to the Sun.

\subsubsection{White-light synoptic maps of the streamer belt}
\label{subsubsec:intro_WLmaps}

White-light (and EUV) Carrington maps have been widely used throughout this thesis and their building process will be introduced in more detail in chapter \ref{cha:stationnary}. In essence, the Sun's rotation on itself (at an average 27 day rotation period) is exploited to stack bands of coronagraph images taken off the solar limb into a synoptic (Mercator-type) map of the full corona at a fixed altitude. \\

This technique has been widely used from the 1970s on the first images taken by  the space-based \emph{CORONASCOPE II} and \emph{SOLWIND} coronagraphs \citep{Bohlin1970,Wang1992} as well as from the Mauna Loa Solar Observatory in Hawaï \citep{Hansen1976}. Later on, synoptic observations from the white-light coronagraph onboard \emph{Skylab} were exploited to model the large-scale coronal density structures by assuming that the peaks in brightness mark the location of the HCS \citep{Guhathakurta1996}. The continuous monitoring of the solar corona by \textit{SoHO} has then enabled a more systematic comparison between the location of streamers with the magnetic topology of the solar corona derived from PFSS calculations \citep{Wang1998,Wang2000,Wang2007} and global coronal models \citep{Gibson2003,Thernisien2006,dePatoul2015,PintoRouillard2017}. Rotational tomography techniques have been developed recently to correct for line-of-sight effects and convert WL observations into synoptic density maps \citep{Morgan2020}. Other techniques have involved the combination of coronagraph images obtained from multiple vantage points (e.g. \emph{SoHO} and \emph{STEREO}) simultaneously to derive synchronic circumsolar maps of helmet streamers \citep{Sasso2019}.\\

\begin{figure*}[]
\centering
\includegraphics[width=0.95\textwidth]{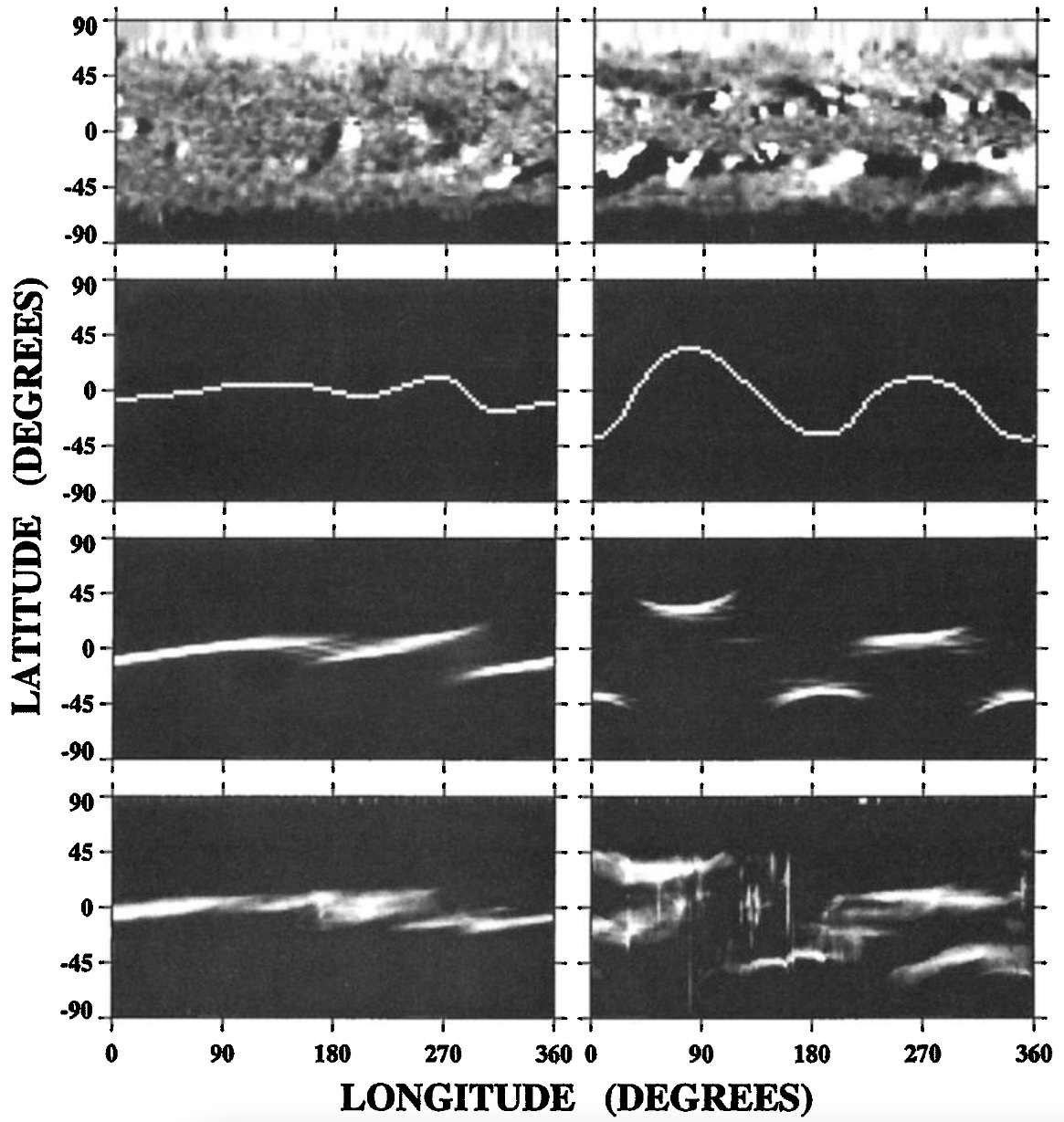}
\caption{Latitude-longitude Carrington maps of the photospheric magnetic field (1st row), of the modeled HCS (2nd row) and white-light emissions (3rd row), and of the real white-light observations taken by \textit{SoHO LASCO-C2}, over Carrington rotation CR1919 (left) and CR1935 (right) that are typical of a solar minimum (left) and maximum (right) coronal configuration. Figure taken from \citet[][Figure 2]{Wang2000}.
\label{fig:Wang2000_fig2}}
\end{figure*}

An example of a white-light synoptic map taken from \citet[][Figure 2]{Wang2000} is shown in the bottom panel of Figure \ref{fig:Wang2000_fig2}. The streamer stalks that we introduced at the beginning of this chapter, form a continuous band of bright emission in the synoptic map that is commonly called the streamer belt. The HPS, also introduced earlier, stands somewhere inside the streamer belt probably where the plasma is densest and scatters photospheric light significantly. White-light synoptic maps are therefore very convenient to visualize the global structure of the solar corona. \\

Still, there remain portions of the streamer belt that are poorly visible or not detected at all. This is a well understood effect and is inherent to the line-of-sight integration of white light scattered by the corona. Regions were the streamer belt is not well observed correspond to locations where the belt undergoes latitudinal excursions. In these regions the belt is inclined with respect to the imager's line-of-sight which means that a smaller section of the streamer belt is contributing to its brightness in the imager. This effect is less pronounced near solar minimum because the streamer belt remains more or less parallel to the equatorial plane. And since (up to now) all images have been taken from inside the ecliptic plane, significant portions of the streamer contributes to its brightness. \\

This distribution changes dramatically near solar maximum when the streamer belt is highly warped covering a broad range of latitudes as shown in the right-hand side panels of Figure \ref{fig:Wang2000_fig2}. At this level of activity, the solar corona becomes highly structured and the streamers appear as separated stripes in a Carrington map rather than a continuous belt. This complex structure is typically induced by the emergence of active regions and the formation of low-latitude coronal holes which force the HPS to warp away from the equator. This is illustrated in the 1st and 2nd rows of Figure \ref{fig:Wang2000_fig2} where an expected shape of the HCS (and therefore the streamer belt) is computed from a 3-D magnetostatic current free reconstruction (i.e. PFSS) of the solar corona that will be introduced later.\\

Some bright excursions of the streamer belt can extend towards higher latitudes even at solar minimum, these arch-light features are typically pseudo-streamers. But other inverted arch features, called "bananas" in \citet{Gibson2003}, can also be observed in white-light Carrington maps from the projection of non-equatorial streamers moving in the coronagraphic images \citep[see e.g.][]{Wang1992,Wang2000,Wang2007}. As we shall see in this thesis, this additional projection effect is particularly strong in observatories that are located much closer to the Sun such as \textit{WISPR} on board \textit{PSP} as discussed in chapter \ref{cha:stationnary}, and in \citet{Liewer2019} and \citet{Poirier2020}.\\

In this thesis we exploit white-light Carrington maps at multiple occasions, to constrain coronal and heliospheric models in a systematic manner (see section \ref{sec:WL_opti}), to track the origin of the slow solar wind measured in situ at \textit{PSP} (see section \ref{sec:dynamics_insitu}) or to study the fine structure of coronal rays observed remotely at \emph{PSP} (see section \ref{sec:stationnary_poirier2020}).

\subsubsection{Spectroscopy of the low solar atmosphere}

The EUV imaging spectrometer (\textit{EIS}) on board the \textit{Hinode} spacecraft has provided a wealth of precious information on the potential sources of the slow solar wind, through composition diagnostics of the low solar atmosphere. \\

\begin{figure*}[]
\centering
\includegraphics[width=0.6\textwidth]{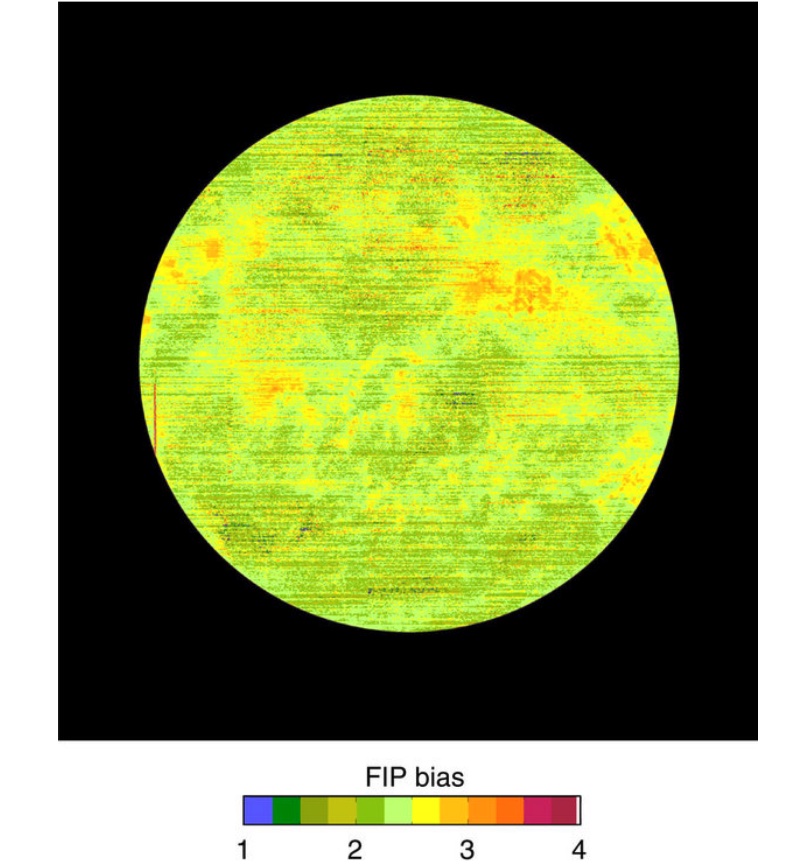}
\caption{2-D map of Silicium (Si) over Surfur (S) abundances ratio on the solar disk. Figure taken from \citet[][Figure 3]{Brooks2015}.
\label{fig:Brooks2015_fig3}}
\end{figure*}

\citet{Brooks2015} have developed an inversion technique to derive the relative abundances of Silicium (Si, low FIP) over Sulfur (S, intermediate FIP) based on spectral line intensities of the solar-disk observed remotely by \textit{Hinode-EIS}. Their 2-D map of Si/S ratio of the solar-disk is given in Figure \ref{fig:Brooks2015_fig3} and is typical of rather high solar activity with several active regions that are distributed over the surface. They have identified the edges of some of these active regions as potential sources of the slow solar wind that was measured in situ at near 1 AU by the \textit{Advanced Composition Explorer} (\textit{ACE}). \\

More generally, spectroscopic signatures of active regions show a quasi systematic enrichment in low-FIP elements and therefore support a slow solar wind that is partially made up of closed-field plasma from active regions \citep{Meyer1985,Neugebauer2002,Liewer2004,Brooks2011,Doschek2019}. A higher charge state is also measured in solar winds that originate from active regions rather than from coronal holes, that is explained by a higher temperature in the corona that boosts up the ionization. The comprehensive remote-sensing suite of the \textit{SoHO} spacecraft allowed for the first time to link spectroscopic diagnostics with white-light signatures of the solar corona. A direct association has been made between coronal loops that are anchored in active regions and enriched in low-FIP elements, and the associated slow solar wind stream seen in white-light coronagraph \citep{Ko2002,Uzzo2004}. In contrast, such enrichment has not been found (or to a much lower extent) above coronal holes \citep{Feldman1998b,Stansby2020a}.

\clearpage
\subsection{Properties of the slow solar wind} 
\label{subsec:intro_SSW}

\subsubsection{The intermittent nature of the slow wind}
\label{subsubsec:intro_SSW_intermittency}

The \textit{Large-Angle and Spectrometric Coronagraph} \citep[LASCO: ][]{Brueckner1995} on board \textit{SoHO} marked a revolution in white-light images of streamers and of the slow solar wind, unveiling their fine structure and dynamism. \\

\begin{figure*}[]
\centering
\includegraphics[width=0.6\textwidth]{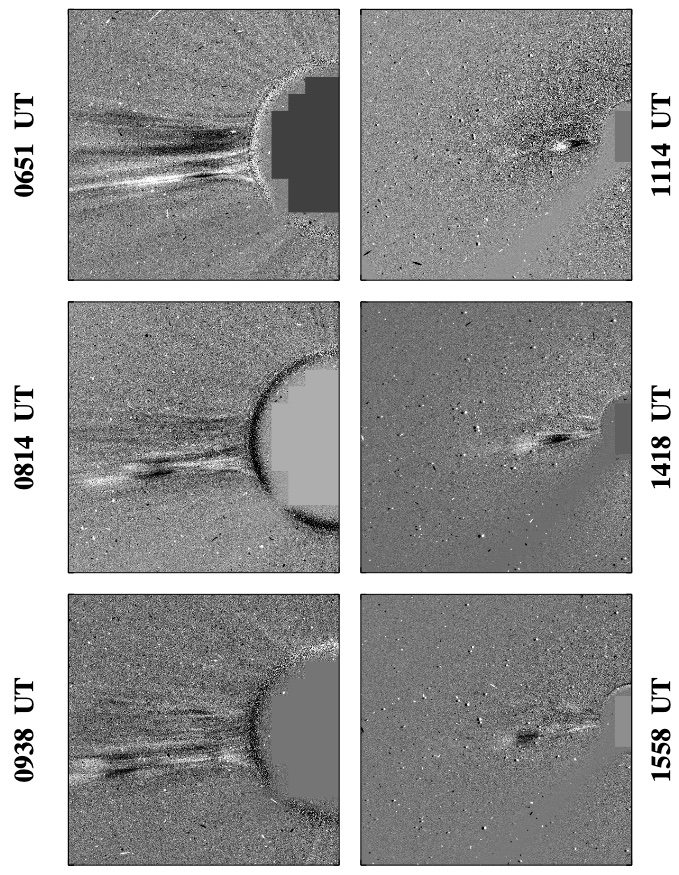}
\caption{ Example of streamer blobs observed by \textit{SoHO-LASCO}, using a running difference technique to enhance density fluctuations in the corona where bright (or dark) regions are indicative of an enhanced (or depleted) density respectively. Figure taken from \citet[][Figure 1]{Sheeley1997}.
\label{fig:Sheeley1997_fig1}}
\end{figure*}

By exploiting these novel data, \citet{Sheeley1997} discovered a plethora of small-scale density perturbations that they called streamer "blobs" and that are shown in Figure \ref{fig:Sheeley1997_fig1}. The faint brightness fluctuations induced by the motion of "streamer blobs" was enhanced using the "running-difference" technique that consists in subtracting subsequent images. By convention bright and dark regions in these images denote respectively an increase or a decrease in the brightness, and likely in the local density of electrons in the solar atmosphere as well. These streamer "blobs" tend to appear from the tip of helmet streamers and propagate along their stalk to move in concert with the slow solar wind. \\

Periodicities in the ambient solar wind impinging the Earth's magnetosphere have been noted in a subsequent study from \citet{Kepko2002}. The tracking of streamer blobs in the remote-sensing and in situ observations from the \textit{STEREO} mission revealed propagating density structures at periodicities varying from $\approx 90-180\rm{min}$ to $\approx 8-16\rm{hr}$ \citep{Viall2010,Viall2015,Kepko2016,Sanchez-Diaz2017a}, which were found in recent \textit{PSP} observations as well \citep{Rouillard2020a}. Some streamer blobs have also been associated to the passage of corotating interacting regions \citep[CIRs: ][]{Pizzo1978}, where fast winds catch up slow winds that generate a compression of the solar wind plasma \citep[see e.g.][]{Rouillard2009,Sheeley2010,Plotnikov2016}. \\

\begin{figure*}[]
\centering
\includegraphics[width=0.6\textwidth]{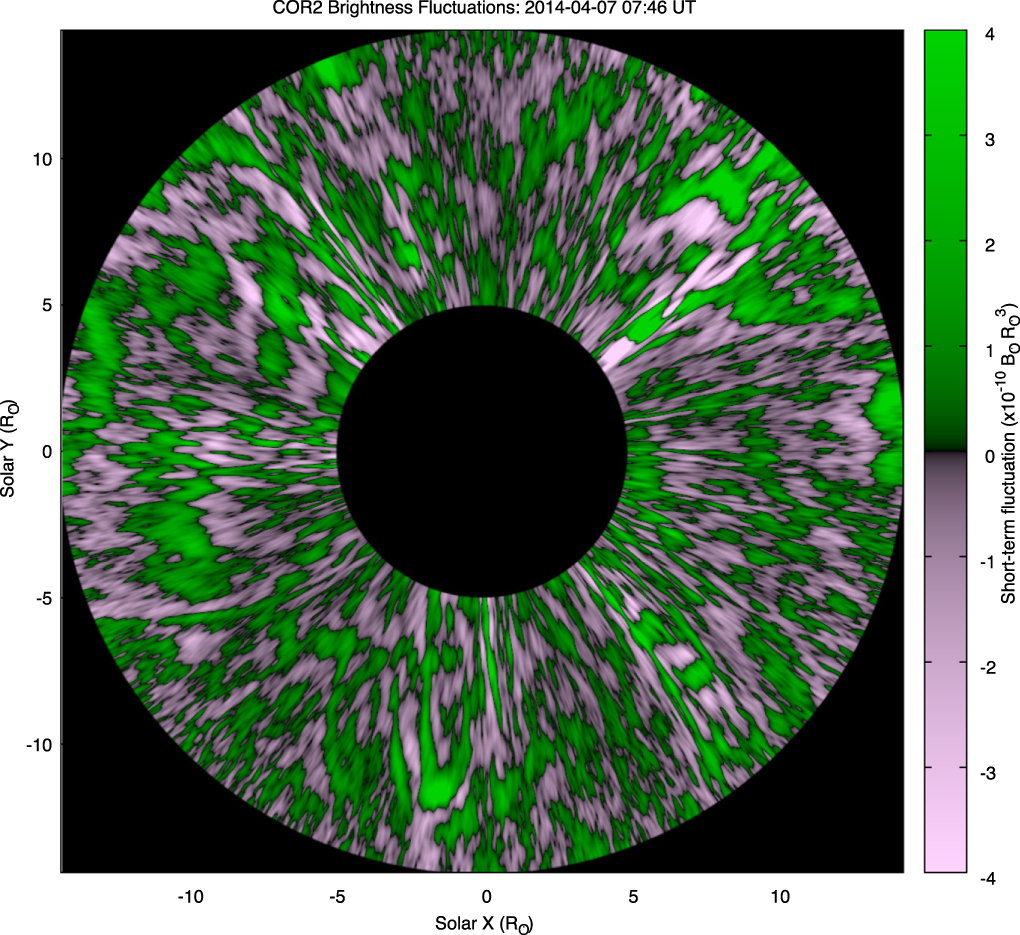}
\caption{ Brightness fluctuations in the upper solar corona detected during a high-cadence campaign of the \textit{STEREO-A COR2} coronagraph. Figure taken from \citet[][Figure 12]{DeForest2018}.
\label{fig:DeForest2018_fig12}}
\end{figure*}

A recent (deep-field) high-cadence campaign of the \textit{STEREO-A COR2} coronagraph dramatically improved observations and revealed the omnipresence of density structures propagating in the slow wind \citep{DeForest2018}. Using sophisticated processing techniques, these authors unveiled a highly structured solar corona with the ubiquitous release of density perturbations with various sizes (see Figure \ref{fig:DeForest2018_fig12}), the largest of which likely being the "streamer blobs" of \citet{Sheeley1997}. Due to line of sight effects, it remains unclear as to whether all the density fluctuations observed by \citet{DeForest2018} propagate inside the streamer belt or if some of the density fluctuations are released from a broader region of the corona. \citet{Griton2020} showed for instance that some of these density fluctuations could be induced low in the solar corona, at the base of the open flux tubes that channel the solar wind, following sudden reconnection events associated with coronal bright points. The latter are observed as enhanced emissions in extreme-ultraviolet and X-ray observations \citep{Madjarska2019} that have been interpreted as hot magnetic loops that form when the magnetic field emerges, interacts and reconnects with pre-existing coronal magnetic field \citep{Kwon2012}. My contribution to the \citet{Griton2020} study will be discussed in section \ref{sec:dynamics_griton2020}. \\

\begin{figure*}[]
\centering
\includegraphics[width=0.8\textwidth]{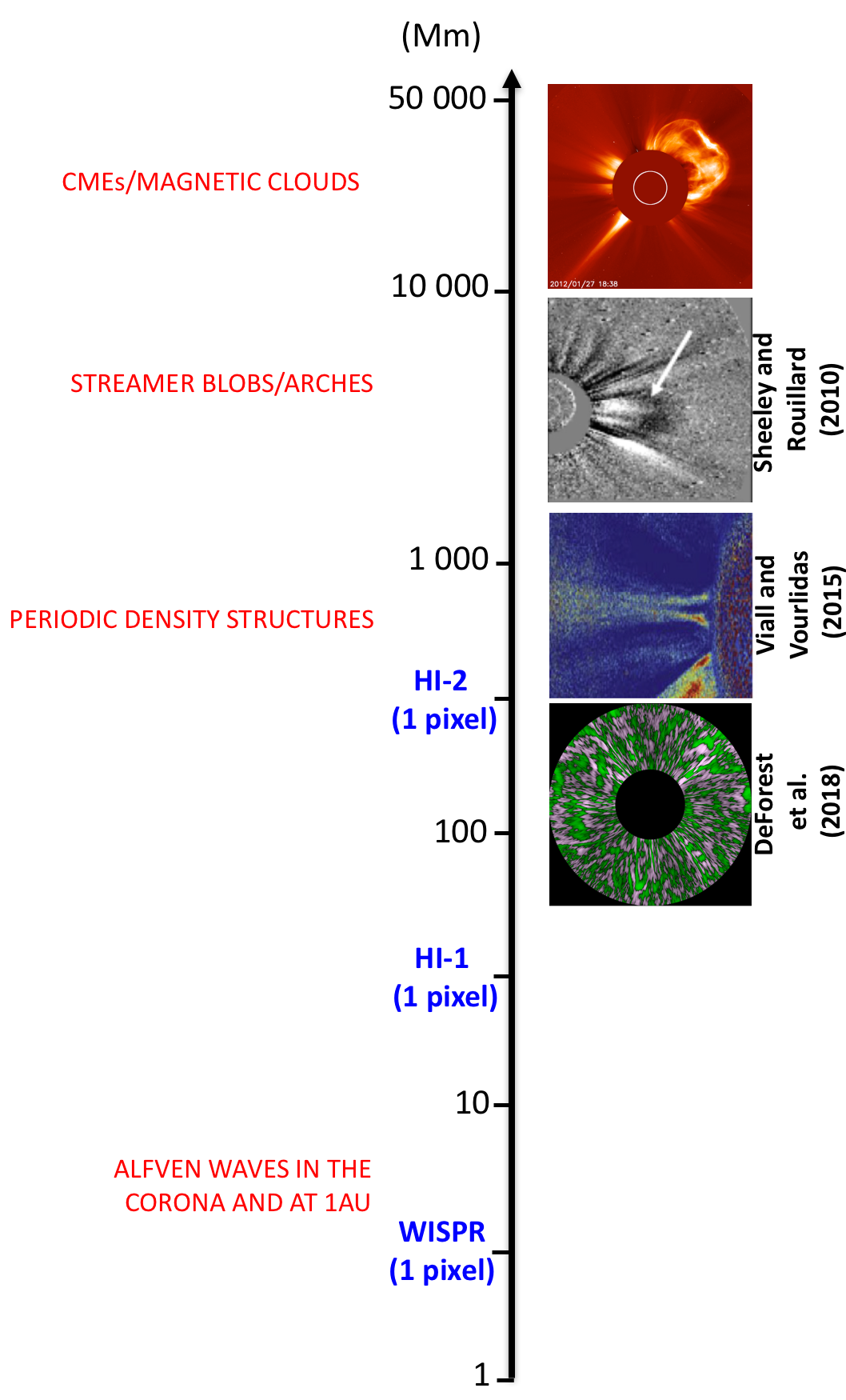}
\caption{ Overview of dynamic structures in the slow solar wind from the large (top) to the small (bottom) spatial scales with their associated references mentioned in the text.
\label{fig:transient_scales}}
\end{figure*}

It is now widely accepted that the slow solar wind is highly dynamic in nature. Figure \ref{fig:transient_scales} summarises the variability of the slow solar wind at various spatial scales, where several of the references mentioned above are represented. At the largest scales significant deflections of the streamers have been observed during the eruption of coronal mass ejections and the passage of their associated shock that propagates upstream \citep[see e.g.][]{Kouloumvakos2020a,Kouloumvakos2020b}. So far, the smallest periodic density structures that have been unveiled from remote-sensing observations are those of \citet{DeForest2018} during a high-cadence \textit{STEREO-A COR2} campaign. The unprecedented close-up distance of \textit{PSP-WISPR} to the Sun is likely to extend this picture beyond to even smaller scales, as we shall illustrate throughout this thesis.

\subsubsection{Composition of the slow wind}
\label{subsubsec:intro_SSW_composition}

Further clues on the origin of the different solar winds can be obtained from composition diagnostics of the solar wind derived from the abundance of their minor constituents such as heavy ions. This is made possible by the fact that the abundance and charge states of heavy ions measured in the solar wind are regulated low in the solar corona and become invariant upwards of the upper corona  \citep{Geiss1982}.  \\

The \textit{Ulysses-SWICS} instrument provided precious composition diagnostics of the solar wind at all latitudes. These measurements showed that the abundance of certain elements is not only different in the fast and slow winds but a particular element can have different ionisation levels \citep{Steiger1996}.

\begin{figure*}[]
\centering
\includegraphics[width=0.95\textwidth]{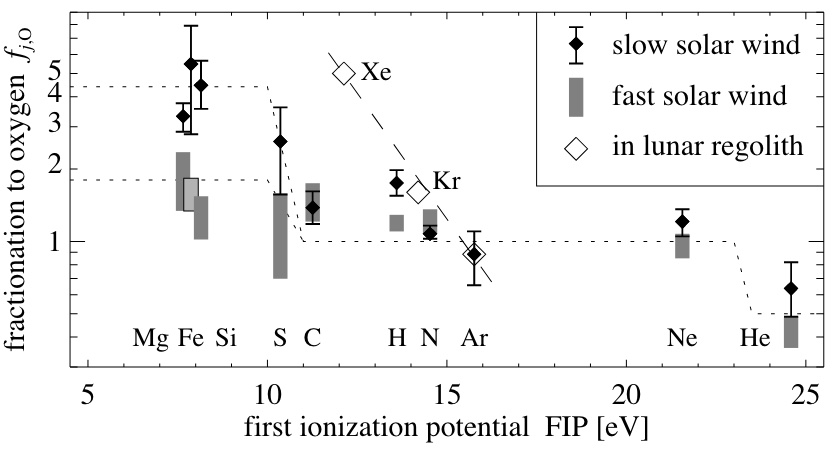}
\caption{Variation of abundances for the main constituents measured in the solar wind as a function of the FIP. Abundances are relative to the abundance of Oxygen. Figure taken from \citet[][Figure 1]{Peter1998b}.
\label{fig:Peter1998_fig1}}
\end{figure*}

\begin{figure*}[]
\centering
\includegraphics[width=0.95\textwidth]{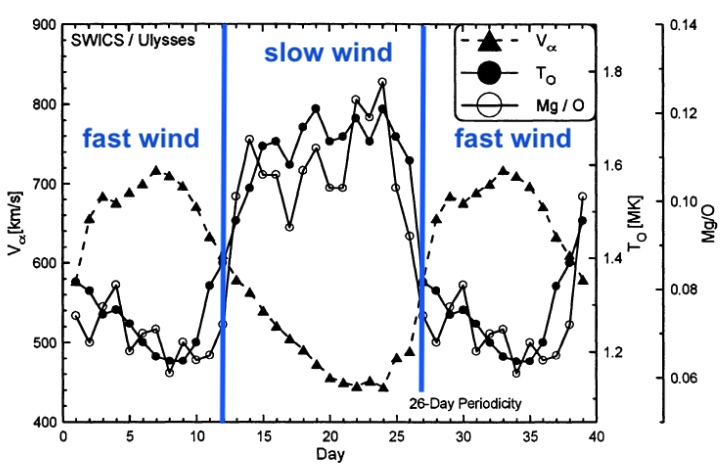}
\caption{Temporal variation of the low FIP Magnesium (Mg) abundance with respect to Oxygen (O), of the solar wind speed (here taken as the speed of doubly ionized Helium), and of the Oxygen temperature. Figure taken from \citet[][Figure 7]{Geiss1995}.
\label{fig:Geiss1995_fig7}}
\end{figure*}

Figure \ref{fig:Peter1998_fig1}
illustrates first the dichotomy in elemental abundance between the slow and fast solar winds, the former being enriched by a factor $\approx3-5$ in heavy ions that have a low first ionization potential (FIP, i.e. the energy required to ionize the first electron) such as Magnesium (Mg), Iron (Fe) and Silicium (Si). Heavy ions with a higher FIP $> 11\ \rm{eV}$ do not show significant deviations from the photospheric abundances measured lower down in the solar atmosphere, except for Helium (He) that tends to be depleted in the solar wind and especially in the fast solar wind where a factor $\approx 1/2$ is measured. 
Figure \ref{fig:Geiss1995_fig7} complements this picture with an abundance of low FIP elements (here Mg) that is inversely (or directly) correlated with the solar wind speed (or temperature). An additional and essential observed property is that the fractionation appears to proceed independently of the mass of the particles, which can be much heavier than the main proton constituents though. We will see in section \ref{subsec:intro_FIP} that this observational fact can have direct implications on the possible physical processes that may control the extraction of heavy ions in the corona. \\

\begin{figure*}[]
\centering
\includegraphics[width=0.8\textwidth]{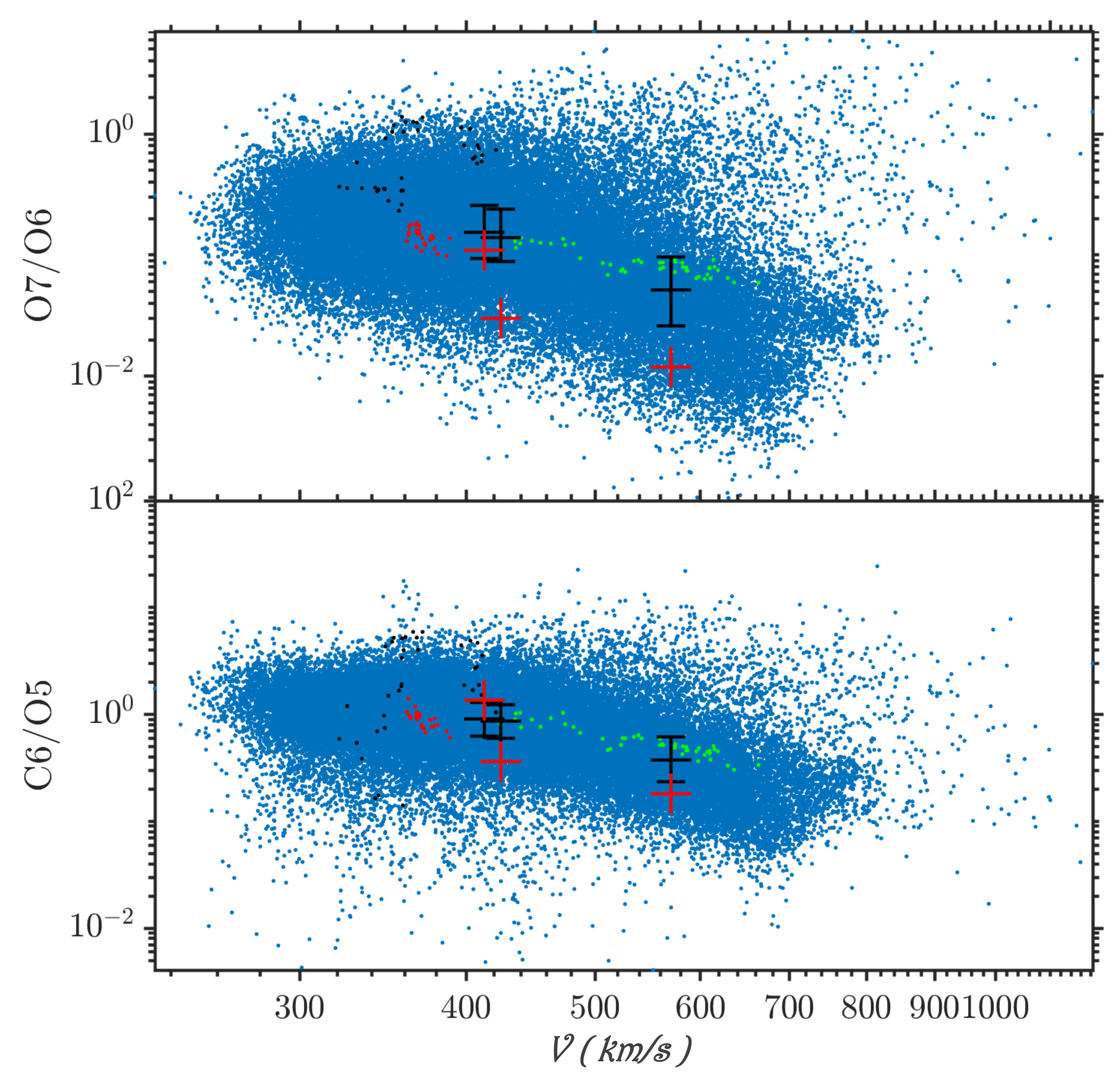}
\caption{In situ measurements from \textit{ACE} of the Oxygen $O^{7+}/O^{6+}$ and Carbon $C^{6+}/C^{5+}$ charge state ratios function of the solar wind speed (blue dots). Data corresponding to fast, slow Alfvénic and slow non-Alfvénic winds as identified by \citet{DAmicis2015} are indicated by green, red and black dots respectively. Simulations from the Irap Solar Atmosphere Model (ISAM) for these three solar wind regimes are shown as red crosses along with their associated uncertainties in black \citep[see][for further details]{Lavarra2022}. Figure taken from \citet[][Figure 1]{Lavarra2022}.
\label{fig:Lavarra2022_fig1}}
\end{figure*}

Furthermore, composition diagnostics of the solar wind made in situ can also provide information about the charge state (or ionization level) of an element. The charge state of an element can be computed by comparing the number densities of its various ions, namely the charge state ratios. The charge state ratios are generally highly variable in both the slow and fast wind and a tendency for the charge state to be more elevated in the slow solar wind (or as we shall see a fraction of that SSW) compared with the fast wind (Figure \ref{fig:Lavarra2022_fig1}). The charge state is inherently linked to the properties of the source region low in the solar atmosphere, at heights where the ionization is still not fully established between the upper chromosphere and low corona (that is discussed later in section \ref{subsec:intro_chromo}). For instance, \citet{Neugebauer2002} and \citet{Liewer2004} have observed enhanced charge state ratios in the slow wind that come from hot active regions. In contrast, they found charge state ratios that tend to be lower in the fast wind, and that are associated to source regions (e.g. coronal holes) that are typically cooler than active regions. \citet{Kepko2016} also showed a similarity between the variability of the charge state ratios measured in situ in the slow wind and the short, hourly time scale of the quasi-periodic structures observed remotely in WL streamers. \\

\subsubsection{Variability of the SSW}
\label{subsubsec:intro_SSW_variability}

In section \ref{subsubsec:intro_SSW_intermittency} and \ref{subsubsec:intro_SSW_composition} we have depicted a SSW that is likely multi-faceted beyond the simplistic bimodal kinetic classification established before. The diversity of observed slow wind regimes is indicative of a multitude of possible candidates of slow wind sources which may differ significantly from one to another. \\ 

While the SSW is systematically enriched in low-FIP elements, a variable abundance in alpha particles (high FIP) has been measured that changes not only on hourly/daily time scales but also through the solar cycle. Two distinct subcategories of SSW have been identified with different abundances in alpha particles \citep{Kasper2007,McGregor2011}. One SSW exhibits a quasi steady depletion in alpha particles that is similar to the one measured in the fast wind that emerges near the center of coronal holes. Whereas the other SSW has a higher but variable abundance in alpha particles that is typical of streamer flows. 

\begin{figure*}[]
\centering
\includegraphics[width=0.7\textwidth]{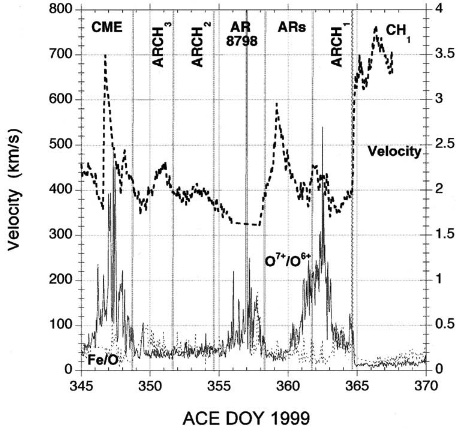}
\caption{Variation of the Oxygen $O^{7+}/O^{6+}$ charge state ratio and of the solar wind speed over time from \textit{ACE} in situ measurements taken at near 1 AU. Figure taken from \citet[][Figure 6]{Liewer2004}.
\label{fig:Liewer2004_fig6}}
\end{figure*}

A similar identification has been made from the measurements of charge state ratios in the SSW \citep{Neugebauer2002,Liewer2004,Stakhiv2015,Stakhiv2016}, with a "streamer-like" SSW that has high charge ratios typical of the hot active regions, and a "coronal hole-like" SSW that has lower charge state ratios as in the fast wind. That is illustrated in Figure \ref{fig:Liewer2004_fig6}. 

Subsequent measurements taken at \textit{PSP} \citep{Rouillard2020a,Griton2021} and at near 1 AU \citep{DAmicis2019} looked at the bulk properties of these two SSW states in detail. The "streamer-like" SSW was generally measured slower, denser and more variable than the "coronal hole-like" SSW. The later is commonly termed the Alfvénic SSW because it can host Alfvénic fluctuations that are as large as those measured in the fast wind \citep{DAmicis2019}. \\

These results suggest two SSW states that are likely generated from different sources and through different mechanisms. The "streamer-like" SSW which is highly variable may be formed by intermittent processes that occur at the tip of streamers and/or at the open-closed boundaries between streamers and coronal holes, that is discussed further in section \ref{sec:intro_dynamic}. We will see in section \ref{sec:intro_stationnary} that the more steady Alfvénic SSW likely originates from open field lines that are rooted closer to the center of coronal holes, and that this SSW can be well described with a quasi-stationary theory \citep[see also][]{PintoRouillard2017,Lavarra2022}.

\clearpage
\section{Basics of the low solar atmosphere}
\label{sec:intro_low_atmosphere}

In this section I introduce the physical ingredients that form a basis to better understand the dynamic of the low solar atmosphere. I start by a short presentation in section \ref{subsec:intro_RT} of the different layers that constitute the low solar atmosphere. Then I discuss in section \ref{subsec:intro_heating} the heating mechanisms that may contribute to the energy transfer in the low solar atmosphere. Furthermore, the major specificities of the chromosphere where most of the \emph{extraction} processes of heavy ions are supposed to take place, are described in section \ref{subsec:intro_chromo}. Finally, I present in section \ref{subsec:intro_FIP} a short review on the physical processes that might operate to enrich the corona with certain heavy ions, and therefore that could constitute building blocks of the quasi-stationnary and dynamic theory of the solar wind that are discussed later on in section \ref{sec:intro_stationnary} and \ref{sec:intro_dynamic}.

\subsection{The chromosphere, transition region and corona}
\label{subsec:intro_RT}

\begin{figure*}[]
\centering
\includegraphics[width=1.\textwidth]{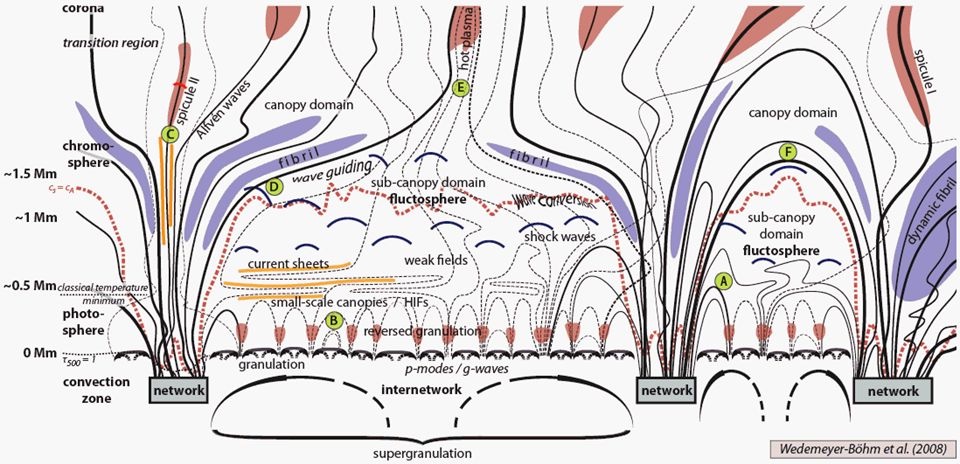}
\caption{Global picture of the structure of the solar atmosphere. Figure taken from \citet[][Figure 16]{Wedemeyer2009}.
\label{fig:Wedemeyer2009_fig16}}
\end{figure*}

\begin{figure*}[]
\centering
\includegraphics[width=1.\textwidth]{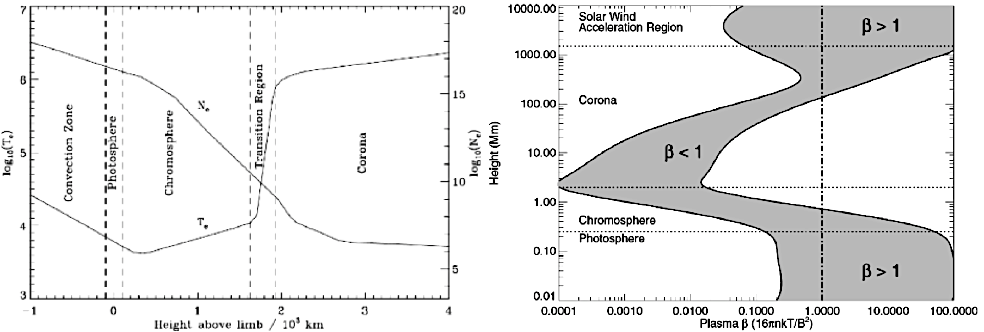}
\caption{Left panel: Electron density and temperature in the different layers of the solar atmosphere. Right panel: Typical range of the plasma beta parameter. Figure taken from \citet{Aschwanden2005}.
\label{fig:T_beta}}
\end{figure*}

The different layers of the solar atmosphere host a broad range of physical processes that are illustrated in Figure \ref{fig:Wedemeyer2009_fig16}. The solar corona situated at the top of this figure is heated to temperatures reaching $\approx 1-3\ \rm{MK}$ through yet undetermined processes. A very thin interface of about $\approx 100\ \rm{km}$ thick, called the transition region (TR), separates the hot corona and the cooler/denser chromosphere. Typical electron density and temperature profiles are shown in the left-hand side panel of Figure \ref{fig:T_beta}. The TR being so thin compared to the magnetic scale height we can assume that the magnetic field strength does not change significantly across this region. Consequently the altitude of the TR which is quite variable is controlled by an equilibrium reached between the relative thermal pressures of the chromosphere and the corona. A significant amount of energy passes through the TR by the downward conductive heat flux brought down from the hot corona to the cooler chromosphere. In closed-field plasmas, most of the energy that is deposited in the corona is convected downward to the upper chromosphere where it is dissipated in the form of radiative emissions (often called radiative cooling), that is discussed in great detail in section \ref{subsec:intro_chromo}. \\

The transition region is a dynamic interface between two regions that are driven by different physical processes and that obey different constraints. The solar corona is mostly controlled by the magnetic field which is continuously perturbed by the effects of magnetic flux emergence and convective motions driven from below. These perturbations can drive complex structures that store magnetic free energy, which will often be removed by transient magnetic reconfigurations that will allow the coronal magnetic field to retrieve a lower energy state. In contrast a significant part of the chromosphere is controlled by plasma processes that result from convective motions transmitted from the convection zone through the photosphere. This dichotomy is usually well described by the plasma beta parameter that is the ratio of the thermal $n k_b T$ to magnetic $B^2/(2\mu_0)$ pressure, and of which typical ranges are plotted on the right-hand side of Figure \ref{fig:T_beta}. The height in the chromosphere where the plasma beta is equal to 1 is called the equipartition layer. The height of this layer therefore marks where the magnetic field begins to influence the dynamics of the chromosphere. The position of this layer is shown as a red dotted line in Figure \ref{fig:Wedemeyer2009_fig16}. \\

\begin{figure*}[]
\centering
\includegraphics[width=0.9\textwidth]{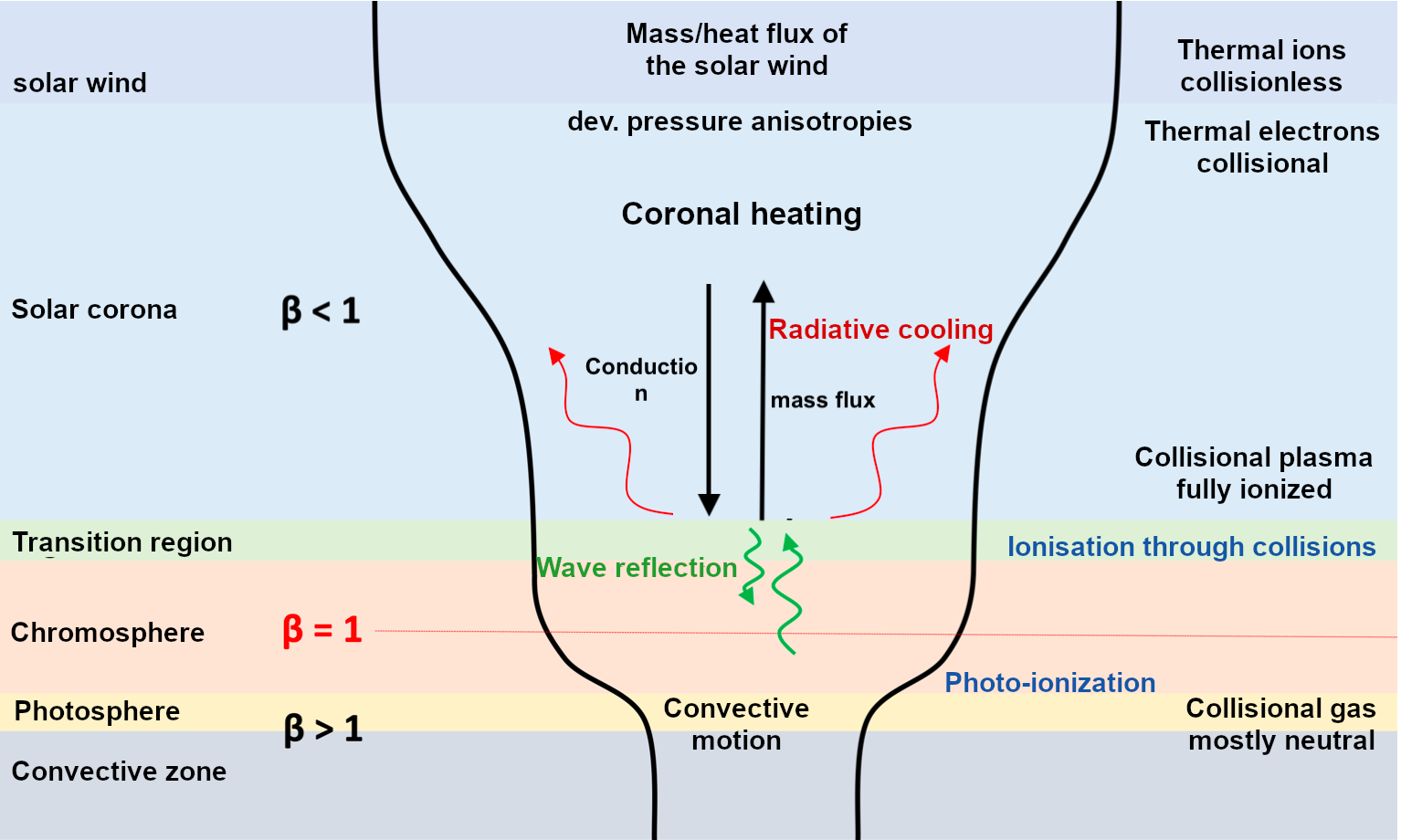}
\caption{Overall structure of the solar atmosphere with its associated physical properties and dominant processes. Figure adapted from the habilitation thesis of Dr. Alexis Rouillard (2021).
\label{fig:Alexis_HDR}}
\end{figure*}

As the electron density decreases with altitude, one progressively transitions from a chromosphere dominated by collisions to a corona where collisions become too scarce to influence energy transfers. The sudden rise in temperature and the drop in density at the TR already greatly reduces the collision frequency ($\nu \propto n T^{-3/2}$) between charged particles which interact via Coulomb interactions. This has a number of profound effects on particle transport and will be of importance for the results discussed in chapter \ref{cha:ISAM_results} because we will see that Coulomb collisions play a major role in the FIP effect. 
For completeness we present in Figure \ref{fig:Alexis_HDR} a summary picture of the different regions of the solar atmosphere and of the physics that rules these regions. The physical processes that have not been discussed yet are introduced in the subsequent paragraphs.

\subsection{Heating of the solar atmosphere}
\label{subsec:intro_heating}

If the coronal heating problem remains highly debated in the scientific community, several promising energy sources have been identified that are able to sustain the corona at several millions degrees. \\

In open-field regions, a significant part of the input energy is converted into kinetic energy by accelerating  coronal plasma to solar wind speeds. There is no such energy conversion in closed-field regions, and most of the energy must be conducted downward to the chromosphere where it is dissipated by electromagnetic radiation. The total energy flux that is necessary to compensate for the combined radiative and conductive energy losses ranges from $\approx 10^7\ \rm{erg.cm^{-2}.s^{-1}}$ in regions with strong magnetic fields (active regions), to $\approx 3 \times 10^5\ \rm{erg.cm^{-2}.s^{-1}}$ in quiet Sun regions \citep{Withbroe1977,Klimchuk2006}.\\

As already mentioned the mechanical displacement of the magnetic field at the photosphere or below is likely a main source of energy that can fuel the corona with the required energy flux \citep{Klimchuk2006}. These displacements are mostly induced by convective motions of the plasma that rises from the solar interior up to the photosphere. These displacements trigger the generation of a myriad of acoustic and magneto-acoustic waves that propagate throughout the chromosphere and eventually reach the solar corona. Waves may undergo mode conversion as they propagate to the upper chromosphere \citep[see e.g.][]{Khomenko2006,Khomenko2008}. The magnetic field lines that reach the upper chromosphere can expand significantly to merge into large-scale flux tubes, this can be seen from point D to E in the illustration of Figure \ref{fig:Wedemeyer2009_fig16} \citep[see also][]{Cranmer2005}. Mode conversion is also expected to occur somewhere in the chromosphere where the sound speed $c_s=\sqrt{k_b T/m}$ equals the Alfvén speed $c_A=B/\sqrt{\mu_0 \rho}$ \citep{Khomenko2006,Carlsson2007}, typically at the already mentioned equipartition layer (see Figure \ref{fig:Wedemeyer2009_fig16}). \\

Pure acoustic waves have been suggested as potential contributors to the heating of the chromosphere and corona \citep{Biermann1948,Schwarzschild1948}. As they propagate towards the upper layer of the chromosphere, acoustic waves steepen into shocks where they eventually dissipate their energy to heat up the plasma \citep{Schrijver1995,Carlsson2007}. Chromospheric simulations that include acoustic shocks and a detailed radiative transfer treatment \citep[see e.g.][and references therein]{Carlsson2002a} have the advantage of matching some spectroscopic observations, such as the chromospheric emission line of singly ionized Calcium (CaII). However they often fail at reproducing correctly the intensities of other emission lines of the mid/upper chromosphere, and subsequent studies have shown that acoustic waves alone probably cannot account for the total heating required to heat the chromosphere \citep{Carlsson2007}. In addition it is generally thought that the contribution of acoustic waves to coronal heating is likely negligible as most of their energy is dissipated in the upper chromosphere through shock formation and ultimately in the transition region due to steep temperature and density gradients \citep{Klimchuk2006}.\\

The magnetic field has therefore been pointed out as playing a key role in the transport of energy from the photosphere to the upper chromosphere and corona. Subsequent magneto-hydrodynamics simulations have shown that the braiding of magnetic field lines, that is induced by shear photospheric motions at the granular scale, can dissipate enough energy in the corona to maintain coronal temperatures at $\approx 1\ \rm{MK}$ \citep[see e.g.][]{Galsgaard1996,Gudiksen2005}. While nano-flare heating through magnetic reconnection appears to contribute significantly to coronal heating in these simulations, a significant portion of the energy is also transported by a mechanical flux throughout the corona. It is now widely accepted that shear Alfvén waves can provide the required mechanical flux to heat up the corona \citep{Carlsson2007}. In contrast to magnetoacoustic waves such as slow and fast modes, shear Alfvén waves remain incompressible during most of their transit through the chromosphere and are thought to penetrate the corona. \\

Since the energy is primarily transported by low-frequency Alfvén waves, an additional process is required that transfers the energy to higher frequencies where it can effectively be given to the plasma through wave-particle interactions. A turbulence cascade can provide this energy transfer to the high frequencies at the condition that "something" triggers the cascade. Many studies have then formulated a coronal heating theory where the turbulence cascade is generated through the non-linear interaction between counterpropagating waves at low frequency \citep[see e.g.][]{Zhou1990,Tu1995,LieSvendsen2001,Dmitruk2002,Cranmer2005,Verdini2009,Chandran2011,Verdini2019,Reville2020a}.  \\ 

In this thesis we exploit heating models that assume the dissipation of shear Alfvén-waves because their propagation can be described by a simple transport equation as discussed in section \ref{subsec:ISAM_Aw}. We also resort to ad-hoc heating functions that can well approximate the mechanical flux that is required to heat up the corona as discussed in section \ref{subsec:ISAM_heating_adhoc}. 

\subsection{The partially ionized chromosphere}
\label{subsec:intro_chromo}

In closed-field regions, all the energy that is given to the plasma in the corona must be conducted downward where it is dissipated in the form of radiative emissions. \\

The photosphere is usually defined as the height from which the medium becomes transparent and where photons that were bound in the solar interior through many collisions (absorptions and re-emissions) can escape into the solar atmosphere and beyond. But cooling through radiative emissions becomes efficient only in the mid/upper chromosphere where the decrease in plasma density allows for a significant fraction of the plasma radiative emissions to escape through the optically thin corona. \\

Because the chromosphere is highly dynamic, optically thick in EUV and hence only visible in several chromospheric lines, and can only been observed throughout emissions that are integrated along the line-of-sight, therefore observations can not provide a global picture of the chromosphere but only spare information. Therefore semi-empirical models of the chromosphere have been built to alleviate these limitations, of which the well-known VAL3 \citep{Vernazza1981}, FAL \citep{Fontenla2002} and lastly AL \citep{Avrett2008} chromospheric models. 

\begin{figure*}[]
\centering
\includegraphics[width=0.7\textwidth]{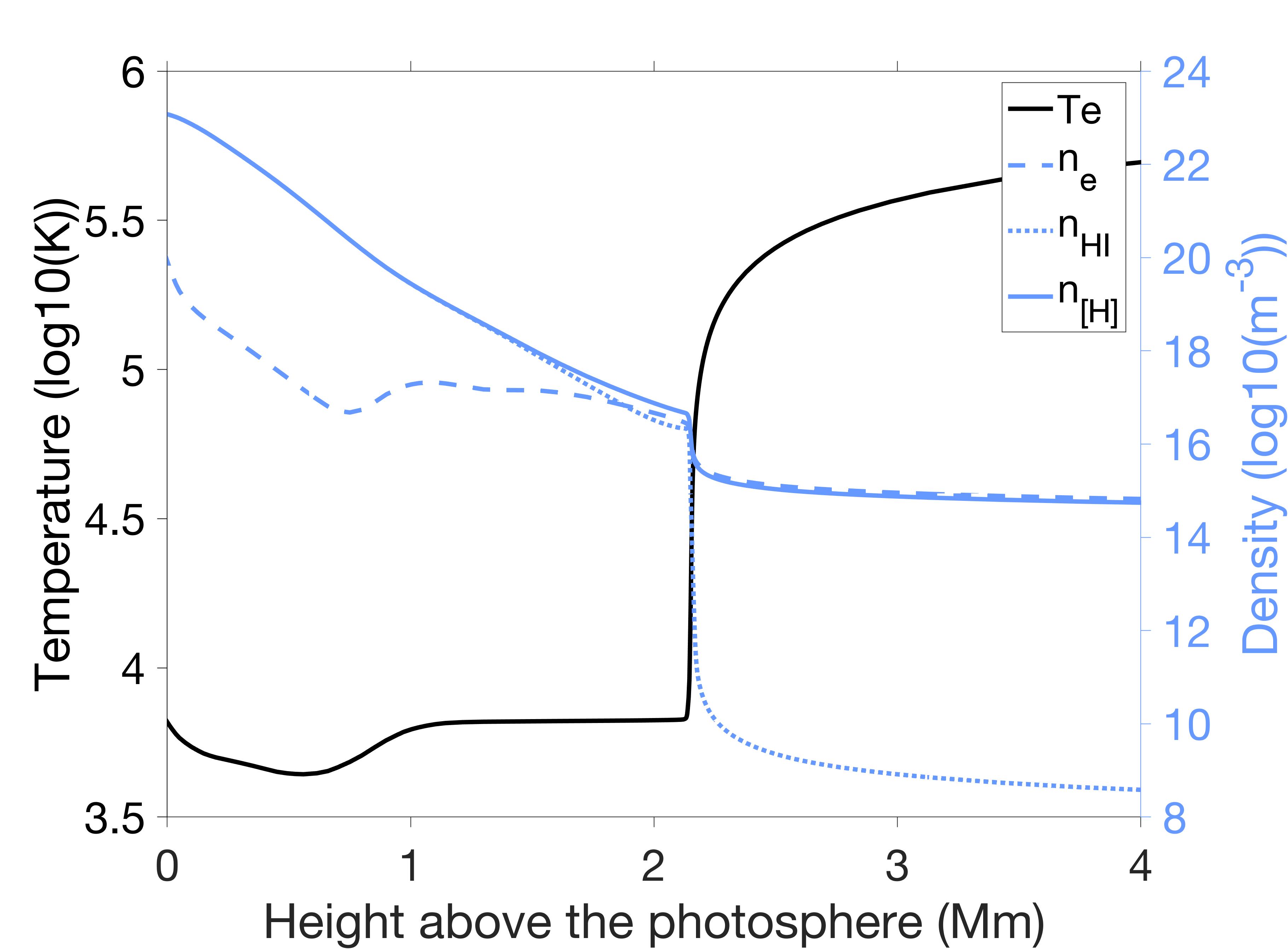}
\caption{Model of the solar chromosphere and low corona for the averaged quiet Sun, from the AL-C7 model of \citet{Avrett2008}. Both the total density of Hydrogen $n_{[H]}$, and the density of neutral Hydrogen alone $n_{HI}$ are plotted. 
\label{fig:AL_profil}}
\end{figure*}

They all consist in sophisticated inversion techniques that convert spectroscopic diagnostics into hydrostatic profiles of the entire chromosphere. In practice, the temperature-height distribution in the chromosphere is adjusted through a trial and error approach until the simulated (synthetic) emission spectra matches the observations. An example is shown in Figure \ref{fig:AL_profil} for the AL-C7 model of \citet{Avrett2008} that corresponds to the averaged quiet Sun. The transition region marks the separation between a chromosphere that is partially ionized with $n_{HI}/n_{[H]} \lesssim 1$ and a corona that is fully ionized in protons where $n_{HI}/n_{[H]} \ll 1$. 

Although these models have been found to reproduce very well the averaged quiet Sun, their relevance is questioned in lights of recent radiative-hydrodynamic simulations that depict a highly dynamic chromosphere \citep{Carlsson2002a,Carlsson2007}. Nonetheless, these semi-empirical models have been of precious help to calibrate our model of the solar atmosphere (ISAM) that is introduced in chapter \ref{cha:ISAM}, and where a quasi-stationnary approach is followed at first. Conversely, we show in this thesis that ISAM may shed new lights on these semi-empirical models by providing an improved treatment of the mass and energy transfer from the upper chromosphere to the corona. \\

The corona itself is fully ionized (see Figure \ref{fig:AL_profil}) and the lack of collisions with electrons rapidly "freezes" the charge states of coronal species. Therefore if the corona must be enriched or depleted in certain heavy ions according to their FIP, then it is likely to occur in the upper chromosphere where all neutrals are not fully ionized yet.

Ionization is facilitated for elements having a low FIP which ionize early on in the chromosphere and become fully ionized in the upper chromosphere. In comparison only $\approx 30\%$ of Hydrogen, with an intermediate FIP ($\simeq 13.6\ \rm{eV}$), is ionized at the top of the chromosphere. In contrast the ionization of high FIP elements such as Helium ($\simeq 24.6\ \rm{eV}$) occurs mostly in the transition region and very low corona. Elements heavier than Hydrogen and Helium can occur in varying charge states according to the number of electrons that their nucleus can host. For instance, the 20-electron Calcium atom while having a low FIP of $\simeq 6.1\ \rm{eV}$ can barely reach the 14th charge state at coronal temperatures of $1-3\ \rm{MK}$, whereas the 2-electron Helium atom with a high FIP of $\simeq 24.6\ \rm{eV}$ will be fully (doubly) ionized in the corona. \\

The chromosphere and solar corona also differ in the physical processes that contribute to the ionization balance. In the chromosphere ionization occurs primarily through photoionization from the incident photospheric radiation field, that is counterbalanced by radiative recombination \citep{Carlsson2002a}. In the transition region and low corona ionization is mostly driven by collisional impact with electrons, which can either proceed directly (direct ionization) or in two phases through an intermediate excited (autoionizing) state. We further discuss the specificity of each of these ionization/recombination processes in section \ref{sec:ISAM_ioniz}. \\

Hydrogen is peculiar in the sense that its ionization balance may proceed more slowly than the typical hydrodynamic time scales in the chromosphere \citep{Kneer1980}. Radiative-hydrodynamic simulations have shown that Hydrogen is likely not in ionization equilibrium with the local thermal equilibrium, with a tendency for Hydrogen ionization to be boosted up near the peak temperature of acoustic shocks that traverse in the upper chromosphere \citep{Carlsson2002a}. This further questions the validity of an average "static" chromosphere as given by the semi-empirical models discussed above, because ionization of Hydrogen is likely not in equilibrium and depends on the previous history of the chromosphere. Our approach to address this difficulty in our model of the solar atmosphere is discussed in section \ref{subsec:ISAM_ioniz_future}.

\subsection{Extraction of heavy ions from the chromosphere}
\label{subsec:intro_FIP}

The composition in heavy ions that is measured in situ in the solar wind is likely settled low in the solar atmosphere before the ionization balance "freezes" in the corona where collisions are scarcer. Furthermore, since relative abundances are uniformly distributed at the photosphere \citep{Asplund2009}, the separation between heavy ions must then occur somewhere between the photosphere and the corona, and hence most likely in the chromosphere. More precisely, most of the fractionation is probably established in the upper chromosphere where neutrals become ionized and where neutrals having a low FIP start to ionize first. That way one can possibly explain the separation between low and high FIP elements if there exists an outward force that selectively pulls off ions from the chromosphere. Without this hypothetical force, heavier elements would fall faster than lighter elements due to gravitation, and therefore would produce a separation of elements according to their mass and not their FIP. \\

Several forces have been suggested in the literature which may correspond to this hypothetical force and can produce a separation between low and high FIP elements \citep[see][for a review on the FIP effect]{Henoux1998,Laming2015}. They can be sorted into two main groups whether the magnetic field plays a role or not in the separation process. Many studies of the FIP effect show that pure diffusion effects can induce a fractionation according to the FIP and not mass. Some of these studies consider the effect of frictional coupling of particles having different velocities \citep{Wang1996,Peter1996,Marsch1995,Peter1998b,Bo2013}, alone or coupled with thermal diffusion effects along the magnetic field \citep{Geiss1986,Hansteen1997,Killie2005,Killie2007}. While others have studied the effect of diffusion across the magnetic field \citep{Steiger1989,Antiochos1994}. \\

\citet{Vauclair1996} formulated a scenario where the observed coronal abundances would result from the emergence of a magnetic flux tube from the photosphere into the corona, and that only low-FIP elements that ionize early on may be able to catch up with the convective motion of the rising flux tube. Recent observations from \textit{Hinode} suggest nonetheless that when emerging fluxes reconnect with pre-existing coronal fields, that opens up new channels where photopheric and coronal abundances are likely mixed up \citep{Baker2015}.

\citet{Schwadron1999,Laming2004,Laming2009} have found an intimate connection between heating processes that are based on wave-particle interactions, and the FIP effect that may result from the same interactions. It is now widely accepted that low-frequency Alfvén waves can carry the necessary mechanical energy from the chromosphere to the corona, where the energy is eventually dissipated at small scales through wave-particle interactions (see section \ref{subsec:intro_heating}). It is however uncertain how such high-frequency waves could survive in the chromosphere, nonetheless \citet{Laming2004,Laming2009,Laming2015} formulated a theory where heavy ions may interact with low-frequency Alfvén waves through the ponderomotive force. In essence the ponderomotive force corresponds to a time-averaged description of the Lorentz forces acting in an oscillating electromagnetic field, that results in a net force that is directed from low to high wave-energy densities \citep{Lundin2006,Laming2015}. \\

The frictional coupling of heavy ions with the protons (proton drag) alone has been found to be very efficient to separate heavy ions from the main neutral Hydrogen gas, that occurs mostly in the ionization layer of Hydrogen in the upper chromosphere.

In the hydrodynamic case modeled by \citet{Wang1996}, that requires however the existence of an ambipolar flow in the upper chromosphere where protons are drifting upward with respect to neutral Hydrogen. Low-FIP elements that ionize early on in the chromosphere are then carried out by the proton flow through frictional coupling and the Coulomb interaction. This ambipolar flow is likely transitory in closed-field regions but they suggest that it can be temporaly sustained during phases of chromospheric evaporation when an enhanced heating is applied at the base of the corona. Periodic signatures of chromospheric evaporation have been observed in EUV loops, which may not necessarily require an external input of energy via a sudden reconnection event for instance but could just result from thermal non equilibrium (TNE) cycles in coronal loops \citep[see e.g.][]{Auchere2016}. Under some conditions coronal loops can enter in TNE and undergo alternating phases of evaporative upflows from the chromosphere and condensation downflows from the corona \citep[see also][]{Johnston2017,Johnston2019}, and hence can possibly sustain the above mentioned ambipolar flow. We will discuss further about TNE cycles in section \ref{subsec:ISAM_results_H_thermodynamics} using our model of the solar atmosphere called ISAM. Therefore this separation process would be the most efficient in closed-field plasmas where the energy deposited in the corona is primarily convected downward to the chromosphere, and not converted into kinetic energy as in open-field plasmas. \citet{Bo2013} also showed the importance of frictional coupling with protons to prevent heavy elements from settling in the upper chromosphere, and that even in an hydrostatic chromosphere where neutral Hydrogen and protons are supposed at rest. 

\begin{figure*}[]
\centering
\includegraphics[width=0.85\textwidth]{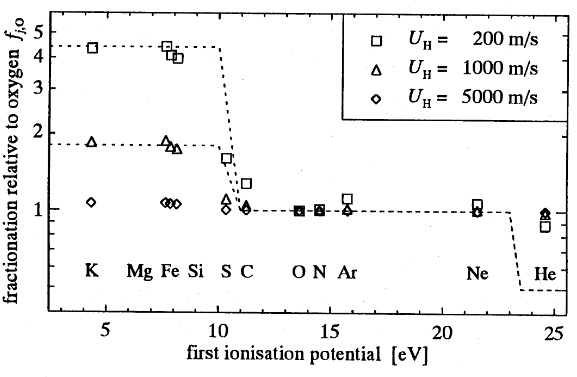}
\caption{Variation of abundances relative to Oxygen as a function of the FIP, given by the velocity-drift model of \citet{Peter1996}. Figure taken from \citet[][Figure 1]{Peter1996}.
\label{fig:Peter1996_fig1}}
\end{figure*}

\citet{Peter1996,Peter1998b} studied the case of open-field plasmas. By varying the mean flow velocity of Hydrogen in the chromosphere, they argue that frictional coupling with protons can reproduce in overall the typical FIP fractionation that is measured in situ in the slow and fast solar wind, as shown in Figure \ref{fig:Peter1996_fig1}. \\

All frictional models presented above only consider the upper chromosphere in the separation process, and hence they do not address the exchange of matter throughout the transition region and corona.
By including the corona and thermal diffusion effects, namely the thermal force, \citet{Killie2007} showed a different picture that seems in conflict with some past studies, especially for the case of closed-field plasmas. Because they have a downward (and not outward) flow of protons, their proton flow acts as a barrier to the extraction of heavy ions with low FIP from the chromosphere. However, the thermal diffusion effects that they introduce produce a net force that is directed upward and which pulls low-FIP elements towards the hotter corona. Nonetheless, they argue that low FIP elements such as Silicium or Iron still are unable to flow into the corona due to a "proton barrier" that is stronger than the thermal force. Therefore their model tends to produce an inverse FIP effect which is generally not observed in the solar corona but in active stellar coronae \citep[see the review from][]{Laming2015}. In order to alleviate this paradox, they conclude that this enriched material in low-FIP elements that is trapped in the corona, may initially originate from a (partially) stratified chromosphere. And that this enriched chromospheric material would be naturally lifted up as a loop emerges from the photosphere into the corona \citep[see e.g.][]{Vauclair1996}. \\

Many possible contributors to the FIP effect have been discussed throughout this section. The collisional coupling with protons is commonly found to contribute significantly to the separation between low and high FIP elements in the upper chromosphere. However diffusion effects are slow and as a result the FIP effect can take up to several days or weeks to settle in a non-perturbed chromosphere \citep[see e.g.][]{Killie2007}, that is consistent with enhanced FIP biases measured with the aging of active regions \citep{Widing2001}. Furthermore some authors have noticed the necessity of having an external mixing mechanism of the chromosphere that could prevent some heavy elements from being too severely depleted or enriched due to gravitational stratification \citep{Hansteen1997,Killie2007,Bo2013}. \citet{Laming2009} supplemented this point by arguing that in the absence of wave-particle interactions with Alfvén waves (through the ponderomotive force), hydrodynamic turbulence may provide a chromospheric mixing on time scales that are shorter than gravitational stratification to operate, but still sufficiently long so that the FIP effect can establish. The emergence of magnetic flux from the photosphere may be another major source of chromospheric mixing with photospheric abundances \citet{Baker2015}. As discussed in section \ref{subsec:intro_chromo}, the chromosphere is inherently dynamic and as a consequence it is much unlikely to have a chromosphere that remains stratified over time. \\

Although all studies introduced above are able to produce a FIP effect, the modeling framework and the required conditions differ from one study to another, where in most studies the coupling between the transition region and the corona is not considered. All forces that are considered in these studies likely contribute to the FIP effect, however it remains to quantify their relative contribution in a self-consistent manner, in a comprehensive framework that spans from the chromosphere, transition region and corona. For that purpose we develop a model of the solar atmosphere that is described in chapter \ref{cha:ISAM}. First applications of the model to the case of closed-field plasmas are presented in chapter \ref{cha:ISAM_results} where the diffusion effects discussed above can be tested right away.



\section{Theories on the origin of the slow solar wind}
\label{sec:intro_theories}

\subsection{The quasi-stationnary theory}
\label{sec:intro_stationnary}

\begin{figure*}[]
\centering
\includegraphics[width=0.5\textwidth]{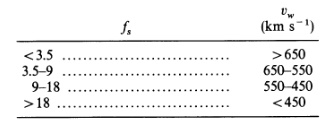}
\caption{Correspondence between expansion factors (computed between the photosphere and a coronal height of $2.5\ R_\odot$) and solar wind speeds measured in situ at 1 AU. Figure taken from \citet[][Table 1]{Wang1990}.
\label{tab:Wang1990_tab1}}
\end{figure*}

Before the \textit{Ulysses} mission, \citet{Wang1990} collected 22 years of solar wind speed data from the \textit{Vela}, \textit{IMP}, and \textit{ISEE 3/ICE} missions to demonstrate a long-term inverse correlation between the solar wind bulk velocity measured in situ at 1 AU, and the divergence rate of the coronal magnetic field of the estimated source region. In Table \ref{tab:Wang1990_tab1}, a correspondence is shown between expansion rates and the typical associated solar wind speeds. \\

\begin{figure*}[]
\centering
\includegraphics[width=0.95\textwidth]{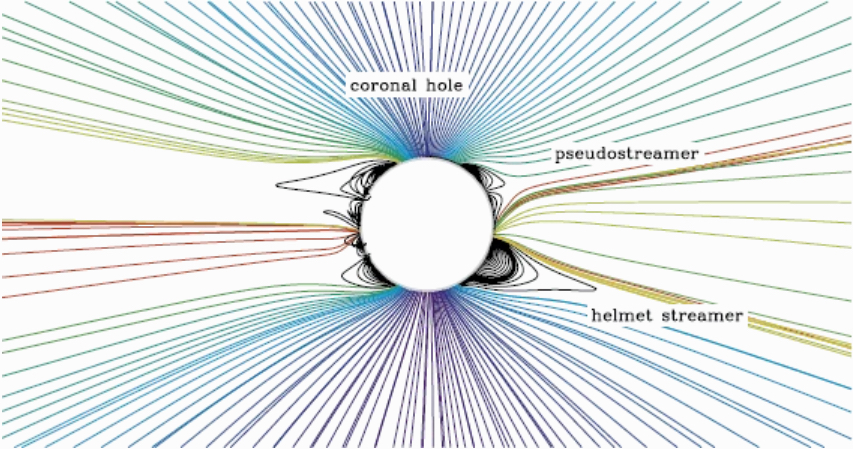}
\caption{Closed (black) and open (multi-color) magnetic field lines traced from a time-steady solution of the polytropic MHD conservation equations, computed by the Magnetohydrodynamics Around a Sphere (MAS) code \citep{Linker1999,Riley2001}. Photospheric boundary conditions were from Carrington Rotation 2058 (June-July 2007). Colors of open field lines correspond to the \citet{Wang1990} expansion factor $f_{ss}=(R_\odot/R_{ss})^2(B_\odot/B_{ss})$: $f_{ss}\leq 4$ (purple), $f_{ss}\simeq 6$ (blue), $f_{ss}\simeq 10$ (green), $f_{ss}\simeq 15$ (gold), $f_{ss}\geq 40$ (red). Figure taken from \citet[][Figure 1]{Cranmer2017}.
\label{fig:Cranmer2017_fig1}}
\end{figure*}

An inspection of 3-D reconstructions of the coronal magnetic field then suggests that the SSW originates from flux tubes that are contiguous to closed field regions (e.g. the coronal loops beneath helmet streamers) where the magnetic field undergoes a large expansion across the corona as illustrated in Figure \ref{fig:Cranmer2017_fig1}. At opposite, large coronal holes such as those situated near the poles typically produce a fast solar wind near their center because the magnetic field expands much less. For smaller coronal holes found at equatorial latitudes where the expansion factor is not as low, one may find winds with intermediate speeds or even slow Alfvénic winds as discussed in section \ref{sec:dynamics_insitu} and in \citet{Griton2021}. At rare occasions large coronal holes have been observed near the equator, which can produce a very fast wind even at equatorial latitudes and also drastically affect the overall shape of the solar corona with a HPS that is nearly vertical \citep[see e.g.][]{Sanchez-Diaz2017b}. \\

The multiple flux tubes model MULTI-VP introduced in section \ref{subsec:MULTI-VP} and described in \citet{PintoRouillard2017} successfully reproduces a solar wind driven by the expansion rate and is exploited at several occasions in this thesis. In MULTI-VP, the generation of different solar wind regimes is built in the prescription that is adopted to heat up the plasma. In essence the expansion factor controls the heights at which the energy is deposited, that affects in turn the amount of chromospheric plasma that is pushed out to the corona via chromospheric evaporation \citep[see also][]{Hansteen2012}. Thanks to its versatility, MULTI-VP is therefore able to reproduce the overall structure of the solar corona as seen in WL coronagraphs \citep{PintoRouillard2017}, with even finer details that are discussed further in section \ref{sec:stationnary_poirier2020}. \\

At greater heights in the corona, the input energy given to the heating of the plasma is progressively converted into kinetic energy to accelerate the solar wind \citep{PintoRouillard2017}. If one also consider pressure anisotropies that cannot be neglected in the high corona within the acceleration region of the solar wind, then the mirror force also plays a major role in the acceleration process \citep{Lavarra2022}. \\

Recently, \citet{Lavarra2022} have shown with the Irap Solar Atmosphere Model (ISAM, see also chapter \ref{cha:ISAM}) that the quasi-stationnary theory can also account for the charge states measured in situ in both the fast and slow winds, and even in the slow Alfvénic wind as illustrated in Figure \ref{fig:Lavarra2022_fig1}. Whether this model can also reproduce the variations in the abundances of alpha particles and minor ions measured in situ in these various solar wind regimes will be investigated in a future study. \\

As introduced in section \ref{subsec:intro_FIP}, diffusion effects such as frictional coupling with the mean proton flow may explain the measured enhanced abundances of low-FIP elements for different solar wind speeds, but fail at predicting the variable abundances in alpha particles \citep{Peter1996,Peter1998b}. We have also introduced in section \ref{subsec:intro_FIP} other processes such as wave-particle interactions that may also play a role in the quasi-stationnary theory. However, we also pointed out in section \ref{subsec:intro_FIP} that the fractionation processes remain inherently limited in open-field media described by the quasi-stationnary theory.

\subsection{Dynamic theories of the slow solar wind}
\label{sec:intro_dynamic}

As discussed in section \ref{subsec:intro_FIP}, closed-field regions are propitious environment which can reach fractionation levels that are as high as those measured in situ in the slow solar wind. Therefore we invoke the necessity of a dynamic theory that complements the quasi-stationnary theory, where the slow solar wind is partially made up of closed-field plasma that is expelled in the slow wind through magnetic reconnection processes that are discussed throughout this section.

\subsubsection{Magnetic reconnection at the tip of streamers}
\label{subsec:intro_dynamics_streamertip}

It is likely that the blobs observed above streamers (see section \ref{subsubsec:intro_SSW_intermittency}) are produced at the HCS through magnetic reconnection known to occur in many regions of the solar corona. \\

\begin{figure*}[]
\centering
\includegraphics[width=0.4\textwidth]{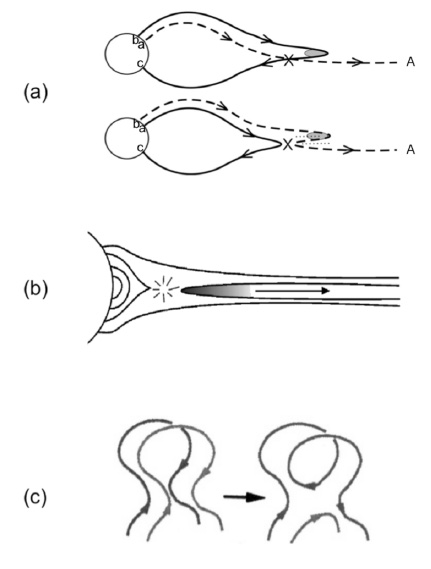}
\caption{Sketch of different reconnection scenarios for the release of closed-field plasma in the slow solar wind near the tip of streamers by: (a) interchange reconnection \citep{Crooker2003}, (b) reconnection in the HCS \citep{Wang1998}, and (c) reconnection between a pair of coronal loops \citep{Gosling1995}. Figure taken from \citet[][Figure 1.10]{Sanchez-Diaz2017_PHD}.
\label{fig:Sanchez2017_fig1.10}}
\end{figure*}

\citet{Wang2000} suggested that such intermittent outflows could be induced by magnetic reconnection near the cusp of streamers when coronal loops rising in the solar atmosphere get stretched enough to trigger magnetic reconnection between magnetic fields of opposite direction, as illustrated in Figure \ref{fig:Sanchez2017_fig1.10}b. This picture has the advantage to be also consistent with the observation of associated plasma inflows in \textit{LASCO} \citep{Wang2000} and associated for the first time to outflowing blobs in \citet{Sanchez-Diaz2017a}. The tearing instability that develops during the expansion of loops has been proposed as the mechanism that triggers the generation of streamer blobs via magnetic reconnection at the HCS \citep{Reville2020b,Reville2022}. We investigate this process and its expected remote-sensing signatures in chapter \ref{sec:dynamics_tearing} where we exploit the magneto-hydrodynamic WindPredict-AW model \citep{Reville2020a}. \\

\subsubsection{Interchange reconnection at open-closed boundaries}
\label{subsec:intro_interchange}

Another possible formation mechanism for transient structures is magnetic interchange reconnection in open-closed boundaries where magnetic loops can interact with open magnetic fields in regions of high magnetic shear, this is also illustrated in Figure \ref{fig:Sanchez2017_fig1.10}a. An import driver of interchange reconnection is certainly the rate at which open and closed field regions are susceptible to drift towards each other. \\

Sunspot observations indicate that the solar surface or photosphere is mostly in differential rotation, meaning that equatorial regions rotate faster than polar ones \citep{Scheiner1630,Bumba1969}. However, the photospheric differential rotation gets rapidly restrained at greater heights in the solar atmosphere, by the well-established corona that exhibits a more rigid rotation \citep{Fisher1984,Hoeksema1987,Bird1990}. These different rotation rates probably lead to the inter-penetration of closed and open fields, a propitious environment for interchange reconnection \citep[see e.g.][]{Fisk1996}. \\

The open magnetic field lines inside coronal holes tend to rotate rigidly with the solar corona \citep{Lionello2005}, and hence should drift with respect to closed-field regions that rotate with the photosphere. For instance the open-closed boundary between polar holes and coronal loops situated beneath helmet streamers is a favoured environment where interchange reconnection might occur continuously \citep{Crooker2003,Pinto2021}. Such a scenario may also explain the numerous magnetic field reversals, or 'switchbacks', that have been recently measured in situ at \textit{PSP} \citep{Bale2021}. \\

\begin{figure*}[]
\centering
\includegraphics[width=0.5\textwidth]{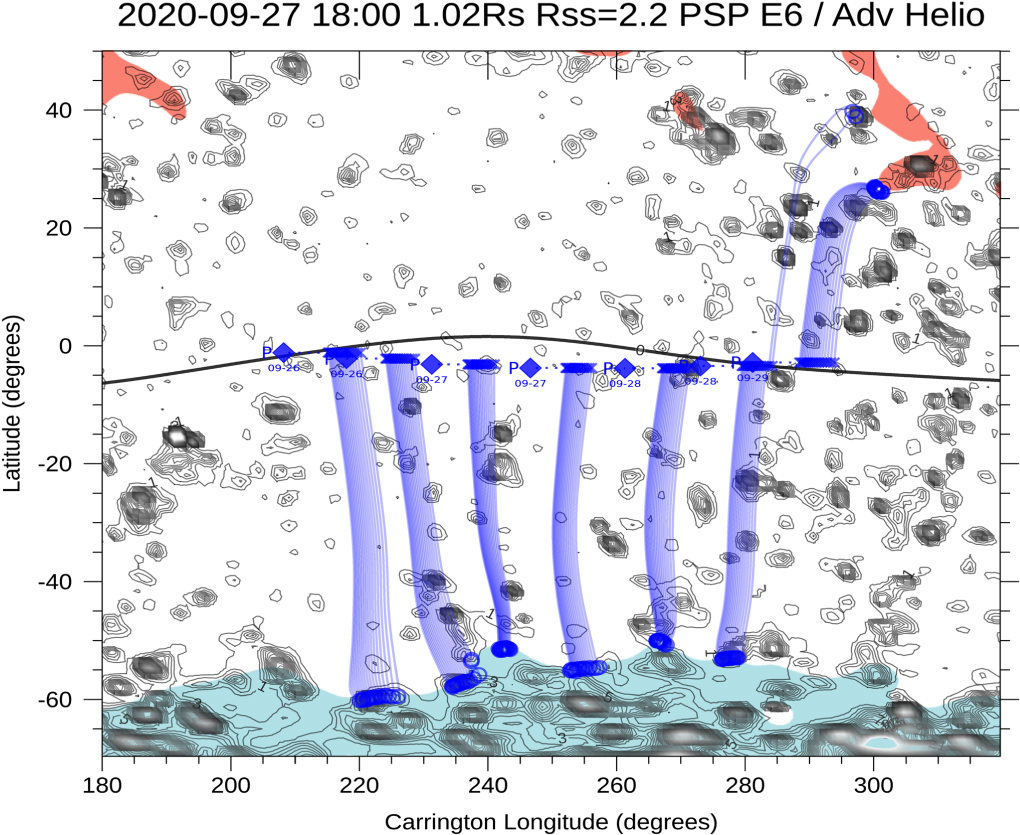}
\caption{Connectivity map of \textit{PSP} during its 6th passage close to the Sun (26-29 of September 2020). The northern and southern polar coronal holes are represented as red and cyan shaded areas respectively. The magnetic pressure $B^2/(2\mu_0)$ is plotted as dark contours at $0.02\ R_\odot\simeq 14\ \rm{Mm}$ above the photosphere, and represent the open-closed boundaries associated with the supergranular network. The \textit{PSP} trajectory (blue diamonds) is projected ballistically to a height of $1.2\ R_\odot$ (blue crosses), and then tracked down to $0.02\ R_\odot$ (blue circles) above the photosphere using a 3-D reconstruction of the magnetic field. The position of the HCS predicted by the 3-D reconstruction model is plotted as a solid black line at $1.2\ R_\odot$. Figure taken from \citet[][Figure 1]{Bale2021}.
\label{fig:Bale2021_fig1}}
\end{figure*}

\citet{Bale2021} draw a more detailed picture of interchange reconnection that occurs much lower in the solar atmosphere (near the transition region), and at a much smaller spatial scale. These interchange-like reconnection events could take place in open-closed boundaries that correspond with the supergranular network with typical scale of $\approx\ 15-20\ \rm{Mm}$, which is illustrated in Figure \ref{fig:Bale2021_fig1} as dark contours. They also notice a slight enrichment in the alpha particles (i.e. doubly ionized Helium) measured in situ in the slow solar wind when \textit{PSP} passes over these regions. Their work shows that the release of closed-field plasma in the slow solar wind through interchange reconnection occurs systematically and at even smaller scales than previously thought. In addition, the statistical analyses of \textit{PSP} data carried out by \citet{Fargette2021,Fargette2022} found that the scales and occurrence rate of switchbacks are compatible with processes occurring at the granular and mesogranular scale. They proposed that magnetic loops that tied in to the differentially-rotating photosphere may reconnect with the more rigidly rotating coronal magnetic field. This will be discussed in section \ref{subsec:intro_Sweb}. \\

We shall note that magnetic reconnection events associated with strong jets can also be generated in pre-existing null-point topologies within unipolar regions, if some twisting of the field lines is enforced at the base \citep{Pariat2009}. Alternative scenarios for the origin of switchbacks also involve interchange-like reconnection, not with pre-existing loops as discussed previously, but with loops that emerge on the photosphere in otherwise unipolar regions thereby triggering microjets \citep{Sterling2015}. More generally, jet-like features or spicules have been observed multiple times in chromospheric spectral lines as short-lived $\approx1 - 10\ \rm{min}$ and small-scale $\lesssim 300-1500\ \rm{km}$ intensity enhancements associated to upward velocities of about $25\ \rm{km/s}$, and where their temperature of about $5000-15000\ \rm{K}$ indicate a material of chromospheric origin \citep{Sterling2000}. Coronal jets are also signatures of magnetic reconnection that are frequently observed in the solar corona \citep{Raouafi2016}. The recent high-resolution remote-sensing observations of \textit{Solar Orbiter} further unveiled the existence of very small-scale "campfires" popping out frequently of the chromosphere, which for now are considered as new elements in the flare-microflare-nanoflare family \citep{Berghmans2021}. All together, small-scale reconnection events may contribute significantly at heating the chromosphere and transition region, and probably at mixing up the chromosphere with photospheric abundances as well. The wealth of sudden brightenings detected in EUV depicts a highly dynamic solar chromosphere and corona where magnetic reconnection events are ubiquitous, and hence where closed-field material can be expelled into the solar wind through interchange reconnection. \\

As we shall see in section \ref{subsec:intro_Sweb}, it is now widely accepted in the literature that a significant part of the slow solar wind is born at open-closed boundaries where interchange reconnection is found to occur systematically. That process allows especially to explain why signatures of the slow wind (which for recall is enriched in low-FIP elements as in coronal loops, see section \ref{subsubsec:intro_SSW_composition}) are also detected far from the HCS from composition measurements taken in situ \citep{Zurbuchen2007}. Using high-resolution and three-dimensional MHD simulations, \citet{Higginson2017} show that a slow wind can form far from the HCS at the open-closed boundaries of polar coronal hole extensions, which under some circumstances can create a narrow open-field corridor from polar to mid latitudes as illustrated in Figure \ref{fig:Antiochos2011_fig4} (see also section \ref{subsubsec:intro_streamers}). More generally, we will see in section \ref{subsec:intro_Sweb} that these open-field boundaries are ubiquitous in the solar atmosphere and that form a wide web of separatrices \citep[called the S-web: ][]{Antiochos2011} and from which the mysterious slow wind that is detected far away from the HCS is supposed to be born. \\

\subsubsection{Release of magnetic flux ropes from the tip of streamers}
\label{subsec:intro_dynamics_fluxropes}

Magnetic reconnection may also occur between adjacent coronal loops beneath a streamer \citep{Gosling1995}, as depicted in Figure \ref{fig:Sanchez2017_fig1.10}c. Such scenario would lead to the release of plasma outflows transported by magnetic flux ropes in the slow wind. In fact, the panels b and c of Figure \ref{fig:Sanchez2017_fig1.10} may represent two aspects of the same release process, where the formation of a flux rope (panel c) is accompanied by a density enhancement (i.e. a "streamer blob", panel b) at the same time. \\

The continuous tracking of blobs expelled from the tip of helmet streamers all the way to points of in situ measurements reveals that blobs indeed transport helical magnetic fields \citep{Rouillard2009,Rouillard2011a,Rouillard2011b}, that is further supported by recent \textit{PSP} observations \citep{Lavraud2020,Rouillard2020a}. More systematic statistical analyses of blobs observed in \textit{STEREO} images, and of transient structures measured in situ inside the HPS revealed that the topology of blobs is consistent with magnetic flux ropes \citep{Sanchez-Diaz2019} that form via magnetic reconnection at the tip of helmet streamers \citep{Sanchez-Diaz2017b}. \\

\begin{figure*}[]
\centering
\includegraphics[width=0.6\textwidth]{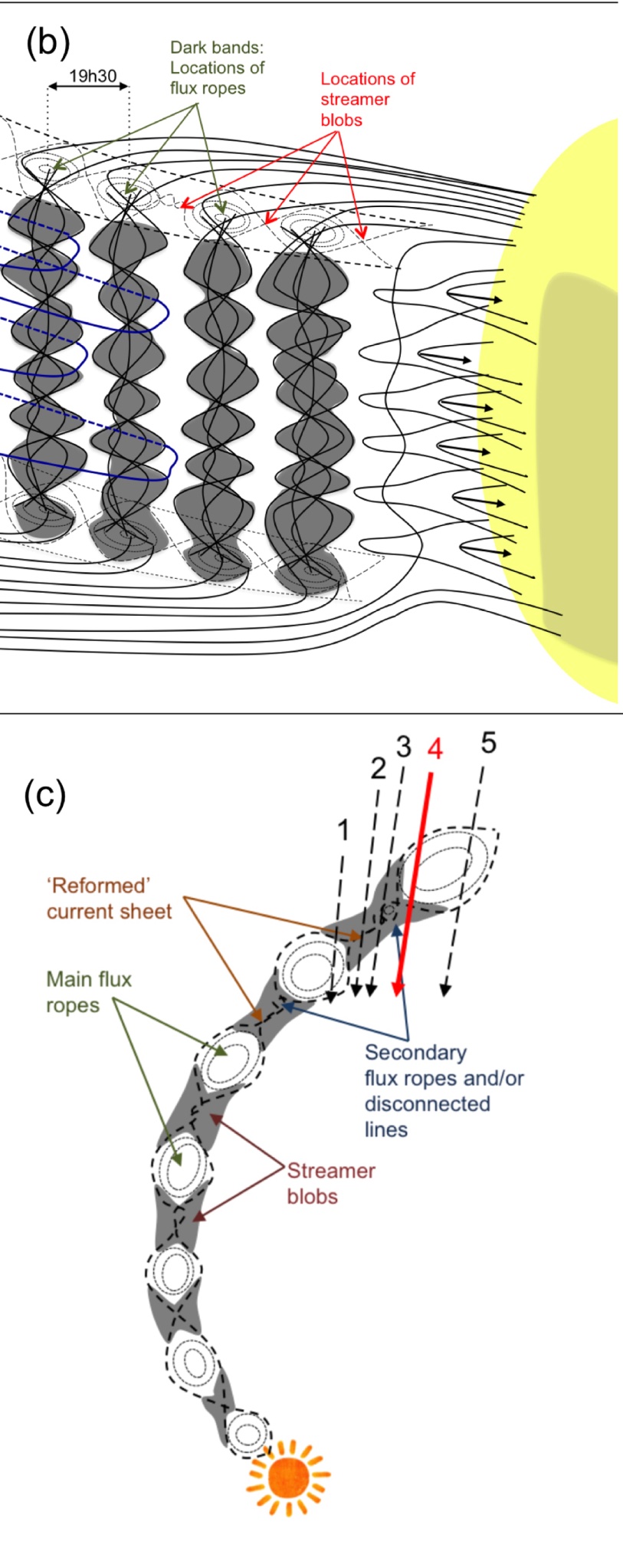}
\caption{ Schematic of the generation of magnetic flux ropes and streamer blobs in a HCS that is seen face-on (panel b) and edge-on (panel c). The grey shaded areas depict either the flux ropes in panel b, or the streamer blobs in panel c. Figure extracted from \citet[][Figure 12]{Sanchez-Diaz2017b}.
\label{fig:Sanchez2017b_fig12}}
\end{figure*}

An illustration is given in Figure \ref{fig:Sanchez2017b_fig12} where flux ropes are generated along a HCS that is seen face-on in this case. In this picture, the bright "blobs" seen in WL imagery correspond to density structures that are located at the interstices of successive flux ropes. The HCS that reforms in between flux ropes may be favorable to the generation of sub-density structures at smaller scales and with higher periodicities up to the hourly time scales that have been measured in situ \citep{Viall2010,Viall2015,Kepko2016}. We will see in section \ref{subsec:dynamics_tearing_3D} that this picture is consistent with the recent 3-D MHD modelling work of \citet{Reville2022}, where flux ropes are produced by the reconnection between coronal loops that have been stretched up in the atmosphere, and that fast quasi-periodic structures can form from subsequent magnetic reconnections at the HCS through the tearing instability. \\

All the processes discussed above involve magnetic reconnection that liberates into the slow solar wind, plasma that was initially confined to the corona along magnetic loops. As we shall see, magnetic loops are particularly enriched in elements with low FIP and this reconnection could contribute to enriching the slow solar wind in low-FIP elements as discussed in section \ref{subsubsec:intro_SSW_composition}. 

\subsubsection{Open-closed boundaries as potential source regions of the slow wind}
\label{subsec:intro_Sweb}

In light of the previous sections we now consider further open-closed boundaries as potential source regions of the slow wind, and discuss processes that could allow that wind to form far away from bipolar streamers. The idea that magnetic reconnection is likely to occur continually in certain regions of the solar corona dates back to the first explanations of the quasi-rigid rotation of coronal holes \citep{Nash1988,Wang1988}. This quasi-rigid rotation was interpreted as the effect of a wave of magnetic reconnection occurring between the open magnetic field threading the coronal hole and the closed loops anchored in the differentially-rotating photosphere situated below \citep{Wang1988}. This was later confirmed by full 3-D MHD simulations of the solar corona \citep{Lionello2005}. \\

It was further demonstrated that magnetic reconnection can occur even in the absence of anti-parallel magnetic field lines, that is in other layers than for instance the standard HCS that are called quasi-separatrix layers (QSLs) \citep{Priest1995,Demoulin1996}. As we will see in the next paragraph, QSLs are thin layers that connect open field lines of the same polarity but of different origin at the solar surface. Therefore QSLs can form above the cusp of pseudo-streamers for instance and be associated with their stalk as seen in WL (see section \ref{subsubsec:intro_streamers}). Therefore, while QSLs do not have a net inversion of the magnetic field as at the HCS, they still exhibit strong gradients in their magnetic connectivity. As in the HCS, electric currents have also been found in QSLs that trigger magnetic reconnection \citep{Aulanier2005,Aulanier2006}. If the magnetic reconnection in QSLs fails at justifying the enrichment of the SSW in closed-field material, it does explain part of the observed high variability of the SSW, even away from the HCS and the regular streamers. \\

\begin{figure*}[]
\centering
\includegraphics[width=0.95\textwidth]{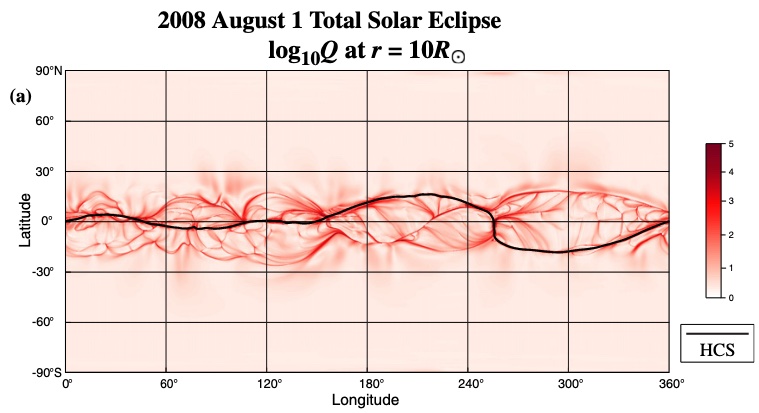}
\caption{Latitude-longitude Carrington map of the squashing factor (color plotted in a logarithmic scale) at $10\ R_\odot$ derived from 3-D magnetohydrodynamic simulation run with the MAS code \citep[see][and references therein]{Antiochos2011}. Figure taken from \citet[][Figure 7]{Antiochos2011}.
\label{fig:Antiochos2011_fig7}}
\end{figure*}

Later, \citet{Antiochos2011} argued that open-closed boundaries must be ubiquitous and weave in what they call the S-web (or web of separatrix and quasi-separatrix layers), that are observed even during periods of low solar activity as shown in Figure \ref{fig:Antiochos2011_fig7} for the solar cycle 23 minimum. The S-web has been mapped by computing from 3-D magnetic vector fields, such as the output of 3-D MHD or magnetostatic models described in section \ref{sec:modeling}, a parameter called the squashing factor for each field line. This factor quantifies how much magnetic field lines that are contiguous at a reference altitude in the upper corona diverge from each other towards the solar surface at their footpoint \citep{Titov2007}. A high squashing factor will tend to mean that two field lines that are adjacent high in the corona are widely separated at their photospheric footpoint. Since coronal loops tend to force a significant separation of its overlying open field lines, these systems will be marked by high squashing factors and the latter parameter is a good estimate of the location of open-closed field boundaries. In addition these systems tend to be associated with higher magnetic shears known as quasi-separatrix layers \citep{Demoulin1997} and the squashing factor gives topological information on where magnetic shears are likely to occur in the coronal field \citep{Titov2011}. The coronal loops observed below both bipolar and unipolar (pseudo) streamers will tend to force a significant separation of open field lines from streamer tops where they are adjacent to their footpoint at the solar surface which will be reflected as high squashing factors. \\ 

Consequently there is a clear association between the bright streamers seen in white-light Carrington maps and regions where high squashing factors are likely to occur. The highest squashing factors (dark red colors) in Figure \ref{fig:Antiochos2011_fig7} correspond to the tip of helmet streamer and generally follow the shape of the streamer belt. Intermediate squashing factors (reddish colors) occur more along pseudo-streamers, they appear as arches in the white-light Carrington map and cover a broader region of the corona than the streamer belt. \\

\citet{Crooker2012} carried out a ballistic mapping of the slow wind measured in situ back to the upper corona, and compared the estimated source regions to maps of the squashing factor. They found a good association between slow wind source regions and regions of high squashing factor that can occur well beyond the width of the streamer belt. These associations suggest that the slow wind originates from coronal regions where magnetic loops are adjacent to open magnetic fields and tend to develop elevated magnetic shears perhaps prone to the occurrence of magnetic reconnection. That was further supported by \citet{Baker2009} who observed in \textit{Hinode-EIS} data ubiquitous outflows along QSLs that form above active regions. \\

\begin{figure*}[]
\centering
\includegraphics[width=0.7\textwidth]{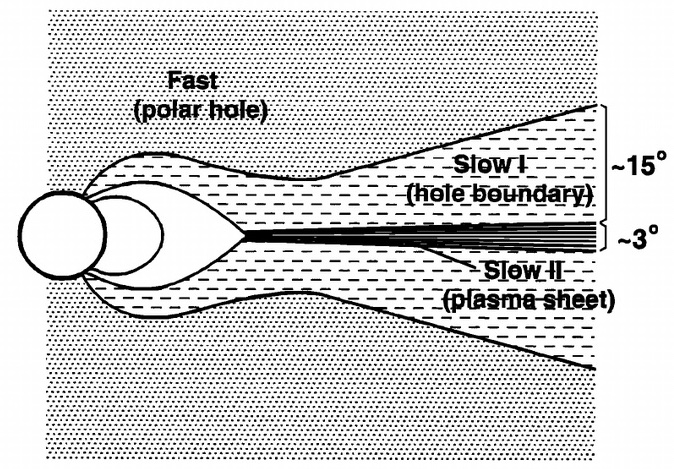}
\caption{Schematic of the distribution of fast and slow solar winds in the meridional plane. Figure taken from \citet[][Figure 12]{Wang2000}.
\label{fig:Wang2000_fig12}}
\end{figure*}

It is likely that the heliosphere is filled up with dense plasma to some degrees away from the HCS as illustrated in Figure \ref{fig:Wang2000_fig12}. This interpretation complements the basic picture portrayed at the beginning of this chapter of a slow wind that comes from the tip of bipolar streamers and that form the dense HPS. It is also coherent with a HPS that represents only the core and brightest portion of coronal observations in white-light from coronagraphs or during total solar eclipses, the rest of the emissions being due to slightly less dense plasmas that probably come from open-closed interfaces.

%% file: chapters/Introduction_FR.tex
\chapter{Introduction (français)}
\label{cha:intro_FR}

\minitoc

Dès la seconde moitié du XIXe siècle, on soupçonnait les aurores boréales d'être induites par des flux occasionnels de particules solaires heurtant l'atmosphère terrestre. Vers le milieu du vingtième siècle, l'astronome allemand Ludwig Biermann \citep{Biermann1951} a conclu, à partir d'observations de queues de comètes, que les particules solaires interagissant avec les comètes faisaient probablement partie d'un flux continu (et non sporadique) de particules expulsées de l'atmosphère solaire. Sydney Chapman, astronome et géophysicien britannique, soutient alors que l'atmosphère solaire pourrait s'étendre bien au-delà de la couronne, probablement jusqu'à au moins l'orbite de la Terre \citep{Chapman1957}. Ce n'est qu'en 1958 que \citet{Parker1958} montre théoriquement que le Soleil peut produire un flux supersonique de particules chargées qu'il appelle alors le vent solaire. Cette découverte est rapidement validée par les premières missions spatiales (\textit{Luna-1,2,3}: voir e.g. \citet{Gringauz1961}, \textit{Venera-1}: \citet{Gringauz1964}, et \textit{Mariner-2}: \citet{Neugebauer1962}) et pose les bases d'une demi-décennie de recherches en physique solaire. \\

\begin{figure*}[]
\centering
\includegraphics[width=0.85\textwidth]{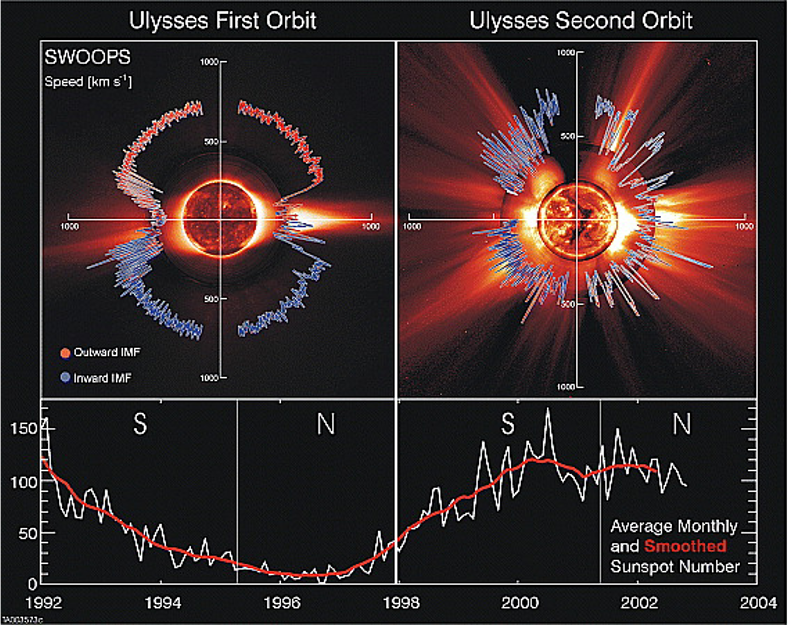}
\caption{ La vitesse du vent solaire telle que mesurée par le satellite \textit{Ulysses} lors de son premier (panneau de gauche) et deuxième (panneau de droite) passage. Les grandes structures de la couronne solaire sont illustrées par des images composites assemblées à partir: d'images du disque solaire prises par le télescope \textit{SoHO-EIT} dans l'ultraviolet extrême à $195$\si{\angstrom}, d'images de la couronne interne par le K-coronamètre de Mauna Loa ($7000$-$9500$\si{\angstrom}), et d'images dans le domaine visible provenant du coronagraphe \textit{SoHO} \textit{LASCO-C2}. Le panneau inférieur montre l'évolution du nombre de taches solaires qui est un bon indicateur du niveau d'activité solaire. Figure adaptée de \citet[][figure 1]{McComas2003}.
\label{fig:Ulysses_FR}}
\end{figure*}

\begin{figure*}[]
\centering
\includegraphics[width=0.95\textwidth]{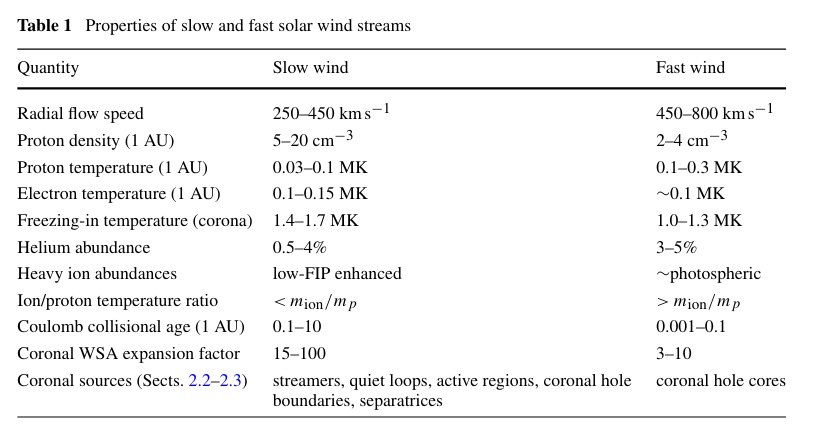}
\caption{Propriétés du vent solaire lent et rapide. Figure prise de \citet[][Table 1]{Cranmer2017}.
\label{tab:Cranmer2017_tab1_FR}}
\end{figure*}

À la fin du XXe siècle, la mission \textit{Ulysses} développée conjointement par l'ESA et la NASA a été la première à scruter le vent solaire hors du plan de l'écliptique solaire (jusqu'à près de $80^\circ$) en recueillant des mesures in situ de la couronne solaire et de l'héliosphère aux hautes latitudes. La première orbite de \textit{Ulysses} proche du minimum solaire a révélé un vent solaire de nature bimodale avec un vent lent présent principalement aux basses latitudes et un vent solaire rapide prédominant dans les régions polaires (voir le panneau de gauche de la figure \ref{fig:Ulysses_FR}) \citep{McComas2003}. En effet, le vent solaire a été classé en deux grands régimes, lent ou rapide selon que sa vitesse moyenne est inférieure ou supérieure à $450\ \rm{km/s}$. Une revue substantielle des principales différences entre les vents solaires lents et rapides a été réalisée par \citet{Cranmer2017} pour laquelle un résumé de leurs propriétés est donné dans la Table \ref{tab:Cranmer2017_tab1_FR}. Une étude récente de \citet{Sanchez-Diaz2016} a même identifié un régime de vent solaire très lent avec des vitesses moyennes typiques inférieures à $300\ \rm{km/s}$.

\section{Considérations générales sur l'origine énigmatique du vent lent}

S'il existe désormais un consensus sur la région source du vent solaire rapide, l'origine du vent lent reste très débattue. Les missions \textit{Parker Solar Probe} \citep[PSP: ][]{Fox2016} et \textit{Solar Orbiter} \citep[SolO: ][]{Muller2013,Muller2020} lancées respectivement en août 2018 et février 2020, ont été spécifiquement conçues pour répondre à cette question. En mesurant le vent solaire in situ et en imageant l'atmosphère solaire plus près du Soleil que jamais atteint auparavant, ces missions fournissent une pléthore de nouvelles informations sur les vents solaires naissants et de leurs origines possibles. \\

Le vent solaire est principalement constitué d'Hydrogen ionisé (protons) à $\approx 95\%$ et d'Helium doublement ionisé (appelées particules alpha) à $\approx 4\%$ où les $\approx 1\%$ restants incluent une myriade d'ions mineurs plus lourds \citep{Rouillard2021}. En plus de la distribution bimodale de la vitesse du vent solaire illustrée par les mesures de \textit{Ulysses}, des données supplémentaires in situ provenant du \textit{Solar Wind Ion Composition Spectrometer} \citep[SWICS: ][]{Gloeckler1992} à bord d'\textit{Ulysses} et du \textit{Advanced Composition Explorer} (\textit{ACE}) a montré une variabilité significative dans les abondances des particules alpha \citep[voir e. g.][]{Kasper2007,McGregor2011} et des ions mineurs dans le vent solaire \citep{Geiss1995,Steiger1996}, ainsi que dans les états de charge des ions mineurs \citep[voir e.g.][]{Stakhiv2015,Stakhiv2016,Neugebauer2002,Liewer2004}, ce qui est discuté plus en détail dans la section \ref{subsubsec:intro_SSW_composition_FR}. Comme la composition du vent solaire ne change pas entre la couronne solaire et les points de mesures in situ dans l'héliosphère, les différentes compositions des vents rapides et lents ont été reliées à différentes sources locales. En particulier, les abondances ioniques mesurées in situ dans les vents lents et rapides ont été associées à celles mesurées par spectroscopie dans les boucles coronales des régions actives \citep [voir e.g.][]{Ko2002,Brooks2011,Doschek2019} et les trous coronaux respectivement \citep{Feldman1998b}. \\

Deux théories se distinguent pour la formation du vent solaire lent. Les prémisses de l'imagerie en lumière blanche coronale et héliosphérique avec le \textit{Solar and Heliospheric Observatory} \citep[SoHO: ][]{Domingo1995} ont mis en évidence un vent lent très variable \citep{Sheeley1997}, ce qui a été confirmé plus tard par les observations du \textit{Solar-TErrestrial RElations Observatory} \citep[STEREO: ][]{Kaiser2008} \citep[voir e.g.][]{Rouillard2009,Rouillard2011a,Plotnikov2016,Sanchez-Diaz2017a,DeForest2018} et récemment par \textit{PSP} \citep{Rouillard2020a}, et a donc conduit à une myriade de théories dynamiques où le vent lent peut être alimenté sporadiquement par du plasma initialement confiné le long des boucles coronales. Cette théorie explique tout naturellement l'abondance ionique du vent lent. En parallèle, une théorie quasi-stationnaire a été proposée qui unifie le vent lent et le vent rapide comme différentes manifestations du plasma accéléré continûment le long des lignes de champ magnétique ouvertes. Bien que cette théorie quasi-stationnaire peine à réconcilier les mesures de composition du vent lent in situ avec les observations spectroscopiques des boucles, elle a l'avantage de fournir des prédictions théoriques quantitatives pour les propriétés globales du vent solaire rapide et lent. Les théories quasi-stationnaires et dynamiques ont donc leurs propres avantages et inconvénients, qui seront présentés dans les sections \ref{sec:intro_stationnary_FR} et \ref{sec:intro_dynamic_FR} et discutés tout au long de cette thèse. \\

Un enrichissement similaire en certains ions mineurs, ceux qui ont un faible potentiel de première ionisation (FIP) comme le Fer et le Magnésium, mesuré in situ dans le vent lent, a également été observé dans les boucles coronales qui constituent la couronne dite "fermée". Les processus physiques qui enrichissent la couronne en éléments à faible potentiel d'ionisation, appelé "effet FIP", font encore l'objet de débats (voir la section \ref{subsec:intro_FIP_FR}). En plus d'examiner les processus susceptibles d'expulser ce plasma enrichi dans le vent lent (appelé processus d'\emph{expulsion}), nous devons également nous demander comment ces ions mineurs, beaucoup plus lourds que les principaux constituants protoniques, parviennent à s'échapper de la basse atmosphère solaire où ils sont formés pour s'extraire de la couronne (appelé processus d'\emph{extraction}). Concernant les origines possibles du vent solaire lent, la modélisation de ce processus d'extraction (ou comme nous le verrons le fractionnement des ions lourds par leur premier potentiel d'ionisation) peut-elle aider à faire la différence entre les théories quasi-stationnaires et dynamiques du vent lent? \\

La théorie quasi-stationnaire est élégante dans le sens où elle peut reproduire de nombreuses propriétés globales du vent lent et rapide, dont la variation bimodale de la vitesse du vent solaire mais aussi sa température et densité moyennes. À cette fin, le modèle MULTI-VP \citep{PintoRouillard2017} qui est introduit dans la section \ref{subsec:MULTI-VP} est exploité dans la section \ref{sec:dynamics_insitu} pour analyser la structure du vent lent et des streamers d'un point de vue quasi-stationnaire. Une analyse rapprochée des récentes observations de télédétection par la sonde \textit{Parker Solar Probe} confirme le comportement quasi-stationnaire du vent lent à des échelles encore plus petites, qui est discuté dans le chapitre \ref{cha:stationnary}. Dans la section \ref{sec:dynamics_griton2020}, nous montrons que la théorie quasi-stationnaire peut être étendue pour expliquer certaines des caractéristiques intermittentes relevées dans les observations de télédétection en lumière blanche du vent lent. Enfin, nous verrons que cette théorie peut être une clé pour expliquer les abondances mesurées du vent lent en ions lourds, en considérant des processus quasi-stationnaires qui impliquent des processus de diffusion/collision qui sont introduits dans la section \ref{subsec:intro_FIP_FR} et analysés en détail dans le chapitre \ref{cha:ISAM_results}. \\

Seules les théories dynamiques du vent lent impliquant le phénomène de reconnexion magnétique peuvent permettre le transfert de plasma des boucles coronales vers le vent lent. Elles semblent donc naturellement plus adaptées pour expliquer pourquoi le vent lent présente des abondances similaires à celles observées dans les boucles coronales. Pourtant, il reste très difficile d'identifier précisément les régions sources du vent lent car la reconnexion magnétique avec les boucles coronales peut se produire à différents endroits de l'atmosphère solaire. Comme nous le verrons dans la section \ref{sec:intro_dynamic_FR}, une grande partie du vent lent se forme au dessus des streamers et des grandes boucles coronales proche de ce qu'on appelle la couche de plasma héliosphérique (HPS), mais d'autres sources montrant des compositions typiques d'un vent lent ont aussi été identifiées loin de la HPS \citep{Zurbuchen2007}. La nature intermittente du vent lent observée dans les images en lumière blanche peut s'expliquer en partie par la reconnexion magnétique à l'extrémité des streamers, qui est étudiée plus en détail dans la section \ref{sec:dynamics_tearing} avec un modèle magnéto-hydrodynamique appelé WindPredict-AW \citep{Reville2020a}. Une idée commune est que plus un plasma reste piégé longtemps le long des boucles coronales, plus il est susceptible de subir des processus de fractionnement, tels que les effets de diffusion, la stratification gravitationnelle et les interactions onde-particule, qui peuvent ainsi enrichir l'atmosphère solaire en éléments à faible FIP. Par exemple, une augmentation de l'abondance coronale des ions mineurs ayant un faible FIP a été observée pendant le vieillissement des régions actives à partir des données \textit{Skylab} \citep{Widing2001}. Ces résultats ont été nuancés par les observations récentes de la mission \textit{Hinode} qui suggèrent que le champ magnétique joue un rôle majeur dans l'évolution de l'abondance dans les régions actives où la reconnexion du champ préexistant avec le flux magnétique émergeant de la photosphère peut mélanger les abondances coronales et photosphériques \citep{Baker2015}. Un certain nombre d'études ont abordé les processus susceptibles de séparer les ions mineurs en fonction de leur FIP qui sont présentés dans la section \ref{subsec:intro_FIP_FR}, mais à ce jour, il n'y a pas de consensus sur les rôles relatifs des différents processus invoqués. Afin de progresser vers une évaluation systématique des mécanismes proposés, je présente dans le chapitre \ref{cha:ISAM} un modèle de l'atmosphère solaire que j'ai spécifiquement adapté pour étudier les processus de fractionnement des ions mineurs dans les boucles coronales. Les premières applications de ce modèle sont présentées dans le chapitre \ref{cha:ISAM_results} et peuvent apporter un nouvel éclairage sur les processus physiques qui contrôlent la composition des ions mineurs dans les boucles coronales et par conséquent sur les régions sources du vent lent. \\

Avant de présenter plus en profondeur les théories dynamiques et quasi-stationnaires du vent lent dans les sections \ref{sec:intro_stationnary_FR} et \ref{sec:intro_dynamic_FR} respectivement, je présente dans la section \ref{sec:intro_general_FR} quelques propriétés générales connues de l'atmosphère solaire, du vent solaire lent et de ses régions sources possibles. Je commence par présenter la structure globale de l'atmosphère solaire dans la section \ref{subsec:intro_high_atmosphere_FR}. Les principales caractéristiques du vent lent sont ensuite discutées dans la section \ref{subsec:intro_SSW_FR} où les propriétés globales et les aspects de composition sont considérés. J'aborde ensuite la dynamique de la basse atmosphère solaire dans la section \ref{sec:intro_low_atmosphere_FR} où je présente les ingrédients physiques qui constituent une base pour mieux comprendre le plasma et les transferts d'énergie depuis la chromosphère jusqu'à la couronne solaire.

\section{Observation du vent lent et de ses régions sources potentielles}
\label{sec:intro_general_FR}

\subsection{Structure de l'atmosphère solaire}
\label{subsec:intro_high_atmosphere_FR}

\subsubsection{La couche de courant et de plasma héliosphérique}
\label{subsubsec:intro_HCS_HPS_FR}

\begin{figure*}[]
\centering
\includegraphics[width=0.95\textwidth]{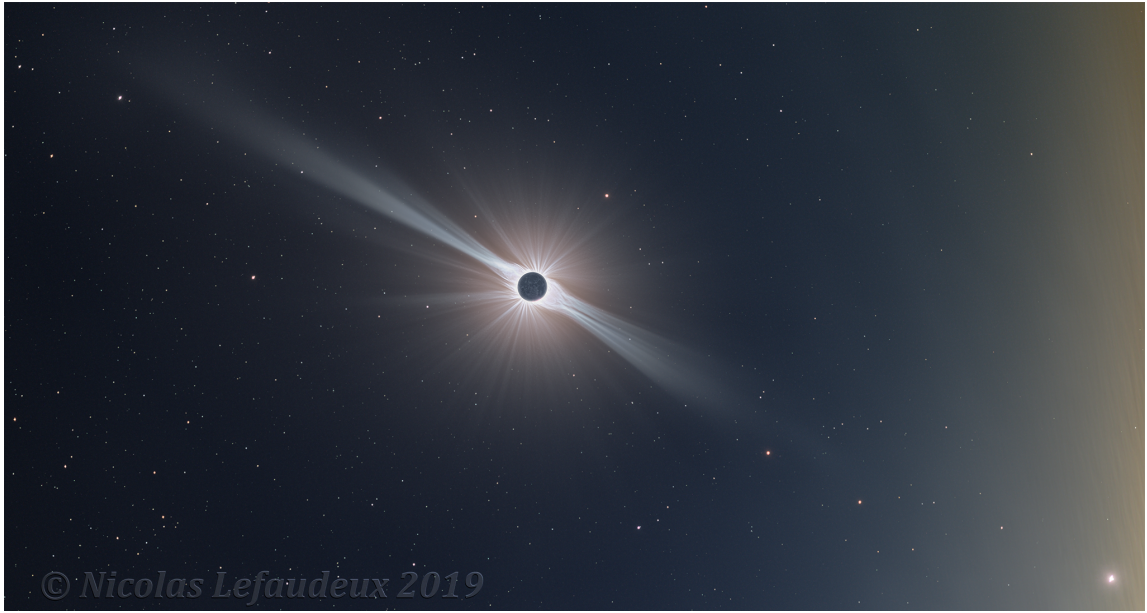}
\caption{Photographie (retravaillée) de la couronne solaire lors de l'éclipse solaire totale de 2019. Crédits: Nicolas Lefaudeux - \url{https://hdr-astrophotography.com}.
\label{fig:eclipse_FR}}
\end{figure*}

Le vent lent n'est pas seulement plus lent, il a aussi tendance à être plus dense, avec une densité de plasma à une unité astronomique (UA) d'environ $5-20\ \rm{cm^{-3}}$, contre $2-4\ \rm{cm^{-3}}$ mesurée dans le vent rapide. Cela signifie qu'il y a globalement plus d'électrons dans le vent lent, ce qui fait qu'il est généralement bien observé dans les images en lumière blanche. En effet, les électrons sont les principaux diffuseurs de la lumière émise depuis la photosphère par le biais du processus connu sous le nom de diffusion de Thomson. L'image coronographique présentée en arrière plan dans le panneau de gauche de la figure \ref{fig:Ulysses_FR} illustre cette relation en montrant que les régions les plus brillantes des images se situent près de l'équateur solaire, où le vent lent a été principalement observé par \textit{Ulysses}. Ces structures brillantes sont appelées streamers et à leur extrémité se forme une tige qui s'étend loin dans la couronne, comme le montre la photographie prise par Nicolas Lefaudeux lors de l'éclipse totale du 2 juillet 2019 (voir figure \ref{fig:eclipse_FR}). Cette image montre à quel point le vent lent, dense et brillant, structure l'héliosphère. Comme nous le verrons en détail tout au long de cette thèse, la couche de plasma héliospérique est logée à l'intérieur de ces régions denses de l'atmosphère solaire. \\

Les missions \textit{Interplanetary Monitoring Platform} (\textit{IMP}) de la NASA ont révélé l'existence d'une division de l'héliosphère en deux hémisphères magnétiques, sous la forme d'une ceinture qui est structurée mais qui s'étend de manière continue en entourant le Soleil \citep{Ness1964,Wilcox1965}. Ces observations ont été confirmées par les mesures in situ effectuées ultérieurement par \textit{Pioneer-11}  \citep{Smith1978} et expliquées par des théories magnéto-hydrodynamiques \citep{Schulz1973}, cette structure a été appelée la couche de courant héliosphérique (HCS). Des analyses approfondies des traversées de la HCS par les sondes héliosphériques \textit{IMP} ont montré une corrélation persistante entre les inversions de champ magnétique mesurées in situ et une densité de plasma accrue \citep{Wilcox1967}. Il a alors été suggéré que la source des HCS mesurées dans l'héliosphère pourrait correspondre aux extrémités des streamers brillants observés dans les coronographes \citep{Howard1974,Hundhausen1977}. \\

Les trois missions spatiales de la série \textit{International Sun-Earth Explorer} conçues conjointement par l'ESA et la NASA ont ensuite effectué des mesures à plus haute résolution temporelle du plasma héliosphérique. Ils ont dévoilé que les inversions du champ magnétique ont lieu à l'intérieur d'une couche de plasma mince et dense appelée couche de plasma héliosphérique \citep{Winterhalter1994}, et suggèrent donc que la HPS et la HCS proviennent probablement de la partie la plus brillante des streamers observés en lumière blanche \citep[voir aussi][]{Guhathakurta1996}. Dans cette thèse, nous étayons cette affirmation en interprétant les nouvelles observations en lumière blanche réalisées à l'intérieur de la couronne solaire par le télescope \textit{Wide-Field Imager for Parker Solar Probe} (\textit{WISPR}) à bord de \textit{PSP}.

\subsubsection{Le Soleil calme et actif}

Les taches solaires, observées à la surface du Soleil comme des régions sombres en raison de leur température locale plus froide, sont observées depuis plus de deux millénaires. Le nombre de taches solaires à la surface du soleil varie considérablement en fonction du niveau d'activité solaire, suivant un cycle solaire bien connu de 11 ans. Lors des périodes de forte activité, le nombre de taches solaires peut atteindre des valeurs supérieures à 100, avec une valeur maximale qui varie d'un cycle à l'autre. Un exemple de l'évolution des taches solaires au cours des cycles solaires 22 et 23 est illustré dans le panneau inférieur de la figure \ref{fig:Ulysses_FR}. \\

\begin{figure*}[]
\centering
\includegraphics[width=0.5\textwidth]{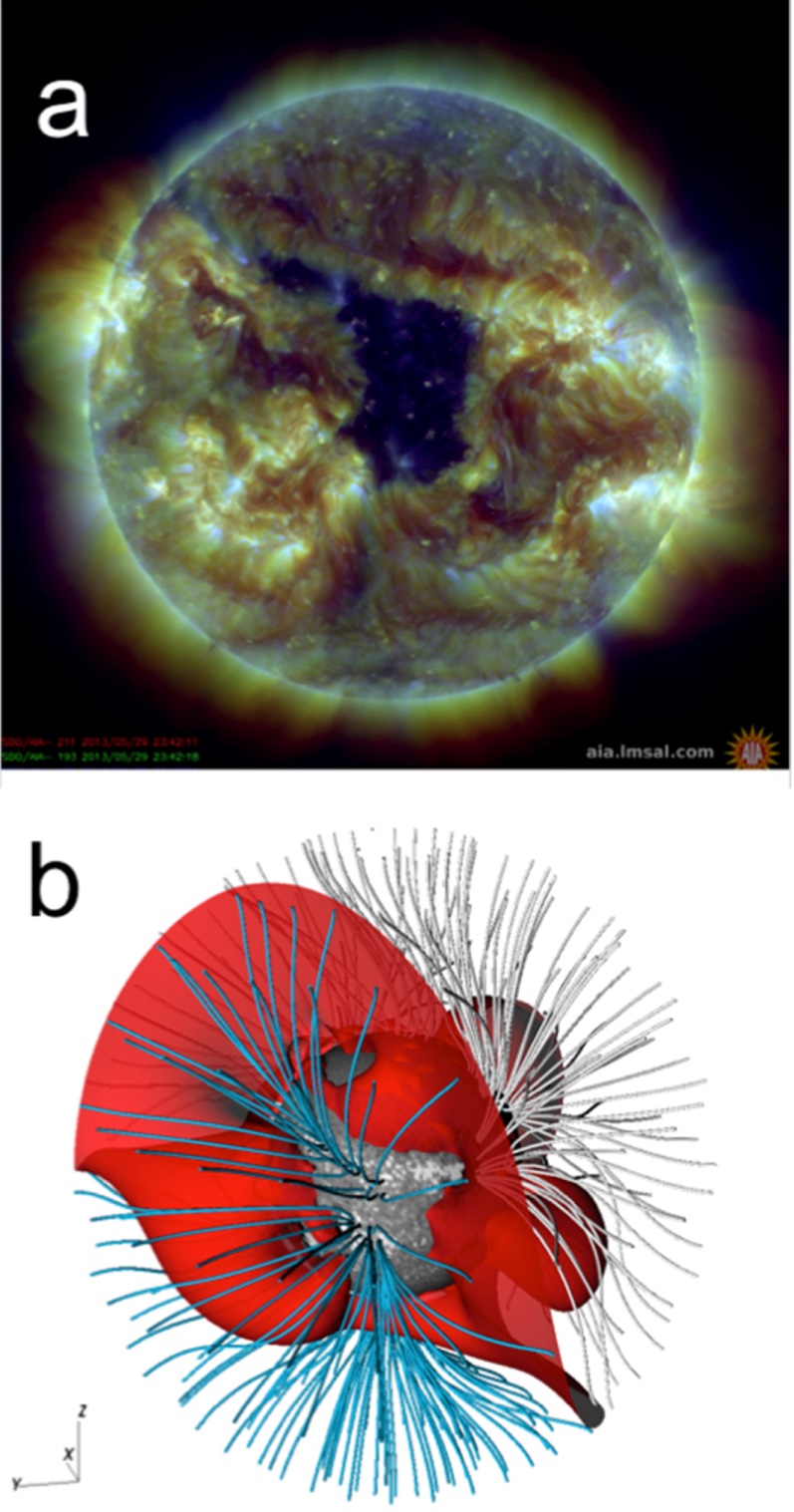}
\caption{ Illustration d'une couronne solaire hautement structurée durant une période de forte activité. Panneau supérieur: images en EUV prises à $193$\si{\angstrom} et $211$\si{\angstrom} par \textit{SDO-AIA} le 29 Mai 2013. Panneau inférieur: une reconstruction à champ potentiel (PFSS, see section \ref{subsec:PFSS}) du champ magnétique coronal pour la rotation Carrington 2137, où les lignes de champ ouvertes sont colorées en bleu et gris suivant leur polarité, et où la HCS est affichée comme une surface rouge. Figure prise de \citet[][Figure 1]{Sanchez-Diaz2017a}.
\label{fig:Sanchez2017a_fig1_FR}}
\end{figure*}

Durant les périodes de faible activité solaire avec seulement quelques taches solaires, la HPS et la HCS restent situées près du plan équatorial. Au cours du second passage de \textit{Ulysses} près du maximum solaire, une configuration beaucoup plus complexe de la couronne solaire peut être observée dans le panneau de droite de la figure \ref{fig:Ulysses}. Cette image reflète une couronne solaire hautement structurée dans les images en lumière blanche ainsi qu'une héliosphère constituée de vents solaires lents et rapides présents à toutes les latitudes \citep{McComas2003}. Dans de tels cas, la HCS et aussi la HPS sont considérablement déformées et peuvent être comparées à la robe d'une ballerine plutôt qu'à un tapis plat. Un exemple d'une telle configuration de haute activité est donné en figure \ref{fig:Sanchez2017a_fig1_FR} pour un cas rare où un très grand trou coronal est présent aux basses latitudes, lequel dévie amplement la HCS vers les régions polaires. \\

La structure du champ magnétique coronal est continuellement reconfigurée à l'échelle globale au cours du cycle solaire, ce qui est couplé à l'émergence de régions actives et de trous coronaux aux basses latitudes. 

Les trous coronaux sont des régions plus froides de la couronne solaire que l'on peut observer dans les images en ultraviolet extrême (EUV) comme des régions plus sombres de l'atmosphère \citep{Waldmeier1981}. Leurs températures plus basses sont induites par les effets du vent solaire, qui en s'échappant de ces régions, entraîne la matière et l'énergie hors du trou \citep{Aschwanden2014}.

Les régions actives sont des régions de l'atmosphère solaire dominées par des boucles coronales qui confinent le plasma et où le champ magnétique photosphérique peut atteindre environ $\approx 100\ \rm{G}$. Les températures coronales élevées ainsi que la forte densité atteinte dans ces boucles les rendent facilement détectables dans les imageurs en ultraviolet extrême comme des structures très brillantes. \\

Si les trous coronaux sont principalement concentrés dans les régions polaires lors du minimum solaire, des trous coronaux isolés mais plus petits peuvent être trouvés errant aux latitudes équatoriales pendant les périodes de forte activité. Inversement, des régions plus actives peuvent être trouvées à des latitudes plus élevées avec l'augmentation de l'activité solaire, qui autrement restent concentrées près de l'équateur dans un rayon de $\approx 20^\circ$ de latitude solaire. Les trous coronaux et les régions actives contribuent tous deux de manière significative à façonner le champ magnétique coronal, et donc plus globalement, la HCS et la HPS.

\subsubsection{Streamers et pseudo-streamers}
\label{subsubsec:intro_streamers_FR}

\begin{figure*}[]
\centering
\includegraphics[width=0.95\textwidth]{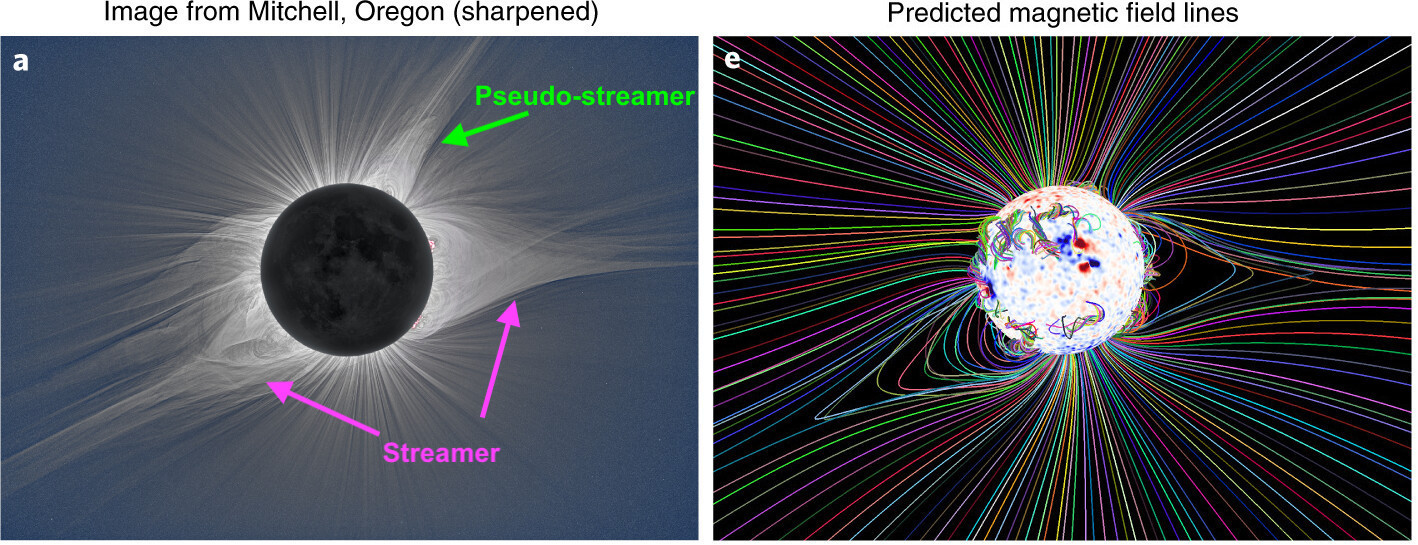}
\caption{Comparaison entre une image réelle de la couronne (gauche) prise lors de l'éclipse solaire totale d'Août 2017, et une reconstruction associée (droite). Figure adapatée de la figure 1 de \citet{Mikic2018}. Crédits pour l'image a: \copyright 2017 Miloslav Druckmüller, Peter Aniol, Shadia Habbal. \label{fig:Mikic2018_fig1_FR}}
\end{figure*}

Comme nous l'avons déjà illustré, d'excellentes conditions d'observation de la couronne solaire sont certainement réunies lors des éclipses totales de soleil, lorsque la Lune cache le disque solaire et dévoile les faibles émissions coronales. Équipés de caméras et de télescopes modernes, les astronomes amateurs peuvent réaliser des images hautement résolues de l'atmosphère solaire. Une photographie traitée de l'éclipse solaire totale du 21 août 2017 est présentée dans la figure \ref{fig:Mikic2018_fig1_FR} et révèle la structure de la couronne solaire avec un haut niveau de détail \citep [voir aussi][]{November1996,Druckmuller2014}. Une modélisation tridimensionnelle (3-D) du champ magnétique de la couronne est présentée à titre de comparaison dans le panneau latéral droit. Elle a été produite par l'équipe \textit{Predictive Science} à l'aide de techniques de modélisation magnéto-hydrodynamique avancées \citep{Mikic2018}. Les "helmet streamer" brillants indiqués par les flèches roses dans le panneau de gauche enferment des régions dominées par des champs magnétiques fermés, tandis que les tiges des streamers sont principalement constituées de champs magnétiques qui sont connectés au champ magnétique interplanétaire. La frontière entre ces deux parties est appelée le point de culminement, où les boucles coronales les plus hautes sont souvent considérablement étirées par l'expansion du vent solaire. \\

\begin{figure*}[]
\centering
\includegraphics[width=0.85\textwidth]{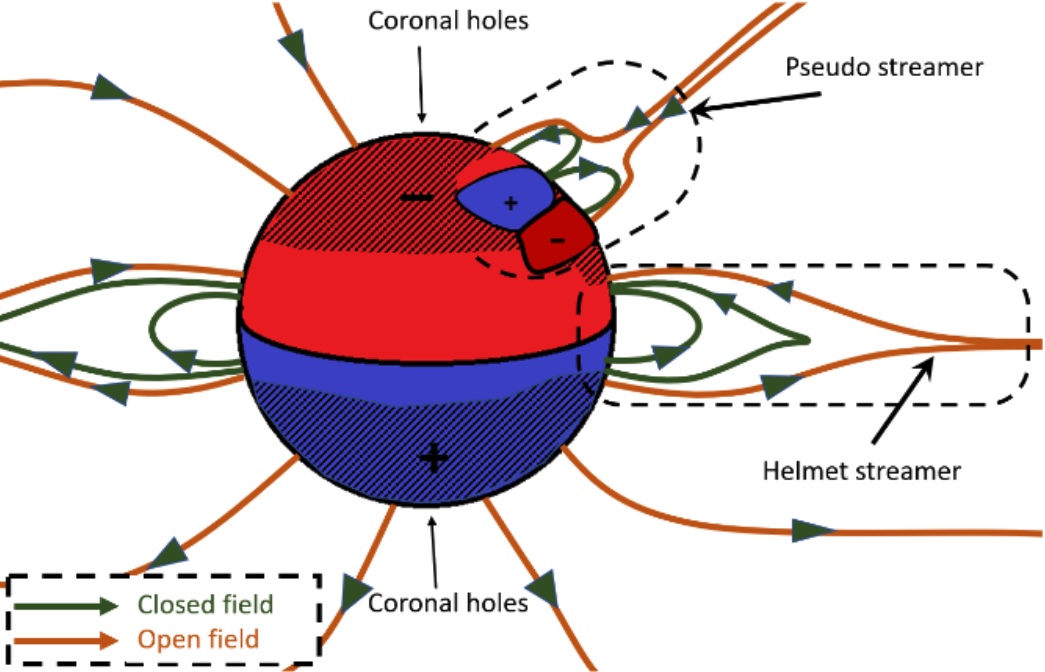}
\caption{ Schéma de la structure du champ magnétique coronal. Figure prise de \citet{Pellegrin2021}, avec permission de l'auteur.
\label{fig:Pellegrin2021_FR}}
\end{figure*}

Les streamers bipolaires sont définis comme des structures coronales brillantes constituées de boucles qui relient des polarités opposées comme le montre la figure \ref{fig:Pellegrin2021_FR}, leur extension dans la couronne est formée par des champs magnétiques ouverts de polarités opposées qui participent à la formation de la HCS et HPS. Le streamer présent dans la portion nord de la figure \ref{fig:Mikic2018_fig1_FR}, bien que plus petit ressemble à un streamer bipolaire, mais il est différent par nature car il se forme dans des régions qui sont unipolaires de chaque coté et donc où les deux côtés du streamer ont la même polarité comme schématisé sur la figure \ref{fig:Pellegrin2021_FR}. Ce type particulier de streamer est appelé pseudo-streamer et son extension dans la couronne et l'héliosphère n'abrite pas de HCS. Tout au long de cette thèse, nous utiliserons simplement le terme "streamer" pour désigner les streamers bipolaires classiques associés à la HCS et HPS. Bien que les pseudo-streamers soient différents des streamers bipolaires, ils produisent clairement un vent solaire plus dense que celui observé dans le vent rapide, ce qui les fait ressortir sur les images d'éclipse (figure \ref{fig:Mikic2018_fig1_FR}). \\

\begin{figure*}[]
\centering
\includegraphics[width=0.8\textwidth]{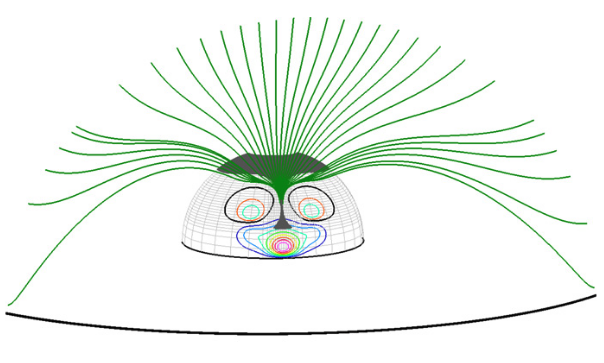}
\caption{Exemple d'un étroit corridor de champ ouvert (lignes vertes) se formant le long d'une extension d'un trou coronal polaire (zone grise). Les contours de l'amplitude du champ magnétique sont tracés en couleur au niveau de la photosphère. Figure prise de \citet[][Figure 4]{Antiochos2011}.
\label{fig:Antiochos2011_fig4_FR}}
\end{figure*}

En pratique, les pseudo-streamers peuvent se former à tout endroit où un fort bipôle isolé émerge dans une région globalement unipolaire de la photosphère, comme le montre la figure \ref{fig:Pellegrin2021_FR}. La formation de trous coronaux isolés ou même d'extensions des trous coronaux polaires déclenche également la formation de pseudo-streamers à grande échelle. Un exemple est schématisé sur la figure \ref{fig:Antiochos2011_fig4_FR} où les champs ouverts d'un trou coronal polaire, représentés par les zones grises au niveau de la photosphère \citep[voir aussi][]{Antiochos2011}, s'étendent jusqu'aux basses latitudes. Les lignes de champ ouvertes (en vert) qui sont associées à l'extension étroite du trou coronal, se connectent à l'héliosphère à une distance angulaire notable de l'HCS, qui ici repose sur le plan équatorial et est tracée comme une ligne continue noire. Cette configuration idéalisée nécessite la présence de deux bipôles de chaque côté de l'extension polaire \citep{Antiochos2011}, qui créent deux lignes d'inversion de polarité supplémentaires qui sont tracées comme des lignes continues noires au niveau de la photosphère. \\

\begin{figure*}[]
\centering
\includegraphics[width=0.95\textwidth]{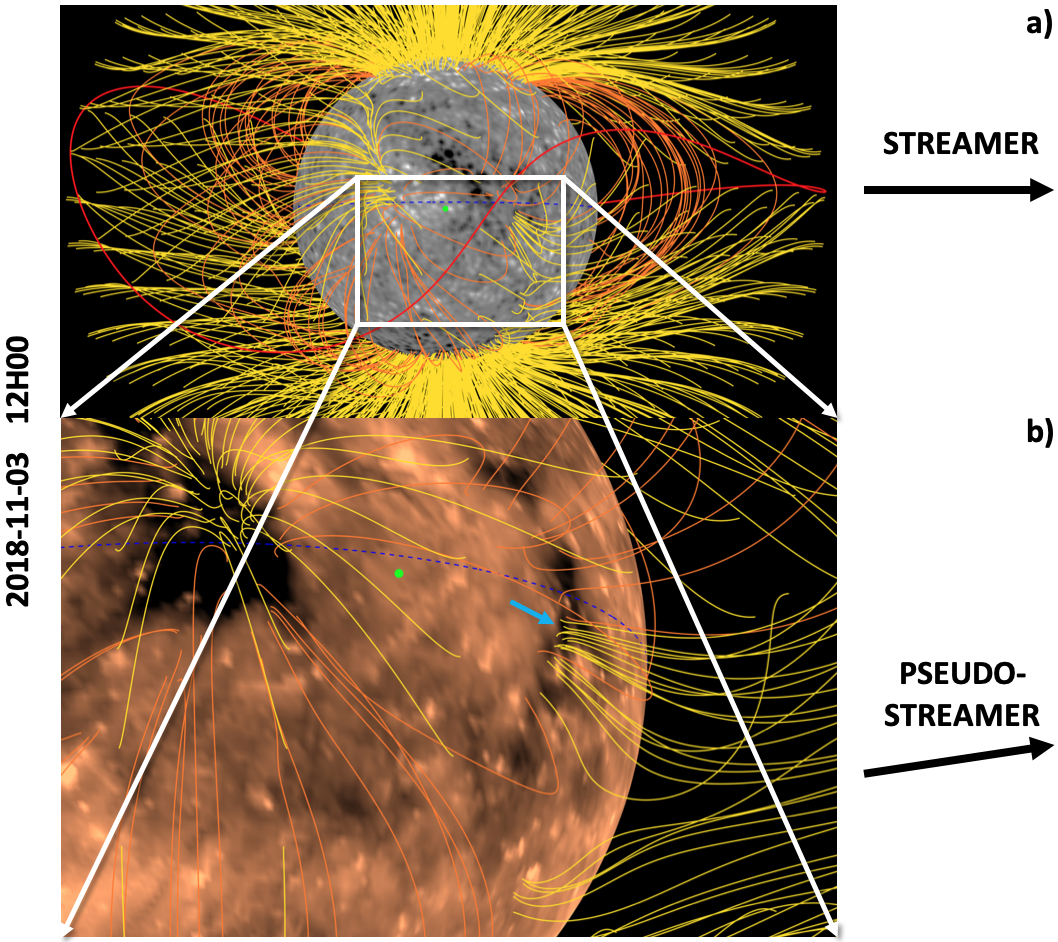}
\caption{ Panneau (a) : une reconstruction 3-D du champ magnétique coronal du 5 Novembre 2018 à l'aide d'une méthode d'extrapolation PFSS (voir section \ref{subsec:PFSS}). La carte magnétique utilisée comme entrée pour l'extrapolation PFSS est affichée en échelle de gris à la surface du Soleil. Les lignes de champ magnétique ouvertes et fermées sont représentées en jaune et en orange respectivement. La ligne d'inversion de polarité (c'est-à-dire la ligne de base de la HCS) est représentée par une ligne rouge à une hauteur de $2.1\ R_\odot$. Panneau (b) : une vue agrandie d'un pseudo-streamer ancré dans un petit trou coronal équatorial isolé (flèche bleue). Est montré également une vue projetée sur la surface d'observations réalisées en ultraviolet extrême dans la longueur d'onde de $193$\si{\angstrom}.
\label{fig:Poirier2020_fig7_FR}}
\end{figure*}

Les trous coronaux isolés bien développés et les extensions des trous coronaux polaires sont en général bien détectés dans les images du disque solaire en ultraviolet extrême, un exemple est présenté en figure \ref{fig:Poirier2020_fig7_FR}. L'extension du trou coronal polaire est souvent, mais pas toujours, le précurseur d'un trou coronal isolé mais plus petit qui se développe aux basses latitudes. La configuration 3-D associée des lignes de champ ouvertes (en jaune) est représentée dans le panneau inférieur de la figure \ref{fig:Poirier2020_fig7_FR} que j'ai produite pour l'article \citet{Poirier2020}. Dans cette étude, le pseudo-streamer issu de ce petit trou coronal équatorial était responsable de l'apparition de rayons brillants supplémentaires dans les images en lumière blanche prises par \textit{WISPR}, ce qui est discuté plus en détail dans la section \ref{sec:stationnary_poirier2020}. Une autre étude de \citet{Griton2021}, à laquelle j'ai contribué, a identifié un petit trou coronal équatorial comme l'une des régions sources du vent solaire lent prélevé par \textit{PSP} lors de son deuxième passage près du Soleil.

\subsubsection{Cartes synoptiques en lumière blanche des streamers}
\label{subsubsec:intro_WLmaps_FR}

Les cartes de Carrington en lumière blanche (WL) (et EUV) ont été largement utilisées tout au long de cette thèse et leur processus de construction sera présenté plus en détail dans le chapitre \ref{cha:stationnary}. En résumé, la rotation du Soleil sur lui-même (à une période de rotation moyenne de 27 jours) est exploitée pour empiler des bandes d'images coronographiques prises au niveau du limbe solaire, pour former une carte synoptique (de type Mercator) de la couronne complète à une altitude fixe. \\

Cette technique a été largement utilisée dès les années 1970 sur les premières images prises par les coronographes spatiaux \emph{CORONASCOPE II} et \emph{SOLWIND} \citep{Bohlin1970,Wang1992} ainsi que par l'observatoire solaire de Mauna Loa à Hawaï \citep{Hansen1976}. Plus tard, les observations synoptiques du coronographe en lumière blanche à bord de \emph{Skylab} ont été exploitées pour modéliser les structures de densité coronale à grande échelle en supposant que les pics de luminosité marquent l'emplacement de la HCS \citep{Guhathakurta1996}. La surveillance continue de la couronne solaire par \textit{SoHO} a ensuite permis une comparaison plus systématique entre la localisation des streamers et la topologie magnétique de la couronne solaire dérivée des calculs PFSS \citep{Wang1998,Wang2000,Wang2007} et des modèles coronaux globaux \citep{Gibson2003,Thernisien2006,dePatoul2015,PintoRouillard2017}. Des techniques de tomographie rotationnelle ont été développées récemment pour corriger les effets de la ligne de visée et convertir les observations WL en cartes de densité synoptiques \citep{Morgan2020}. D'autres techniques ont impliqué la combinaison d'images prises par des coronographes situés à diverses points d'observation (par exemple \emph{SoHO} et \emph{STEREO}) simultanément pour dériver des cartes synchrones circumsolaires des streamers \citep{Sasso2019}. \\

\begin{figure*}[]
\centering
\includegraphics[width=0.95\textwidth]{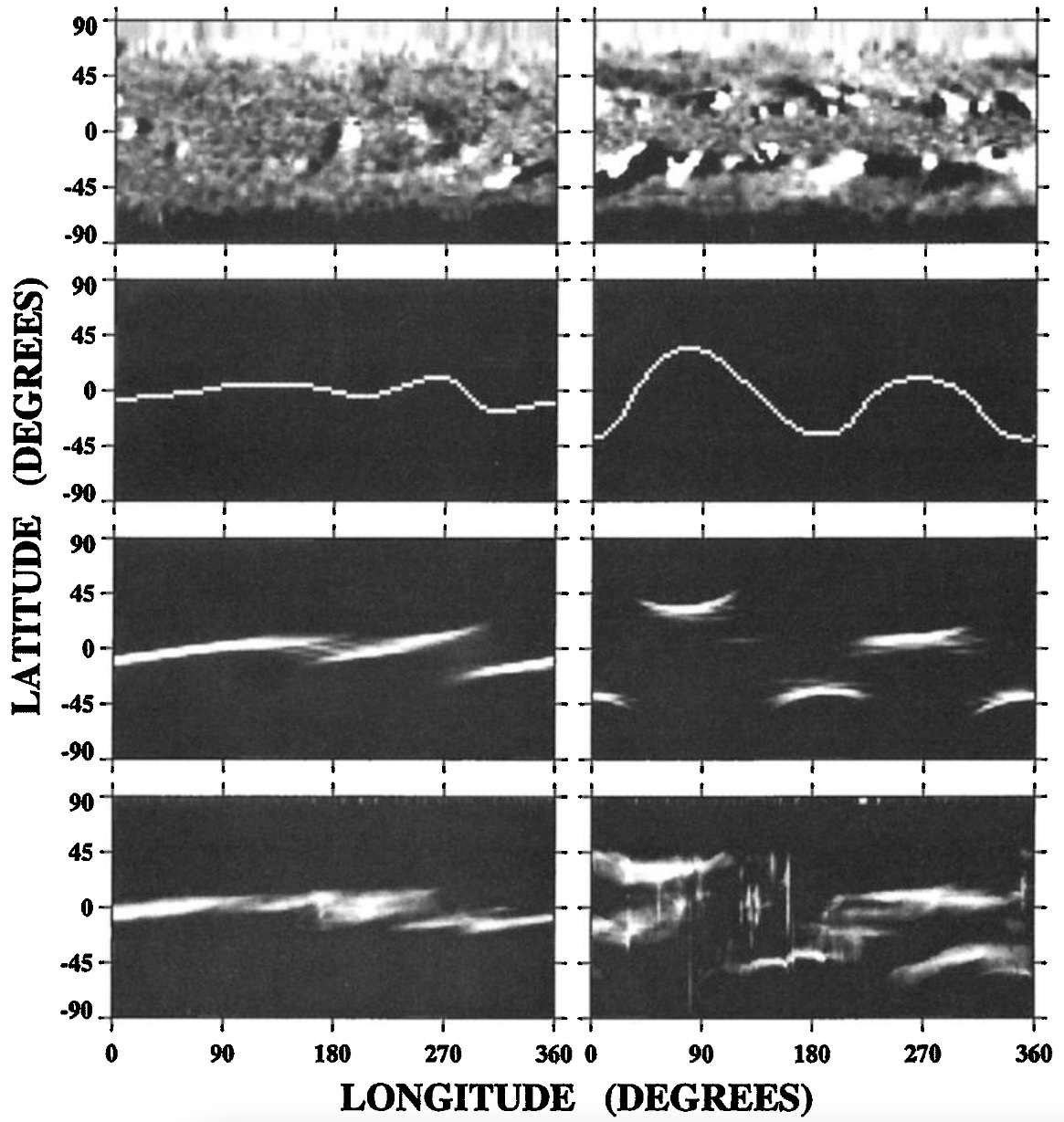}
\caption{ Cartes de Carrington en latitude et en longitude du champ magnétique photosphérique (1ère ligne), de la HCS prédite (2ème ligne) et des émissions de synthèse en WL (3ème ligne), ainsi que des observations réelles en WL réalisées par \textit{SoHO LASCO-C2}, sur la rotation de Carrington CR1919 (gauche) et CR1935 (droite) qui sont typiques d'une configuration coronale minimale (gauche) et maximale (droite) du Soleil. Figure prise de \citet[][Figure 2]{Wang2000}.
\label{fig:Wang2000_fig2_FR}}
\end{figure*}

Un exemple de carte synoptique en lumière blanche tirée de \citet[][Figure 2]{Wang2000} est présenté dans le panneau inférieur de la figure \ref{fig:Wang2000_fig2_FR}. Les tiges des streamers, que nous avons présentées au début de ce chapitre, forment une bande continue d'émission brillante sur la carte synoptique, communément appelée ceinture de streamers. La HPS, également présentée précédemment, se trouve quelque part à l'intérieur de la ceinture de streamers, probablement là où le plasma est le plus dense et diffuse la lumière photosphérique de manière significative. Les cartes synoptiques en lumière blanche sont donc très pratiques pour visualiser la structure globale de la couronne solaire. \\

Pourtant, il reste des portions de la ceinture de streamers qui sont peu visibles ou pas du tout détectées. Il s'agit d'un effet bien compris, inhérent à l'intégration le long de la ligne de visée de la lumière blanche diffusée par la couronne. Les régions où la ceinture de streamers n'est pas bien observée correspondent aux endroits où la ceinture subit des excursions latitudinales. Dans ces régions, la ceinture est inclinée par rapport à la ligne de visée de l'imageur, ce qui signifie qu'une plus petite section de la HPS est capturée par l'imageur. Cet effet est moins prononcé à l'approche du minimum solaire car la HPS reste plus ou moins parallèle au plan équatorial. Et puisque (jusqu'à présent) toutes les images ont été prises depuis l'intérieur du plan écliptique, des portions significatives de la HPS contribuent à sa luminosité. \\

Cette distribution change radicalement proche du maximum solaire, lorsque la ceinture de streamers est fortement déformée et couvre une large gamme de latitudes, comme le montrent les panneaux latéraux de droite de la figure \ref{fig:Wang2000_fig2_FR}. À ce niveau d'activité, la couronne solaire devient très structurée et les streamers apparaissent comme des bandes séparées dans une carte de Carrington plutôt que comme une ceinture continue. Cette structure complexe est typiquement induite par l'émergence de régions actives et la formation de trous coronaux à basse latitude qui forcent la HPS à s'éloigner de l'équateur. Ceci est illustré dans les 1ère et 2ème lignes de la figure \ref{fig:Wang2000_fig2_FR} où une forme attendue de la HCS (et donc de la ceinture de streamers) est calculée à partir d'une reconstruction magnétostatique tridimensionnelle sans courant (i.e. PFSS) de la couronne solaire qui sera présentée plus tard. \\

Certaines excursions brillantes de la ceinture de streamers peuvent s'étendre vers des latitudes plus élevées même au minimum solaire, ces formes en arc lumineux sont typiquement des pseudo-streamers. Mais d'autres formes en arc inversé, appelées "bananes" dans \citet{Gibson2003}, peuvent également être observées dans les cartes de Carrington en lumière blanche à partir de la projection de streamers non-équatoriaux se déplaçant dans les images coronographiques \citep [voir e.g.][]{Wang1992,Wang2000,Wang2007}. Comme nous le verrons dans cette thèse, cet effet de projection supplémentaire est particulièrement fort dans les observatoires qui se situent beaucoup plus près du Soleil, tels que \textit{WISPR} à bord de \textit{PSP} comme discuté dans le chapitre \ref{cha:stationnary}, ainsi que dans \citet{Liewer2019} et \citet{Poirier2020}. \\

Dans cette thèse, nous exploitons les cartes de Carrington en lumière blanche à de multiples occasions, pour contraindre les modèles coronaux et héliosphériques de manière systématique (voir section \ref{sec:WL_opti}), pour suivre l'origine du vent solaire lent mesuré in situ à \textit{PSP} (voir section \ref{sec:dynamics_insitu}) ou pour étudier la structure fine des rayons coronaux observés par télédétection à \emph{PSP} (voir section \ref{sec:stationnary_poirier2020}).

\subsubsection{Spectroscopie de la basse atmosphère solaire}

Le spectromètre en EUV (\textit{EIS}) à bord du vaisseau spatial \textit{Hinode} a fourni une pléthore d'informations précieuses sur les sources potentielles du vent solaire lent, grâce à des diagnostics de composition de la basse atmosphère solaire. \\

\begin{figure*}[]
\centering
\includegraphics[width=0.6\textwidth]{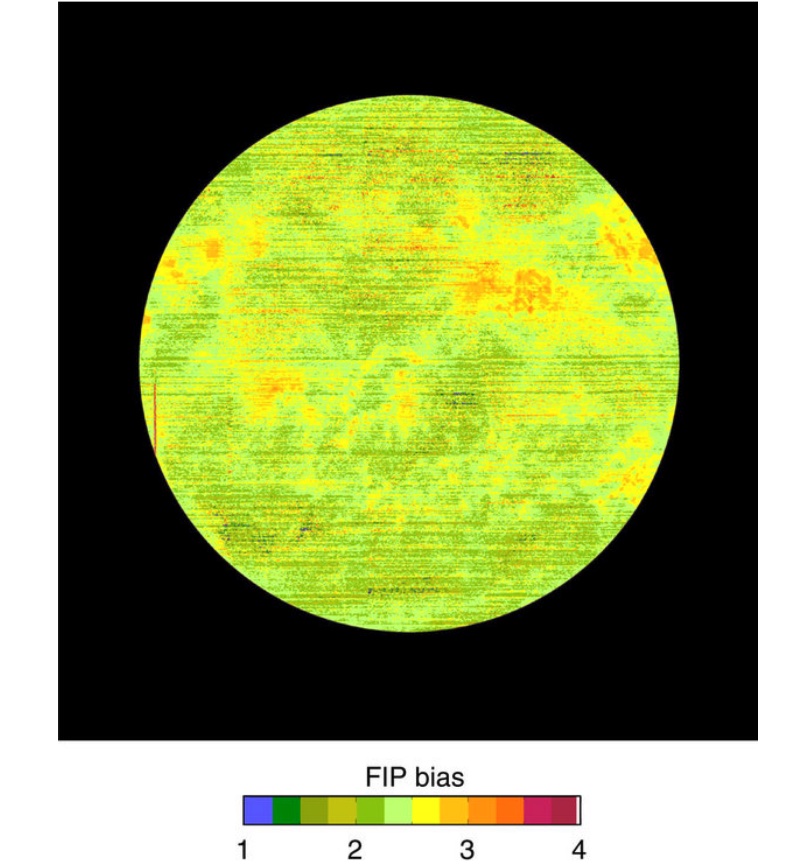}
\caption{Carte 2-D du rapport d'abondance Silicium (Si) sur Soufre (S) du disque solaire. Figure prise de \citet[][Figure 3]{Brooks2015}.
\label{fig:Brooks2015_fig3_FR}}
\end{figure*}

\citet{Brooks2015} ont développé une technique d'inversion pour dériver les abondances relatives du Silicium (Si, FIP faible) par rapport au Soufre (S, FIP intermédiaire) à partir des intensités des raies spectrales du disque solaire observé à distance par \textit{Hinode-EIS}. Leur carte bidimensionnelle du rapport Si/S du disque solaire est présentée dans la figure \ref{fig:Brooks2015_fig3_FR} et est typique d'une activité solaire plutôt élevée avec plusieurs régions actives réparties à la surface. Ils ont identifié les bords de certaines de ces régions actives comme sources potentielles d'un vent solaire lent qui a été mesuré in situ à près de 1 UA par \textit{ACE}. \\

Plus généralement, les signatures spectroscopiques des régions actives montrent un enrichissement quasi systématique en éléments à faible FIP et soutiennent donc l'hypothèse d'un vent solaire lent qui est partiellement constitué de plasma provenant de champs fermés des régions actives \citep{Meyer1985,Neugebauer2002,Liewer2004,Brooks2011,Doschek2019}. Un état de charge plus élevé est également mesuré dans les vents solaires provenant des régions actives plutôt que dans ceux provenant des trous coronaux, ce qui s'explique par une température plus élevée dans la couronne qui stimule l'ionisation. La suite complète de télédétection de la sonde \textit{SoHO} a permis pour la première fois de relier les diagnostics spectroscopiques aux signatures en lumière blanche de la couronne solaire. Un lien direct a été établi entre les boucles coronales ancrées dans les régions actives et enrichies en éléments à faible FIP, et l'écoulement de vent solaire lent associé, observé dans le coronographe en lumière blanche \citep{Ko2002,Uzzo2004}. En revanche, un tel enrichissement n'a pas été trouvé (ou dans une bien moindre mesure) au-dessus des trous coronaux \citep{Feldman1998b,Stansby2020a}.

\subsection{Propriétés du vent solaire lent} 
\label{subsec:intro_SSW_FR}

\subsubsection{La nature intermittente du vent lent}
\label{subsubsec:intro_SSW_intermittency_FR}

Le coronagraphe \textit{LASCO} à bord de \textit{SOHO} \citep{Brueckner1995} a marqué une révolution dans les images en lumière blanche des streamers et du vent solaire lent, dévoilant leur structure fine et leur dynamisme. \\

\begin{figure*}[]
\centering
\includegraphics[width=0.6\textwidth]{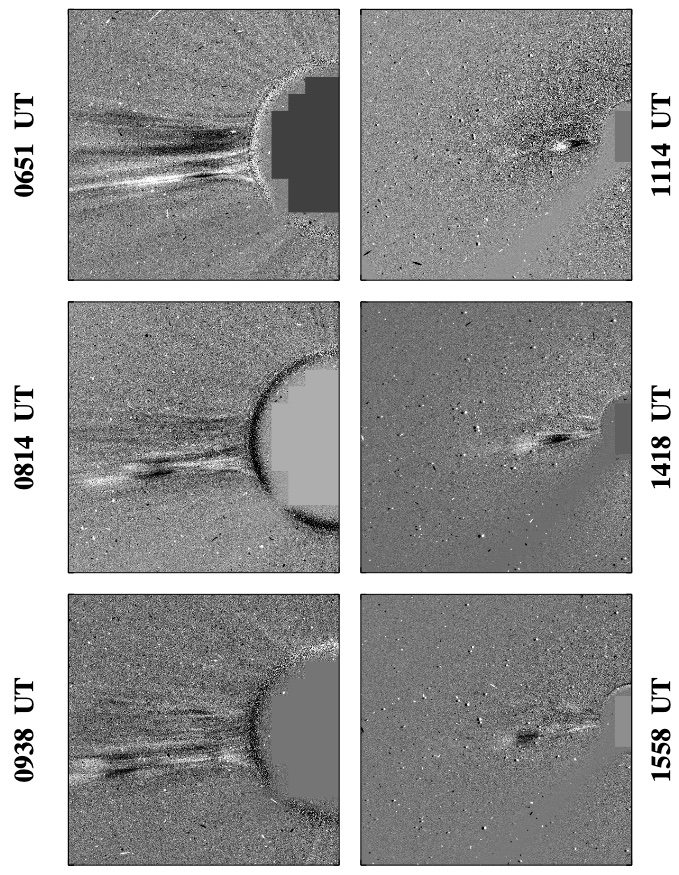}
\caption{ Exemple de blobs de streamers observés par \textit{SoHO-LASCO}, en utilisant une technique de "running difference" (voir texte) pour mettre en valeur les fluctuations de densité dans la couronne où les régions brillantes (ou sombres) indiquent respectivement une densité accrue (ou plus faible). Figure tirée de \citet[][Figure 1]{Sheeley1997}.
\label{fig:Sheeley1997_fig1_FR}}
\end{figure*}

En exploitant ces nouvelles données, \citet{Sheeley1997} ont découvert une pléthore de perturbations de densité à petites échelles qu'ils ont appelées "blobs" de streamers et qui sont montrés dans la figure \ref{fig:Sheeley1997_fig1_FR}. Les faibles fluctuations de luminosité induites par le mouvement des "streamer blobs" ont été mises en évidence à l'aide de la technique "running-difference" qui consiste à soustraire des images successives. Par convention, les régions claires et sombres de ces images indiquent respectivement une augmentation ou une diminution de la luminosité et probablement de la densité locale d'électrons dans l'atmosphère solaire. Ces "blobs" de streamers ont tendance à apparaître à l'extrémité des streamers et à se propager le long de leur tige pour se déplacer de concert avec le vent solaire lent. \\

Des périodicités dans le vent solaire ambiant qui frappe la magnétosphère terrestre ont été notées dans une étude ultérieure de \citet{Kepko2002}. Le suivi des streamers blobs dans les observations de télédétection et in situ de la mission \textit{STEREO} a révélé des structures de densité se propageant à des périodicités variant de $\approx 90-180\rm{min}$ à $\approx 8-16\rm{hr}$ \citep{Viall2010,Viall2015,Kepko2016,Sanchez-Diaz2017a}, qui ont aussi été retrouvées dans les observations récentes de \textit{PSP} \citep{Rouillard2020a}. Certains streamers blobs ont également été associés au passage de régions d'interaction corotatives \citep[CIRs: ][]{Pizzo1978} où des vents rapides rattrapent des vents lents génèrant ainsi une compression du plasma du vent solaire \citep[voir e.g.][]{Rouillard2009,Sheeley2010,Plotnikov2016}. \\

\begin{figure*}[]
\centering
\includegraphics[width=0.65\textwidth]{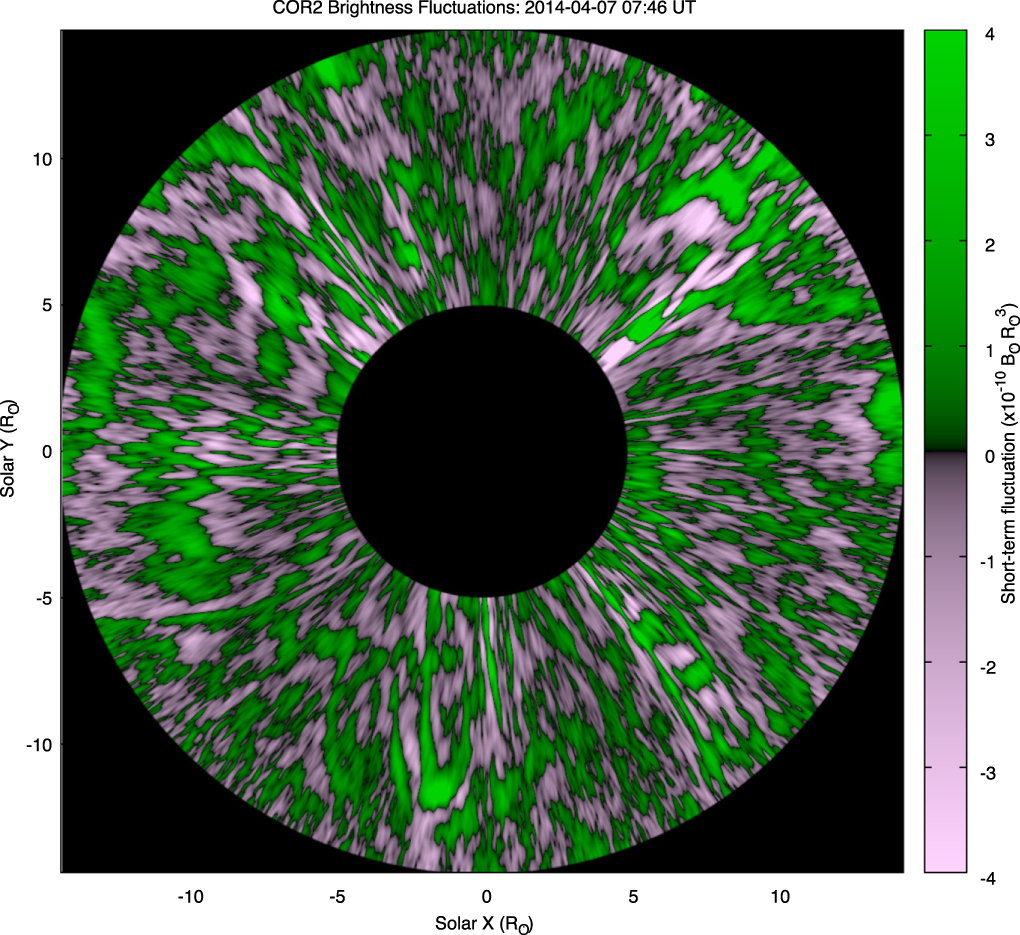}
\caption{ Fluctuations de luminosité dans la haute couronne solaire détectées lors d'une campagne à haute cadence du coronographe \textit{STEREO-A COR2}. Figure tirée de \citet[][Figure 12]{DeForest2018}.
\label{fig:DeForest2018_fig12_FR}}
\end{figure*}

Une récente campagne à haute cadence (en champ profond) menée par le coronographe \textit{STEREO-A COR2} a considérablement amélioré les observations et révélé l'omniprésence des structures de densité se propageant dans le vent lent \citep{DeForest2018}. En utilisant des techniques de traitement sophistiquées, ces auteurs ont dévoilé une couronne solaire hautement structurée avec la libération omniprésente de perturbations de densité de différentes tailles (voir Figure \ref{fig:DeForest2018_fig12_FR}), dont les plus grandes sont probablement les "streamer blobs" de \citet{Sheeley1997}. En raison des effets de ligne de visée, il n'est pas clair si toutes les fluctuations de densité observées par \citet{DeForest2018} se propagent à l'intérieur de la ceinture de streamers ou si certaines des fluctuations de densité sont libérées à partir d'une région plus large de la couronne. \citet{Griton2020} ont montré par exemple que certaines de ces fluctuations de densité pouvaient être induites dans la couronne solaire, à la base des tubes de flux ouverts qui canalisent le vent solaire, et à la suite d'événements de reconnexion brefs associés à des points lumineux dans la couronne. Ces derniers sont observés comme des émissions accrues en ultraviolet extrême et en rayons X \citep{Madjarska2019} qui ont été interprétées comme des boucles magnétiques chaudes qui se forment lorsque le champ magnétique émerge, intéragit et se reconnecte avec le champ magnétique coronal préexistant \citep{Kwon2012}. Ma contribution à l'étude de \citet{Griton2020} sera discutée dans la section \ref{sec:dynamics_griton2020}. \\

\begin{figure*}[]
\centering
\includegraphics[width=0.8\textwidth]{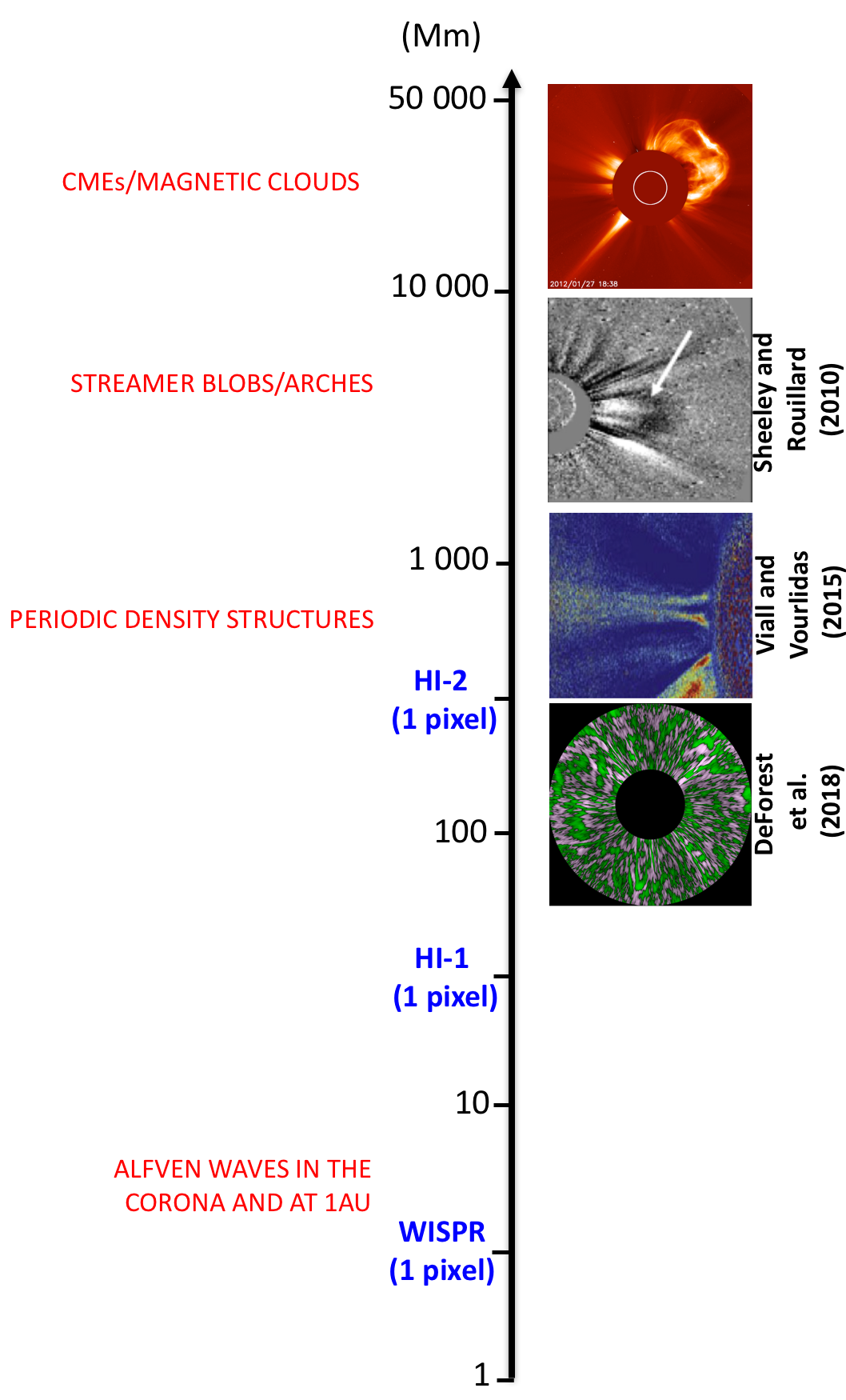}
\caption{Aperçu des structures dynamiques dans le vent solaire lent, depuis les grandes (en haut) jusqu'aux petites (en bas) échelles spatiales, avec leurs références associées qui ont été mentionnées dans le texte.
\label{fig:transient_scales_FR}}
\end{figure*}

Il est maintenant largement admis que le vent solaire lent est hautement dynamique par nature. La figure \ref{fig:transient_scales_FR} résume la variabilité du vent solaire lent à diverses échelles spatiales, où plusieurs des références mentionnées ci-dessus sont représentées. Aux plus grandes échelles, des déflexions significatives des streamers ont été observées lors de l'éruption d'éjections de masse coronale et du passage de leur choc associé qui se propage en amont \citep [voir e.g.][]{Kouloumvakos2020a,Kouloumvakos2020b}. Jusqu'à présent, les plus petites structures périodiques de densité qui ont été dévoilées à partir d'observations de télédétection sont celles de \citet{DeForest2018} au cours d'une campagne \textit{STEREO-A COR2} à haute cadence. La distance rapprochée sans précédent de \textit{PSP-WISPR} au Soleil est susceptible d'étendre cette frise à des échelles encore plus petites, comme nous l'illustrerons tout au long de cette thèse.

\subsubsection{Composition du vent lent}
\label{subsubsec:intro_SSW_composition_FR}

D'autres indices sur l'origine des différents vents solaires peuvent être obtenus à partir des diagnostics de composition du vent solaire dérivés depuis l'abondance de leurs constituants mineurs tels que les ions lourds. Ceci est rendu possible par le fait que l'abondance et les états de charge des ions lourds mesurés dans le vent solaire sont régulés dans les basses couches de la couronne solaire et deviennent ensuite invariants dans les parties supérieures de la couronne \citep{Geiss1982}.  \\

L'instrument \textit{Ulysses-SWICS} a fourni de précieux diagnostics de la composition du vent solaire à toutes les latitudes. Ces mesures ont montré que l'abondance de certains éléments est non seulement différente dans les vents rapides et lents mais qu'un élément particulier peut avoir des niveaux d'ionisation différents \citep{Steiger1996}.

\begin{figure*}[]
\centering
\includegraphics[width=0.95\textwidth]{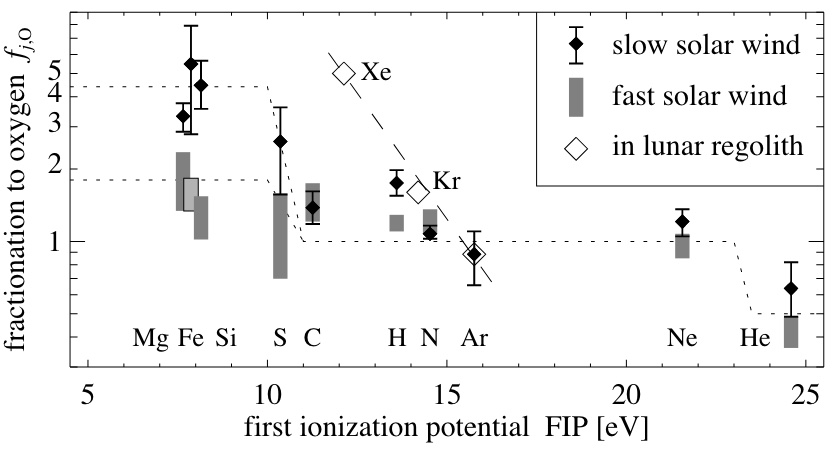}
\caption{Variation des abondances des principaux constituants mesurés dans le vent solaire en fonction du FIP. Les abondances sont relatives à l'abondance de l'Oxygène. Figure tirée de \citet[][Figure 1]{Peter1998b}.
\label{fig:Peter1998_fig1_FR}}
\end{figure*}

La figure \ref{fig:Peter1998_fig1_FR} illustre d'abord la dichotomie de l'abondance élémentaire entre les vents solaires lents et rapides, les premiers étant enrichis d'un facteur $\approx 3-5$ en ions lourds qui ont un faible potentiel de première ionisation (FIP, c'est-à-dire l'énergie requise pour ioniser le premier électron) tels que le Magnésium (Mg), le Fer (Fe) et le Silicium (Si). Les ions lourds dont le potentiel d'ionisation est supérieur à $11 \rm{eV}$ ne présentent pas d'écarts significatifs par rapport aux abondances photosphériques mesurées plus bas dans l'atmosphère solaire, à l'exception de l'Hélium (He) qui a tendance à être appauvri dans le vent solaire, et plus particulièrement dans le vent solaire rapide où un facteur d'environ 1/2 est mesuré. 

\begin{figure*}[]
\centering
\includegraphics[width=0.95\textwidth]{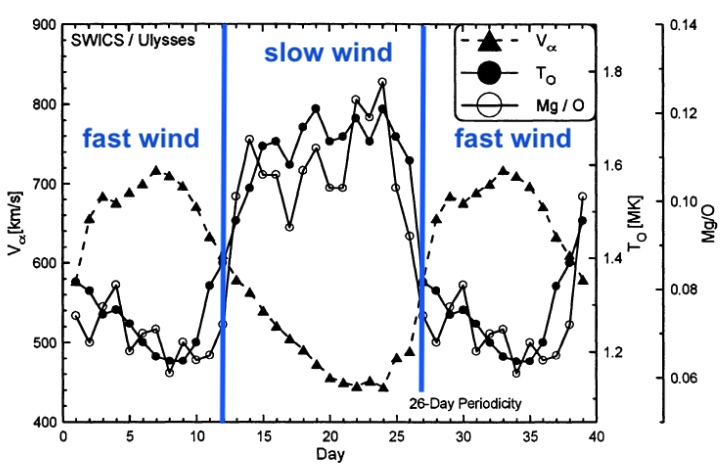}
\caption{Variation temporelle de l'abondance en Magnésium (Mg, bas FIP) par rapport à l'Oxygène (O), de la vitesse du vent solaire (prise ici comme la vitesse de l'Hélium doublement ionisé), et de la température de l'Oxygène. Figure tirée de \citet[][Figure 7]{Geiss1995}.
\label{fig:Geiss1995_fig7_FR}}
\end{figure*}

La figure \ref{fig:Geiss1995_fig7_FR} complète cette image avec une abondance d'éléments à faible FIP (ici Mg) qui est inversement (ou directement) corrélée avec la vitesse (ou la température) du vent solaire. Une propriété supplémentaire et essentielle observée est que le fractionnement semble s'effectuer indépendamment de la masse des particules, qui peuvent cependant être beaucoup plus lourdes que les principaux constituants protoniques. Nous verrons dans la section \ref{subsec:intro_FIP_FR} que ce fait observationnel peut avoir des implications directes sur les processus physiques possibles qui peuvent contrôler l'extraction des ions lourds dans la couronne. \\

\begin{figure*}[]
\centering
\includegraphics[width=0.8\textwidth]{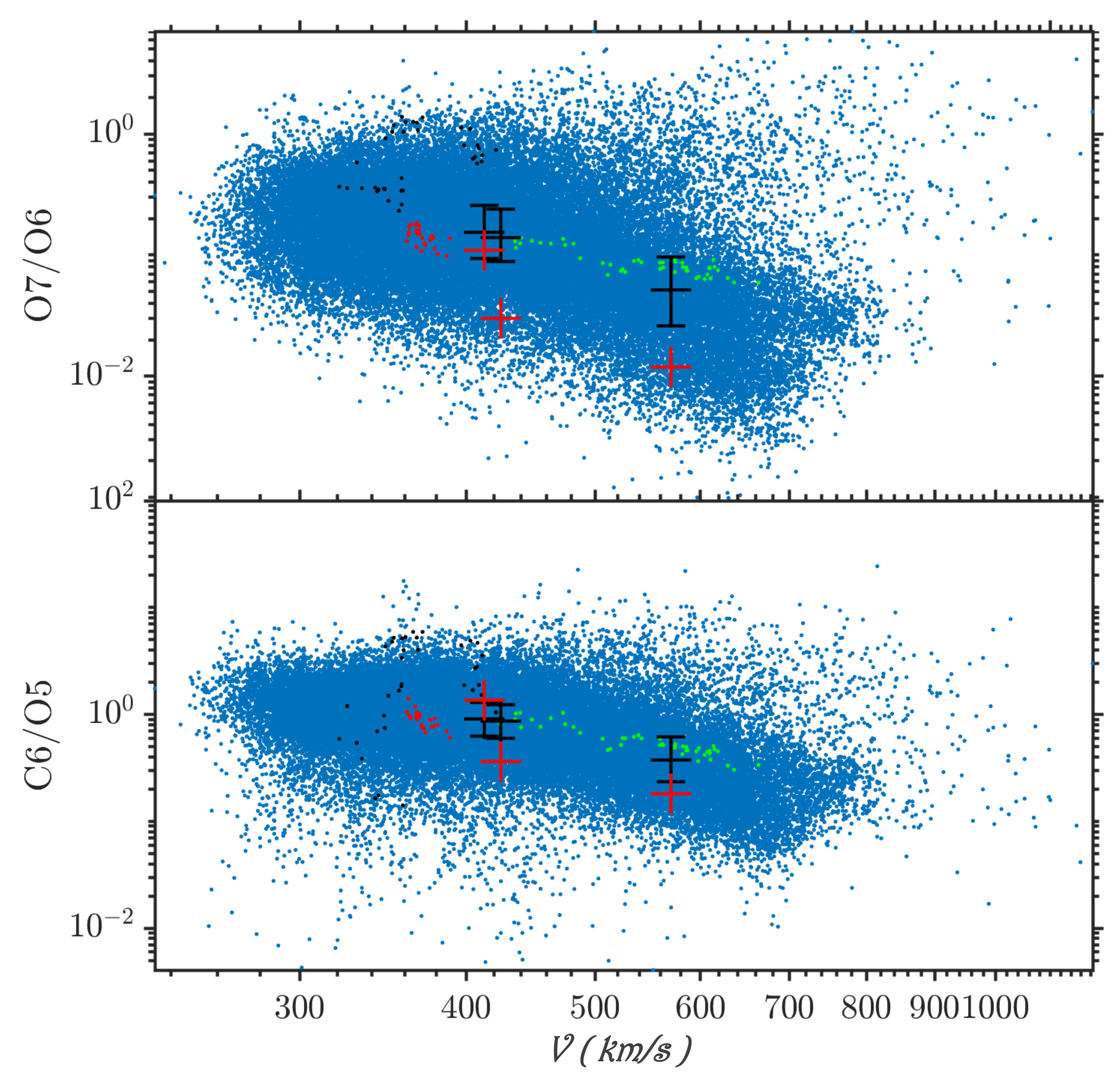}
\caption{Mesures in situ prises par \textit{ACE} du rapport d'état de charge de l'Oxygène $O^{7+}/O^{6+}$ et du Carbone $C^{6+}/C^{5+}$ en fonction de la vitesse du vent solaire (points bleus). Les mesures correspondant à des vents rapides, lents Alfvéniques et lents non Alfvéniques identifiés par \citet{DAmicis2015} sont indiquées par les points verts, rouges et noirs respectivement. Des simulations de l'état de charge pour ces trois régimes de vents, obtenues à partir du modèle d'atmosphère solaire de l'IRAP (ISAM), sont montrées par les croix rouges avec leurs incertitues associées en noir \citep[voir ][pour plus de détails]{Lavarra2022}. Figure prise de \citet[][Figure 1]{Lavarra2022}.
\label{fig:Lavarra2022_fig1_FR}}
\end{figure*}

En outre, les diagnostics de composition du vent solaire réalisés in situ peuvent également fournir des informations sur l'état de charge (ou niveau d'ionisation) d'un élément. L'état de charge d'un élément peut être calculé en comparant les densités de ses différents ions, à savoir les rapports d'état de charge. Les rapports d'état de charge sont généralement très variables dans le vent lent et le vent rapide, et l'état de charge a tendance à être plus élevé dans le vent solaire lent (ou, comme nous le verrons, dans une fraction de ce vent) par rapport au vent rapide (figure \ref{fig:Lavarra2022_fig1_FR}). L'état de charge est intrinsèquement lié aux propriétés de la région source dans la basse atmosphère solaire, à des hauteurs où l'ionisation n'est pas encore complètement établie entre la haute chromosphère et la basse couronne (ce qui est discuté plus loin dans la section \ref{subsec:intro_chromo_FR}). Par exemple, \citet{Neugebauer2002} et \citet{Liewer2004} ont observé des rapports d'états de charge plus élevés dans le vent lent provenant de régions actives chaudes. En revanche, ils ont trouvé des rapports d'état de charge qui ont tendance à être plus faibles dans le vent rapide, et qui sont associés à des régions sources (par exemple, des trous coronaux) qui sont généralement plus froides que les régions actives. \citet{Kepko2016} ont également montré une similitude entre la variabilité des rapports d'état de charge mesurés in situ dans le vent lent et la courte périodicité temporelle des structures quasi-périodiques observées par télédétection en WL dans les streamers. \\

\subsubsection{Variabilité du vent lent}
\label{subsubsec:intro_SSW_variability_FR}

Dans les sections \ref{subsubsec:intro_SSW_intermittency_FR} et \ref{subsubsec:intro_SSW_composition_FR}, nous avons décrit un vent solaire lent qui présente probablement de multiples facettes au-delà de la classification cinétique bimodale simpliste établie auparavent. La diversité des régimes de vent lent observés est révélatrice d'une multitude de candidats possibles de sources de vent lent qui peuvent se différencier significativement les uns des autres. \\ 

Alors que le vent lent est systématiquement enrichi en éléments à faible FIP, une abondance variable en particules alpha (FIP élevé) a été mesurée qui change non seulement sur des échelles de temps qui s'étendent sur plusieurs heures ou jours mais aussi au cours du cycle solaire. Deux sous-catégories distinctes de vents lents ont été identifiées avec des abondances différentes en particules alpha \citep{Kasper2007,McGregor2011}. L'un des vents lents présente un appauvrissement quasi constant en particules alpha qui est similaire à celui mesuré dans le vent rapide qui émerge près du centre des trous coronaux. Tandis que l'autre vent lent a une abondance plus élevée mais variable en particules alpha qui est typique des écoulements provenant des streamers. 

\begin{figure*}[]
\centering
\includegraphics[width=0.7\textwidth]{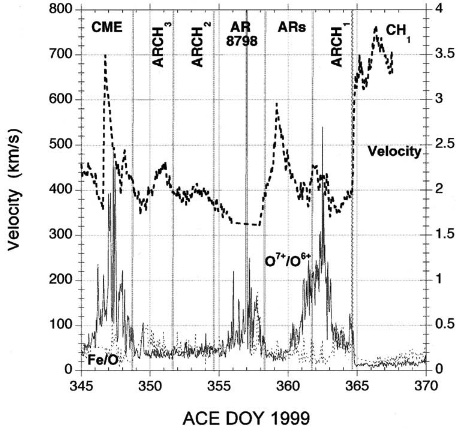}
\caption{Variation temporelle du rapport d'état de charge $O^{7+}/O^{6+}$ et de la vitesse du vent solaire dans des mesures in situ prises par \textit{ACE} depuis 1 UA. Figure prise de \citet[][Figure 6]{Liewer2004}.
\label{fig:Liewer2004_fig6_FR}}
\end{figure*}

Une identification similaire a été faite à partir des mesures des rapports d'état de charge dans le vent solaire lent \citep{Neugebauer2002,Liewer2004,Stakhiv2015,Stakhiv2016}, avec un vent lent de streamers qui présente des rapports de charge élevés typiques des régions actives chaudes, et un vent lent de trous coronaux qui a des rapports d'état de charge plus faibles comme dans le vent rapide. Ceci est illustré dans la figure \ref{fig:Liewer2004_fig6_FR}. 

Des mesures ultérieures effectuées à \textit{PSP} \citep{Rouillard2020a,Griton2021} et à proximité de 1 UA \citep{DAmicis2019} ont examiné en détail les propriétés globales de ces deux états de vent lent. Les mesures du vent lent typique de streamers montraient généralement un vent plus lent, plus dense et plus variable que celui des trous coronaux. Ce dernier est communément appelé le vent solaire lent Alfvénique car il peut héberger des fluctuations Alfvéniques aussi importantes que celles mesurées dans le vent rapide \citep{DAmicis2019}. \\

Ces résultats suggèrent deux états de vent lent qui sont probablement générés par des sources différentes et par des mécanismes différents. Le vent lent des streamers qui est très variable peut être formée par des processus intermittents qui se produisent à l'extrémité des streamers et/ou aux frontières ouvertes-fermées entre les streamers et les trous coronaux, ce qui est discuté plus en détail dans la section \ref{sec:intro_dynamic_FR}. Nous verrons dans la section \ref{sec:intro_stationnary_FR} que le vent lent alfvénique qui est plus stable provient probablement des lignes de champ ouvertes qui sont enracinées plus près du centre des trous coronaux, et que ce vent lent peut être bien décrit avec une théorie quasi-stationnaire \citep[voir aussi][]{PintoRouillard2017,Lavarra2022}.

\clearpage
\section{Principes fondamentaux de la basse atmosphère solaire}
\label{sec:intro_low_atmosphere_FR}

Dans cette section, je présente les ingrédients physiques qui constituent une base pour mieux comprendre la dynamique de la basse atmosphère solaire. Je commence par une brève présentation dans la section \ref{subsec:intro_RT_FR} des différentes couches qui constituent la basse atmosphère solaire. Ensuite, je discute dans la section \ref{subsec:intro_heating_FR} des mécanismes de chauffage qui peuvent contribuer au transfert d'énergie dans la basse atmosphère solaire. De plus, les principales spécificités de la chromosphère, où la plupart des processus d'extraction des ions lourds sont supposés avoir lieu, sont décrites dans la section \ref{subsec:intro_chromo_FR}. Enfin, je présente dans la section \ref{subsec:intro_FIP_FR} une brève revue des processus physiques qui pourraient opérer pour enrichir la couronne avec certains ions lourds, et donc qui pourraient constituer des éléments de construction des théories quasi-stationnaires et dynamiques du vent solaire qui sont discutées plus loin dans la section \ref{sec:intro_stationnary_FR} et \ref{sec:intro_dynamic_FR}.

\subsection{La chromosphere, région de transition et couronne}
\label{subsec:intro_RT_FR}

\begin{figure*}[]
\centering
\includegraphics[width=1.\textwidth]{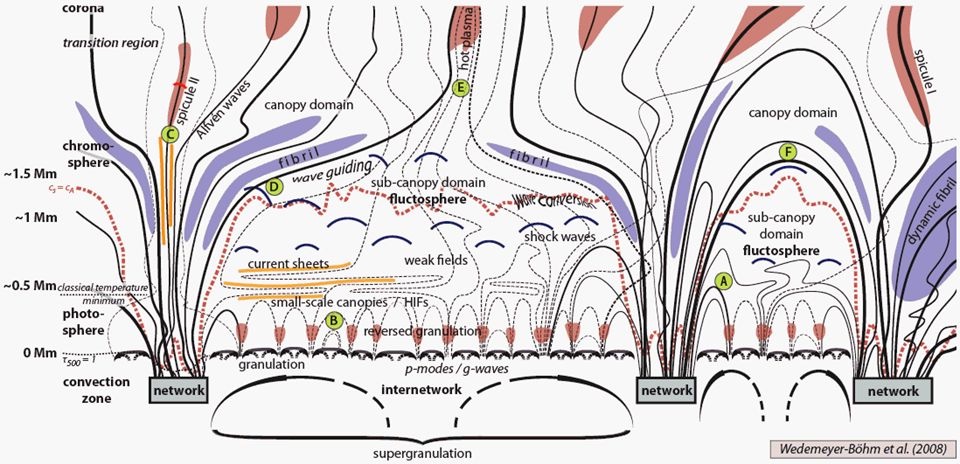}
\caption{Structure globale de l'atmosphère solaire. Figure prise de \citet[][Figure 16]{Wedemeyer2009}.
\label{fig:Wedemeyer2009_fig16_FR}}
\end{figure*}

\begin{figure*}[]
\centering
\includegraphics[width=1.\textwidth]{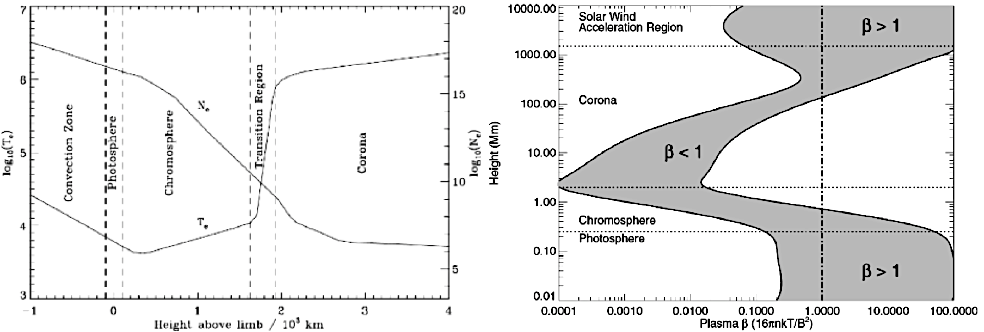}
\caption{Panneau de gauche : densité et température des électrons dans les différentes couches de l'atmosphère solaire. Panneau de droite : plage typique de variation du paramètre bêta associé au plasma. Figure tirée de \citet{Aschwanden2005}.
\label{fig:T_beta_FR}}
\end{figure*}

Les différentes couches de l'atmosphère solaire abritent un large éventail de processus physiques qui sont illustrés dans la figure \ref{fig:Wedemeyer2009_fig16_FR}. La couronne solaire située en haut de cette figure est chauffée à des températures atteignant environ $1-3\ \rm{MK}$ par des processus encore indéterminés. Une interface très fine d'environ $\approx 100\ \rm{km}$ d'épaisseur, appelée région de transition (TR), sépare la couronne chaude de la chromosphère plus froide/dense. Des profils typiques de densité électronique et de température sont représentés dans le panneau latéral gauche de la figure \ref{fig:T_beta_FR}. La TR étant si mince par rapport aux échelles caractéristiques du champ magnétique, nous pouvons supposer que l'intensité du champ magnétique ne change pas de manière significative dans cette région. Par conséquent, l'altitude de la TR, qui est assez variable, est contrôlée par un équilibre atteint entre les pressions thermiques relatives de la chromosphère et de la couronne. Une quantité importante d'énergie traverse la TR par le flux de chaleur conductif descendant de la couronne chaude vers la chromosphère plus froide. Pour un plasma confiné dans un champ fermé, la majeure partie de l'énergie qui est déposée dans la couronne est convectée vers le bas jusqu'à la chromosphère supérieure où elle est dissipée sous forme d'émissions radiatives (souvent appelé refroidissement radiatif), ce qui est discuté en détail dans la section \ref{subsec:intro_chromo_FR}. \\

La région de transition est une interface dynamique entre deux régions qui sont dirigées par des processus physiques différents et qui obéissent à des contraintes différentes. La couronne solaire est principalement contrôlée par le champ magnétique, qui est continuellement perturbé par les effets d'émergence du flux magnétique et des mouvements convectifs provenant du bas. Ces perturbations peuvent entraîner des structures complexes qui stockent de l'énergie libre magnétique, laquelle sera souvent libérée par des reconfigurations magnétiques transitoires qui permettront au champ magnétique coronal de retrouver un état de moindre énergie. En revanche, une partie importante de la chromosphère est contrôlée par des processus plasma qui résultent de mouvements convectifs transmis depuis la zone de convection et à travers la photosphère. Cette dichotomie est généralement bien décrite par le paramètre bêta du plasma qui est le rapport entre la pression thermique $n k_b T$ et la pression magnétique $B^2/(2\mu_0)$, et dont les valeurs typiques sont représentées dans la partie droite de la figure \ref{fig:T_beta_FR}. La hauteur dans la chromosphère où le bêta du plasma est égal à 1 est appelée la couche d'équipartition. La hauteur de cette couche marque donc l'endroit où le champ magnétique commence à influencer la dynamique de la chromosphère. La position de cette couche est représentée par une ligne pointillée rouge dans la figure \ref{fig:Wedemeyer2009_fig16_FR}. \\

\begin{figure*}[]
\centering
\includegraphics[width=0.95\textwidth]{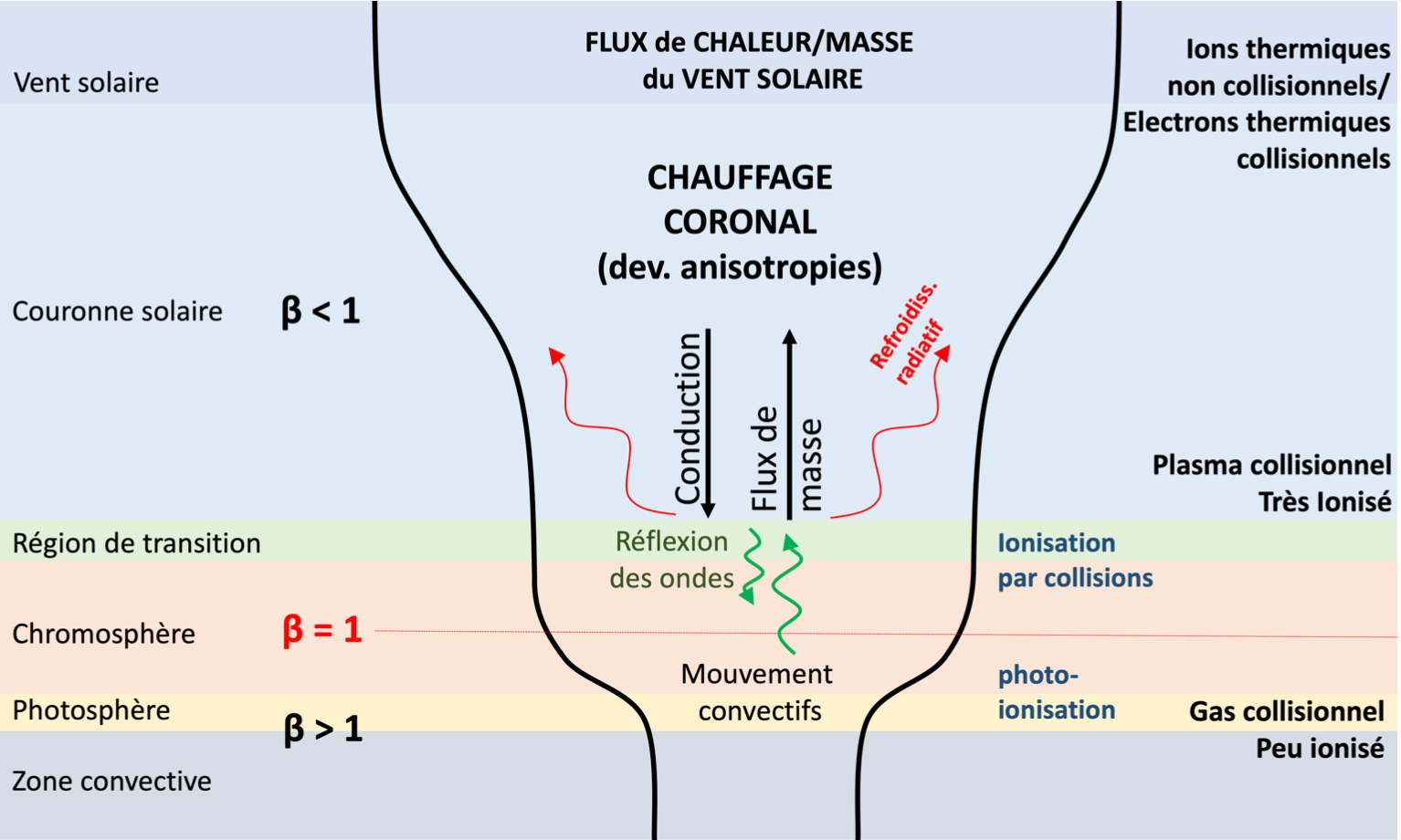}
\caption{Structure globale de l'atmosphère solaire avec ses propriétés physiques associées et ses processus dominants. Figure prise de l'habilitation à diriger des thèses de Dr. Alexis Rouillard (2021).
\label{fig:Alexis_HDR_FR}}
\end{figure*}

Comme la densité électronique diminue avec l'altitude, on passe progressivement d'une chromosphère dominée par les collisions à une couronne où les collisions deviennent trop rares pour influencer les transferts d'énergie. L'augmentation soudaine de la température et la baisse de la densité au niveau de la TR réduisent déjà fortement la fréquence de collision ($\nu \propto n T^{-3/2}$) entre les particules chargées qui interagissent via des interactions de Coulomb. Ceci a un certain nombre d'effets profonds sur le transport des particules et sera important pour les résultats discutés dans le chapitre \ref{cha:ISAM_results} car nous verrons que les collisions de Coulomb jouent un rôle majeur dans l'effet FIP. 
De manière plus exhaustive, nous présentons également en figure \ref{fig:Alexis_HDR_FR} un aperçu sommaire des différentes régions de l'atmosphère solaire et de la physique qui régit ces régions. Les processus physiques qui n'ont pas encore été abordés sont présentés dans les paragraphes suivants.

\subsection{Chauffage de l'atmosphère solaire}
\label{subsec:intro_heating_FR}

Si le problème du chauffage de la couronne solaire reste très débattu dans la communauté scientifique, plusieurs sources d'énergie prometteuses ont été identifiées, capables de maintenir la couronne à plusieurs millions de degrés. \\

Dans les régions à champ ouvert, une partie importante de l'énergie donnée au plasma est convertie en énergie cinétique pour accélérer le plasma jusqu'à la vitesse du vent solaire. Cette conversion d'énergie n'existe pas dans les régions à champ fermé, et la majeure partie de l'énergie doit être conduite vers le bas, vers la chromosphère, où elle est dissipée sous forme de rayonnement électromagnétique. Le flux d'énergie total nécessaire pour compenser les pertes d'énergie radiatives et conductives combinées varie entre $\approx 10^7\ \rm{erg.cm^{-2}.s^{-1}}$ dans les régions à fort champ magnétique (régions actives) et $\approx 3 \times 10^5\ \rm{erg.cm^{-2}.s^{-1}}$ dans les régions calmes du Soleil \citep{Withbroe1977,Klimchuk2006}. \\

Comme nous l'avons déjà mentionné, le déplacement mécanique du champ magnétique au niveau de la photosphère ou en dessous est probablement une source principale d'énergie qui peut alimenter la couronne avec un flux d'énergie suffisant \citep{Klimchuk2006}. Ces déplacements sont principalement induits par les mouvements convectifs du plasma qui s'élève de l'intérieur du soleil jusqu'à la photosphère. Ces déplacements déclenchent la génération d'une myriade d'ondes acoustiques et magnéto-acoustiques qui se propagent dans la chromosphère et atteignent finalement la couronne solaire. Les ondes peuvent subir une conversion de mode lorsqu'elles se propagent vers la chromosphère supérieure \citep[voir e.g.][]{Khomenko2006,Khomenko2008}. Les lignes de champ magnétique qui atteignent la chromosphère supérieure peuvent s'étendre considérablement pour fusionner en tubes de flux à grande échelle, comme on peut le voir du point D à E dans l'illustration donnée en figure \ref{fig:Wedemeyer2009_fig16_FR} \citep[voir aussi][]{Cranmer2005}. La conversion de mode devrait également se produire quelque part dans la chromosphère où la vitesse du son $c_s=\sqrt{k_b T/m}$ égalise la vitesse d'Alfvén $c_A=B/\sqrt{\mu_0 \rho}$ \citep{Khomenko2006,Carlsson2007}, typiquement au niveau de la couche d'équipartition déjà mentionnée (voir figure \ref{fig:Wedemeyer2009_fig16_FR}). \\

Les ondes acoustiques pures ont été proposées comme contributeurs potentiels au chauffage de la chromosphère et de la couronne \citep{Biermann1948,Schwarzschild1948}. En se propageant vers la couche supérieure de la chromosphère, les ondes acoustiques se transforment en chocs où elles finissent par dissiper leur énergie pour chauffer le plasma \citep{Schrijver1995,Carlsson2007}. Les simulations chromosphériques qui incluent des chocs acoustiques et un traitement détaillé du transfert radiatif \citep [voir e.g.][et les références qui y sont mentionnées]{Carlsson2002a} ont l'avantage de correspondre à certaines observations spectroscopiques, comme la raie d'émission chromosphérique du Calcium simplement ionisé (CaII). Cependant, elles ne parviennent souvent pas à reproduire correctement les intensités d'autres raies d'émission de la chromosphère moyenne et supérieure, et des études ultérieures ont montré que les ondes acoustiques ne permettraient pas à elles seules de chauffer la chromosphère \citep{Carlsson2007}. En outre, on pense généralement que la contribution des ondes acoustiques au chauffage coronal est probablement négligeable, car la majeure partie de leur énergie est dissipée dans la chromosphère supérieure par la formation de chocs et, ultimement, dans la région de transition en raison des forts gradients de température et de densité \citep{Klimchuk2006}. \\

Le champ magnétique a donc été désigné comme jouant un rôle clé dans le transport de l'énergie depuis la photosphère vers la chromosphère supérieure et la couronne. Des simulations magnéto-hydrodynamiques ultérieures ont montré que le tressage des lignes de champ magnétique, qui est induit par les mouvements photosphériques de cisaillement à l'échelle granulaire, peut dissiper suffisamment d'énergie dans la couronne pour maintenir des températures coronales à $\approx 1\rm{MK}$ \citep [voir e.g.][]{Galsgaard1996,Gudiksen2005}. Alors que les nano éruptions ("nano-flares") issues de la reconnexion magnétique semblent contribuer de manière significative au chauffage de la couronne dans ces simulations, une partie importante de l'énergie est aussi transportée par un flux mécanique à travers la couronne. Il est maintenant largement admis que les ondes d'Alfvén de cisaillement peuvent fournir le flux mécanique nécessaire au chauffage de la couronne \citep{Carlsson2007}. Contrairement aux ondes magnétoacoustiques telles que les modes lents et rapides, les ondes d'Alfvén de cisaillement restent incompressibles pendant la majeure partie de leur transit à travers la chromosphère et on pense qu'elles pénètrent très largement dans la couronne. \\ 

Comme l'énergie est principalement transportée par des ondes d'Alfvén de basse fréquence, il faut un processus supplémentaire qui transfère l'énergie vers des fréquences plus élevées où elle peut effectivement être transmise au plasma par des interactions onde-particule. Une cascade de turbulence peut assurer ce transfert d'énergie vers les hautes fréquences à la condition que "quelque chose" déclenche cette cascade. De nombreuses études ont alors formulé une théorie du chauffage coronal où la cascade de turbulence est générée depuis les basses fréquences par l'interaction non linéaire entre des ondes se propageant dans des directions opposées \citep[voir e.g.][]{Zhou1990,Tu1995,LieSvendsen2001,Dmitruk2002,Cranmer2005,Verdini2009,Chandran2011,Verdini2019,Reville2020a}. \\ 

Dans cette thèse, nous exploitons des modèles de chauffage qui supposent la dissipation des ondes d'Alfvén de cisaillement car leur propagation peut être décrite par une simple équation de transport comme discuté dans la section \ref{subsec:ISAM_Aw}. Nous avons également recours à des fonctions de chauffage ad-hoc qui peuvent bien approximer le flux mécanique nécessaire pour chauffer la couronne comme discuté dans la section \ref{subsec:ISAM_heating_adhoc}. 

\subsection{La chromosphère partiellement ionisée}
\label{subsec:intro_chromo_FR}

Dans les régions à champ fermé, toute l'énergie donnée au plasma dans la couronne doit être conduite vers le bas où elle est dissipée sous forme d'émissions radiatives. \\

La photosphère est généralement définie comme la hauteur à partir de laquelle le milieu devient transparent et où les photons qui étaient auparavent coincés à l'intérieur du soleil par de nombreuses collisions (absorptions et réémissions) peuvent s'échapper dans l'atmosphère solaire et au-delà. Mais le refroidissement par émissions radiatives ne devient efficace que dans la chromosphère moyenne/supérieure, où la diminution de la densité du plasma permet à une fraction significative des émissions radiatives du plasma de s'échapper à travers la couronne qui est optiquement mince. \\

La chromosphère étant très dynamique, optiquement épaisse en ultraviolet extrême et donc uniquement visible dans quelques raies chromosphériques, et ne pouvant être observée qu'à travers des émissions intégrées le long de la ligne de visée, les observations ne peuvent donc pas fournir un profil global de la chromosphère mais seulement des informations éparses. C'est pourquoi des modèles semi-empiriques de la chromosphère ont été construits pour contourner ces limitations, dont les célèbres modèles chromosphériques VAL3 \citep{Vernazza1981}, FAL \citep{Fontenla2002} et enfin AL \citep{Avrett2008}. 

\begin{figure*}[]
\centering
\includegraphics[width=0.7\textwidth]{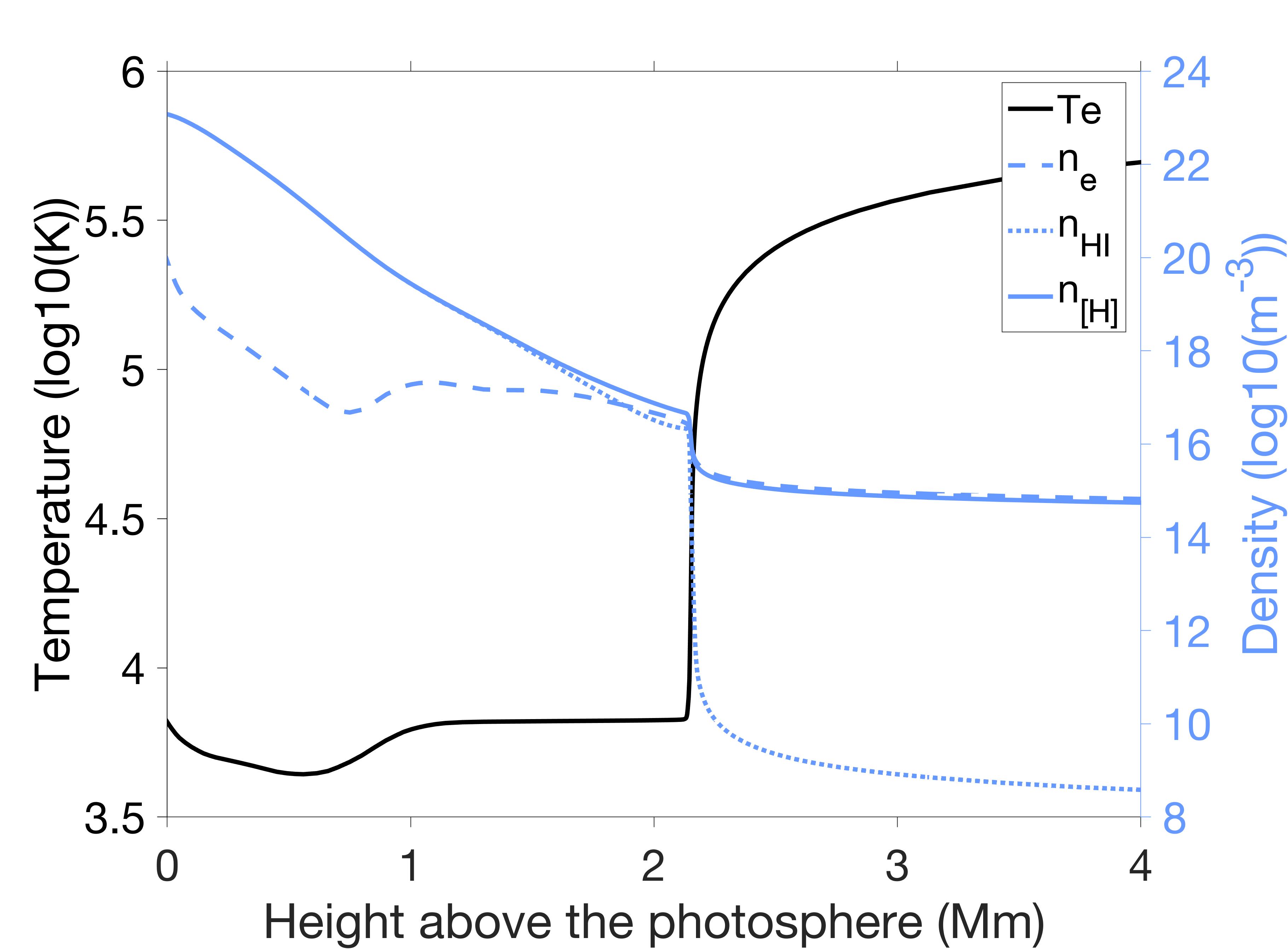}
\caption{Modèle de la chromosphère solaire et de la basse couronne pour le Soleil calme moyenné, à partir du modèle AL-C7 de \citet{Avrett2008}. La densité totale d'Hydrogène $n_{[H]}$, et celle de l'Hydrogène neutre seul $n_{HI}$ sont représentées. 
\label{fig:AL_profil_FR}}
\end{figure*}

Ils consistent tous en des techniques d'inversion sophistiquées qui convertissent les diagnostics spectroscopiques en des profils hydrostatiques de toute la chromosphère. En pratique, la distribution température-hauteur dans la chromosphère est ajustée par une approche itérative jusqu'à ce que le spectre d'émission simulé (synthétique) corresponde aux observations. Un exemple est illustré en figure \ref{fig:AL_profil_FR} pour le modèle AL-C7 de \citep{Avrett2008} qui correspond au Soleil calme moyenné. La région de transition marque la séparation entre une chromosphère partiellement ionisée avec $n_{HI}/n_{[H]} \lesssim 1$ et une couronne entièrement ionisée en protons où $n_{HI}/n_{[H]} \ll 1$. 

Bien que ces modèles reproduisent très bien le Soleil calme moyenné, leur pertinence est remise en question à la lumière des récentes simulations radiatives-hydrodynamiques qui décrivent une chromosphère hautement dynamique \citep{Carlsson2002a,Carlsson2007}. Néanmoins, ces modèles semi-empiriques ont été d'une aide précieuse pour calibrer notre modèle de l'atmosphère solaire (ISAM) qui est introduit dans le chapitre \ref{cha:ISAM}, et où une approche quasi-stationnaire est suivie dans un premier temps. Inversement, nous montrons dans cette thèse que ISAM peut apporter un éclairage nouveau sur ces modèles semi-empiriques en fournissant un traitement amélioré du transfert de masse et d'énergie depuis la haute chromosphère vers la couronne. \\

La couronne elle-même est entièrement ionisée (voir figure \ref{fig:AL_profil_FR}) et le manque de collisions avec les électrons "gèle" rapidement les états de charge des espèces coronales. Par conséquent, si la couronne doit être enrichie ou appauvrie en certains ions lourds en fonction de leur FIP, il est probable que cela se produise dans la chromosphère supérieure où tous les neutres ne sont pas encore totalement ionisés.

L'ionisation est facilitée pour les éléments ayant un faible FIP qui s'ionisent tôt dans la chromosphère et deviennent ainsi complètement ionisés dans la chromosphère supérieure. En comparaison, seul environ $30\%$ de l'Hydrogène, dont le FIP est intermédiaire ($\simeq 13.6\ \rm{eV}$), est ionisé au sommet de la chromosphère. En revanche, l'ionisation d'éléments à FIP élevé tels que l'Hélium ($\simeq 24.6\ \rm{eV}$) se produit principalement dans la région de transition et la très basse couronne. Les éléments plus lourds que l'Hydrogène et l'Hélium peuvent se présenter sous différents états de charge en fonction du nombre d'électrons que leur noyau peut accueillir. Par exemple, l'atome de Calcium à 20 électrons, bien qu'ayant un faible FIP de $\simeq 6.1\ \rm{eV}$ peut à peine atteindre le 14e état de charge à des températures coronales de $1-3\ \rm{MK}$, alors que l'atome d'hélium à 2 électrons avec un FIP élevé de $\simeq 24.6\ \rm{eV}$ sera entièrement (doublement) ionisé dans la couronne. \\

La chromosphère et la couronne solaire diffèrent également par les processus physiques qui contribuent à l'équilibre d'ionisation. Dans la chromosphère l'ionisation se produit principalement par photoionisation à partir du rayonnement incident d'origine photosphérique, une ionisation qui est contrebalancée par la recombinaison radiative \citep{Carlsson2002a}. Dans la région de transition et basse couronne l'ionisation provient surtout de l'impact collisionnel avec les électrons, qui peut se faire soit directement (ionisation directe), soit en deux phases à travers un état intermédiaire excité (auto-ionisation). Nous discutons plus en détail de la spécificité de chacun de ces processus d'ionisation/recombinaison dans la section \ref{sec:ISAM_ioniz}. \\

L'Hydrogène est particulier dans le sens où son équilibre d'ionisation peut s'établir plus lentement que les échelles de temps hydrodynamiques typiques de la chromosphère \citep{Kneer1980}. Les simulations radiatives-hydrodynamiques ont montré que l'hydrogène n'est probablement pas en équilibre d'ionisation avec l'équilibre thermique local, avec une tendance pour l'ionisation de l'Hydrogène à être stimulée lors du passage de chocs acoustiques qui traversent la chromosphère supérieure \citep{Carlsson2002a}. Ceci remet en question la validité d'une chromosphère moyenne "statique" telle que donnée par les modèles semi-empiriques discutés ci-dessus, car l'ionisation de l'Hydrogène n'est probablement pas en équilibre et dépend de l'historique de la chromosphère. Notre approche pour résoudre cette difficulté dans notre modèle de l'atmosphère solaire est discutée dans la section \ref{subsec:ISAM_ioniz_future}.

\subsection{Extraction des ions lourds depuis la chromosphère}
\label{subsec:intro_FIP_FR}

La composition en ions lourds qui est mesurée in situ dans le vent solaire s'établit probablement dans les basses couches de l'atmosphère solaire avant que l'équilibre d'ionisation ne se "gèle" dans la couronne où les collisions sont plus rares. De plus, puisque les abondances relatives sont uniformément distribuées à la photosphère \citep{Asplund2009}, la séparation entre les ions lourds doit alors se produire quelque part entre la photosphère et la couronne, et donc très probablement dans la chromosphère. Plus précisément, la majeure partie du fractionnement s'établit probablement établit dans la chromosphère supérieure, où les neutres deviennent ionisés et où les neutres ayant un faible FIP commencent à s'ioniser en premier. De cette façon, on peut éventuellement expliquer la séparation entre les éléments à faible et à fort FIP s'il existe une force extérieure qui attire sélectivement les ions de la chromosphère. Sans cette force hypothétique, les éléments plus lourds tomberaient plus vite que les éléments plus légers en raison de la gravitation, et produiraient donc une séparation des éléments en fonction de leur masse et non de leur FIP. \\

Plusieurs forces ont été suggérées dans la littérature qui peuvent correspondre à cette force hypothétique et peuvent produire une séparation entre les éléments à faible et à fort FIP \citep[voir][pour une revue sur l'effet FIP]{Henoux1998,Laming2015}. On peut les classer en deux groupes principaux selon que le champ magnétique joue un rôle ou non dans le processus de séparation. De nombreuses études de l'effet FIP montrent que les effets de diffusion pure peuvent induire un fractionnement selon le FIP et non la masse. Certaines de ces études considèrent l'effet du couplage frictionnel de particules ayant des vitesses différentes \citep{Wang1996,Peter1996,Marsch1995,Peter1998b,Bo2013}, seul ou couplé à des effets de diffusion thermique le long du champ magnétique \citep{Geiss1986,Hansteen1997,Killie2005,Killie2007}. Alors que d'autres ont étudié l'effet de la diffusion à travers le champ magnétique \citep{Steiger1989,Antiochos1994}. \\

\citet{Vauclair1996} ont formulé un scénario selon lequel les abondances coronales observées résulteraient de l'émergence d'un tube de flux magnétique de la photosphère vers la couronne, et que seuls les éléments à faible FIP qui s'ionisent tôt pourraient être capables de rattraper le mouvement convectif du tube de flux ascendant. Des observations récentes de \textit{Hinode} suggèrent néanmoins que lorsque des flux émergents se reconnectent avec les champs coronaux préexistants, cela ouvre de nouveaux canaux où les abondances photophériques et coronales sont probablement mélangées \citep{Baker2015}. 

\citet{Schwadron1999,Laming2004,Laming2009} ont trouvé une connexion intime entre les processus de chauffage qui sont basés sur les interactions onde-particule, et l'effet FIP qui peut résulter de ces mêmes interactions. Il est maintenant largement admis que les ondes d'Alfvén de basse fréquence peuvent transporter l'énergie mécanique nécessaire depuis la chromosphère vers la couronne, où l'énergie est finalement dissipée aux petites échelles par des interactions onde-particule (voir section \ref{subsec:intro_heating_FR}). On ne sait toutefois pas comment de telles ondes à haute fréquence peuvent survivre dans la chromosphère, mais \citet{Laming2004,Laming2009,Laming2015} ont formulé une théorie selon laquelle les ions lourds peuvent interagir avec les ondes d'Alfvén à basse fréquence par le biais de la force pondéromotrice. Essentiellement, la force pondéromotrice correspond à une description temporelle moyennée des forces de Lorentz agissant dans un champ électromagnétique oscillant, qui résultent en une force nette dirigée depuis les régions à densité d'onde faible vers élevée \citep{Lundin2006,Laming2015}. \\

Le couplage par friction des ions lourds avec les protons (force de traînée des protons) s'est avéré très efficace pour séparer les ions lourds du gaz d'Hydrogène neutre principal, qui se produit principalement dans la couche d'ionisation de l'Hydrogène dans la chromosphère supérieure.

Dans le cas hydrodynamique modélisé par \citet{Wang1996}, cela nécessite cependant l'existence d'un flux ambipolaire dans la chromosphère supérieure où les protons dérivent vers le haut par rapport à l'hydrogène neutre. Les éléments à faible FIP qui s'ionisent tôt dans la chromosphère sont alors entraînés par le flux de protons par couplage frictionnel et par l'interaction coulombienne. Ce flux ambipolaire est probablement transitoire dans les régions à champ fermé mais ces auteurs suggèrent qu'il peut être temporairement soutenu pendant les phases d'évaporation chromosphérique lorsqu'un chauffage accru est appliqué à la base de la couronne. Des signatures périodiques d'évaporation chromosphérique ont été observées dans les boucles en EUV, ce qui ne nécessite pas nécessairement un apport externe d'énergie via un événement de reconnexion soudain par exemple, mais pourrait simplement résulter de cycles de non-équilibre thermique (TNE) dans les boucles coronales \citep[voir e.g.][]{Auchere2016}. Dans certaines conditions, les boucles coronales peuvent entrer en TNE et subir des phases alternées de flux ascendants d'évaporation depuis la chromosphère et de flux descendants de condensation depuis la couronne \citep[voir aussi][]{Johnston2017,Johnston2019}, et peuvent donc éventuellement soutenir le flux ambipolaire mentionné ci-dessus. Nous discuterons plus en détail des cycles TNE en section \ref{subsec:ISAM_results_H_thermodynamics} en utilisant notre modèle de l'atmosphère solaire appelé ISAM. Par conséquent, ce processus de séparation serait le plus efficace dans les plasmas à champ fermé où l'énergie déposée dans la couronne est principalement convectée vers la chromosphère, et non convertie en énergie cinétique comme dans les plasmas à champ ouvert. \citet{Bo2013} ont également montré l'importance du couplage par friction avec les protons pour empêcher la stratification des éléments lourds dans la chromosphère supérieure, et ce même dans une chromosphère hydrostatique où l'Hydrogène neutre et les protons sont supposés au repos. 

\begin{figure*}[]
\centering
\includegraphics[width=0.85\textwidth]{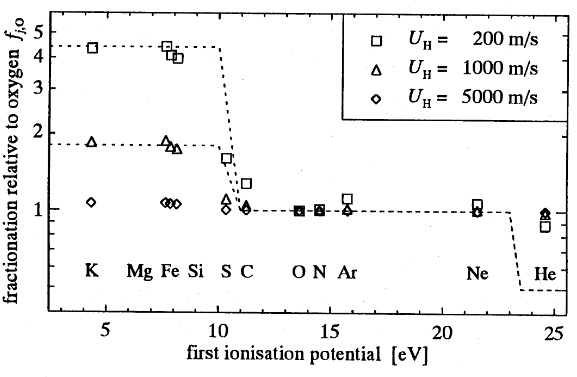}
\caption{Variation de l'abondance relative d'Oxygen en fonction du FIP, générée depuis le modèle à vitesse différentielle de \citet{Peter1996}. Figure prise de \citet[][Figure 1]{Peter1996}.
\label{fig:Peter1996_fig1_FR}}
\end{figure*}

\citet{Peter1996,Peter1998b} ont étudié le cas des plasmas à champ ouvert. En faisant varier la vitesse moyenne d'écoulement de l'Hydrogène dans la chromosphère, ils montrent que le couplage frictionnel avec les protons peut reproduire globalement le fractionnement typique du FIP qui est mesuré in situ dans le vent solaire lent et rapide, comme illustré en figure \ref{fig:Peter1996_fig1_FR}. \\

Tous les modèles de friction présentés ci-dessus ne prennent en compte que la chromosphère supérieure dans le processus de séparation, et ne traitent donc pas de l'échange de matière dans la région de transition et la couronne. En incluant la couronne et les effets de diffusion thermique, à savoir la force thermique, \citet{Killie2007} ont montré une image différente qui semble être en conflit avec certaines études antérieures, en particulier pour le cas des plasmas à champ fermé. Comme ils présentent un flux de protons descendant (et non ascendant), leur flux de protons agit comme une barrière à l'extraction depuis la chromosphère des ions lourds à faible FIP. Cependant, les effets de diffusion thermique qu'ils introduisent produisent une force nette qui est dirigée vers le haut et qui tire les éléments à faible FIP vers la couronne plus chaude. Néanmoins, ils soutiennent que les éléments à faible FIP, tels que le Silicium ou le Fer, sont quand même incapables de pénétrer dans la couronne en raison d'une "barrière de protons" qui est plus forte que la force thermique. Par conséquent, leur modèle tend à produire un effet FIP inverse qui n'est généralement pas observé dans la couronne solaire mais dans les couronnes stellaires actives \citep[voir la revue de][]{Laming2015}. Pour pallier à ce paradoxe, ils concluent que cette matière enrichie en élements à faible FIP qui se retrouve piégée dans la couronne, proviendrait initialement d'une chromosphère (partiellement) stratifiée. Et que cette matière enrichie chromosphérique pourrait naturellement être transportée vers le haut lors de l'émergence des boucles depuis la photosphère vers la couronne \citep [voir e.g.][]{Vauclair1996}. \\

De nombreux facteurs pouvant contribuer à l'effet FIP ont été examinés tout au long de cette section. On pense généralement que le couplage collisionnel avec les protons contribue de manière significative à la séparation entre les éléments à faible et à fort FIP dans la chromosphère supérieure. Cependant, les effets de diffusion sont lents et, par conséquent, l'effet FIP peut prendre jusqu'à plusieurs jours ou semaines pour s'installer dans une chromosphère non perturbée \citep[voir e.g.][]{Killie2007}, ce qui est cohérent avec les biais FIP accrus mesurés durant le vieillissement des régions actives \citep{Widing2001}. De plus, certains auteurs ont remarqué la nécessité d'avoir un mécanisme de mélange externe de la chromosphère qui pourrait empêcher certains éléments lourds d'être trop sévèrement appauvris ou enrichis en raison de la stratification gravitationnelle \citep{Hansteen1997,Killie2007,Bo2013}. \citet{Laming2009} ont complété ce point en avançant qu'en l'absence d'interactions onde-particule avec les ondes d'Alfvén (par le biais de la force ponderomotrice), la turbulence hydrodynamique peut fournir un mélange chromosphérique sur des échelles de temps plus courtes que celles de la stratification gravitationnelle pour opérer, mais qui restent suffisamment longues pour que l'effet FIP puisse s'établir. L'émergence de flux magnétique depuis la photosphère peut être une autre source majeure de mélange chromosphérique aux abondances photosphériques \citet{Baker2015}. Comme nous l'avons vu dans la section \ref{subsec:intro_chromo_FR}, la chromosphère est intrinsèquement dynamique et, par conséquent, il est très peu probable que la chromosphère reste stratifiée au cours du temps. \\

Bien que toutes les études présentées ci-dessus soient capables de produire un effet FIP, le cadre de modélisation et les conditions requises diffèrent d'une étude à l'autre, où dans la plupart des études le couplage entre la région de transition et la couronne n'est pas considéré. Toutes les forces qui sont considérées dans ces études contribuent probablement à l'effet FIP, cependant il reste à quantifier leur contribution relative d'une manière auto-consistante, dans un cadre global qui couvre la chromosphère, la région de transition et la couronne. À cette fin, nous développons un modèle de l'atmosphère solaire qui est décrit dans le chapitre \ref{cha:ISAM}. Les premières applications du modèle au cas des plasmas en champ fermé sont présentées dans le chapitre \ref{cha:ISAM_results} où les effets de diffusion discutés ci-dessus peuvent être testés immédiatement.


\section{Théories sur l'origine du vent solaire lent}
\label{sec:intro_theories_FR}

\subsection{La théorie quasi-stationnaire}
\label{sec:intro_stationnary_FR}

\begin{figure*}[]
\centering
\includegraphics[width=0.5\textwidth]{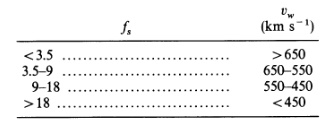}
\caption{Correspondances entre les facteurs d'expansion (calculés entre la photosphère et une hauteur coronale de $2.5\ R_\odot$) et la vitesse du vent solaire mesurée in situ à 1 UA. Figure prise de \citet[][Tableau 1]{Wang1990}.
\label{tab:Wang1990_tab1_FR}}
\end{figure*}

Avant la mission \textit{Ulysses}, \citet{Wang1990} a rassemblé 22 ans de données sur la vitesse du vent solaire provenant des missions \textit{Vela}, \textit{IMP}, et \textit{ISEE 3/ICE} pour démontrer une corrélation inverse à long terme entre la vitesse globale du vent solaire mesurée in situ à 1 UA, et le taux de divergence du champ magnétique coronal de la région source estimée. Dans le tableau \ref{tab:Wang1990_tab1_FR}, une correspondance est montrée entre les taux d'expansion et les vitesses du vent solaire typiquement associées. \\

\begin{figure*}[]
\centering
\includegraphics[width=0.95\textwidth]{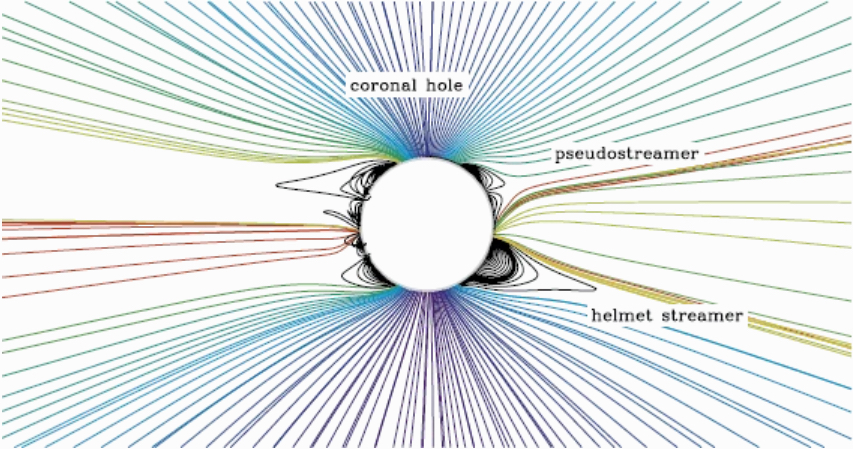}
\caption{Lignes de champ magnétique fermées (noir) et ouvertes (multicolore) tracées à partir d'une solution stationnaire aux équations de conservation MHD polytropiques, calculée par le code "Magnetohydrodynamics Around a Sphere" (MAS) \citep{Linker1999,Riley2001}. La condition limite au niveau de la photosphère corresponds à la rotation Carrington 2058 (Juin-Juillet 2007). Les couleurs des lignes de champ ouvertes correspondent au facteur d'expansion de \citet{Wang1990} pour des valeurs: $f_{ss}=(R_\odot/R_{ss})^2(B_\odot/B_{ss})$: $f_{ss}\leq 4$ (violet), $f_{ss}\simeq 6$ (bleu), $f_{ss}\simeq 10$ (vert), $f_{ss}\simeq 15$ (or), $f_{ss}\geq 40$ (rouge). Figure prise de \citet[][Figure 1]{Cranmer2017}.
\label{fig:Cranmer2017_fig1_FR}}
\end{figure*}

Une inspection des reconstructions 3-D du champ magnétique coronal suggère alors que le vent solaire lent provient des tubes de flux qui sont contigus aux régions à champ fermé (par exemple, les boucles coronales sous les streamers) où le champ magnétique subit une grande expansion à travers la couronne, comme illustré en figure \ref{fig:Cranmer2017_fig1_FR}. À l'opposé, les grands trous coronaux tels que ceux situés près des pôles produisent généralement un vent solaire rapide près de leur centre car le champ magnétique se dilate beaucoup moins. Pour les trous coronaux plus petits situés à des latitudes équatoriales où le facteur d'expansion n'est pas aussi faible, on peut trouver des vents de vitesse intermédiaire ou même des vents Alfvéniques lents comme discuté dans la section \ref{sec:dynamics_insitu} et dans \citet{Griton2021}. En de rares occasions, de grands trous coronaux ont été observés près de l'équateur, ce qui peut produire un vent très rapide même aux latitudes équatoriales et également affecter de manière drastique la forme globale de la couronne solaire avec une HPS qui est presque verticale \citep[voir e.g.][]{Sanchez-Diaz2017b}. \\

Le modèle à tubes de flux multiples MULTI-VP introduit dans la section \ref{subsec:MULTI-VP} et décrit dans \citet{PintoRouillard2017} reproduit avec succès un vent solaire induit par le taux d'expansion et est exploité à plusieurs reprises dans cette thèse. Dans MULTI-VP, la génération de différents régimes de vent solaire est intégrée dans la prescription qui est adoptée pour chauffer le plasma. Essentiellement, le facteur d'expansion contrôle les hauteurs où l'énergie est déposée, ce qui affecte ensuite la quantité de plasma chromosphérique qui est extrait vers la couronne via l'évaporation chromosphérique \citep[voir aussi][]{Hansteen2012}. Grâce à sa polyvalence, MULTI-VP est donc capable de reproduire la structure globale de la couronne solaire telle qu'elle est observée par les coronographes en lumière visible \citep{PintoRouillard2017}, avec des détails encore plus fins qui sont discutés plus loin dans la section \ref{sec:stationnary_poirier2020}. \\

À des hauteurs plus importantes dans la couronne, l'énergie donnée en entrée pour chauffer le plasma est progressivement convertie en énergie cinétique pour accélérer le vent solaire \citep{PintoRouillard2017}. Si l'on considère également les anisotropies de pression qui ne peuvent pas être négligées dans la haute couronne au sein de la région d'accélération du vent solaire, alors la force miroir joue également un rôle majeur dans le processus d'accélération \citep{Lavarra2022}. \\

Récemment, \citet{Lavarra2022} ont montré avec le modèle d'atmosphère solaire de l'IRAP (ISAM, voir aussi le chapitre \ref{cha:ISAM}) que la théorie quasi-stationnaire peut également rendre compte des états de charge mesurés in situ dans les vents rapides et lents, et même dans le vent lent Alfvénique, comme l'illustre la figure \ref{fig:Lavarra2022_fig1_FR}. Une étude future permettra de déterminer si ce modèle peut également reproduire les variations d'abondance des particules alpha et des ions mineurs mesurées in situ dans ces différents régimes de vent solaire. \\

Comme présenté dans la section \ref{subsec:intro_FIP_FR}, les effets de diffusion tels que le couplage frictionnel avec le flux moyen de protons peuvent expliquer les abondances accrues mesurées des éléments à faible FIP pour différentes vitesses du vent solaire, mais ne parviennent pas à prédire les abondances variables dans les particules alpha \citep{Peter1996,Peter1998b}. Nous avons également présenté dans la section \ref{subsec:intro_FIP_FR} d'autres processus tels que les interactions onde-particule qui peuvent également jouer un rôle dans la théorie quasi-stationnaire. Cependant, nous avons également souligné dans la section \ref{subsec:intro_FIP_FR} que l'efficacité des processus de fractionnement reste intrinsèquement limitée dans les milieux à champ ouvert tels que décrits par la théorie quasi-stationnaire. 

\subsection{Théories dynamiques du vent solaire lent}
\label{sec:intro_dynamic_FR}

Comme discuté dans la section \ref{subsec:intro_FIP_FR}, les régions à champ fermé sont des environnements propices qui peuvent atteindre des niveaux de fractionnement aussi élevés que ceux mesurés in situ dans le vent solaire lent. Par conséquent, nous invoquons la nécessité d'une théorie dynamique qui complète la théorie quasi-stationnaire, et où le vent solaire lent est partiellement constitué de plasma en champ fermé qui se retrouve expulsé dans le vent lent par les processus de reconnexion magnétique qui sont discutés tout au long de cette section. 

\subsubsection{Reconnection magnétique à l'extrémité des streamers}
\label{subsec:intro_dynamics_streamertip_FR}

Il est probable que les blobs observés au-dessus des streamers (voir section \ref{subsubsec:intro_SSW_intermittency}) sont produits au niveau de la HCS par la reconnexion magnétique connue pour se produire dans de nombreuses régions de la couronne solaire. \\

\begin{figure*}[]
\centering
\includegraphics[width=0.4\textwidth]{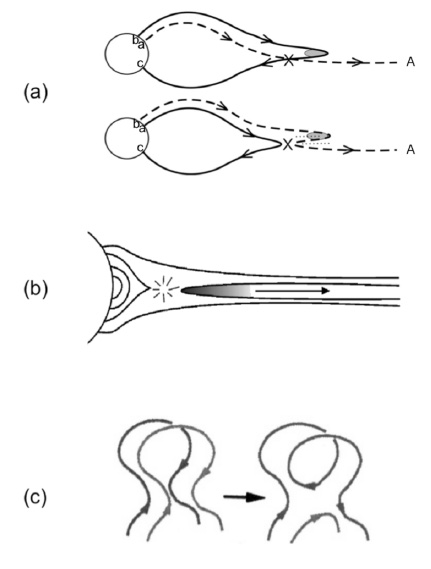}
\caption{Esquisse de différents scénarios de reconnexion pour la libération de plasma en champ fermé dans le vent solaire lent à l'extrémité des streamers par: (a) la reconnexion d'interchange entre lignes ouvertes et fermées \citep{Crooker2003}, (b) la reconnexion dans la HCS \citep{Wang1998}, et (c) la reconnexion entre pairs de boucles coronales \citep{Gosling1995}. Figure tirée de \citet[][Figure 1.10]{Sanchez-Diaz2017_PHD}.
\label{fig:Sanchez2017_fig1.10_FR}}
\end{figure*}

\citet{Wang2000} ont suggéré que de tels écoulements intermittents pourraient être induits par la reconnexion magnétique à l'extrémité des streamers lorsque les boucles coronales s'élevant dans l'atmosphère solaire sont suffisamment étirées pour déclencher la reconnexion magnétique entre des champs magnétiques de direction opposée, comme illustré en figure \ref{fig:Sanchez2017_fig1.10_FR}b. Cette image a l'avantage d'être également cohérente avec l'observation des flux de plasma descendants dans \textit{LASCO} \citep{Wang2000}, lesquels ont été associés pour la première fois à des blobs de streamers dans \citet{Sanchez-Diaz2017a}. L'instabilité de déchirement qui se développe pendant l'expansion des boucles a été proposée comme le mécanisme déclencheur de la génération de blobs de streamers via la reconnexion magnétique au niveau de la HCS \citep{Reville2020b,Reville2022}. Nous étudions ce processus et ses signatures attendues par télédétection dans le chapitre \ref{sec:dynamics_tearing} où nous exploitons le modèle magnéto-hydrodynamique WindPredict-AW \citep{Reville2020a}. \\

\subsubsection{Reconnection d'interchange aux interfaces ouvert-fermé}
\label{subsec:intro_interchange_FR}

Un autre mécanisme de formation possible pour les structures transitoires est la reconnexion magnétique d'interchange aux frontières ouvert-fermé où les boucles magnétiques peuvent interagir avec les champs magnétiques ouverts dans les régions de fort cisaillement magnétique, ce qui est également illustré dans la figure \ref{fig:Sanchez2017_fig1.10_FR}a. Un facteur important de la reconnexion d'interchange est certainement la vitesse à laquelle les régions de champ ouvert et fermé sont susceptibles de dériver l'une par rapport à l'autre. \\

Les observations des taches solaires indiquent que la surface solaire ou photosphère est principalement en rotation différentielle, ce qui signifie que les régions équatoriales tournent plus vite que les régions polaires \citep{Scheiner1630,Bumba1969}. Cependant, la rotation différentielle de la photosphère est rapidement freinée plus haut dans l'atmosphère solaire, par la couronne bien établie qui présente une rotation plus rigide \citep{Fisher1984,Hoeksema1987,Bird1990}. Ces différents taux de rotation conduisent probablement à l'interpénétration de champs fermés et ouverts, un environnement propice à la reconnexion d'interchange \citep[voir e.g.][]{Fisk1996}. \\

Les lignes de champ magnétique ouvertes à l'intérieur des trous coronaux ont tendance à tourner de manière rigide avec la couronne solaire \citep{Lionello2005}, et devraient donc dériver par rapport aux régions à champ fermé qui tournent avec la photosphère. Par exemple, la frontière ouvert-fermé entre les trous polaires et les boucles coronales situées sous les streamers est un environnement privilégié où la reconnexion d'interchange pourrait se produire en permanence \citep{Crooker2003,Pinto2021}. Un tel scénario pourrait également expliquer les nombreuses inversions de champ magnétique, ou "switchbacks", qui ont été récemment mesurées in situ à \textit{PSP} \citep{Bale2021}. \\

\begin{figure*}[]
\centering
\includegraphics[width=0.7\textwidth]{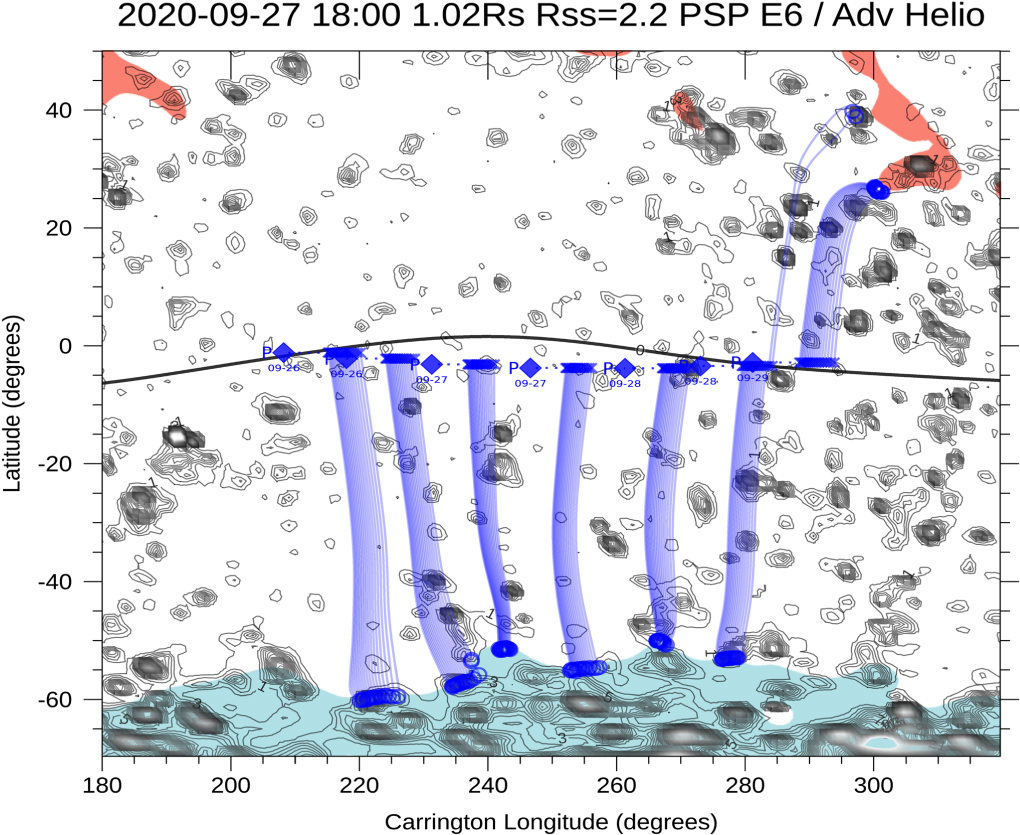}
\caption{Carte de connectivité de \textit{PSP} lors de son 6ème passage près du Soleil (26-29 septembre 2020). Les trous coronaux polaires nord et sud sont représentés respectivement par des zones de couleur rouge et cyan. La pression magnétique $B^2/(2\mu_0)$ est représentée par des contours noirs à $0.02\ R_\odot\simeq 14\ \rm{Mm}$ au-dessus de la photosphère, et représente les frontières ouvert-fermé associées au réseau supergranulaire. La trajectoire de \textit{PSP} (losanges bleus) est projetée de manière balistique à une hauteur de $1.2\ R_\odot$ (croix bleues), puis tracée jusqu'à $0,02\ R_\odot$ (cercles bleus) au-dessus de la photosphère en utilisant une reconstruction 3-D du champ magnétique. La position de la HCS prédite par le modèle de reconstruction 3-D est représentée par une ligne noire pleine à $1,2\ R_\odot$. Figure tirée de \citet[][Figure 1]{Bale2021}.
\label{fig:Bale2021_fig1_FR}}
\end{figure*}

\citet{Bale2021} dessinent une image plus détaillée de la reconnexion d'interchange qui se produit beaucoup plus bas dans l'atmosphère solaire (près de la région de transition), et à une échelle spatiale beaucoup plus petite. Ces événements de reconnexion d'interchange pourraient avoir lieu dans des frontières ouvert-fermé qui correspondent au réseau supergranulaire avec une échelle typique de $\approx 15-20\ \rm{Mm}$, ce qui est illustré en figure \ref{fig:Bale2021_fig1_FR} sous la forme de contours noirs. Ces auteurs remarquent également un léger enrichissement des particules alpha (c'est-à-dire de l'Hélium doublement ionisé) mesurées in situ dans le vent solaire lent lorsque \textit{PSP} passe au-dessus de ces régions. Leur travail montre que la libération de plasma en champ fermé dans le vent solaire lent par la reconnexion d'interchange se produit systématiquement et à des échelles encore plus petites que ce que l'on pensait auparavant. De plus, les analyses statistiques des données \textit{PSP} effectuées par \citet{Fargette2021,Fargette2022} ont montré que les échelles et le taux d'occurrence des reconnexions d'interchange sont compatibles avec des processus se produisant à l'échelle granulaire et mésogranulaire. Ils ont proposé que les boucles magnétiques qui restent ancrées à la photosphère en rotation différentielle puissent se reconnecter au champ magnétique coronal en rotation plus rigide. Ce point sera abordé dans la section \ref{subsec:intro_Sweb_FR}. \\

Nous noterons également que les événements de reconnexion magnétique associés à de forts jets peuvent être générés dans des topologies de points nuls préexistants au sein de régions unipolaires, si une certaine torsion des lignes de champ est imposée à la base \citep{Pariat2009}. Des scénarios alternatifs pour l'origine des "switchbacks" impliquent également une reconnexion d'interchange, non pas avec des boucles préexistantes comme discuté précédemment, mais avec des boucles qui émergent à la photosphère dans des régions autrement unipolaires, déclenchant ainsi des micro-jets \citep{Sterling2015}. Plus généralement, des structures semblables à des jets ont été observées à plusieurs reprises dans les raies spectrales de la chromosphère, sous la forme de sursauts d'intensité de courte durée (environ $1$ à $10\ \rm{min}$), associés à des petites tailles (environ $300$ à $1500\ \rm{km}$) et à des vitesses ascendantes d'environ $25\ \rm{km/s}$, et où leur température $\approx 5000-15000\ \rm{K}$ indique de la matière d'origine chromosphérique \citep{Sterling2000}. Les jets coronaux sont également des signatures de la reconnexion magnétique qui sont fréquemment observées dans la couronne solaire \citep{Raouafi2016}. Les récentes observations de télédétection à haute résolution de \textit{Solar Orbiter} ont en outre révélé l'existence de "feux de camp" à très petites échelles surgissant fréquemment dans la chromosphère, qui sont pour l'instant considérés comme de nouveaux éléments de la famille des éruptions, micro-éruptions et nano-éruptions \citep{Berghmans2021}. Dans l'ensemble, les événements de reconnexion à petites échelles peuvent contribuer de manière significative au chauffage de la chromosphère et de la région de transition, et probablement aussi au mélange de la chromosphère aux abondances photosphériques. La profusion de sursauts détectés en EUV dépeint une chromosphère et une couronne solaire hautement dynamiques où les événements de reconnexion magnétique sont omniprésents, et donc où la matière en champ fermé peut être expulsée dans le vent solaire par reconnexion d'interchange. \\

Comme nous le verrons dans la section \ref{subsec:intro_Sweb_FR}, il est maintenant largement accepté dans la littérature qu'une partie significative du vent solaire lent naît aux frontières ouvert-fermé où l'on constate que la reconnexion par échange se produit systématiquement. Ce processus permet notamment d'expliquer pourquoi des signatures du vent lent (qui pour rappel est enrichi en éléments à faible FIP comme dans les boucles coronales, voir section \ref{subsubsec:intro_SSW_composition_FR}) sont également détectées loin de la HCS à partir de mesures de composition effectuées in situ \citep{Zurbuchen2007}. À l'aide de simulations MHD tridimensionnelles et à haute résolution, \citet{Higginson2017} montrent qu'un vent lent peut se former loin de l'HCS aux frontières ouvert-fermé des extensions des trous coronaux polaires, ce qui, dans certaines circonstances, peut créer un étroit couloir de champ ouvert depuis les latitudes polaires jusqu'aux latitudes moyennes, comme l'illustre la figure \ref{fig:Antiochos2011_fig4_FR} (voir également la section \ref{subsubsec:intro_streamers_FR}). Plus généralement, nous verrons dans la section \ref{subsec:intro_Sweb_FR} que ces frontières ouvert-fermé sont omniprésentes dans l'atmosphère solaire et qu'elles forment un large réseau de séparatrices \citep[appelé S-web: ][]{Antiochos2011} et d'où est censé naître le mystérieux vent lent détecté loin des streamers et de la HCS. \\

\subsubsection{Libération de champs torsadés à l'extrémité des streamers}
\label{subsec:intro_dynamics_fluxropes_FR}

La reconnexion magnétique peut également se produire entre des boucles coronales adjacentes sous un streamer \citep{Gosling1995}, comme le montre la figure \ref{fig:Sanchez2017_fig1.10_FR}c. Un tel scénario conduirait à la libération dans le vent lent, d'écoulements de plasma transportés par des champs magnétiques torsadés. En fait, les panneaux b et c de la figure \ref{fig:Sanchez2017_fig1.10_FR} peuvent représenter deux aspects du même processus de libération, où la formation d'un champ torsadé (panneau c) s'accompagne d'une augmentation de la densité à sa périphérie (c'est-à-dire d'un "streamer blob", panneau b) au même moment. \\

Le suivi continu des blobs expulsés à l'extrémité des streamers jusqu'aux points de mesures in situ révèle que les blobs transportent effectivement des champs magnétiques hélicoïdaux \citep{Rouillard2009,Rouillard2011a,Rouillard2011b}, comme le révèlent également les observations récentes de \textit{PSP} \citep{Lavraud2020,Rouillard2020a}. Des analyses statistiques plus systématiques des blobs observés dans les images de \textit{STEREO}, et des structures transitoires mesurées in situ à l'intérieur de la HPS ont révélé que la topologie des blobs est cohérente avec des champs torsadés magnétiques \citep{Sanchez-Diaz2019} qui se forment par reconnexion magnétique. \\

\begin{figure*}[]
\centering
\includegraphics[width=0.6\textwidth]{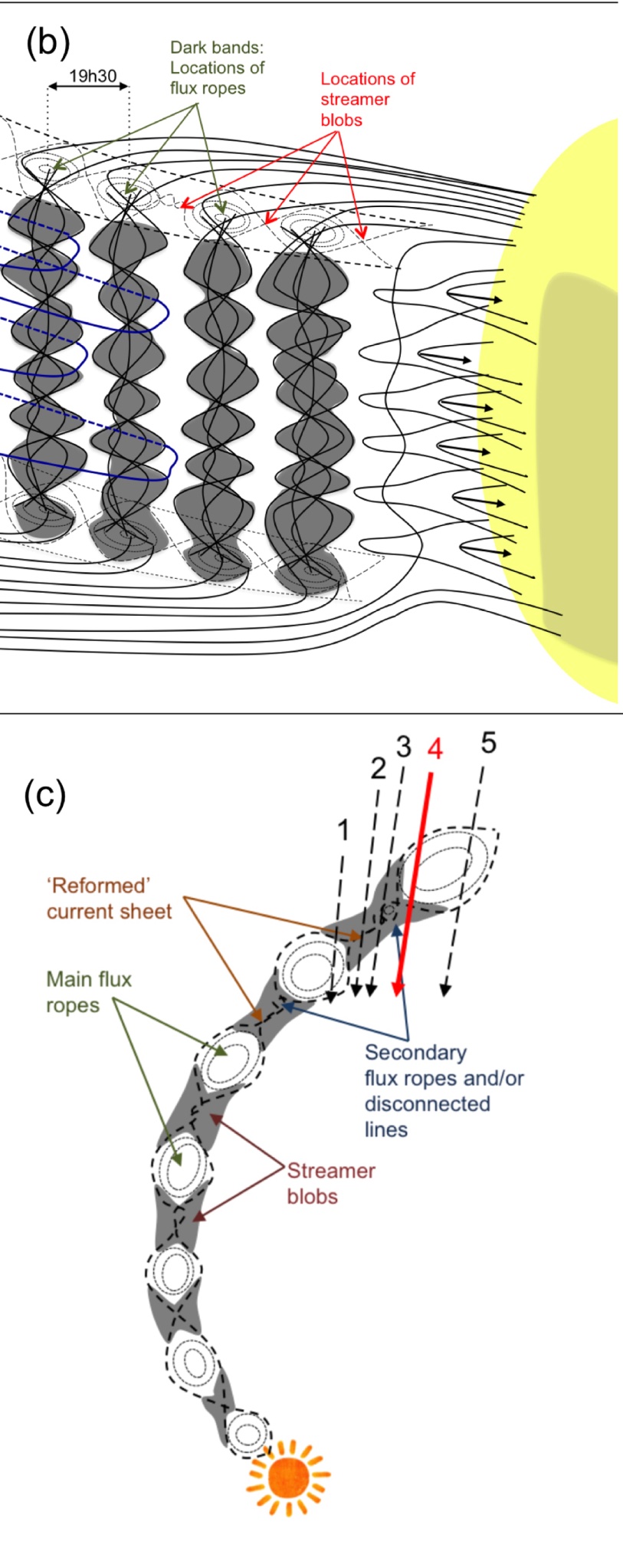}
\caption{Schéma de la génération de champs torsadés et de blobs de streamers dans une HCS vue de face (panneau b) et de côté (panneau c). Les zones grisées représentent soit les champs torsadés dans le panneau b, soit les blobs de streamer dans le panneau c. Figure extraite de \citet[][Figure 12]{Sanchez-Diaz2017b}.
\label{fig:Sanchez2017b_fig12_FR}}
\end{figure*}

Une illustration est donnée dans la figure \ref{fig:Sanchez2017b_fig12_FR} où des champs torsadés sont générées le long d'une HCS qui est vue de face dans ce cas. Dans cette image, les " blobs " brillants observés en imagerie WL correspondent à des structures de densité situées aux interstices de champs torsadés successifs. Les couches de courant qui se forment entre les champs torsadés peuvent être favorables à la génération de structures transitoires supplémentaires à des échelles plus petites et avec des périodicités plus élevées, jusqu'aux échelles de temps qui ont été mesurées in situ \citep{Viall2010,Viall2015,Kepko2016}. Nous verrons dans la section \ref{subsec:dynamics_tearing_3D} que cette interprétation est cohérente avec les récents travaux de modélisation MHD 3-D de \citet{Reville2022}, où les champs torsadés magnétiques sont produits par la reconnexion entre les boucles coronales qui ont été étirées dans l'atmosphère, et que les structures transitoires plus petites et plus fréquentes se forment par une reconnexion magnétique supplémentaire par le biais de l'instabilité de déchirement. \\

Tous les processus discutés ci-dessus impliquent une reconnexion magnétique qui libère dans le vent solaire lent le plasma qui était initialement confiné dans la couronne le long des boucles magnétiques. Comme nous le verrons, les boucles magnétiques sont particulièrement enrichies en éléments à faible FIP et cette reconnexion pourrait contribuer à enrichir le vent solaire lent en éléments à faible FIP comme discuté dans la section \ref{subsubsec:intro_SSW_composition_FR}. 

\subsubsection{Les frontières ouvert-fermé comme régions sources potentielles du vent lent}
\label{subsec:intro_Sweb_FR}

Suite aux sections précédentes, nous considérons maintenant d'autres frontières ouvert-fermé comme des régions sources potentielles du vent lent, et nous discutons des processus qui pourraient permettre à ce vent de se former loin des streamers bipolaires. L'idée que la reconnexion magnétique serait susceptible de se produire continuellement dans certaines régions de la couronne solaire remonte à la première explication de la rotation quasi-rigide des trous coronaux \citep{Nash1988, Wang1988}. Cette rotation quasi-rigide a été interprétée comme l'effet d'une vague de reconnexion magnétique se produisant entre le champ magnétique ouvert du trou coronal et les boucles fermées ancrées dans la photosphère située en dessous et qui est en rotation différentielle \citep{Wang1988}. Ce phénomène a ensuite été confirmé par des simulations MHD 3-D complètes de la couronne solaire \citep{Lionello2005}. \\

Il a également été démontré que la reconnexion magnétique peut se produire même en l'absence de lignes de champ magnétique antiparallèles, c'est-à-dire dans d'autres couches que, par exemple, les HCS standard, appelées couches quasi-séparatrices (QSL) \citep{Priest1995,Demoulin1996}. Comme nous le verrons dans le paragraphe suivant, les QSL sont des couches minces qui relient des lignes de champ ouvertes de même polarité mais d'origines différentes à la surface solaire. Les QSL peuvent donc se former au-dessus des pseudo-streamers par exemple (voir section \ref{subsubsec:intro_streamers_FR}). Par conséquent, bien que les QSLs ne présentent pas une inversion nette du champ magnétique comme dans la HCS, elles présentent toujours de forts gradients dans leur connectivité magnétique. Comme dans la HCS, des courants électriques ont également été détectés dans les QSLs où ils déclenchent la reconnexion magnétique \citep{Aulanier2005,Aulanier2006}. La reconnexion magnétique dans les QSLs explique au moins en partie la variabilité observée du vent solaire lent, même loin de la HCS et des streamers. \\

\begin{figure*}[]
\centering
\includegraphics[width=0.95\textwidth]{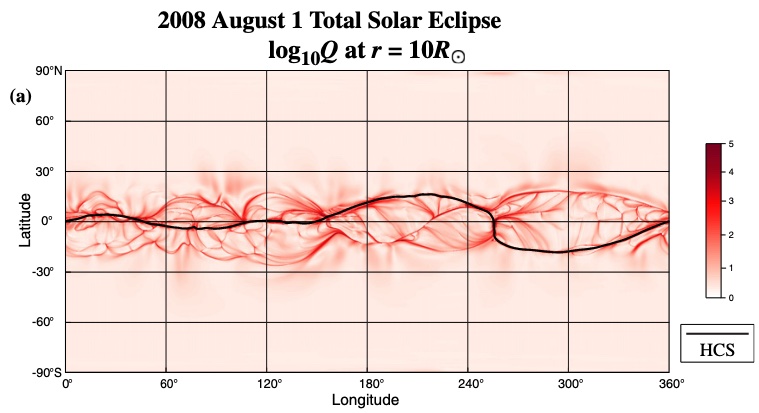}
\caption{Carte Carrington en latitude-longitude du squashing factor (couleur tracée sur une échelle logarithmique) à $10\ R_\odot$ dérivée à partir de la simulation magnéto-hydrodynamique 3-D du code MAS \citep{Antiochos2011}. Figure prise de \citet[][Figure 7]{Antiochos2011}.
\label{fig:Antiochos2011_fig7_FR}}
\end{figure*}

Plus tard, \citet{Antiochos2011} ont soutenu que les frontières ouvert-fermé doivent être omniprésentes et former un réseau appelé toile S (ou toile de couches séparatrices et quasi séparatrices, en anglais S-web), des frontières qui sont observées même pendant les périodes de faible activité solaire comme le montre la figure \ref{fig:Antiochos2011_fig7_FR} pour le minimum du cycle solaire 23. La S-web a été cartographiée en calculant à partir des champs magnétiques vectoriels 3-D, tels que les résultats des modèles MHD ou magnétostatiques 3-D décrits dans la section \ref{sec:modeling}, un paramètre appelé facteur d'écrasement (ou "squashing factor" en anglais) pour chaque ligne de champ. Ce facteur quantifie à quel point des lignes de champ magnétique contiguës à une altitude de référence dans la haute couronne, divergent les unes des autres vers la surface solaire au niveau de leur point d'ancrage photosphérique \citep{Titov2007}. Un squashing factor élevé signifie donc que deux lignes de champ adjacentes dans la haute couronne sont largement séparées à leur point d'ancrage photosphérique. Puisque les boucles coronales ont tendance à forcer une séparation significative des lignes de champ ouvertes qui les recouvrent, ces systèmes seront marqués par des squashing factors élevés et ce dernier paramètre est une bonne estimation de l'emplacement des frontières ouvert-fermé. De plus, ces systèmes ont tendance à être associés à des cisaillements magnétiques plus importants, connus sous le nom de couches quasi-séparatrices \citep{Demoulin1997} et le squashing factor donne des informations topologiques sur l'endroit où les cisaillements magnétiques sont susceptibles de se produire dans le champ coronal \citep{Titov2011}. Les boucles coronales observées sous les streamers bipolaires et unipolaires (pseudo-streamers) auront tendance à forcer une séparation significative des lignes de champ ouvertes des streamers sous forme d'arcs dans les cartes de Carrington en lumière blanche et couvriront une région de la couronne plus large que la ceinture de streamers.\\

Par conséquent, il existe une association claire entre les streamers brillants observés dans les cartes de Carrington en lumière blanche et les régions où des squashing factors élevés sont susceptibles d'appraître. Les squashing factors les plus élevés (couleurs rouge foncé) sur la figure \ref{fig:Antiochos2011_fig7_FR} correspondent à l'extrémité des streamers bipolaires et suivent généralement la forme de la ceinture de streamers. Les squashing factors intermédiaires (couleurs rougeâtres) se produisent davantage le long des pseudo-streamers et couvrent une région bien plus large que la ceinture de streamers. 

\citet{Crooker2012} ont réalisé une cartographie balistique du vent lent mesuré in situ jusqu'à la haute couronne et ont comparé les régions sources estimées aux cartes du squashing factor. Ils ont trouvé une bonne association entre les régions sources du vent lent et les régions de squashing factor élevé qui peuvent se produire bien au-delà de la ceinture de streamers. Ces associations suggèrent que le vent lent provient de régions coronales où les boucles magnétiques sont adjacentes à des champs magnétiques ouverts et tendent à développer des cisaillements magnétiques élevés, peut-être propices à l'apparition de la reconnexion magnétique. Cela a été confirmé par \citet{Baker2009} qui a observé dans les données \textit{Hinode-EIS} des écoulements omniprésents le long des QSL qui se forment au-dessus des régions actives. \\

\begin{figure*}[]
\centering
\includegraphics[width=0.7\textwidth]{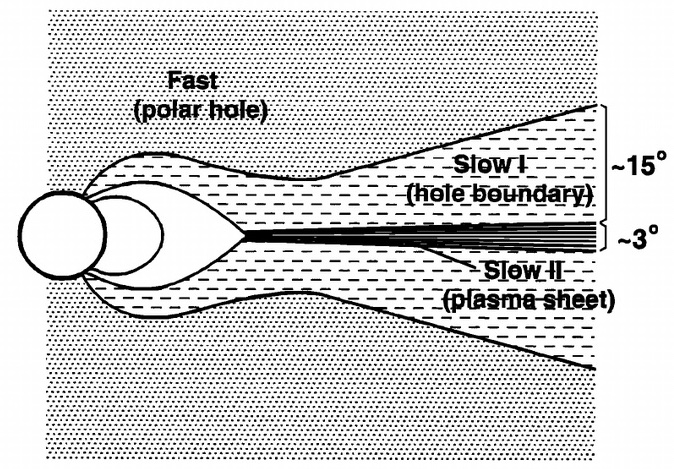}
\caption{Schéma de la distribution des vents solaires rapides et lents dans le plan méridional. Figure tirée de \citet[][Figure 12]{Wang2000}.
\label{fig:Wang2000_fig12_FR}}
\end{figure*}

L'héliosphère est donc probablement alimentée en plasma dense même à une certaine distance de la HCS, comme l'illustre la figure \ref{fig:Wang2000_fig12_FR}. Cette interprétation complète le scénario initial décrit au début de ce chapitre d'un vent lent qui provient de l'extrémité des streamers bipolaires et qui forme la dense HPS. Ceci est également cohérent avec une HPS qui ne représente qu'une sous-partie très brillante des observations de la couronne en lumière blanche réalisées par les coronographes ou pendant les éclipses solaires totales, le reste des émissions étant dû à des plasmas légèrement moins denses qui proviennent probablement d'interfaces ouvert-fermé.

%% file: chapters/Instruments_tools.tex
\chapter{Instrumentation and tools}
\label{cha:instruments_tools}

\minitoc

\section{Instrumentation}

Getting a complete picture of the Sun from its surface to the solar wind is a challenging task for astronomers and instrument designers. Detectors designed to observe the solar disk can be ineffective at capturing the much fainter emission from the upper corona. The faint solar corona is commonly revealed by using an artificial occulter placed ahead of the detector that blocks emissions from the solar disk. This technique usually induces some stray light which can be partially removed depending on the occulter/detector distance and the quality of the optical alignment. \\

For white-light observations some ideal conditions are certainly met when the Moon acts as a natural occulter of the solar disk during total solar eclipses. Such events were already exploited in the late 1800's where first images of the global shape of the solar corona were produced \citep[see e.g.][]{Maunder1899}. Now many astronomers continue to take advantage of these rare events to take highly detailed pictures of the solar corona which can be highly valuable to study the magnetic field topology \citep[see e.g. ][]{Boe2020}. The first ground-based coronagraphs made of an artificial occulter were designed by \citet{Lyot1930} and \citet{Evans1948} followed by subsequent space-based coronagraphs (e.g. \textit{CORONNASCOPE-II}, \textit{SOLWIND} and \textit{Skylab}) that allowed for more systematic observations of the solar corona beside the rare eclipses \citep[see e.g.][]{Bohlin1970,Wang1992,Guhathakurta1996}. \\

In the past decades, the \textit{Solar and Heliospheric Observatory} \citep[SoHO: ][]{Domingo1995} has offered a continuous monitoring of the solar corona in white-light, through a set of two coronagraphs (C2 and C3), part of the \textit{Large-Angle Spectrometric Coronograph} \citep[LASCO: ][]{Brueckner1995} complemented by many other remote-sensing and in situ instruments that will not be presented here. The \textit{SoHO} observations gave the opportunity to study the evolution of the global shape of the solar corona over several solar cycles and has greatly contributed to the evaluation of coronal models from the simple PFSS reconstructions \citep[see e.g. ][]{Wang1998,Wang2000,Wang2007} to more global coronal models \citep[see e.g.][]{Gibson2003,Thernisien2006,dePatoul2015,PintoRouillard2017}. Later on, the \textit{Solar TErrestrial RElations Observatory} \citep[STEREO: ][]{Kaiser2008} space mission composed of two identical payloads placed on twin spacecrafts revolutionized heliospheric science by imaging simultaneously the solar wind from several vantage points. The mission was designed to offer a stereoscopic view of the Sun when the two satellites (\textit{STEREO-A} and \textit{STEREO-B}) orbit temporarily the Lagrange points L4 and L5 located at 60 degrees from the Earth viewpoint. Very unfortunately the contact was lost with the \textit{STEREO-B} spacecraft during superior solar conjunction in the year 2014. A better tracking of Coronal Mass Ejections (CMEs) and the solar wind was therefore possible during their propagation in the heliosphere until they reach the Earth's magnetosphere. Combining coronagraphic images from \textit{STEREO} and \textit{SoHO} taken at multiple vantage points also allowed to update more frequently the mapping of helmet streamers in white-light synoptic maps, and to derive "synchronic maps" of the solar atmosphere \citep[see ][]{Sasso2019}. \\

The recently launched \textit{Parker Solar Probe} \citep[PSP: ][]{Fox2016} mission started to provide unprecedented close-up views of the solar wind and CMEs with the inner and outer telescopes of the \textit{Wide-Field Imager for Parker Solar Probe} \citep[WISPR: ][]{Vourlidas2016}. The \textit{Solar Orbiter} \citep[SolO: ][]{Muller2013,Muller2020} mission will supplement the \textit{PSP} observations with a comprehensive set of remote-sensing instruments comprising both of full-disk imagers and coronagraphs. Upcoming missions are also about to pursue the \textit{SoHO} and \textit{STEREO} missions, of which the \textit{Polarimeter to UNify the Corona and Heliosphere} (\textit{PUNCH}\footnote{\textit{PUNCH} website: \url{https://punch.space.swri.edu/}}). \\

In the following paragraphs of this section, I focus on the instruments that are exploited in this thesis. The coronographs on board the \textit{SoHO} and \textit{STEREO} missions have been used in several applications and are introduced in section \ref{subsec:inst_LASCO} and \ref{subsec:inst_STEREO} respectively. Of particular interest here, they have been helpful to support in situ measurements of streamer blobs (see chapter \ref{cha:dynamics}), to constrain coronal and heliospheric models in a systematic manner (see section \ref{sec:WL_opti}) and to study the shape of the streamer belt (see chapter \ref{cha:stationnary}). The unprecedented images taken by \textit{WISPR} which is introduced in section \ref{subsec:inst_WISPR} have been critical to carry out this thesis, for studying the inner structure of the streamer belt (see chapter \ref{cha:stationnary}), and for understanding the temporal and spatial characteristics of streamer blobs (see chapter \ref{cha:dynamics}).

\subsection{The \textit{SoHO-LASCO} imaging suite}
\label{subsec:inst_LASCO}

The \textit{Solar and Heliospheric Observatory} \citep[SoHO: ][]{Domingo1995} is a collaborative ESA-NASA mission launched on December 2, 1995. The spacecraft is equipped with the \textit{Large-Angle and Spectrometric Coronagraph} \citep[LASCO: ][]{Brueckner1995}, comprising three white-light coronagraphs but only \textit{LASCO-C2} and \textit{LASCO-C3} are still operating nominally. The near Earth orbit of \textit{SoHO} at Lagrange point L1 allows a continuous monitoring of the Sun by providing near real time imaging that is essential for space weather applications and as a support to operations for other missions. \\

\begin{figure*}[]
\centering
\includegraphics[width=0.8\textwidth]{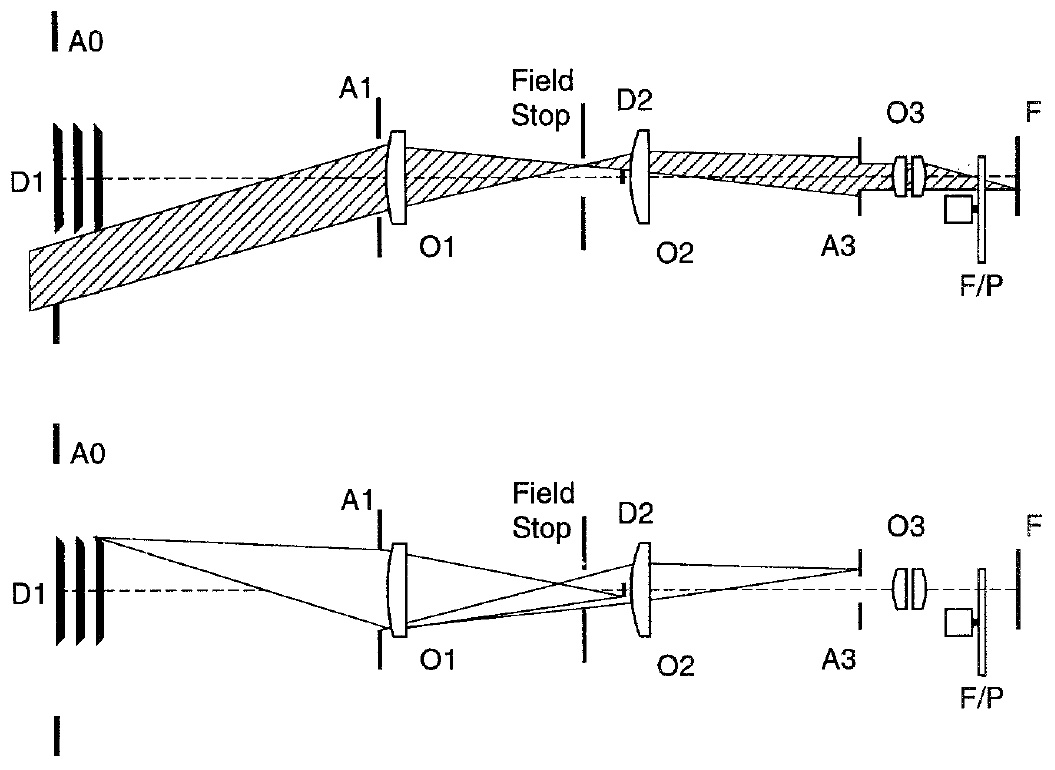}
\caption{Schematics of the \textit{LASCO-C2} and \textit{LASCO-C3} coronagraphs. The top panel shows the path followed by the incoming light from the target until it reaches the detector. The bottom panel illustrates how the residual light that is diffracted at the edges of the external occulter D1 and apertures is deviated to not reach the detector. Figure taken from \citet[][Figure 3]{Brueckner1995})
\label{fig:LASCO_instrument}}
\end{figure*}

The two \textit{LASCO-C2} and \textit{LASCO-C3} coronagraphs combined cover a wide portion of the solar corona with their respective field-of-view (FOV) ranging from $1.5\ R_\odot$ to $6\ R_\odot$ and from $6\ R_\odot$ to $30\ R_\odot$ \citep{Brueckner1995}. As opposite to the \textit{LASCO-C1} coronagraph, \textit{LASCO-C2} and \textit{LASCO-C3} are occulted by an external occulter (labelled D1 in Figure \ref{fig:LASCO_instrument}). They are both based on the design invented by \citet{Lyot1930} to remove scattered light from an ordinary objective lens telescope. 

The incident light is recorded by a 1024x1024 pixel CCD camera at the image plane F. Due to the fall of the corona's brightness with height, the two telescopes have different sensitivities which range $2\times10^{-7}$ to $5\times10^{-10}$ for \textit{LASCO-C2} and $3\times10^{-9}$ to $1\times10^{-11}$ for \textit{LASCO-C3} (in solar brightness $B_\odot$ unit). The approximate spatial resolution (taken as 2 pixels) is approximately $23$ and $112$ arc second for \textit{LASCO-C2} and \textit{LASCO-C3} respectively. The 1024x1024 pixel images are first compressed on board by the flight software before being sent to the ground to meet the limitations from the telemetry.

\subsection{The \textit{STEREO-SECCHI} imaging suite}
\label{subsec:inst_STEREO}
The \textit{Solar TErrestrial RElations Observatory} \citep[STEREO: ][]{Kaiser2008} is a NASA mission launched on October 25, 2006 on a Delta II rocket. It was a revolutionary mission in the sense that it was composed of twin spacecrafts - one ahead of Earth in its orbit, the other trailing behind, giving a first 3-D view of coronal mass ejections. \\

\begin{figure}
	\centering
	\includegraphics[width=0.8\textwidth]{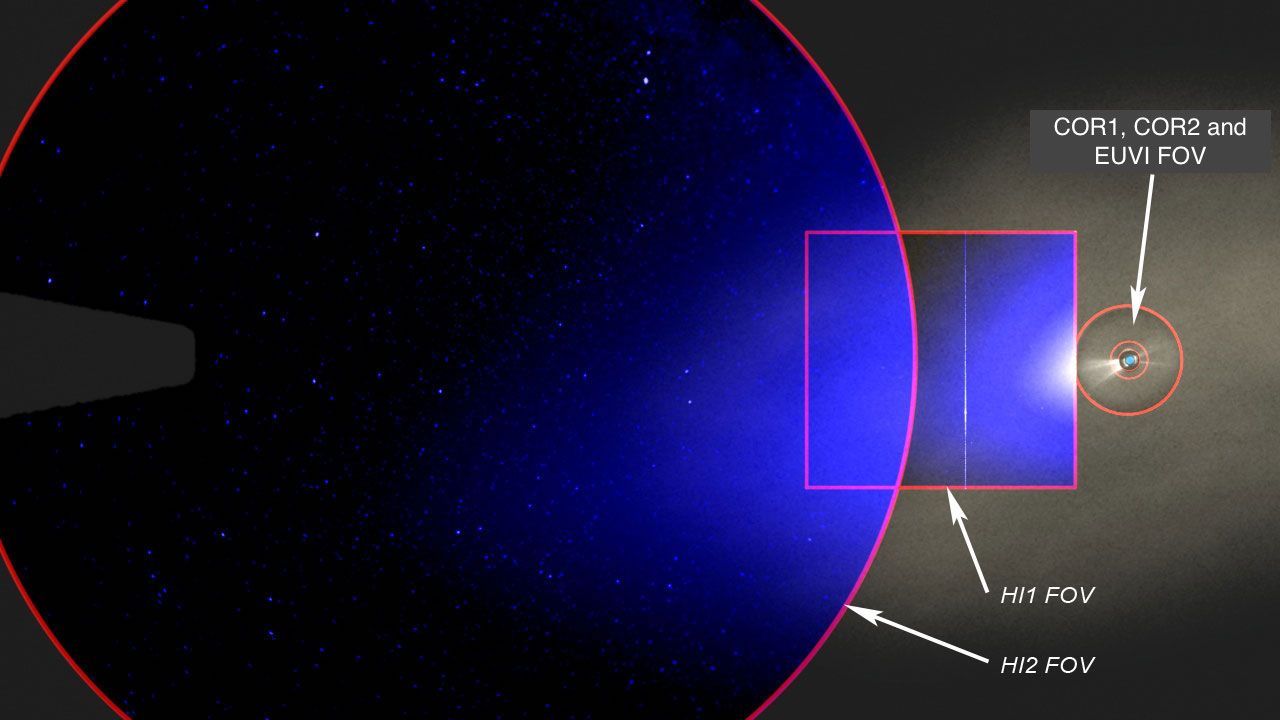}
	\caption{Combined field-of-view of the \textit{SECCHI} telescopes on board \textit{STEREO-A}. (Credit: \textit{STEREO/GSFC}) }
	\label{fig:secchi_fov}
\end{figure}

The two spacecrafts \textit{STEREO-A} and \textit{STEREO-B} are equipped with the \textit{Sun Earth Connection Coronal and Heliospheric Investigation} \citep[SECCHI: ][]{Howard2008}, an observatory comprising: an extreme ultraviolet imager (\textit{EUVI}), two coronagraphs (inner: \textit{COR1} and outer: \textit{COR2}) and two heliospheric imagers (inner: \textit{HI-1} and outer: \textit{HI-2}). The heliospheric imagers offer an uninterrupted view of the solar wind from the Sun to Earth-like distances. The \textit{HI-1} and \textit{HI-2} cameras unlike coronagraphs are not Sun-centered and instead the center of their field-of-view points respectively $14^\circ$ and $50^\circ$ azimuth (or elongation) angle away from the Sun. Furthermore the combined \textit{HI-1} and \textit{HI-2} FOV provide a coverage over solar elongation angles from $4.0^\circ$ to $88.7^\circ$ at the viewpoints of the two spacecrafts \citep{Eyles2009}. Since the two spacecrafts are nearly at $1\ AU \approx 215$ solar radii, \textit{HI-1} and \textit{HI-2} then provide a very wide coverage from $\approx 15$ to $\approx 333$ solar radii in the plane of the sky. Figure \ref{fig:secchi_fov} illustrates the wide coverage of the \textit{SECCHI} telescopes on board \textit{STEREO-A}. Thanks to the large field-of-view of \textit{HI-1} and \textit{HI-2}, density perturbations in the solar wind can be tracked all the way to spacecrafts situated in the inner heliosphere or even to the Earth. Unfortunately the \textit{STEREO-B} spacecraft is not operating anymore due to a loss of contact from October 1st 2014.

\subsection{The \textit{PSP-WISPR} white-light telescopes}
\label{subsec:inst_WISPR}

The NASA's \textit{Parker Solar Probe} \citep[PSP: ][]{Fox2016} mission was launched on 2018 August 12. One of its main objectives is to shed new light on the heating and acceleration processes of the solar wind. For this purpose the mission was designed to fill the data gap between the corona that can be observed during total solar eclipses and by coronagraphs (i.e. below $\approx 10\ R_\odot$), and the in situ measurements taken by the past Ulysses and Marineer II missions (i.e. above $0.3\ AU$). The spacecraft follows elliptical orbits where perihelion is progressively brought closer to the Sun thanks to subsequent gravitational assists with Venus. After five Venus flyby, the \textit{PSP} closest approach point has been reduced from $31\ R_\odot$ down to $10\ R_\odot$, to reach its final objective at $8\ R_\odot$ by 2024. \\

\begin{figure*}[]
\centering
\includegraphics[width=0.6\textwidth]{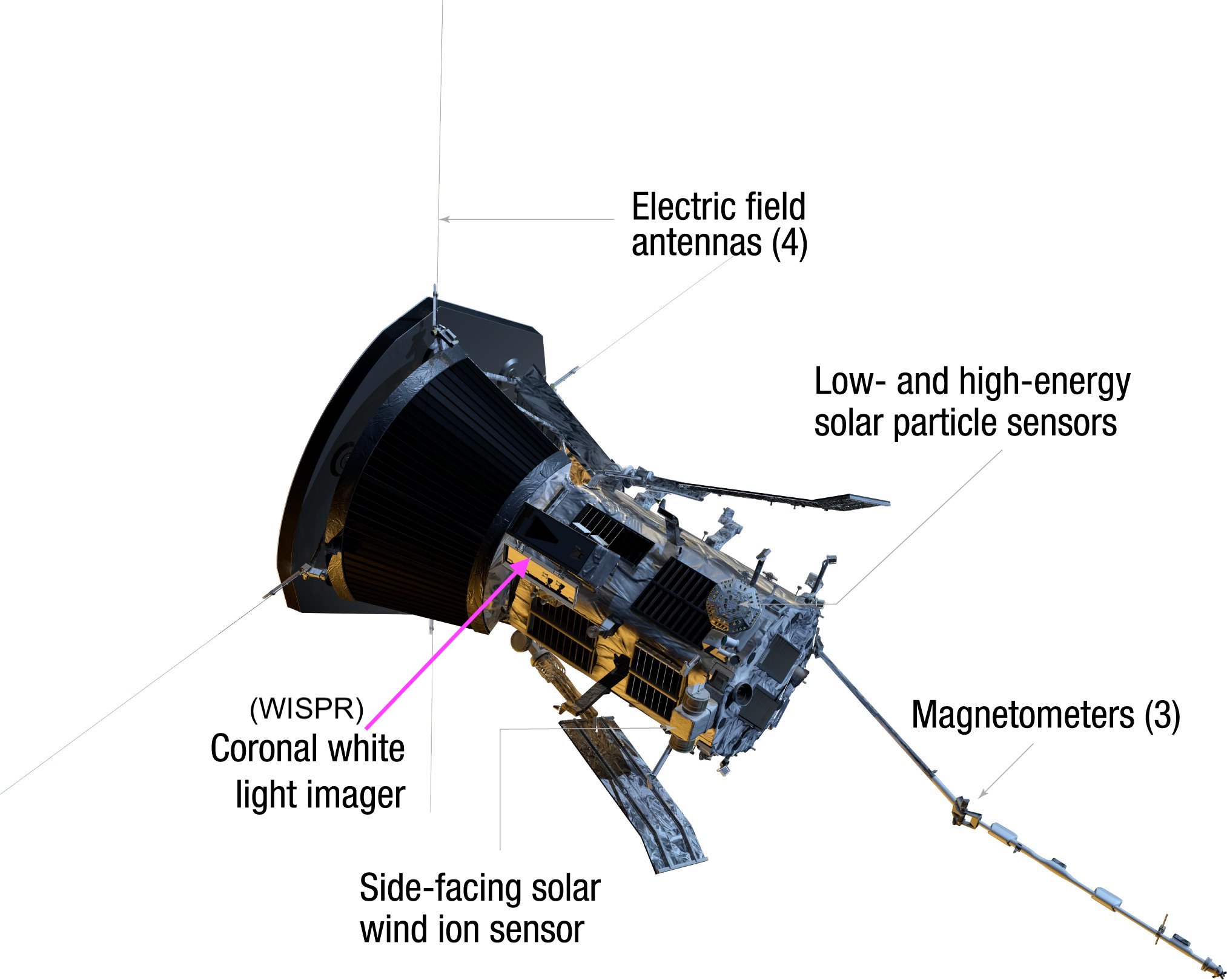}
\caption{Position of the \textit{WISPR} white-light imager on the ram side of \textit{PSP} (NASA/JHUAPL).
\label{fig:PSP_3D}}
\end{figure*}

\begin{figure*}[]
\centering
\includegraphics[width=0.8\textwidth]{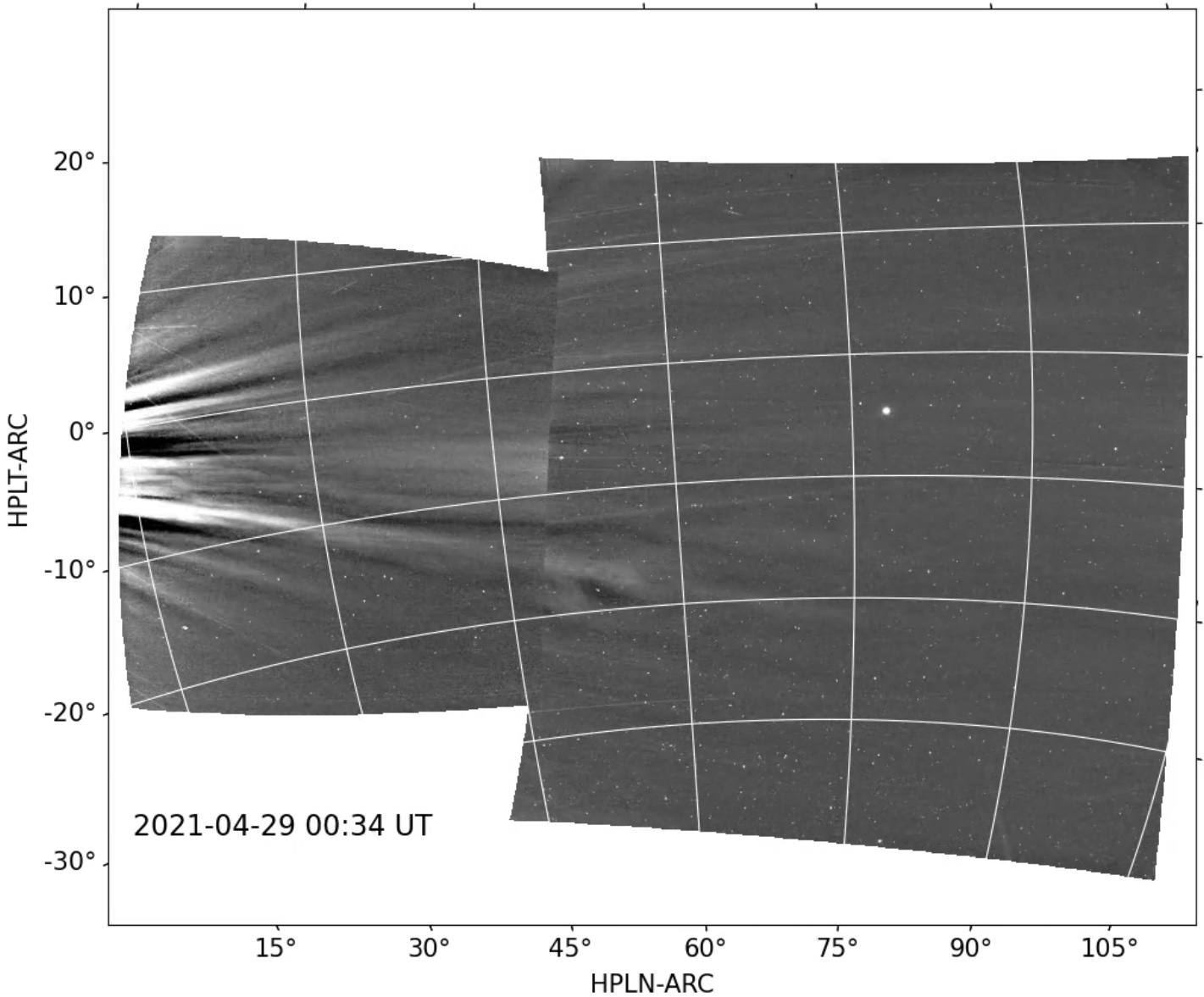}
\caption{Merged image from the inner \textit{WISPR-I} and \textit{WISPR-O} outer telescopes around the 8th perihelion on 2021-04-29 at 00:34 UT. \textit{PSP} was located at a heliocentric distance of $16.5\ R_\odot$. The helioprojective grid (centered at \textit{PSP}) is plotted with elongation angles given as the abscissa. Image produced by the \textit{WISPR} team.
\label{fig:WISPR_merged_img}}
\end{figure*}

\textit{PSP} is equipped with a heliospheric imager that records the brightness of the corona from a vantage point situated in the corona. The \textit{Wide-Field Imager for Solar PRobe} \citep[WISPR: ][]{Vourlidas2016} is mounted on the ram side of the spacecraft (see Figure \ref{fig:PSP_3D}), so the solar wind structures can be imaged prior to their in situ measurement. As opposite to a coronagraph, \textit{WISPR} is not facing the Sun directly so there is no need for an occulter to hide the bright solar disk. The \textit{WISPR} field-of-view is radially offset from the Sun and is centered $10^\circ$ below the ecliptic plane. \textit{WISPR} offers a large FOV thanks to its two telescopes which combined, cover a range of heliocentric distances from $2.2\ R_\odot$ to $20\ R_\odot$ at closest approach ($0.044\ AU$) and from $9.5\ R_\odot$ to $83\ R_\odot$ at $0.25\ AU$. The inner (\textit{WISPR-I}) telescope extends in elongation angles (azimuthal angle away from the Sun) from $13.5^\circ$ to $53^\circ$ and the outer telescope (\textit{WISPR-O}) extends from $50^\circ$ to $108^\circ$ (illustrated in Figure \ref{fig:WISPR_merged_img}). At closest approach \textit{WISPR} can reach a spatial resolution of $17$ arc second, about the same as the \textit{LASCO-C2} coronagraph (in 1 AU equivalent quantities) but with a wider FOV \citep[see][Table 1]{Vourlidas2016}. At further distances from the Sun ($0.25\ AU$), \textit{WISPR} performances are similar to those of the heliospheric imager \textit{HI-1} of the \textit{STEREO-A} spacecraft with a spatial resolution of $94$ arc second (again in 1 AU equivalent quantities) and a total FOV of $9.5\ R_\odot$ to $83\ R_\odot$. These similar performances with 1 AU observatories applies only for targets far from the telescopes, but thanks to the unprecedented closeup distance of \textit{PSP} to the Sun \textit{WISPR} can image coronal structures in much greater detail. \\

\begin{figure*}[]
\centering
\includegraphics[width=0.95\textwidth]{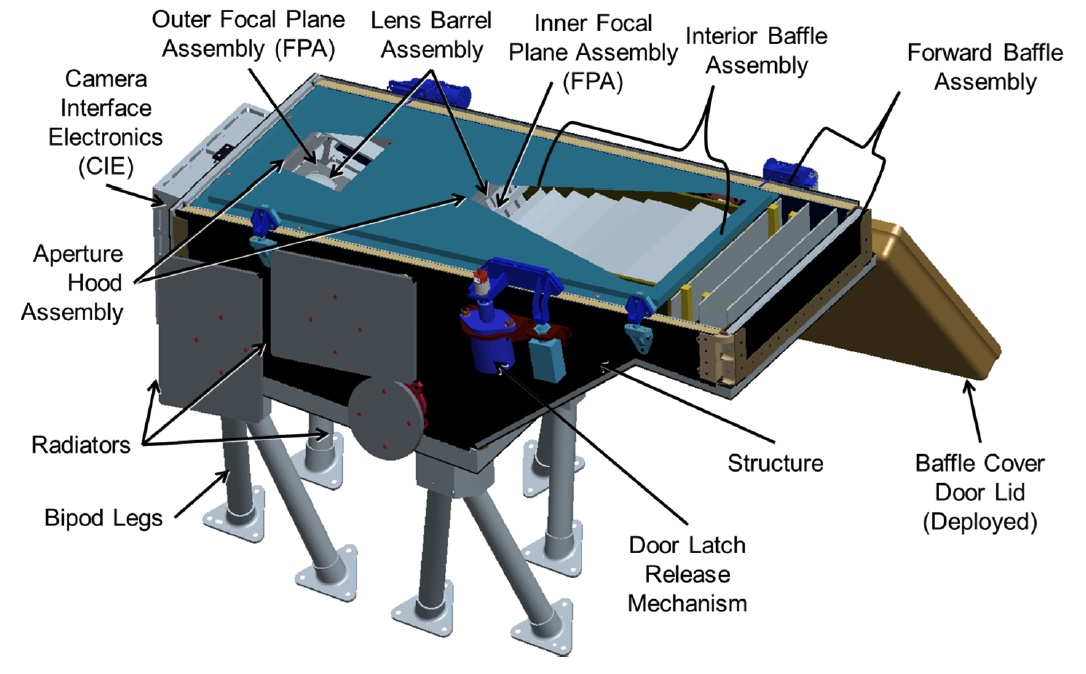}
\includegraphics[width=0.95\textwidth]{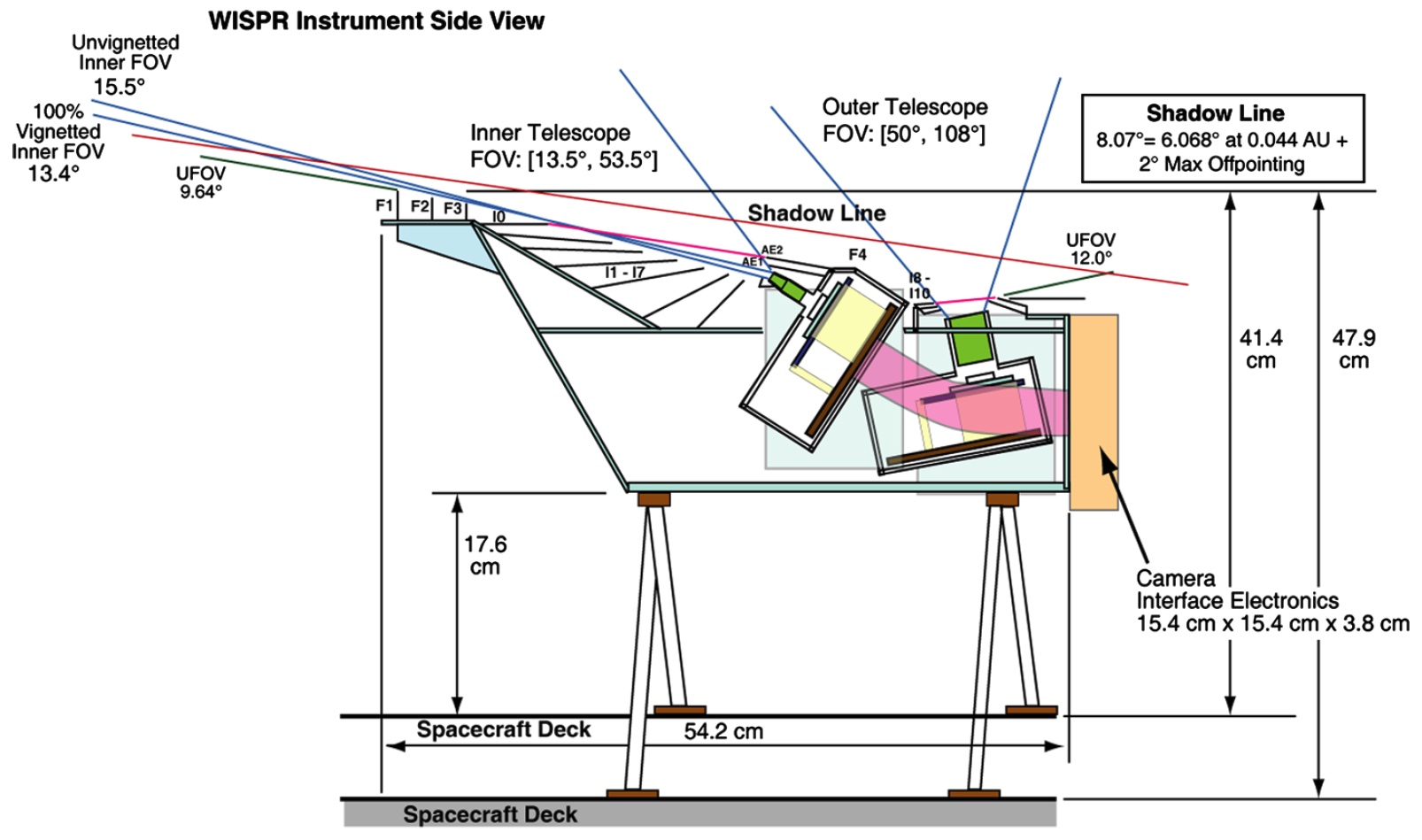}
\caption{Panel a: 3-D rendering of \textit{WISPR} (taken from \citet[][Figure 9]{Vourlidas2016}). Panel b: schematic of \textit{WISPR} (taken from \citet[][Figure 15]{Vourlidas2016}).
\label{fig:WISPR_instrument}}
\end{figure*}

The inner telescope required a complex assembly of baffles to prevent stray light from entering the telescope (see Figure \ref{fig:WISPR_instrument}a). The inner and outer telescopes are equipped with two separated large-format (2K $\times$ 2K) APS CMOS detectors (see Figure \ref{fig:WISPR_instrument}b) which limit damaging from dust particles. As expected many dust particle impacts have been recorded during the first \textit{PSP} encounters \citep{Szalay2020}. The heat shield in front of the spacecraft deviates dust particles whose trajectories then appear in the images. Sometimes images are too deteriorated by dust particle rays to be exploitable, a drawback that becomes more frequent as \textit{PSP} is getting closer to the Sun. Hopefully most \textit{WISPR} images are not too polluted by dust particles and remain highly valuable for science applications. \\

A great challenge has been the calibration of \textit{WISPR} images which is explained in details in \citet{Hess2021}. Level-2 \textit{WISPR} images undergo a series of corrections that are similar to those currently used for other space-based telescopes. They include corrections for: the detector (bias removal, linearity, exposure time normalization), the optical system (stray-light, vignetting) and photometry (calibration factor). For most science applications, level-2 images can not be directly exploited because they are saturated by the so-called F-corona, the light scattered by dust particles. To study the corona one needs to reveal the light scattered by coronal electrons, the K-corona which appears much fainter in coronagraphs and heliospheric imagers. The transformation from level-2 to level-3 then involves the subtraction of a F-corona background to the raw images. For the \textit{SoHO} and \textit{STEREO} observatories orbiting $1\ AU$, background images are constructed at a weekly or monthly cadence to derive an average brightness for the F-corona \citep{Morrill2006}. This technique had to be revisited for space-based observatories that were rapidly moving along their orbit. A more sophisticated method has recently been developed which allows the construction of a F-corona model from a single image \citep{Stenborg2018}. They exploited four years of the heliospheric imager \textit{HI-1} on board \textit{STEREO-A} to parametrize the shape of the dust cloud according to orbital parameters. The technique remains more or less the same for \textit{WISPR}, while some tuning of the F-corona model was needed (a future paper will be released by the \textit{WISPR} team). \textit{WISPR} level-2 and level-3 images and movies are provided by the National Research Laboratory\footnote{\textit{WISPR} data access: \url{https://wispr.nrl.navy.mil/wisprdata}}.

\subsection{Synoptic photospheric magnetic maps}
\label{subsec:magmaps}
The performances of coronal and heliospheric models can be seriously affected by the choice of the surface magnetograms that are specified at the inner boundary of the simulated domain. In this thesis, a considerate effort has been dedicated to benchmark different sources of surface magnetic maps, in order to improve the accuracy of the models introduced in section \ref{sec:modeling}. This is discussed in details in section \ref{sec:WL_opti} as well as in \citet{Poirier2021} and \citet{Badman2022}. \\

Synoptic maps of the surface magnetic field are basically built by combining full disk magnetograms data over a full solar rotation. Magnetograms are obtained by magnetographs which measure the amplitude of the magnetic field along the line-of-sight mostly using the Zeeman effect. Basically the Zeeman effect relates the strength of the magnetic field with an alteration of the light emitted by the photospheric plasma. Different photospheric spectral lines can be used for that purpose, such as the Nickel (Ni) $6768$ Angström (\si{\angstrom}) line as done by the Global Oscillation Network Group (GONG).

In the following paragraph I introduce the different sources of synoptic magnetic maps that have been used throughout this thesis.

\subsubsection{The Wilcox Solar Observatory}
\label{subsubsec:wso}

The Wilcox Solar Observatory (WSO) has provided daily ground-based observation of the mean photospheric magnetic field since 1976. The WSO comprises of one single ground based observatory located near the Stanford University in California, therefore WSO operations are limited by the diurnal cycle. They provide synoptic maps of the photospheric magnetic field averaged over a full solar rotation or Carrington rotation, which we exploit in this thesis. The magnitude of the magnetic field is determined with a precision of $0.1$ Gauss (G), and is mapped on a heliographic latitude versus longitude map of $5^\circ$ angular resolution \citep{Duvall1977}. They have been widely used in the scientific community and precious to study the Sun evolution over more than four solar cycles to this date. WSO synoptic maps are available at the Stanford University WSO online archive \footnote{WSO data access: \url{http://wso.stanford.edu/forms/prsyn.html}}.

\subsubsection{The Global Oscillation Network Group}
\label{subsubsec:gong}

The Global Oscillation Network Group (GONG), funded by the National Solar Observatory (NSO) integrated synoptic program, provides synoptic surface magnetic maps at a 1-hour cadence. The continuous monitoring of the Sun's disk is possible thanks to their various ground observatories spread out over the Earth \citep[see][for further details]{Hill2018}. Approximately 10,000 full-disk 10-min average magnetograms acquired from the different sites are remapped and merged together to produce a hourly synoptic map in heliographic coordinates. We used the zero-corrected GONG data products where the open magnetic flux at the photosphere has been rebalanced between the northern and southern polar regions. A major issue in the previous GONG maps was a quasi systematic shift in latitude of the neutral line with respect to the expected location deduced from white-light observations. In the new product, called hereafter GONG-z, the instrumental bias has been smoothed out among the various sites to reduce the uncertainty on the background magnetic field from $10\ \rm{G}$ to $0.1\ \rm{G}$ \citep{Hill2018}. The zero-corrected GONG-z synoptic maps are provided via the GONG archive \footnote{GONG archive: \url{https://gong2.nso.edu/archive/patch.pl}} with a $1^\circ$ angular resolution in both heliographic longitude and latitude. 

\subsubsection{The Air force Data Assimilative Flux Transport model}
\label{subsubsec:adapt}

Synoptic maps of the surface magnetic field often suffer from a lack of observations of the solar disk on the far side as viewed from Earth. That means only half of the solar surface can be imaged at a given instant and that a full solar rotation period (27 days as viewed from Earth) is required to fully update a synoptic map. This delay can result in erroneous reconstructions of the coronal magnetic field when there are strong reconfigurations of the photospheric magnetic field on the far side that cannot be observed by Earth's ground-based observatories. Such large-scale reconfigurations of the photospheric magnetic field can be induced for instance by the emergence of an active region. \\

The Air force Data Assimilative Flux Transport (ADAPT) model tackles this issue by simulating the transport of magnetic flux across the solar surface (see \citet{Arge2010,Arge2011,Arge2013} or this document \footnote{\url{https://www.swpc.noaa.gov/sites/default/files/images/u33/SWW_2012_Talk_04_27_2012_Arge.pdf}}). The model is based on the work of \citet{Worden2000} which includes various physical mechanisms such as differential rotation, meridional flow, supergranular diffusion and random flux emergence. 

The standard synoptic maps (e.g. from WSO or GONG) are static in the sense that the acquired surface data remains in solid rotation with the Sun so that even differential rotation is not accounted for. Similarly to the standard products the ADAPT synoptic maps are continuously updated as soon as new full disk observations are available. However the novelty is in using sophisticated data assimilation techniques to combine in a coherent manner the new data with the results from the flux transport forward modeling. 

A recent improvement in the ADAPT maps has been the assimilation of additional information about the far side magnetic structures by exploiting helioseismological observations of the near side. There are many uncertainties coming from the observations and the parameters of the flux transport model itself. Therefore a large ensemble of ADAPT models are run but in practice only 12 solutions (or realizations) are given at the end in the final products. Our bench-marking of source magnetic maps discussed in section \ref{sec:WL_opti} and in \citet{Poirier2021} considers separately each of the 12 ADAPT realizations. \\ 

We exploited in this thesis synoptic maps processed by the ADAPT model from GONG data (hereafter called GONG-ADAPT), which are provided by the NSO integrated synoptic program \footnote{GONG-ADAPT maps are available here: \url{https://gong.nso.edu/adapt/maps/}}.

\section{Supporting tools}
\label{sec:tools}
In this section I introduce some of the tools that I used the most during my thesis. As already illustrated, the interpretation of in situ measurements from \textit{PSP} and \textit{SolO} can benefit greatly from a connection between those data with remote-sensing observations. This naturally led me to contribute to the development of tools that aim at fulfilling this objective. The Magnetic Connectivity Tool is the result of a large team effort to build a tool that supports coordinated campaigns during the operations of \textit{Solar Orbiter}, and by exploiting \textit{SolO}'s extended remote-sensing suite through the search of relevant targets on the solar surface (see section \ref{subsec:Connectivity_tool}). The AMDA and CLWeb tools are helpful to quickly visualize in situ data and are briefly introduced in section \ref{subsec:AMDA_CLweb}. Finally I give in section \ref{subsec:SolarSoft} some details about how I used the SolarSoft/IDL package to process remote-sensing observations and to produce synthetic images relevant to the studies that I present below in chapter \ref{cha:stationnary} and \ref{cha:dynamics}.

\subsection{The Magnetic Connectivity Tool}
\label{subsec:Connectivity_tool}
\begin{figure*}[]
\centering
\includegraphics[width=0.8\textwidth]{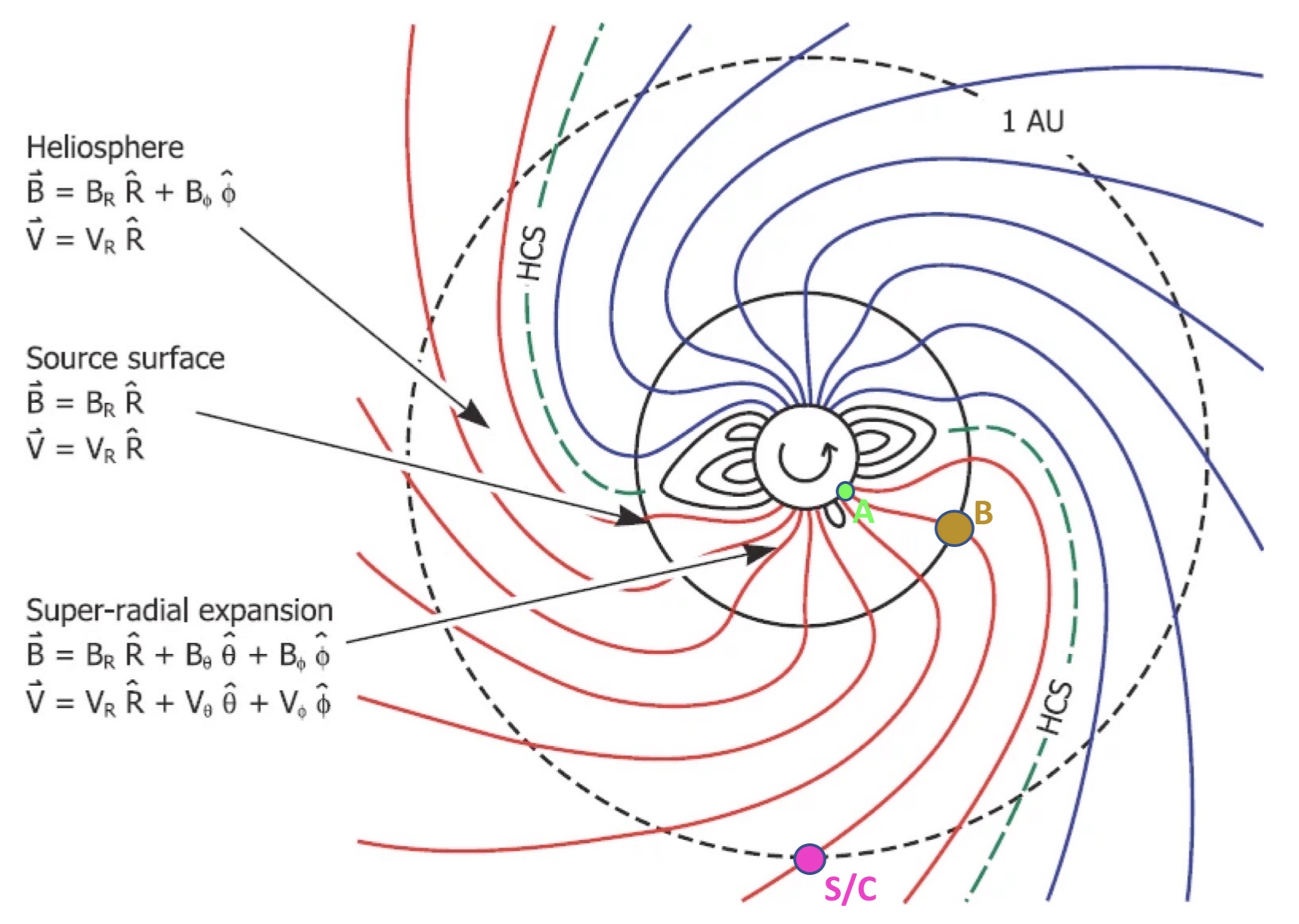}
\caption{Schematic illustrating how magnetic connectivity is established in the MCT, showing the location of a spacecraft (S/C), the intersection of the Parker Spiral (B) and the photospheric/low corona magnetic footpoint of field line (A). Schematic taken from \citet[][Figure 1]{Owens2013}.
\label{fig:MCT_diagram}}
\end{figure*}

The Magnetic Connectivity Tool (MCT) is an initiative from the Modelling and Data Analysis Working Group (MADAWG) that aims at supporting the \textit{Solar Orbiter} mission with modeling and data analysis tools \citep{Rouillard2020}. The MCT is designed to identify the possible connections of any spacecraft (S/C) in the heliosphere back to the Sun surface (see point "S/C" and "A" in Figure \ref{fig:MCT_diagram}). In science mode, the user can retrieve an estimate of the source region which generated the plasma that has been measured in situ at the spacecraft of interest. In forecast mode, the user can estimate the likely solar source region of the plasma measured in situ a posteriori. The forecast mode is particularly relevant for mission operations by providing (ground and space based) remote-sensing observatories with a list of relevant targets on the solar disk. This fulfils one of the main objective of the \textit{Solar Orbiter} mission, imaging the nascent solar wind prior to its measurement in the heliosphere. \\

\begin{figure*}[]
\centering
\includegraphics[width=0.9\textwidth]{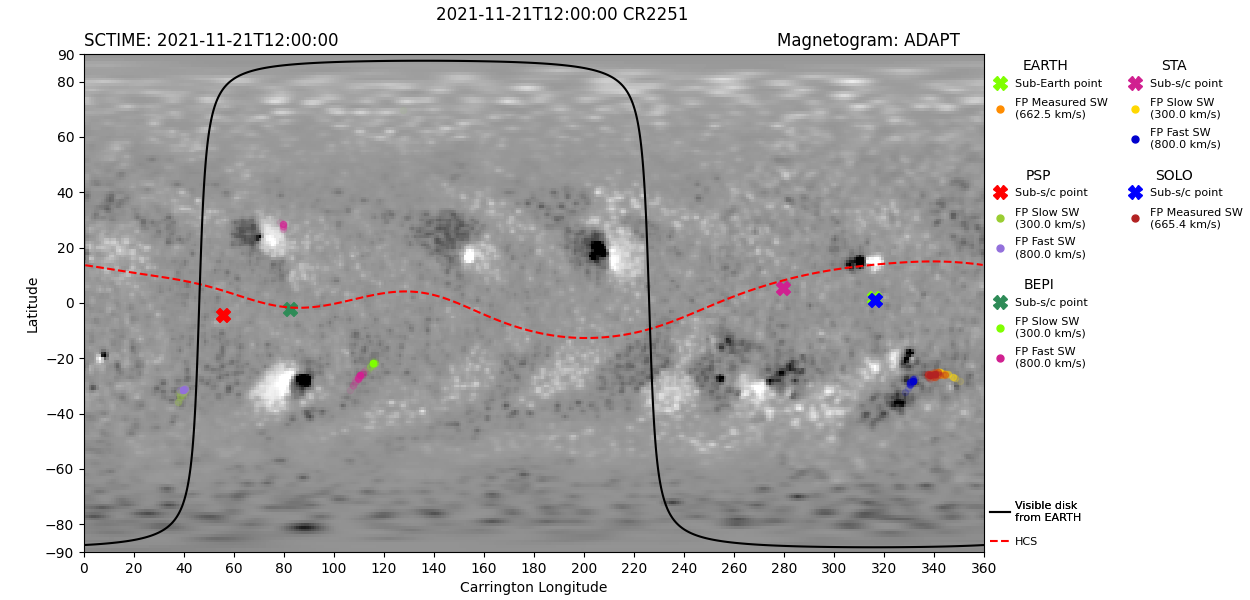}
\includegraphics[width=0.9\textwidth]{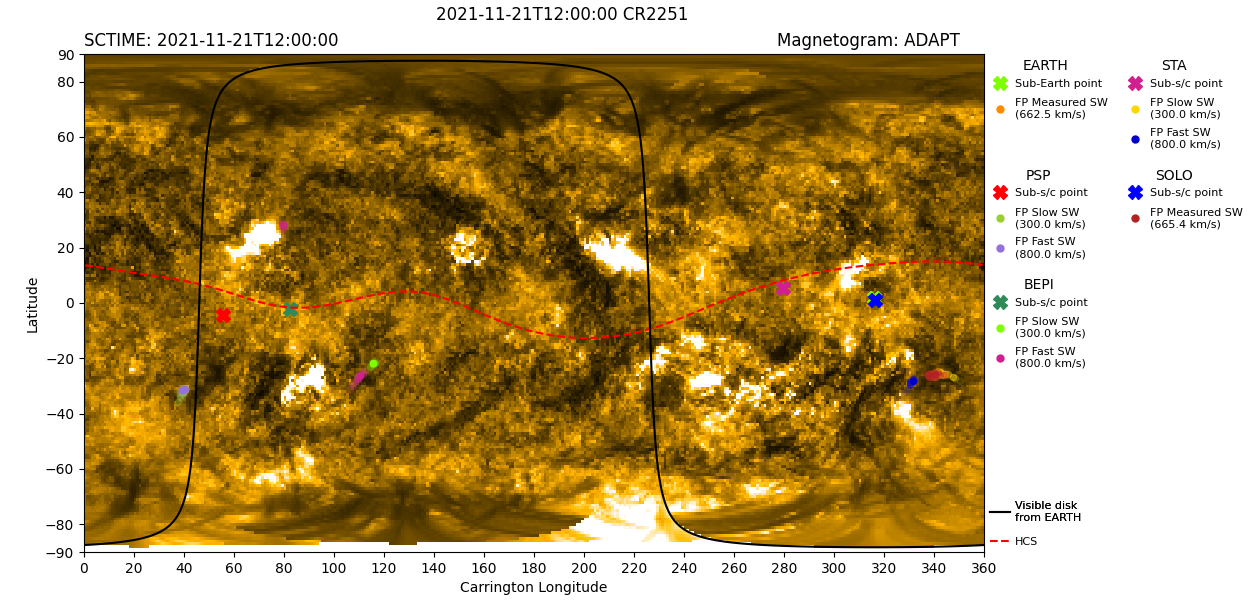}
\includegraphics[width=0.9\textwidth]{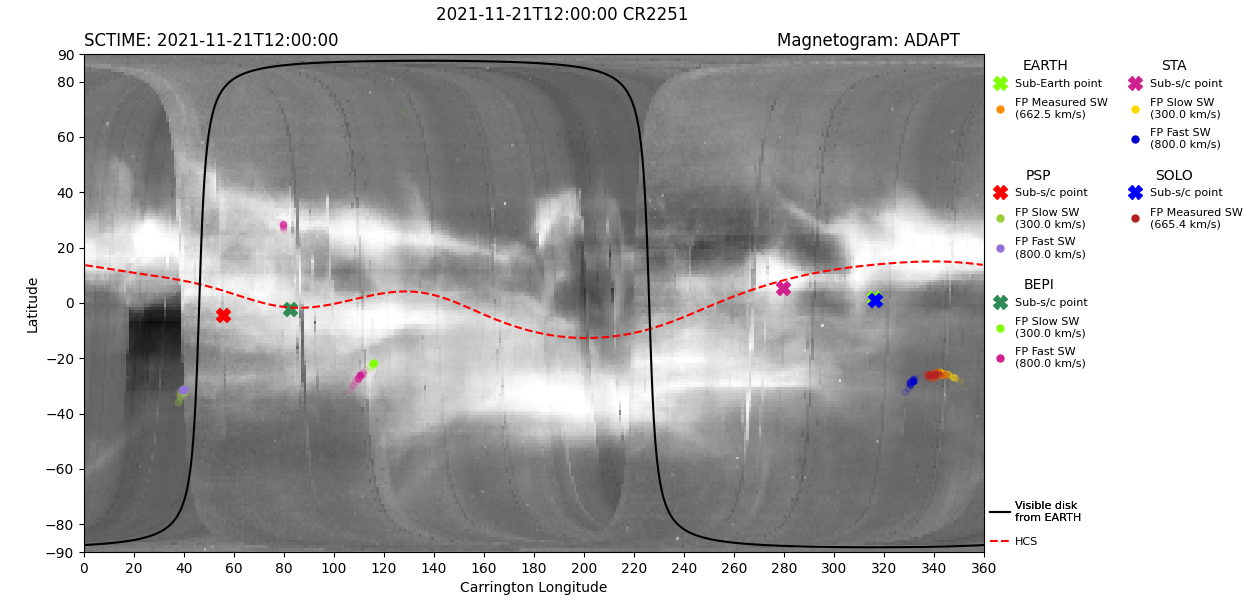}
\caption{Snapshots from the MCT website with input date 2021-11-21 at 12:00 UT. The polarity inversion line which is obtained from the coronal model is plotted as a dashed red line. The dark line denotes the visible portion of the solar disk from the Earth perspective. Colored dots represent the likelihood of spacecraft connectivity to the surface. Colored crosses are vertical projections of the spacecraft location onto the solar disk. Top panel: GONG-ADAPT (2nd realization) synoptic map of the surface magnetic field which has been used for the PFSS reconstruction of the coronal magnetic field. Middle panel: EUV synoptic map at $171$ \si{\angstrom} from the \textit{SDO-AIA}. Bottom panel: white-light synoptic map at $2.5\ R_\odot$ from the \textit{SoHO LASCO-C2} coronagraph. \label{fig:MCT_snapshots}}
\end{figure*}

The MCT is a webservice tool that is available to the public here \footnote{Public access to the Magnetic Connectivity Tool: \url{http://connect-tool.irap.omp.eu/}}. An illustration of the online tool is given in Figure \ref{fig:MCT_snapshots} where the colored patches depict the likelihood of the spacecraft connectivity low in the solar atmosphere or photosphere. The top and middle panels are helpful to relate such regions to the magnetic activity near the surface while the bottom panel gives an insight on the global shape of the solar corona. \\

The MCT relies on models of the magnetic field for two regions, one for the solar corona and one for the heliospheric region situated between the outer boundary of the corona and the point of in situ measurements. In the heliosphere the solar wind is assumed to be frozen-in the interplanetary magnetic field which is in radial expansion. The latter is rooted in the rotating solar corona and hence forms the so-called Parker Spiral (see Figure \ref{fig:MCT_diagram}), along which the solar wind propagates in the heliosphere \citep{Parker1958}. In this region of the heliosphere the solar wind has likely completed its initial acceleration phase and hence an uniform velocity is supposed. When no in situ measurements are available, the solar wind is either assumed to be slow ($\approx 300\ km/s$) or fast ($\approx 800\ km/s$). \\

Lower down in the corona the magnetic field is no longer under pure radial expansion. In practice any reconstruction of the coronal magnetic field could be used, such as from MHD or PFSS models (see section \ref{sec:modeling} for an introduction to these models). The current version of MCT includes PFSS reconstructions that are computed from distinct sources of surface magnetic maps (see section \ref{subsec:magmaps}), and for several source surface heights (see section \ref{subsec:PFSS}). To account for uncertainties in the models, an ensemble of connections is calculated from an extended region spanning $5^\circ$ in heliocentric longitude and latitude around the spacecraft. This is supposed to mimic the effect of slight deviations from the Parker spiral model in response to magnetic turbulence. Evaluating the performance of these models with respect to observations is essential to improve estimates of magnetic connectivity. My main contribution to this tool consisted in developing a pipeline to benchmark all coronal models using white-light synoptic maps of the streamer belt (see section \ref{subsec:poirier2021}). Future improvements of the MCT will include other types of observations: e.g. of the solar wind in situ and coronal holes from EUV solar disk images. The difficulty remains in bringing together different observations and metrics in a consistent manner (see section \ref{subsec:badman2022}).

\subsection{The propagation tool}
\label{subsec:propagation_tool}
The propagation tool is a desktop Java-based application\footnote{The propagation tool is provided by the Centre de Données de la Physique des Plasmas (CDPP): \url{http://propagationtool.cdpp.eu/}} born from the desire to connect coronal structures imaged near the Sun by remote-sensing observatories and their in situ counterpart in the heliosphere. A great novelty of the tool is to provide an interactive access to J-maps which are helpful to track brightness variations of the corona and solar wind continuously with elongation angle away from the Sun. These brightness variations can include white-light signatures of transients such as coronal mass ejections, corotating interaction regions and solar wind perturbations. J-maps are projections over elongation (ordinate axis) and time (abscissae axis) of the brightness extracted along a specific direction in the images, commonly at the ecliptic.

\subsection{AMDA and CLWeb}
\label{subsec:AMDA_CLweb}
AMDA is a public web-based interface\footnote{AMDA is provided by the CDPP: \url{http://amda.irap.omp.eu/}} that allows the user to visualize in situ data from most of the past and on going space-based heliospheric missions \citep{Genot2021}. AMDA is a powerful tool that, in place of the user, deals with the harsh task of combining heterogeneous datasets provided by different data centres. In addition AMDA includes helpful features to assist the user in searching and manipulating the multidatasets: e.g. data mining, event catalogues, ephemeris and models. Recently a python package \textit{speasy} has been developed that provides python users with most of the AMDA functionalities\footnote{\textit{speasy} python package access: \url{https://speasy.readthedocs.io/en/stable/}}. \\

In the same vein as AMDA, CLWeb is an online quick visualizer tool\footnote{CLWeb access: \url{http://clweb.irap.omp.eu/}} that provides access to in situ data from many space-based missions. While CLWeb is mostly exploited in-house at IRAP, access can be granted to external users upon request. In addition to time series, CLWeb can make 2-D plots of the velocity distribution function as measured by the \textit{Proton Alpha Sensor} on board \textit{Solar Orbiter}.

\subsection{SolarSoft and Astropy}
\label{subsec:SolarSoft}
SolarSoft is a library developed by the scientific community on the Interactive Data Language (IDL)\footnote{Solarsoft online documentation: \url{https://sohowww.nascom.nasa.gov/solarsoft/}}. It contains data bases, routines and system utilities providing a common programming and data analysis environment for solar physics \citep{Freeland1998}. SolarSoft is commonly used to process and visualize images of the solar disk and corona such as those taken by \textit{SoHO LASCO}, \textit{STEREO SECCHI} and \textit{PSP WISPR}. In this thesis we mainly exploited the World Coordinate System (WCS) routines built into SolarSoft to extract accurate pointing information for the \textit{WISPR} instrument (a detailed documentation on these routines is available on the GSFC/NASA website\footnote{\url{https://hesperia.gsfc.nasa.gov/ssw/gen/idl/wcs/wcs_tutorial.pdf}}). The WCS is a standard representation of pixel coordinates in the Flexible Image Transport System (FITS) which is an image format widely used in astronomy \citep{Greisen2002,Thompson2006,Thompson2010}. That step was crucial to built synthetic white-light \textit{WISPR} images afterwards (see chapters \ref{cha:stationnary} and \ref{cha:dynamics}). \\

\textit{Astropy} basically offers most of the SolarSoft functionalities on the Python programming language. \textit{Astropy} comes from a community effort to build a long-term data analysis environment for astrophysics that aims at being comprehensive, open and interlinked with other astronomy Python packages\footnote{\textit{Astropy} main web-page: \url{https://www.astropy.org/index.html}} \citep{astropy2013,astropy2018}. \textit{Astropy} has been used by Dr. Athanasios Kouloumvakos to process raw images and build white-light synoptic maps from \textit{SoHO LASCO} observations. These synoptic maps have been extensively exploited for most of the works presented in this thesis.

\section{External models}
\label{sec:modeling}

Numerical models have become essential tools in heliospheric research, and in particular to support the operations of space missions. This is of critical importance for the \textit{PSP} and \textit{SolO} missions whose primary objectives are to characterize the solar sources of the solar winds, storms and energetic particles measured in situ. Coronal magnetic field models are for example necessary to estimate how a spacecraft is magnetically connected to the Sun, which provides an estimate of the path followed by the plasma of the solar wind or even energetic particles. \\

Potential Field Source Surface (PFSS) models constitute the simplest representations of the large-scale coronal magnetic field and are discussed in section \ref{subsec:PFSS}. In this thesis, I will exploit PFSS extrapolations to search for the potential sources of the plasma measured in situ at \textit{PSP} (see section \ref{sec:dynamics_insitu}), to interpret the origin of coronal rays observed by \textit{WISPR} (see chapter \ref{cha:stationnary}), and for systematic comparisons with streamer belt observations (see section \ref{sec:WL_opti}). \\

More realistic reconstructions of the coronal magnetic field can be obtained with magneto-hydrodynamics (MHD) models at the cost of significant computational resources. However MHD models have the benefits to simulate, in addition to the magnetic field topology, the plasma bulk properties that are highly valuable to interpret remote-sensing and in situ observations. Large-scale 3-D MHD simulations such as the WindPredict-AW model (discussed in section \ref{subsec:WindPredict}) are particularly relevant to determine the overall structure of the corona and heliosphere and hence simulate the plasma environment at different probes or in the vicinity of the Earth. Another major interest in using 3-D global simulations such as the WindPredict-AW model is its capability to simulate large-scale transients such as streamer blobs while preserving their magnetic topology as they propagate in the corona and heliosphere (see section \ref{sec:dynamics_tearing}). \\

The Multiple Flux tube Solar Wind Model (MULTI-VP) (introduced in section \ref{subsec:MULTI-VP}) is an alternative to full-fledged 3-D MHD models by constructing a 3-D datacube from multiple 1-D solutions. The benefit in computational tractability from using such multi 1-D models is however at the expense of not accounting for possible reconfigurations of the magnetic field in three dimensions or for the retroaction of the plasma on the magnetic field. One should note however that while 3-D interactions are naturally accounted for in global 3-D simulations, these complex processes can be poorly resolved due to the scales resolved and the inherent physical assumptions of the MHD approach (such as in current sheets). In the study I carried out that is presented in section \ref{sec:stationnary_poirier2020}, I show that MULTI-VP is particularly suited to study the fine structure of streamer rays as seen in \textit{WISPR} observations because it can be run at high resolution. \\

Furthermore, for the sake of computational tractability most multi 1-D and 3-D models assume a single or two fluid plasma of Hydrogen protons and electrons. As introduced in section \ref{subsec:intro_FIP}, the underlying physics of the FIP effect requires a comprehensive description of how minor elements interact with the background hydrogen solar wind plasma, which entails the development of proper multi-species models. In chapter \ref{cha:ISAM} I give a comprehensive description of the Irap Solar and Atmospheric Model (ISAM) that I have co-developed to study the composition of coronal loops and for which a considerable attention has been given to the physics inserted into the model.

\subsection{The Potential Field Source Surface (PFSS) model}
\label{subsec:PFSS}
The Potential Field Source Surface (PFSS) models basically extrapolate a 3-D magnetic field assuming a potential field without any electric currents $\nabla \times B=\mu_0 J=0$. The inner boundary is specified by direct observations in the form of a given synoptic map of the photospheric magnetic field (see section \ref{subsec:magmaps}). At the outer boundary or source surface, the magnetic field is assumed to be purely radial and fully open. This equipotential condition set at the source surface allows to mimic the formation of the solar wind which connects the coronal magnetic field to the interplanetary medium. It is assumed that the dynamic pressure of the solar wind is sufficiently strong to force the coronal magnetic field to be radial. Such condition is not necessary in MHD models where the solar wind plasma is solved together with the magnetic field.

\subsubsection{Solving the magnetostatic equation}

The height of the source surface is typically set at a height of $2.5\ R_\odot$ \citep{Hoeksema1984}. The 3-D coronal magnetic field is obtained by simply solving the current free magnetostatic equation $\nabla \times B=\mu_0 J=0$ in addition to the divergence free condition $\nabla \cdot B=0$. There are different numerical approaches to solve this equation. A common method (which we use in our in-house solver at IRAP) computes the spherical harmonics of the magnetic scalar potential $\Phi$. Indeed the magnetic field can be written in the form of a scalar potential $B=-\nabla\Phi$ while automatically satisfying the relation $\nabla\times B=\nabla\times \nabla \Phi=0$. The divergence free relation $\nabla \cdot B=\nabla^2 \Phi=0$ then requires to find the $\Phi$ function having a null laplacian. Such functions are well known as spherical harmonic functions and can be directly expressed as functions of the spherical coordinates $(r,\theta,\phi)$ and the associated Legendre polynomials $P_l^m(x)$ by the relation:

\begin{equation}
    \Phi=\sum_{l=0}^{+\infty} r^{-(l+1)}\sum_{m=-l}^{+l} f_{l,m} Y_{l,m}(\theta,\phi)
\end{equation}
where $Y_{l,m}(\theta,\phi)=k_{l,m} P_l^m(cos\theta)e^{im\phi}$.

\subsubsection{Reconstruction of the coronal magnetic field}

\begin{figure*}[]
\centering
\includegraphics[width=0.8\textwidth]{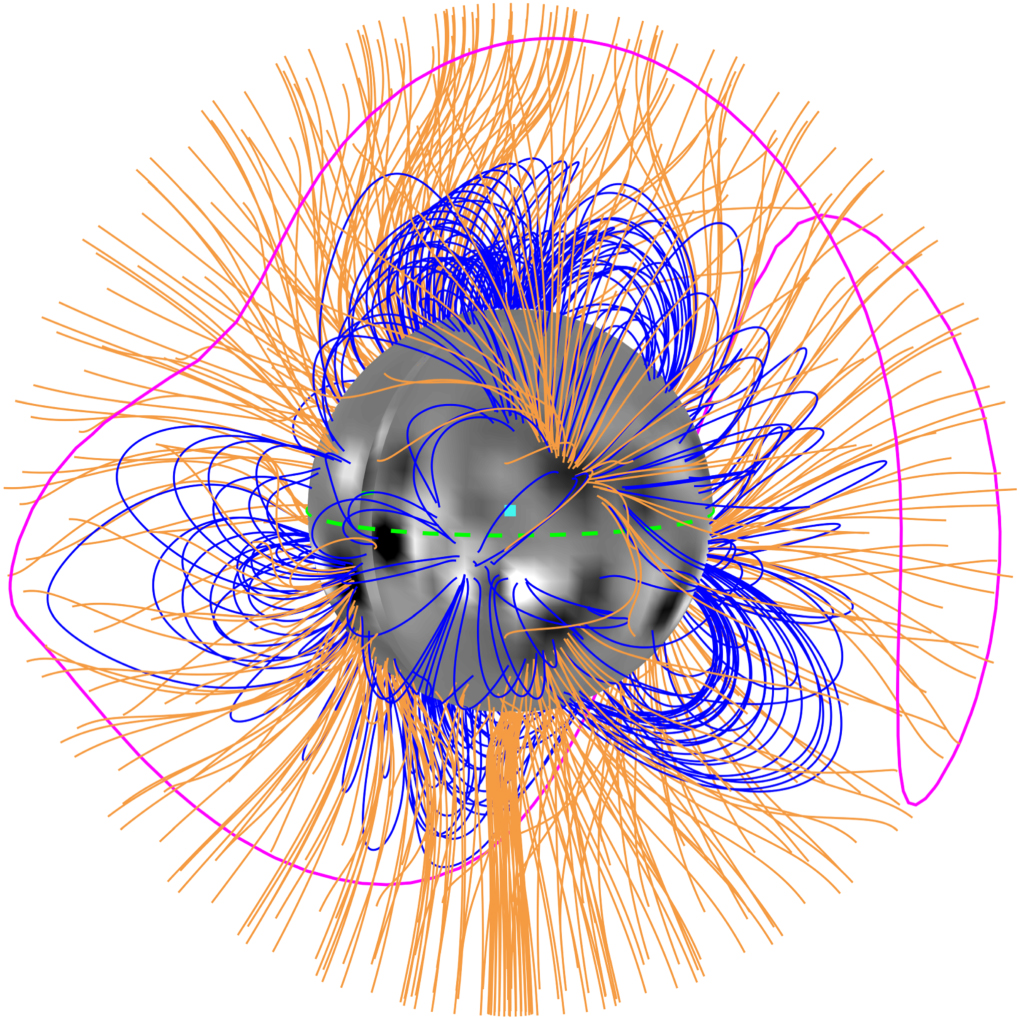}
\caption{PFSS reconstruction of the coronal magnetic field using a photospheric magnetic field map from WSO (Carrington rotation CR2149) and a source surface height of $2.5\ R_\odot$. Open and closed magnetic field lines are depicted in orange and blue respectively. The solar equator is represented by a dashed green line. The line of polarity inversion or neutral line is traced in purple and is the baseline of the heliospheric current sheet. Figure 7 that I produced for \citet{Griton2020}.
\label{fig:Griton2020_fig7}}
\end{figure*}

A PFSS reconstruction of the coronal magnetic field is given in Figure \ref{fig:Griton2020_fig7} during a period of high solar activity that I produced for the \citet{Griton2020} paper. The neutral line or polarity inversion line (purple line) is extracted at the outer boundary (source surface) and represents the baseline of the HCS. The HCS is significantly warped and extends far above and below the solar equator (green dashed line). One can see that there is no more closed magnetic field (in blue) at the source surface where the open field (in orange) becomes purely radial. The surface magnetic field map that has been specified at the inner boundary is color plotted in a gray scale where white (or black) spots represent regions of strong magnetic field of positive (or negative) polarity. Most of the large-scale coronal loops shown in blue have one of their footprint anchored into these strong unipolar regions. Some of the largest loops beneath the neutral line are typically identified as helmet (or base) of streamers in eclipse images (see Figure \ref{fig:Mikic2018_fig1}) while other loops can be located below pseudo-streamers (an example of a pseudo-streamer is shown more clearly in Figure \ref{fig:Poirier2020_fig8}).

\subsubsection{Limitations of PFSS models}

Recent studies have demonstrated that the standard height of $2.5\ R_\odot$ for the source surface often leads to a poorer representation of the coronal magnetic field when compared with observations. For instance, \citet{Boe2020} have shown from total eclipse observations that most of the magnetic field becomes radial only above $3.0\ R_\odot$. \citet{Riley2006} and \citet{Panasenco2020} have highlighted the variation of this height from place to place around the Sun, and hence pointed out the limitations of using a spherical source surface with a constant height. Finally a recent collaborative effort \citep{Badman2022} that I contributed to significantly, evaluated how much this height would vary to match different remote-sensing and in situ observations. We concluded from this study that the typical $2.5\ R_\odot$ source surface height cannot satisfy all observations simultaneously, and that for most of the time a compromise is needed. We also highlighted the impact on the PFSS models performance due to the choice of the source magnetic map. Further details about my contribution to this work are discussed in section \ref{subsec:badman2022}.

\subsection{The WindPredict-AW model}
\label{subsec:WindPredict}
The WindPredict-AW model is a magneto-hydrodynamic model driven by Alfvén waves \citep{Reville2020a,Reville2020b,Reville2022} and based on the PLUTO numerical solver \citep{Mignone2007}. The 3-D MHD equations for a fully ionized single-fluid plasma are solved in a conservative form \citep{Jacques1977}:
\begin{subequations}
\label{eq:PLUTO_eq}
\begin{align}
    &\frac{\partial}{\partial t} \rho + \nabla \cdot \rho \mathbf{u} =0 \label{eq:PLUTO_rho}\\
    &\frac{\partial}{\partial t}\rho \mathbf{u}+\nabla \cdot \left( \rho \mathbf{u}\mathbf{u}-\mathbf{B}\mathbf{B}+\mathbf{I}p\right) =-\rho\nabla \Phi \label{eq:PLUTO_u}\\
    &\frac{\partial}{\partial t}\left(E+\epsilon+\rho \Phi \right)+\nabla \cdot \left[\left( E+p+\rho \Phi\right)\mathbf{u} -\mathbf{B\left(\mathbf{u}\cdot\mathbf{B} \right)}+(\mathbf{u}+\mathbf{v_A})\epsilon^+ +(\mathbf{u}-\mathbf{v_A})\epsilon^-\right] =Q_{tot} \label{eq:PLUTO_E}\\
    &\frac{\partial}{\partial t} \mathbf{B} + \nabla \cdot \left( \mathbf{uB}-\mathbf{Bu}\right) =0 \label{eq:PLUTO_B}\\
    &\nabla \cdot B =0
\end{align}
\end{subequations}
where $\rho = m_p n_p$ is the mass density, $\mathbf{u}$ is the bulk velocity, $\mathbf{B}$ the magnetic field, $\mathbf{I}$ the identity matrix, $\Phi = -GM_\odot/r$ the gravitational potential, $E\equiv \rho e + \rho u^2/2 + B^2/2$, $\rho e$ the internal energy, $p=p_{th}+B^2/2$ the total (thermal and magnetic) pressure. The source term $Q_{tot}=Q_h+Q_w-Q_c-Q_r$ includes thermal conduction, ad-hoc and Alfvén wave heating, and radiative cooling. These terms are discussed in the following paragraphs. The set of equations is closed by the typical adiabatic relation for an ideal gas $\rho e=p_{th}/(\gamma-1)$ where the adiabatic index $\gamma$ equals $5/3$ for a fully ionized hydrogen gas. \\

The heat conducted downward from the corona is lost by radiative cooling in the transition region and at the top of the chromosphere. For computational tractability, WindPredict-AW relies on an approximation for the radiative losses in an optically thin medium based on \citet{Athay1986}'s formulation that is given in section \ref{subsec:ISAM_radloss_MVP}. \\

In contrast to the high-order moment model ISAM (presented in chapter \ref{cha:ISAM}) that solves the heat flux explicitly, WindPredict-AW assumes a form of the heat flux in order to close its set of coupled conservation equations. To model the solar atmosphere at great extent, two distinct heat fluxes are adopted to account for both collisionaly dominated and collisionless media. In a medium dominated by the collisions, the classical Spitzer-Härm heat conduction term $\mathbf{q_s=-\mathcal{K}_0 T^{5/2}\nabla T}$ is generally adopted to a good approximation with a conductivity coefficient set to $\mathcal{K}_0=9\times 10^{-7}\ \rm{(erg.cm^{-1}.s^{-1}.K^{-7/2})}$. In contrast, the electron collisionless heat flux $\mathbf{q_p}=3/2p_{th}\mathbf{u_e}$ of \citet{Hollweg1986} is adopted in the higher corona. The total heat flux $\mathbf{q_{tot}}=\alpha \mathbf{q_s}+(1-\alpha)\mathbf{q_p}$ is then defined so that there is a smooth transition between the two regimes where $\alpha=1/(1+(r-R_\odot)^4/(r_{coll-R_\odot})^4)$ \citep{Reville2020a}. That corresponds to the term $Q_c=\nabla \cdot \mathbf{q_{tot}}$ appearing in equation \ref{eq:PLUTO_E}. \\

Contribution from Alfvén waves to the plasma momentum equation is accounted for by an additional term $p_w=\epsilon/2$ to the total pressure $p$ in equation \ref{eq:PLUTO_u}. The transport of the wave energy $\epsilon=\epsilon^+ +\epsilon^-$ is computed following the Wentzel\textendash Kramers\textendash Brillouin (WKB) approximation \citep[see e.g.][]{Alazraki1971,Hollweg1974,Tu1995} and is given by equation \ref{eq:Ew} that is introduced latter on in section \ref{subsubsec:ISAM_Aw_WKB}. \\

In the WindPredict-AW model, Alfvén waves dissipate their energy to the plasma through a turbulence cascade that is driven by non-linear interaction between counter-propagating Alfvén waves. This framework is introduced in detail in section \ref{subsubsec:ISAM_Aw_Qw} where each wave population (inward and outward propagating) dissipates their energy at a rate:
\begin{equation}
    Q_w^\pm=\frac{\rho}{8}\frac{\left|z^\pm \right|^2}{\mathcal{L}}\left|z^\mp \right| 
    \label{eq:PLUTO_Qw_final}
\end{equation}
where $\mathcal{L}=\mathcal{L}_\odot/\sqrt{B}$ is a characteristic transverse length of the turbulence cascade, that is usually set to match the typical size of supergranules in the low corona $\mathcal{L}_\odot=0.022\ R_\odot\sqrt{B_\odot}\approx15,000\ km\sqrt{B_\odot}$ (with $B_\odot$ in Gauss unit) \citep{Verdini2007}. In the above formulation, each Alfvén wave population ($+,-$) gives a fraction of their energy to the turbulence cascade that provide a total plasma heating rate $Q_w=Q_w^+ + Q_w^-$. Although counter-propagating waves naturally coexist in closed-field geometries, it is not the case in open solar wind solutions for which a specific strategy is adopted and is presented in section \ref{subsubsec:ISAM_Aw_Qw}. \\

\subsection{The Multiple Flux tube Solar Wind Model (MULTI-VP)}
\label{subsec:MULTI-VP}

MULTI-VP is a 3-D multi-tube magneto-hydrodynamic code that solves for the MHD properties of the solar wind such as speed, density, and temperature by solving a set of 1-D MHD conservation equations along individual flux tubes \citep[see][]{PintoRouillard2017}. The model is typically run on thousands of magnetic flux tubes to simulate the entire solar wind escaping the solar atmosphere. The geometry of these flux tubes is an input of the model and can be either idealized or extracted from global reconstructions of the coronal magnetic field, that can be given by for instance full 3-D MHD (see e.g. section \ref{subsec:WindPredict}) or PFSS extrapolations (see \ref{subsec:PFSS}). The plasma is supposed to be fully ionized, one-fluid, quasi neutral and isothermal. The energy equation includes the effect of heating, thermal conduction, and radiative cooling, which are essential in order to simulate a realistic solar wind mass flux \citep{Hansteen1995,PintoRouillard2017}. \\

The 1-D conservative equations solved in MULTI-VP can be expressed as follows along the curvilinear abscissa $s$ of the magnetic flux tube \citep{PintoRouillard2017}:

\begin{subequations}
\begin{align}
    & \frac{\partial}{\partial t}\rho + \nabla_\parallel \cdot \left(\rho u\right) =0 \label{eq:MVP_rho}\\
    & \frac{\partial}{\partial t}u + u \nabla_\parallel u +\frac{\nabla_\parallel p_{th}}{\rho} + \frac{GM_\odot}{r^2}cos(\alpha) = \nu \nabla_\parallel^2 u \label{eq:MVP_u}\\
    & \frac{\partial}{\partial t}T + u \nabla_\parallel T + (\gamma-1)T\nabla_\parallel \cdot u +\frac{\gamma-1}{\rho}\nabla_\parallel \cdot F_c = -\frac{\gamma-1}{\rho}\left[\nabla_\parallel \cdot F_h + \rho^2\Lambda(T)\right] \label{eq:MVP_T}
\end{align}
\end{subequations}
where $\rho = m_p n_p$ is the plasma mass density, $u$ the plasma bulk velocity along the flux tube, $T$ the total plasma temperature, $p_{th}=nk_bT$ the plasma thermal pressure, $k_b$ the Boltzmann constant, $G$ the gravitational constant, $M_\odot$ the Sun mass, $\alpha=$ the angle between the magnetic field $\hat{\mathbf{b}}$ and the vertical direction $\hat{\mathbf{r}}$, $\nu$ the kinematic viscosity, $\gamma=5/3$ the adiabatic index or ratio of specific heats, $F_h$ and $F_c$ the mechanical heating and conductive heat fluxes, and $\Lambda(T)$ the radiative losses. Compared to the full 3-D MHD equations (eq \ref{eq:PLUTO_eq}) described in \ref{subsec:WindPredict}, the 1-D set here does not include explicit terms nor a specific equation for the magnetic field since it is specified as an input to the model. But the magnetic field still plays a role on the plasma via the divergence operator defined as:

\begin{equation}
    \nabla_\parallel \cdot (*)=\frac{1}{A}\nabla_\parallel(A*) = \frac{1}{A} \frac{\partial}{\partial s}(A*)=B\frac{\partial}{\partial s}(*/B)
    \label{eq:MVP_div}
\end{equation}
where $\mathbf{B}=B(s)\hat{\mathbf{b}}$. This relation accounts for the effect of the expansion of the flux tube's cross-sectional area ($A(s)\propto 1/B(s)$) on the background plasma, that is sometimes called the geometric effect.

The thermal energy of the plasma that is primarily heated in the corona is conducted downward by the conductive heat flux $F_c$ until the heat gets eventually evacuated in the lower and denser layer of the solar atmosphere by strong radiative emissions. As in WindPredict-AW, the conductive heat flux is not solved explicitly and hence must be approximated. The usual Spitzer\textendash Härm approximation is retained in the collisional part of the domain:
\begin{equation}
    F_c=-\kappa_0T^{5/2}\nabla_s T
\end{equation}
with $\kappa_0=9\times 10^{-7}\ erg.cm^{-1}.s^{-1}.K^{-7/2}$. \\

The dynamics of the winds simulated in MULTI-VP are highly dependent on the balance between the energy inputs and losses along the flux tubes. As introduced in section \ref{subsec:intro_heating}, MULTI-VP assumes that all the energy is transported from the base of the flux tube to the corona in the form of a mechanical flux. In the vein of the quasi-stationnary theory introduced in section \ref{sec:intro_stationnary}, MULTI-VP adopts an ad-hoc formulation of the heating flux $F_h$ that is inspired from the work of \citet{Withbroe1988} and where the topology of the magnetic flux controls the dynamics of the simulated winds as described in section \ref{subsec:stationnary_poirier2020_numerical}. \\

Similarly to the WindPredict-AW model introduced in section \ref{subsec:WindPredict}, the usual formulation of \citet{Athay1986} is used to estimate radiative losses in an optically thin medium, which is given in section \ref{subsec:ISAM_radloss_MVP}. But the MULTI-VP implementation includes an additional correction term to account for re-absorption of the radiated emissions (radiative heating) by the optically thick plasma in the chromosphere \citep{PintoRouillard2017}. \\

\section{Constraining coronal and heliospheric models of the solar wind}
\label{sec:WL_opti}

We have introduced in section \ref{sec:tools} and \ref{sec:modeling} tools and models that were exploited in this thesis by providing context as well as critical information necessary to better understand observations of the solar wind as it propagates from the inner corona to the heliosphere. Conversely, we show in this section that some well-known features of the solar corona and the solar wind can be valuable assets to improve and constrain these tools and models. As we gain insight and a better understanding of the features that are observed remotely and measured in situ we can place ever more stringent tests to the models. \\

In this context, a collaborative project was initiated between IRAP and external collaborators, throughout the MADAWG and an international working group entitled \textit{"Exploring The Solar Wind In Regions Closer Than Ever Observed Before"} led by Dr. Louise Harra and financed by the International Space Science Institute (ISSI). My contribution to these projects were to develop comparative procedures to evaluate in a systematic manner, coronal and heliospheric models using white-light observations. These procedures are discussed in section \ref{subsec:poirier2021} and in greater detail in \citet{Poirier2021}. 

In \citet{Badman2022} we show that other types of observations, including magnetic sectors measured in situ as well as the size and location of coronal holes observed in ultraviolet imaging, are essential ingredients to benchmark all aspects of the models. I have significantly contributed to this work by providing the evaluation pipeline for the white-light observations. In section \ref{subsec:badman2022} I only give a summary of the main outcomes from this study while the full published version can be found online \citep[see][]{Badman2022}.

\subsection{Exploiting White-Light Observations to Improve Estimates of Magnetic Connectivity (Poirier et al., 2021)}
\label{subsec:poirier2021}

\begin{figure*}[!h]
\centering
\includegraphics[width=0.8\textwidth]{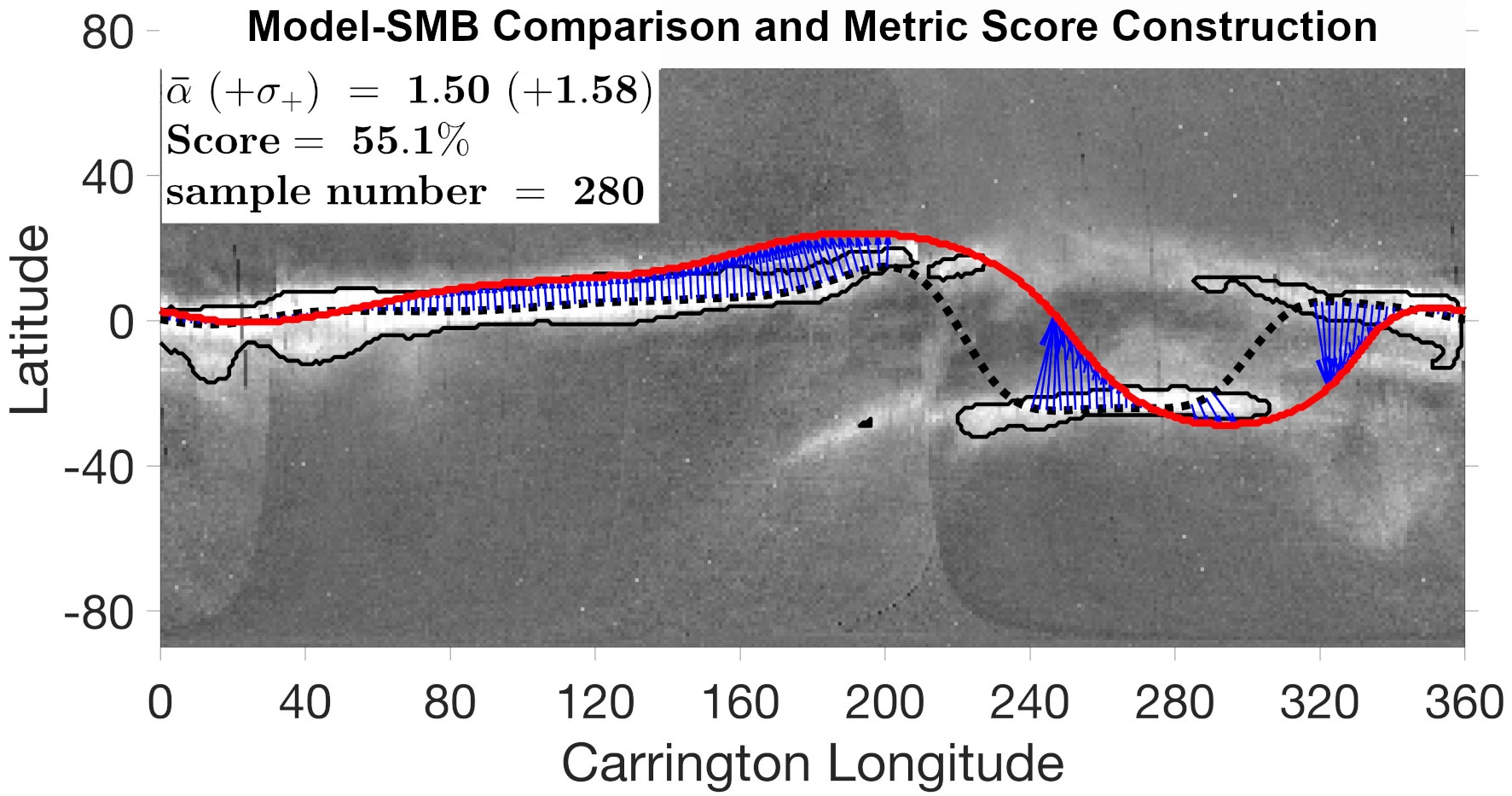}
\caption{Comparison of the HCS predicted by a model (solid red line) with the HCS derived from white-light observations of the streamer belt (dashed black line) taken by \textit{SoHO LASCO-C2} between 2018, October 30 and 2018, November 13. The angular deviations between the two lines that are used to compute the WL metric are represented by the blue arrows. Results from the WL metric are given in the legend, of which the WL global score (in \%). Figure 2 that I produced for \citet{Badman2022}. \label{fig:Badman2022_fig2}}
\end{figure*}

As introduced in section \ref{subsec:intro_high_atmosphere}, WL observations contain a wealth of valuable information about the solar wind and the overall structure of the solar corona and heliosphere, which can be compared with models in many different ways. For the purpose of supporting the \textit{Solar Orbiter} and \textit{Parker Solar Probe} missions, we based our benchmarking on the position of the HCS which draws an overall picture of the structure of the heliosphere by marking the separation between the positive and negative magnetic sectors. We exploit white-light synoptic maps of the streamer belt (see section \ref{subsubsec:intro_WLmaps}) and we extract an estimate position of the HCS where the emissions are the brightest in the core of the streamer belt, that is likely associated to the very dense plasma that constitutes the HPS (see section \ref{subsubsec:intro_HCS_HPS}).

On the modeling side, the HCS can be fetched from any models that provide a 3-D reconstruction of the magnetic field, as the surface where the magnetic polarity switches sign. In practice we extract this boundary or neutral line at a specific distance from the Sun, at the outer boundary of MHD models or at the source surface height of PFSS models where the magnetic field becomes purely radial.

In a study that I carried out with IRAP colleagues and which is described in detail in \citet{Poirier2021}, I define a comparative procedure to compare the HCS extracted from the models and the WL observations of the streamer belt. Basically I detect the core of the streamer rays and trace the line where the brightness is maximum. I then compare this line with the polarity inversion line (or neutral line, solid red line) that is extracted from the models. At last I define a function that attributes a score according the related positions of the two lines \citep[see][eq 1-3]{Poirier2021}. An example is shown in Figure \ref{fig:Badman2022_fig2} for a PFSS model with a source surface height set at $2.5\ R_\odot$. \\

One of the most relevant applications of this work is the ability to better estimate how a spacecraft is magnetically connected to the Sun. Accurate estimates of magnetic connectivities are essential to provide context to in situ measurements and to link those data with their possible counterpart in the lower solar atmosphere. The white-light comparative procedure has therefore been implemented into the MCT introduced in section \ref{subsec:Connectivity_tool}.

We have also highlighted the importance of using white-light observations to better constrain the models. The comparative procedure allows us to select an optimal photospheric map which is a critical parameter for most coronal models. Almost all studies presented in this thesis have benefited from this functionality.

\subsection{Constraining Global Coronal Models with Multiple Independent Observables (Badman et al., 2022)}
\label{subsec:badman2022}

\begin{figure*}[!h]
\centering
\includegraphics[width=0.8\textwidth]{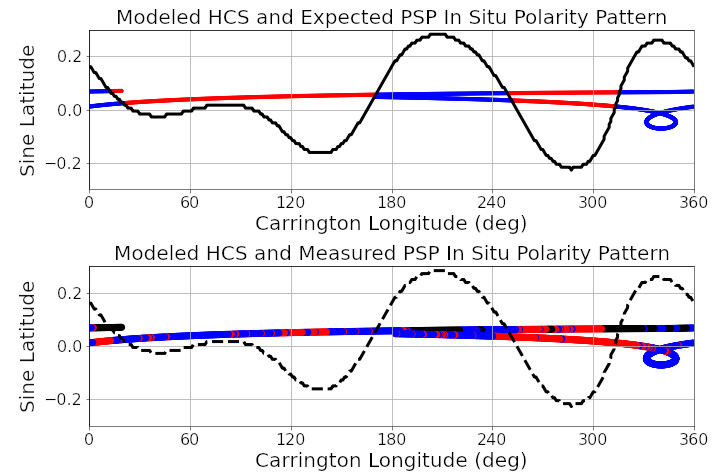}
\caption{ Top panel: Prediction of magnetic polarity measurements at \textit{PSP} from a PFSS model. The simulated polarity switch sign when the spacecraft moves across the modeled HCS (dark solid line). Bottom panel: Actual magnetic polarity measured by \textit{PSP} during its first passage to the Sun (October/November 2018). In both real and virtual modes, the \textit{PSP}'s trajectory is projected closer to the Sun (at around $2.0-2.5\ R_\odot$) by following the Parker spiral (also known as a ballistic model). That way the measured magnetic polarities can be compared with the magnetic sectors predicted by PFSS models.  Figure adapted from \citet[][Figure 3]{Badman2022}. \label{fig:badman_fig3}}
\end{figure*}

\begin{figure*}[!h]
\centering
\includegraphics[width=0.7\textwidth]{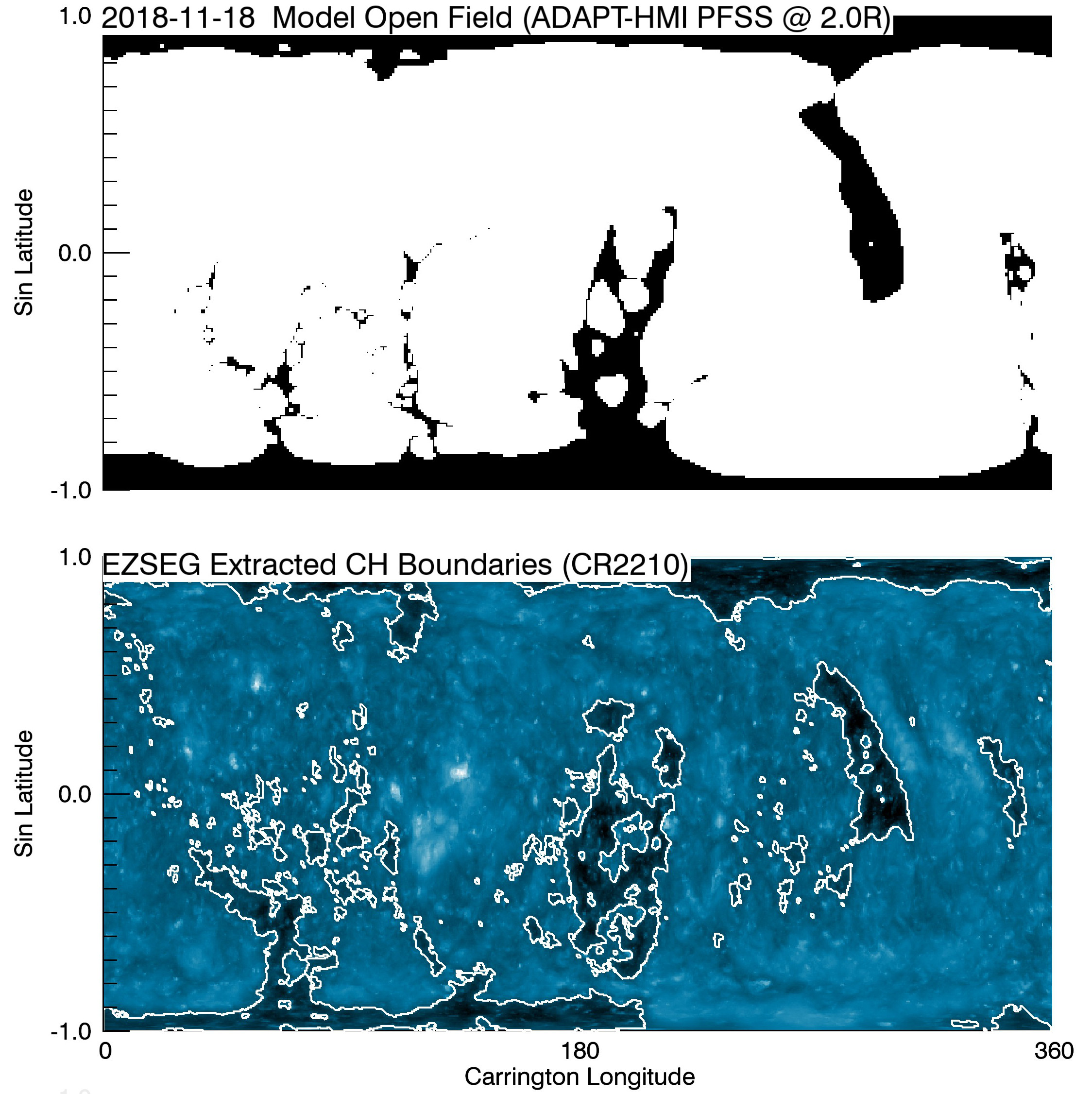}
\caption{Top panel: Coronal holes distribution (dark shaded areas) as predicted by a PFSS model.
Bottom panel: EUV synoptic map at $193$ \si{\angstrom} built from \textit{AIA-EUVI} observations during Carrington rotation CR2210. Darker regions in the EUV map are identified as coronal holes with a white contour. Figure adapted from \citet[][Figure 1]{Badman2022}. \label{fig:badman_fig1}}
\end{figure*}

We have seen in section \ref{subsec:poirier2021} that WL observations can be exploited to evaluate coronal and heliospheric models based on the identification of the HCS. However observations of the solar corona and heliosphere are not limited to WL coronagraph images. \\

In situ measurements of the solar wind bulk properties and of the interplanetary magnetic field are available at various places in the heliosphere and distances to the Sun. Spacecrafts that cross the HCS show clear inversion of the magnetic polarity, that is of the radial component of the interplanetary magnetic field. When located at different points in the heliosphere, space-based missions provide direct measurements, although localized, of the spatial distribution of magnetic sectors over time. Such temporal measurements can be simulated by inserting virtual probes into 3-D reconstructions of the interplanetary magnetic field. An example is given in Figure \ref{fig:badman_fig3} for a PFSS model that is compared against real \textit{PSP} measurements.

Much closer in, EUV observations of the solar disk provide a distribution of the magnetically closed or open regions around the Sun. Open regions directly connect to the interplanetary magnetic field and supply the heliosphere with solar material. Those regions are cooler due to the energy that is evacuated by the escaping plasma and usually appear much darker than magnetic closed regions in EUV. Such dark regions are commonly called coronal holes and can be directly compared to the open regions given by 3-D reconstructions of the coronal magnetic field as illustrated in Figure \ref{fig:badman_fig1}. \\

To summarize, one can combine these different sets of observations to evaluate various aspects of coronal and heliospheric models. Magnetic sectors measured in situ allow to evaluate the models capability to capture HCS crossing with a good timing, which can be a critical aspect for space weather applications. Coronal hole observations are helpful to check whether a model can reproduce or not the magnetic topology close to the surface, which can be essential for connectivity science and as a support to the \textit{Solar Orbiter} and \textit{Parker Solar Probe} operations. WL synoptic maps (see section \ref{subsubsec:intro_WLmaps}) provide an additional constraint at an intermediate height, between the inner corona and heliosphere, when the coronal magnetic field stabilizes into a purely spherical expansion.

The challenge addressed by \citet{Badman2022} then consists in defining a general framework to compare the models with these three different sets of observations. I contributed to this work by providing the evaluation pipeline based on the white-light observations that is almost identical to the one already presented in \citet{Poirier2021} and introduced in section \ref{subsec:poirier2021}. The comparative procedures based on coronal hole observations and magnetic sectors measured in situ are also explained in detail in \citet{Badman2022}. The reader is invited to refer to the full published version that is available online. \\

In \citet{Badman2022} we applied our comparative framework to various models of which: two PFSS models based on the open source package \textit{pfsspy}\footnote{\textit{pfsspy: \url{https://pfsspy.readthedocs.io/en/stable/}}} and on the standard implementation from GONG \citep{Harvey1996}, the Wang-Sheeley-Arge (WSA) model that alleviates the current-free limitation assumed in the classical PFSS reconstructions with a Schatten Current Sheet (SCS) model \citep{Schatten1972}, and the Magnetohydrodynamics Around a Sphere (MAS) model that is a full-fledged 3-D MHD model that solves the magnetic field and the thermodynamics of the plasma together \citep{Linker1999,Riley2001} \citep[see][for further details on the modeling framework]{Badman2022}. \\

In the following I summarize some of the main outcomes from this study, the reader is also invited to refer to \citet{Badman2022} for further details.

\begin{itemize}
    \item Most of the time a compromise is needed to satisfy simultaneously the WL, coronal hole and magnetic sector observations.
    \item Although the typical $2.5\ R_\odot$ source surface height for PFSS models does not always give the best scores, it is an optimal choice that performs reasonably well in general.
    \item Coronal hole observations favor PFSS models with low source surface heights whereas high source surface heights match better with WL observations of the streamer belt.
    \item As expected, the WSA model circumvents this limitation thanks to a correction that accounts for the effect of the transverse plasma pressure on the coronal magnetic field, and hence encourages a more pronounced flattening of the HCS in accordance with WL observations.
    \item The MAS MHD model performs in general the best and for all three types of observations. That highlights the advantage of inserting more physics into the models at the cost of much heavier computations.
\end{itemize}

%% file: chapters/Stationnary.tex
\chapter{The fine structure of streamer rays imaged by \textit{PSP-WISPR}}
\label{cha:stationnary}

\minitoc

\textit{(This chapter contains material from the \citet{Poirier2020} paper.)} \\

The appearance of coronal rays in white-light imagery taken from Earth's heliocentric distances evolves gradually over daily timescales even without transient releases. This daily evolution is primarily due to the effect of solar rotation that brings rays located at different heliocentric longitudes (and latitudes) into the plane-of-the-sky of the observing telescope. On longer timescales, the positions and brightness of coronal rays respond to changes in the topology of the coronal magnetic field during the solar cycle \citep{Golub2009}. Despite decades of observations, the physical mechanisms that produce these rays are still debated.\\

A source of difficulty resides in the nature of the observations themselves; any WL image of the solar corona results necessarily from the integration of sunlight that has been scattered by electrons situated along each line-of-sight (LOS) of each pixel in the image. This observational constraint complicates any interpretation of the 3-D structure of streamer rays and the determination of their source closer to the surface of the Sun. \\

Because of the LOS effects, it is generally difficult to analyze the detailed streamer topology from 1 AU. For this reason, the \textit{PSP} is equipped with a heliospheric imager \textit{WISPR} that records the brightness of the corona from a vantage point that is situated for the first time in the corona itself. According to the Thomson scattering theory, as an imager gets closer to the Sun, it becomes sensitive to plasma located over a more narrow region of the solar atmosphere, acting as a microscope scrutinizing the fine-structure coronal rays compared with instruments based at near 1 AU, that is discussed further in section \ref{sec:stationnary_theory}. \\

Since a component at least of the slow solar wind (SSW) appears to originate in streamer rays, the next sections aim at evaluating whether the quasi-stationary theory of the SSW (see section \ref{sec:intro_stationnary}) can explain the appearance of coronal rays in \textit{WISPR}. We first introduce in section \ref{sec:stationnary_theory} the Thomson scattering theory necessary to interpret \textit{WISPR} observations. The novelty of \textit{WISPR} observations compared to typical 1 AU observatories is then discussed in section \ref{sec:stationnary_comp}. We focus on the 1st \textit{PSP} encounter that I studied for the \citet{Poirier2020} paper, and of which the main outcomes are summarized in section \ref{subsec:stationnary_poirier2020_conclusion}. Finally we take a tour of the subsequent \textit{WISPR} observations taken over PSP's first nine solar encounters, in section \ref{sec:stationnary_WISPR}.

\section{The Thomson scattering theory}
\label{sec:stationnary_theory}

Because of the LOS effects, WL remote-sensing observations only provide limited information about the 3-D structure of the solar corona. Additional inputs from the theory are often necessary to interpret WL observations in greater detail. \\

The scattering of sunlight by coronal electrons has been formulated in the Thomson scattering theory that has been reviewed recently in \citep{Howard2009,Howard2012}. Following this theory, the total intensity received along the LOS of a detector from scattered electrons is as follows:
\begin{equation}
\label{eq:Thomson_scattering}
    I_{tot} = \int_0^\infty n_e z^2 G dz
\end{equation}
where $I_{tot}$ is the total intensity expressed in solar brightness unit $B_\odot$, $n_e$ the electron density, $z$ the path length along the LOS. The scattering factor $G$ can be expressed as a function of the van de Hulst coefficients \citep[see][eq 25-28]{Howard2009} in which geometric effects and the fall off of incident sunlight are encoded. Those geometric effects include for instance the collimation of sunlight with distance from the Sun, where the Sun is not considered as a point-source. \\

\begin{figure*}[]
\centering
\includegraphics[width=0.85\textwidth]{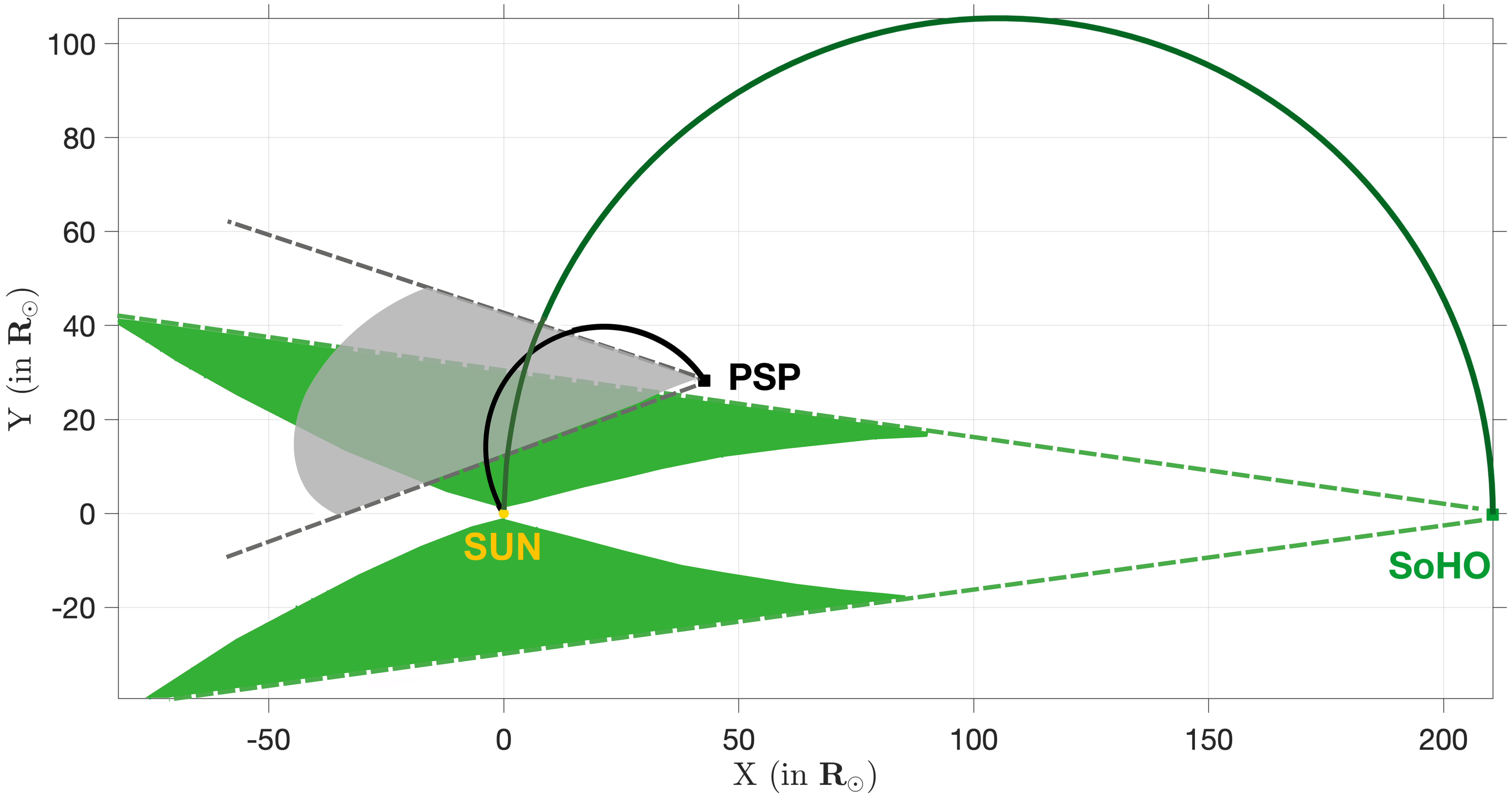}
\caption{ Region of sensitivity of \textit{WISPR-I} compared to \textit{LASCO-C2}, defined as the LOS portions that contain $99\%$ of the total light received by the detectors. Figure 14 that I produced for the \citet{Poirier2020} paper.
\label{fig:Poirier2020_fig14}}
\end{figure*}

In the \citet{Poirier2020} paper, we argued that the brightness in \textit{WISPR-I} images originates from light mostly scattered by electrons situated close to a surface called the "Thomson sphere" and that is located ahead of \textit{PSP} \citep{Vourlidas2006}. In fact, this effect is further described in \citet{Howard2009} and \citet{Howard2012} who show that while the scattering efficiency is minimal at the Thomson sphere, it is largely offset by the combined fall off of the incident sunlight and of the electron density with distance from the Sun (both as $1/r^2$). The Thomson sphere is therefore the locus of points along all LOS where density and incident brightness from the Sun are maximised and therefore produce a total scattered intensity ($I_{tot}$) that is \emph{maximum} in this region. However, \citet{Howard2012} warn of the fact that this \emph{maximum} at the Thomson sphere is greatly smeared out. Therefore, a detector such as \textit{WISPR} would not be only sensitive to electrons that are concentrated near the Thomson sphere but rather, to a broader region on either side of the Thomson sphere that is called the "Thomson plateau" within which local intensities contribute almost equally to the total light received by the detector. That is illustrated in Figure \ref{fig:Poirier2020_fig14} where the shaded areas account for $99\%$ of the total brightness received by \textit{PSP WISPR-I} and the \textit{SOHO LASCO-C2} coronagraph. \\

Because brightness contributions can be significant along an extended portion of the LOS within the Thomson plateau and even beyond for particularly dense structures, it is often difficult to segregate which density structure contributes to the total brightness. To alleviate this difficulty I adopt for the following analyses a forward modeling strategy, where I exploit global 3-D MHD simulations of the solar corona and wind to produce synthetic WL observations. Then I carry out an in-depth analysis of each local contribution to the total brightness, where one method consists in separating the foreground and background contributions from either side of the Thomson sphere and in generating a synthetic image for each one of these sub-domains as done in \citet{Poirier2020}.

\section{Producing \textit{WISPR} synthetic images}
\label{sec:stationnary_synthetic}

Two methods are generally employed to interpret the shape and origin of the structures imaged by WL instruments, that are often based on the Thomson scattering theory introduced in section \ref{sec:stationnary_theory}. \\

First, inversion techniques derive the 3-D structure of the corona directly from the imagery, they can infer the electron density by rotational tomography \citep[see e.g.][]{Kramar2016b,Morgan2020} and even the magnetic field direction via spectropolarimetry \citep[see e.g][]{Kramar2016a}.

The second method, which we follow throughout this thesis, is the forward modeling technique that consists in producing synthetic WL observations from a given 3-D coronal structure. A comparison of the synthetic versus real products then provide a baseline to interpret the imaged structures. A great benefit of using this method is that we start from a modelled (known) coronal structure and we can exploit it to understand how local density structures may contribute to the total light received by a detector. In the \citet{Poirier2020} paper we show that even simplistic PFSS reconstructions of the coronal magnetic field alone (see section \ref{subsec:PFSS} for an introduction on PFSS models) already give valuable insight to understand WL observations. We also resort to extensive 3-D simulations such as MULTI-VP (see section \ref{subsec:MULTI-VP}) and WindPredict-AW (see section \ref{subsec:WindPredict}) to reproduce a solar corona from density cubes that are influenced by the forming solar wind. The Thomson scattering theory (see section \ref{sec:stationnary_theory}) is then followed to compute a synthetic image where the simulated electron density is integrated along each LOS of an instrument. \\

For the construction of the synthetic WL images we use the following procedure. For each \textit{WISPR-I} image, we calculate the heliographic coordinates of all pixels situated on a reference surface that we choose to be the Thomson sphere. We define LOSs that start from \textit{PSP}, pass through and beyond the Thomson surface. Along each LOS, we interpolate the electron density of the simulated 3-D datacube produced by any global MHD model such as MULTI-VP or WindPredict-AW, and we use the Thomson scattering theory to calculate a total intensity using equation \ref{eq:Thomson_scattering}. There are three major time-dependent effects that are involved in this process. 

First, because \textit{PSP-WISPR} is a detector placed on a rapidly moving observatory sweeping extended regions of the solar corona in only a few days, together with a rapid variation of its distance to the Sun, the FOV of \textit{WISPR} must be updated very regularly. In order to carry out an accurate comparison between observed and simulated \textit{WISPR} images it is therefore critical to keep an accurate tracking of \textit{WISPR}'s pointing through usage of the World Coordinate System (WCS).

Second, there is the time-dependency associated with the photospheric magnetic map that is used as the inner boundary to all 3-D simulations of the solar corona. For the sake of computational tractability we can generally afford to run only one single full 3-D MHD simulation within an observing window of \textit{WISPR}. Such simulations are commonly termed as "stationary" or ''static'' because in that case only one photospheric magnetic map is set at the inner boundary, that is maintained unchanged over the whole simulated time interval. 

However, time-dependent simulations such as Multi-VP and WindPredict involve a "dynamic" phase during which they evolve from an initial state until a permanent (stationary) regime is reached. The latter constitutes the third time-dependency that needs to be considered throughout the building process of \textit{WISPR} synthetic images. \\

In the case treated in \citet{Poirier2020} and in section \ref{sec:stationnary_poirier2020} where we focus on the quasi-stationary behavior of the solar wind, we run our forward modelling technique on a single simulation snapshot but with LOSs that are continuously updated throughout the observing window of \textit{WISPR-I}. \\

To address the variability of the solar wind as done in \citet{Griton2020} and in section \ref{sec:dynamics_griton2020} and \ref{sec:dynamics_tearing}, there remains an essential task that consists in calibrating the timeline of the simulations that are used as inputs to build the synthetic images. Because the cadence of the simulation outputs does not necessarily match the one of \textit{WISPR}, I select for each synthetic requested \textit{WISPR} image, a simulation snapshot that is the closest to the real observation time. I prefer this method rather than performing time interpolation of the 3-D datacubes to avoid introducing non-physical structures in the simulations, especially when simulating the propagation of density perturbations in the solar wind (see section \ref{sec:dynamics_griton2020} and \ref{sec:dynamics_tearing}). Therefore when the cadence of the simulation outputs is lower than \textit{WISPR} ($\simeq 13-16\ \rm{min}$ for the cases considered in this thesis) the same simulation snapshot can be re-used several times in the building of subsequent synthetic images.

\section{Analysis of the first observations of \textit{WISPR} with the MULTI-VP model (Poirier et al. 2020)}
\label{sec:stationnary_poirier2020}

In this section, I focus on the first \textit{WISPR-I} observations that I analysed in detail in \citet{Poirier2020}. Although \textit{PSP} approached the Sun at much closer distances during the next encounters, we show in \citet{Poirier2020} that the \textit{WISPR} observations taken during the first passage already enrich our knowledge of the nature and structure of the slow solar wind. In the following I only give a summary of the \citet{Poirier2020} study, the reader is invited to refer to the full version of the paper published online for more detail. \\

\subsection{Comparison with 1 AU observatories}
\label{sec:stationnary_comp}

We exploit level-3 \textit{WISPR} images that capture primarily the K-corona which is the photospheric light scattered by coronal electrons (see section \ref{subsec:inst_WISPR}). We focus on the \textit{WISPR-I} inner telescope where the WL signature from streamers is the most distinguishable (see Figure \ref{fig:WISPR_merged_img} for a comparison of the \textit{WISPR-I} and \textit{WISPR-O} FOV). \\

Comparing images from different instruments can be a challenging task, especially from separate observatories. As seen in section \ref{sec:stationnary_theory}, remote-sensing instruments are in practice sensitive to local emissions from a broad region that is called the Thomson plateau \citep{Howard2012}. This is particularly true for observations taken from near 1 AU. Figure \ref{fig:Poirier2020_fig14} shows a comparison of the regions likely observed by \textit{WISPR-I} and \textit{LASCO-C3} on 2018 November 1 as predicted by the Thomson scattering theory. It shows that \textit{WISPR-I} and \textit{LASCO C3} imaged a similar coronal region. As we can see in Figure \ref{fig:Poirier2020_fig14}, the brightness recorded by \textit{WISPR-I} originates from a broad region along the LOS that extends from \textit{PSP} to well behind the Thomson sphere. As \textit{PSP} gets closer to the Sun, this region shrinks. Nevertheless, Figure \ref{fig:Poirier2020_fig14} shows that \textit{WISPR-I} already records plasma brightness from a smaller region than \textit{LASCO C3}. \\

\begin{figure*}[]
\centering
\includegraphics[width=0.95\textwidth]{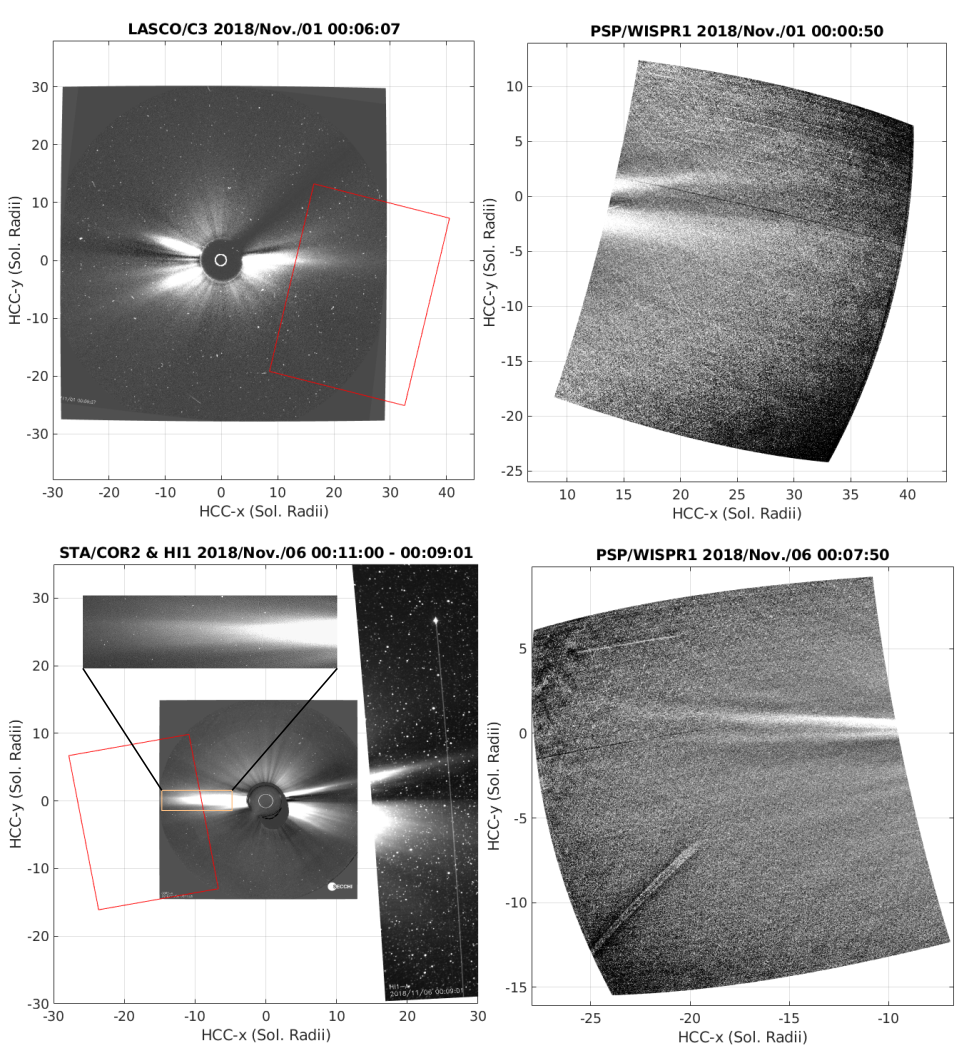}
\caption{ A comparison between \textit{WISPR-I} images with \textit{LASCO-C3} images at around 00:06 UT on 2018 November 1 (top panels) and \textit{STA COR-2A} and \textit{HI1} images at around 00:11 UT on 2018 November 6 (bottom panels). The red boxes overplotted at the left panels represent the \textit{WISPR-I} field of view for comparison with those observations. Figure taken from \citet[][Figure 4]{Poirier2020}.
\label{fig:Poirier2020_fig4}}
\end{figure*}

Considering the above complications, we give in Figure \ref{fig:Poirier2020_fig4} an overview of the zoomed-in view offered by \textit{WISPR-I} on the typical streamer structures observed from near 1~AU observatories. In Figure \ref{fig:Poirier2020_fig4}, we compare \textit{WISPR-I} with \textit{LASCO C3} and \textit{STEREO COR-2A} observations for two selected days when the comparison was optimal. On 2018 November 1 (top panels), the two brightest rays situated just a few degrees north and south of the equatorial plane are imaged by both spacecraft. In addition to these two bright features, a number of much fainter rays are also visible at \textit{PSP}, unveiling an apparent complex structuring of the corona, which is not resolvable in \textit{LASCO-C3} images. Besides, we note that, on 2018 November 1, \textit{WISPR-I} observations had not reached their highest resolution for this encounter yet, as \textit{PSP} was located at $\simeq 51\ R_\odot$, e.g., $\simeq 16\ R_\odot$ away from perihelion. \\

On 2018 November 6, the best alignment was with \textit{STA}, and Figure \ref{fig:Poirier2020_fig4} (bottom panels) shows a comparison of \textit{COR-2A} and \textit{WISPR-I} images. The pair of bright rays located just above the equatorial plane is not clearly distinguishable in \textit{COR-2A} despite the adequate resolution of the instrument to resolve such structures. This is likely a LOS integration effect, suggesting that 1 AU observatories may be not capturing the fine coronal structure. The zoomed-in view that \textit{WISPR-I} offers and the shrinking of the Thomson plateau, make it possible to observe coronal rays in finer detail compared with near 1 AU observations. \\

\subsection{WL synoptic map of the first \textit{WISPR-I} observations}
\label{subsec:stationnary_poirier2020_WLmap}

\begin{figure*}[]
\centering
\includegraphics[width=0.95\textwidth]{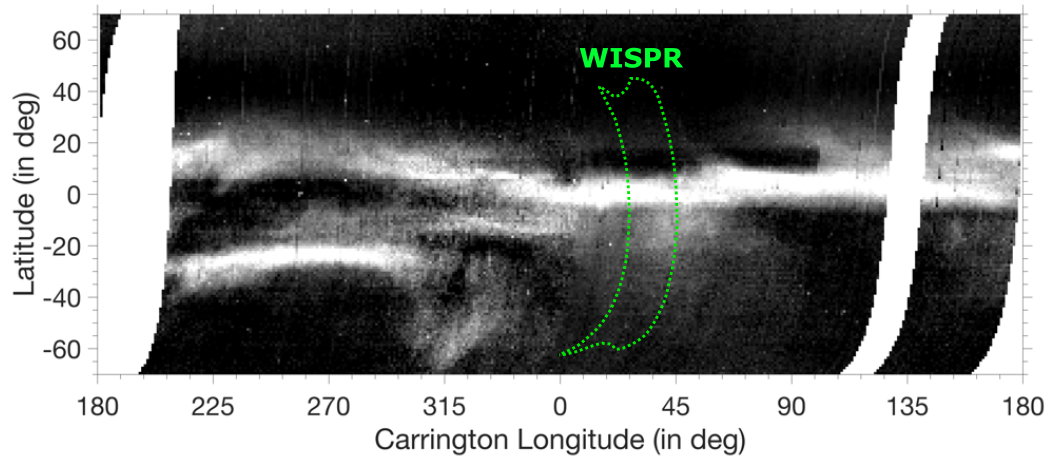}
\caption{ Carrington map from \textit{LASCO-C3} WL observations during the time interval 2018 October 20 to November 14. The likely scanned region by \textit{WISPR} during its first observing window is represented by the green dotted line. Figure taken from \citet[][Figure 5]{Poirier2020}.
\label{fig:Poirier2020_fig5}}
\end{figure*}

Following a similar approach as the one employed for years to exploit 1 AU coronagraphic images (see section \ref{subsubsec:intro_WLmaps}), I built synoptic WL maps of \textit{WISPR-I} images. However, because \textit{PSP} is a fast-moving observatory that is sometimes in corotation with the rotating solar corona, a latitude versus time format was found in general to be more convenient than the usual latitude versus Carrington longitude format. To illustrate this point, a visual comparison is given in Figure \ref{fig:Poirier2020_fig5} between a \textit{WISPR-I} synoptic map that combines images taken during the entire 1st encounter (i.e. about 10 days of coverage) and the associated \textit{LASCO-C2} map assembled over half a synodic period (about $27/2=13.5$ days). Therefore \textit{WISPR} mostly scanned the same region of the solar corona during the whole observing window. That is however less pronounced in the recent closest approaches of \textit{PSP}, where the quasi-corotating phase is greatly reduced thanks to reduced periapsis distances. \\

The building process of the \textit{WISPR-I} maps is detailed in \citet{Poirier2020} and is illustrated in Figure 16 of this paper. In summary I extract in each \textit{WISPR-I} image all pixels that are located at a specific distance to the Sun, and then I map those pixels in a latitude versus time format. In this process, one has firsthand to project the image on a specific plane of reference. Motivated by the Thomson scattering theory introduced in section \ref{sec:stationnary_theory}, a natural choice is the Thomson sphere at the core of the broad Thomson plateau where WL emissions are roughly expected to occur. For 1 AU observatories and targets that are close to the Sun, one could simply use the plane-of-the-sky to a good approximation. \\

\begin{figure*}[]
\centering
\includegraphics[width=0.95\textwidth]{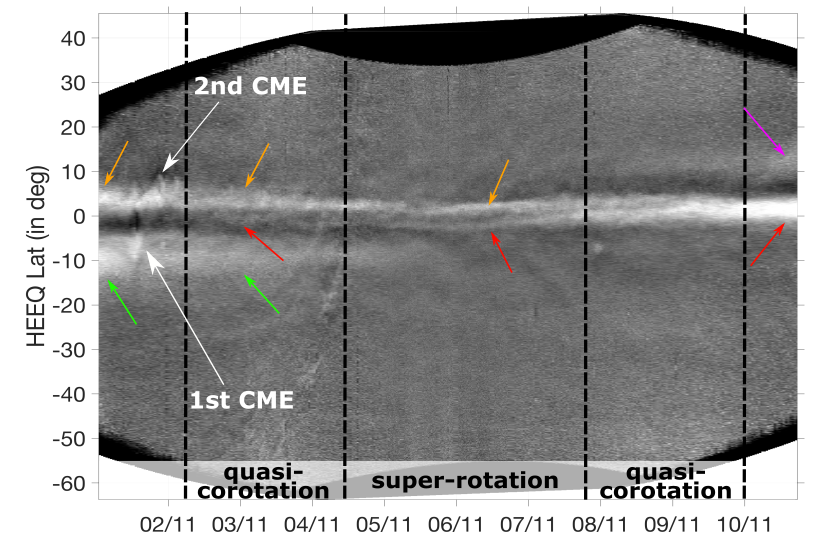}
\caption{ Latitude vs. time map from \textit{WISPR-I} images taken during the first encounter: from 2018 November 1, 00:45~UT, to 2018 November 10, 17:29~UT. The y-axis is the heliographic latitude in degrees. Figure 6 that I produced for the \citet{Poirier2020} paper.
\label{fig:Poirier2020_fig6}}
\end{figure*}

Figure \ref{fig:Poirier2020_fig6} presents a \textit{WISPR-I} latitude versus
time synoptic map for the first encounter, combining about 10 days of observations. At first sight one can distinguish many interesting features. We give a close-up look at those features in the next sections and we investigate their possible origin.

\subsection{Interpreting the large-scale structures}
\label{subsec:stationnary_poirier2020_largescale}

As mentioned earlier, PFSS reconstructions of the coronal magnetic field are rather simplistic but valuable as they already provide some clues to interpret the large-scale structures observed by \textit{WISPR}.

In Figure \ref{fig:Poirier2020_fig7}(a), we present the results of a PFSS extrapolation based on the GONG-ADAPT magnetogram produced for the 1st \textit{PSP} encounter on 2018 November 5 at 12:00~UT. In this view, \textit{WISPR-I} was observing plasma that originated near the west limb of the Sun (i.e. on the right-hand side of the image). The open and closed magnetic field lines are shown in yellow and orange, respectively, and the position of the coronal neutral line, i.e. the origin of the HCS, at the source surface is given by the red line. 

The streamer stalk denoted by the black arrow is associated to the bright band (orange arrows) located at $\approx 5\ \rm{deg}$ north in the \textit{WISPR-I} map shown in Figure \ref{fig:Poirier2020_fig6}.

Figure \ref{fig:Poirier2020_fig7}(b) presents a zoomed-in view of panel (a) centered to the south of the streamer. The EUV map reveals the presence of an isolated coronal hole (indicated by a blue arrow) at the southern edge of the regular streamer, and that is associated with a low-lying cusp-like structure. This is a typical pseudo-streamer, and it likely produces the bright band (green arrows) located approximately at $\approx 10\ \rm{deg}$ south of the equator in the \textit{WISPR-I} map shown in Figure \ref{fig:Poirier2020_fig6}. \\

From November 3rd, the northern streamer starts to split in two individual thin strands (red arrow in Figure \ref{fig:Poirier2020_fig6}) that persist until November 9th. The splitting of streamer rays has already been observed from 1 AU observatories but to a larger extent, and was generally associated to an inclination of the HPS with the LOS of the observer (see section \ref{subsubsec:intro_WLmaps}). Here the splitting produce two strands of $\approx 5\ \rm{deg}$ each, that is about the typical width of the HPS (see section \ref{subsubsec:intro_HCS_HPS} and \citet{Winterhalter1994}). Hence \textit{WISPR} may be the first heliospheric imager to unveil the thin and dense HPS that is confined within the core of streamers. A visual inspection of the PFSS reconstruction shown in Figure \ref{fig:Poirier2020_fig7}(a) reveals a HPS that is slightly warped on the right-hand side of the figure, and that may produce the splitting of the streamer rays observed by \textit{WISPR}. To examine the origin of this feature in detail, we exploit in the next sections additional inputs from high-resolution 3-D MHD simulations. \\

On November 1st and 2nd, large-scale fluctuations of the northern streamer have been detected by \textit{WISPR-I}.  They were found to be associated to the deflection of the streamer induced by the passage of two coronal mass ejections (CMEs). These CMEs were slow and dragged along in the slow solar wind \citep{Hess2020,Rouillard2020b}. I contributed significantly to the \citet{Rouillard2020b} paper in which I tracked the slow CME imaged by \textit{WISPR} on November, 1-2 2018 as it propagated throughout the corona and heliosphere, using both direct triangulation methods and a physical based model. \\

Corrugations at the edge of the northern streamer can also be seen while being much smaller at other times in the \textit{WISPR-I} map, especially on November 3rd. It is important to notice that \textit{PSP} was in a quasi-corotation phase during this period and therefore these perturbations should be transient structures propagating along with the slow solar wind. Some of them have been associated to density fluctuations expelled along the edge of streamers \citep{Rouillard2020a} that have been measured later on at \textit{PSP} in situ. The presence of such perturbations may be the signature of magnetic reconnection processes occurring at lower heights and which generate density fluctuations that are dragged in the slow solar wind, as stated in the dynamic theory introduced in section \ref{sec:intro_dynamic} and that is further discussed in section \ref{sec:stationnary_WISPR} in lights of the recent \textit{WISPR} observations.

\subsection{Numerical setup}
\label{subsec:stationnary_poirier2020_numerical}

To interpret \textit{WISPR-I} observations, I exploit the forward modeling method already discussed together with a high-resolution simulation of the solar corona and of the solar wind using the MULTI-VP MHD model introduced in section \ref{subsec:MULTI-VP} \citep[see also][]{PintoRouillard2017}. For this purpose, MULTI-VP was run on thousands of magnetic flux tubes to simulate the entire solar wind escaping the solar atmosphere. The inner boundary of the simulation domain is at the photosphere and extends typically to about $30\ R_\odot$. For the purpose of this study, the outer boundary of the MULTI-VP simulation was set to $90\ R_\odot$ in order to include the brightness contribution of electrons situated far behind the Thomson sphere. \\

The MULTI-VP model aims at simulating the solar wind as it forms along an expanding magnetic flux tube. Therefore the heating flux $F_h$ that appears in the energy equation (eq \ref{eq:MVP_T}) is a function dependent of the flux tube's cross sectional area $A(s)$ and of the magnetic field amplitude at the base of the flux tube $|B_\odot|$ \citep{PintoRouillard2017}, with a formulation inspired from \citet{Withbroe1988}:
\begin{equation}
    F_h=12\times 10^5 |B_\odot| \left(\frac{A_\odot}{A}\right)exp\left[-\frac{s-R_\odot}{H_f}\right]\ \ erg.cm^{-2}.s^{-1}\ (\text{with }|B_\odot|\text{ in }G) \label{eq:MVP_Q}
\end{equation}
where $H_f=5f_{ss}^{-1.1}R_\odot$ is the heating scale height. The expansion factor $f_{ss}=A_{ss}/A_\odot(r_\odot/r_{ss})^2$ is computed between the base of the flux tube at $R=R_\odot$ and the height $R=R_{ss}$ in the corona from which the magnetic field becomes purely radial. In the case where PFSS reconstructions are used to specify the magnetic field topology to MULTI-VP, this height is the height of the PFSS model's outer boundary called the source surface (see section \ref{subsec:PFSS}). \\

The MULTI-VP solar wind model runs on coronal magnetic fields that can be derived by potential field source-surface (PFSS, see section \ref{subsec:PFSS}) extrapolations of magnetograms measured by different observatories. In this study, we used photospheric magnetic field maps from WSO (see section \ref{subsubsec:wso}) and those computed by the ADAPT model (see section \ref{subsubsec:adapt}) based on GONG magnetograms (see section \ref{subsubsec:gong}). Hereafter we will refer to these two sets of simulations as MVP-WSO and MVP-ADAPT. \\

The MULTI-VP simulations have a grid adapted to the input magnetogram resolution. The MVP-WSO run has a grid of 5$^\circ$ resolution in both latitude and longitude, and the MVP-ADAPT run is at a higher grid resolution of 2$^\circ$. The MVP-ADAPT simulation provides 2.5 times finer resolution compared with the MVP-WSO run; this provides a significant impact on the synthetic images which will be discussed later on. \\

\subsection{\textit{WISPR-I} synthetic images}
\label{subsec:stationnary_poirier2020_synthetic}

MULTI-VP provides densities in all regions of space occupied by open magnetic fields. Using those cubes, we produce synthetic WL imagery by applying the theory of Thomson scattering \citep[see][]{Howard2009} introduced in section \ref{sec:stationnary_theory} and the method described in section \ref{sec:stationnary_synthetic}. These synthetic images are compared with \textit{WISPR} observations, providing a baseline to interpret the imaged structures. \\

\begin{figure*}[ht!]
\centering
\includegraphics[scale=0.55]{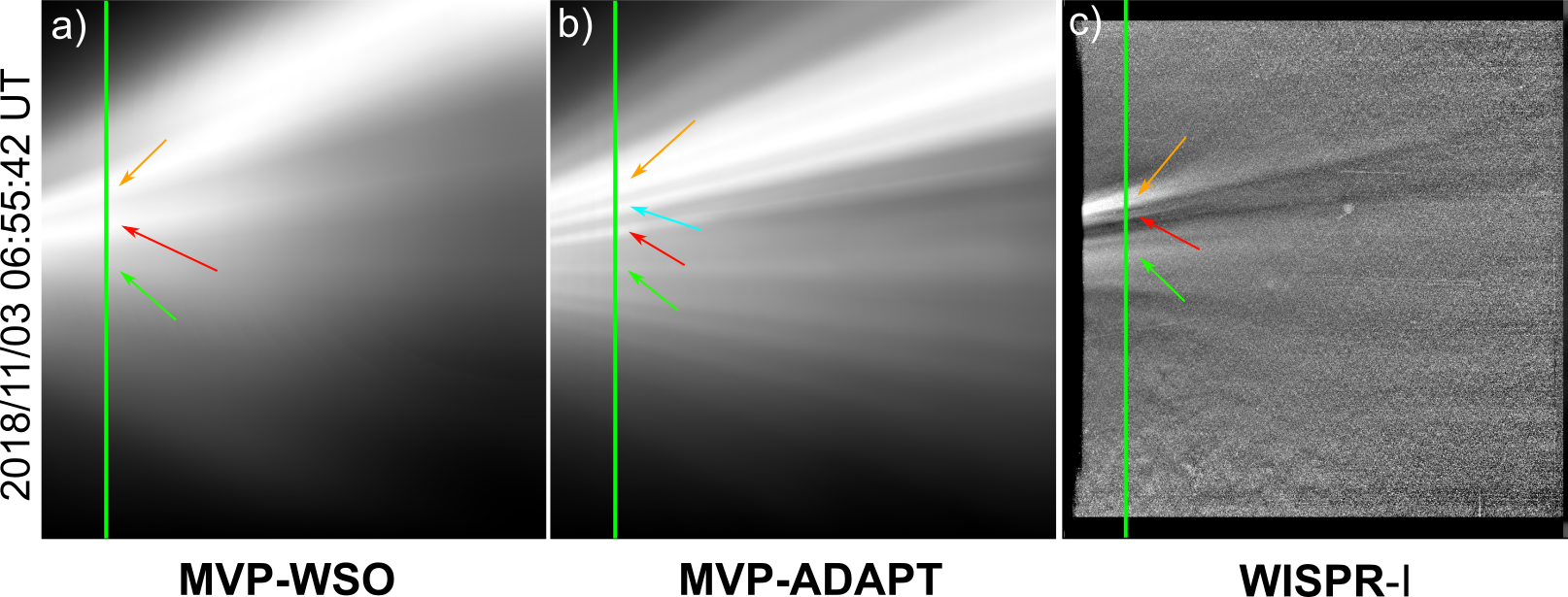}
\caption{ A comparison between synthetic WL images (panels a and b) and a \textit{WISPR-I} image (panel c) on 2018 November 3, at 06:55~UT. The synthetic images produced by MHD data from the MULTI-VP model and two different source magnetograms; WSO magnetogram of Carrington rotation CR2210 (panel a) and GONG-ADAPT magnetogram of 2018 November 5, 12:00~UT (panel b). The arrows superimposed at the images are the same as in Figure \ref{fig:Poirier2020_fig6}, and are indicative of the position of the features discussed in the text. Figure 8 that I produced for the \citet{Poirier2020} paper. \label{fig:Poirier2020_fig8}}
\end{figure*}

Figure \ref{fig:Poirier2020_fig8} presents examples of synthetic WL images produced from the MVP-WSO (panel a) and MVP-ADAPT (panel b) runs. Panel (c) shows the corresponding \textit{WISPR-I} image on 2018 November 3 at 06:55~UT for comparison with the simulated images. For illustration purposes, I rescaled the intensity for both synthetic WL images to enhance the visibility of the streamer rays. As a consequence, in the following sections I only perform a qualitative comparison of the features aspect and position between synthetic products and real observations. \\

Comparing the synthetic images between the two runs, I find, as explained earlier, similarities in the features observed near the equator, but there are also some striking differences that are presented in detail in \citet{Poirier2020}. 

Overall, the high-resolution MVP-ADAPT simulation produces a more detailed view of the streamers and reveals substructures that are absent in the MVP-WSO simulation. The streamer rays (orange arrows) are reproduced by both MVP-WSO and MVP-ADAPT. This is also the case for the pseudo-streamer identified in Figure \ref{fig:Poirier2020_fig7}(b) and here pointed with the green arrows. 

The splitting of the northern streamer into two lanes observed in the \textit{WISPR-I} map (orange and red arrows in Figure \ref{fig:Poirier2020_fig6}) is also well reproduced in both simulations but it appears at a slightly different latitude compared to the observations. It is probably related to the inherent uncertainties in polar field measurement. At first sight, it can be hard to identify the two-lane splitting within the multiple adjacent thin rays visible in the MVP-ADAPT image. Section \ref{subsec:stationnary_poirier2020_smallscale} will remove this ambiguity and confirm the identification made here. \\

Most of the differences between the two synthetic WL images of Figure \ref{fig:Poirier2020_fig8} appear in the northern streamer, where the higher-resolution MVP-ADAPT simulation shows an additional subdivision of the northern streamer into at least two separate rays (orange and blue arrows). However, this subdivision is not discernible in the \textit{WISPR} image (panel c). The more diffuse streamer rays that appear in the MVP-WSO image could result from multiple rays unresolved in this lower resolution simulation but clearly seen in the higher-resolution MVP-ADAPT run (panel b). Overall, the MVP-ADAPT simulation captures better the finer-scale structures of coronal rays that are also observed by \textit{WISPR-I} in the streamer and pseudo-streamer. However, the MVP-ADAPT synthetic image tends to show additional streamer rays in the northern streamer that are not seen clearly in the \textit{WISPR-I} image, and that we discuss in more detail in the next section. 

\subsection{Interpreting the small-scale structures}
\label{subsec:stationnary_poirier2020_smallscale}

In the following, we further exploit the simulations to interpret the properties of coronal rays observed by \textit{WISPR-I}. The analysis of all the features appearing in the real/synthetic images and maps has proven to be a complex task. Mainly, the LOS effects make it difficult when we need to identify the source regions responsible for the different rays visible in the synthetic maps or images. \\ 

\begin{figure*}[]
\centering
\includegraphics[width=0.95\textwidth]{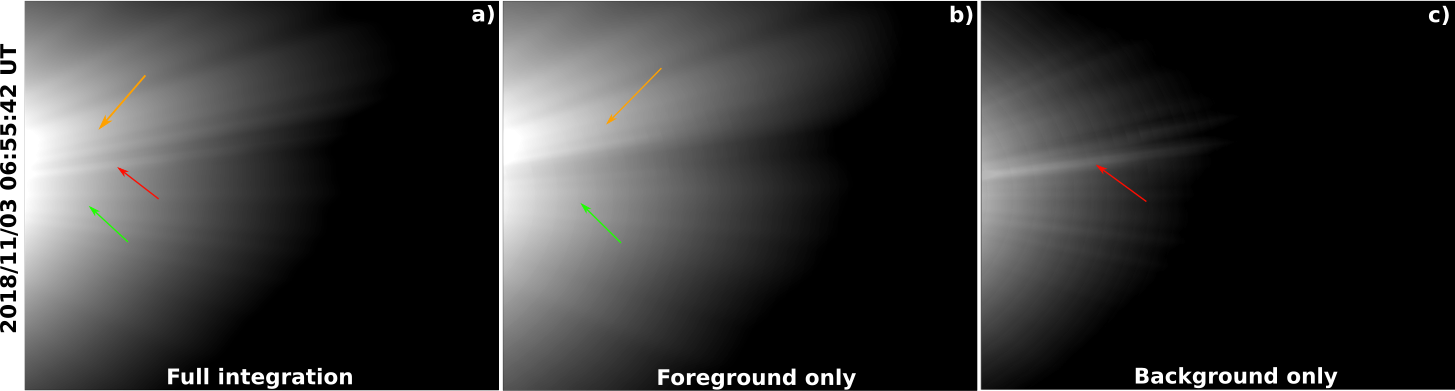}
\caption{ Three WISPR WL synthetic images on 2018 November 3 at 06:55~UT from the MVP-ADAPT run. They correspond to integration along the LOS over (a) the full span, (b) the foreground only, and (c) the background only. The arrows' color coding is the same as in the previous figures. Figure 10 that I produced for the \citet{Poirier2020} paper.
\label{fig:Poirier2020_fig10}}
\end{figure*}

To get a better insight of where the bright rays are situated relative to the Thomson sphere, I recomputed the synthetic images by splitting the integration path along each LOS in two separate domains. The first domain covers only the region from the observer (\textit{PSP}) up to the Thomson sphere (``foreground'' region), while the second extends far out and beyond the Thomson sphere (``background'' region). In Figure \ref{fig:Poirier2020_fig10}, the initial WL synthetic image from the MVP-ADAPT run (Figure \ref{fig:Poirier2020_fig8}(b)) is again plotted in panel (a) along with its associate foreground (panel b) and background (panel c) subimages. The foreground subimage (panel b) looks very similar to the full image (panel a) and contributes to most of the diffuse brightness of the broad northern and southern rays (annotated by the orange and green arrows). In contrast, the background subimage (panel c) only reveals the thin and bright central ray at a few degrees south (marked by the red arrow), which is not visible in the foreground subimage. This is a clear hint that the full synthetic images consist of rays located over an extended region in front of and beyond the Thomson sphere. As already discussed in section \ref{sec:stationnary_theory}, there are indeed non-negligible contributions to the total brightness on both sides of the Thomson surface that are included in the LOS integration domain. \\

\begin{figure*}[]
\centering
\includegraphics[width=0.95\textwidth]{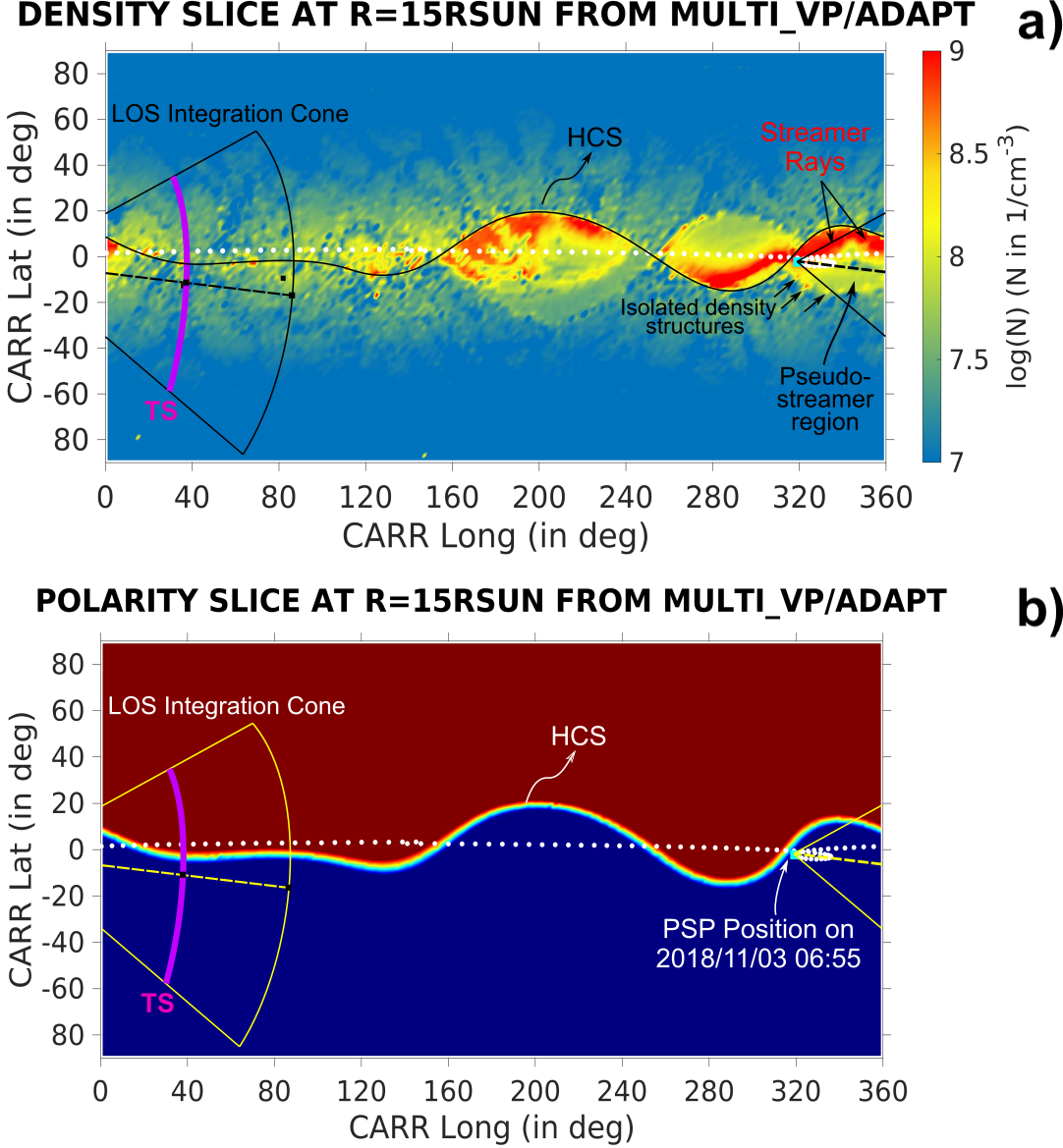}
\caption{ Panels (a) and (b) show, respectively, the Carrington maps of the simulated density and magnetic field polarity at $15\ R_\odot$ from the MVP-ADAPT simulation. The magenta line traces the intersection of the Thomson sphere with the Carrington map. The cone of integration, defined by the intersection between the field of view of \textit{WISPR-I} and the map, is shown in black (yellow) in panel a (b). Figure 11 from the \citet{Poirier2020} paper.
\label{fig:Poirier2020_fig11}}
\end{figure*}

In order to understand these subimages in more detail, I investigated further the 3-D topology of the corona. For that we use the density in the simulations as a proxy to visualize in 3-D these bright structures. Figure \ref{fig:Poirier2020_fig11}(a) shows a Carrington map of the simulated density at $15\ R_\odot$ for the MVP-ADAPT run. The white dashed line represents the \textit{PSP} projected trajectory, and the cyan square is the \textit{PSP} position on 2018 November 3 at 06:55 UT. The magenta line separates the foreground from the background integration domain. Figure \ref{fig:Poirier2020_fig11}(b) shows the Carrington map of the magnetic polarity from the MVP-ADAPT run. Comparing both Carrington maps, we can see that the densest solar wind forms in this simulation around the HCS and likely constitutes the HPS as introduced in section \ref{subsubsec:intro_HCS_HPS}. Dense wind also forms along arcs that connect different parts of the HCS; these correspond to the cusp of pseudo-streamers (see section \ref{subsubsec:intro_streamers}). There are also patches of slightly more tenuous slow winds extending away from the HCS. These are likely the product of intermediate expansion factors of the coronal magnetic field that lead to winds with intermediate speeds according to the quasi-stationnary theory introduced in section \ref{sec:intro_stationnary}. \\

One can see that the foreground domain is dominated by an intense and extended density enhancement associated with the northern streamer. This is consistent with the foreground subimage (Figure \ref{fig:Poirier2020_fig10}(b)) as \textit{PSP} is close in space to this high-density region. The instrument records a significant increase in brightness over a broad region extending northwards from near the equator. \textit{PSP} is therefore imaging different regions of the streamer from a vantage point that is just below the HPS.

We confirm that the pseudo-streamer rays (green arrow in Figure \ref{fig:Poirier2020_fig8}) originate near the unipolar cusp identified in Figure \ref{fig:Poirier2020_fig7}. This region is located south of the HCS and well in front of the Thomson sphere. Figure \ref{fig:Poirier2020_fig11}(a) reveals that the wind forming in that region is not as dense as the simulated streamer flows and would indeed appear less bright in the images. Consequently, this less dense region appears in the foreground subimage as a much fainter diffuse region in the lower half of the image (see the green arrow in Figure \ref{fig:Poirier2020_fig10} panel b).

On the contrary, the background integration domain covers a region of much lower density with a thin and flat layer of local density enhancement associated with the HPS. Imaging this east-west oriented structure from a larger distance explains why the streamer appears this time as a very thin and bright streamer ray visible in the background subimage (indicated by a red arrow in Figure \ref{fig:Poirier2020_fig10}(c)). \\

Therefore, the full synthetic image (Figure \ref{fig:Poirier2020_fig10}(a)) shows both the broad and diffuse light scattering emission of the foreground as well as the thin ray of the same northern streamer that flattens at lower latitudes behind the Thomson sphere. We must conclude that comparing \textit{WISPR-I} images by simply taking slices of simulations near the Thomson sphere is inadequate, and a complete analysis of \textit{WISPR-I} images requires an analysis that integrates foreground and background features. \\ 

\begin{figure}[ht!]
\centering
\includegraphics[scale=0.48]{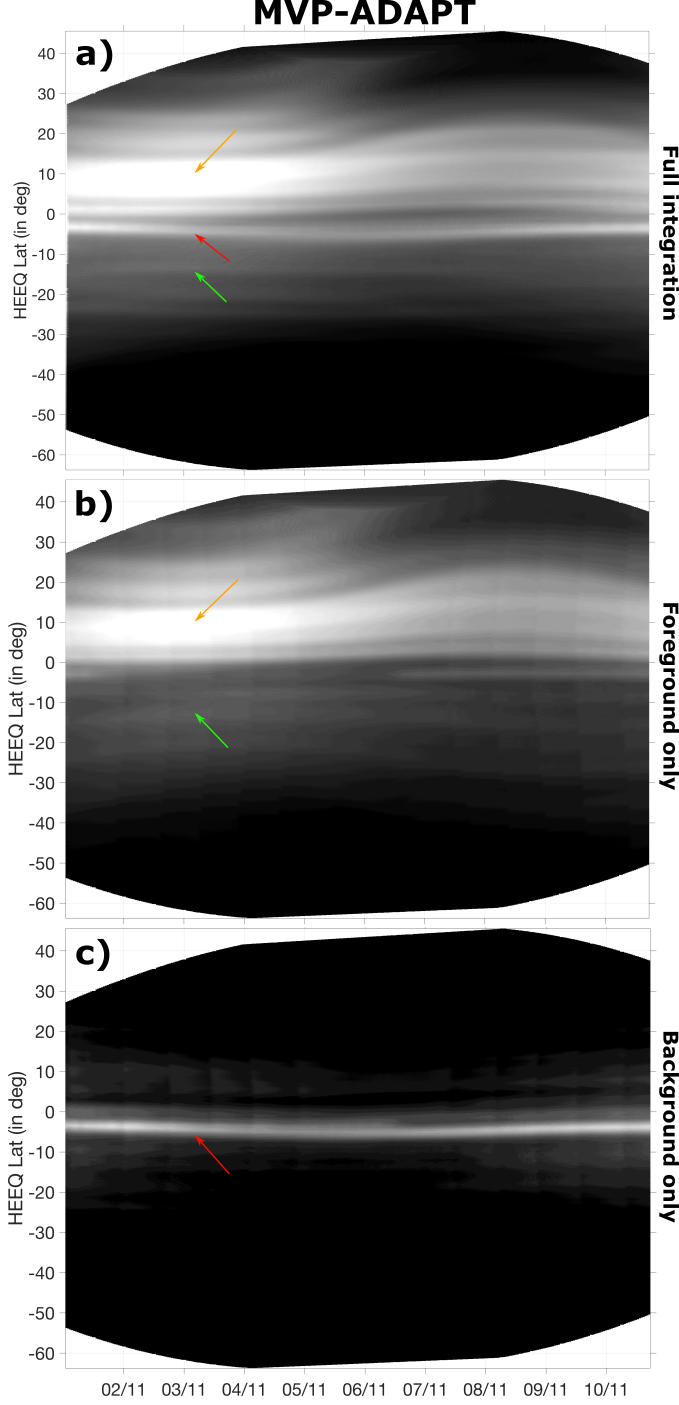}
\caption{Synthetic latitude vs. time maps from the MVP-ADAPT run. They have been generated from (a) the full synthetic images, (b) the foreground subimages only, and (c) the background subimages only. The arrows' color coding is the same as in the previous figures. Figure 12 that I produced for the \citet{Poirier2020} paper. \label{fig:Poirier2020_fig12}}
\end{figure}

Similarly to the \textit{WISPR-I} map shown in Figure \ref{fig:Poirier2020_fig6}, we can build synoptic maps from individual synthetic images generated from the MVP-ADAPT simulation. That is shown in Figure \ref{fig:Poirier2020_fig12}(a) where the associated contributions from the foreground and background are also given. Similarly to the analysis done on the WL synthetic subimages, the foreground submap (panel b) contributes to most of the bright structures seen in the full map (panel a). An exception is the thin southern bright stripe already mentioned and located in the background submap (red arrow in panel c). These submaps give us the last hint to understand the origin of the apparent two-lane splitting of the northern streamer ray that we identified earlier, which is annotated by the orange and red arrows and visible from 2018 November 3 to 9 in the \textit{WISPR-I} map (Figure \ref{fig:Poirier2020_fig6}).

From the decomposition of an MVP-ADAPT WL synthetic image into two subimages (foreground and background as shown in Figure \ref{fig:Poirier2020_fig10}) as well as the analysis of the polarity and density slices of the MVP-ADAPT run, we interpret this splitting as the result of an LOS integration effect from two very distinct regions. The initial northern streamer, slightly folded, visible in the foreground and located at a few degrees above the equator flattens further in the background to remain flat at a few degrees below the equator. This flat part of the streamer in the background is highly visible in the polarity slice (see Figure~\ref{fig:Poirier2020_fig11}(b)) from $\sim$30$^\circ$ to $\sim$100$^\circ$ Carrington longitude. Therefore, from the modeling, we can identify the apparent streamer splitting in the \textit{WISPR-I} map to be the result of an LOS integration effect of a slightly folded HPS extending at and beyond the Thomson sphere. \\

The density and polarity maps (Figure \ref{fig:Poirier2020_fig11}(a) and (b)) also reveal that smaller folds of the HCS situated in the foreground could create the main subdivision of the northern streamer seen in the MVP-ADAPT synthetic image (see e.g. the blue arrow in Figure \ref{fig:Poirier2020_fig8}(b)). This may also include the rays located at the highest latitudes, which drift towards higher latitudes as the effect of \textit{PSP} plunging into these structures. Such features are hardly distinguishable in the \textit{WISPR-I} map shown in Figure \ref{fig:Poirier2020_fig6}, but they appear much more clearly in the recent encounters (see Figure \ref{fig:WISPR_tour}). \\

Several dense flux tubes can also be observed in the southern region in the density map between Carrington longitudes 320$^\circ$ and 340$^\circ$. They are isolated flux tubes with higher densities that produce additional thin rays in the southern part of the synthetic foreground image (Figure \ref{fig:Poirier2020_fig10}). Others similar patches can be seen all over the Sun but they are hardly distinguishable in the dense northern streamer due to the saturation of the color scale. These dense flux tubes are likely the product of different magnetic conditions at the base of the flux tubes and that in turn affect the mass flux of the winds simulated in MULTI-VP. This is an innate characteristic of MULTI-VP that we already discussed at several occasions and that is built in the heating prescription adopted in MULTI-VP. Therefore the fine pattern of the photospheric magnetic map offered by the GONG-ADAPT magnetogram is reflected in MULTI-VP by a myriad of winds with various mass fluxes. These individual slow solar wind streams are, nonetheless, not clearly visible in the \textit{WISPR-I} observations and may be mixed up with the rays induced by the small warps of the HPS discussed in the previous paragraph. Still, we shall point out that similar fine-scale striations of the streamers can be seen in the more recent encounters where the \textit{PSP-WISPR} "microscope" effect was enhanced by taking images from a vantage point that was well inside the solar corona. Whether or not these individual slow solar wind streams predicted by the quasi-stationnary theory and simulated by MULTI-VP are realistic or not will have to be investigated in detail in lights of the more recent \textit{WISPR-I} observations, that is left for future studies. \\

\subsection{Conclusion}
\label{subsec:stationnary_poirier2020_conclusion}

The previous section was rich in details, therefore we shall give a summary of the main outcomes and limitations from this study and how these results may affect our understanding of the slow solar wind. The reader is invited to refer to \citet{Poirier2020} for an extensive discussion on the limitations of this study. \\

Modeling of the solar wind and corona has been extensively used in this study not to perform direct comparisons with \textit{WISPR-I} images but to help us understand the origin of the different coronal rays observed by \textit{WISPR-I} during the first encounter of \textit{PSP}. We showed the need of having a fine-enough simulation (e.g. MVP-ADAPT) in order to reproduce even the smallest features observed by \textit{WISPR-I}. \\

The high-resolution MVP-ADAPT simulation (with 2$^\circ$ angle resolution) allowed us to give further context and potentially explain the apparent splitting of the brightest streamer rays seen by \textit{WISPR-I}. Our results suggest that this originates from the LOS integration along an extended region where the HPS undergoes a latitude change. Our model shows that the HPS latitude changes from $\sim$10$^\circ$ to $\sim$-5$^\circ$ over a $\sim$60$^\circ$ Carrington longitude span at the region where \textit{WISPR-I} observations were made. The effect of such folds in the HPS have been known since \textit{LASCO} observations to produce separated streamer rays as discussed in section \ref{subsubsec:intro_WLmaps} \citep[see also][]{Sheeley1997,Wang1998}. The novelty in \textit{WISPR-I} observations is to act as a microscope to catch even small latitudinal changes in the HPS, allowing a more detailed evaluation of current coronal models. \\ 

This microscope effect manifests in \textit{WISPR-I} images by revealing very thin streamer rays of about $3-5^\circ$ in angular width. That is about the typical width of the HPS measured in situ \citep{Winterhalter1994}, and hence \textit{WISPR-I} may be the first WL heliospheric imager to unveil the thin HPS. As we shall see in section \ref{sec:stationnary_WISPR}, that is even more remarkable in the recent observations performed by \textit{WISPR-I}.

\section{A quick tour of subsequent \textit{WISPR-I} observations}
\label{sec:stationnary_WISPR}

\begin{figure*}[]
\centering
\includegraphics[width=0.95\textwidth]{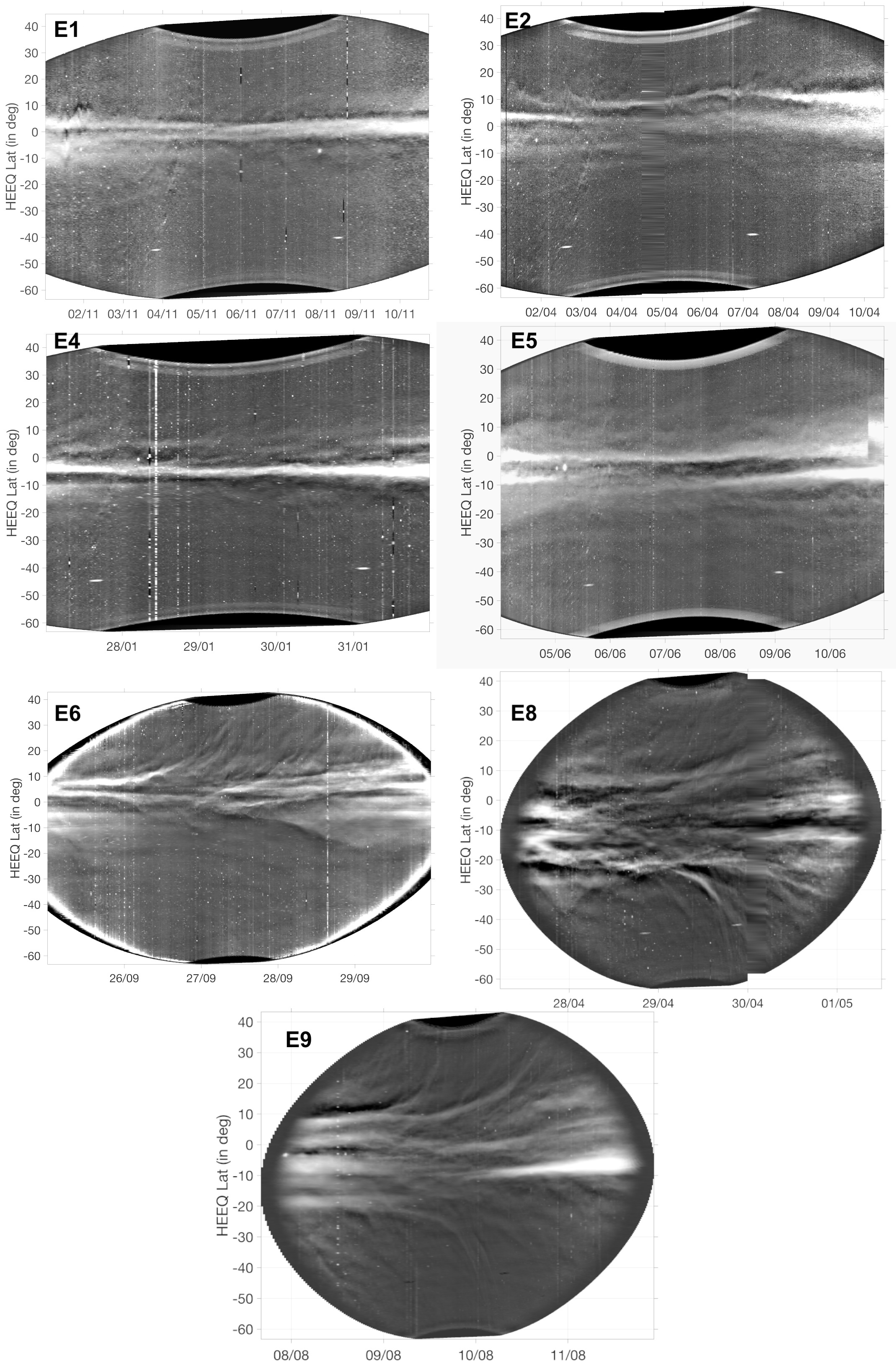}
\caption{ \textit{WISPR-I} WL synoptic maps for encounters 1, 2, 4, 5, 6, 8 and 9.
\label{fig:WISPR_tour}}
\end{figure*}

In figure \ref{fig:WISPR_tour} we show at a glance many \textit{WISPR-I} observations represented as latitude versus time maps for seven of the first ninth passages of \textit{PSP} to the Sun (same format as for Figure \ref{fig:Poirier2020_fig6}). 
A general observation is that the streamer belt remained more or less in the equatorial plane for the first seventh orbits, that is typical of a period with low solar activity. This was also noted in the plasma measurements taken in situ at \textit{PSP} which show a high occurrence of slow and dense solar winds coming primarily from the edge of streamers, as discussed in section \ref{sec:dynamics_insitu} for encounter 2. \\

Encounters 8 and 9 reveal a much more structured corona with multiple striations in the coronal rays. That may be the result of moving towards the ascending phase of the solar cycle with a HPS that becomes significantly warped. Whether all these striations are due to folding in the HPS is a question that still remains to be elucidated and that we partially addressed in section \ref{subsec:stationnary_poirier2020_smallscale}. In lights of the forward modeling that we performed for the 1st encounter, we also suggest that some of these striations may correspond to individual slow wind streams of distinct origins and hence to winds that have a different history in their formation (e.g. for the deposit of energy). We will see in section \ref{subsec:dynamics_tearing_3D} that at high solar activity, the solar atmosphere does not solely consists of a HPS but rather forms a complex network of dense layers that are associated with pseudo-streamers and more generally the QSLs (S-web, see section \ref{subsec:intro_Sweb}). Therefore, \textit{WISPR-I} may capture some of these QSLs that complicates the interpretation of the observations. A modeling of encounter 9 has been attempted in section \ref{subsec:dynamics_tearing_3D} using a 3-D MHD simulation from the WindPredict-AW model. We note that a fine level tuning of the model would be required to match the actual \textit{WISPR-I} observations. The exploitation of \textit{WISPR-I} images together with the use of synthetic imagery methods constitute a baseline to improve the accuracy of the current models of the solar atmosphere to an unprecedented level of detail. \\

Furthermore, one can notice a feature that is present in almost all \textit{WISPR-I} maps shown in Figure \ref{fig:WISPR_tour}. I refer to the dark bands that split the bright streamers in two separate rays. That feature generally results from a slight inclination of the streamer belt (or HPS) with respect to the line-of-sight of the observer, that has been explained in section \ref{subsubsec:intro_WLmaps} and examined in detail during the 1st encounter in section \ref{sec:stationnary_poirier2020}. \\

One can also see rays that drift towards high latitudes. Such drifting rays were poorly visible during the 1st encounter (Figure \ref{fig:Poirier2020_fig6}) but they become well visible from encounter 6th, and is a 3-D perspective effect of \textit{PSP} plunging directly into the streamers \citep[see also][]{Liewer2019}. Furthermore, this feature will be more visible in the synthetic \textit{WISPR-I} imagery performed in section \ref{subsec:dynamics_tearing_3D}, from two 3-D MHD simulations of encounter 5 and 9 based on the WindPredict-AW model.\\

Occasionally, one can see some corrugations in the streamer belt. As noted for the 1st encounter (see section \ref{subsec:stationnary_poirier2020_largescale}), some are related to the deflection of streamers induced by the passage of CMEs. Other corrugations in the streamers seen in \textit{WISPR} have been associated to density fluctuations expelled along the edge of streamers and measured in situ at \textit{PSP} \citep{Rouillard2020a}, and that present flux rope signatures as discussed in section \ref{sec:dynamics_WISPR}. These small-scale perturbations of the streamer become more and more preponderant in the subsequent encounters as \textit{PSP} gets closer to the Sun and as the microscope effect of \textit{WISPR} accentuates. That favors a theory of the formation of the slow solar wind that is ruled by systematic magnetic reconnection events as formulated in the dynamic theory introduced in section \ref{sec:intro_dynamic} and that is discussed in more detail in chapter \ref{cha:dynamics} in light of the recent \textit{PSP} observations. \\

\section{Conclusion}

In the present chapter, novel remote-sensing observations from \textit{PSP-WISPR} were exploited showing a structuring of the solar wind at much smaller scales than what has typically been achieved from 1 AU observatories. Because of the peculiarity of \textit{PSP} that dives deeply in the corona with unprecedented speeds, I had to develop new methods to interpret the WL images taken by \textit{WISPR}. At several occasions I pointed out the importance of considering the Thomson scattering theory (introduced in section \ref{sec:stationnary_theory}) to interpret the nature of the WL emissions captured by \textit{WISPR}. I showed that even for \textit{WISPR} that images at the corona up close, it remains difficult to locate precisely the regions where WL emissions occur. Fortunately, the closer-up \textit{PSP} approaches the Sun, the thinner is its area of sensitivity. The microscope effect of \textit{WISPR} was even more remarkable in the recent observations presented in section \ref{sec:stationnary_WISPR}, that reveal a fine-striation of the SSW at even smaller scales compared to the first observations that we analysed in detail in \citet{Poirier2020}  (section \ref{sec:stationnary_poirier2020}). \\

I  developed a new method to facilitate the interpretation of \textit{WISPR} images. By stacking images over time while accounting for \textit{PSP}'s position and \textit{WISPR}'s pointing, I built synoptic maps that show at a glance the structure of the corona imaged by \textit{WISPR}. By identifying phases where \textit{PSP} is in quasi-corotation with the rotating solar corona, we showed that the time-dependent and spatial-dependent components of the SSW can be disentangled. That way I could focus in this chapter on the analysis of the fine-structure of a likely quasi-stationnary SSW component imaged by \textit{WISPR}, while the more dynamic component SSW is discussed in detail in chapter \ref{cha:dynamics}. \\

Using a high-resolution MHD simulation from the MULTI-VP model, patches of dense solar winds have been found to form over the entire source region of the slow solar wind roughly below $40^\circ$ of heliographic latitude. This points towards a "texture" of the solar wind that finds its root in the highly structured nature of the coronal magnetic field. That was achieved following the quasi-stationnary theory (section \ref{sec:intro_stationnary}) where the energy source that drives the solar wind is controlled by the topology of the coronal magnetic field. By employing my forward modeling technique, I produced synthetic \textit{WISPR} images where these patches reflected in a fine striation of the streamer rays. \\

At the largest scales, this was interpreted as a small latitudinal shift of the location of the HPS between the foreground and background. Although this effect has long been known since the early stage of  \textit{SoHO} which orbits at near 1 AU, \textit{WISPR} is able to capture irregularities in the shape of the HPS with unprecendented details. \\

However, synthetic \textit{WISPR} images built from the MULTI-VP simulations showed a fine-structuring of the SSW at even smaller scales. We concluded that this is likely the result of two distinct contributions: sub-foldings of the HPS at smaller scales ($\lesssim 5^\circ$), and individual denser slow wind streams that take root in regions of distinct origins. The presence of such fine rays has been shown in highly processed eclipse images (see e.g. Figure \ref{fig:Mikic2018_fig1} or also \citet{November1996,Druckmuller2014}) and dominates significantly the recent images taken by \textit{WISPR} (see section \ref{sec:stationnary_WISPR}). More generally, this fine striation could be the manifestation of a SSW that forms along a highly-structured S-web (section \ref{subsec:intro_Sweb}), where WL emissions are expected not only from the HPS but from the QSLs at a much larger extent. The last observation carried on by \textit{WISPR} will have to be investigated in detail to identify precisely the sources of the observed fine-structuring of the SSW, that is left to future studies. \\

%% file: chapters/Dynamics.tex
\chapter{Variability of the slow solar wind}
\label{cha:dynamics}

\minitoc

In this chapter we will first show in section \ref{sec:dynamics_insitu} how the recent in situ measurements taken by \textit{PSP} support both the quasi-stationnary and dynamic theories of the SSW formation introduced in section \ref{sec:intro_stationnary} and \ref{sec:intro_dynamic} respectively. Then we inspect in section \ref{sec:dynamics_WISPR} the intermittent nature of the SSW (see section \ref{subsubsec:intro_SSW_intermittency}) from the recent close-up perspective offered by \textit{WISPR}. The WindPredict-AW MHD model introduced in section \ref{subsec:WindPredict} is exploited in section \ref{sec:dynamics_tearing} to investigate the formation process of density perturbations in the SSW through dynamic reconnection processes at the HCS, within the observational context of \textit{WISPR}. Finally, we show in section \ref{sec:dynamics_griton2020} that other transient processes occurring lower in the solar corona are likely to generate density fluctuations at higher frequencies.

\section{Variability of the SSW as measured in situ at \textit{PSP}}
\label{sec:dynamics_insitu}

\begin{figure*}[]
\centering
\includegraphics[width=0.8\textwidth]{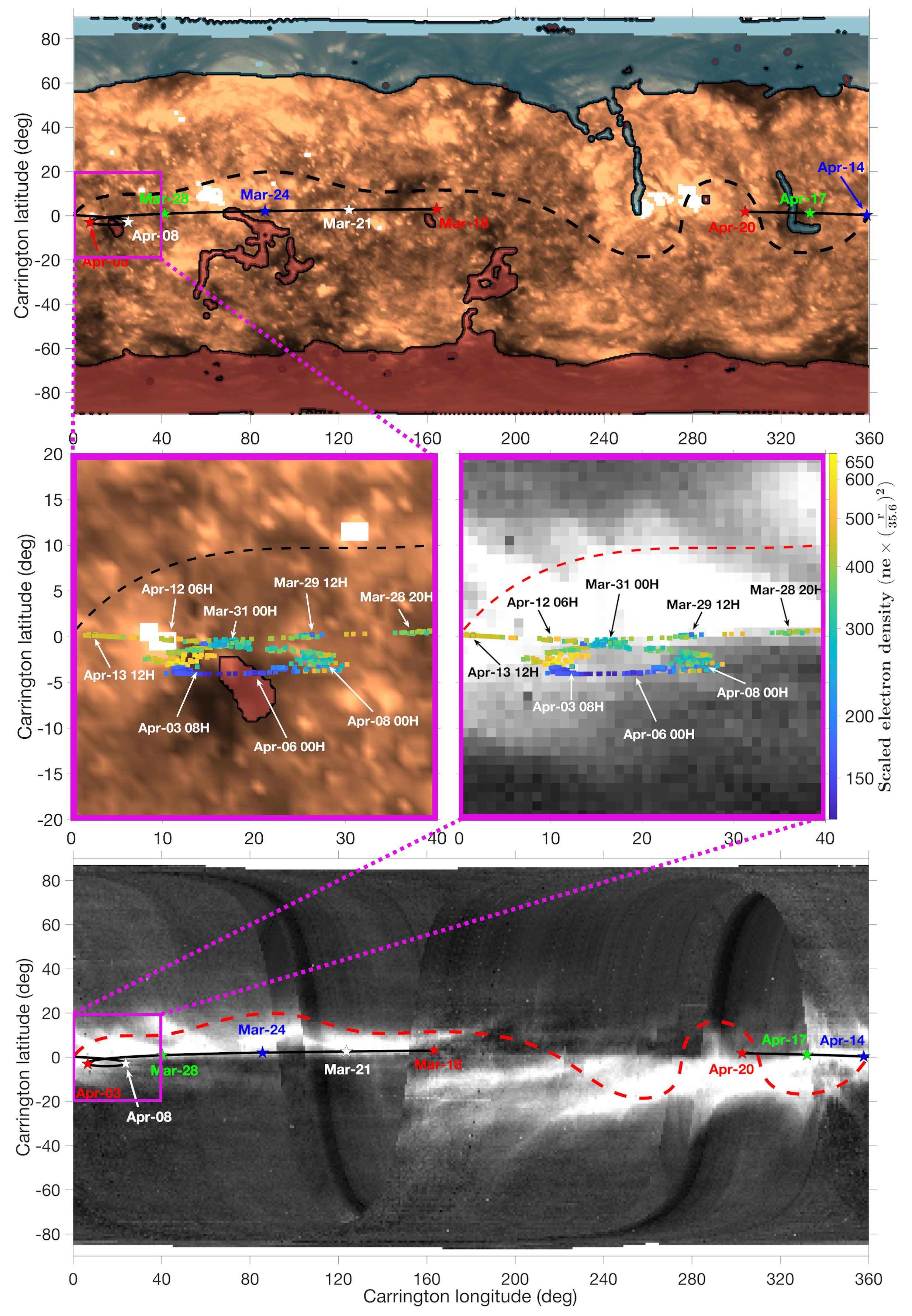}
\caption{Top panel: Carrington map from \textit{SDO}-\textit{AIA} 193\si{\angstrom} EUV observations, on March 31, 2019, at 00:00 UT. Bottom panel: Carrington map from \textit{SOHO} \textit{LASCO-C3} WL observations, on April 3, 2019, at 12:00 UT and at a height of 5 $R_\odot$. The \textit{PSP} trajectory (dark solid-line) is projected at the height of the Carrington maps (and at 2.1 $R\odot$ for the \textit{AIA} EUV map), using the Parker Spiral \citep{Parker1958} and the solar wind speed measured in situ at \textit{PSP}. Along the \textit{PSP} projected trajectory, some \textit{PSP} projected positions for several dates and times are depicted with colored stars and arrows. On the zoom-in views in the middle panels, the plasma electron density $n_{e}(35.6 R\odot)$ (in $cm^{-3}$) taken from quasi-thermal-noise (QTN) data of the \textit{FIELDS}-\textit{RFS} instrument \citep[see][for a description of the method]{Moncuquet2020} is plotted along the PSP projected trajectory with a logarithmic color scale. The red and dark dashed-lines represent the neutral line from an optimized PFSS reconstruction with a source surface height of $2.1\ R_\odot$ and a GONG-ADAPT photospheric magnetic map (2nd realization) of March 24, 2019, at 00:00 UT. Open field regions of positive (negative) polarity from the PFSS model are represented at the photosphere by a semi-transparent blue (red) layer. Figure 8 that I produced for \citet{Griton2021}.
\label{fig:griton2021_fig8}}
\end{figure*}

During its first solar encounters, the \textit{Parker Solar Probe} mostly measured SSW with speeds less than $500\ km/s$. The examination of streamer rays in WL observations by \citet{Rouillard2020a} during the second orbit of \textit{PSP} revealed that \textit{PSP} mainly scanned dense and SSW plasma that originated from the edge of streamers. However near perihelion, \textit{PSP} exited temporarily the dense streamer flows to sample a more tenuous SSW that likely emerged from deeper inside a coronal hole. A thorough study of this same time interval by \citet{Griton2021} revealed that the SSW measured in situ by \textit{PSP} comes from two different sources that have a distinct magnetic topology. At closest approach \textit{PSP} temporary exited the dense streamer flows to intercept magnetic field lines that are rooted in the center of a small equatorial coronal hole. This configuration is illustrated in Figure \ref{fig:griton2021_fig8} that I produced for the \citet{Griton2021} paper. The small equatorial coronal hole is visible as a small darker region in the EUV synoptic map (top and middle panels). \\

My contribution to this paper consisted in providing a detailed analysis of the \textit{PSP} connectivity to the Sun. For this purpose I basically followed the method presented in section \ref{subsec:poirier2021} to select the most realistic PFSS reconstruction of the coronal magnetic field, by comparing the modelled HCS (or neutral line) with the WL observations of the streamer belt derived from \textit{SOHO} \textit{LASCO}-C2 coronagraphic images. An optimal source surface height was found at $2.1\ R_\odot$ that captured both the streamer belt structure and the small equatorial coronal hole. This value is however not optimal to reproduce at best the streamer belt shape alone. This dichotomy between the source surface height optimized from EUV or WL observations has already been discussed in section \ref{subsec:badman2022} and is a shortcoming of PFSS models that lack some physics compared to more realistic MHD models. The connectivity of \textit{PSP} to the Sun was then estimated following the method of the Magnetic Connectivity Tool described in section \ref{subsec:Connectivity_tool} and assuming that \textit{PSP} connects to the source surface with a Parker spiral (ballistic model). \\

Around closest approach, from March 28th to April 13th 2019, \citet{Griton2021} identified two intervals of SSW that appeared to exhibit two distinct states that differed by their densities, magnetic field properties and plasma beta (the ratio between thermal to magnetic pressure, see section \ref{subsec:intro_RT}). State 1 was the SSW interval associated to the edge of the streamer and was found to be denser with a higher thermal pressure than the plasma coming from deeper inside the small equatorial coronal hole, or state 2. The magnitude of the magnetic field is also smaller near the streamer than in the center of the coronal hole that results in a higher beta parameter in the streamer flow (state 1). That is illustrated in the middle panel of Figure \ref{fig:griton2021_fig8} where the electron density has been plotted on top of a WL map of the streamer belt. In this plot, the state 1 (and state 2) SSW that has a high (and low) electron density can be identified to the greenish/yellowish (and blueish) colors. \citet{Griton2021} showed using the MULTI-VP model that the quasi-stationnary theory is able to reproduce the bulk properties of these two SSW states, this is discussed in more detail below. \\

The study focused on time intervals of very calm solar winds that were detected in \emph{both} state 1 and 2. While the SSW appeared overall much more perturbed in the streamer flows than from the coronal hole flow, some very quiet intervals were also measured inside the streamer flows (corresponding to state 1) meaning that a transient origin of these streamer flows may not always be necessary. \\

\begin{figure*}[]
\centering
\includegraphics[width=0.95\textwidth]{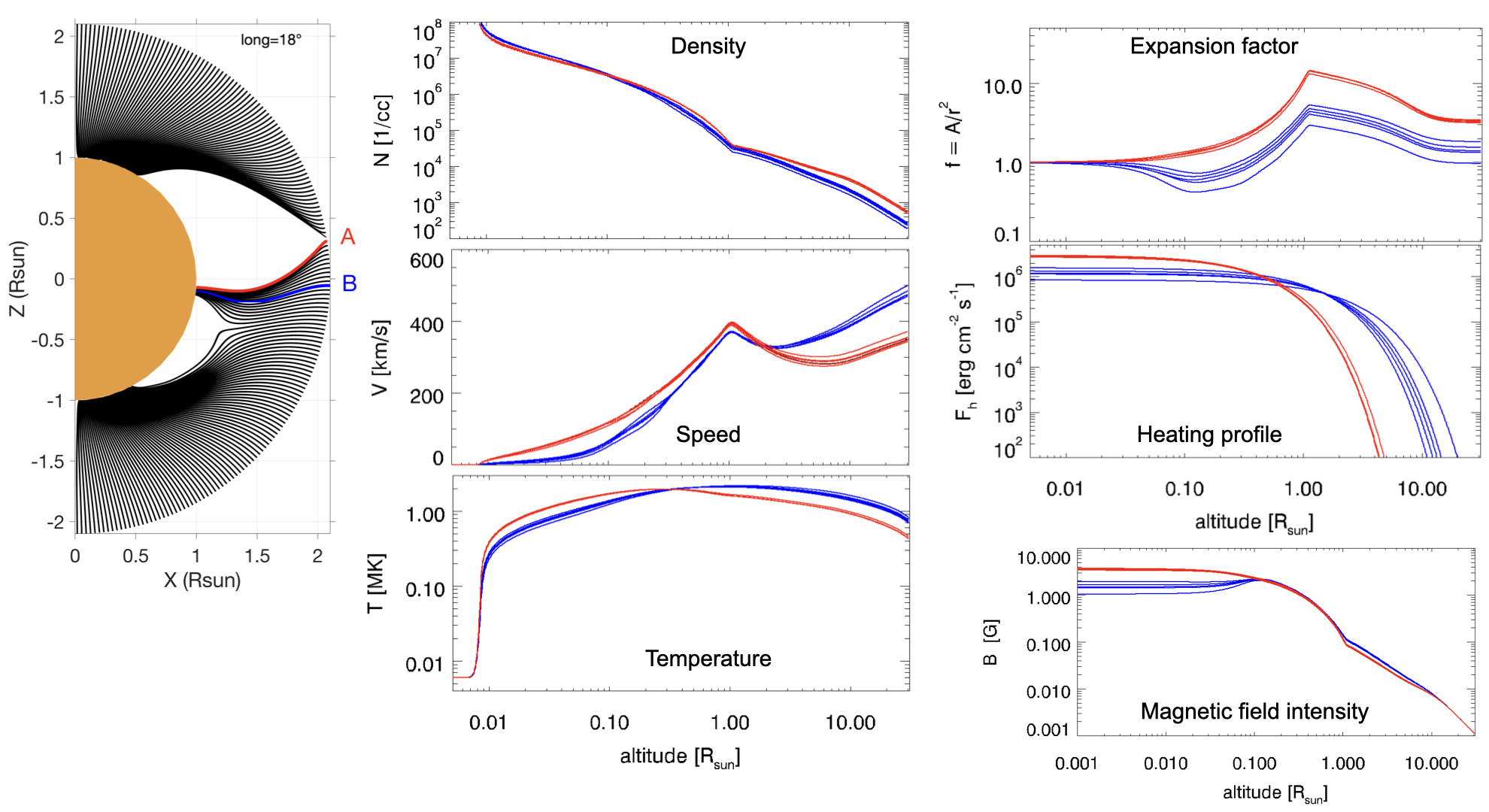}
\caption{ Left column: Samples of magnetic field lines from the optimized PFSS magnetic field model with the source surface at $2.1\ R_\odot$, in the plane cut through $18^\circ$ longitude. \textit{PSP} certainly crossed magnetic field lines connected to field lines A (state 1) and B (state 2). Middle column: plasma density (top), speed (middle) and temperature profiles (bottom) computed by MULTI-VP along 5 flux tubes around field line A (in red) and around field line B (in blue). Right column: same as the middle column but for the expansion factor (top), heating profile (middle) and magnetic field intensity profiles (bottom). Figure taken from \citep[][Figure 9]{Griton2021}.
\label{fig:griton2021_fig9}}
\end{figure*}

The MULTI-VP model (introduced in section \ref{subsec:MULTI-VP}) was exploited to check whether the quasi-stationnary theory could reproduce the two quiet SSW states measured at \textit{PSP}, which may result from the difference in the magnetic geometry of the two distinct source regions. As introduced in section \ref{sec:intro_stationnary}, the various solar wind speeds produced by MULTI-VP directly reflect the formulation that is adopted for the coronal heating (see section \ref{subsec:stationnary_poirier2020_numerical}) where the expansion factor controls the height at which the heat is deposited in the corona and hence affects the density and speed of the solar wind \citep{PintoRouillard2017}. MULTI-VP has been run on two sets of magnetic flux tubes associated with the SSW states 1 and 2 defined in \citet{Griton2021}, of which the solutions are given in Figure \ref{fig:griton2021_fig9} where the red (and blue) color denotes the SSW state 1 (and 2). The corresponding topology of the coronal magnetic field is traced in the left-hand side panel of Figure \ref{fig:griton2021_fig9} that I produced for the \citet{Griton2021} paper. One can notice a strong expansion of the flux tubes that are contiguous to the streamer (red lines) which have the effect in MULTI-VP to increase the amount of energy that is deposited low in the corona. As introduced in section \ref{subsec:intro_RT}, this pushes the transition region downward at a height where the chromosphere can balance the enhanced thermal pressure of the low corona. This results in the evaporation of chromospheric material into the corona and solar wind. The outcome is a  slow and dense solar wind analogous to SSW state 1 measured by \textit{PSP} at the streamer edge. In contrast the magnetic field expansion is less strong along the flux tubes that are rooted close to the center of equatorial coronal holes producing a more tenuous and slightly more rapid SSW (state 2). The latter remains nonetheless much slower than the typical fast winds that forms above polar coronal holes. Because the equatorial hole considered here is much smaller than typical polar coronal holes, and because it is enclosed by two dominating closed-field regions (north: a streamer and south: a pseudo-streamer) this configuration induces high-expansion factors along the field lines associated with both states of SSW. \\

These recent \textit{PSP} measurements illustrate nicely that the SSW can take different forms characterised by different densities, plasma beta as well as magnetic fields. While the SSW from streamers appears overall to be denser and much more variable (as introduced in section \ref{subsubsec:intro_SSW_variability}) than the SSW released from closer to the center of coronal holes, a quiet SSW appears to also form inside streamers. Numerical modelling with MULTI-VP shows that the different densities of the SSW from streamers and coronal holes could result from distinct heating depositions with height that are controlled by the different expansion factors. A theory explaining the properties of streamer flows must therefore explain why it can be at times very calm and at other times very variable. We investigate in section \ref{sec:dynamics_tearing} and \ref{sec:dynamics_griton2020} using numerical modelling how variability can be produced inside streamer flows, and what that variability should look like in coronal imagery and from the \textit{WISPR} perspective. \\

\section{Variability of the SSW as observed by \textit{PSP-WISPR}}
\label{sec:dynamics_WISPR}

\begin{figure*}[]
\centering
\includegraphics[width=0.95\textwidth]{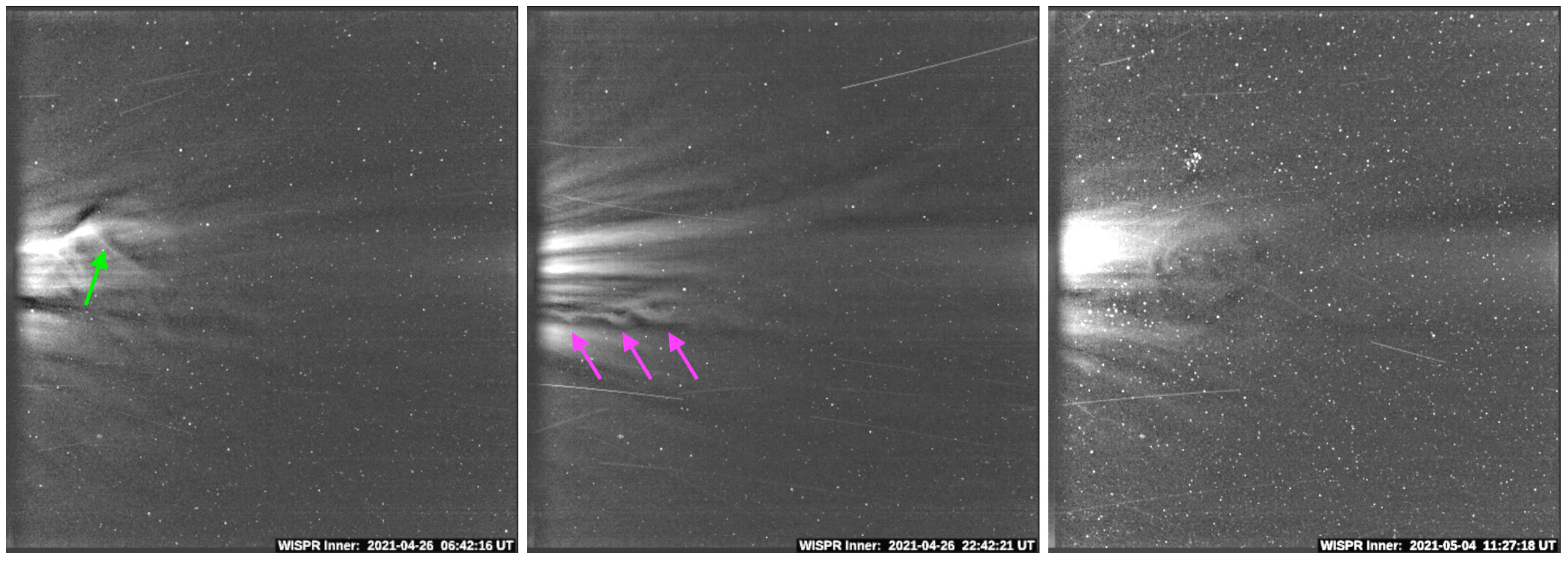}
\caption{ A sample of several density fluctuations observed by \textit{WISPR-I}.
\label{fig:WISPR_blobs}}
\end{figure*}

The first observations provided by the heliospheric WL imager \textit{WISPR} on board \textit{PSP} (see section \ref{subsec:inst_WISPR}) have been rich in structures that were often unresolved from typical 1 AU observatories. In addition to revealing the fine-structure of streamers as discussed in chapter \ref{cha:stationnary}, \textit{WISPR} have also detected many density perturbations propagating along the streamers of which, some of them are shown in Figure \ref{fig:WISPR_blobs}. In overall we detect two different topologies in the WL signatures that suggest either structures of different shapes or different viewing conditions. \\

The left-hand and right-hand side panels of Figure \ref{fig:WISPR_blobs} depict two large-scale structures that are likely situated close to \textit{PSP}. They look very similar to the signature of a flux rope seen edge-on, as in other events of pristine slow CMEs imaged in great detail by \textit{WISPR} \citep{Hess2020,Korreck2020} and for which I had tracked the flux rope associated to these CMEs throughout the FOV of \textit{WISPR} and other space-based WL imagers \citep{Korreck2020,Kouloumvakos2020b,Rouillard2020b}. These flux rope signatures often appear as a circular annulus of increased density that marks the envelope of their cross-section, that is more or less deformed depending on the viewing angle. Their trailing edge is particularly well visible as a bright "u-shape" or "v-shape" even from 1 AU observatories \citep{Rouillard2009b}. When seen face-on, the global shape of flux ropes is generally identified as long arches which have long been observed from 1 AU \citep[see e.g.][]{Sheeley2010}.

The middle panel of Figure \ref{fig:WISPR_blobs} presents a track of three small-scale blobs. Similarly to the two other structures discussed in the previous paragraph, these blobs appear as bright annulus that suggest a flux rope structure seen edge-on. In this case however, the WL signature is significantly smaller and smeared out. At this point it is hard to say whether these structures are actually small-scale flux ropes or if that is because they are located far away from \textit{WISPR}. \\

\begin{figure*}[]
\centering
\includegraphics[width=0.7\textwidth]{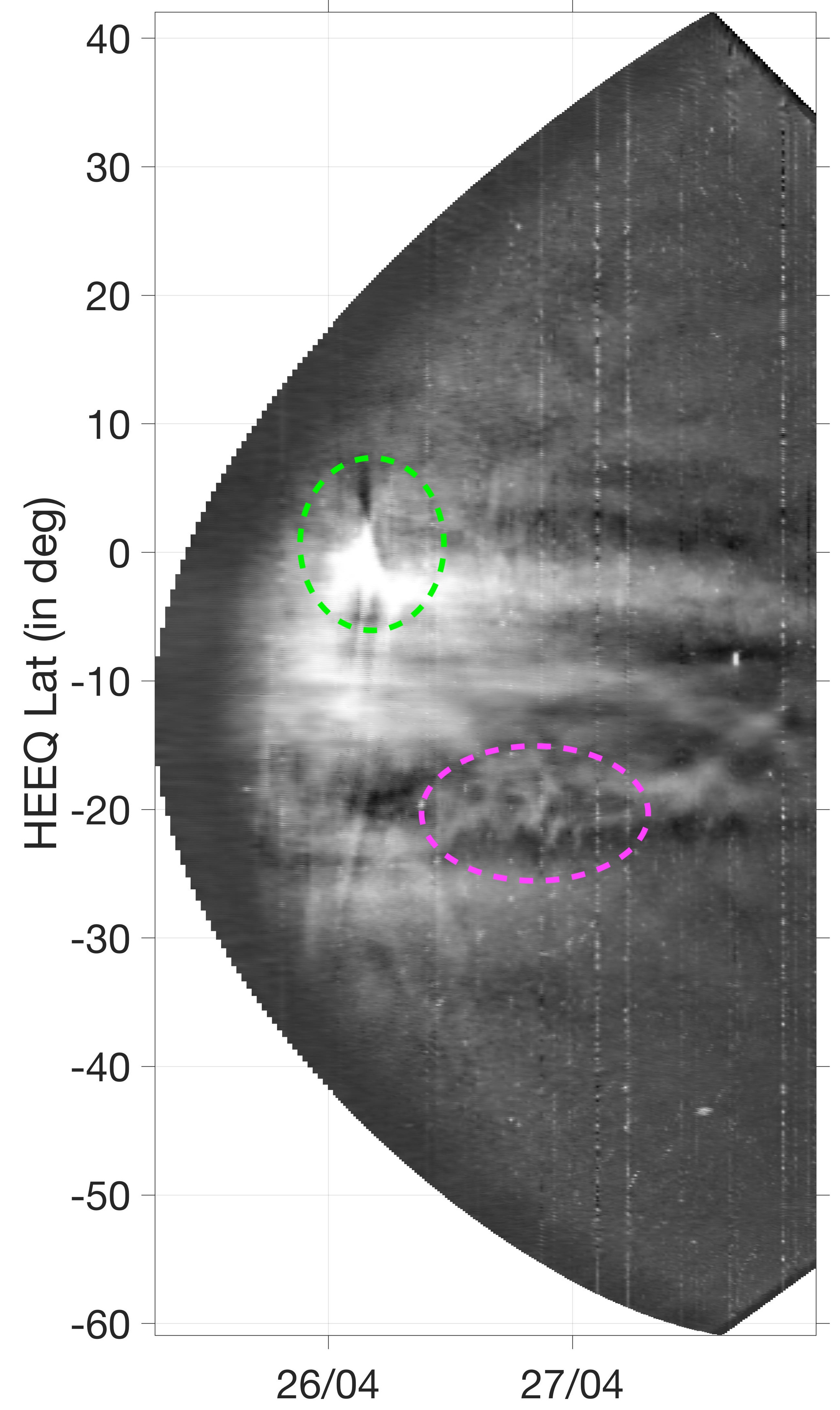}
\caption{ A \textit{WISPR-I} synoptic map extracted at $10\ R_\odot$ and focused on the inbound of the 8th passage of \textit{PSP} on April 25-27, 2021. The density perturbations that have been identified in Figure \ref{fig:WISPR_blobs} are denoted here using the same colors.
\label{fig:WISPR_map_E8_zoomed_10Rs}}
\end{figure*}

The \textit{WISPR-I} synoptic maps that I produced have been found very convenient to track both small-scale and large-scale density perturbations over long time intervals, as discussed in section \ref{sec:stationnary_WISPR} and shown in Figure \ref{fig:WISPR_tour}. The large-scale flux rope identified in Figure \ref{fig:WISPR_blobs} (green color) is also visible in the associated \textit{WISPR-I} synoptic map shown in Figure \ref{fig:WISPR_map_E8_zoomed_10Rs}. That is also true for the trail made of three small-scale flux ropes (pink color), which show a periodicity of $\approx 3-4\rm{hr}$. \\

To summarize, the two large-scale flux ropes shown in Figure \ref{fig:WISPR_blobs} might belong to the "streamer blob" family that we discussed in section \ref{subsubsec:intro_SSW_intermittency} and that is represented in Figure \ref{fig:transient_scales}. In contrast, the trail of small-scale flux ropes shown in the middle panel of Figure \ref{fig:WISPR_blobs} might be associated to quasi-periodic density structures with a higher periodicity, which have already been measured in situ at \textit{PSP} but at another epoch \citep{Rouillard2020a}. Fluctuations at even smaller scales can be seen in the \textit{WISPR-I} movies as corrugations in the streamers but would be hardly distinguishable in individual images. Fortunately, the \textit{WISPR-I} synoptic maps that I produced provide a nice view of these corrugations as presented in Figure \ref{fig:WISPR_tour}. They might fall in the small-scale fluctuations unveiled from 1 AU by \citet{DeForest2018} (see also Figure \ref{fig:transient_scales}) or even correspond to structures of smaller sizes. \\

In this section we only presented a reduced sample of the myriad of blobs observed by \textit{WISPR}. A future study that assembles a large catalogue of these structures based on their shape and periodicity would be highly constructive. In the following sections we employ synthetic imagery methods combined with global MHD simulations to get additional insights on the nature of the perturbations observed by \textit{WISPR}.

\section{Quasi-periodic density structures generated by magnetic reconnection at the HCS}
\label{sec:dynamics_tearing}

We examine in this section the release of density perturbations from the tip of streamers, and more specifically through the process of magnetic reconnection at the HCS (see section \ref{subsec:intro_dynamics_streamertip}), including the generation of magnetic flux ropes  (see section \ref{subsec:intro_dynamics_fluxropes}). To investigate the nature and origin of these transient features observed in WL by \textit{WISPR} (see section \ref{sec:dynamics_WISPR}), we exploit the 3-D MHD Alfvén-wave driven solar wind model WindPredict-AW  already introduced in section \ref{subsec:WindPredict} in two different setups. We first use in section \ref{subsec:dynamics_tearing_2p5D} an idealized 2.5-D high-resolution resistive simulation that is similar to the one presented in \citet{Reville2020b} and that simulates the development of magnetic reconnection at the HCS through the tearing instability, and consequently of the formation of quasi-periodic density structures in the streamers. We then test in section \ref{subsec:dynamics_tearing_3D} how this process may hold in two realistic 3-D simulations of the corona and of the solar wind, of which one has already been proven to generate 3-D magnetic flux ropes in similar regions of the solar atmosphere as those observed in situ by the \textit{Parker Solar Probe} and \textit{Solar Orbiter} \citep{Reville2022}.

\subsection{Idealized 2.5-D setup: the tearing instability}
\label{subsec:dynamics_tearing_2p5D}

We exploit in this section an idealized 2.5-D version of the WindPredic-AW model introduced in section \ref{subsec:WindPredict} and that is described in detail in \citet{Reville2020b}. The motivation for this idealized setup is to analyse the formation of magnetoplasma perturbations released from the tip of streamers as suggested by the early observations of \citet{Sheeley1997} (see section \ref{subsec:intro_dynamics_streamertip}). \citet{Reville2020b} showed with this type of simulation that the generation of "streamer blobs" is two-fold, first a pressure instability forces the outward stretching and expansion of coronal loops that begin to thin out along what will become the heliospheric plasma sheet. This results eventually into the onset of a tearing-mode instability that triggers magnetic reconnection at the HCS with an associated release of a 2-D plasmoids separated by density perturbations that propagate along the streamer. \\

\citet{Reville2020b} have identified two different periodicities in the generated structures. They show that the stretching of coronal loops due to coronal heating is further enhanced by the SSW that is accelerated along the adjacent open field lines that form the cusp of the streamer. Overall this stretching process occurs over a  $\simeq 30\ \rm{hr}$ period and, as already mentioned, is interrupted by the onset of the tearing-mode. In 2.5-D MHD this process ejects plasmoids along the heliospheric plasma sheet, in 3-D MHD large-scale magnetic flux ropes can be released through this process as discussed in the next section. This $\simeq 30\ \rm{hr}$ periodicity roughly corresponds to the 16-20 hour periodicity derived from \textit{STEREO HI-1} heliospheric imagery by \citet{Sanchez-Diaz2017b}. The region situated by subsequent plasmoid/flux ropes where the tearing instability develops marks the onset of smaller reconnection events that form smaller scale quasi-periodic structures (see also Figure \ref{fig:Sanchez2017b_fig12}). The periodicities of these additional quasi-periodic structures were found to be consistent with the hourly time scale of those observed in WL coronal and heliospheric imagers \citep{Viall2010,Viall2015} and detected in situ \citep{Kepko2016}.  \\

\citet{Reville2020b,Reville2022} argue that the 30-hour cycles correspond to the time for the streamer loops to refill in heated coronal material and recover from its exhaust phase induced by the reconnection at the HCS. In essence the recovery time of the streamer is linked to the processes that heat up the coronal part of the loops that constitute the streamer. Streamers can become unstable \citep{Higginson2018} and grow if their internal pressure exceeds the confining force induced by the magnetic field. For the largest flux ropes produced this way in the streamers, a toroidal geometry (with a poloidal and toroidal field) could develop during the magnetic reconnection process. In this case additional toroidal forces could also contribute to the outward acceleration of these magnetic flux ropes as studied in the \citet{Rouillard2020b} paper and for which I contributed significantly. For this paper I integrated to a toy flux rope model a fully 3-D geometry as well as a calculation of the shifting of flux surfaces known as the Shafranov shift. \\

During their replenishing growth phase, coronal loops stretch until they break up by reconnection, this is accompanied by a significant transfer of matter where the plasma that was initially confined in coronal loops gets expelled into the SSW. Throughout this process, the slow wind should then be fueled with low-FIP elements that have been fractionated from other elements beforehand in the coronal loops following processes that have been discussed in section \ref{subsec:intro_FIP}. As the tearing instability develops the matter gets concentrated primarily at the interstices between the generated fast quasi-periodic structures mentioned above, where the resulting density enhancement actually correspond to the "streamer blobs" observed in WL. That may also explain the variability of the composition measured in situ in the SSW precisely in these quasi-periodic structures \citep[see e.g.][]{Kepko2016}. The magnetic reconnection that develops near the tip of streamers as produced in the WindPredict-AW simulation provides an attractive explanation for the intermittency of the streamer flows (see section \ref{subsubsec:intro_SSW_intermittency}), its peculiar composition typical of coronal loops (see section \ref{subsubsec:intro_SSW_composition}), and the variability of its composition (see section \ref{subsubsec:intro_SSW_variability}). \\

\begin{figure*}[]
\centering
\includegraphics[width=0.5\textwidth]{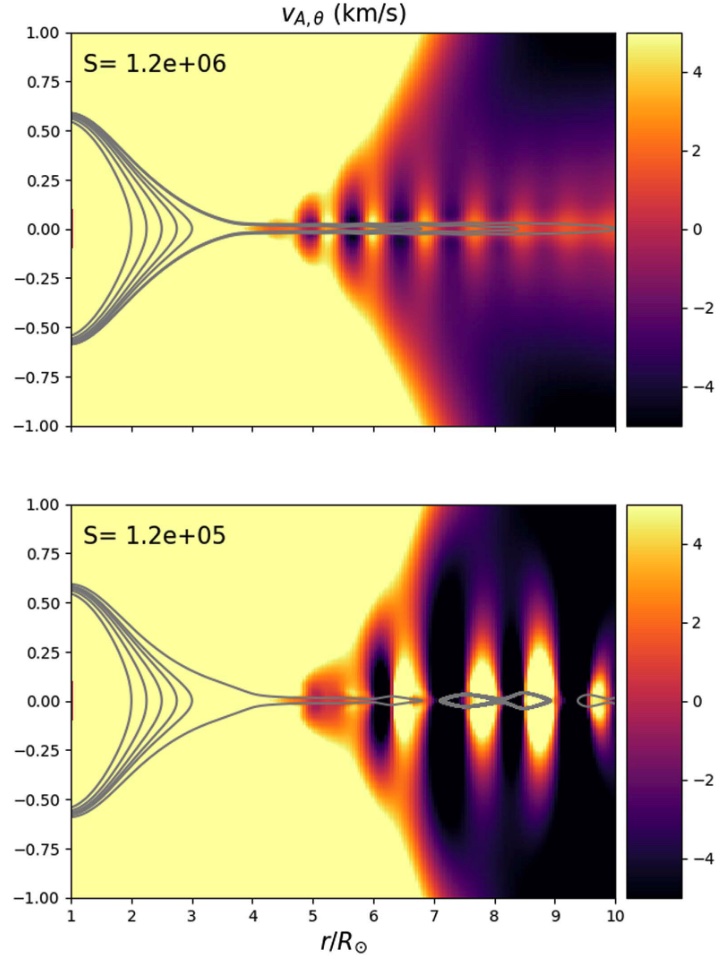}
\caption{ Snapshots of two WindPredict-AW 2.5-D simulations of \citet{Reville2020b} illustrating the generation of quasi-periodic structures from the tearing instability, with two distinct Lundquist numbers $S$. The latitudinal component of the Alfvén speed is color plotted with a sample of magnetic field lines in grey.
\label{fig:Reville2020b_fig3}}
\end{figure*}

To favor an environment where the tearing instability can fully occur, the choice is made of a 2.5-D setup that permits sufficiently high numerical resolution near the HCS to model accurately the development of the tearing mode. By doing so the magnetic reconnection at the HCS can be computed through an explicit resistive term in the MHD equations. Additionally, the model exploits a Harten\textendash Lax\textendash van Leer Discontinuities (HLLD) Riemann solver \citet{Miyoshi2005} that can handle sharp gradients near the HCS. This setup allows high values of the Lundquist number $S=Lv_A/\eta \geq 10^4$, an indicator of the ability of the medium to trigger the tearing-mode instability. Although this value of $S$ remains much lower than what is actually measured in situ in the HCS ($S\approx 10^{14}$), \citet{Reville2020b} show that periodic structures still arise and propagate with timescales coherent with those typically measured in situ in the SSW. This is illustrated in Figure \ref{fig:Reville2020b_fig3} for two different Lundquist numbers, where the top panel depicts a situation that is more favorable to the full development of the tearing instability thanks to a Lundquist number that is higher than in the bottom panel. \\

\subsection{Realistic 3-D setup: comparison with \textit{PSP-WISPR} observations}
\label{subsec:dynamics_tearing_3D}

The idealized 2.5-D setup presented in section \ref{subsec:dynamics_tearing_2p5D} allows us to generate plasmoïd-like structures that propagate in the streamer. In reality, these structures may consist of 3-D global structures in the form of magnetic flux ropes as measured in situ in the SSW \citep{Sanchez-Diaz2019}. In this sense, the two reconnection processes suggested by \citet{Sheeley1997} and \citet{Gosling1995} and that are depicted in the panels b and c of Figure \ref{fig:Sanchez2017_fig1.10} may be interpreted as two manifestations of the same reconnection process at the tip of streamers, where the panel c gives a more precise 3-D description of the topology of the magnetic field. As introduced in section \ref{subsec:intro_dynamics_fluxropes}, such scenario would lead to the generation of successive 3-D flux rope structures that are punctuated by streamer blobs at their interstice as illustrated in Figure \ref{fig:Sanchez2017b_fig12}. \\

This motivated a subsequent study from \citet{Reville2022} where global 3-D simulations of the solar wind using the WindPredict-AW model introduced in section \ref{subsec:WindPredict}, have been exploited to study the generation and dynamics of flux ropes propagating in the streamer and along the HCS. By a qualitative comparison with real \textit{PSP} and \textit{SolO} in situ measurements taken on June 2020, \citet{Reville2022} produced flux rope releases in roughly the same heliospheric region as the flux ropes detected in situ. I contributed to this study by helping to select an optimal photospheric magnetic map to be given as an inner boundary condition to the WindPredict-AW model, using the WL remote-sensing observations and following the methodology presented in section \ref{subsec:poirier2021}. In the same vein as the extended benchmarking framework presented in section \ref{subsec:badman2022}, \citet{Reville2022} performed an additional selection of the photospheric magnetic map based on in situ measurements of the magnetic polarity to ensure a correct modeling of the magnetic sectors at \textit{PSP} and \textit{SolO}. \\

\begin{figure*}[]
\centering
\includegraphics[width=0.95\textwidth]{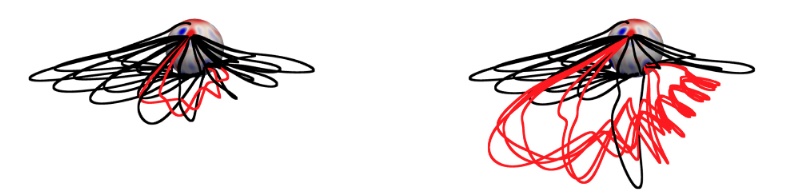}
\caption{ 3-D structure of a streamer flux rope generated in the WindPredict-AW 3-D simulation of June 14, 2020 described in \citet{Reville2022}. This figure has been extracted from \citet[][Figure 7]{Reville2022}.
\label{fig:Reville2022_fig7}}
\end{figure*}

In contrast to the 2.5-D setup, the 3-D setup cannot afford such a level of refinement and therefore it is the numerical diffusion induced by the numerical scheme itself that allows for reconnection to happen. This numerical diffusion can be of several order of magnitudes larger than what is actually permitted by the physics. Therefore, the magnetic reconnection process is bound in the 3-D setup by the actual numerical size of the mesh near the HCS. That constrains the 3-D setup with a lower Lundquist number $S\approx500-1000$ that is not optimal for the full development of the tearing instability as in the 2.5-D simulation. Despite this limitation, the 3-D simulation does produce streamer blobs or flux ropes but only at the longest periodicity of $\simeq 20\ \rm{h}$ (as those measured in situ by \citet{Sanchez-Diaz2019}) and not the fast quasi-periodic structures of about hour-long timescales achieved in the 2.5-D simulation (and which have been detected both remotely \citep{Viall2010,Viall2015} and in situ \citep{Kepko2016}). An example of a magnetic flux rope structure is illustrated in the right panel of Figure \ref{fig:Reville2022_fig7}, after the main reconnection event at the HCS. \\

Following the Thomson scattering theory (see section \ref{sec:stationnary_theory}) and the same method as in section \ref{sec:stationnary_synthetic}, I here build \textit{WISPR-I} synthetic images from the 3-D datacubes of plasma density simulated by the WindPredict-AW model. In particular, I exploit the June 14, 2020 simulation presented in \citet{Reville2022} for a subset of the simulation snapshots between time $t=60\ \rm{h}$ and $t=180\ \rm{h}$ that were found to contain the flux ropes structures. However, the cadence of the simulation outputs (i.e. $\simeq 2.2\ \rm{h}$) is not as high as for \textit{WISPR-I} (i.e. $\simeq 16\ \rm{min}$), hence a simulation snapshot is used several times ($7.5$ in average) before using the next one. In another 3-D WindPredict-AW simulation that we present later in this section, we directly fixed the cadence of the simulation outputs to be the same as \textit{WISPR-I} so that the tracking of propagating structures in the \textit{WISPR} FOV is more accurate. We stress out, nonetheless, that the FOV of \textit{WISPR-I} remains almost unchanged over a time period of two hours during which \textit{PSP} moves only about $1.5^\circ$ in Carrington longitude with almost no variation in distance to the Sun. \\

Furthermore, we need to calibrate the timeline of the simulation snapshots with the observation times of \textit{WISPR-I}. In \citet{Reville2022}, a post-event photospheric map (of June 14) has been used as the inner boundary to the WindPredict-AW model for the reasons mentioned at the beginning of this section. This particular map was updated to account for an active region that emerged on the far-side of the Sun and hence was not captured by ground-based observatories during the \textit{WISPR-I} observation window (June 4-10). The timeline is hence shifted so that the 5 days ($120\ \rm{h}$) of coverage of the WindPredict-AW simulation are centered at the 5th perihelion of \textit{PSP}, leading to a simulation that spans a physical time interval from June 5 00:00 to June 9 23:59 with a cadence of $\simeq 2.2\ \rm{h}$. For the rest of the \textit{WISPR-I} observing window (June 4 and June 10), the \textit{WISPR-I} synthetic images are produced using either the $t=60\ \rm{h}$ (or $t=180\ \rm{h}$) simulation snapshot. \\

\begin{figure*}[]
\centering
\includegraphics[width=0.6\textwidth]{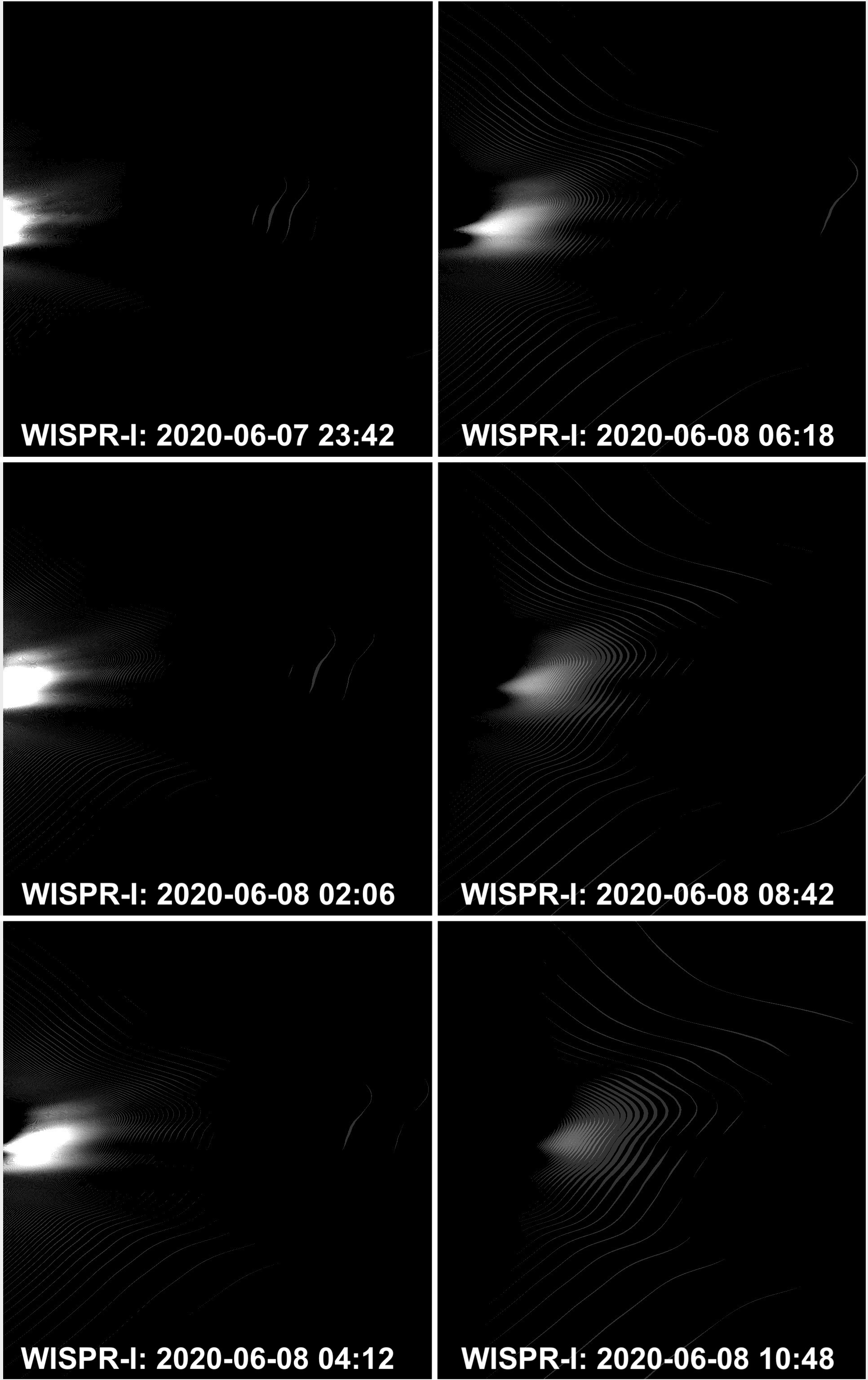}
\caption{ Synthetic \textit{WISPR-I} running difference images of the formation and propagation of a "streamer blob" associated to a flux rope topology from the WindPredict-AW 3-D simulation of June 14, 2020 described in \citet{Reville2022}.
\label{fig:Synthetic_WISPR_E5_rdiff}}
\end{figure*}

\begin{figure*}[]
\centering
\includegraphics[width=0.95\textwidth]{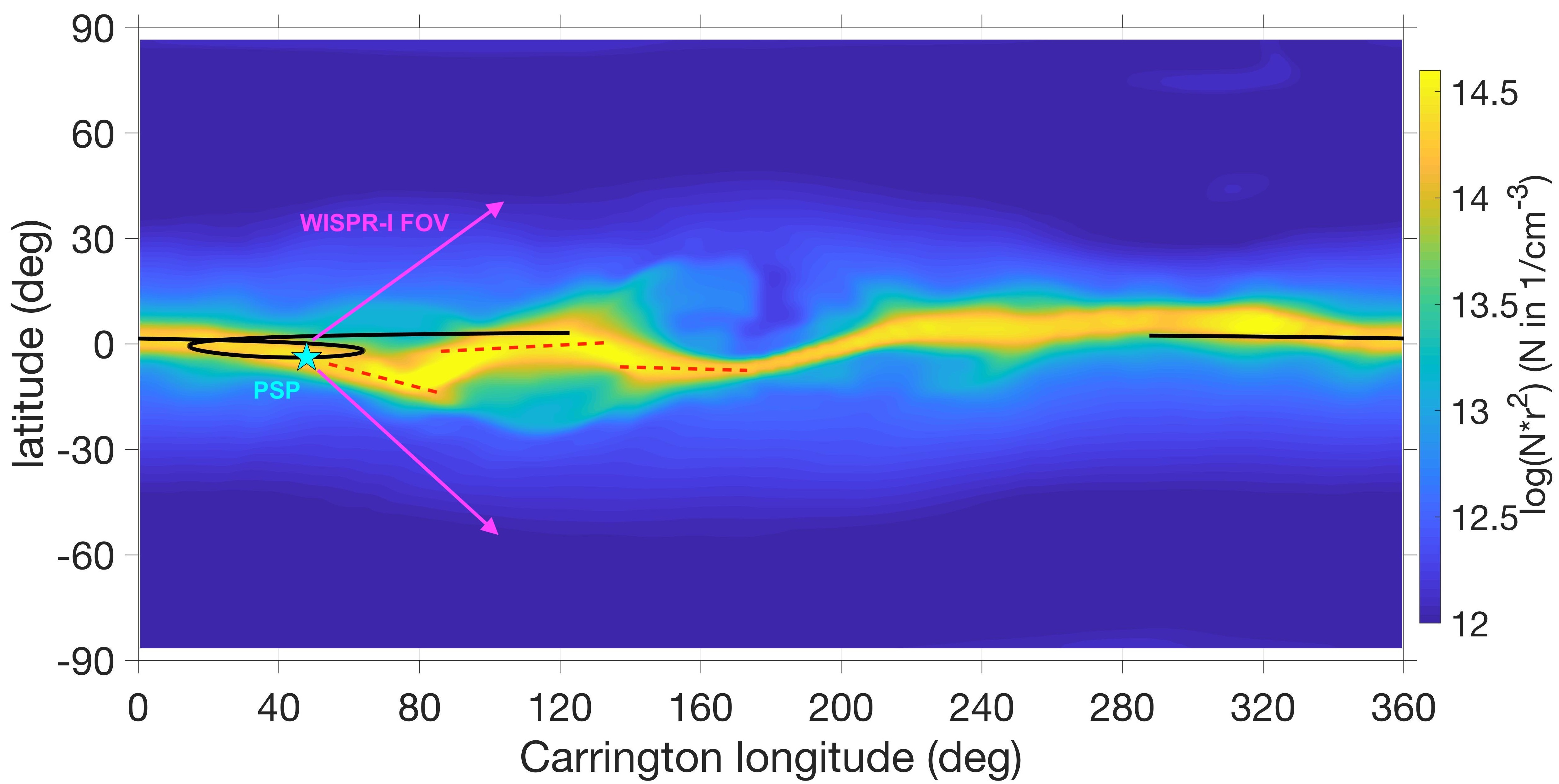}
\caption{ Carrington map of the simulated density at $12\ R_\odot$ from the June 14, 2020 simulation. The position of \textit{PSP} on June 7 23:42 is marked by the cyan star symbol with a rough estimate of the \textit{WISPR-I} FOV  indicated by the magenta arrows. At that time, \textit{PSP} was located at $\simeq 28.4\ R_\odot$ in heliocentric distance. The dashed red lines indicate the portions of the HPS that are likely captured by \textit{WISPR-I}.
\label{fig:Carrmap_dens_E5}}
\end{figure*}

Synthetic \textit{WISPR-I} images of the formation and propagation of a streamer blob as generated in the WindPredict-AW 3-D simulation are shown in Figure \ref{fig:Synthetic_WISPR_E5_rdiff}, where subsequent images are subtracted following the running difference technique to enhance the visibility of the WL signature. We do not find clear similarities with the subset of flux ropes that we have identified in section \ref{sec:dynamics_WISPR} from \textit{WISPR-I} images. That may be clarified by constructing an extended observational dataset of blobs seen by \textit{WISPR} and that is left for future studies. Looking at Figure 8 in \citet{Reville2022}, this flux rope likely originates from a portion of the HCS that is within $150-250^\circ$ of Carrington longitude among four others that are seen in the same region. I only show one of the five flux ropes in Figure \ref{fig:Synthetic_WISPR_E5_rdiff} but I could distinctly identify all of them in the \textit{WISPR-I} synthetic images. These flux ropes are located far away from \textit{PSP} which was located between $\simeq 15-60^\circ$ of Carrington longitudes as illustrated in Figure \ref{fig:Carrmap_dens_E5}. Therefore the structure of these flux ropes is not well captured compared to the case shown later on in Figure \ref{fig:Synthetic_WISPR_E9_rdiff7}. \\

\begin{figure*}[]
\centering
\includegraphics[width=0.8\textwidth]{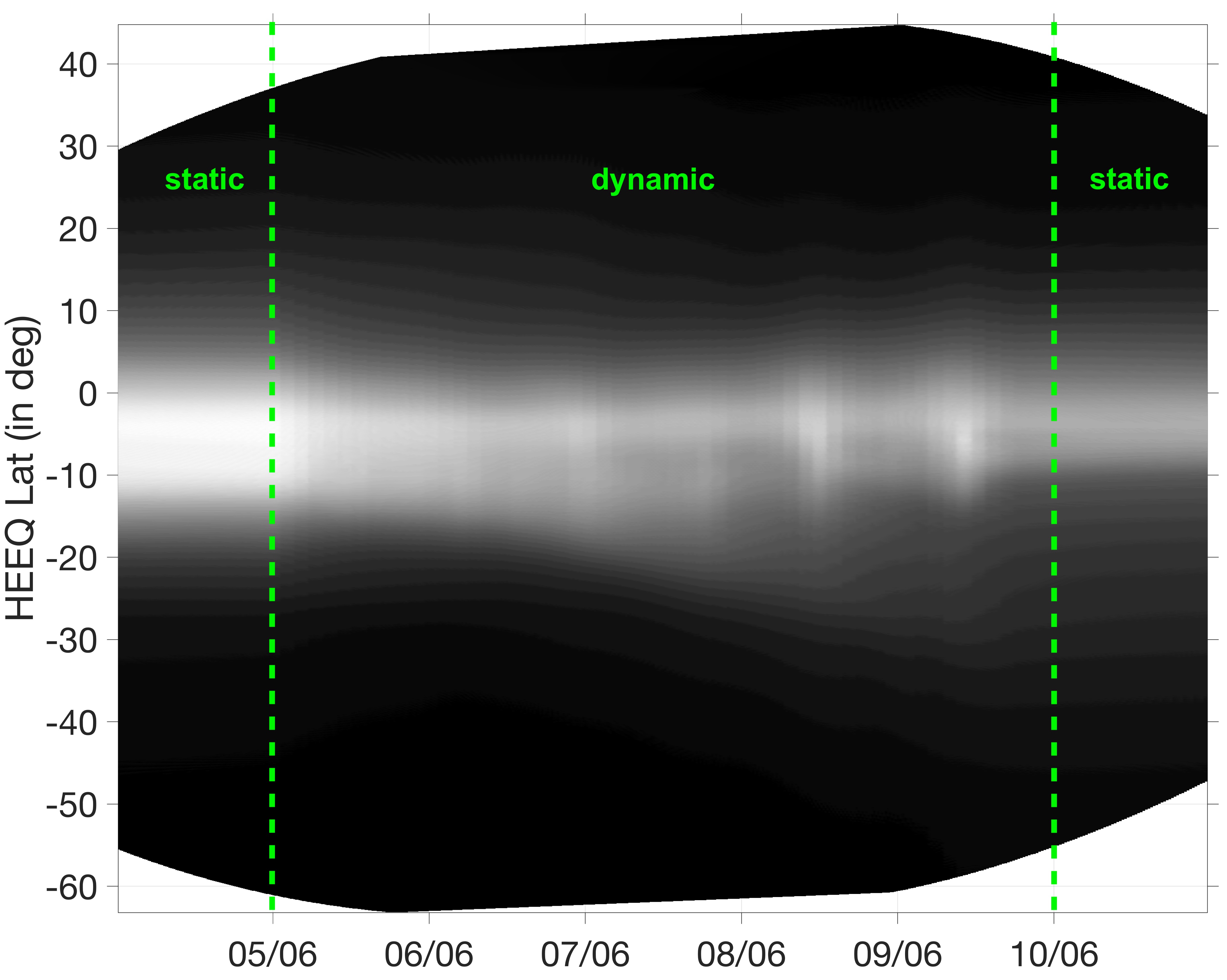}
\caption{ Synthetic \textit{WISPR-I} map for the 5th passage of \textit{PSP} to the Sun from the WindPredict-AW 3-D simulation of June 14, 2020 described in \citet{Reville2022}. The map is extracted at $12 R_\odot$.
\label{fig:Synthetic_WISPRmap_E5_12Rs}}
\end{figure*}

\begin{figure*}[]
\centering
\includegraphics[width=\textwidth]{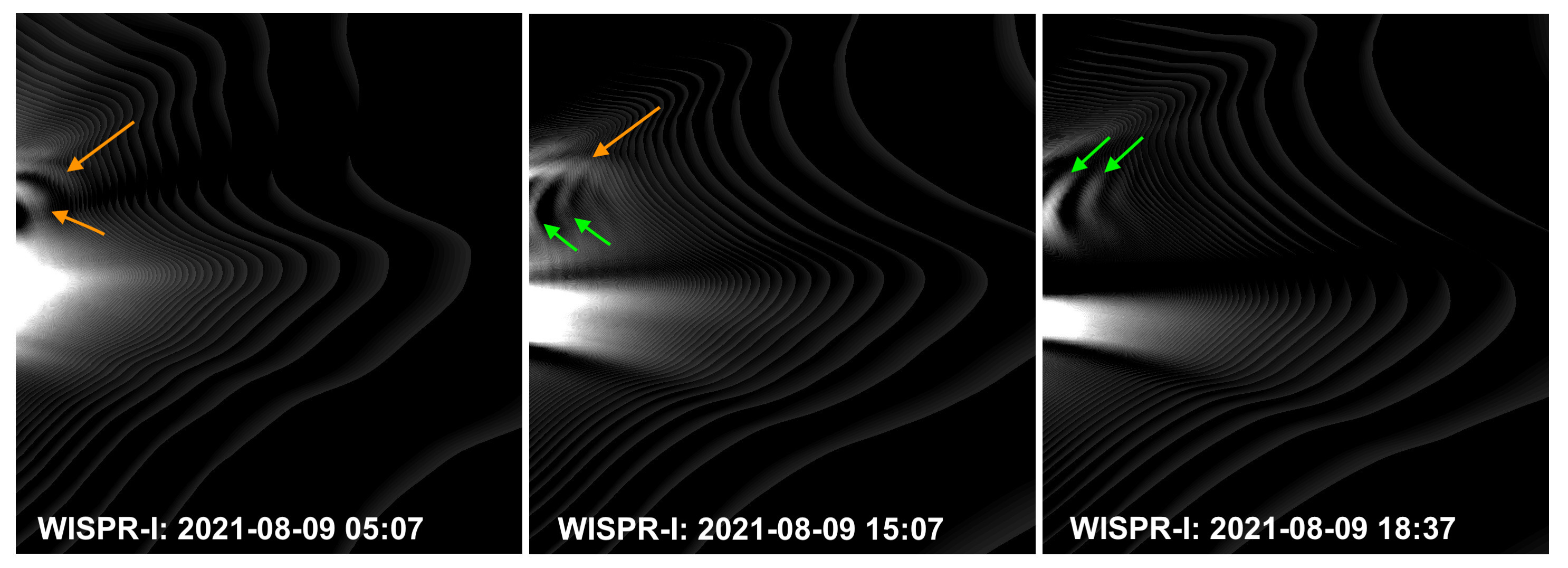}
\caption{ Same as figure \ref{fig:Synthetic_WISPR_E5_rdiff} but for a flux rope generated in the simulation of August 14 2021.
\label{fig:Synthetic_WISPR_E9_rdiff7}}
\end{figure*}

Similarly to the first observations of \textit{WISPR-I} that we analyse in detail in section \ref{sec:stationnary_poirier2020}, we can build a synthetic \textit{WISPR-I} synoptic map that stacks synthetic \textit{WISPR-I} images over time (see section \ref{subsec:stationnary_poirier2020_WLmap} for a description of the method). The result of this approach is shown in Figure \ref{fig:Synthetic_WISPRmap_E5_12Rs} where the corresponding observed \textit{WISPR-I} synoptic map was already presented in Figure \ref{fig:WISPR_tour}. We can draw a similar conclusion to that of section \ref{sec:stationnary_poirier2020} where the apparent splitting of the streamer rays in two bands is again the result of the HPS changing latitude between the foreground and background portion of the LOS of \textit{WISPR-I}. This effect is depicted in Figure \ref{fig:Carrmap_dens_E5} by the two dashed red lines between $85-130^\circ$ and $140-180^\circ$ of Carrington longitude that represent two bundles of LOSs along which a large portion of the dense HPS is integrated and produces the two bright stripes visible in both the real and synthetic \textit{WISPR-I} maps. As discussed in section \ref{sec:stationnary_WISPR}, one can also see a clear drifting of the southern streamer ray away from the equator and that is an effect of \textit{PSP} plunging into the streamer. That corresponds to a portion of the HPS that is in the direct vicinity of \textit{WISPR-I} (red dashed line between $50-85^\circ$ of Carrington longitude) at a time where \textit{PSP} crosses the HCS as shown in Figure \ref{fig:Carrmap_dens_E5}. 
In addition, one can see clear signatures of five distinct features in the streamer ray that appear every $\approx 20\ \rm{h}$ and which correspond to the passage of the five flux ropes mentioned above across the FOV of \textit{WISPR-I}. Finally, the flux ropes appear quite fragmented in the synthetic \textit{WISPR-I} map along the time axis (abscissa axis). That is the effect of updating the simulated density every $\simeq 2.2\ \rm{h}$ whereas \textit{WISPR-I} images are cadenced at $\simeq 16\ \rm{min}$. \\

We repeat the analysis for the ninth \textit{PSP} solar encounter where \textit{WISPR-I} observed a HCS that was much more warped. Since streamer flux ropes are usually aligned with the streamer belt \citep{Sheeley2010, Rouillard2011a}, a warped streamer means that \textit{WISPR} will observe different WL signatures of these flux ropes that will either appear edge-on or with a high inclination angle. The WindPredict-AW setup is similar to the previous one but here the inner boundary is set with the GONG-ADAPT (11th realization) magnetogram of August 14, 2021 00:00. Just as for the simulation of June 14 2020, the magnetogram was selected among many different sources and dates to best match the observations. The latter included WL streamer belt observations taken from $1\ AU$ (using the method presented in section \ref{subsec:poirier2021}) complemented with the magnetic sectors and timing of HCS crossings measured in situ by \textit{PSP}. Once the simulation reached a permanent regime, outputs of the entire 3-D simulated domain have been extracted every $\simeq 13\ \rm{min}$ to match the cadence of \textit{WISPR-I}. For the sake of memory space, we limit ourselves to $\approx 100$ outputs nonetheless, that cover a time interval of $\simeq 22\ \rm{hr}$ around perihelion. \\

\begin{figure*}[]
\centering
\includegraphics[width=1.\textwidth]{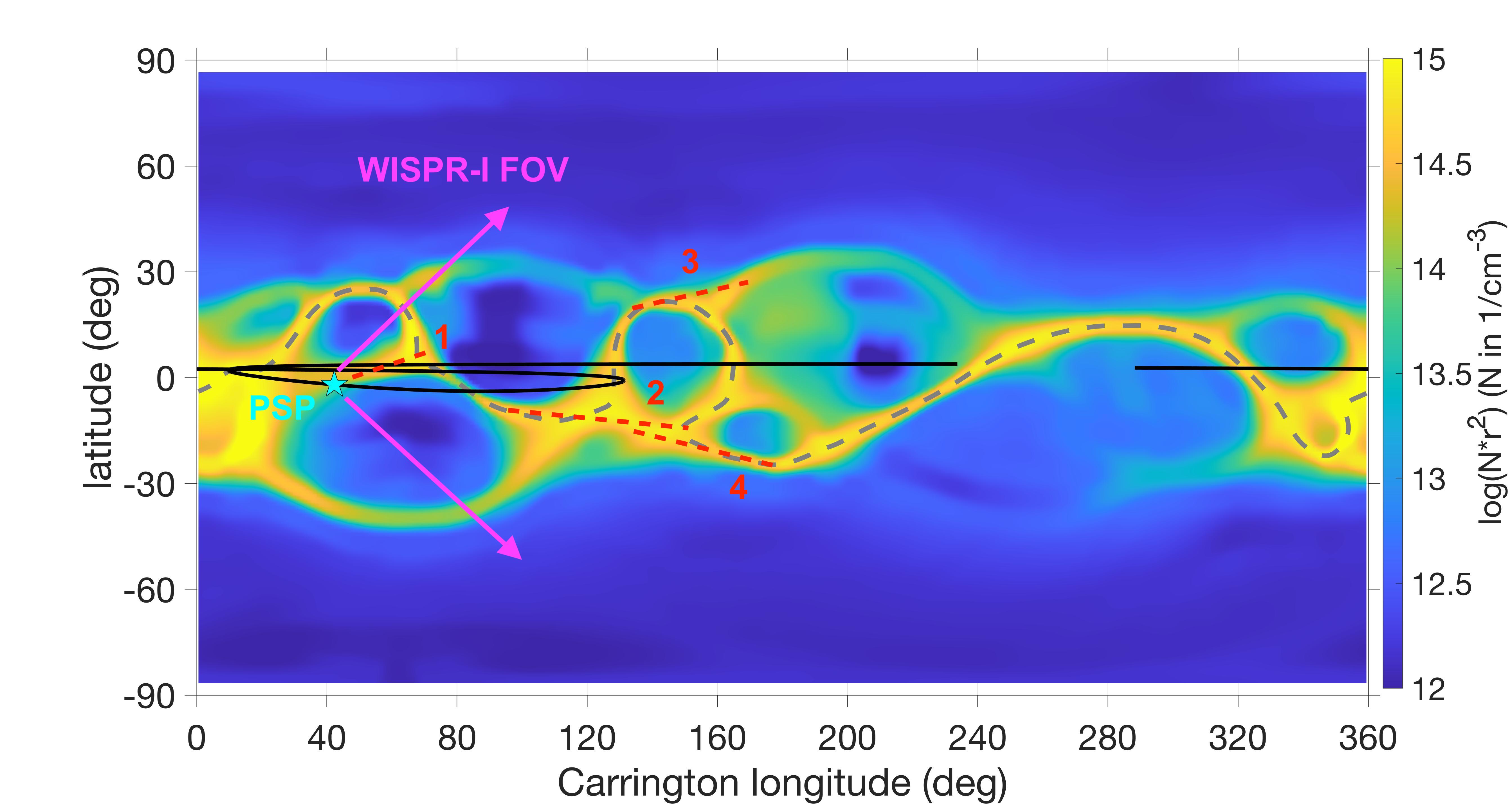}
\caption{ Carrington map of the simulated density at $6.5\ R_\odot$ from the August 14 2021 simulation. The position of \textit{PSP} on August 9 05:07 is marked by the cyan star symbol with a rough estimate of the \textit{WISPR-I} FOV indicated by the magenta arrows. At that time, \textit{PSP} was located at $\simeq 17.5\ R_\odot$ in heliocentric distance. The dashed red lines indicate the portions of the HPS that are likely captured by \textit{WISPR-I}. The grey dashed line denotes the HCS where magnetic polarity switches sign in the simulation.
\label{fig:Carrmap_dens_E9}}
\end{figure*}

Following the same procedure as before, we show in Figure \ref{fig:Synthetic_WISPR_E9_rdiff7} synthetic \textit{WISPR-I} running-difference images of a flux rope produced by reconnection at the HCS in the August 14 2021 simulation. The overall shape of the solar corona (from the same simulation) is given in Figure \ref{fig:Carrmap_dens_E9} and appears, as expected for the higher level of activity, much more structured and complex than in the previous simulation. The flux rope seen in the synthetic \textit{WISPR-I} images (see Figure \ref{fig:Synthetic_WISPR_E9_rdiff7}) comes from a portion of the HCS that is within $\approx70^\circ$ and $\approx 100^\circ$ of Carrington longitude. The WL signature of the flux rope appears much more complex than in the previous case (figure \ref{fig:Synthetic_WISPR_E5_rdiff}) and is very similar to actual flux ropes observed by \textit{WISPR-I} (see section \ref{sec:dynamics_WISPR}) where \textit{WISPR-I} is very close to flux ropes that are mostly seen edge-on. The leading edge of the flux rope appears as arches (orange arrows) whereas the trailing edge has a peculiar "v-shape" (green arrows). The growth and expansion of the flux rope compresses the solar wind plasma that stacks on the front and hence produces the bright arches. As discussed in section \ref{sec:dynamics_WISPR}, the "v-shape" could correspond to a LOS effect, but here this also likely a remnant feature of the magnetic reconnection process itself which disconnected the coronal loops as depicted in Figure \ref{fig:Sanchez2017_fig1.10}(c). At the trailing edge the reconnection accelerates and pushes the coronal loop plasma to the back of the flux rope, that is also accompanied by an injection of poloidal flux into the flux rope as we discussed in \citet{Rouillard2020b}. \\ 

\begin{figure*}[]
\centering
\includegraphics[width=0.8\textwidth]{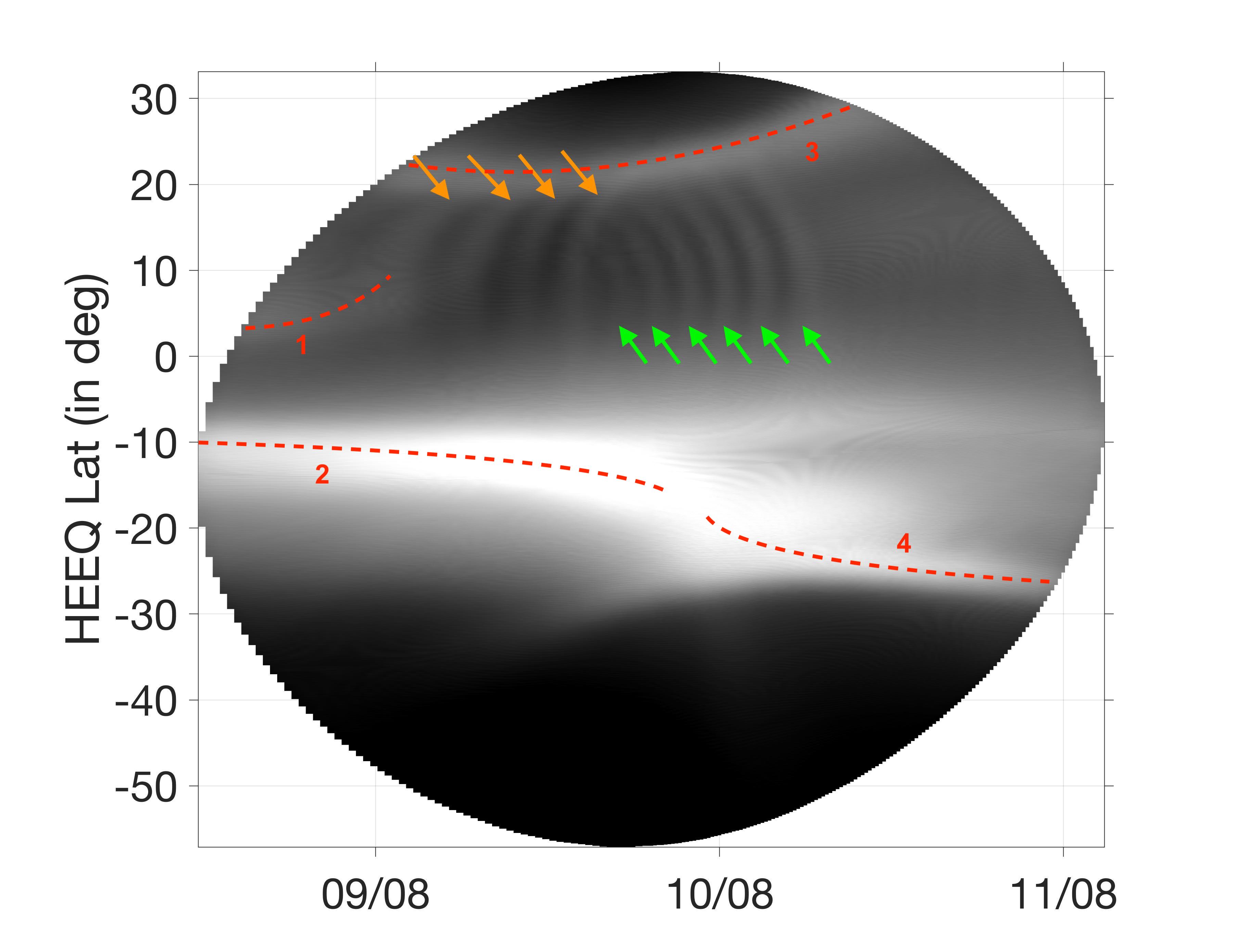}
\caption{ Synthetic \textit{WISPR-I} map for the 9th passage of \textit{PSP} close to the Sun, from the WindPredict-AW 3-D simulation of August 14 2021. The map is extracted at $5\ R_\odot$. To ease the interpretation each streamer ray is identified by a dashed red line and a number where the associated portion of the HPS producing the brightness is identified in Figure \ref{fig:Carrmap_dens_E9} using the same numbers.
\label{fig:Synthetic_WISPRmap_E9_5Rs}}
\end{figure*}

The HCS (see the dashed grey line in Figure \ref{fig:Carrmap_dens_E9}) is significantly warped, this is a consequence of a much more active Sun compared to the June 14 2020 simulation. This complicates the interpretation of the WL signature in recent \textit{WISPR-I} observations. The \textit{WISPR-I} synoptic map shown in Figure \ref{fig:WISPR_tour} for the ninth encounter (E9) now appears much more striated by streamer rays. As discussed in section \ref{sec:stationnary_WISPR}, these fine structures can be the result of multiple folds in the HPS and other high-density plasma layers formed by pseudo-streamers or more generally the complex network of QSLs (see section \ref{subsec:intro_Sweb}). For illustration purposes we show in Figure \ref{fig:Synthetic_WISPRmap_E9_5Rs} the synthetic \textit{WISPR-I} map built from the August 14 2021 simulation that looks very different to the real map at first sight. At this stage we do not aim at reproducing the actual position of each ray observed by \textit{WISPR-I} because this would require further iterations of model optimisation that is left for future study. However, this simulation provides additional insight on the interpretation of WL signatures in a highly-structured solar atmosphere. For that purpose we denoted several bright rays in Figure \ref{fig:Synthetic_WISPRmap_E9_5Rs} (dashed red lines) for which we could identify an associated source region in the density map shown in Figure \ref{fig:Carrmap_dens_E9}. The plunging effect of \textit{WISPR-I} into the corona is even more pronounced (ray 1) than in the previous simulation, this orbital effect is a recurring feature of the latest \textit{WISPR-I} observations (see Figure \ref{fig:WISPR_tour}). The multiple source regions identified in Figure \ref{fig:Carrmap_dens_E9} and \ref{fig:Synthetic_WISPRmap_E9_5Rs} further strengthen our previous observation that very distant regions can contribute equally to the brightness received by \textit{WISPR-I} and that a significant part of the WL signal comes from high-density regions that are not always associated with the bipolar streamer. This should be investigated further in future studies, \textit{WISPR} is here likely providing detailed information on the network of pseudo-streamers and perhaps more generally the S-web (see section \ref{subsec:intro_Sweb}). \\

Both the leading and trailing edge of the flux rope that we have tracked in the running difference images (see Figure \ref{fig:Synthetic_WISPR_E9_rdiff7}) can also be seen distinctly in the synthetic \textit{WISPR-I} map and produce highly inclined stripes that spread over a wide range of latitudes. The trailing edge produces vertical stripes (orange arrows) that are similar to that of an actual flux rope observed by \textit{WISPR-I} and that we identified in Figure \ref{fig:WISPR_map_E8_zoomed_10Rs} by a green ellipse. Unfortunately the leading edge was not clearly visible in this particular event, and the other flux rope shown in the right-hand side panel of Figure \ref{fig:WISPR_blobs} is too faint to be properly tracked in a \textit{WISPR-I} synoptic map. That should be clarified in subsequent observations as \textit{WISPR} will likely reveal more events of this kind by getting closer and closer to the observed flux rope structures.

\section{Propagating density structures triggered by transient heating at the coronal base}
\label{sec:dynamics_griton2020}
\textit{(This section contains some material from the \citet{Griton2020} paper.)} \\

We have already highlighted the successes of MULTI-VP to model the fine-structure of the upper corona and nascent solar wind imaged by \textit{WISPR-I} (see section \ref{sec:stationnary_poirier2020}), including the SSW variability measured in situ at \textit{PSP} (see section \ref{sec:dynamics_insitu}). We see in this section that the MULTI-VP model can be also exploited to interpret the origin of the transient structures generally observed in WL coronagraphs. \\

The main motivation of the \citet{Griton2020} paper to which I contributed significantly, was to understand the origin of some of the density fluctuations observed by \citet{DeForest2018} during a recent (deep-field) high-cadence campaign of the \textit{STEREO-A COR-2} coronagraph. We already introduced their observations in section \ref{subsubsec:intro_SSW_intermittency} and in Figure \ref{fig:DeForest2018_fig12}. This figure depicts a highly structured corona typical of high solar activity where the HPS is largely inclined with respect to the equator, this is illustrated in Figure \ref{fig:Griton2020_fig7} which I produced for the \citet{Griton2020} paper. The largest fluctuations observed by \citet{DeForest2018} likely belong to the "streamer blobs" family observed earlier on by \citet{Sheeley1997}, which likely originate from magnetic reconnection at the tip of streamers in the form of magnetic flux ropes (as shown in the previous section). However, most fluctuations observed by \citet{DeForest2018} are likely to be smaller transient structures. Their presence at all position angles in the \textit{STEREO} images could be the result of the streamers being highly warped during that period, meaning that the density fluctuations are released at high latitudes as well. The origin of these small-scale density fluctuations is still debated and they could come also from other coronal regions than just the streamers. We have found no evidence from the dynamic WindPredict-AW simulations that they could originate in magnetic reconnection from streamer tops. Therefore in \citet{Griton2020} we investigated instead whether these density fluctuations could originate from lower heights in the solar corona during impulsive heating events.  \\

In particular we wanted to check whether these density fluctuations are generated low in the corona during transient heating events detected as local brightenings in EUV and X-rays, known as coronal bright points (CBPs) \citep{Madjarska2019}. These CBPs appear as hot loops (log$T$[K]$\simeq 6.2$) that have received a significant amount of energy through what are likely to be localised magnetic reconnection events \citep{Kwon2012}. The combined analysis of intense CBPs observed in EUV and WL images of the upper corona have revealed that outflowing density structures known as "coronal jets" often originate as CBPs low in the corona \citep{Wang1998}. Recent numerical simulations also corroborate this view \citep[e.g.][]{Singh2019}. These coronal jets become part of the outflowing solar wind outstreaming from coronal holes \citep{Wang1998}. Hence an association exists between the occurrence of density structures in the outflowing solar wind and the occurrence of CBPs low in the corona. Furthermore, based on a statistical analysis of the occurrence of CBPs carried out by \citet{Alipour2015}, \citet{Griton2020} suggest that on average 1.6 CBPs should be expected for each $1^\circ$ band of elevation angle in Figure \ref{fig:DeForest2018_fig12}, and producing brightness fluctuations in WL every $\simeq 15$ hours. \\

An alternative setup of the MULTI-VP model introduced in section \ref{subsec:MULTI-VP} was developed to simulate the outward propagation of density fluctuations originating in impulsive heating events at CBPs. For that an additional transient heating term is added to the background coronal heating (eq \ref{eq:MVP_Q}) so that the total heating rate $Q_h=-\nabla_\parallel \cdot F_h$ (in $erg.cm^{-3}.s^{-1}$) is \citep{Griton2020}:
\begin{equation}
\label{eq:MVP_Q2}
    Q_h=-12\times 10^5 |B_\odot|\left(\frac{A_\odot}{A}\right)\left(\frac{1}{H_f}exp\left[-\frac{s-R_\odot}{H_f}\right] + \frac{1}{r_p}\mathcal{H}(t)a_p exp\left[-\frac{(s-R_p)^2}{r_p^2}\right]\right)
\end{equation}
This perturbed component in $Q_h$ is a Gaussian function whose center, width and amplitude is parametrized by $R_p$, $r_p$ and $a_p$. The time dependency of the impulsive heating is introduced with the time dependent $\mathcal{H}(t)$ factor. 

A series of MULTI-VP runs was performed with this setup, first on a simple idealized geometry and then a more realistic three-dimensional version, in order to test the effect of each of the heating parameters on the generated density fluctuations. The parameters assumed for the spatial distribition (heliocentric latitude/longitude), frequency of occurrence, amplitude and height deposition of the energy to the plasma were taken from observations of CBPs \citep{Alipour2015, Madjarska2019}. These transient heating events induce perturbations that have a significant impact on both the simulated background solar wind and the generated density fluctuations \citep[see][]{Griton2020}. In a nutshell, the sudden deposition of energy in the low corona triggers slow-mode (compressive) waves that propagate outwards at the local sound speed in the simulated solar wind stream. The associated compression fronts of these waves produce density enhancements that are found sufficiently strong to be clearly visible in synthetic imagery and with intensity variations that are comparable with \textit{STEREO-A COR-2} observations. The reader is invited to refer to the \citet{Griton2020} paper for more details. \\

\begin{figure*}[]
  \centering
  \includegraphics[width=\textwidth]{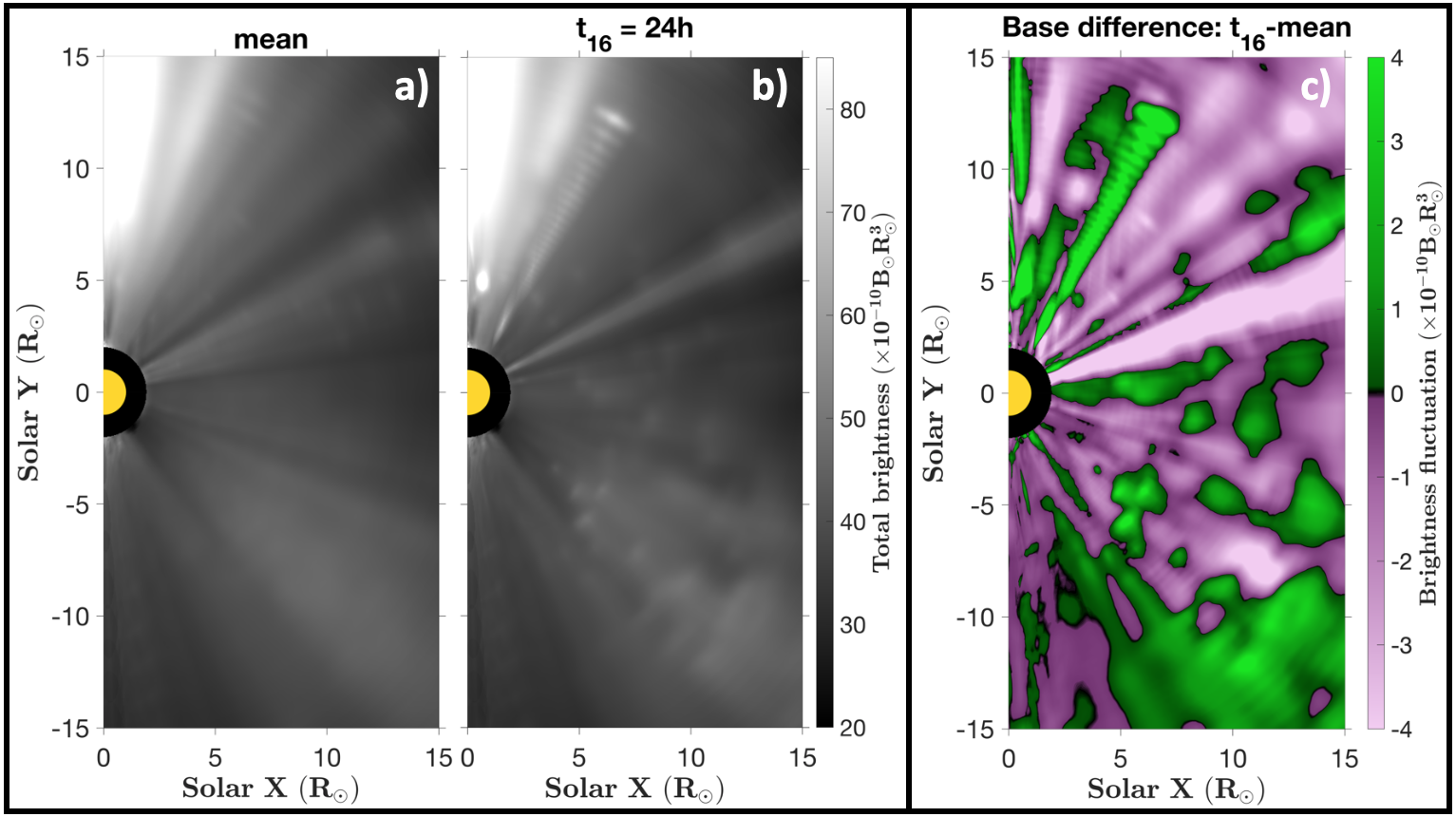}
  \caption{
        Panel a: mean (averaged) synthetic image over times $t=15\mathrm{h}$ to $t=30\mathrm{h}$ from simulation 9 \citep[see][]{Griton2020}. 
        Panel b: synthetic image at time $t=24\mathrm{h}$.
        Panel c: synthetic base difference image at time $t=24\mathrm{h}$ with respect to the mean background image shown in panel a.} 
  \label{fig:Griton2020_fig10}
\end{figure*}

For this paper I performed WL synthetic imagery from the realistic three-dimensional case mentioned above and which includes randomly distributed heating events based on the occurrence rate of CBPs (simulation 9 in \citet{Griton2020}). I followed the same method as the one described in section \ref{sec:stationnary_synthetic} based on the Thomson scattering theory introduced in section \ref{sec:stationnary_theory}. Figure \ref{fig:Griton2020_fig10} shows a result from this process where the brightness is multiplied by $r^3$ (where $r$ is the heliocentric distance) to enhance the visibility of the fluctuations as done in \citet{DeForest2018}. To allow for a visual comparison with their figure (Figure \ref{fig:DeForest2018_fig12}), I then employed a base-difference method where a background synthetic image representative of the mean solar wind flow (panel a) is subtracted from a raw synthetic image (panel b) to enhance the visibility of brightness fluctuations (panel c). The color scale is similar to the one used in Figure \ref{fig:DeForest2018_fig12} so that the reader can directly compare our simulated density fluctuations with the ones observed by \citet{DeForest2018} from \textit{STEREO-A COR-2} images. \\

Figure \ref{fig:Griton2020_fig10}(c) shows several propagating structures characterised by increases in total brightness $\lesssim4\times10^{-10}\ B_{\odot}R_{\odot}^3$ or alternatively $\sim11\%$ with respect to the mean background image. The shape and order of magnitude are consistent with the ubiquitous brightness fluctuations observed by \textit{STEREO COR2} and analyzed in \citet{DeForest2018}. Synthetic brightness fluctuations show different shapes, scales and intensities. Regarding shapes, one can see patches (or blobs) alongside elongated features. Elongated features are mainly visible due to the saturation of the color scheme, where neighboring fluctuations coalesce. They may also result from the superposition of background and foreground contributions. \\

In conclusion, this study reveals that at least a fraction of the brightness fluctuations observed in the corona could be caused by heating events due to CBP occurring below, near the coronal base. Since CBPs occur over the entire solar surface, the formation and outward propagation of these density fluctuations is not limited to the streamer belt and can appear both in the slow and fast solar wind regions. A process that we neglected and is known to occur during CBPs is magnetic reconnection. In particular CBPs likely result from reconnection between coronal magnetic fields. At the source of the solar wind, interchange reconnection between the pre-existing open-field of coronal holes and the highly dynamic closed-field of small-scale loops is likely to occur and generate complex, probably twisted, magnetic fields that are released as jets or even microjets. As discussed in section \ref{subsec:intro_interchange} and \ref{subsec:intro_Sweb}, many mechanisms can favor such interchange reconnection in coronal holes, of which the emergence of magnetic flux, the convective motions at the photosphere, and the differential versus rigid rotation. We have seen in this section that, given a source of energy provided by dynamic processes, the quasi-stationnary theory can explain part of the fluctuations observed in the solar wind. In the \citet{Griton2020} study the simulated fluctuations were the most visible in the high speed and tenuous solar wind streams that originate far from the streamer, but they were also seen in the dense slow wind that typically forms further away from the center of coronal holes or near the center of small equatorial coronal holes (such as the one discussed in section \ref{sec:dynamics_insitu}).

\section{Conclusion}

We showed in the previous chapter \ref{cha:stationnary} that coronal and solar wind imagery can be used to investigate the SSW's spatial and temporal variability with greater ease than with the in situ measurements considered in this chapter (see section \ref{sec:dynamics_insitu}). This is because time-dependent variability tends to manifest itself as the outward propagation of bright (high density) structures in the images whereas spatial variability induces brightness variations that evolve more slowly in the images even for the rapidly-moving \textit{Parker Solar Probe}. In chapter \ref{cha:stationnary} we depicted SSW source regions that are highly structured spatially in the form of highly striated streamer rays observed in great detail by \textit{WISPR}.  \\

In the present chapter, we looked at whether this spatial variability could to some extent be isolated in in situ measurements thanks to the \textit{PSP}'s novel orbit. We therefore carried out a more in-depth analysis (section \ref{sec:dynamics_insitu}) of how the variability of the SSW measured in situ by \textit{PSP} results from both spatial and time-dependent effects on a broad range of scales. We focused first on a time period when \textit{PSP} transitioned from what was interpreted as streamer flows to coronal hole flows. While streamer belt flows appear much more variable, exhibiting strong variations in plasma and magnetic field parameters \citep{Reville2022}, we found that a quiet solar wind component also exists in this streamer flow \citep{Griton2021} and that its mean properties differ from those measured in the quiet component of the coronal hole flow. \\

We interpreted these differences as a \emph{spatial-dependent} effect of the probe moving across distinct SSW that form along magnetic field lines with different magnetic topologies at their source. The quasi-stationary theory was found to provide a quantitative assessment of what we categorised as a spatial variability in the SSW. Just as we did in chapter \ref{cha:stationnary} to interpret the fine structure of streamer rays observed in \textit{PSP-WISPR}, we used the MULTI-VP MHD model to study quantitatively the differences between the quiet solar wind measured in streamers and coronal hole flows. To do that we kept the coronal heating term (eq \ref{eq:MVP_Q}) dependent on the expansion factor of flux tubes channelling the solar wind. A transition of what appears to be two states of the SSW measured in situ by \textit{PSP} during its second encounter is naturally interpreted as the transition between field lines that have different expansion factors at their source. This topological change in magnetic field is related to \textit{PSP}'s magnetic connectivity shifting from the edge of the streamer belt to the center of a small equatorial coronal hole. If the first in situ measurements taken by \textit{PSP} were mostly constituted of streamer-like SSW, \textit{PSP} has now collected over the recent orbits a wealth of in situ data of solar wind plasmas that originate from a diversity of source regions. A systematic comparison between the magnetic topology of the source regions and the bulk properties of the SSW measured in situ at \textit{PSP} could provide further tests to the quasi-stationary theory of the SSW. \\

We then examined how the various physical processes proposed in the dynamic theory of the SSW could explain the \emph{time-dependent} variability of the SSW. For that we analysed and interpreted further  \textit{WISPR} observations by exploiting advanced 3-D time-dependent numerical modelling. As already noted in chapter \ref{cha:stationnary}, WL emissions from the stationary part of the corona detected in the foreground and background of \textit{WISPR-I} images can complicate their interpretation. However we show that \textit{WISPR-I} can capture perturbations in the SSW with fine details that are otherwise smeared out at further distances to the Sun, this includes what appear to be very small plasmoids outflowing along streamer rays. I exploited the WindPredict-AW MHD simulations performed by Victor Réville and described in \citet{Reville2020b,Reville2022} to examine how well they can explain the transients observed in \textit{WISPR}. Because this model solves the coupled transport of the solar wind plasma with the magnetic field, dynamic processes such as the generation of periodic structures from the tip of streamers could be investigated in detail. The large plasmoids and flux ropes released from the streamers result from the extension and thinning of streamer loops. This process is driven by pressure imbalance between the loops and the surrounding medium and that then provides favorable conditions for magnetic reconnection to occur at the HCS through the tearing instability. The reconnection between coronal loops that have been stretched during this process leads to the formation of flux ropes that propagate along with the SSW, whose signatures have been detected in situ in past studies \citep{Rouillard2009, Sanchez-Diaz2019}. \\

The 2.5-D MHD numerical setup of WindPredict-AW that was introduced in section \ref{subsec:dynamics_tearing_2p5D}, reproduces nicely the density structures that occur with periodicities consistent with past in situ and remote-sensing observations. In order to fully address the spatial and temporal variability in these images I had to exploit the 3-D setup of WindPredict-AW. The 3-D MHD setup cannot resolve the scales necessary to model accurately the tearing mode instability but it does produce (streamer-induced) outflowing 3-D flux ropes, albeit with greater size and slightly longer periodicities that deduced from both in situ measurements and remote-sensing observations by \textit{SolO} and \textit{PSP} \citep{Reville2022}.  I used my forward modeling technique to simulate \textit{WISPR} images of the quasi-periodic structures generated in WindPredict-AW. That includes a close-up view of a flux rope simulated in WindPredict-AW that looks similar to other flux ropes imaged by \textit{WISPR-I} and which have been presented in section \ref{sec:dynamics_WISPR}. Future studies should extend this type of analysis by searching and cataloguing more small flux rope like structures observed in \textit{WISPR} images in order to include many more cases. \\

Section \ref{sec:dynamics_griton2020} extends the analysis made in section \ref{sec:dynamics_tearing} to address the origin of even smaller density structures that appear ubiquitous in coronal imagery from \textit{STEREO}. In contrast to the formation process discussed in section \ref{sec:dynamics_tearing}, we examine in section \ref{sec:dynamics_griton2020} the generation of density perturbations at coronal heights situated well below the tip of streamers, at the base of the corona. The MULTI-VP model was exploited to simulate how the quasi-stationnary solar wind may be altered when subjected to a varying input of energy. As discussed in section \ref{subsec:intro_interchange} and \ref{subsec:intro_Sweb}, magnetic reconnection is thought to be ubiquitous in the low corona. The heating associated with these reconnection events was considered in MULTI-VP by an additional impulsive heating component that triggered periodic pressure pulses that induced high-density structures rising up in the corona. By supposing that these heating events follow the statistics of CBPs which are interpreted as electromagnetic signatures of magnetic reconnection events happening in the low corona, we showed that periodic heating can generate density perturbations that fill out a significant portion of the corona typically observed by WL coronagraphs. However, it remains to assess how much of these perturbations relate to those actually observed by \textit{WISPR}, that is left for a future study. \\

%% file: chapters/ISAM_v2.tex
\chapter{The Irap Solar Atmosphere Model (ISAM)}
\label{cha:ISAM}
\minitoc

ISAM (for Irap Solar Atmosphere Model) is a multi-specie kinetic-fluid model that solves for the transport of neutrals and charged particles along a magnetic field line and from the chromosphere to the corona. The model assumes gyrotropy of the velocity distribution function (VDF) and solves a set of transport equations for 16 moments of the VDF ( $n,\ u,\ T^\parallel,\ T^\perp,\ q^\parallel,\ q^\perp$ ) (see section \ref{sec:ISAM_origin} and \ref{sec:ISAM_transport_eq}). The transport equations are solved for neutral atoms, ions and electrons. ISAM solves for transport equations that are concretely in 1-D but include the effects of pressure anisotropy. The predominant ionization and recombination processes are accounted for in order to model partial ionization in the chromosphere (see section \ref{sec:ISAM_ioniz}), this is essential to address the FIP effect. Thermal and momentum exchanges between species through collisions are also accounted for (see section \ref{sec:ISAM_collisions}). The solution of the model will depend strongly on the chosen heating profile which can be either prescribed with an ad-hoc heating function or through an explicit calculation of the dissipation of Alfvén waves (see section \ref{sec:ISAM_heating}). Near the top of the chromosphere and at the transition region, the energy balance includes cooling/heating terms that account for the radiative emissions/absorptions in the optically thick and thin regions (see section \ref{sec:ISAM_radloss}). This comprehensive set of physical processes solved by ISAM allows us to compute dynamically the different layers of the solar atmosphere: chromosphere, transition region and corona.

\section{Legacy}

The structure of ISAM is based on an ionospheric code called IPIM \citep{Blelly1993,Marchaudon2015}. The model has been adapted to the solar atmosphere by my colleague Michael Lavarra who included the coupling of neutrals to charged particles and exploited the model to study the properties of the solar wind in open magnetic field geometries \citep{Lavarra2022}. In parallel I developed a different setup to simulate closed-field plasmas confined in coronal loops.

ISAM is a model that follows the footsteps of previous implementations of high-order moment approaches \citep{Schunk1977,Demars1979,Blelly1993,LieSvendsen2001,Killie2004,Janse2006}. They all aim at describing in detail the transport of matter and energy with a self-consistent resolution of the heat flux that seeks to be valid for both collision-dominated and collisionless plasmas.

\section{Origin: the Boltzmann equation}
\label{sec:ISAM_origin}

High-order approaches all start from a kinetic description of the plasma. Because a peculiar description of each particle that constitutes the plasma is computationally prohibitive for typical densities around $10^{10}-10^{20}$ in the solar atmosphere, therefore a statistical description using the velocity distribution function $f(\mathbf{r},\mathbf{v},t)$ (VDF) is more appropriated. In other words, $f(\mathbf{r},\mathbf{v},t)$ is the particle density at time $t$ in the volume defined by $\mathbf{r}-d\mathbf{r} \leq \mathbf{r}\leq \mathbf{r}+d\mathbf{r}$ with a peculiar velocity $\mathbf{v}-d\mathbf{v} \leq \mathbf{v}\leq \mathbf{v}+d\mathbf{v}$. 

The dynamical evolution of such a system that is out of balance due to external interactions is described by the Boltzmann equation. For an ensemble $s$ of particles (species) with mass $m_s$ and VDF $f_s$ undergoing a force $F_s$, the Boltzmann equation reads:
\begin{equation}
    \frac{D}{Dt}f_s=\frac{\partial}{\partial t}f_s + \mathbf{v_s} \cdot \mathbf{\nabla}_r f_s + \frac{1}{m_s}\mathbf{F_s}\cdot\mathbf{\nabla}_v f_s = \frac{\delta f_s}{\delta t}
    \label{eq:ISAM_boltzmann}
\end{equation}
where $\mathbf{\nabla}_r$ and $\mathbf{\nabla}_v$ are the gradients in the coordinate and velocity space. Collisions of a specie $s$ with other species will affect the VDF throughout the right-hand side collision term $\frac{\delta f_s}{\delta t}$. Solving directly for the Boltzmann equation is usually not appropriated to describe the macroscopic state of a system. Only in rare occasions the Boltzmann equation has been fully solved numerically, as in the work of \citet{Spitzer1953} who gave their names to the well-known Spiter-Härm classical heat flux prescription that is widely used by the scientific community to describe heat flux in collisional plasmas. Such approaches are usually referred as the classical theories.

Since we are interested in the macroscopic properties of a plasma, it is more convenient to represent its statistical state through the moments of the VDF, assuming that the macroscopic scales are sufficiently large for the notion of density and temperature to be meaningful. For plasmas typically the particles exhibit collective behaviour mediated by collisions and/or the electromagnetic field. In a gas composed of charged particles such as a plasma, a convenient scale can be introduced above which the electric field induced by for instance an ion in the gas will be screened by the surrounding electrons, the time required for this screening to take place in a weakly collisional plasma is related to the transit time of a thermal particle to cross the Debye length:
\begin{equation}
    \lambda_D=\sqrt{\frac{\epsilon_0 k_b T_e}{e^2 n_e}}\frac{1}{\sqrt{1+\sum_j Z_j^2 \frac{n_j}{n_e}\frac{T_e}{T_j}}}
\end{equation}
where $\epsilon_0$ is the vacuum permittivity, $k_B$ the Boltzmann constant, $e$ the electron elementary charge, $Z_j$ the ion charge state, $n_{e,j}$ and $T_{e,j}$ the electron and ion density and total temperature. The Debye sphere of radius $D$ defines the region of maximum influence of a particle in the plasma. The typical Debye length reaches macroscopic scales of about $10^{-6}-10^{-2}\ \rm{m}$ in the solar chromosphere and corona, and $\approx 10\ \rm{m}$ in the solar wind but in any case it remains much smaller than the minimum grid cell size in ISAM (i.e. $15\ \rm{km}$ at the transition region for the runs presented in chapter \ref{cha:ISAM_results}). \\

The first 16 moments of the VDF solved in ISAM are obtained by subsequent averaging of the VDF:
\begin{subequations}
\label{eq:ISAM_moments}
\begin{align}
    n_s &= \int f_s d\mathbf{v_s} \ \ &\text{density}\\
    \mathbf{u_s} &= \left< \mathbf{v_s}-\mathbf{t_s}\right> = \left< \mathbf{v_s}\right> \ \ &\text{drift velocity}\\
    T_s^\parallel &= m_s/k_b \left< \mathbf{t_s}^{\parallel 2} \right> \ \ &\text{parallel temperature}\\
    T_s^\perp &= m_s/(2k_b) \left< \mathbf{t_s}^{\perp 2} \right> \ \ &\text{perpendicular temperature}\\
    \mathbf{q_s^\parallel} &= n_s m_s \left< \mathbf{t_s}^{\parallel 2} \mathbf{t_s} \right> \ \ &\text{heat flow vector for parallel energy}\\
    \mathbf{q_s^\perp} &= n_s m_s/2 \left< \mathbf{t_s}^{\perp 2} \mathbf{t_s} \right> \ \ &\text{heat flow vector for perpendicular energy}\\
    \mathbf{P_s} &= n_s m_s \left< \mathbf{t_s}\mathbf{t_s}\right> \ \ &\text{pressure tensor}\\
    \mathbf{\tau_s} &= \mathbf{P_s} - p_s^\perp\mathbf{I}-(p_s^\parallel-p_s^\perp)\mathbf{e_3}\mathbf{e_3}\ \ &\text{stress tensor}
\end{align}
\end{subequations}
where the averaging operator is defined as:
\begin{equation}
        \left<\mathbf{A}(\mathbf{r},\mathbf{v_s},t)\right> = \frac{1}{n_s(\mathbf{r},t)}\int f_s(\mathbf{r},\mathbf{v_s},t)\mathbf{A}(\mathbf{r},\mathbf{v_s},t)d\mathbf{v}_s
\end{equation}
and $\mathbf{t_s=v_s-u_s}$ is the random (thermal) speed, $p_s^\parallel=n_s k_b T_s^\parallel$ and $p_s^\perp=n_s k_b T_s^\perp$ the parallel and perpendicular thermal pressures, $\mathbf{I}=\mathbf{e_1}\mathbf{e_1}+\mathbf{e_2}\mathbf{e_2}+\mathbf{e_3}\mathbf{e_3}$ is the unit dyadic tensor (or identity matrix). Note that the $\parallel$ and $\perp$ subscripts for the heat flow vectors indicate whether the parallel or perpendicular energy is transported. In all other cases the $\parallel$ and $\perp$ subscripts refer to the vector component parallel and perpendicular to the magnetic field. \\

By forming the relevant moments listed in equation \ref{eq:ISAM_moments} from the Boltzmann equation, one can derive a complex set of transport equations. This system of equations needs to be closed though because the transport equation for the $n^{th}$ moment introduces the moment of order $n+1$. A closed-set of transport equations can be obtained by approximating the shape of the VDF. \\

An isotropic plasma in thermodynamic equilibrium can be described by a Maxwell distribution function of the form:
\begin{equation}
\label{eq:ISAM_VDF_M}
    f_{s0}^M=n_s\left(\frac{\beta_s}{2\pi}\right)^{3/2}\exp\left(\frac{-\beta_s \mathbf{t_s}^2}{2}\right)
\end{equation}
with $\beta_s=m_s/(k_b T_s)$. However such approximation is not appropriated in a plasma where collisions are too scarce to allow an efficient redistribution of the energy, as in the solar corona. A bi-Maxwellian is therefore recommended to account for the plasma anisotropy in the direction parallel $T_s^\parallel$ and perpendicular $T_s^\perp$ to the magnetic field:
\begin{equation}
    f_{s0}^{BM}=n_s\frac{\beta_s^\perp}{2\pi}\sqrt{\frac{\beta_s^\parallel}{2\pi}}\exp\left(-\frac{\beta_s^\perp \mathbf{t_s}^{\perp 2}}{2}-\frac{\beta_s^\parallel \mathbf{t_s}^{\parallel 2}}{2}\right)
\end{equation}
where $\mathbf{t_s}^\parallel$ and $\mathbf{t_s}^\perp$ denote the thermal velocity vector $\mathbf{t_s}$ splitted in a part that is aligned or perpendicular to the magnetic field. Since both the Maxwellian and bi-Maxwellian functions are symmetric, the associated heat flux and stress moments are automatically null. \\

Most high-order approaches consist in expanding the above zeroth-order Maxwellian or bi-Maxwellian VDF in an orthogonal series of the form:
\begin{equation}
    f_s =f_{s0}(1+\chi_s)
\end{equation}
where $\chi_s \ll 1$ comprises a series of higher order terms.

Different set of transport equations can then be derived depending of the number of high order terms retained in the series expansion of the VDF. The usual 5-moment, 8-moment, 10-moment and 13-moment sets all assume a zeroth-order Maxwellian VDF $f_{s0}=f_{s0}^M$ but do not retain the same higher order terms. For instance the 5-moment set only keeps the zeroth-order term $f_s=f_{s0}=f_{s0}^M$, and then describes an isotropic plasma with no heat flux nor stresses. ISAM solves for the 16-moment set of transport equations where the zeroth-order bi-Maxwellian VDF is expanded to first order \citep{Demars1979}:
\begin{align}
\label{eq:ISAM_VDF16}
    f_s &=f_s^{BM}(1+\chi_s)\\
    \text{where:}\quad \chi_s &=-\frac{\beta_s^\perp}{p_s^\perp}\left(1-\frac{\beta_s^\perp \mathbf{t_s}^{\perp 2}}{4}\right)\mathbf{q_s^\perp}\cdot \mathbf{t_s^\perp} \notag\\
    &-\frac{\beta_s^\perp}{p_s^\parallel}\left(1-\frac{\beta_s^\perp \mathbf{t_s}^{\perp 2}}{2}\right)\mathbf{q_s^\perp}\cdot \mathbf{t_s^\parallel} \notag\\
    &-\frac{\beta_s^\parallel}{2p_s^\perp}\left(1-\beta_s^\parallel \mathbf{t_s}^{\parallel 2}\right)\mathbf{q_s^\parallel}\cdot \mathbf{t_s^\perp} \notag\\
    &-\frac{\beta_s^\parallel}{p_s^\perp}\left(1-\frac{\beta_s^\parallel \mathbf{t_s}^{\parallel 2}}{3}\right)\mathbf{q_s^\parallel}\cdot \mathbf{t_s^\parallel} \notag\\
    &+\frac{\beta_s^\parallel}{p_s^\perp}(\mathbf{\tau_s}:\mathbf{t_s^\perp}\mathbf{t_s^\parallel}) \notag\\
    &+\frac{\beta_s^\perp}{2p_s^\perp}(t_{s1}^2-t_{s2}^2)(\mathbf{\tau_s}:\mathbf{e_1}\mathbf{e_1}) \notag\\
    &+\frac{\beta_s^\perp}{p_s^\perp}(\mathbf{\tau_s}:\mathbf{e_1}\mathbf{e_2}) \notag
\end{align}
with $\beta_s^{\parallel,\perp}=m_s/(k_b T_s^{\parallel,\perp})$ and where $p_s^\parallel=n_s k_b T_s^\parallel$ and $p_s^\perp=n_s k_b T_s^\perp$ are the parallel and perpendicular thermal pressures. For recall the $\mathbf{e_1}$, $\mathbf{e_2}$ and $\mathbf{e_3}$ unit vectors define an orthogonal base where $\mathbf{e_3}$ is tangent to the magnetic field line, $t_{s1}$ and $t_{s2}$ are the components of the vector $\mathbf{t_s}$ expressed in this base. 

As we will see in section \ref{sec:ISAM_collisions}, the assumption on the VDF will also have an impact on the collision terms. \\

The validity of the 16-moment approach is conditioned by a small deviation to the zeroth-order VDF  $\chi_s \ll 1$ that translates into the following conditions:
\begin{enumerate}
    \item small stresses (or pressure anisotropy):
    \begin{equation}
    \label{eq:ISAM_col_cond2}
        |p_s^\perp-p_s^\parallel|/p_s\ll 1
    \end{equation}
    \item small heat flows:
    \begin{equation}
    \label{eq:ISAM_col_cond3}
        \gamma_s^\parallel=\frac{q_s^\parallel}{p_s^\parallel c_s^\parallel}\ll 1
    \end{equation}
    \begin{equation}
    \label{eq:ISAM_col_cond4}
        \gamma_s^\perp=\frac{q_s^\perp}{p_s^\perp c_s^\parallel}\ll 1
    \end{equation}
\end{enumerate}
where $c_s^\parallel=\sqrt{k_b T_s^\parallel/m_s}$ is the thermal (or sound) speed associated to the parallel temperature. 

\section{The 1-D transport equations}
\label{sec:ISAM_transport_eq}

ISAM solves the 16-moment set of transport equations described in \citet{Demars1979} and \citet{Blelly1993}) for neutrals, ions and electrons. The conservative equations are similar to the 1-D MHD MULTI-VP model introduced in section \ref{subsec:MULTI-VP} and to a larger extent to the 3-D MHD WindPredict-AW model presented in section \ref{subsec:WindPredict}. The main differences with ISAM come from additional terms for particle collisions and chemical reactions, pressure anisotropy, and an explicit resolution of the heat flux transport thanks to a high-order closure of the transport equations. \\

The 16-moment set of transport equations projected along the magnetic field direction can be expressed in a general formulation for neutrals, charged particles and electrons as follows \citep{Lavarra2022}:
\begin{subequations}
\begin{flalign}
    \label{eq:ISAM_ns}
    & \frac{\partial}{\partial t}\rho_s+u_s \nabla_\parallel \rho_s +\rho_s \frac{1}{A}\nabla_\parallel(A u_s) = m_s\frac{\delta n_s}{\delta t}\\
    \label{eq:ISAM_us}
    & \frac{\partial }{\partial t} u_s + u_s \nabla_\parallel u_s + \frac{\nabla_\parallel n_s k_b T_s^\parallel}{\rho_s} + \frac{k_b}{m_s}\left(T_s^\parallel - T_s^\perp \right)\frac{1}{A}\nabla_\parallel A + \frac{GM_\odot}{r^2}cos(\theta) - \frac{1}{m_s n_s}F_s = \frac{\delta u_s}{\partial t}\\
    \label{eq:ISAM_Tps}
& \begin{aligned}
    \frac{\partial }{\partial t}T_s^\parallel + u_s \nabla_\parallel T_s^\parallel + 2T_s^\parallel \nabla_\parallel u_s + \frac{1}{n_s k_b} \nabla_\parallel q_s^\parallel + \frac{q_s^\parallel - 2q_s^\perp}{n_s k_b A} \nabla_\parallel A &= \frac{1}{n_s k_b}\left(Q_{h,s}^\parallel - \mathcal{R}_s^\parallel \right) \\
    &+ \frac{\delta T_s^\parallel}{\delta t} - \frac{T_s^\parallel}{n_s}\frac{\delta n_s}{\delta t}
\end{aligned}\\
\label{eq:ISAM_Tts}
& \begin{aligned}
    \frac{\partial }{\partial t}T_s^\perp + u_s \nabla_\parallel T_s^\perp + \frac{1}{n_s k_b} \nabla_\parallel q_s^\perp +\left( \frac{2q_s^\perp}{n_s k_b} + u_s T_s^\perp\right) \frac{1}{A} \nabla_\parallel A &= \frac{1}{n_s k_b}\left(Q_{h,s}^\perp - \mathcal{R}_s^\perp \right) \\
    &+ \frac{\delta T_s^\perp}{\delta t} - \frac{T_s^\perp}{n_s}\frac{\delta n_s}{\delta t}
\end{aligned}\\
\label{eq:ISAM_qps}
& \begin{aligned}
    \frac{\partial }{\partial t}q_s^\parallel + u_s \nabla_\parallel q_s^\parallel + 4q_s^\parallel \nabla_\parallel u_s + \frac{3 n_s k_b^2 T_s^\parallel}{m_s} \nabla_\parallel T_s^\parallel &+ u_s q_s^\parallel\frac{1}{A}\nabla_\parallel A \\
    &= \frac{\delta q_s^\parallel}{\delta t} - 3 n_s k_b T_s^\parallel \frac{\delta u_s}{\delta t}
\end{aligned}\\
\label{eq:ISAM_qts}
& \begin{aligned}
    \frac{\partial }{\partial t}q_s^\perp + u_s \nabla_\parallel q_s^\perp + 2 q_s^\perp \nabla_\parallel u_s + \frac{n_s k_b^2 T_s^\parallel }{m_s}\nabla_\parallel T_s^\perp &+\left(\frac{n_s k_b^2 T_s^\perp }{m_s} (T_s^\parallel - T_s^\perp ) +2 u_s q_s^\perp\right)\frac{1}{A}\nabla_\parallel A \\
    &= \frac{\delta q_s^\perp}{\delta t} - n_s k_b T_s^\perp \frac{\delta u_s}{\delta t} 
\end{aligned}
\end{flalign}
\end{subequations}
where $G$ is the gravitational constant, $M_\odot$ the Sun mass, $\theta$ the inclination angle of the magnetic field with respect to the vertical/radial direction. The gradient along the direction parallel to the magnetic field is defined as $\nabla_\parallel(*)=\partial/\partial c (*)$ where $c$ is the curvilinear abscissa. Similarly to the 1-D MULTI-VP model (see eq \ref{eq:MVP_div}), the divergence operator along the direction parallel to the magnetic field is defined as $\nabla_\parallel \cdot (*)=1/A \nabla_\parallel(A*)$ where the cross-sectional area $A$ of the considered flux tube is inversely proportional to the magnetic field strength ($A(c)\propto 1/B(c)$).

Depending on the heating model (see section \ref{sec:ISAM_heating}), the energy given to the plasma can be distributed differently among species, and to the parallel $Q_{h,n}^\parallel$ and perpendicular $Q_{h,n}^\perp$ thermal energy. The same goes for radiation cooling/heating through the terms $\mathcal{R}_n^\parallel$ and $\mathcal{R}_n^\perp$. The right-hand side terms $\delta*/\delta t$ account for the interactions between species via ionization/recombination processes (see section \ref{sec:ISAM_ioniz}) and collisions (see section \ref{sec:ISAM_collisions}). \\

For ions, the force $F_s$ includes the interaction with an electrostatic polarization field of the form:
\begin{equation}
    F_s=\frac{Z_i}{m_i} e E=-\frac{Z_i}{m_i}\frac{1}{n_e}\left[\nabla_\parallel (n_e k_b T_e^\parallel) + k_b(T_e^\parallel-T_e^\perp)\frac{1}{A} \nabla_\parallel A\right] + Z_i\frac{m_e}{m_i}\frac{\delta u_e}{\delta t} \label{eq:ISAM_Fpol}
\end{equation}
with $e$ the elementary charge. \\

Similarly to the ions, the parallel and perpendicular energy and heat flow equations are solved for the electrons. However the continuity and momentum equations are directly deduced from the ion populations assuming quasi-neutrality and ambipolar flow with no aligned current $\mathbf{J}=\mathbf{0}$:
\begin{subequations}
\label{eq:ISAM_eq_e}
\begin{align}
    n_e &=\sum_{i=ions}(Z_i  n_i)\label{eq:ISAM_ne}\\
    u_e &=\frac{\sum_{i=ions}(Z_i n_i u_i)}{n_e}\label{eq:ISAM_ue}
\end{align}
\end{subequations}
\newline

An important if not mandatory step in any numerical model is to scale or normalize the physical quantities prior their numerical resolution. The main motivation is to prevent for numerical truncation errors by scaling the physical quantities by characteristic parameters of the system. For completeness, the normalization procedure is given in Appendix \ref{sec:ISAM_normalization}. Once normalized, the transport equations of the 16-moment set are solved numerically by the LCPFCT algorithm described in section \ref{subsec:ISAM_lcpfct}. \\

We have already discussed in the previous section the underlying assumptions of the 16-moment approach that are listed in equations \ref{eq:ISAM_col_cond2}-\ref{eq:ISAM_col_cond4}. Similarly, the unicity of the solution given by the transport equations is conditioned by the criteria of hyperbolicity. Our transport equations are hyperbolic in the sense that they are "wave-like" as opposite to elliptical and parabolic equations. That means that a perturbation can not be perceived on the other side of an hyperbolic system until it actually reaches this point, whereas a perturbation is perceived at once in the entire domain in elliptical and parabolic systems. Therefore the hyperbolicity criteria consists in ensuring that these waves or perturbations do not travel faster than permitted by the physics. That is checked in ISAM by adjusting the numerical time step until the conditions described in Appendix \ref{sec:ISAM_hyperbol} are respected.

\section{Partial ionization}
\label{sec:ISAM_ioniz}
The multi-specie structure of ISAM allows us to simulate partially ionized plasma. Modeling a realistic partial ionization is a real challenge especially in the solar chromosphere. First the chromosphere is highly dynamic and likely evolves on short timescales before Hydrogen can actually reach an ionization equilibrium \citep{Carlsson2002a}. Second there is a strong coupling in the chromosphere between radiative and hydrodynamic processes which would require a detailed treatment of radiative transfer and of the energy levels. Ideally one would need a self-consistent radiation-hydrodynamic modeling of the chromosphere that we cannot afford here to keep ISAM tractable at the computational level. The treatment of ionization in ISAM is therefore a compromise between the expensive full non-LTE approaches and the crude LTE approximation where ionization is in statistical equilibrium at the local thermodynamic plasma properties. In contrast to LTE, ISAM still solves the rate equations for ionization which can proceed at different timescales than the hydrodynamic ones, but we rely on tabulated rates for ionization and recombination processes. \\

I carried out an extensive survey of the literature to gather a comprehensive and reliable set of rates for ionization/recombination and charge-exchange processes. These rates (the $\alpha$ terms defined later on in equations \ref{eq:PI_rate}, \ref{eq:RR_rate}, \ref{eq:DR_rate}, \ref{eq:DI_rate}, \ref{eq:CE_rate}) are included through the right-hand side terms $\delta n_s/\delta t$ of the continuity equations (see eq \ref{eq:ISAM_ns}) either as production or loss terms. In the next paragraphs I review and discuss the reaction rates that are currently used in ISAM.

\subsection{Photoionization}
Photoionization (PI) occurs when an incident photon has sufficient energy to extract a bound electron from a neutral or ionized atom:
\begin{equation}
    X^{(n-1)+}+h\nu \rightarrow X^{n+} + e^-
\end{equation}
The ejected (free) electron has an energy equal to the energy of the incident photon minus the minimum required energy for ionization. The ionized atom (or residual ion) can also be left at an excited state. Since we do not model energy level populations in ISAM for computational tractability, we sum up contributions over all possible excited states of the residual ion. Resonant photoionization can also occur through intermediate excited states. \\

I use the analytical fits from \citet{Verner1996b} for the partial photoionization cross sections of ground state atoms and ions. All ionization stages for all atoms of interest to model the solar atmosphere are available, from Hydrogen to Zinc. Resonances have been smoothed out, therefore resonant photoionization is not accounted in these fits. The partial photoionization cross sections are summed up over all possible excited states of the residual ion to get a total photoionization cross section $\sigma^{PI}_{X^{(n-1)+}}$. The total photoionization rate can then be computed by integration over the frequency of the incident electromagnetic spectrum \citep{Mihalas1978}:
\begin{equation}
    \alpha^{PI}_{X^{(n-1)+}}=\int_{\nu_0}^\infty \frac{4\pi J_\nu}{h\nu}\sigma^{PI}_{X^{(n-1)+}}(\nu)d\nu
    \label{eq:PI_rate}
\end{equation}
where $J_\nu$ is the spectral exitance (or radiance) of the solar spectrum. The PI $n_{X^{(n-1)+}}\alpha^{PI}_{X^{(n-1)+}}$ productions (and losses) are then directly inserted into the continuity equations (see eq \ref{eq:ISAM_ns}) through the right-hand side terms $\delta n_{X^{n+}}/\delta t$ (and $\delta n_{X^{(n-1)+}}/\delta t$). \\

For simplicity I approximate the solar spectrum to a black body emission at a photospheric temperature of $T_0=5800\ K$ following a Planck law:
\begin{equation}
    J_\nu=\frac{2\pi h\nu^3}{c^2}\frac{1}{\exp\left(\frac{h\nu}{k_b T_0}\right)-1}
\end{equation}
where $h$ is the Planck constant, $c$ the light speed, $\nu$ the frequency. The incident radiation field is then supposed to not be altered throughout the different layers of the solar atmosphere. This rough approximation allows us to decouple the photoionization processes and the hydrodynamic evolution of the chromospheric plasma, hence without any impact on the computational cost. \\

The photoionization cross sections from \citet{Verner1996b} have been exploited in many studies \citep[see e.g.][]{Ferland1998,Mazzotta1998,Avrett2008,Ferland2017}. While these cross sections have been determined carefully within the framework of the Opacity Project \citep{Opacity1995}, more recent calculations and measurements are now available. For instance, an extensive and collaborative effort is on-going that proposes a unified treatment for photoionization, radiative and dielectronic recombinations using the R-matrix method \citep{Nahar2004}. The NORAD-Atomic-Data database\footnote{The NORAD-Atomic-Data database: \url{https://norad.astronomy.osu.edu/##AtomicDataTbl1}} \citep{Nahar2020} is under construction and likely contains the most accurate cross sections to date. 

There are many limitations with our current implementation of photoionization which we discuss further in section \ref{subsec:ISAM_ioniz_future}.

\subsection{Radiative recombination and dielectronic recombination}
Photorecombination between an electron and an ion can be seen as the reverse process of photoionization:
\begin{equation}
    X^{n+}+e^- \rightarrow X^{(n-1)+} + h\nu
\end{equation}
As for photoionization this process may involve resonance and intermediate excited states. As such photorecombination is often split into its non-resonant part called radiative recombination (RR) and its resonant part called dielectronic recombination (DR). \\

I use the analytical fits from \cite{Verner1996a} which directly give the total rate for radiative recombination:
\begin{equation}
    \alpha^{RR}_{X^{n+}}=A\left[\sqrt{\frac{T_e}{T_0}}\left(1+\sqrt{\frac{T_e}{T_0}}\right)^{1-b}\left(1+\sqrt{\frac{T_e}{T_1}}\right)^{1+b}\right]
    \label{eq:RR_rate}
\end{equation}
where $T_e$ is the electronic temperature. Their fits include all fitting coefficients $A$, $b$, $T_0$ and $T_1$ for all ions from Hydrogen to Zinc. I use in ISAM a slightly different version that has been improved later on by \citet{Mazzotta1998}. \\

Dielectronic recombinations rates are fetched from \citet[][eq 7]{Mazzotta1998} who fitted an extensive set of collected data with the analytical relation:
\begin{equation}
    \alpha^{DR}_{X^{n+}}=\frac{1}{(k_b T_e)^{3/2}}\sum_{i=1}^4 c_j \exp\left(-\frac{E_i}{k_b T_e}\right)
    \label{eq:DR_rate}
\end{equation}
The fits coefficients $E_i$ and $c_j$ are provided in a VizieR online data catalog \citep[see][]{Mazzotta1998VizieR} for all ions from Helium to Nickel.

The RR and DR $n_{X^{n+}}n_{e^-}\alpha^{RR-DR}_{X^{n+}}$ productions (and losses) are directly inserted into the continuity equations (see eq \ref{eq:ISAM_ns}) through the right-hand side terms $\delta n_{X^{(n-1)+}}/\delta t$ (and $\delta n_{X^{n+}}/\delta t$). \\

These RR and DR rates have been widely used in the scientific community \citep[see e.g.][]{Dere2007} and are now provided through the well-known CHIANTI atomic database\footnote{The CHIANTI Atomic database: \url{http://www.chiantidatabase.org/}} \citep{Dere1997,DelZanna2021}. The RR and DR which have long been treated separately can now be unified with the self-consistent R-matrix approach \citep{Nahar2004}. As for photoionization, the NORAD-Atomic-Data database\footnote{The NORAD-Atomic-Data database: \url{https://norad.astronomy.osu.edu/##AtomicDataTbl1}} \citep{Nahar2020} likely gives the most up-to-date data for photorecombination and will be exploited in future ISAM versions.

\subsection{Collisional and auto-ionization}
An ion can be ionized by collision with an electron provided that the electron has sufficient energy. The ionization can be direct (i.e. direct ionization (DI)):
\begin{equation}
    X^{(n-1)+}+e^- \rightarrow X^{n+} + e^- + e^-
\end{equation}
or can proceed in two phases with an intermediate autoionizing state. The latter is commonly referred to as excitation-autoionization (EA):
\begin{subequations}
\begin{align}
    X^{(n-1)+}+e^- \rightarrow X^{(n-1)+*} + e^-\ \ &\text{: collisional excitation}\\
    X^{(n-1)+*} \rightarrow X^{n+} + e^-\ \ &\text{: autoionization}
\end{align}
\end{subequations}

The current ISAM version exploits the fitting formula from \citet[][eq 1 to 3]{Mazzotta1998} which is based on the previous works from \citet{Arnaud1985} and \citet{Arnaud1992}:
\begin{equation}
    \alpha^{DI}_{X^{(n-1)+}}=\frac{6.69\times 10^7}{(k_b T_e)^{3/2}}\sum_j\frac{e^{-x_j}}{x_j}F(x_j)\quad \rm{(in\ cm^3.s^{-1})}
    \label{eq:DI_rate}
\end{equation}
where
\begin{subequations}
\begin{align}
    F(x_j)&=A_j[1-x_jf_1(x_j)]+b_j[1+x_j-x_j(2+x_j)f_1(x_j)]C_jf_1(x_j)+D_jx_jf_2(x_j) \\
    f_1(x)&=\int_0^\infty\frac{dt}{1+t}e^{-tx}\\
    f_2(x)&=\int_0^\infty\frac{dt}{1+t}e^{-tx}ln(t)
\end{align}
\end{subequations}
and $x_j=I_j/(k_b T_e)$ (see \citet{Mazzotta1998} and references therein for the values of $A_j$, $b_j$, $C_j$ and $D_j$). Contributions from all subshells $j$ of the ionizing ion are summed up. The actual fits that are currently used in ISAM also include autoionization contributions for ground state ions.

The DI and EA $n_{X^{(n-1)+}}n_{e^-}\alpha^{DI-EA}_{X^{(n-1)+}}$ productions (and losses) are directly inserted into the continuity equations (see eq \ref{eq:ISAM_ns}) through the right-hand side terms $\delta n_{X^{n+}}/\delta t$ (and $\delta n_{X^{(n-1)+}}/\delta t$). \\

More recent DI and EA fitted rates have been computed by \citet{Dere2007}. They compared their ionization equilibria (computed with the new rates) with previous ones from \citet{Arnaud1985} and \citet{Arnaud1992}, and they only got negligible discrepancies except for specific elements which are not solved in ISAM yet. The unified R-matrix approach already mentioned above also provides new DI and EA rates through the NORAD-Atomic-Data database\footnote{The NORAD-Atomic-Data database: \url{https://norad.astronomy.osu.edu/##AtomicDataTbl1}} \citep{Nahar2020}. Future versions of ISAM may use these new data but in this first implementation I employ the fits from \citet{Mazzotta1998} which are sufficiently accurate for the applications presented in chapter \ref{cha:ISAM_results}.

\subsection{Charge exchange reactions}
Ion-neutral interactions can result in a transfer of an electron through resonant charge exchange. Charge-exchange (CE) reactions may occur when an ion collides with its parent neutral, or accidentally between unlike ions and neutrals:
\begin{equation}
    X^{p+}+Y^{n+} \rightarrow X^{(p+1)+}+Y^{(n-1)+}
\end{equation}
Charge exchanges between alike ions and neutrals as between neutral Hydrogen and protons: $H+H^+ \rightarrow H^++H$ do not play a role in the ionization balance as the gain and loss in $H$ and $H^+$ cancel out exactly in the continuity equation. However these charge exchange reactions will have an impact on the momentum and energy transfer (see section \ref{sec:ISAM_collisions}). In contrast, accidental charge-exchange reactions such as $O+H^+ \rightarrow O^++H$ (back and forth) have a non-null contribution to the density budget and must be included in the right-hand side of the continuity equations. \\

Ionization and recombination rates for accidental charge exchange are taken from \citet[][Table IIIA and IIIB]{Arnaud1985}:
\begin{equation}
    \alpha^{CE}_{X^{p+}-Y^{n+}}= \begin{dcases*} 
    a(T^*_4)^b\left(1+c\exp(T^*_4 d)\right) & for $\begin{bmatrix}H\\He\end{bmatrix}+Y^{n+} \rightarrow \begin{bmatrix}H^+\\He^+\end{bmatrix}+ Y^{(n-1)+}$\\
    a\exp\left(-\frac{\Delta E}{k_b T^*}\right)\left(1-0.93\exp(-T^*_4 c)\right) & for $O+H^+ \rightarrow O^++H$\\
    a(T^*_4)^b \exp(-T^*_4 c)\exp\left(-\frac{\Delta E}{k_b T^*}\right) & for $\begin{bmatrix}H^+\\He^+\end{bmatrix}+Y^{(n-1)+} \rightarrow \begin{bmatrix}H\\He\end{bmatrix}+Y^{n+}$
\end{dcases*}
\label{eq:CE_rate}
\end{equation}
where $T^*_4=T^*/10^4$ and $T^*$ is the reduced temperature defined in the center-of-mass of the two colliding particles:
\begin{equation}
    T^*=\frac{m_2T_1+m_1T_2}{m_1+m_2}
\end{equation}
Fits coefficients $a$, $b$, $c$, $d$ and $\Delta E$ are given in Tables IIIA and IIIB in \citet{Arnaud1985} who include both Hydrogen and Helium CE reactions with most of the minor species present in the solar atmosphere. These rates have been used in other studies of the FIP effect \citep[see e.g.][]{Steiger1989,Killie2007}. Future versions of ISAM will likely exploit the new rates from \citet{Kingdon1996} who extended the work of \citet{Arnaud1985} to all CE reactions between Hydrogen and the first 30 elements. Since we limit ourself to modeling Hydrogen, Helium and only a few minor species in this thesis, the rates from \citet{Arnaud1985} are sufficient for a start.

The CE $n_{X^{p+}}n_{Y^{n+}}\alpha^{CE}_{X^{p+}-Y^{n+}}$ productions (losses) are directly inserted into the continuity equations (see eq \ref{eq:ISAM_ns}) through the right-hand side terms $\delta n_{X^{(p+1)+}}/\delta t$ and $\delta n_{Y^{(n-1)+}}/\delta t$ ($\delta n_{X^{p+}}/\delta t$ and $\delta n_{Y^{n+}}/\delta t$). 

\subsection{Limitations and future improvements}
\label{subsec:ISAM_ioniz_future}

Partial ionization of Hydrogen in the upper part of the chromosphere is a delicate topic that has been addressed only by a few studies and is still not fully understood. Although our approach may be reasonable to model partial ionization of minor species, it is likely not realistic for Hydrogen. \\

The difficulty lies mostly in the fact that Hydrogen is not found in statistical equilibrium in the upper chromosphere \citep{Carlsson2002a}, and therefore it becomes necessary to look in detail at the populations of the excited levels of Hydrogen, and not only the ground state as in our current approach. Indeed they show that Hydrogen is predominantly photoionized not from the ground state ($n=1-\infty$, Lyman continuum) but from the first excited level ($n=2-\infty$, Balmer continuum) in most of the chromosphere. They conclude that the ionization degree of Hydrogen in the chromosphere is therefore directly dependent on the proportion of Hydrogen that is in the excited $n=2$ level. Radiative recombination from the upper levels followed by bound-bound radiative deexcitation down to the $n=2$ level is one of the two major processes that regulate the $n=2$ level population. The other process is by collision with electrons which can become a rapid means of production of $n=2$ excited Hydrogen atoms from the ground state as soon as the electron temperature increases. For temperatures above $\approx 1-2\times 10^4\ K$ as in the transition region, ionization of Hydrogen becomes dominated by collisional ionization from the ground state and our approach is much more accurate in this case. \\

According to these observations, a realistic modeling of the Hydrogen ionization balance in the chromosphere requires at least, to model a two-level Hydrogen atom. That would be an affordable task for a future short-term development in ISAM. Basically we will have to solve an additional specie for the $n=2$ level of Hydrogen which can in principle be already handled in the structure of the code. The main difficulty however will be to fetch a new set of rates to control the $n=2$ level of which, photoionization, collisional excitation and radiative deexcitation rates. For the results presented in this thesis, the current implementation of photoionization is accurate enough to match reasonably well the proton density in the upper chromosphere as predicted by semi-empirical models as we shall see in section \ref{subsec:ISAM_results_H_chromo}. \\

The photoionization rates also depend on the assumed incident radiation field which we take as the spectrum of a perfect emitter at the photosphere. The solar spectrum emitted at the photosphere is in reality transformed throughout the different layers of the low solar atmosphere as photons are continually absorbed and re-emitted. The resulting spectrum at the top of the chromosphere can therefore significantly differ from the one assumed at the photosphere. Another simple approach would be to consider the actual solar spectrum observed in the corona where the medium becomes optically thin. In any case, our current approach that assumes a constant photospheric radiation field may remain appropriated to model photoionization of neutral Hydrogen in the Balmer continuum, because the Balmer continuum radiation field is likely set at the photosphere \citep{Carlsson2007}. \\

Partial ionization of Hydrogen could be resolved in detail by using a proper radiative transfer code. While it is nowadays computationally prohibitive to fully solve for radiative transfer in 3-D MHD simulations, it is accessible for 1-D simulations. The RADYN code is one of them that solves the hydrodynamic equations together with rate equations for the excitation level populations and a non-LTE treatment of radiative transfer \citep[see][and references therein]{Carlsson2002a}. However their hydrodynamic approach is limited to a 5-moment set where the plasma is considered as isotropic and where heat flows are not solved explicitly. Including a complete radiative transfer solution in ISAM may be highly challenging due to the already complex high-order approach and is left for potential future developments. \\

An extensive set of external data is used in ISAM to compute the ionization/recombination rates for Hydrogen, Helium and minor species. Although I have already made a step forward at reviewing the validity of these datasets, there is still room for improvement. The exploitation of the NORAD atomic database which promotes accuracy and self-consistency is likely the next objective to pursue.

\section{Collisions between species}
\label{sec:ISAM_collisions}

The transition from the collisional chromosphere and the collisionless mid/high corona is dynamically solved in ISAM thanks to a comprehensive treatment of collisions between particles. By combining the later with the heat flow equation that is solved explicitly in our high-order approach, a great strength of ISAM is in solving the detailed transport of energy across the thin transition region. Modeling the heat flux accurately is of critical importance for simulating the thermal force which plays a major role in the FIP effect (see section \ref{subsec:intro_FIP}). Momentum transfer between species is also consistently implemented so that both the impact of the friction and thermal forces on the FIP fractionation can be assessed precisely. In the next sections I want to provide the user with a comprehensive description of the current treatment of collisions in ISAM. It has been a demanding task that I carried out to revise the physics and optimize the code when solving for minor species.

\subsection{General collision terms}

The collision terms $\delta */\delta t$ that appear on the right-hand side of the transport equations (see section \ref{sec:ISAM_transport_eq}) are derived similarly as the left-hand side terms, that is by forming the subsequent moments of the VDF collision transfer rate $\delta f_s/\delta t$ (see section \ref{sec:ISAM_origin}):
\begin{subequations}
\begin{align}
    \frac{\delta n_s}{\delta t} &= \int\left( \frac{\delta f_s}{\delta t}\right)d\mathbf{t_s} \\
    \frac{\delta \mathbf{u_s}}{\delta t} &= \frac{1}{n_s m_s} \frac{\delta \mathbf{M_s}}{\delta t} = \frac{1}{n_s m_s} \int \left(\frac{\delta f_s}{\delta t}m_s\mathbf{v_s}\right)d\mathbf{t_s} \\
    \frac{\delta T_s^\parallel}{\delta t} &= \frac{m_s}{k_b n_s}\int \left(\frac{\delta f_s}{\delta t}\mathbf{t_s}^{\parallel 2}\right)d\mathbf{t_s} \\
    \frac{\delta T_s^\perp}{\delta t} &= \frac{m_s}{2 k_b n_s}\int \left(\frac{\delta f_s}{\delta t}\mathbf{t_s}^{\perp 2}\right)d\mathbf{t_s} \\
    \frac{\delta \mathbf{q_s^\parallel}}{\delta t} &= m_s\int \left(\frac{\delta f_s}{\delta t}\mathbf{t_s}^{\parallel 2}\mathbf{t_s} \right)d\mathbf{t_s} \\
    \frac{\delta \mathbf{q_s^\perp}}{\delta t} &= \frac{m_s}{2}\int \left(\frac{\delta f_s}{\delta t}\mathbf{t_s}^{\perp 2}\mathbf{t_s} \right)d\mathbf{t_s}
\end{align}
\end{subequations}

All collisions treated in ISAM can be described as elastic collisions, in the sense that the kinetic energy and momentum are conserved in a collision. Elastic collisions can be described by the Boltzmann collision integral \citep[][eq 2]{Demars1979}:
\begin{equation}
\label{eq:Boltzmann_col_int}
    \frac{\delta f_s}{\delta t} = \sum_{t\neq s}\int g_{st}\sigma_{st}(g_{st},\theta)\left[f_s' f_t' - f_s f_t\right]d\mathbf{v_t}d\Omega
\end{equation}
where $g_{st}$ is the relative velocity between the two colliding particles $t$ and $s$, $\sigma_{st}(g_{st},\theta)$ the differential cross-section, $\Omega$ the solid angle in the frame moving with particle $s$, $\theta$ the scattering angle and the prime subscripts indicate quantities evaluated after collision.\\

The derivation beyond this step can be involved depending on the interaction type and the approximation chosen for the VDF of the two colliding particles. In most high-order approaches the Boltzmann collision integral must be approximated. In addition to the underlying assumptions associated with the 16-moment approach (see section \ref{sec:ISAM_origin}) of which small heat fluxes and stresses, it is further assumed that the velocity differences between the colliding particles are small compared to the averaged thermal velocity:
\begin{equation}
\label{eq:ISAM_col_cond1}
    \frac{|u_s-u_t|}{(c_s+c_t)/2}\ll 1
\end{equation}
where $c_s=\sqrt{k_b T_s/m_s}$ is the total thermal (or sound) speed of specie $s$ where $T_s$ is the total temperature which can be expressed as $T_s=(T_s^\parallel+2T_s^\perp)/3$. \\

These conditions correspond to the "semilinear" approximation of \citet{Burgers1969}. The first condition is a usual minimum requirement to permit sufficient simplifications of the Boltzmann collision integral. The second and third condition is enforced by the choice of the VDF $f_s=f_s^{BM}(1+\chi_s)$ which is by definition a small deviation to the pure bi-Maxwellian. The direct implication of $\chi_s\ll 1$ is that we can omit the cross-product terms $\chi_s \chi_t$ appearing in $f_s f_t$ of eq \ref{eq:Boltzmann_col_int}. 

We systematically checked these three conditions for the results presented in this thesis. In the case of significant differences in the drift velocity, a correction term (in $(u_t-u_s)^2$) can be added in the collisions terms of the energy equations that accounts for frictional heating \citep{Blelly1993}. \\

The collision terms solved in ISAM have been derived from the 16-moment formulas given by \citet[][]{Demars1979} which describe all interactions in a general formulation:
\begin{subequations}
\begin{flalign}
\label{eq:Demars1979_dudt}
&\begin{aligned}
    \frac{\delta u_s}{\delta t}=\frac{1}{n_s m_s}\sum_t \frac{3}{4\pi\Gamma[(5-n)/2]}n_s m_s \nu_{st}\Biggl[ & 2 I_{002}\frac{\sigma_{st}^\perp}{\sigma_{st}^\parallel}(u_t-u_s) \\
    &+\frac{2}{k_b \sigma_{st}^\parallel}(2 I_{202}-I_{002})\left(\frac{q_t^\perp}{n_t m_t}-\frac{q_s^\perp}{n_s m_s} \right) \\
    &+ \frac{\sigma_{st}^\perp}{k_b \sigma_{st}^{\parallel 2}}\left(\frac{2}{3}\frac{\sigma_{st}^\perp}{\sigma_{st}^\parallel} I_{004}-I_{002}\right)\left(\frac{q_t^\parallel}{n_t m_t}-\frac{q_s^\parallel}{n_s m_s} \right)\Biggr]
\end{aligned}\\
\label{eq:Demars1979_dTpardt}
&\begin{aligned}
     \frac{\delta T_s^\parallel}{\delta t}=\sum_t \frac{3}{2\pi\Gamma[(5-n)/2]}\frac{m_s}{m_s+m_t}\nu_{st}\Biggl[ & \frac{2\pi}{3}m_t(u_t-u_s)^2 \\ 
     &+ \frac{A_2(a)}{A_1(a)} m_t\sigma_{st}^\perp(I_{200}-I_{002}) + 2\frac{\sigma_{st}^\perp}{\sigma_{st}^\parallel}I_{002}(T_t^\parallel-T_s^\parallel) \Biggr]
\end{aligned}\\
\label{eq:Demars1979_dTperpdt}
&\begin{aligned}
     \frac{\delta T_s^\perp}{\delta t}=\sum_t \frac{3}{4\pi\Gamma[(5-n)/2]}\frac{m_s}{m_s+m_t}\nu_{st}\Biggl[ & \frac{4\pi}{3}m_t(u_t-u_s)^2 \\ 
     &+ \frac{A_2(a)}{A_1(a)} m_t\sigma_{st}^\perp(I_{002}-I_{200}) + 4I_{200}(T_t^\perp-T_s^\perp) \Biggr]
\end{aligned}\\
\label{eq:Demars1979_dqpardt}
&\begin{aligned}
    \frac{\delta q_s^\parallel}{\delta t}=\sum_t \frac{3}{2\pi\Gamma[(5-n)/2]}\nu_{st}\Biggl[ & k_b n_s m_s \sigma_{st}^\perp R_{st}^{(2)}(u_t-u_s) \\
    & + \frac{\sigma_{st}^\perp}{\sigma_{st}^\parallel}\left(- R_{st}^{(4)}q_s^\perp - R_{st}^{(8)}q_s^\parallel + \frac{n_s m_s}{n_t m_t}\left(R_{st}^{(6)}q_t^\perp + R_{st}^{(10)}q_t^\parallel\right) \right) 
&\end{aligned}\\
\label{eq:Demars1979_dqperpdt}
&\begin{aligned}
    \frac{\delta q_s^\perp}{\delta t}=\sum_t \frac{3}{2\pi\Gamma[(5-n)/2]}\nu_{st}\Biggl[ & k_b n_s m_s \sigma_{st}^\perp S_{st}^{(2)}(u_t-u_s) \\
    & + \frac{\sigma_{st}^\perp}{\sigma_{st}^\parallel}\left(- S_{st}^{(4)}q_s^\perp - S_{st}^{(8)}q_s^\parallel + \frac{n_s m_s}{n_t m_t}\left(S_{st}^{(6)}q_t^\perp + S_{st}^{(10)}q_t^\parallel\right) \right) 
\end{aligned}
\end{flalign}
\end{subequations}
where $\sigma_{st}^{\parallel, \perp}=(m_t T_s^{\parallel, \perp}+m_s T_t^{\parallel, \perp})/(m_s m_t)$. The terms $I_{202}$, $I_{202}$, $I_{202}$, $I_{202}$, $R_{st}^{(2)}$, $R_{st}^{(4)}$, $R_{st}^{(6)}$, $R_{st}^{(8)}$, $R_{st}^{(10)}$, $S_{st}^{(2)}$, $S_{st}^{(4)}$, $S_{st}^{(6)}$, $S_{st}^{(8)}$, $S_{st}^{(10)}$ are functions of the species temperature and mass, and of the type of the interaction. These terms together with the gamma function $\Gamma[(5-n)/2]$ (where $n=4/(a-1)-1$) are all given in \citet{Demars1979} for a general inverse-power force ($\propto r^{-a}$). The $A_2(a)$, $A_1(a)$ quantities are pure numbers given for several inverse-power laws in \citet{Chapman1970}. \\

The collision terms of \citet{Demars1979} repeated above (eq \ref{eq:Demars1979_dudt}\textendash\ref{eq:Demars1979_dqperpdt}) are valid for any inverse-power interaction and then can be applied to Coulomb, non-resonant ion-neutral (Maxwell) and neutral-neutral interactions. They also derived a separate set of collision terms for resonant ion-neutral charge-exchange reactions. However in ISAM we resort to alternative approaches to treat neutral-neutral and resonant ion-neutral charge-exchange reactions which we discuss further in section \ref{subsec:neutral_neutral_col} and \ref{subsec:ion_neutral_col}. \\

The actual collision terms solved in ISAM are a simplified version of eq \ref{eq:Demars1979_dudt}\textendash\ref{eq:Demars1979_dqperpdt} where we further assume a small temperature difference between the colliding species, that falls in the "linear" approximation of \citet{Burgers1969}. The obtained collision terms are then comparable to the 13-moment set of \citet[][eq 50a to 50e]{Schunk1977} but with a distinction on the transport of parallel $q_s^\parallel$ and perpendicular $q_s^\perp$ energy. The collision terms as actually solved in ISAM for all interactions are given below:
\begin{subequations}
\begin{flalign}
\label{eq:ISAM_col_u}
    & \frac{\delta u_s}{\delta t} = \frac{1}{n_s m_s}\left[\underbrace{\sum_{t \neq s} n_s m_s\nu_{st}(u_t-u_s)}_\textrm{Friction force} + \underbrace{\sum_{t \neq s} \nu_{st}\frac{z_{st}\mu_{st}}{k_b T_{st}}\left(\frac{q_s^{\parallel} + 2q_s^{\perp}}{2}-\frac{q_t^{\parallel} + 2q_t^{\perp}}{2}\frac{n_s m_s}{n_t m_t}\right)}_\textrm{Thermal force}\right] \\
&\begin{aligned}
    \frac{\delta T^{\parallel}_s}{\delta t} &= \sum_{t \neq s} \frac{\nu_{st}}{m_s+m_t}\left[2m_s(T^{\parallel}_t-T^{\parallel}_s) + \frac{4}{5}m_t(T^{\perp}_s-T^{\parallel}_s) + \frac{4}{5}m_s(T^{\perp}_t-T^{\parallel}_t)\right] \\
    &+ \frac{4}{5}\nu_{ss}(T^{\perp}_s-T^{\parallel}_s)
\end{aligned} \\
&\begin{aligned}
    \frac{\delta T^{\perp}_s}{\delta t} &= \sum_{t \neq s} \frac{\nu_{st}}{m_s+m_t}\left[2m_s(T^{\perp}_t-T^{\perp}_s) + \frac{2}{5}m_t(T^{\parallel}_s-T^{\perp}_s) + \frac{2}{5}m_s(T^{\parallel}_t-T^{\perp}_t)\right] \\
    &+ \frac{2}{5}\nu_{ss}(T^{\parallel}_s-T^{\perp}_s)
\end{aligned} \\
&\begin{aligned}
    \frac{\delta q_s^{\parallel}}{\delta t} &= T_s k_b n_s\sum_{t \neq s} \nu_{st}A^{*0p}_{st}(u_t-u_s)\\
    &+ n_s\sum_t \nu_{st}\left(D^{*1pt}_{st}\frac{q_s^{\perp}}{n_s}+D^{*4pt}_{st}\frac{q_t^{\perp}}{n_t}+D^{*1pp}_{st}\frac{q_s^{\parallel}}{n_s}+D^{*4pp}_{st}\frac{q_t^{\parallel}}{n_t}\right)\\
    &- 3n_s k_b T_s^{\parallel}\frac{\delta U_s}{\delta t}\\
\end{aligned} \\
&\begin{aligned}
    \frac{\delta q_s^{\perp}}{\delta t} &= T_s k_b n_s\sum_{t \neq s} \nu_{st}A^{*0t}_{st}(u_t-u_s)\\
    &+ n_s\sum_t \nu_{st}\left(D^{*1tt}_{st}\frac{q_s^{\perp}}{n_s}+D^{*4tt}_{st}\frac{q_t^{\perp}}{n_t}+D^{*1tp}_{st}\frac{q_s^{\parallel}}{n_s}+D^{*4tp}_{st}\frac{q_t^{\parallel}}{n_t}\right)\\
    &- n_s k_b T_s^{\parallel}\frac{\delta U_s}{\delta t}\\
\end{aligned} \\
\end{flalign}
\end{subequations}
where $\mu_{st}=(m_s m_t)/(m_s + m_t)$ is the reduced mass and $T_{st}=(m_t T_s+m_s T_t)/(m_s + m_t)$ the temperature expressed in the center-of-mass of the two colliding particles. The coefficients $z_{st}$, $A^{*0p}_{st}$, $A^{*0t}_{st}$, $D^{*1pt}_{st}$, $D^{*4pt}_{st}$, $D^{*1pp}_{st}$, $D^{*4pp}_{st}$, $D^{*1tt}_{st}$, $D^{*4tt}_{st}$, $D^{*1tp}_{st}$ and $D^{*4tp}_{st}$ only depend on the type of the interaction and on the mass of the two colliding particles, they are given in appendix \ref{sec:mass_ratios}. \\

For ions, we also consider the contribution from the electrons $\delta u_e/\delta t$ on the ion momentum transport equation via the electrostatic polarization field (see eq \ref{eq:ISAM_Fpol}). This implies a modification of the total friction force and thermal force undergone by the ion $i$ of charge state $Z_i$:
\begin{align}
\label{eq:ISAM_col_ui}
    \frac{\delta u_i}{\delta t}+\frac{m_e Z_i}{m_i}\frac{\delta u_e}{\delta t} = \frac{1}{n_i m_i} & \Biggl[ \sum_{t \neq i} n_i m_i\nu_{it}(u_t-u_i) +\frac{Z_i n_i}{n_e}\sum_{t \neq e} n_e m_e\nu_{et}(u_t-u_e) \\
    &+ \sum_{t \neq i} \nu_{it}\frac{z_{it}\mu_{it}}{k_b T_{it}}\left(\frac{q_i^{\parallel} + 2q_i^{\perp}}{2}-\frac{q_t^{\parallel} + 2q_t^{\perp}}{2}\frac{n_i m_i}{n_t m_t}\right) \notag\\
    &+ \frac{Z_i n_i}{n_e} \sum_{t \neq e} \nu_{et}\frac{z_{et}\mu_{et}}{k_b T_{et}}\left(\frac{q_e^{\parallel} + 2q_e^{\perp}}{2}-\frac{q_{t}^{\parallel} + 2q_t^{\perp}}{2}\frac{n_e m_e}{n_t m_t}\right)\Biggr] \notag
\end{align}
where $n_e$ and $u_e$ are given by eq \ref{eq:ISAM_ne} and \ref{eq:ISAM_ue}. \\

The remaining unknowns in these equations are the collision frequencies $\nu_{st}$ which are discussed in the next paragraphs for each of the interactions considered in ISAM. Since all collisions are treated as elastic where the kinetic energy and momentum are conserved through the collision, the collision frequencies for the back and forth interactions are then linked by the relation $m_s n_s \nu_{st}=m_t n_t \nu_{ts}$. \\ 

Elastic interactions are often described as collisions between hard-sphere particles where the diameter of the colliding particles is set as the characteristic range of the interaction. This approach is for instance convenient to derive simpler expressions for the collision frequencies. We will see in the next paragraphs that this approach is appropriated for short-range interactions such as neutral-neutral or ion-neutral interactions but not for the long-range Coulomb interaction.

\subsection{Neutral-neutral collisions}
\label{subsec:neutral_neutral_col}

The collisions terms for Maxwell type interactions are used in ISAM rather than the hard-sphere collision terms derived by \citet{Demars1979}. The reason behind this choice is that Maxwell type interactions eliminate the dependency of the collision terms on the energy while remaining reasonably representative of a short-range interaction. As a consequence the mass ratios given in Appendix \ref{sec:mass_ratios} and appearing in the heat flow collision terms will be the same as those for ion-neutral Maxwell type interactions. We plan to improve our treatment of neutral-neutral collisions by deriving the mass ratios for a more realistic potential, the Lennard-Jones potential (12-6) discussed just below. \\

The collision terms having been discussed, we now explain our choice for the neutral-neutral collision frequency. In ISAM we exploit a more realistic collision frequency than the one typically assumed for hard-sphere interactions  \citep[e.g.][eq C4]{Schunk1977}, where in the former neutrals interact with each other via a Lennard-Jones (12-6) repulsive-attractive potential. The attraction potential results from the creation of induced dipoles as in van der Waals interactions whereas the short-range $r^{-12}$ repulsion arises when the two neutrals get too close to each other so that the two electron clouds do not overlap. The associated Lennard-Jones (12-6) potential can be expressed as:
\begin{equation}
    \Phi=4\epsilon\left[\left(\frac{\sigma}{r} \right)^{12} -\left(\frac{\sigma}{r} \right)^6 \right]
\end{equation}
where $\sigma=(\sigma_s + \sigma_t)/2$ is the closest possible distance between the two particles and $\epsilon=\sqrt{\epsilon_s \epsilon_t}$ is the characteristic energy. These parameters have been fitted for many atoms by \citet{Rappe1992}. The corresponding neutral-neutral collision frequency is \citep[][eq 2.29]{Boqueho2005Thesis}:
\begin{equation}
    \nu_{st}=0.579\frac{8\sqrt{\pi}}{3}\frac{m_t}{m_s+m_t}n_t\left(\frac{K_{st}}{\mu_{st}} \right)^{1/3}\left(\frac{2 k_b T_{st}}{\mu_{st}} \right)^{1/6}\quad \rm{(in\ s^{-1})}
\end{equation}
where $K_{st}=20\epsilon\sigma^6$. As an example for atomic Hydrogen (HI) and Helium (HeI) $\epsilon$ and $\sigma$ equal $\epsilon_{HI}\simeq22.14\ K$, $\epsilon_{HeI}\simeq28.18\ K$, $\sigma_{HI}=2.57$ \si{\angstrom} and $\sigma_{HeI}=2.10$ \si{\angstrom}.

\subsection{Collisions between charged particles: Coulomb collisions}
\label{subsec:coulomb_col}
Charged particles (ion-ion and ion-electron) interact through the well-known long-range Coulomb interaction:
\begin{equation}
    F_{st}=\frac{Z_S Z_t e^2}{4\pi\epsilon_0 r^2}
\end{equation}
with $\epsilon_0$ the vacuum permittivity. The corresponding collision frequency is \citep[][eq 9.114]{Banks1973}:
\begin{equation}
    \nu_{st}=\frac{4}{3}\sqrt{2\pi}\frac{m_t}{m_s+m_t}\frac{n_t}{(k_b T_{st})^{3/2}}\frac{Z_s^2 Z_t^2 e^4}{\sqrt{\mu_{st}}} ln(\Lambda) \quad \rm{(in\ s^{-1})}
\end{equation}
where the Coulomb logarithm is taken from \citet[][eq 5.5]{Schunk1975} \citep[see also][eq 22.8]{Burgers1969}:
\begin{equation}
    \Lambda = 12\pi n_e\left(\frac{\epsilon_0 k_b T_e}{e^2 n_e}\right)^{3/2}
\end{equation}
In these formula the elementary charge is expressed in Coulomb unit $e=1.60217662\times10^{-19}\ C$ and all other quantities in SI units. The Coulomb logarithm represents the screening of the Coulomb interaction range beyond the Debye sphere by surrounding electrons, that translates mathematically into a finite expression of the collision frequency.

\subsection{Ion-neutral collisions}
\label{subsec:ion_neutral_col}

Ion-neutral interactions can be either resonant or non-resonant. The resonant interaction occurs when an ion collides with its parent neutral or accidentally as in the case of $H^+\ +\ O \rightarrow \ H\ +\ O^+$ whereas the non-resonant one happens for collisions between unlike ions and neutrals. \\

The non-resonant ion-neutral interaction is well-defined by elastic collisions between Maxwellian-type molecules. In this case the interaction can be described by an induced dipole attraction \citep[][eq 9.69]{Banks1973}:
\begin{equation}
    \Phi=-\frac{\alpha e^2}{2 r^4}
\end{equation}
where $\alpha$ is the neutral atom polarizability. In reality the attraction force becomes countered by a short-range quantum mechanical repulsion above $300\ K$ \citep[see][section 9.6]{Banks1973}. But because there is only a few data available for ion-neutral interactions at high temperature, the short-range effects are omitted in the collision terms. The resulting non-resonant (Maxwell molecule) ion-neutral collision frequency is then \citep[][eq 9.73]{Banks1973}:
\begin{equation}
    \nu_{st}=2.6\times 10^{-9}\frac{m_t}{m_s+m_t}n_t\sqrt{\frac{\alpha_t}{\mu_{st}}}\quad \text{(in }s^{-1}\text{)}
\end{equation}
where $s$ and $t$ here stand for ion and neutral respectively. The neutral atom polarizabilities $\alpha_t$ can be found in \citet[][Table 9.10, p219]{Banks1973} or \citet[][Table 1., p267]{Marsch1995}. The latter have been used in another study on the FIP effect by \citet{Bo2013}. For instance we take $\alpha_H=0.667$ \si{\angstrom}$^3$ and $\alpha_{He}=0.205$ \si{\angstrom}$^3$ for neutral Hydrogen and Helium respectively. A great advantage of Maxwell molecule-type interactions is that the dependency on the temperature is eliminated, hence the collision frequency is directly determined once the neutral identity is known. For temperatures greater than $1000\ K$, the non-resonant polarization interaction becomes negligible and can be replaced by the resonant charge-exchange reaction \citep[see][Figure 9.26]{Banks1973}. Therefore, in ISAM we replace the non-resonant collision frequency by the resonant charge-exchange one whenever applicable. \\

\citet{Demars1979} derived a separate set of collision terms for the resonant ion-neutral charge-exchange reactions which are not yet implemented in ISAM. As a consequence ISAM solves the same collision terms for both non-resonant and resonant ion-neutral interactions but consider different collision frequencies. We claim that by simply changing the collision frequency, the overall dynamic of the system should still be realistic. We note however that the thermal force as derived by \citet{Demars1979} for resonant charge-exchange interaction is no longer zero which might affect our results on the FIP effect. As for now, we let this point for future improvements of ISAM. We take the collision frequencies from \citet[][Table 5., p823]{Schunk1980} for the resonant charge-exchange interaction between alike ions and neutrals, and collision frequencies from \citet[][Table IIIA-B., p445-446]{Arnaud1985} for the accidental charge-exchange interactions between unlike ions and neutrals. In both cases we have a dependency of the collision frequency on the ion-neutral reduced temperature $T_{st}$ and on the neutral density. \\

\citet{Kingdon1996} provide an extended dataset with an end-to-end formalism to get collision frequencies between neutral Hydrogen and all ions of charge $Z_i=1-4$ for the first 30 elements. We will likely exploit their fits in future version of ISAM. However, for the dominant resonant interactions involving Hydrogen and Helium we might rely for future ISAM version, on the recent work of \citet{Vranjes2013} who exploited the most accurate (at least at the date of their publication) dataset from \citet{international1999iaea}. We further comment that recent modeling of ion-neutral interactions \citep[e.g.][]{Martinez2020} still exploit the collision frequencies from \citet{Vranjes2013}. \\

For electron-neutral interactions, we assume a two-body elastic interaction where we fetch the collision frequencies from \citet[][Table 3., p822]{Schunk1980}. They are dependent on the neutral density and electron temperature.

\section{Heating models}
\label{sec:ISAM_heating}
Depending on the numerical setup and application, different heating terms and models will be adopted and combined to address heating processes at different heights in the solar atmosphere from the chromosphere, through the transition region and into the corona. For certain applications one could perform cross combinations of different heating models whereby one model is used to specify the total amount of energy given to the plasma, and another that specifies how this heat is distributed among the different species and when relevant, in the directions parallel and perpendicular to the magnetic field.

As expected, the plasma remains collision-dominated even at coronal heights in the simulations of small-scale loops that I present in this thesis (see chapter \ref{cha:ISAM_results}). Therefore the choice of how energy is deposited relative to the magnetic field and among the species will not be of critical importance and should not affect our results. However, future studies on loops that extend much further up in the corona should consider these aspects.

\subsection{Ad-hoc heating prescription}
\label{subsec:ISAM_heating_adhoc}

Ad-hoc heating terms have been extensively used in ISAM because their simple form offers great flexibility by avoiding additional coupling in the transport equations. One successful form of such ad-hoc heating terms supposes a strong dependence on the input magnetic field geometry, a common formulation (inspired from e.g. \citet[][]{Withbroe1988}) involves for instance the variation of the cross sectionnal area $A(c)\propto 1/B(c)$ of the magnetic flux tube along the curvilinear coordinate $c$:
\begin{equation}
\label{eq:ISAM_Qh}
    Q_h=\frac{F_\odot}{H_f}\left(\frac{A_0}{A}\right)\exp \left[ -\frac{c-R_\odot}{H_f}\right]
\end{equation}
where $F_\odot$ is the heating flux at the base and $H_f$ the heating-scale height. A similar formulation as in the MULTI-VP model supposes a dependence of $F_\odot$ and $H_f$ on the basal magnetic field and expansion factor \citep{PintoRouillard2017}. Both formulations are implemented in ISAM but the results presented in this thesis have mostly made use of the former version where $F_\odot$ and $H_f$ are fixed as constants. Several of the results presented in this thesis use two combinations of this ad-hoc law, one dedicated to the high chromosphere and transition region, and another to the coronal part of the loop. \\

The distribution of the total heating flux among the species and the parallel/perpendicular direction can be decided at a second stage. In most implementations where I used the above ad-hoc function (\ref{eq:ISAM_Qh}), I distributed all the heating on protons and in the perpendicular direction only $Q_{h,p}^\perp=Q_h$. The motivation behind this choice is that Hydrogen is mostly ionized in regions where I apply the ad-hoc heating. The choice of the direction along which the heat is deposited depends on the physics assumed for the heating process which is actually highly debated in the community. For instance, ion heating by particle resonance with ion-cyclotron waves is expected to be strong in the direction perpendicular to the magnetic field. These specific heating mechanisms that occur at the microscopic kinetic scales are not addressed in a comprehensive manner in this thesis but could be an interesting future application of the model. In these first applications, heating is applied systematically in the perpendicular direction supposing that wave-particle resonance interaction is mediated by ion-cyclotron waves.

\subsection{Heating based on the dissipation of Alfvén waves following the WKB approximation}
\label{subsec:ISAM_Aw}

Coronal heating is still largely debated in the scientific community, but one mechanism that has been addressed extensively is heating via dissipation of transverse Alfvén waves. Many modelers have used this form of heating during the past two decades \citep[see e.g.][]{LieSvendsen2001,Chandran2011,Reville2020a,Reville2020b,Reville2022} of which the WindPredict-AW MHD model presented in section \ref{subsec:WindPredict}. In this section, we start by introducing the basic equations for the transport of transverse Alfvén waves in the Wentzel-Krimers-Brillouin (WKB) approximation. Then we discuss the damping of Alfvén waves through a non-linear turbulence cascade in section \ref{subsubsec:ISAM_Aw_Qw}. We finally present in section \ref{subsubsec:ISAM_Aw_heating} some processes that can convert Alfvén-wave energy into an effective energy that heats up the plasma.

\subsubsection{Alfvén-wave propagation in the WKB approach}
\label{subsubsec:ISAM_Aw_WKB}

Let us start from the ideal single-fluid MHD equations:
\begin{subequations}
\begin{align}
    \frac{\partial}{\partial t}\rho + \nabla \cdot (\rho \mathbf{u}) &=0 \\
    \rho\frac{\partial}{\partial t}\mathbf{u} + \rho \mathbf{u}\cdot \nabla \mathbf{u} &= -\nabla p_{th}+\mathbf{j}\times \mathbf{B} \\
    \frac{\partial}{\partial t}\mathbf{B} &= \nabla \times (\mathbf{u} \times \mathbf{B}) \\
    \nabla \times \mathbf{B} &= \mu_0 \mathbf{j} \\
    \nabla \cdot \mathbf{B} &= 0
\end{align}
\end{subequations}
where $p_{th}=n k_b T$ is the plasma thermal pressure. With some algebra and integrating the divergence free condition of the magnetic field, these equations can be reduced to:
\begin{subequations}
\begin{align}
    \frac{\partial}{\partial t}\rho + \nabla \cdot (\rho \mathbf{u}) &=0\\
    \rho\frac{\partial}{\partial t}\mathbf{u} + \rho \mathbf{u}\cdot \nabla \mathbf{u} &= -\nabla P + \frac{\mathbf{B}\cdot \nabla \mathbf{B}}{\mu_0} \label{eq:MHD_u}\\
    \frac{\partial}{\partial t}\mathbf{B} + \mathbf{u}\cdot \nabla \mathbf{B} - \mathbf{B}\cdot \nabla \mathbf{u} + (\nabla \cdot \mathbf{u})\mathbf{B} &= \mathbf{0} \label{eq:MHD_B}
\end{align}
\end{subequations}
with $P=p_{th}+\mathbf{B}^2/(2\mu_0)$ the total pressure.\\

A convenient formulation of the ideal MHD equations can be derived by defining the Elssasër variables \citep{Elsasser1950}:
\begin{equation}
    \mathbf{Z}^\pm = \mathbf{u} \mp \frac{\mathbf{B}}{\sqrt{\mu_0 \rho}}
\end{equation}
where the second right-hand side term is the magnetic field expressed in velocity unit and that is commonly called the Alfvén speed $\mathbf{v_A}=\mathbf{B}/\sqrt{\mu_0 \rho}$. By adding (or subtracting) the induction equation (eq \ref{eq:MHD_B}) divided by $\sqrt{\mu_0\rho}$ to the momentum equation (eq \ref{eq:MHD_u}) one can derive a simple MHD formula for the Elsassër variables \citep{Magyar2019}:
\begin{subequations}
\label{eq:MHD_z}
\begin{align}
    \frac{\partial}{\partial t}\mathbf{Z}^\pm + \mathbf{Z}^\mp \cdot \nabla \mathbf{Z}^\pm &= - \nabla P \\
    \nabla \cdot \mathbf{Z}^\pm &= 0
\end{align}
\end{subequations}

We now consider transverse perturbations of arbitrary amplitudes over the mean flow, that can be expressed in terms of Elssasër variables as follows:
\begin{equation}
    \mathbf{Z}^\pm = z_0^\pm \mathbf{e^\parallel} + z^\pm \mathbf{e^\perp} = (u_0 \mp v_{A0})\mathbf{e^\parallel} +\left(\delta u \mp \frac{\delta B}{\sqrt{\mu_0 \rho_0}}\right)\mathbf{e^\perp}
\end{equation}
where $v_{A0}=B_0/\sqrt{\mu_0 \rho_0}$ is the Alfvén speed of the mean flow, and $\mathbf{e^{\parallel,\perp}}$ denote the unit vectors tangent and perpendicular to the mean magnetic field. By injecting this decomposition into equation \ref{eq:MHD_z} and keeping only the first order terms, one can obtain the transport equation for the perturbed Elsassër variables \citep[][eq 32]{Cranmer2005}:
\begin{subequations}
\begin{align}
\label{eq:Cranmer2005_eq32}
    \frac{\partial}{\partial t}z^\pm + (u_0 \pm v_{A0})\frac{\partial}{\partial c}z^\pm = (u_0 \mp v_{A0})\left( \frac{z^\pm}{4}\frac{\partial}{\partial c}\ln \rho+\frac{z^\mp}{2}\frac{\partial}{\partial c}\ln v_A\right)
\end{align}
\end{subequations}
with $c$ the curvilinear abscissa along the magnetic field line.

In this framework the Elsassër variables represent two counter-propagating Alfvén waves, where $z^+$ and $z^-$ correspond to the waves that propagate forward and backward respectively with respect to the mean magnetic field. Linear interactions between the inward and outward propagating waves are accounted for in the $z^\mp$ term on the right-hand side of equation \ref{eq:Cranmer2005_eq32}, that represents the reflection of Alfvén waves in regions with strong homogeneities. \\

Similarly to previous works, I resort to an integrated approach where the monochromatic equations for $z^\pm$ are summed up over the full spectrum. One can then define the frequency-averaged wave energy density as $\epsilon^\pm = \rho \left<z^\pm\right>^2/4$. 

In ISAM we follow the Wentzel\textendash Kramers\textendash Brillouin (WKB) approximation whereby
the mean flow is supposed homogeneous at the scale of the fluctuations:
\begin{equation}
    \frac{\lambda}{L}\ll 1
\end{equation}
where $\lambda$ is the Alfvén-wave wavelength and $L$ the characteristic length scale of the mean flow (e.g. the gravity scale height). The WKB approximation is then equivalent to having a null linear reflection term in equation \ref{eq:Cranmer2005_eq32}.

In the WKB approximation, the transport equation for the Alfvén-wave energy can then be expressed as \citep[see e.g.][]{Alazraki1971,Hollweg1974,Tu1995}: 
\begin{equation}
    \frac{\partial}{\partial t}\epsilon^\pm + \nabla_\parallel \cdot \left( (u_0\pm v_{A0})\epsilon\pm \right) = -\frac{\epsilon^\pm}{2}\nabla_\parallel \cdot u_0 - Q_h^\pm
    \label{eq:Ew}
\end{equation}
with an extra dissipation (sink) term $Q_h^\pm$ that accounts for the wave energy lost to heat up the plasma, and where the divergence operator is defined as $\nabla_\parallel \cdot (*)=1/A \partial (A*)/\partial c$. This equation is solved in ISAM in addition to the hydrodynamic transport equations introduced in section \ref{sec:ISAM_transport_eq}. \\

The framework described above should be adapted to multi-specie plasmas that is left for future improvements of ISAM. For simplicity I consider here that the Alfvén wave energy is primarily transported by the protons and hence I take $\rho_0=m_p n_p$ and $u_0=u_p$. \\

Similarly to the transport equations of the 16-moment set, I solve the propagation of the Alfvén-wave energy using the LCPFCT algorithm that is presented in section \ref{subsec:ISAM_lcpfct}. The actual quantity that is solved by LCPFCT is $\mathcal{E}=\epsilon/\rho$ which is normalized using the scaling parameters defined in Appendix \ref{sec:ISAM_normalization}.

\subsubsection{Alfvén-wave dissipation}
\label{subsubsec:ISAM_Aw_Qw}

Alfvén waves moving through an inhomogeneous plasma can be reflected in regions marked by significant changes in Alfvén speed such as the transition region and certain regions of the solar corona. This reflection will produce two populations of counter-propagating waves that are likely to interact \citep{Dmitruk2002} and feed a turbulence cascade where low-frequency waves are converted into higher frequency waves at smaller scales. This cascade is expected to increase the population of waves that can effectively interact with and heat charged particles. While the entire wave-wave and subsequent wave-particle interactions is highly complex, various proxies of different complexity have been derived to estimate the amount of Alfvén-wave energy that is given to the plasma through this turbulence cascade. We discuss several of these approaches in this section, which all assume the same transport equation for the mother Alfvén-wave given by eq \ref{eq:Ew}. \\

All these approaches are based on a phenomenological expression of the Alfvén-wave dissipation rate that depends only on the large-scale properties and not on a full description of the turbulent cascade. Since this turbulence cascade requires the interaction between two counter-propagating Alfvén waves, the dissipation rate $Q_h^\pm$ that appears in the WKB transport equation (eq \ref{eq:Ew}) is non-linear and reads \citep[][eq 57]{Dmitruk2002}:
\begin{equation}
\label{eq:Qw}
    Q_h^\pm=\frac{\epsilon^\pm}{2\mathcal{L}}|\left<z^\mp\right>|
\end{equation}
with $\mathcal{L}$ a transverse length scale or correlation length of the turbulence. The $\mathcal{L}$ length scale is commonly assumed to scale as the cross-sectional width of magnetic flux-tubes as they expand in the corona and hence $\mathcal{L}=\mathcal{L}_\odot\sqrt{A/A_\odot}$. \\

The characteristic length $\mathcal{L}_\odot$ is a free parameter that has to be adjusted to the magnetic topology of the simulated region and to the height of the inner boundary. For instance in the WindPredict-AW model (see section \ref{subsec:WindPredict}), $\mathcal{L}_\odot$ is usually set to match the typical size of supergranules in the low corona $\mathcal{L}_\odot=0.022 R_\odot \sqrt{B_\odot}\approx 15000\ \rm{km} \sqrt{B_\odot}$ with $B_\odot$ expressed in Gauss unit \citep{Verdini2007}, which is associated to a transverse velocity fluctuation at the base that is fixed at $36\ \rm{km.s^{-1}}$. For the case studies of small-scale loops presented in this thesis, we usually fix the length scale to be around the typical size of granules around $\approx 3500\ km$ together with an Alfvén-wave amplitude at the base of $\approx 20-30\ km/s$. \\

The total amount of the wave energy that is given to the plasma is $Q_h=Q_h^+ + Q_h^-$. As discussed above, dissipation of the wave energy in equation \ref{eq:Qw} is ensured if both the inward and outward propagating waves coexist. \\

For open solar wind solutions where only outward propagating waves are generated from the inner boundary, there is no means within the WKB framework to convert the outward into inward propagating waves because of the lack of reflection terms in equation \ref{eq:Ew}. This limitation is often bypassed by assuming a constant refection rate $\mathcal{R}$ where the reflected population is instantly dissipated:
\begin{equation}
\label{eq:WindPredict_Qw}
    Q_h^\pm =\frac{\epsilon^\pm}{2\mathcal{L}}\left(|\left<z^\mp\right>| + \mathcal{R}|\left<z^\pm\right>| \right) 
\end{equation}
The WindPredict-AW model (see section \ref{subsec:WindPredict}) and the heating model of \citet{LieSvendsen2001} essentially follow the same approach by taking $\mathcal{R}=0.1$ and $\mathcal{R}=1.0$ respectively. 

\citet{Chandran2011} suggest another approach where the counter-propagating waves result from reflections in regions with large-scale inhomogeneities such as in the transition region: 
\begin{equation}
    |z^-|=\frac{\mathcal{L}(u_0+v_{A0})}{v_{A0}}\left| \frac{\partial v_{A0}}{\partial c}\right|
\end{equation}
which inserted into equation \ref{eq:Qw} leads to a total dissipation rate:
\begin{equation}
    Q_h=c_d\frac{(u_0+v_{A0})}{v_{A0}}\left| \frac{\partial v_{A0}}{\partial c}\right| \epsilon^+
\end{equation}
where it is supposed that $z^- \ll z^+$ and hence the $\epsilon^-$ term is neglected in equation \ref{eq:Qw}. \\

This requirement on $|z^-|$ is alleviated in closed-field geometries such as coronal loops where Alfvén waves can more easily propagate back and forth on both sides of the loop, through reflection at the transition region for instance. Coronal loops are therefore much more propitious for initiating a turbulence cascade than in open solar wind solutions where the counter-propagating population is less prevalent. The validity of \citet{Chandran2011}'s model is therefore questionable in coronal loops because the counter-propagating wave is likely as much dominant as the mother wave population.

The WindPredict-AW model takes a zero reflection rate $\mathcal{R}=0$ in closed-field geometries, and hence equation \ref{eq:WindPredict_Qw} reduces exactly to equation \ref{eq:Qw}. I basically follow the same approach in ISAM.

\subsubsection{Plasma heating}
\label{subsubsec:ISAM_Aw_heating}

Once the turbulence cascade is initiated, low-frequency Alfvén waves progressively convert into higher frequency waves which at some point can interact with the charged particles that constitute the plasma.\\

At first approach, we can follow the same method as for the ad-hoc heating prescription introduced in section \ref{subsec:ISAM_heating_adhoc}, that is applying the heating along the perpendicular direction only. This approach supposes that the wave-particle interaction occurs primarily through ion-cyclotron resonance on protons.

A more complete approach consists in distributing the energy over the parallel and perpendicular directions according to the quasi-linear theory of ion-cyclotron resonance \citep[][eq 28-30]{LieSvendsen2001}: 
\begin{subequations}
\begin{align}
    Q_{h,p}^\parallel &= 2 Q_h \frac{F}{F+1}\\
    Q_{h,p}^\perp     &=   Q_h \frac{1}{F+1}\\
    F &= \frac{T_p^\perp-\eta T_p^\parallel}{\frac{m_p}{k_b}v_A^2+\eta\left(\frac{\eta+1}{\eta}T_p^\parallel-T_p^\perp\right)}
\end{align}
\end{subequations}
where $\eta=5/3$ corresponds to homogeneous Kolmogorov turbulence. \\

\citet{Chandran2011} suggests a more sophisticated approach by considering an anisotropic Alfvén-wave cascade that is not only transverse but also has a parallel component to the magnetic field. In their model perpendicular heating on protons is ensured by stochastic heating when the Alfvén-wave energy cascades to kinetic scales around the proton gyroradius. They also include a parallel contribution to the proton heating through a Landau damping of Alfvén waves, but that is very weak for the low coronal heights simulated in this thesis. Finally, the remaining Alfvén-wave energy that cascades to lower scales contributes at heating the electrons. For completeness we repeat here their equations adapted to our own notations:
\begin{subequations}
\begin{align}
    Q_{h,e} &= \frac{(1+\gamma_e t_c)Q_h}{1+\gamma_{tot} t_c}\\
    Q_{h,p}^\perp &= \frac{\gamma_s t_c Q_h}{1+\gamma_{tot} t_c}\\
    Q_{h,p}^\parallel &= \frac{\gamma_p t_c Q_h}{1+\gamma_{tot} t_c}
\end{align}
\end{subequations}
where
\begin{subequations}
\begin{align}
    \gamma_s &= 0.18\mathcal{E}_p \Omega_p \exp \left(-\frac{c_2}{\mathcal{E}_p} \right)\\
    \gamma_e &= \frac{1}{t_c}0.01\left(\frac{T_e}{T_p\beta_p}\right)^{1/2}\left[\frac{1+0.17\beta_p^{1.3}}{1+(2800\beta_e)^{-1.25}} \right] \\
    \gamma_p &= \frac{1}{t_c}0.08\left(\frac{T_e}{T_p}\right)^{1/4}\beta_p^{0.7}\exp \left( -\frac{1.3}{\beta_p}\right)
\end{align}
\end{subequations}
where $\beta_{p,e}=8\pi n k_b T_{p,e}/B_\odot^2$, $t_c=\rho\delta v_p^2/Q_h$, $\delta v_p=|z^-|/2(\rho_p/\mathcal{L})^{1/4}$,  $\mathcal{E}_p=\delta v_p/c_p^\perp$ and $c_p^\perp=\sqrt{2k_b T_p^\perp/m_p}$ the proton thermal perpendicular speed. 

This distribution involves an additional free parameter $c_2$ that we set to $0.15$, that is about the original value used in \citet{Chandran2011}. This parameter controls the amount of wave energy that is transferred to the protons. As noted earlier, this parameter will not be critical in coronal loops where collisions efficiently redistribute the thermal energy over the different species.

\subsection{The Shell-Atm turbulence model}
\label{subsec:ISAM_Aw_ShellATM}

To properly resolve Alfvén wave reflection and dissipation near the transition region, one needs to solve the non-WKB equations at the price of higher complexity and computational resources. To treat this more complicated problem, I have made use of a separate existing code called Shell-Atm to properly account for Alfvén waves reflection and dissipation. \\

The Shell-Atm turbulence model takes off the limitations of the WKB approximation by fully resolving the non-WKB coupled transport and dissipation of both inward and outward propagating transverse Alfvén waves. This is achieved by solving the transport equations not only over the spatial coordinates and time, but also over the spectral dimension. The consideration of the spectral dimension therefore allows for a more realistic description of the turbulence cascade which leads up to dissipation. Shell-Atm has already been applied to the case of coronal loops \citep{Buchlin2007} and open solar winds \citep{Verdini2009,Verdini2019}. \citet{Reville2021b} investigated the FIP fractionation generated by the ponderomotive acceleration exploiting the detailed description of the turbulence cascade offered by Shell-Atm. \\

Shell-Atm solves for the incompressible transport of Alfvénic fluctuations from a given background profile of the plasma bulk velocity, temperature and density \citep{Buchlin2007,Verdini2009,Verdini2019}. The usual Elsassër variables (as defined in section \ref{subsubsec:ISAM_Aw_WKB}) are transported via the following equation \citep[][eq 1]{Verdini2009}:
\begin{equation}
\label{eq:ShellATM}
\begin{split}
    \frac{\partial}{\partial t}z^\pm +(u\pm v_A)\frac{\partial}{\partial r}z^\pm=&-\frac{1}{2}(u\mp v_A)\left(\frac{d}{dr}\log(v_A) +\frac{d}{dr}\log(A)\right)z^\pm \\
    &+ \frac{1}{2}(u\mp v_A)\left(\frac{d}{dr}\log(v_A)\right)z^\mp\\
    &-k^2(\nu^+ z^\pm+\nu^- z^\mp) + ik(T^\pm)^*
\end{split}
\end{equation}
where $\nu^+$ and $\nu^-$ include both effects of kinematic viscosity and magnetic resistivity, and with $k=k^\perp$ the wave number of the transverse fluctuations. Non-linear interactions between the inward and outward wave populations are accounted for in the $T^\pm$ term which contains all cross products $z^+ z^-$. 

In practice, equation \ref{eq:ShellATM} is solved on a finite number of wave numbers (called shells) $k_n=k_0 2^n$. That way Shell-Atm can capture in great detail the non-linear and dissipative processes that occur throughout the cascade, from the (large) inertial scales to the (small) dissipation scales. \\

Coupling directly the Shell-Atm and ISAM models can be really challenging so we preferred for a start to proceed with several manual back-and-forth iterations. In this approach ISAM supplies Shell-Atm with hydrodynamic profiles for the plasma density, bulk speed and temperature and in return Shell-Atm provides a total heating rate that is inserted back into the ISAM energy equations (eq \ref{eq:ISAM_Tps}\textendash\ref{eq:ISAM_Tts}). Of interest for the transfer of heavy ions throughout the solar atmosphere as discussed in section \ref{subsec:intro_FIP} and chapter \ref{cha:ISAM_results}, Shell-Atm also provides a total wave pressure and ponderomotive acceleration which are inserted back into the ion momentum equations (eq \ref{eq:ISAM_us}) and that will be described further in a subsequent dedicated paper. The total heating rate is summed up over all shells and combines contributions from both inward and outward wave populations \citep[][eq 7]{Verdini2019}:
\begin{equation}
    \frac{Q_h}{\rho}=\frac{1}{2}\sum_n\nu k_n^2\left(\left|z_n^+\right|^2+\left|z_n^-\right|^2 \right)
\end{equation}
\newline

The heating in ISAM can then be dispatched to protons only or we can follow the same prescriptions as those already discussed in section \ref{subsubsec:ISAM_Aw_heating}.

\subsection{Limitations and future improvements}
\label{subsec:ISAM_heating_future}

Our description of Alfvén-wave heating based on the WKB approximation is highly approximate although it can be extended to include wave reflection. A full-fledged treatment of the non-linear interaction and dissipation is much more appropriated especially in the transition region where the Alfvén speed undergoes sharp variations. Implementing such treatment in ISAM would be highly demanding computationally and at the risk of affecting the stability of the high-order approach which already involves many couplings between the hydrodynamic transport equations. A systematic coupling of ISAM with the Shell-Atm turbulence code (see section \ref{subsec:ISAM_Aw_ShellATM}) would be more achievable provided that enough intermediate sanity checkpoints are set thoroughly. \\

The chromosphere likely hosts other types of waves than pure transverse Alfvén waves, such as slow and fast-mode waves, compressional and torsional Alfvén waves, as well as wave-conversion mechanisms and instabilities. 

Acoustic shocks are also suspected to play a noticeable role in heating and enhancing the ionization level in the chromosphere. The transport equations of ISAM are solved with the LCPFCT algorithm that is specifically designed to propagate shocks. We could exploit this capacity to model the propagation and steepening of acoustic shocks in the chromosphere. Some shocks naturally appeared in ISAM simulations of the open solar wind \citep{Lavarra2022} but no tests have been performed for closed loop geometries yet. The modeling of shocks might completely disturb the stability of our high-order approach in confined environments, this is something that we should be able to test in the future. \\

Acoustic shocks are also suspected to undergo mode conversion into magnetoacoustic waves when they reach the transition region where the plasma beta is about unity. Mode conversion is also facilitated by a rapid change in the magnetic field topology, that expands rapidly to form large-scale flux-tubes in the upper chromosphere and transition region. Unfortunately magnetoacoustic waves can not be self-consistently treated in ISAM since the magnetic field is not solved together with the hydrodynamic equations but is an input to the model. The relevance of magnetoacoustic waves is also inherently limited in 1-D approaches although we could use some proxies or ad-hoc prescriptions.

\section{Radiative cooling}
\label{sec:ISAM_radloss}
A detailed assessment of the energy balance in the low solar atmosphere requires both to consider the energy gain (heating) discussed in section \ref{sec:ISAM_heating} and energy losses. Thermal energy can be converted into kinetic energy or advected during the acceleration phase of the solar wind. For confined medium such as coronal loops, one needs another mechanism to prevent the plasma from heating up to unrealistic temperatures. In such cases, radiative cooling is the dominant process that dissipates the thermal energy accumulated in the corona. Basically radiative cooling occurs in two steps, first an excitation of the atom by impact with an electron that is then followed by a radiative deexcitation with emission of a photon. Radiative cooling starts being really effective in the upper chromosphere and transition region where the plasma becomes optically thin. Deeper in the chromosphere, only a small portion of radiation is transported outwards due to rapid re-absorption of the emitted photons by the dense, optically thick plasma. At greater heights in the corona, the lack of collisions greatly reduces the efficiency of radiative cooling. In the next sections we describe several approaches by order of complexity to account for radiative cooling and heating.

\subsection{Ad-hoc prescription}
\label{subsec:ISAM_radloss_MVP}

The "coronal approximation" has been widely used among the scientific community to estimate radiative losses in the corona, and to some extent in the transition region, because it is a simple function of the electronic temperature and density, and therefore a full-fledged treatment of the radiative transfer can be avoided. The "coronal approximation" includes the main processes that control the ionization equilibria in the solar corona, of which ionization by collisions with electrons and radiative recombination. Radiative losses are then estimated for the major constituants of the solar corona by considering electron-impact excitation that is followed by deexcitation and emission of a photon. \citet{Athay1986} has noted a deviation to the usual "coronal approximation" for temperatures below $\approx 6\times 10^4\ \rm{K}$ if contributions from the Lyman alpha line of neutral Hydrogen are accounted for. \\

The optically thin function of \citet{Athay1986} has been widely used in the scientific community to estimate radiative cooling in the transition region and corona, for temperatures greater than $\approx 2\times 10^4\ \rm{K}$.

In ISAM I have tested the same prescription as in the MULTI-VP model (introduced in section \ref{subsec:MULTI-VP}) which is based on the optically thin function of \citet{Athay1986} but with a correction factor to account for optical thickness in the upper chromosphere for temperatures lower than $\approx 2\times 10^4\ \rm{K}$ \citep{PintoRouillard2017}:
\begin{equation}
\label{eq:ISAM_lambda}
    \Lambda(T) = 10^{-21}10^{[log_{10}(T/T_M)]^2}\chi(T)\quad \rm{(in\ erg.cm^3.s^{-1})}
\end{equation}
where
\begin{equation*}
    \chi(T)=
    \begin{dcases*}
        1 & for $T>T_1$\\
        \frac{T-T_0}{T_1-T_0} & for $T_0<T<T_1$\\
        0 & for $T<T_0$
\end{dcases*}
\end{equation*}
where $T_0$, $T_1$ and $T_M$ are fitting parameters which allow to calibrate the medium opacity and then to adjust radiative losses. The temperature profile in the chromosphere, transition region and low corona is highly impacted by the choice of these fitting parameters. One can get a reasonable temperature profile with a chromospheric plateau at $T_0=6500\ K$ and a transition region with temperatures ranging between $T_1=2\times 10^4\ K$ and $T_M=2\times 10^5\ K$. \\

The total radiative losses are then estimated as:
\begin{equation}
\label{eq:ISAM_radloss_MVP}
    \mathcal{R} = n_e (n_p+n_H)\Lambda(T)
\end{equation}
where $n_p$, $n_H$ and $n_e$ are the particle densities of protons, neutral Hydrogen and electrons respectively. 

Alternative formulations can be found in the literature and in other models, such as in the WindPredict-AW and MULTI-VP models (introduced in section \ref{subsec:WindPredict} and \ref{subsec:MULTI-VP}) where the radiative losses are scaled as $n_e^2$. Because ISAM has been developed to model deeper layer of the solar atmosphere where neutral hydrogen is dominant over protons and electrons, we scale our radiative losses as $n_e (n_p+n_H)$. A similar scaling is used in the well-known \textit{Chianti} spectral code for astrophysical plasmas \citep[see e.g.][]{Chianti09}. \\

\begin{figure*}[]
\centering
\includegraphics[width=0.8\textwidth]{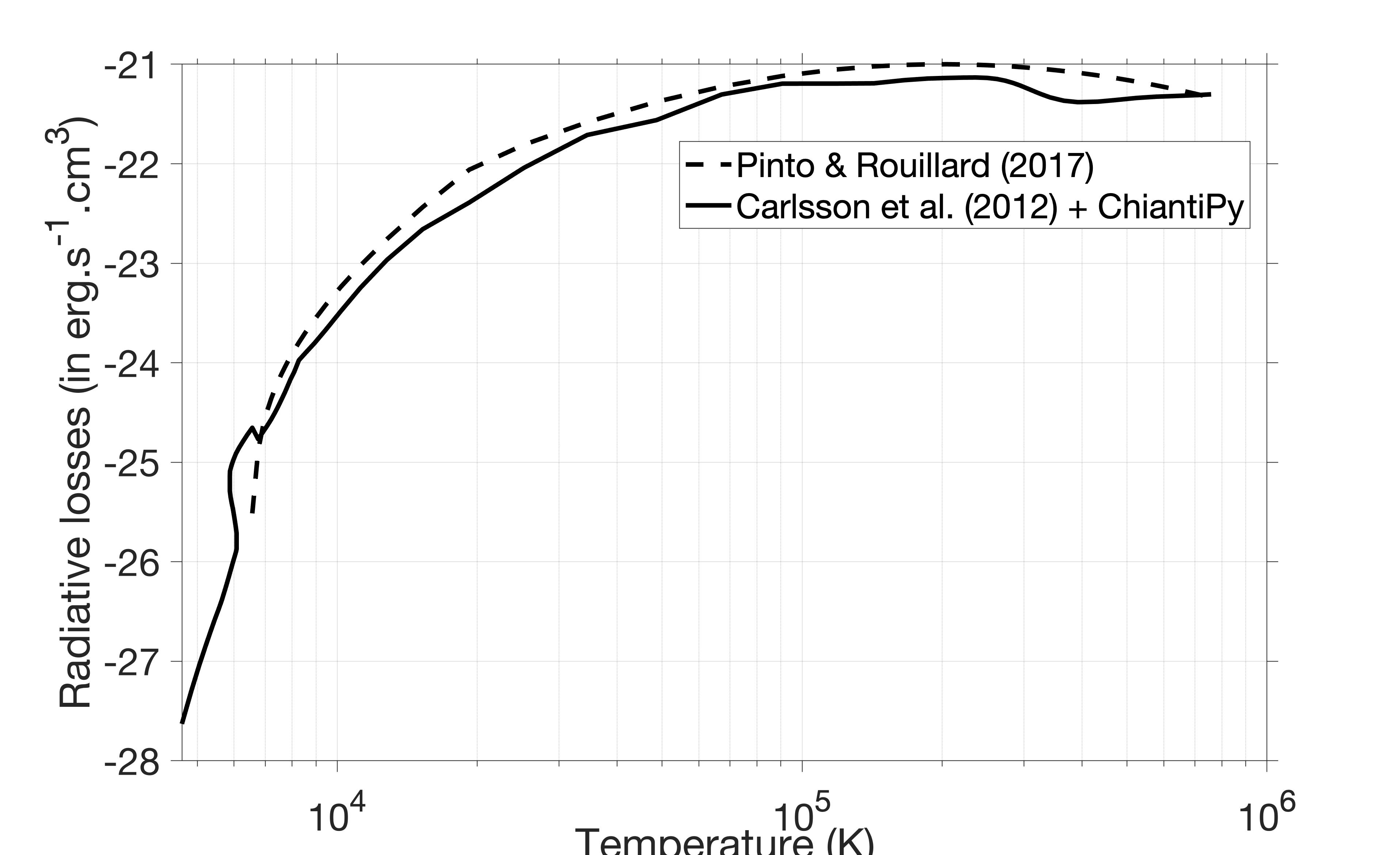}
\caption{ Radiative losses as defined in the MULTI-VP model (dashed black line) \citep{PintoRouillard2017}, and as derived from the improved method (solid black line) that includes both the semi-empirical recipe of \citet{Carlsson2012} (for $2000<T<30000\ \rm{K}$) and \textit{ChiantiPy} (for $T>30000\ \rm{K}$) that are introduced later in section \ref{subsec:ISAM_radloss_Carlsson2012} and \ref{subsec:ISAM_radloss_chianti}. The improved method has also a dependence on the column mass, which here is retrieved from the ISAM run presented in section \ref{subsec:ISAM_results_H_chromo}. 
\label{fig:ISAM_lambda}}
\end{figure*}

The optically thin function of \citet{Athay1986} slightly overestimates the total radiative losses in the corona compared to the more comprehensive treatment of Chianti, that is shown in Figure \ref{fig:ISAM_lambda} and discussed further in section \ref{subsec:ISAM_radloss_chianti}. However its simple implementation in ISAM has proven to be sufficiently accurate and very efficient, while not affecting the stability of the 16-moment approach neither increasing the computational cost. Since contributions from all species are summed up in this radiative loss prescription we chose to distribute the cooling isotropically on electrons only. This is also justified by the fact that radiative cooling takes place in regions where thermal energy is rapidly re-distributed by collisions with the electrons.

\subsection{Improved recipe for radiative cooling/heating in the upper chromosphere}
\label{subsec:ISAM_radloss_Carlsson2012}

The ad-hoc prescription presented in the previous section only aims at giving a rough estimate of the total radiative gains and losses in an optically thin plasma with an arbitrary correction for the opacity. If one wants to properly account for radiative losses and gains in an optically thick plasma, the resort on non local-thermal-equilibrium (non-LTE) simulations is necessary. In contrast to LTE simulations where the thermodynamic properties of the plasma can be decoupled from radiative processes, non-LTE solves simultaneously the (magneto-)hydrodynamic transport equations, the radiative transfer, and the rate equations for level populations. Such simulations are usually limited to one or two dimensions and still must include some approximations for computational tractability. \\

An intermediate approach is to exploit the results from detailed radiative transfer simulations to build a semi-empirical law for radiative cooling and heating. This task has been carried out by \citet{Carlsson2012} who provide a recipe for radiative cooling and heating including the most prevalent spectral chromospheric lines from neutral Hydrogen, singly ionized Calcium and singly ionized Magnesium. 

Essentially they exploited two non-LTE simulations: one based on the RADYN code which solves in 1-D for the full three-fold approach including hydrodynamic transport, radiative transfer and level population rates \citep[][and references therein]{Carlsson2002a}; and a 2-D simulation using the radiation-magneto-hydrodynamics code Bifrost \citep{Gudiksen2011} for which a detailed description of the simulation setup can be found in \citet{Leenaarts2011}. 

A realistic modeling of Hydrogen in the chromosphere requires a detailed treatment of non-equilibrium ionization together with non-LTE hydrogen excitation \citep{Carlsson2002a}. For this reason the RADYN code has been used to produce the recipe for neutral Hydrogen (HI) whereas the recipes for the singly ionized atom of Calcium (CaII) and Magnesium (MgII) have been determined from the 2-D Bifrost simulation \citep[see][for further details]{Carlsson2012}. \\

The total radiative loss (or gain) for an element $X$ including the neutral and all its ionization stages, can be estimated in our standard notations as \citep[][eq 1]{Carlsson2012}:
\begin{equation}
\label{eq:Carlsson2012_recipe}
    \mathcal{R}_{[X]}=-\sum_m L_{X^{+m}}(T)E_{X^{+m}}(\tau)\frac{n_{X^{+m}}}{n_{[X]}}A_{[X]} (n_H+n_p) n_e
\end{equation}
where $n_{X^{+m}}$ is the density of element $X$ in ionization stage $m$ ($m=0$ for neutrals), and $n_{[X]}$ the total density of element $X$. 

The recipe is built in a similar manner as in the ad-hoc prescription (eq \ref{eq:ISAM_lambda} and \ref{eq:ISAM_radloss_MVP}) with an optically thin radiative function $L_{X^{+m}}(T)$ that is corrected by an opacity term $E_{X^{+m}}(\tau)$. The difference with the simple ad-hoc prescription is that these two functions have been derived empirically from detailed radiative transfer simulations and that the opacity function is now dependent on the optical depth or column mass $\tau$. A detailed description of the procedure that I carried out to implement this recipe in ISAM is given in Appendix \ref{sec:ISAM_radloss_detailed}. \\

The above recipe is valid for chromospheric temperatures between $~2000\ K$ and $30 000\ K$. I have implemented in ISAM the summed up contributions of HI, CaII and MgII as a 2-D lookup table that we interpolate with the plasma thermodynamic properties solved in ISAM. An example of application of this recipe is shown in Figure \ref{fig:ISAM_lambda} for a column mass profile given by the ISAM run presented in section \ref{subsec:ISAM_results_H_chromo}. By including a more accurate treatment of the medium opacity, the improved recipe gives a direct estimate of the total radiative losses in the mid-upper chromosphere. That is not the case for the ad-hoc prescription introduced in section \ref{subsec:ISAM_radloss_MVP}, which is switched off as soon as temperatures get lower than a minimum threshold (set at $6500\ \rm{K}$ in this thesis).  \\

For temperatures greater than $30 000\ K$ I made an attempt to compute the optically thin function $L_{X^{+m}}(T)$ following the "coronal approximation" as given in equation 4 of \citet{Carlsson2012}. While I have done it successfully for HI and CaII, it required a significant amount of time to gather all the needed data. As already mentioned in section \ref{subsec:ISAM_radloss_MVP}, many other elements also contribute to the radiative losses in the transition region and corona. Therefore for temperatures greater than $30 000\ K$, I resort on the \textit{Chianti} spectral code and database to estimate the radiative losses for all major and minor constituents of the solar corona.

\subsection{Improved optically thin radiative cooling from \textit{Chianti}}
\label{subsec:ISAM_radloss_chianti}

I exploit the \textit{ChiantiPy} \textit{Python} package distributed on \textit{Github}\footnote{ChiantiPy: \url{https://github.com/chianti-atomic/ChiantiPy/}} and the Chianti database version 10 \citep{DelZanna2021} to better estimate radiative cooling in the transition region and corona for all major and minor constituents of the solar atmosphere. \textit{Chianti} provides a fully integrated environment for spectral diagnostics with a critically evaluated atomic database \citep{Dere1997}. \textit{Chianti} accounts for radiative cooling from bound-bound (radiative deexcitation) and free-bound (photoionization) transitions as well as free-free (Bremsstrahlung) emissions. \\

I have calculated in \textit{ChiantiPy} the total radiative losses for temperatures between $30000\ K$ and $10\ MK$, for which the full procedure is given in Appendix \ref{sec:ISAM_radloss_detailed}. That completes the lookup table established in section \ref{subsec:ISAM_radloss_Carlsson2012}. Since \textit{ChiantiPy} assumes an optically thin plasma which is coherent for such high temperatures in the solar atmosphere, we get rid of the dependence on the optical depth or column mass. An example is shown in Figure \ref{fig:ISAM_lambda}.

\subsection{Limitations and future improvements}
\label{subsec:ISAM_radloss_future}
Future short-term improvements should consider individual computations of the radiative losses for each specie solved in ISAM. Because we consider the total integrated radiative losses in the solar atmosphere, that supposes an underlying state of the ionization fractions and abundances. The improved recipe for chromospheric radiative cooling/heating introduced in section \ref{subsec:ISAM_radloss_Carlsson2012} therefore has been fed with specific ionization fractions simulated in Bifrost and coronal abundances from \citet{Schmelz2012}. Similarly, the optically thin radiative losses that I computed using \textit{ChiantyPy} assume the same coronal abundances but exploit the default ionization equilibria provided in the \textit{Chianti} database \citep{Dere2007,Dere2009}. \\

By implementing a separate recipe for each specie solved in ISAM, we should be able to use the ionization fractions and abundances as computed in ISAM in a self-consistent manner. This is especially important for the abundances since ISAM can provide coronal abundances that account for detailed transport processes including FIP fractionation. The radiative losses given by \textit{ChiantiPy} can already be calculated separately for each specie. The contributions from HI, CaII and MgII as provided by the semi-empirical recipe from \citep{Carlsson2012} could also be splitted and provided as separate lookup tables in ISAM. \\

A more accurate treatment of the energy balance in the optically thick chromosphere would ideally require a full-fledged radiative transfer approach. For the same reasons than those enunciated in section \ref{subsec:ISAM_ioniz_future} this is not a task that can be tackled right away and therefore is left for future long-term developments.

\section{Numerical method}
\label{sec:ISAM_numerical}
In this section we present the numerical specificities of ISAM. The sharp gradients near the transition region are resolved thanks to a Flux-Corrected Transport algorithm and a refined mesh that are presented in section \ref{subsec:ISAM_lcpfct} and \ref{subsec:ISAM_grid}. 
There is a particular numerical treatment of collisions in ISAM which I further improved to maximize the code efficiency when solving for minor species, a description of the method is given in Appendix \ref{sec:ISAM_col_numerical}. Finally in section \ref{subsec:ISAM_BC} I draw a conclusion from all the ISAM runs I performed on the calibration of the boundary conditions for the case of coronal loop geometries.

\subsection{A Flux-Corrected Transport algorithm to solve for conservative equations}
\label{subsec:ISAM_lcpfct}
The transport equations of the 16-moment set solved in ISAM and introduced in section \ref{sec:ISAM_transport_eq} are resolved numerically with a Flux-Corrected Transport algorithm initially developed by J.P. Boris at the Naval Research Laboratory. \citet{Boris1993} provide a fully integrated fortran package to solve for generalized continuity equations of the form:
\begin{equation}
    \frac{\partial}{\partial t}(*) + \mathbf{\nabla} \cdot (* \mathbf{u}) = \mathcal{S}
\end{equation}
where $\mathcal{S}$ contains all other source terms that can not be included in $\mathbf{\nabla} \cdot (* \mathbf{u})$. Flux-Corrected Transport algorithms are particularly suited to simulate conservative physical quantities for systems with steep gradients because they ensure:
\begin{itemize}
    \item \textbf{Positivity}: If the conserved quantity $*$ is positive at time $t$ then it remains positive at time $t+\Delta t$ whatever is the advection velocity $\mathbf{u}$, at the condition that information do not jump more than one cell during $\Delta t$. This is generally expressed as the Courant\textendash Friedrichs\textendash Lewy condition $| \frac{u\Delta t}{\Delta r}|\lesssim 1$.
    \item \textbf{Accuracy}: Numerical diffusion does not smear discontinuities over neighboring cells beyond the physical limit.
    \item \textbf{Monotonicity}: The numerical scheme does not introduce additional new maxima or minima to the ones that arise naturally from physical advection.
\end{itemize}
The basic low-order linear finite methods that are used to approximate gradients, e.g. the upwind or three-point explicit finite-difference scheme, fail at uniting all three conditions \citep{Boris1993}. To ensure positivity one would need a high numerical diffusion which is physically not acceptable to simulate systems with steep gradients. On the other hand one can risk taking a low numerical diffusion coefficient at the price of positivity that is no more preserved and possibly the appearance of unphysical new maxima and/or minima in the solution. Flux-Corrected Transport (FCT) algorithms have been developed to resolve this dilemma and satisfy all three conditions. \\

Basically FCT algorithms solve this dilemma by introducing a non-linear method to determine the numerical diffusion coefficient that is necessary to ensure positivity. The numerical diffusion coefficient is corrected (or limited) whether the evaluated cell is far or not from a sharp discontinuity so that numerical diffusion can be minimized to achieve accuracy while maintaining positivity and monotonicity of the solution. The LCPFCT algorithm developed by \citep{Boris1993} and that we exploit in ISAM uses a flux correction (or limiting) technique similar to other methods proposed by \citet{vanLeer1979} and \citet{Harten1983}. A complete description of the LCPFCT algorithm can be found in \citet{Boris1993}. 

Let us consider a conservative quantity $\rho$ to be solved, below is a summary of the main steps executed by the LCPFCT algorithm.
\begin{description}
\item[1. Convective and diffusive stage: ] The values at time $t$ ($\rho_i^0$) are advanced to time $t+\Delta t$ by the relation:
\begin{subequations}
\begin{align}
    \Tilde{\rho}_i &=\rho_i^0-\frac{1}{\Delta x}\left[f_{i-\frac{1}{2}}-f_{i+\frac{1}{2}} \right] + \mathcal{S}_i \\
    \text{where  }f_{i+\frac{1}{2}} &= \nu_{i+\frac{1}{2}}\Delta x\left(\rho_{i+1}^0-\rho_{i}^0\right)- \epsilon_{i+\frac{1}{2}}\frac{\Delta x}{2}\left(\rho_{i+1}^0+\rho_{i}^0\right) \\
    \text{and    }f_{i-\frac{1}{2}} &=  \nu_{i-\frac{1}{2}}\Delta x\left(\rho_{i}^0-\rho_{i-1}^0\right)- \epsilon_{i-\frac{1}{2}}\frac{\Delta x}{2}\left(\rho_{i}^0+\rho_{i-1}^0\right)
\end{align}
\end{subequations}
with $\nu_{i\pm 1/2}$ and $\epsilon_{i\pm 1/2}=u_{i\pm 1/2}\Delta t/ \Delta x$ the diffusion and advection coefficients at cell interfaces $i\pm 1/2$.

\item[2. Anti-diffusive fluxes: ] The strong numerical diffusion introduced by the diffusive stage is reduced by applying an anti-diffusive stage. Similarly to the diffusive stage we introduce anti-diffusive fluxes:
\begin{subequations}
\begin{align}
    f_{i+\frac{1}{2}}^{ad} &= -\mu_{i+\frac{1}{2}}\Delta x\left(\Tilde{\rho}_{i+1}-\Tilde{\rho}_{i}\right)\\
    f_{i-\frac{1}{2}}^{ad} &= -\mu_{i-\frac{1}{2}}\Delta x\left(\Tilde{\rho}_{i}-\Tilde{\rho}_{i-1}\right)
\end{align}
\end{subequations}

\item[3. Correction of the anti-diffusive fluxes: ] The raw anti-diffusive fluxes can potentially break down the positivity that was ensured by the diffusive stage. A non-linear correction of the anti-diffusive fluxes is necessary to maintain positivity and monotonicity while reducing the numerical diffusion generated by the diffusive stage. This correction simply enforces that the anti-diffusive stage do not create new maxima or minima nor accentuate those already existing. The raw anti-diffusive fluxes $f_{i\pm\frac{1}{2}}^{ad}$ are then corrected as follows:
\begin{subequations}
\begin{align}
    f_{i+\frac{1}{2}}^{c} = S_{i+\frac{1}{2}} \rm{max}\left(0,\rm{min}\left[S\left(\Tilde{\rho}_{i+2}-\Tilde{\rho}_{i+1} \right),|f_{i+\frac{1}{2}}^{ad}|,S\left(\Tilde{\rho}_{i}-\Tilde{\rho}_{i-1} \right) \right] \right) \\
    f_{i-\frac{1}{2}}^{c} = S_{i-\frac{1}{2}} \rm{max}\left(0,\rm{min}\left[S\left(\Tilde{\rho}_{i+1}-\Tilde{\rho}_{i} \right),|f_{i-\frac{1}{2}}^{ad}|,S\left(\Tilde{\rho}_{i-1}-\Tilde{\rho}_{i-2} \right) \right] \right)
\end{align}
\end{subequations}
with $|S_{i\pm\frac{1}{2}}|=1$, $S_{i+\frac{1}{2}}=\rm{sign}(\Tilde{\rho}_{i+1}-\Tilde{\rho}_{i})$ and $S_{i-\frac{1}{2}}=\rm{sign}(\Tilde{\rho}_{i}-\Tilde{\rho}_{i-1})$.

\item[4. Applying the corrected anti-diffusive stage: ]
The final values $\rho_i^n$ advanced at time $t+\Delta t$ are computed from the intermediate $\Tilde{\rho}_i$ values and the corrected anti-diffusive fluxes $f_{i\pm\frac{1}{2}}^{c}$:
\begin{equation}
    \rho_i^n =\Tilde{\rho}_i-\frac{1}{\Delta x}\left[f_{i-\frac{1}{2}}^{c}-f_{i+\frac{1}{2}}^{c} \right]
\end{equation}
\end{description}

The diffusion $\nu_{i\pm 1/2}$ and anti-diffusive $\mu_{i\pm 1/2}$ coefficients are chosen as \citep[][eq 3.19]{Boris1993}:
\begin{subequations}
\begin{align}
    \nu_{i\pm \frac{1}{2}} &=\frac{1}{6}+\frac{1}{3}\epsilon_{i\pm \frac{1}{2}}^2 \\
    \mu_{i\pm \frac{1}{2}} &=\frac{1}{6}-\frac{1}{6}\epsilon_{i\pm \frac{1}{2}}^2
    \end{align}
\end{subequations}

The LCPFCT fortran package includes different options to treat boundary conditions. In the runs presented in this thesis, we set the values at edges so that the diffusive and anti-diffusive stages cancel out so that only pure advection (and sources) are accounted for. This has the advantage to simplify the procedure which otherwise requires to provide values for two cells outside the simulated domain. Although this simplification might compromise the three golden rules enunciated before, it is not troublesome since we enforce new boundary conditions in ISAM afterwards (see section \ref{subsec:ISAM_BC}). \\

Collision terms are not included in the resolution of the transport equations by the LCPFCT algorithm to limit potential stability issues with the 16-moment high-order approach. Collisions likely occur on time scales $1/\nu_{st}$ that are much lower than the hydrodynamic transport. Therefore we start by solving the collision terms and then we linearly superpose the solution with the solution given by LCPFCT for the hydrodynamic transport. The procedure is given in more detail in Appendix \ref{sec:ISAM_col_numerical}. \\

When a heating model based on the dissipation of Alfvén waves is adopted (see section \ref{subsec:ISAM_Aw}), we need to solve an additional transport equation for the Alfvén-wave energy (eq \ref{eq:Ew}). As for the transport equations of the 16-moment, I solve equation \ref{eq:Ew} using the LCPFCT algorithm. And similarly to the collision terms, I proceed with a linear superposition of the pure transport of the Alfvén-wave energy and of the dissipation process, by assuming that they occur at separate time scales. The procedure is given in detail in Appendix \ref{sec:ISAM_AW_numerical}.

\subsection{Boundary conditions}
\label{subsec:ISAM_BC}
One can get two complete opposite solutions for the same physical problem if the boundary conditions are not defined properly. The choice of boundary conditions was less critical for the open solar wind case described in \citet{Lavarra2022} because the open condition at the upper boundary naturally pumps the plasma out of the solar corona. It has been a harsh task to find suitable boundary conditions to model confined plasma in ISAM for closed loop geometries. Below I give a review of all these attempts, by sharing my experience rather than theoretical statements because of the complexity of the 16-moment approach. \\

The 16-moment set can generate significant perturbations and stability issues that arise from the strong coupling in the transport equations. The principal difficulty lies in finding a compromise between a loose boundary condition that ensures that unwanted perturbations exit the system free of any reflection at the edges, and a strict boundary condition that is stable but leads to the reflection or even the growth of these perturbations. Of course the best would be a boundary condition that is completely "invisible" to the plasma inside of the domain. Nevertheless, a minimum of requirements must be imposed at the simulation boundaries in order to model a peculiar physical problem. \\

In my first attempts I was tempted to enforce null mass flows at the bottom boundary to model confined geometries. That way I could control the dynamic evolution of the system and hopefully converge towards a nearly hydrostatic coronal loop. Although this method worked quite well I obtained solutions that were too dependent of the initial state of the loop because the resolved system was over-constrained. \\

Since this is the first time that ISAM is being tested for confined closed loop geometries I turned on choosing boundary conditions that are more flexible. After many tests I found that applying a condition to the neutral density only, provides the most elastic but cleanest solution with a system that is now likely under-constrained. The final set of boundary conditions can be summarized as follows:
\begin{description}
    \item[Mass and heat flows: ]  The plasma can escape but not enter the loop at the bottom boundary:
    \begin{subequations}
    \begin{align}
        u_s &= min(0,u_s) \\
        \gamma_s^{\parallel,\perp} &= min(0,\gamma_s^{\parallel,\perp})
    \end{align}
    \end{subequations}
    where positive velocities indicate that the plasma rises in the solar atmosphere.
    
    \item[Temperature: ] We do not fix a temperature at the bottom boundary but for numerical reasons we maintain it within a "safe" interval:
    \begin{equation}
        T_s^{\parallel,\perp}=min(T_{max},max(T_{min},T_s^{\parallel,\perp}))
    \end{equation}
    where we set $T_{min}=2000\ K$ and $T_{max}=100\ MK$.
    
    \item[Density: ] In most of the runs presented in this thesis and unless specifically mentioned, the neutral Hydrogen density is fixed at the inner boundary to $8.4\times 10^{20}\ \rm{m^{-3}}$ that corresponds with the chromospheric profiles of \citet{Avrett2008}. Then for minor species, I either specify the density of neutrals or ionized species whether a high-FIP (e.g. Helium) or low-FIP (e.g. Magnesium) element is considered, by using the photospheric abundances relative to Hydrogen that I fetch from \citet[][Table 2]{Avrett2008}.
\end{description}

The present version of boundary conditions is likely to be improved in future developments of ISAM. For instance we could use the method of characteristics to derive a set of boundary conditions that are self-consistent, a method that is well suited for systems driven by coupled transport equations.

\subsection{Grid definition}
\label{subsec:ISAM_grid}

The LCPFCT algorithm (see section \ref{subsec:ISAM_lcpfct}) that solves the transport equations is specifically designed to preserve the sharp gradients and discontinuities that are typically observed in the transition region. \\

The requirement of a fine enough mesh resolution in the transition region should therefore not be as critical as for classical numerical approaches. \citet{Johnston2017,Johnston2019} showed that under-resolved loops in the transition region can lead to an underestimation of the heat flux that can affect the whole thermodynamics of the loop. However these authors used the typical Spitzer\textendash Härm prescription for the heat flux ($-K_0 T^{5/2}\partial T/\partial z$), and do not solve an explicit transport equation for the heat flow as in our high-order approach. \\

I realized after many test runs that the thinness of the grid in the transition region was not so decisive in ISAM as long as there are enough points to describe the sharpest gradients. The accuracy of the solutions is then ensured as long as we respect the stability condition for the transport equations (see section \ref{sec:ISAM_origin}) and the condition of small normalized heat fluxes as assumed in the 16-moment approach (see section \ref{sec:ISAM_origin}). However I noticed a direct implication on the amplitude of small oscillations that appear in the upper part of the transition region. For the smallest loop considered in this thesis of height $10\ \rm{Mm}$ I could achieved a mesh size as down as $15\ km$
in the transition region while maintaining enough points in the rest of the loop. By decreasing the mesh size from $\approx 50\ km$ to $15\ km$ I managed to significantly smooth out those oscillations which were troublesome to interpret the results. \\

I have performed numerous tests to optimize the grid points distribution along the loop. The final adopted grid profile is constituted of three separate laws for the chromosphere, transition region and solar corona:
\begin{subequations}
\begin{align}
    dz^i &= (dz_{max}^{CHR}-dz_{min}^{TR})\left( \frac{z^1-z^i}{z^1-z^0}\right)^{\alpha^{CHR}} + dz_{min}^{TR} \quad &\text{for: }z^0\leq z^i<z^1 \\
    dz^i &= dz_{min}^{TR}\quad &\text{for: }z^1\leq z^i\leq z^2\\
    dz^i &= (dz_{max}^{COR}-dz_{min}^{TR})\left( \frac{z^i-z^2}{z^{apex}-z^2}\right)^{\alpha^{COR}} + dz_{min}^{TR} \quad &\text{for: }z^2<z^i\leq z^{apex} 
\end{align}
\end{subequations}
I fix the extent and location of the transition region where the mesh is the most refined with the $z^1$ and $z^2$ altitude parameters. The mesh size is constrained in those three regions with the three parameters $dz_{max}^{CHR}$, $dz_{min}^{TR}$ and $dz_{max}^{COR}$. In the chromosphere, $dz$ progressively decreases from $dz_{max}^{CHR}$ to $dz_{min}^{TR}$ following a power law. In the transition region I keep the mesh size constant at $dz=dz_{min}^{TR}$. In the corona the mesh size increases again following a power law up to $dz=dz_{max}^{COR}$. The exponents $\alpha^{CHR}$ and $\alpha^{COR}$ of the two power laws are optimized by the code itself to match the number of points that I fix in the chromosphere and corona.  \\

\begin{figure*}[]
\centering
\includegraphics[width=0.8\textwidth]{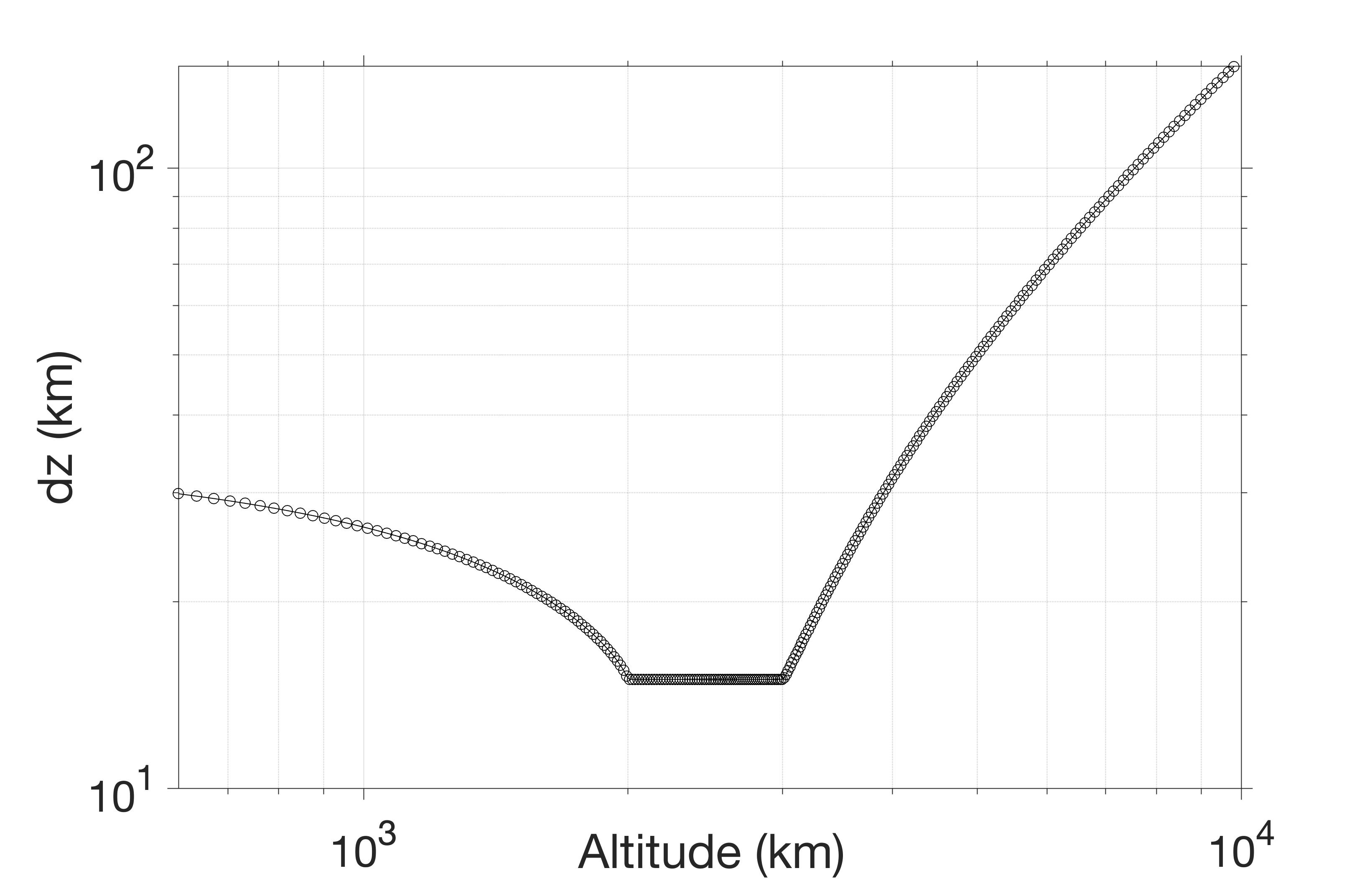}
\caption{Distribution of grid points in ISAM for a small loop of half-length $16\ \rm{Mm}$ and apex height $10\ \rm{Mm}$.
\label{fig:grid}}
\end{figure*}

In the chromosphere I allocate about one-fourth of the total number of points available for half the loop, that is $n^{CHR}\approx 60$ points. The number of points in the transition region is directly deduced from $n^{TR}=\lfloor(z^2-z^1)/dz_{min}^{TR}\rfloor + 1=66$ given that $z^1=2000\ km$, $z^2=3000\ km$ and $dz_{min}^{TR}=15\ km$ (where $\lfloor * \rfloor$ is the truncation or integer part). Finally the left over of grid points is attributed to fill up the coronal part of the loop. An example of the grid points distribution for these parameters is given in Figure \ref{fig:grid}.

%% file: chapters/ISAM_results.tex
\chapter{Separation of heavy ions in coronal loops}
\label{cha:ISAM_results}

\minitoc

As highlighted in section \ref{sec:intro_theories}, several theories for the origin of the slow solar wind (SSW) have been proposed in the last two decades that we have confronted in chapter \ref{cha:stationnary} and \ref{cha:dynamics} with recent modeling efforts and observations from the \textit{Parker Solar Probe}. Besides the structure and bulk properties of the SSW that we discussed in detail in these chapters, another challenging aspect of the SSW that needs to be explained is its composition. \\

As introduced in section \ref{subsubsec:intro_SSW_composition}, heavy ions only constitute about less than $1\%$ of the solar wind. Since their abundance does not change during propagation from the solar corona to points of in situ measurements they represent fossil signatures of the origin and formation process of the SSW. The abundance of heavy ions varies significantly between winds of different regimes (fast and slow) and can also vary for different source regions ("streamer flows" and "coronal hole flows"). The abundance of alpha particles are also highly variable in the SSW with on average a strong depletion in abundance measured in the SSW coming from streamers but not from coronal holes. In contrast the enrichment in low FIP elements is observed in all types of SSW whether it comes from streamer flows or coronal hole flows, albeit with different magnitudes. This FIP effect is however not observed (or to a much less extent) above coronal holes that produce fast winds. It appears therefore that the mechanisms that control the FIP effect may differ or else operate differently than those controlling Helium depletion in the SSW. Clearly a full explanation of the composition of the SSW must address these subtle composition differences and as we discussed in the introduction they will involve probably different mechanisms for Helium and low-FIP elements not only during their extraction from the chromosphere but also during their expulsion in the solar wind. A full modelling of all these processes is beyond the scope of this thesis. 

Dynamic theories are convenient to explain the enrichment of the slow wind in low-FIP elements, by proposing a path way for the plasma that is initially bound into coronal loops to be expelled into the open-field that is connected to the heliosphere (see section \ref{sec:intro_dynamic}). Quasi-stationnary theories are not doomed to failure for all that as a separation of heavy ions is suspected to occur even in the open-field lines alone, through a collisional coupling with the background proton flow (see section \ref{subsec:intro_FIP}). Although these theories provide some hints on the sources of the slow solar wind, there are still two fundamental questions that need to be addressed. Why and how coronal loops get initially enriched in low-FIP heavy ions? And how this confined material happens to be expelled into the slow solar wind? The first question constitutes the backbone of this chapter while the second question has already been partially addressed in chapter \ref{cha:dynamics}.  \\

Since the FIP effect is detected in coronal spectroscopy along magnetic loops, the present chapter focuses on the physical mechanisms that control the extraction of heavy ions along a single magnetic loop extending from the chromosphere to the corona. We first consider the case of Oxygen that has a similar FIP as Hydrogen ($\simeq 13.6\ \rm{eV}$), where the low-FIP Magnesium and high-FIP Helium will be presented in a subsequent work. I exploit the Irap Solar Atmospheric Model (ISAM) that I described in detail in chapter \ref{cha:ISAM} in a framework that simulates plasmas that are confined in coronal loops. The reader is invited to refer to the chapter \ref{cha:ISAM} for a complete description of the improvements I brought to the model as well as for its governing equations. 

Although ISAM has already been successful at reproducing a variety of solar wind regimes including an estimate of their charge states \citep{Lavarra2022}, solar wind solutions that include the coupling of heavy ions with the background protons solar wind self-consistently have not been obtained yet. That will constitute a future study. \\

A simple atmosphere that is only made up of neutral Hydrogen, protons and electrons is first considered in section \ref{sec:ISAM_results_H}, where the physics of coronal loops can already be discussed in detail. This setup will form a baseline to interpret in section \ref{sec:ISAM_results_O} the dynamics of Oxygen in the solar atmosphere. I then conclude and discuss future perspectives to this work in section \ref{sec:ISAM_results_conclusion}.

\section{The background Hydrogen, protons and electrons atmosphere}
\label{sec:ISAM_results_H}

As a preparatory step to build up our expertise for the next section, I consider in this section a pure Hydrogen plasma constituted of neutral and ionized Hydrogen, and electrons. Transport processes can be very slow in coronal regions where the plasma is confined in closed magnetic fields, with time scales of the order of several hours, days or even weeks for the largest coronal loops of streamers seen in WL. To keep the ISAM simulations at a tractable computational time, I consider in this section a small-scale coronal loop of half-length $16\ \rm{Mm}$ and apex height $10\ \rm{Mm}$, that is within the typical range of loops that are observed in EUV in active regions \citep[see e.g.][]{Aschwanden2001}. Therefore I model the plasma composition in a portion of the solar atmosphere where interchange reconnection processes are known to occur (see section \ref{subsec:intro_interchange}), but I will not be able to treat the release of coronal loop material from the tip of streamers (see section \ref{subsec:intro_dynamics_streamertip} and \ref{subsec:intro_dynamics_fluxropes}) that is left for a future study. \\

As introduced in section \ref{sec:intro_low_atmosphere}, the heating of the solar corona is fundamental but remains poorly constrained by the observations. Therefore section \ref{subsec:ISAM_results_H_thermodynamics} constitutes a preliminary work where I investigate how the thermodynamics of the coronal loops simulated in ISAM may be affected when assuming different coronal heating conditions. Furthermore, I address more specifically the physics that is related to the upper chromosphere in section \ref{subsec:ISAM_results_H_chromo}. Then, the transfer of plasma from the upper chromosphere to the corona is analysed in detail in section \ref{subsec:ISAM_results_H_forces} where the different forces at play are compared. 

\subsection{Thermodynamics of coronal loops}
\label{subsec:ISAM_results_H_thermodynamics}

An extensive collection of coronal loop observations has been acquired over the last decades. They were first observed in WL within the bright streamers that are particularly well observed during total solar eclipses as shown in Figure \ref{fig:Mikic2018_fig1}. Spectroscopic diagnostics from X-ray and EUV lines have then provided critical information on the temperature profiles of coronal loops from the early observations of \textit{Skylab} \citep{Rosner1978,Habbal1985}, that were enriched subsequently by a wealth of high-resolution imaging of coronal loops from the \textit{SoHO} \citep{Aschwanden2000a} and \textit{TRACE} \citep{Winebarger2003a} missions. This observational data provided the scientific community with additional constraints to calibrate heating models of coronal loops \citep{Rosner1978,Serio1981,Aschwanden2000b,Aschwanden2001,Winebarger2003b}. By assuming a hydrostatic equilibrium, these authors have built scaling laws that link the thermodynamic properties of loops with their geometric and heating parameters. The earliest laws from \citet{Rosner1978} (also called the RTV laws) assumed an uniform pressure and heating, that was then generalized by \citet{Serio1981} to account for non uniformity. That is motivated by many EUV observations that favor rather a nonuniform heating to explain the quasi isothermal profiles of loop-top temperatures \citep[see also][]{Gudiksen2005}. Later on, \citet{Aschwanden2001} have constructed more rigorous laws to include the variation of scale height in a stratified atmosphere. All the scaling laws mentioned above have in common that they assume pure hydrostatic profiles. \\

However, it was already noted from the \textit{Skylab} observations that not all loops can be fitted with hydrostatic profiles which suggested that some loops are inherently dynamic \citep{Rosner1978,Serio1981,Habbal1985}. The stability of loops have then been further theorized using \textit{SoHO} and \textit{TRACE} diagnostics \citep{Aschwanden2001,Winebarger2003b}. Observational evidences for the occurrence of thermal non-equilibrium (TNE) cycles in coronal loops were then found using time-resolved EUV diagnostics taken by the \textit{SoHO-EIT} \citep{Auchere2016} and \textit{SDO-AIA} \citep{Froment2015} instruments. TNE cycles tend to occur when the heating is concentrated at the base of the corona. In a stratified corona where convection motions are very slow, the plasma is successively transported back and forth between the transition region and the upper most part of the loop where it undergoes heating (evaporation) and cooling (condensation) phases respectively. Many signatures of coronal inflows/rains of cold plasma have been observed in EUV \citep{Schrijver2001,DeGroof2004,Muller2005,Antolin2015}. Recent numerical simulations further support these observations by also arguing that TNE cycles can occur even with a coronal heating rate that is constant over time as long as the deposition of energy is concentrated near the transition region, also known as footpoint heating \citep[see e.g.][]{Lionello2013,Johnston2017,Johnston2019}. \\

I now use the above considerations as a baseline to calibrate our heating parameters in ISAM. For a start I focus on the heating of the corona only, where the inclusion of the chromosphere is discussed later on in section \ref{subsec:ISAM_results_H_chromo}. I assume a perfectly semi-circular coronal loop of half-length $16\ \rm{Mm}$ and height $10\ \rm{Mm}$. To ease the calibration process I use the optically thin prescription described in section \ref{subsec:ISAM_radloss_MVP}, where the radiative losses adjust themselves in the upper chromosphere (via the $\chi$ factor) to maintain a chromospheric plateau at $\simeq 6500\ \rm{K}$. As described in section \ref{subsec:ISAM_heating_adhoc}, simple laws can be used to a good approximation for the coronal heating. To a first approach I rely on these ad-hoc laws so that the heating calibration reduces to only two parameters, the heating flux at the base $F_\odot$ and the heating scale height $H_f$. \\

\begin{figure*}[]
\centering
\includegraphics[width=0.65\textwidth]{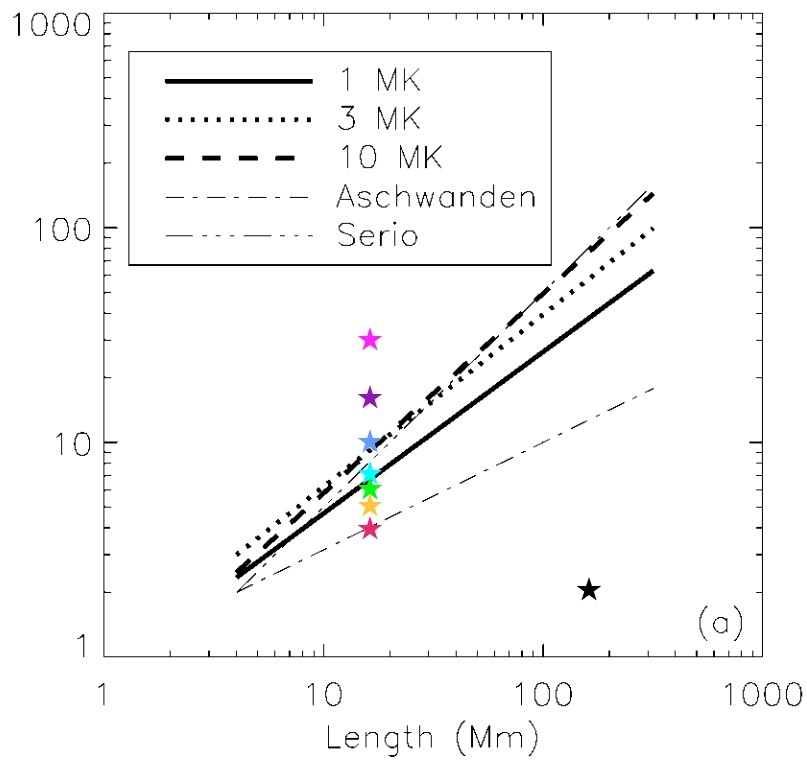}
\caption{ Stability criteria of coronal loops function of their half-length $L$ (x-axis) and heating scale height $H_f$ (y-axis) (both units are in Mm). The 1MK, 3MK and 10MK labelled curves correspond to the criteria from \citet{Winebarger2003a} where the critical heating scale height is $H^*_f(L)\approx A(T)L^\delta(T)$ with $A(T)$ and $\delta(T)$ that are tabulated data. The criteria from \citet{Serio1981} $H^*_f(L)=0.5 L$ and \citet{Aschwanden2001} $H^*_f(L)=\sqrt{L}$ are also plotted. Figure adapted from \citet[][Figure 6]{Winebarger2003a}.
\label{fig:Winebarger2003_fig6}}
\end{figure*}

\begin{figure*}[]
\centering
\includegraphics[width=1.0\textwidth]{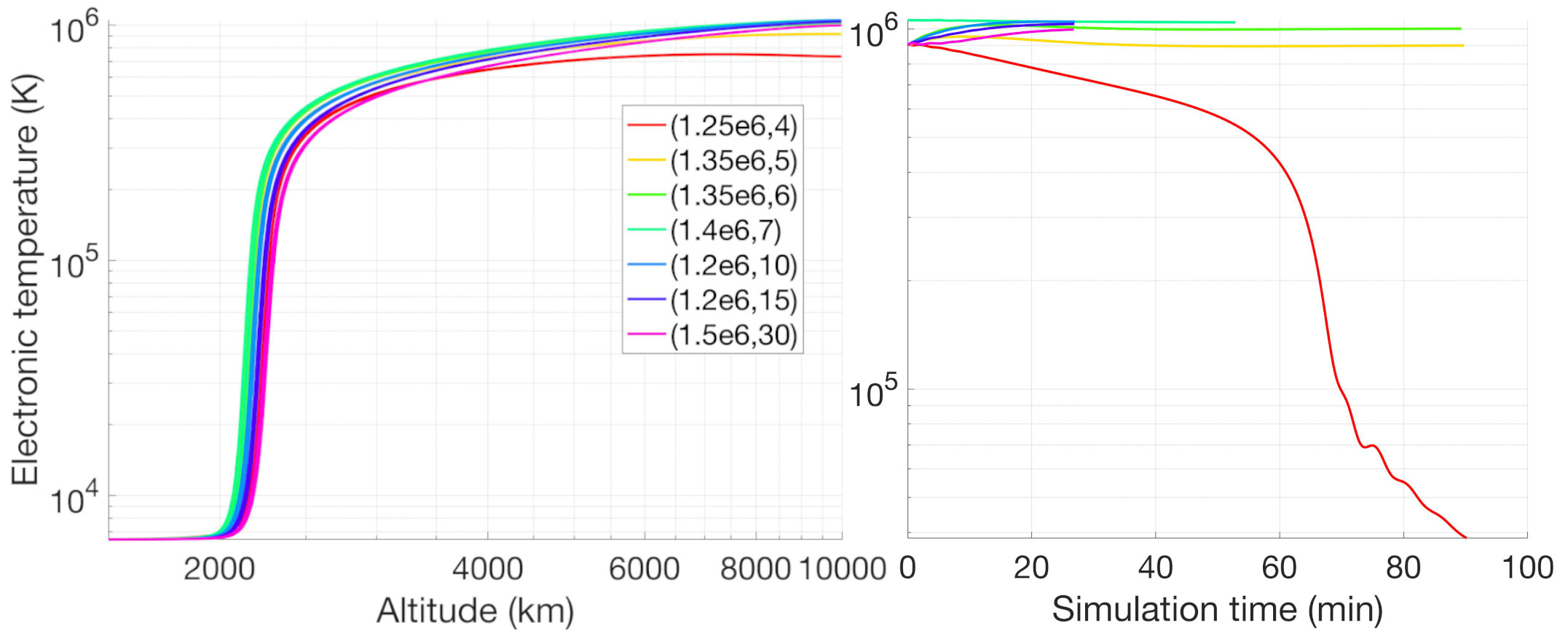}
\caption{ Left panel: comparison of the electronic temperature function of altitude, between ISAM runs that have different coronal heating parameters ($F_\odot,H_f$) and after $26$ min of simulation time. Right panel: same comparison but for the time evolution of the loop-top electronic temperature.
\label{fig:ISAM_Comp_Te}}
\end{figure*}

\begin{figure*}[]
\centering
\includegraphics[width=1.0\textwidth]{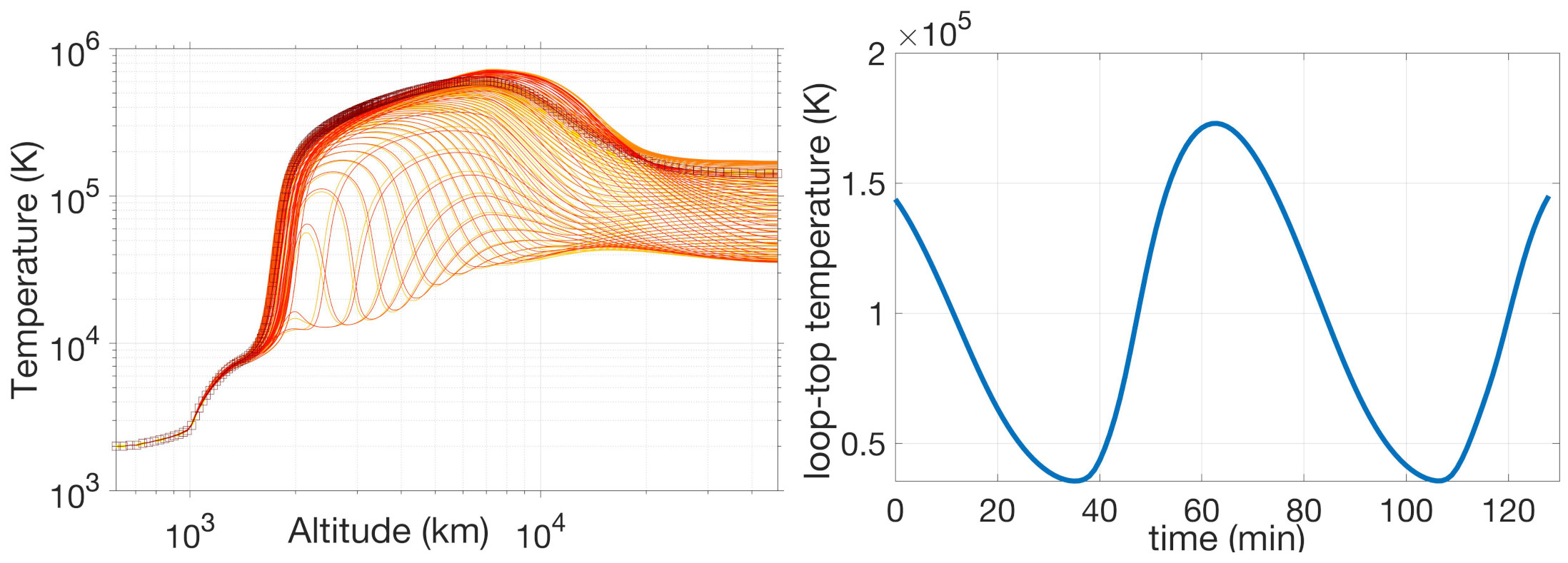}
\caption{Example of a full TNE cycle in a dynamic loop of half-length $153\ \rm{Mm}$, base flux heating $F_\odot=1.e6\ \rm{erg.cm^{-2}.s^{-1}}$ and heating scale height $H_f=2.1\ \rm{Mm}$ following the same format as in Figure \ref{fig:ISAM_Comp_Te}.
\label{fig:ISAM_TNE}}
\end{figure*}

I have performed several ISAM runs with distinct ($F_\odot\ \rm{(erg.cm^{-2}.s^{-1})},H_f\ \rm{(Mm)}$) pair parameters whose values are shown in Figure \ref{fig:Winebarger2003_fig6}. Different stability criteria are also traced in this figure as an indication of the parameter space where quasi-steady (above the curves) or unstable (underneath the curves) solutions of coronal loops are expected. The ISAM runs are all initialized with a quasi-steady state solution that I obtained from the ($1.4e6,7$) set of heating parameters, and that is based on the statistics of steady hydrostatic loops observed by \textit{TRACE} in \citet{Aschwanden2001}. Some pair parameters are well below the criteria established by \citet{Winebarger2003a} for a loop-top temperature of $1\ \rm{MK}$ (solid black line). However, not all ISAM runs maintained a loop-top temperature of $1\ \rm{MK}$ so the 1MK-curve of \citet{Winebarger2003a} should be slightly moved downward to better represent our ISAM simulations. The temperature profiles of all ISAM runs are plotted in Figure \ref{fig:ISAM_Comp_Te}. Among all these ISAM simulations, only the one with the smallest heating scale height of $H_f=4\ \rm{Mm}$ clearly is in an unstable state with a peak temperature that is well below the loop apex. Over time, the coronal part of the loop progressively enters in a critical cooling phase. A much longer simulation time would be required to see the full evolution of this unstable state past this cooling phase. For completeness, I show in Figure \ref{fig:ISAM_TNE} a full TNE cycle obtained in another configuration, for a longer loop with half-length $153\ \rm{Mm}$ and heating parameters ($1.e6,2.1$). In this extreme case, the heating scale height is well below the stability criteria (see black star symbol in Figure \ref{fig:Winebarger2003_fig6}) and hence TNE develops at a greater scale and more rapidly with a period of $\simeq 75\rm{min}$. As we shall see in section \ref{sec:ISAM_results_O}, TNE may play a significant role in the transport of heavy ions through the transition region. The multi-specie architecture of ISAM sounds well appropriate to study how heavy ions may behave in dynamic loops, that is left to future studies. In the following sections, I retain a set of coronal heating parameters ($1.2e6,10$) that allows to maintain a coronal loop in a quasi-steady state and with a loop-top temperature of about $1\ \rm{MK}$. \\

To conclude on this section, the presence of TNE in coronal loops have been observed, predicted theoretically and modeled in past studies, as well as partially addressed in this work with ISAM. That results in a redistribution of the coronal plasma over large scales, hence affecting the mass and heat fluxes significantly. As introduced in section \ref{subsec:intro_FIP}, a separation of certain heavy ions along the magnetic field is expected through friction and thermal diffusion along the magnetic field, which depend directly on the thermodynamic state of the loop. In the following I concentrate on the analysis of coronal loops in a quasi-steady state so that we can draw a first interpretation of the dynamics of heavy ions in the low solar atmosphere. The case of unstable loops will be treated in a future study.

\subsection{Including the upper chromosphere}
\label{subsec:ISAM_results_H_chromo}

As introduced in section \ref{subsec:intro_FIP}, most of the separation processes of heavy ions are expected to establish already in the upper chromosphere and transition region. Thermal diffusion processes, comprising of the thermal forces, are directly impacted by the heat flux profiles and hence by the energy budget in this region. In the following we then improve our energy balance in the upper chromosphere by accounting for more realistic radiative losses. The recipe of \citet{Carlsson2012} and the \textit{ChiantiPy} spectral code introduced in section \ref{subsec:ISAM_radloss_Carlsson2012} and \ref{subsec:ISAM_radloss_Carlsson2012} have been exploited to build a semi-empirical table of which its implementation in ISAM is described in detail in Appendix \ref{sec:ISAM_radloss_detailed}. \\

An additional non-radiative heating of the upper chromosphere is necessary to balance for these new radiative losses. That includes contributions from many different processes that include shock dissipation, waves reflection, dissipation and mode conversion, and more broadly magnetic reconnection. To a first approach, the ad-hoc formulation used for the coronal heating has been found very efficient for sufficiently low-scale height $H_f$ and high base flux $F_\odot$. In contrast to the coronal heating discussed in the previous section, it was much more delicate to calibrate the parameters for chromospheric heating. This is because the amount of energy that is deposited in the upper chromosphere and transition region affects significantly the coronal temperatures by driving more or less mass flow through the transition region. It is the balance between the radiative losses and the downward heat flux that determines the density at the base of the corona \citep{Hansteen2012}. Therefore the aim was to heat sufficiently the upper chromosphere while not depositing too much energy in the transition region that would otherwise induce chromospheric evaporation. A fine tuning has then been performed whereby coronal temperatures were maintained while keeping a heating flux in the corona consistent with the observations and the scaling laws of \citet{Aschwanden2001}. \\

\begin{figure*}[]
\centering
\includegraphics[width=0.95\textwidth]{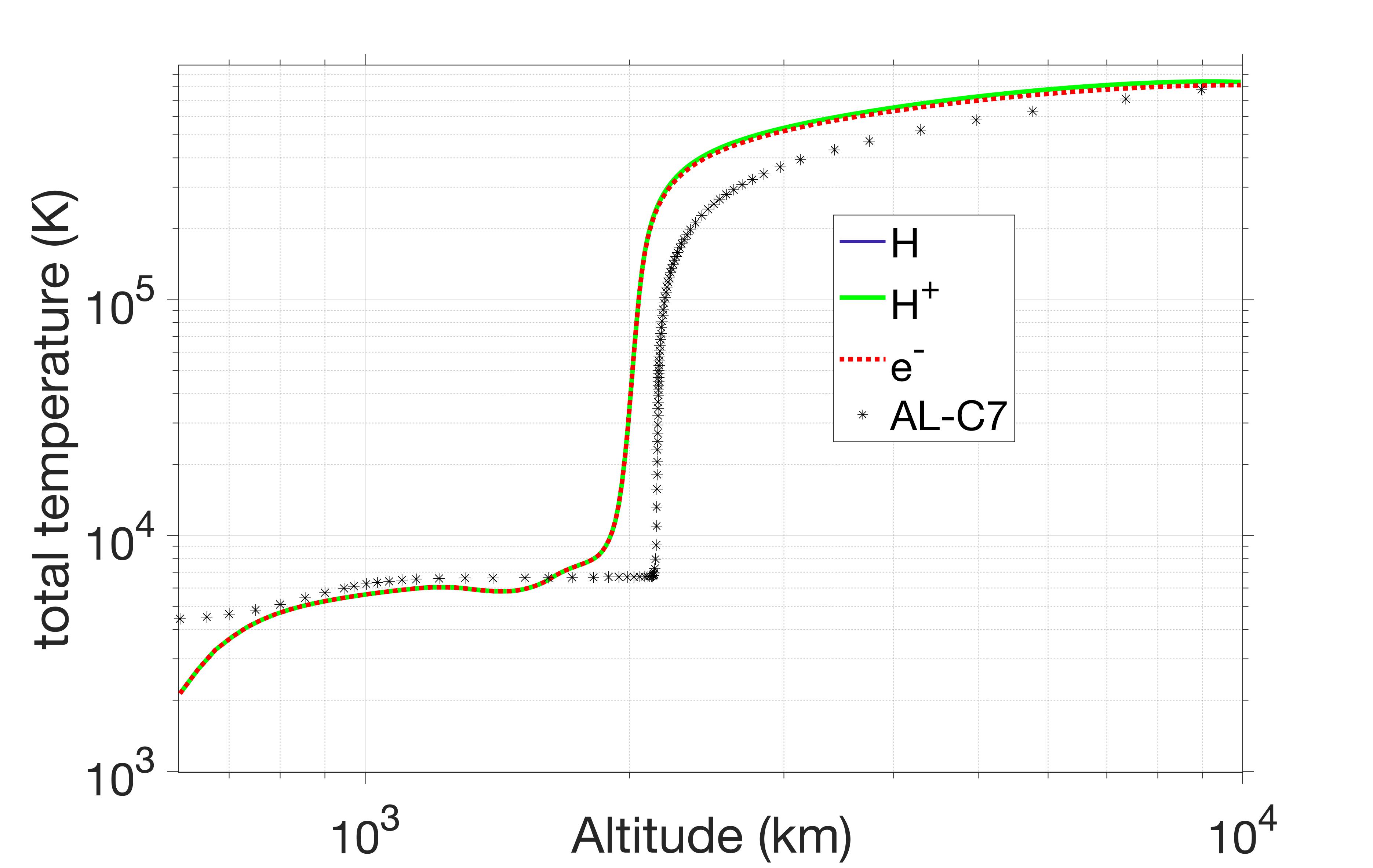}
\caption{ Total temperatures of electron (dashed red line), protons (solid green line) and neutral Hydrogen (solid blue line) computed by ISAM. The neutral Hydrogen and protons temperatures overlap everywhere in the loop which remains highly collisional. The temperature fit from the AL-C7 model is given as black star symbols.
\label{fig:ISAM_H_T}}
\end{figure*}

After many trials, I obtained an optimal set of chromospheric heating parameters ($F_\odot=4.7e6,0.42$) where the average quiet-Sun chromospheric profiles from \citet{Avrett2008} (AL-C7 semi-empirical model) served as a baseline for the calibration. This simulation has been let run until the loop reached a quasi-hydrostatic state with bulk speeds that remain under $\approx 0.3\ \rm{km/s}$. \\

\begin{figure*}[]
\centering
\includegraphics[width=0.85\textwidth]{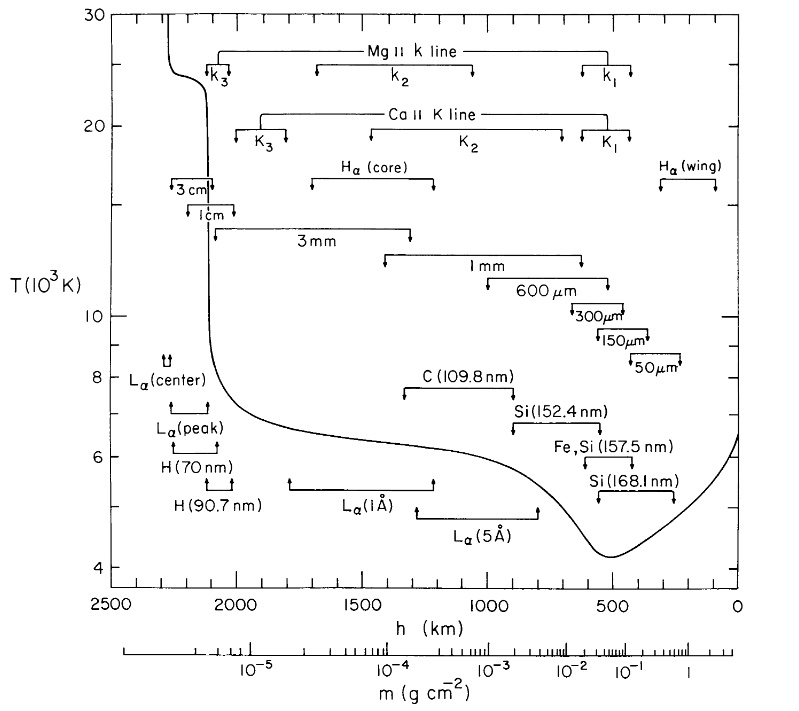}
\caption{ Contributions from the most prevalent chromospheric lines to the average quiet-Sun temperature of the chromosphere. Figure taken from \citet[][Figure 1]{Vernazza1981}.
\label{fig:Vernazza1981_fig1}}
\end{figure*}

The resulting total temperatures for neutral Hydrogen, protons and electrons are shown in Figure \ref{fig:ISAM_H_T}. The loop remains isothermal thanks to the frequent collisions between the species. Slight departures are still visible in the upper most portion of the loop where collisions become more scarce, and hence protons tend to retain the external energy that has been given to them only (this choice is discussed in section \ref{sec:ISAM_heating}). The temperature progressively rises from $\approx 2000\ \rm{K}$ at the bottom boundary ($600\ \rm{km}$) to reach $\approx 6500\ \rm{K}$ at the base of the transition region. This is the result of a continuous decrease of the total radiative losses, where neutral Hydrogen starts to ionize and hence becomes less efficient to radiate the energy. After a short chromospheric plateau at $\approx 6000\ \rm{K}$, there is a first significant rise in the temperature (from $\approx1500\ \rm{km}$) that has been also observed in the advanced hydrodynamic simulations of \citet{Carlsson2002a}. This is the effect of the hot corona radiating on the upper chromosphere. This effect is included in ISAM where the radiative losses given by \citet{Carlsson2012} (see section \ref{subsec:ISAM_radloss_Carlsson2012} and Figure \ref{fig:ISAM_lambda}) account for radiative heating in Lyman-alpha, which is a dominant spectral line in the upper chromosphere and transition region as shown in Figure \ref{fig:Vernazza1981_fig1}. Above the transition region, our solution starts to deviate from the AL-C7 model. As for the scaling laws of coronal heating established by \citet{Aschwanden2001}, the observations of coronal loops in EUV constitute the observational baseline of the AL-C7 model. However, the AL-C7 model assumes a much simpler corona where there is no external heating source and where the coronal temperature has been set so that radiative losses perfectly balance a given downward heat flux. Therefore the AL-C7 model is naturally similar to the solutions with uniform heating determined by \citet{Aschwanden2001} whereas observations rather suggest a nonuniform heating that is concentrated near the transition region and that tends towards a more iso-thermal temperature profile in the corona. \\

As already discussed in section \ref{subsec:intro_chromo}, semi-empirical models of the chromosphere should be used with caution as recent studies based on observations and advanced modelling depict a chromosphere that is highly dynamic in nature and that hence questions the relevance of a quasi-stationnary state \citep{Carlsson2002a,Carlsson2007}. In particular these authors argue that these semi-empirical models may overestimate real chromospheric temperatures, by being biased by their observational base that captures mostly the peak temperature of shocks that propagate in the upper chromosphere. Under such considerations, the AL-C7 chromospheric profile for the temperature has been used as an upper bound throughout the calibration of the chromospheric non-radiative heating. \\

\begin{figure*}[]
\centering
\includegraphics[width=0.85\textwidth]{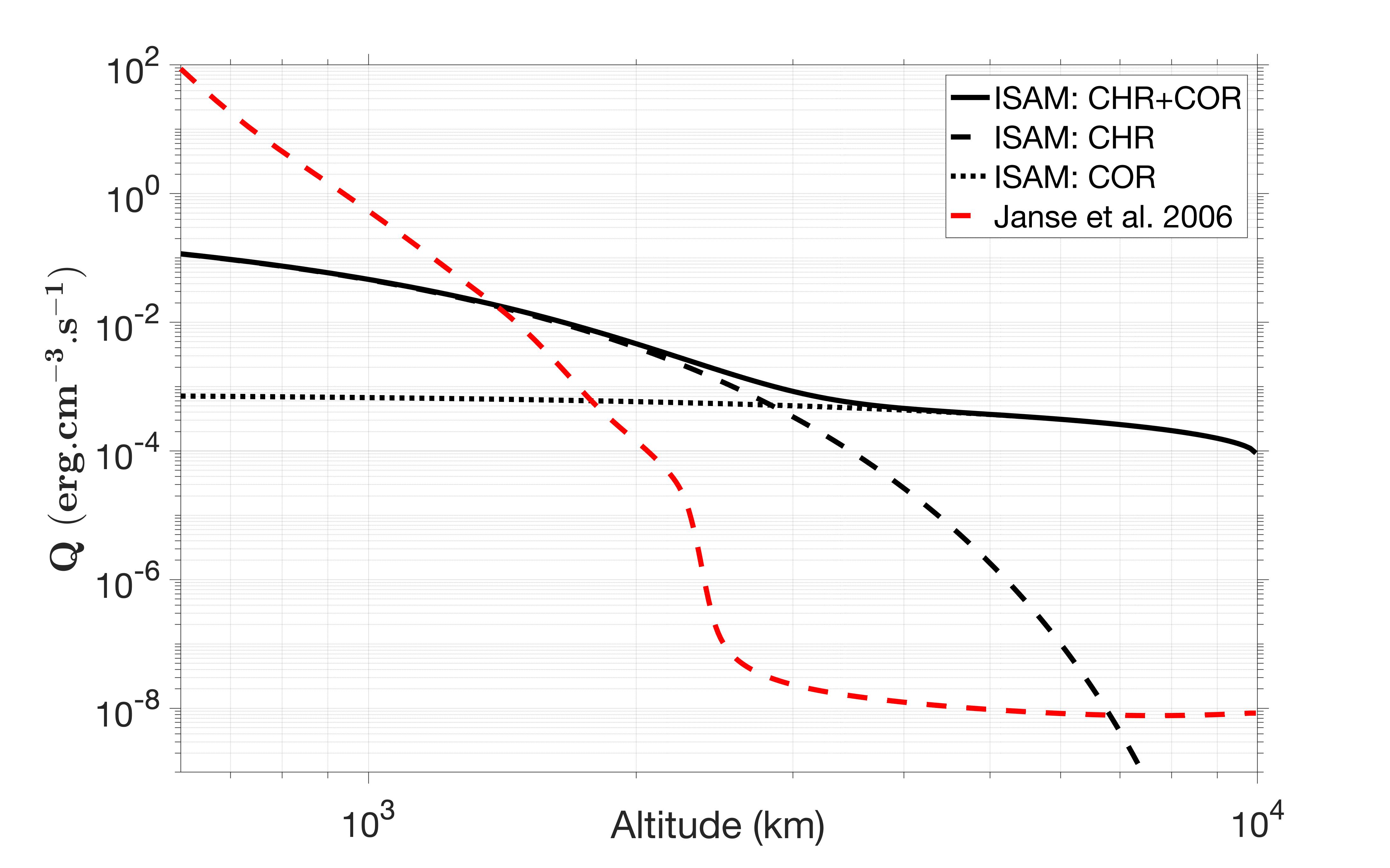}
\caption{ Total (solid black line), chromospheric (dashed black line) and coronal (dotted black line) ah-hoc heating laws adopted in ISAM. The chromospheric heating prescription of \citet{Janse2006} is plotted as a dashed red line.
\label{fig:ISAM_H_Q}}
\end{figure*}

The optimal chromospheric heating profile is traced in Figure \ref{fig:ISAM_H_Q} together with the one derived for the corona in the previous section. The chromospheric heating model of \citet{Janse2006} is also plotted as a comparison, which is much greater than the chromospheric heating rate adopted in ISAM. In fact, their profile has been derived assuming a perfectly optically thin medium and hence they significantly overestimated the total radiative losses. The energy input to heat up the chromosphere in ISAM equals $F_\odot=5\ \rm{kW.m^{-2}}$ that is very close to the $F_\odot=4.2\ \rm{kW.m^{-2}}$ value derived from the VAL3-C semi-empirical chromospheric model of \citet{Vernazza1981} \citep[see also][]{Carlsson2007}. \\

\begin{figure*}[]
\centering
\includegraphics[width=0.85\textwidth]{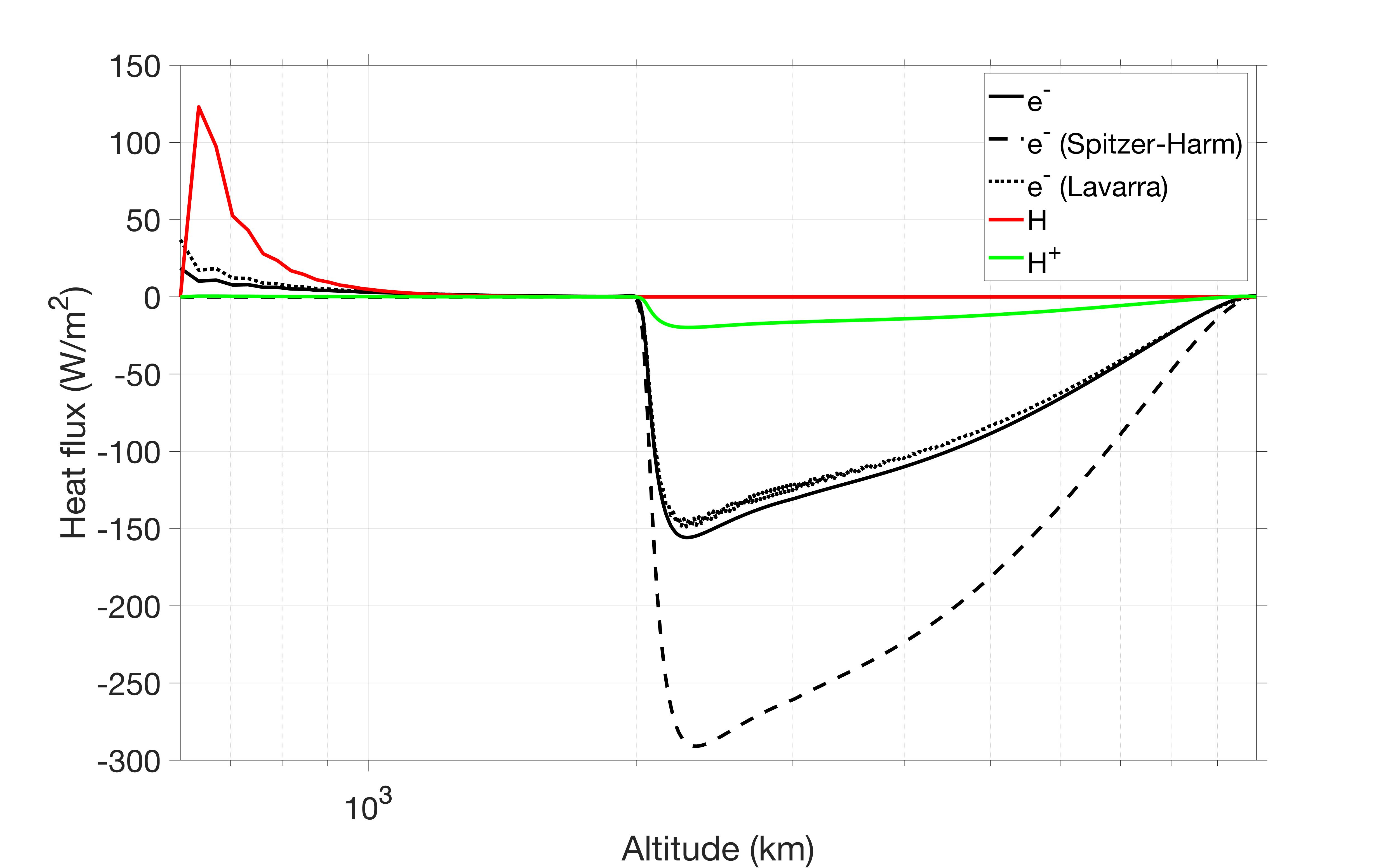}
\caption{ Total heat fluxes of electrons (solid black line), protons (solid green line) and neutral Hydrogen (solid red line). The classical Spitzer-Härm heat flux $F_c=-\kappa_0T^{5/2}\nabla_s T$ with a conductivity $\kappa_0=9.2\times 10^{-7}\ erg.cm^{-1}.s^{-1}.K^{-7/2}$ is plotted as a long dashed black line. The approximation of the total electron heat flux solved in ISAM following \citet[][eq 30-31]{Lavarra2022} is traced as a short dashed black line. Positive heat fluxes are directed outward.
\label{fig:ISAM_H_Fc}}
\end{figure*}

The total heat fluxes given by the optimal chromospheric solution are plotted in Figure \ref{fig:ISAM_H_Fc}. The usual Spitzer\textendash Härm approximation for the collisional regime is also shown for a comparison, that is about twice larger than the one calculated in ISAM. That difference is built in our approximation of the VDF that has been used to derive the transport equations in ISAM (see section \ref{sec:ISAM_origin}). \citet{Killie2004} have shown that this drawback can be prevented by using a higher-order approximation of the VDF, but at the cost of a derivation of the collision terms that becomes much more delicate in the 16-moment approach. This underestimation of the heat flux in ISAM will have an impact on the amplitude of the thermal force (eq \ref{eq:ISAM_col_u}) that pushes heavy ions out of the chromosphere, that is discussed in the next sections. An additional departure to the classical Spitzer\textendash Härm heat flux is also noted in the chromosphere where neutral Hydrogen is dominant and not accounted for in the classical theory. By including the contribution from the collisions of electrons with neutral Hydrogen, a general approximation for the electron heat flux that remains valid throughout the chromosphere to the corona has been given in \citet[][eq 30-31]{Lavarra2022}, that is also plotted in Figure \ref{fig:ISAM_H_Fc}. \\

In coronal loops, most of the energy deposited in the corona is conducted downward through the transition region to be dissipated by radiation. As a consequence, the conductive heat flux is much larger than in open wind solutions where most of the energy is used to accelerate the plasma and is hence dragged along with the solar wind. For the strong conductive heat flux to be dissipated in coronal loops, a higher density is required so that radiative losses can dissipate the energy flowing down from the corona. The collisional coupling of heavy ions with protons is hence much more significant in coronal loops than in open fields, that may explain the high variability of the composition in heavy ions along coronal loops as we shall see in the next sections.

\subsection{Collisional coupling between Hydrogen and protons}
\label{subsec:ISAM_results_H_forces}

As pointed out in past studies that investigated the role of diffusion processes on the FIP effect \citep[see e.g.][]{Peter1998b,Killie2005,Killie2007,Bo2013}, protons have a major impact on the transport of heavy ions through the Coulomb interaction. It has been shown especially that the down-streaming of protons can prevent low-FIP heavy elements (which ionize early in the chromosphere) from reaching the corona. Therefore it is necessary to analyse first the proton dynamics before including heavy ions in the system. As a matter of complexity, I present in this thesis the case of an idealized chromosphere presented in section \ref{subsec:ISAM_results_H_thermodynamics}. The more realistic chromospheric profiles obtained in section \ref{subsec:ISAM_results_H_chromo} will be exploited in a subsequent work. Parameters for coronal heating are chosen as ($F_\odot=1.e6\ \rm{erg.cm^{-2}.s^{-1}}$,$H_f=7\ \rm{Mm}$) for a loop in quasi-steady thermodynamic equilibrium (see section \ref{subsec:ISAM_results_H_thermodynamics}). \\

\begin{figure*}[]
\centering
\includegraphics[width=0.95\textwidth]{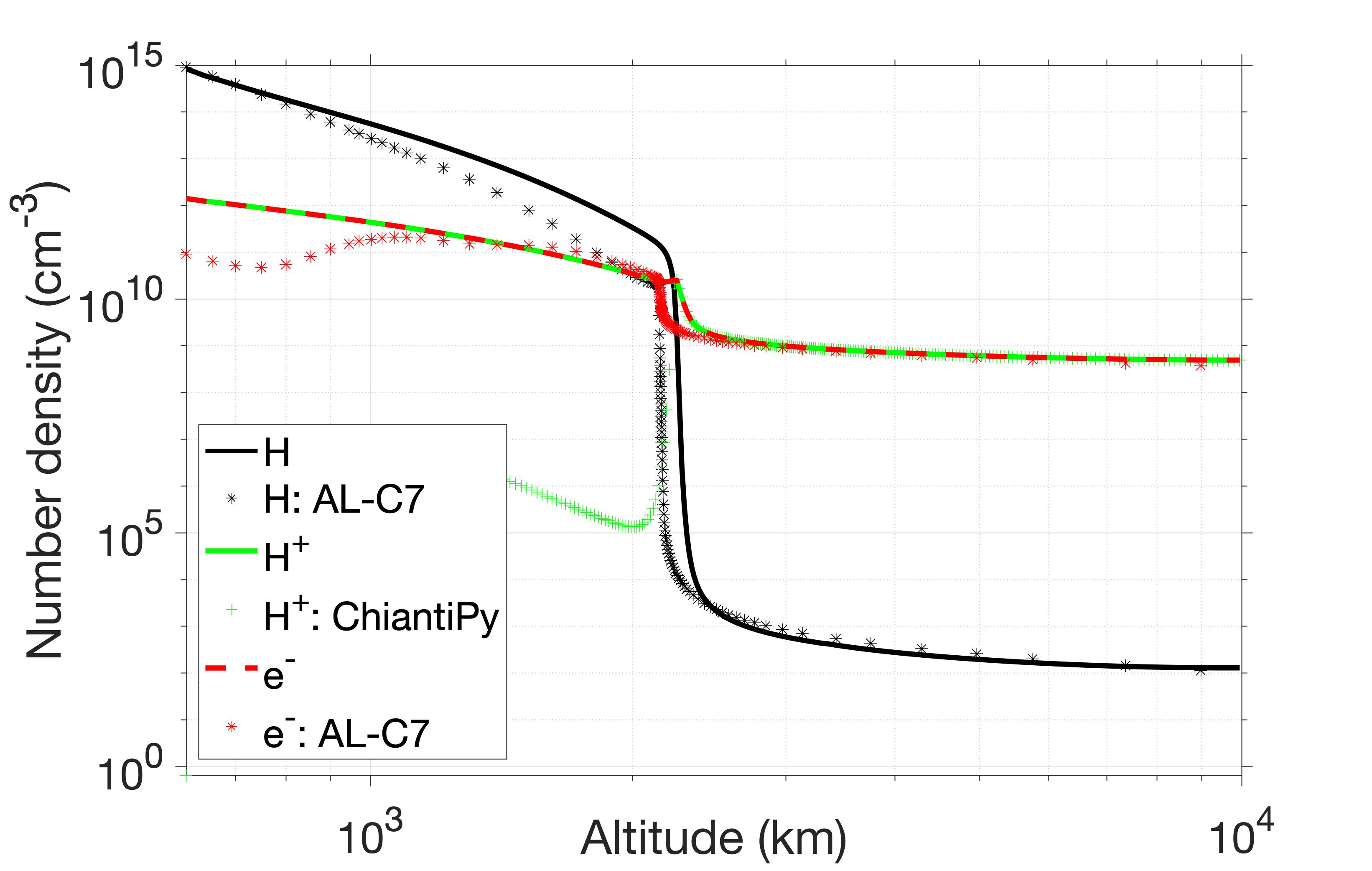}
\caption{ Number densities of electrons (dashed red line), protons (solid green line) and neutral Hydrogen (solid black line) computed by ISAM. The neutral Hydrogen and electron density profiles from the AL-C7 model are marked by black and red star symbols respectively. The proton density computed by the \textit{ChiantiPy} spectral code given the ISAM neutral Hydrogen profile is plotted as green crosses.
\label{fig:ISAM_H_n}}
\end{figure*}

The resulting quasi-steady ISAM solution is shown in Figure \ref{fig:ISAM_H_n} for the number densities, while the temperature profiles are very close to the ($1.4e6,7$) ISAM run already presented in Figure \ref{fig:ISAM_Comp_Te}. In the chromosphere, the neutral Hydrogen density mostly follows a hydrostatic law of which the slope is determined by the local temperature. As we shall see below, low-FIP elements that are already ionized in the chromosphere are significantly coupled with protons through the Coulomb collisions, where the strength of the interaction is highly controlled by the proton density. As already discussed in section \ref{sec:ISAM_ioniz}, the photoionization of neutral Hydrogen that I implemented in ISAM only constitutes a rough estimate of the photoionization rate in the upper chromosphere. Although a proper treatment of radiative transfer would be necessary for a better description of the ionization/recombination processes in the chromosphere, \citet{Carlsson2002a} have shown that the problem can be reduced to a neutral Hydrogen atom with only a few levels and transition rates, that is at reach for future mid-term implementations in ISAM. For the purpose of this thesis where we focus our attention on the upper chromosphere and corona where most of the composition is expected to settle, the current implementation of photoionization in ISAM is accurate enough by providing a mid profile of the proton density to the Al-C7 model. When reaching the transition region and corona, ionization of neutral Hydrogen mostly occurs through collision with electrons that is commonly called the "coronal approximation". In this approximation the ionization/recombination processes are well known and can be described accurately using analytical fits and tabulated data that I implemented in ISAM (see section \ref{sec:ISAM_ioniz}). Similar formulas are used in the \textit{ChiantiPy} spectral code that leads to a proton density in the corona that matches closely the ISAM solution. \\

\begin{figure*}[]
\centering
\includegraphics[width=0.85\textwidth]{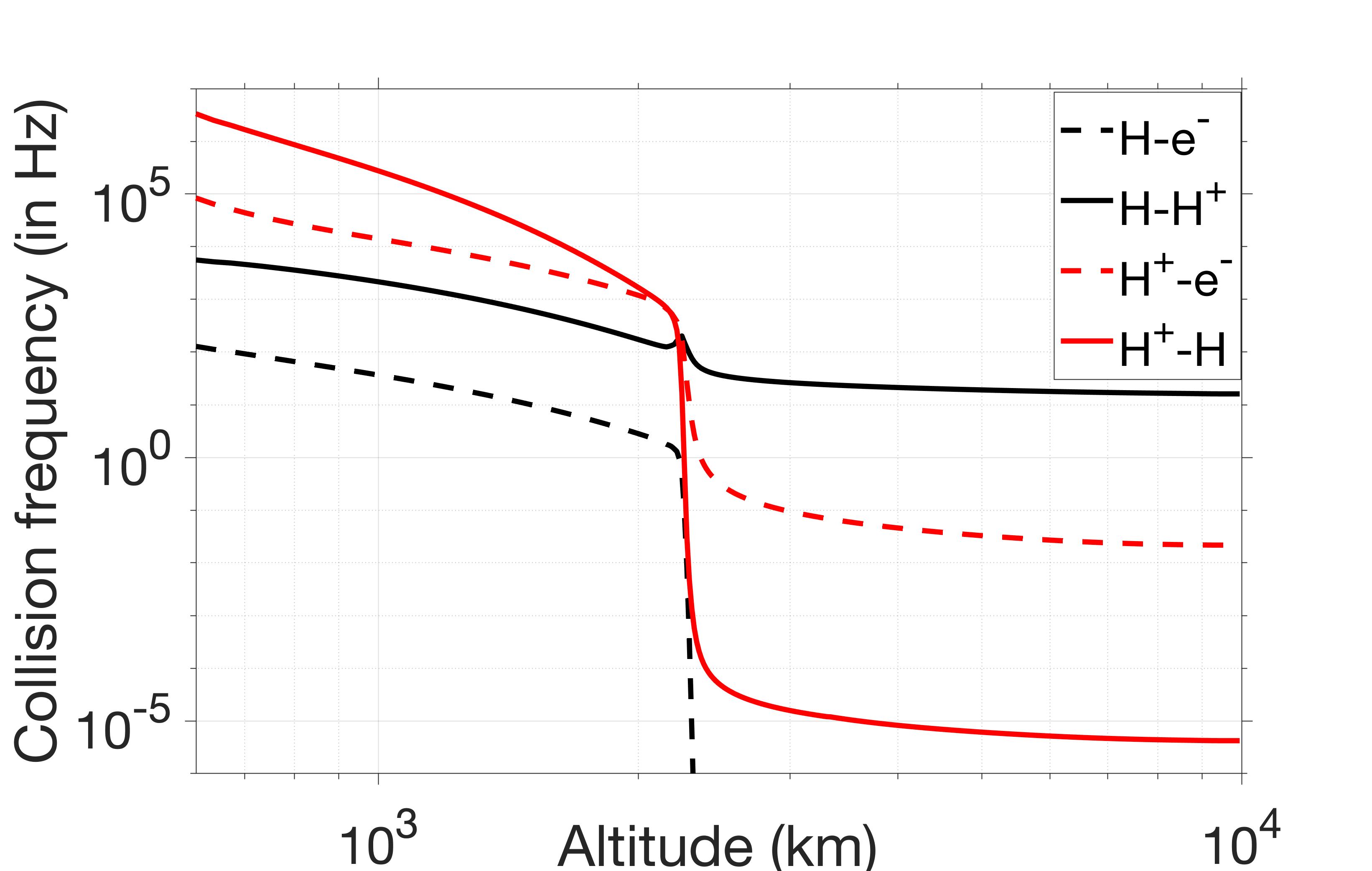}
\caption{ Collision frequencies of neutral Hydrogen with electrons and protons (dashed and solid black lines), and of protons with electrons and neutral Hydrogen (dashed and solid red lines).
\label{fig:ISAM_H_nu}}
\end{figure*}

We present in Figure \ref{fig:ISAM_H_nu} the collisional couplings in the Hydrogen-proton-electron atmosphere solved in ISAM. That includes the charge-exchange resonant interaction between neutral Hydrogen and protons (see section \ref{subsec:ion_neutral_col}), the two-body elastic electron-neutral interaction (see section \ref{subsec:ion_neutral_col}), and the Coulomb interaction between protons and electrons (see section \ref{subsec:coulomb_col}). The few protons that are ionized in the mid/upper chromosphere are closely coupled with the dominant neutral plasma. To first order the resonant collision frequency of protons with neutrals scales as $n_H T^{1/2}$, and is hence mostly controlled by the decrease of the neutral Hydrogen density throughout the chromosphere as ionization proceeds. In the upper chromosphere, the electrons produced from ionization of neutral Hydrogen become dominant and prevail on the collisional coupling with protons through the Coulomb interaction. This proton-electron coupling then rapidly weakens throughout the transition region where the hot coronal temperatures strongly diminish the efficiency of Coulomb collisions which scale as $n_e T^{-3/2}$. \\

\begin{figure*}[]
\centering
\includegraphics[width=1.0\textwidth]{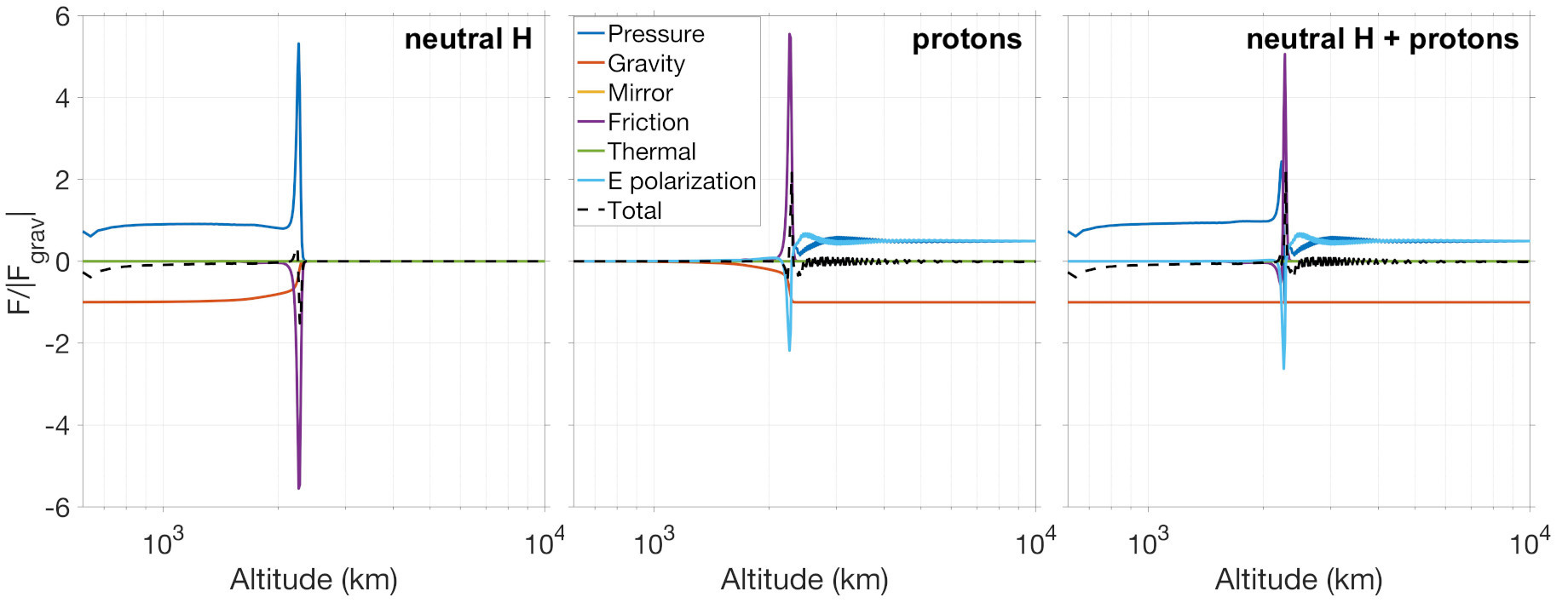}
\caption{ Forces acting on neutral Hydrogen (left), protons (middle) and on both (right). All forces are first multiplied by the ionization fraction ($n_X/n_{[X]}$), and then normalized by the absolute value of gravity. Positive forces are directed outward.
\label{fig:ISAM_H_forces}}
\end{figure*}

After a relaxation time of $\simeq 90\ \rm{min}$, neutral Hydrogen and protons have almost completely settle in the chromosphere and corona respectively, which are in a quasi hydrostatic equilibrium as shown in Figure \ref{fig:ISAM_H_forces}. This equilibrium is reached through a perpetual adjustment of the internal pressure of the system until the gravity pull is balanced. That is fulfilled in the chromosphere where neutrals are dominant, and also in the corona where both the protons, and electrons through the electrostatic field (eq \ref{eq:ISAM_Fpol}), contribute at counterbalancing the gravity. Furhermore, the quasi-neutrality condition assumed in ISAM (eq \ref{eq:ISAM_ne}) imposes that $n_e=n_p$ in a pure Hydrogen plasma. Therefore any discrepancy between the proton and electron contributions to the thermal force is induced by different profiles of proton and electron temperatures. Such discrepancy is visible just above the transition region and is likely the result of the electrons that carry more thermal energy into that region thanks to their enhanced heat flux (see Figure \ref{fig:ISAM_H_Fc}). \\

A junction between these two main hydrostatic equilibria occurs at the transition region. Contrary to neutral Hydrogen, protons face a strong downward pressure force in the transition region that acts as a barrier to the transfer of protons into the corona. Therefore a differential flow appears that generates a net friction force between protons and neutrals, which reads $m_p n_p \nu_{pH}(u_H-u_p)$ for protons or inversely $m_H n_H \nu_{Hp}(u_p-u_H)$ for neutrals where polarization effects are negligible (see eq \ref{eq:ISAM_col_ui}). Note that there is no friction between protons and electrons since we assume a current-free ambipolar flow in ISAM (see eq \ref{eq:ISAM_ue}) and hence $u_e=u_p$ in an atmosphere that is only constituted of protons. \\

A thermal force can arise from thermal diffusion effects between charged species that interact through Coulomb collisions because of the dependency of the collision frequency on the temperature. Basically, charged particles will be tempted to move towards regions of higher temperatures where Coulomb collisions are less frequent, that is from the cool chromosphere to the hot corona. In a loop that remains highly collisional with small pressure anisotropies, the expression of the thermal force that results from the interaction of an ion $i$ with another specie $t$ (i.e. an ion, electron or neutral) can be reduced to (see eq \ref{eq:ISAM_col_ui}):
\begin{subequations}
\begin{align}
    n_i m_i \left[\nu_{it}\frac{z_{it}}{k_b T_{it}(m_i+m_t)}\left(m_t\frac{q_i}{n_i}-m_i\frac{q_t}{n_t}\right) + Z_i\frac{m_e}{m_i}\nu_{et}\frac{z_{et}}{k_b T_{et}(m_e+m_t)}\left(m_t\frac{q_e}{n_e}-m_e\frac{q_t}{n_t}\right) \right]\quad \text{for }t\neq e \\
    n_i m_i \left[\nu_{ie}\frac{z_{ie}}{k_b T_{ie}(m_i+m_e)}\left(m_e\frac{q_i}{n_i}-m_i\frac{q_e}{n_e}\right) + Z_i\frac{m_e}{m_i}\nu_{ei}\frac{z_{ei}}{k_b T_{ei}(m_e+m_i)}\left(m_i\frac{q_e}{n_e}-m_e\frac{q_i}{n_i}\right) \right]\quad \text{for }t=e \label{eq:ISAM_Thermal_b}
\end{align}
\end{subequations}
which accounts for the contributions from the electrostatic polarization field (eq \ref{eq:ISAM_Fpol}). 

The thermal force cancels out exactly in the proton-electron Coulomb interaction ($i=p$ in eq \ref{eq:ISAM_Thermal_b}), since we assume elastic collisions $n_p m_p \nu_{pe}=n_e m_e \nu_{ep}$ and quasi-neutrality $n_p=n_e$. Furthermore, collisions of protons with neutrals are too rare in the transition region to produce a significant thermal force, that is also explained by the fact that protons and neutrals have similar normalized heat fluxes $q_H/n_H$ and $q_p/n_p$ in that region. That justifies our treatment of proton-neutral collisions in ISAM as ideal Maxwell molecule interactions which do not include the thermal force ($z_{it}=0$ in eq \ref{eq:ISAM_col_ui}). Therefore the thermal force likely does not play any role in a solar atmosphere that is only constituted of neutral Hydrogen, protons and electrons. \\

\begin{figure*}[]
\centering
\includegraphics[width=0.85\textwidth]{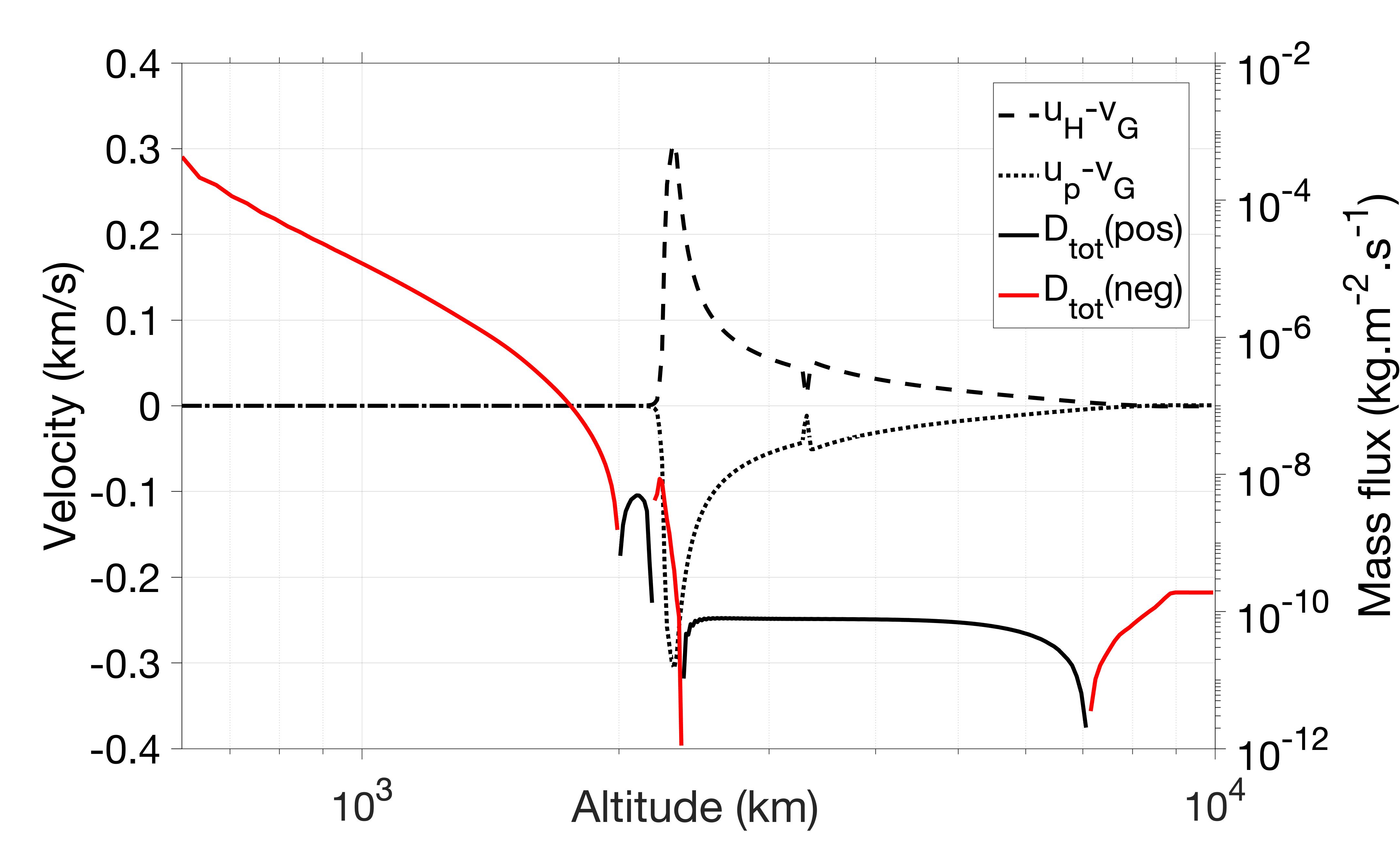}
\caption{ Bulk velocities of neutral Hydrogen and protons (dashed and dotted black lines) expressed in the center of mass $v_G=(m_H u_H+m_p u_p)/(m_H+m_p)$. The total mass flux of the Hydrogen element is plotted on a logarithmic scale that is shown on the right-hand side and where positive and negative values are denoted by black and red colors respectively.
\label{fig:ISAM_H_u}}
\end{figure*}

As noted above, there is an ambipolar flow that forms between neutral Hydrogen and protons that is shown in Figure \ref{fig:ISAM_H_u}. When expressed in the frame associated to the center of mass of neutrals and protons, this ambipolar flow becomes evident with speed profiles that are perfectly opposite. That does not mean that there is no mean flow (plotted on the right-hand side y-axis) despite being close to a quasi-steady state solution. As noted by \citet{Killie2005} and \citet{Killie2007}, this ambipolar flow remains persistent in the ionization layer of Hydrogen as long as the temperature rises at the top of the transition region during chromospheric evaporation phases. The ionization equilibria that was previously established in this region is then perturbed where ionization of neutrals is enhanced due to the higher temperature that hence produces a net outflow to fulfil the increased demand in neutrals. Such evaporation phases have been frequently observed and modeled in dynamic loops that undergo TNE cycles as already discussed in section \ref{subsec:ISAM_results_H_thermodynamics}. Chromospheric evaporation phases can also be initiated by heating pulses near the transition region during magnetic reconnection events for instance. Whenever this ambipolar flow is present, protons flow down the transition region and hence form a "barrier" to any element that is already ionized in this region. That effect is expected to weaken in open solar wind solutions where most of the energy is carried away from the transition region, and hence would explain why closed-field and open-field plasma show different compositions overall. This will have a major impact on the transfer of heavy ions through the transition region where a filtering of heavy elements will naturally occur according to their FIP as we shall see in the next section. 

\begin{figure*}[]
\centering
\includegraphics[width=0.8\textwidth]{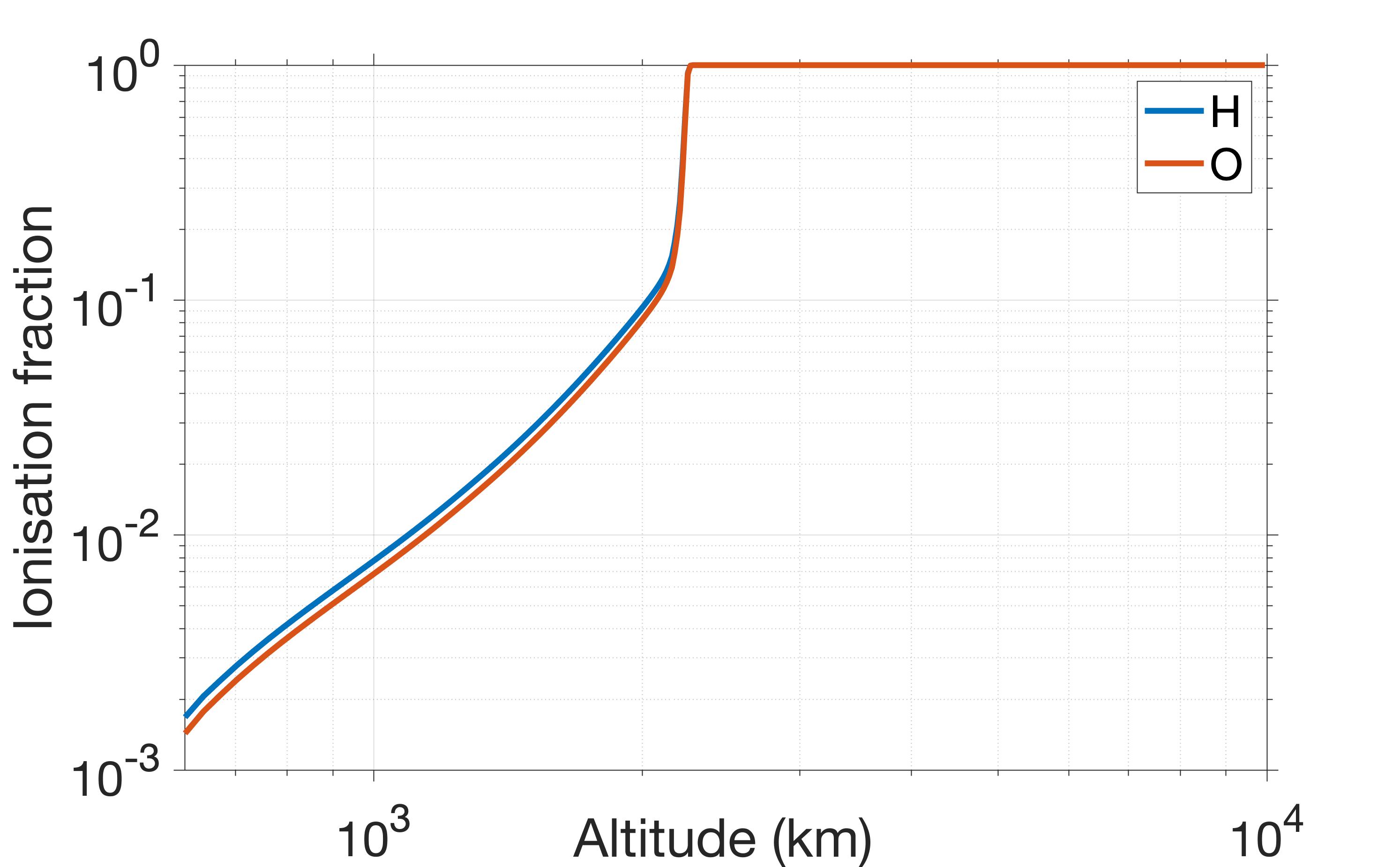}
\caption{ Total ionization fractions of Oxygen and Hydrogen. 
\label{fig:THESE_4_L16_O_A_Frac}}
\end{figure*}

\clearpage
\section{Separation of heavy ions: application to Oxygen}
\label{sec:ISAM_results_O}

Except for Helium that makes about $5-10\%$ of the composition of the solar atmosphere, heavy elements can be considered as minor/trace species that have no feedback on the Hydrogen-proton-electron plasma. In this case, minor species can be solved at a second stage in ISAM for a fixed Hydrogen-proton-electron atmosphere. That allows us to greatly accelerate the simulations by removing the hard time-resolution constrain from the electrons. Although the energy and heat flux transport equations for electrons are not solved in this case, the electron density and bulk velocity are still re-computed at each time step to ensure the condition of quasi-neutrality and ambipolar flows (eq \ref{eq:ISAM_ne}-\ref{eq:ISAM_ue}). In the following, the background atmosphere is taken to be the pure Hydrogen-proton-electron atmosphere presented in section \ref{subsec:ISAM_results_H_forces}. The photospheric abundances relative to Hydrogen from \citet[][Table 2]{Avrett2008} are set at the base of the loop. Practically, that determines at the base of the loop either the density of the neutral or of the first ionization state according to whether the element has a high or low FIP respectively (see also section \ref{subsec:ISAM_BC}). \\

Oxygen is 16 times heavier than Hydrogen but still has a FIP of $\simeq 13.62\ \rm{eV}$ that is very close to that of Hydrogen $13.60\ \rm{eV}$. Another peculiarity of Oxygen is its strong coupling with Hydrogen through the resonant charge-exchange reactions $O + H^+ \leftrightarrow O^+ + H$. Therefore Oxygen has a total ionization fraction in the chromosphere that closely follows that of Hydrogen as shown in Figure \ref{fig:THESE_4_L16_O_A_Frac}, where about $40\%$ of the element is in an ionized state at the base of the transition region at around $2000\ \rm{km}$. For a first application I limit myself to the first four ionized states of Oxygen. In practice, the system could be readily extended to include all ionisation states of Oxygen, but then that would require additional data on the ionization/recombination rates and on the collision frequencies for which I already dedicated a significant time to gather and implement in ISAM (see section \ref{sec:ISAM_collisions} and \ref{sec:ISAM_ioniz}). \\

\begin{figure*}[]
\centering
\includegraphics[width=0.9\textwidth]{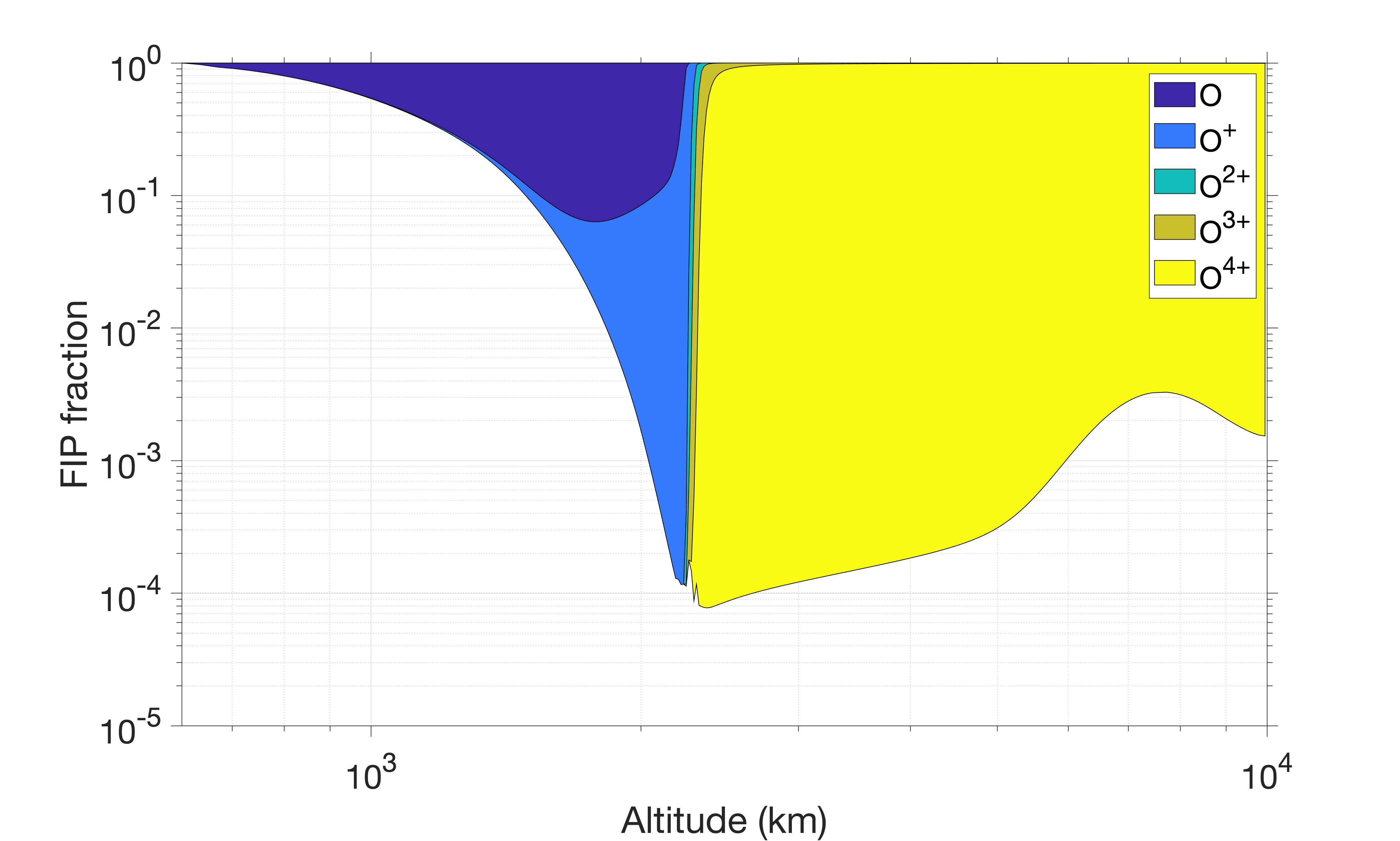}
\caption{ FIP bias of Oxygen relative to Hydrogen $(n_{[O]}/n_{[H]})_{alt}/(n_{[O]}/n_{[H]})_{PHO}$ function of altitude. Colors are indicative of the proportion of each specie (neutral and ions) in the total Oxygen abundance.
\label{fig:ISAM_O_FIP}}
\end{figure*}

As already discussed in section \ref{subsec:intro_FIP}, the level of chromospheric mixing is likely determinant to the amount of heavy ions that can get into the corona through the transition region. I consider first a case of a stratified chromosphere. By starting from a loop that is totally empty in Oxygen, I let the loop fill in Oxygen from the lower boundary until the composition in Oxygen starts to settle. The result is shown in Figure \ref{fig:ISAM_O_FIP} for a loop that run about $11\ \rm{hr}$. The common FIP bias is used as an indicator of the enhancement or depletion of heavy ions with respect to the photospheric abundances, that is defined as the ratio of the relative abundances with respect to Hydrogen between a certain altitude and the base of the loop. According to this definition, the FIP bias equals one at the lower boundary where photospheric abundances are assumed. Then a FIP bias that is either larger or lower than one means either a enrichment or depletion with respect to the photospheric abundances. A first look at Figure \ref{fig:ISAM_O_FIP} indicates a strong depletion of Oxygen in a large portion of the loop with a depletion down to $\approx 10^{-4}$. The fraction of Oxygen that manages to get through the transition region then slightly accumulates higher up in the corona at around $7-8\ \rm{Mm}$. The global depletion of Oxygen is likely a consequence of its mass being 16 times greater than that of Hydrogen and as a result Oxygen is much hardly lifted up in the atmosphere. To better understand which state of Oxygen contribute to the FIP bias locally, the proportion of each specie within the Oxygen element is color plotted under the curve. The thicker the area, the more abundant is the Oxygen specie at that altitude. \\

\begin{figure*}[]
\centering
\includegraphics[width=1.0\textwidth]{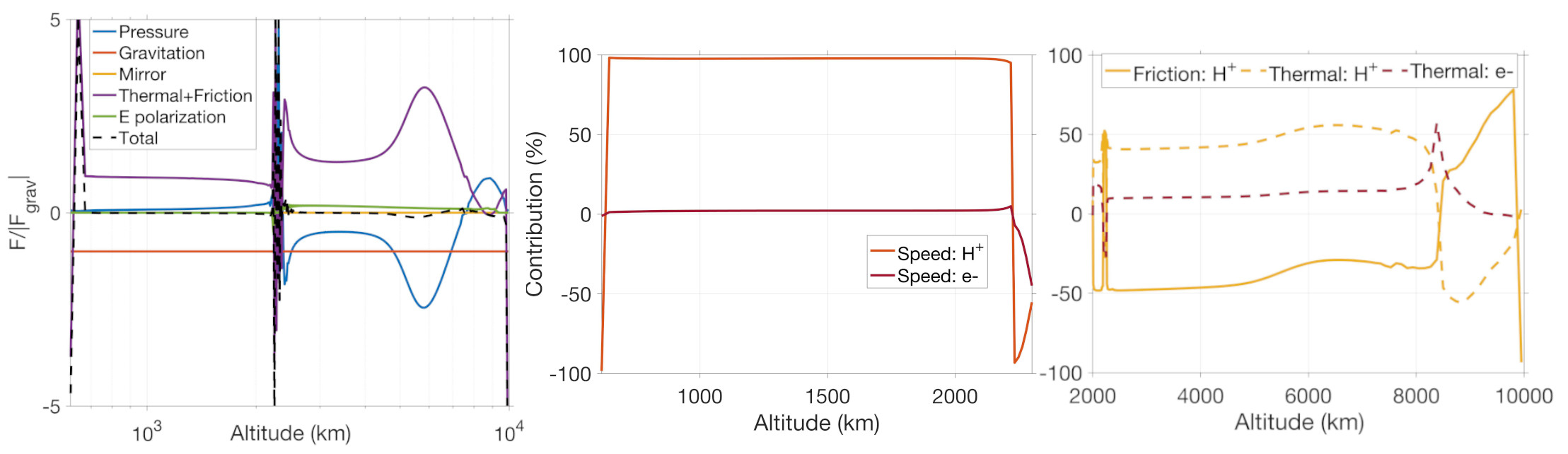}
\caption{ Left panel: summed up contributions of all forces that apply to the Oxygen element using the same normalizing procedure as in Figure \ref{fig:ISAM_H_forces}. The particle-particule interactions that contribute the most to the total friction and thermal forces that apply to neutral Oxygen (middle panel) and $O^{4+}$ (right panel) are also shown. 
\label{fig:ISAM_O_F}}
\end{figure*}

I found this format convenient to analyse the role of each specie in the total fractionation of Oxygen relative to Hydrogen. Therefore I can concentrate my analysis on neutral Oxygen and $O^{4+}$ which constitute most of the chromosphere and corona respectively in the present ISAM run. I should also mention that Oxygen is likely to be ionized at a higher level and that the higher ionization levels of Oxygen will be accounted for in a subsequent paper. The forces at play in the overall Oxygen system are plotted in Figure \ref{fig:ISAM_O_F} using the same normalizing procedure as in Figure \ref{fig:ISAM_H_forces}. Here the thermal and friction forces have been summed up together for a better readability of the graph. Their summed contribution then allows us to better see the influence of external species on the Oxygen system through collisions. For completeness, the separate contributions of friction and thermal effects to the total external collisional force applying to neutral Oxygen and $O^{4+}$  are shown in the middle and right-hand side panels of Figure \ref{fig:ISAM_O_F}. \\

In contrast to the analysis made in section \ref{subsec:ISAM_results_H_forces} for the pure Hydrogen atmosphere, Oxygen is only partially stratified in the chromosphere where the collisional coupling with external species substitutes partially the pressure force to sustain the gravity pull. If at opposite the pressure force had completely fulfilled its role at balancing the gravity, Oxygen would have fully stratified in the chromosphere. According to the scale height $k_b T_s/(m_s g)$ of Oxygen that is $16$ times shorter than that of Hydrogen, that would have led to a depletion of Oxygen relative to Hydrogen of about $8\times 10^{-52}$ at the top of the chromosphere that is much more significant than that simulated in ISAM ($\approx 10^{-4}$, see Figure \ref{fig:ISAM_O_FIP}). The collisional friction of Oxygen with protons is the dominant force that prevents neutral Oxygen from fully stratifying in the chromosphere as shown in the middle panel of Figure \ref{fig:ISAM_O_F}. 

In the corona, thermal diffusion effects becomes dominant through the Coulomb interaction between $O^{4+}$ and protons that generate a net outward thermal force that overcomes the gravity pull. The abundance of Oxygen then progressively increases that reduces the depletion with respect to Hydrogen. This lift up of Oxygen proceeds up to $\simeq 7400\ \rm{km}$ where the total net outward force from the collisional coupling with protons becomes equal to the gravity pull. Past this point the stratification of Oxygen takes over and leads to an even greater depletion at the loop-top. 

The analysis becomes much more complex in the transition region where the ionization of Oxygen proceeds rapidly to the fourth-state. The attribution of more grid points within this thin layer should help at a better description of the forces at play, but at the cost of greater computational times. We can say at first order that Oxygen ions seem to be slightly retained from entering the corona due to the coupling with the down-streaming protons, the so-called proton "barrier" mentioned above and in \citet{Killie2005,Killie2007}), although this barrier is likely overcome here by the thermal diffusion effects that are associated to both protons and electrons. \\

To conclude on the Oxygen balance in this given Hydrogen-proton-electron atmosphere, the strong collisional coupling of neutral Oxygen and $O^{4+}$ through charge-exchange and Coulomb interactions with protons prevents Oxygen from stratifying completely in the chromosphere and corona. That leads up to a significant depletion of Oxygen relative to Hydrogen, but that still remains limited compared to the case of a full stratification where particle-particle interactions would not be accounted for. In the presented ISAM simulation, this stratification of Oxygen takes over $11\ \rm{hr}$ to establish in a non-perturbed chromosphere. Our results agree with a previous study from \citet{Killie2007} who also show significant variations of the Oxygen abundance throughout the upper chromosphere and corona. However, such variations are not supported by the observations and therefore suggest the existence of a mixing process in the chromosphere that counteracts the settling of Oxygen over time. As discussed several times throughout this thesis (see e.g. \ref{subsec:intro_FIP}), the chromosphere is known to be highly dynamic and many sources can provide this mixing including turbulence, wave-particle interactions and flux emergence. The impact of these mixing processes on the abundance of Oxygen will be investigated in a subsequent work.

\section{Conclusion}
\label{sec:ISAM_results_conclusion}

We have presented in this chapter a first application of the Irap Solar Atmospheric Model (ISAM) to the study of composition along coronal loops. That required a significant amount of involvement and determination of myself to get the ISAM model working in such specific cases. The ISAM model has been recently proven to be successful at modeling various solar wind regimes \citep{Lavarra2022}. I brought significant improvements to the initial model to account for the peculiarity of coronal loops, and to allow for a more comprehensive treatment of minor species in the solar atmosphere. All of these changes have been described in detail in chapter \ref{cha:ISAM}. \\

The case of a coronal loop composed of Hydrogen only has been first investigated in section \ref{sec:ISAM_results_H}, as a baseline to build up our expertise on the interpretation of the forces at play in the low solar atmosphere. As introduced in section \ref{sec:intro_low_atmosphere}, we basically found that both the chromosphere and corona satisfy their own hydrostatic equilibria. However, we have seen in section \ref{subsec:ISAM_results_H_thermodynamics} that the coronal part of the loop can be perturbed and enter in a thermal non-equilibrium state where the plasma successively undergoes cooling/heating phases, and bounces back and forth between the transition region and the loop-top. As noted in past studies on the FIP effect \citep{Killie2005,Killie2007}, an increase of the temperature at the base of the corona can convect more thermal energy through the transition region and hence fuels an ambipolar flow between neutral Hydrogen and protons in that region. I have also observed this ambipolar flow in the pure Hydrogen atmosphere presented in section \ref{subsec:ISAM_results_H_forces}, where the down-streaming protons act as a "barrier" for the heavy elements that are already ionized in the transition region (such as low-FIP elements), and hence those heavy ions are more likely to be dragged down by the protons through Coulomb collisions. \\

By including Oxygen in this pure Hydrogen background atmosphere as a trace element, I have shown in section \ref{sec:ISAM_results_O} that Oxygen is able to get through the proton "barrier" where thermal effects are sufficient to overcome the proton drag. That was also noted in  \citet{Killie2007} and is permitted by the fact that Oxygen ionizes at almost the same rate as Hydrogen. This barrier is supposed to be much more effective for low-FIP elements that are completely ionized in the lower transition region however \citep{Killie2007}. As a consequence, an inverse FIP effect across the transition region is favored where low-FIP (and high-FIP) elements are depleted (and enriched) in the corona, an inverse FIP effect that is not observed in the solar corona but in active stellar coronae though \citep[see the review from][]{Laming2015}. This paradox seems inherent to models that consider diffusion effects alone. 

Preliminary tests with the inclusion of low-FIP Magnesium and high-FIP Helium in ISAM tend to supports an inverse FIP effect throughout the transition region as well, where a significant portion of the Magnesium available in the upper chromosphere is prevented from entering the corona. However, these preliminary tests also show that this inverse FIP effect likely does not persist very long, where higher up in the corona Magnesium turns to be well enriched with respect to Hydrogen in the end, thanks to a strong upward thermal force. Therefore the consistency of the modelled abundances is strongly conditioned by the height that is considered, where composition diagnostics of coronal loops done by spectroscopy are usually derived from spectral lines that are relevant to distinct altitudes in the solar atmosphere. \\

Furthermore, the impact of chromospheric mixing on the coronal loop composition has not been investigated in this chapter. The code itself can nonetheless be readily adapted to treat such cases, for which I already performed first attempts. In addition, ISAM could be exploited to generate acoustic shocks at the base of the loop that may then propagate photospheric abundances throughout the chromosphere. As noted by \cite{Laming2009}, wave-particle interactions through the ponderomotive force can also furnish the upper chromosphere in photospheric abundances. The ponderomotive force may also be essential at preventing the inverse FIP effect that tends to be obtained when only diffusion effects are accounted for. The reader may have noted how complex the diffusion effects alone are to interpret. Therefore the results presented in this chapter will serve as a baseline for future applications of ISAM that will consider the contribution from wave-particle interactions. I must highlight nonetheless that I have already made significant preliminary advances towards this goal, including the solving of an additional transport equation in ISAM for the propagation of Alfvén waves (see section \ref{subsubsec:ISAM_Aw_WKB}) and the call to an external turbulence code called Shell-Atm (see section \ref{subsec:ISAM_Aw_ShellATM}).

%% file: chapters/Conclusion_perspectives.tex
\chapter{Conclusion and future perspectives}

Aside from the major breakthroughs that have been made recently from the measurements taken in situ at the \textit{Parker Solar Probe} (\textit{PSP}) of the nascent solar wind, we have shown in chapter \ref{cha:stationnary} that the \textit{WISPR} white-light imager on board \textit{PSP} significantly enriched our knowledge of the structure and formation of the slow solar wind (SSW). By imaging the SSW from a vantage point closer to the source, \textit{WISPR} has depicted a SSW of an unexpected complexity, with small-scale details that are often smeared out from near 1 AU remote observations due to line-of-sight integration effects. While \textit{WISPR} allowed to scrutinize the slow solar wind at scales down to the thin heliospheric plasma sheet (HPS) (i.e. $\lesssim 5^\circ$), \textit{WISPR} has especially portrayed a slow solar wind that emerges from multiple sources distributed further away from the HPS. Therefore \textit{WISPR} corroborates a more general picture where the source regions of the slow solar wind are diverse and are likely dictated by the topology of the magnetic field lower in the corona. In particular, that has been formulated in the quasi-stationnary theory (section \ref{sec:intro_stationnary}) which shows that dense and slow winds should form along magnetic fields that have expanded significantly all the way up from the surface to the corona. This effect has been further quantified in chapter \ref{cha:stationnary} by exploiting the high-resolution MULTI-VP MHD simulations, where the expansion rate of the coronal magnetic field controls the energy that fuels the solar wind. Given a magnetogram of the surface magnetic field that is sufficiently resolved, this modeling framework generated a myriad of streams which in turn produce a fine-structuring of the slow solar wind similarly to the last observations from \textit{WISPR}. However, that may not be the only source of all the fine striation observed by \textit{WISPR} lately, and the contribution from the magnetic field has still to be evaluated precisely. \\

By greatly reducing the gap between the source and the measurement point, the \textit{PSP} mission has already provided a wealth of precious data that incite to rethink the current formation theories of the slow solar wind. Regions of high expansion factors that are favorable to the formation of a quasi-stationary slow solar wind are also propitious environment for magnetic reconnection to occur, these regions being more generally termed the (quasi-)separatrix layers or the S-web. The large affluence of magnetic reconnection events in the corona has long been suspected as a major source of the observed variability of the SSW. In chapter \ref{cha:dynamics}, we considered dynamic theories that may explain part of the variability of the slow wind observed both remotely and in situ at \textit{PSP}. As a step forward to the first "blobs" observed by \textit{SoHO} 25 years ago, we suggested a more general picture of "blobs" that form hand in hand with flux rope structures. The presence of such structures have been suspected from in situ measurements taken at \textit{STEREO} and more recently at \textit{PSP} and \textit{SolO}, but \textit{PSP-WISPR} allowed for the first time to observe remotely these flux ropes in greater depth as shown in section \ref{sec:dynamics_WISPR}. Advanced WindPredict-AW simulations have been exploited in section \ref{sec:dynamics_tearing} to investigate the generation of such flux ropes from reconnection at the tip of streamers. Density structures with multiple periodicities have then been produced with time scales that are coherent with the observations. By simulating their aspect in \textit{WISPR} images I have found a close similarity with actual flux ropes that have been observed by \textit{WISPR}. However, \textit{WISPR} observations unveiled the presence of likely smaller-scale density fluctuations which have also been revealed in a recent deep-field \textit{STEREO-COR2} campaign. The origin of these structures remains largely unknown but we showed in section \ref{sec:dynamics_griton2020} that they can be in part generated by heating pulses at the base of the corona. This constitutes an elegant picture of the formation of the SSW where the dynamic theories provide the necessary energy inputs, through the abundance of magnetic reconnection events that are known to occur in the low corona, and with the associated generation and propagation of density fluctuations into the solar wind. \\

While \textit{WISPR} captured plenty of transient structures, slow solar winds with a low variability have also been observed. That has been further noted in an in-depth analysis of in situ measurements taken at \textit{PSP} combined with a precised assessment of the connectivity that I furnished for this study (section \ref{sec:dynamics_insitu}). Slow winds with different bulk properties have been identified that suggest not a \emph{time-dependent} but \emph{spatial-dependent} variability. Their difference appeared to be inherent to the magnetic topology of their source region and hence corroborates the picture drawn in chapter \ref{cha:stationnary} of a SSW that is textured by the complex structure of the open coronal magnetic field. \\

Although the novel observations made by \textit{PSP} allowed us to pinpoint more precisely the sources of the slow solar wind and of its associated forming mechanisms, subsequent studies should benefit from the large amount of data collected by both \textit{PSP} and now \textit{Solar Orbiter} (\textit{SolO}) to extend the observations that we made on a case by case basis. To benefit the most from these novel measurements, we have highlighted the necessity of developing new methods that employ forward modeling techniques coupled with advanced MHD models of the solar atmosphere, but also novel tools that allow for a more systematic evaluation of the model performances. However, the models employed in this thesis still face significant difficulties at explaining the most recent observations from \textit{PSP} in all their complexity. We especially noted in chapter \ref{cha:stationnary} that \textit{WISPR} now provides more stringent tests to current solar wind models. \\

As introduced in section \ref{subsubsec:intro_SSW_composition} and \ref{subsubsec:intro_SSW_variability}, a variability of the SSW is also observed in term of its composition in heavy ions. As coronal loops have been observed to have similar composition signatures than those measured in situ in the solar wind, we have shown in chapter \ref{cha:ISAM_results} that a high variability of the distribution of heavy ions along coronal loops is also to be expected. I contributed to the development of a multi-specie model of the solar atmosphere called ISAM (presented in chapter \ref{cha:ISAM}), that solves the transport of Hydrogen and minor species in a self-consistent manner from the chromosphere to the corona. ISAM also includes a comprehensive treatment of interactions between species that allowed us to investigate in detail the role of collisional effects in the FIP effect, where I have shown that protons play a significant role in the separation of Oxygen for instance. The strong collisional coupling of Oxygen with Hydrogen through friction and thermal effects prevents the full stratification of Oxygen, that would otherwise be much more marked with a dramatic depletion of Oxygen being 16 times heavier than Hydrogen. Coulomb interactions with protons also tend to favor an inverse FIP effect throughout the transition region where only mid to high FIP elements would have access to the corona by being blind to the "barrier" made by the down-streaming protons in the transition region. This effect is likely to be regulated in dynamic loops that undergo successive evaporation/cooling phases, and to be boosted up during magnetic reconnection events that deposit great amount of energy at the base of the corona. Furthermore, this "local" inverse FIP effect seems to be greatly overcome higher up in the corona as suggested by some preliminary tests that include the low-FIP Magnesium and high-FIP Helium. Although our work only focused on the analysis of the diffusion effects alone onto Oxygen, ISAM has already been found powerful to test some theories of the FIP effect. Therefore I have laid in chapter \ref{cha:ISAM_results} the baseline for subsequent applications of ISAM to study the FIP effect in coronal loops at greater depth, by preparing for the inclusion of wave-particle interactions for instance.

\chapter{Conclusion et perspectives futures}

En plus des percées majeures réalisées récemment grâce aux mesures in situ du vent solaire naissant effectuées par la sonde \textit{Parker Solar Probe} (\textit{PSP}), nous avons montré dans le chapitre \ref{cha:stationnary} que l'imageur en lumière blanche \textit{WISPR} à bord de \textit{PSP} a considérablement enrichi nos connaissances sur la structure et la formation du vent solaire lent (SSW). En observant le SSW depuis un point d'observation plus proche de la source, \textit{WISPR} a dépeint un SSW d'une complexité inattendue, avec des détails à petites échelles qui sont souvent absents dans les observations realisées à une distance de 1 UA en raison des effets d'intégration le long de la ligne de visée. Alors que \textit{WISPR} a permis de scruter le vent solaire lent à des échelles allant jusqu'à la fine couche de plasma héliosphérique (HPS) (i.e. $\lesssim 5^\circ$), \textit{WISPR} a surtout dépeint un vent solaire lent qui émerge de plusieurs sources distribuées plus loin de la HPS. Par conséquent, \textit{WISPR} corrobore une image plus générale où les régions sources du vent solaire lent sont diverses et sont probablement contrôlées par la topologie du champ magnétique plus bas dans la couronne. En particulier, cela a été formulé dans la théorie quasi-stationnaire (section \ref{sec:intro_stationnary_FR}) qui montre que les vents denses et lents devraient se former le long des champs magnétiques qui subissent une forte expansion depuis la surface jusqu'à la couronne. Cet effet a été davantage quantifié dans le chapitre \ref{cha:stationnary} en exploitant des simulations MHD MULTI-VP à haute résolution, où le taux d'expansion du champ magnétique coronal contrôle l'énergie qui alimente le vent solaire. Pour un magnétogramme donné du champ magnétique de surface qui est suffisamment résolu, ce cadre de modélisation a permis de générer une myriade d'écoulements qui, à leur tour, ont produit une structure fine du vent solaire lent comme le révèlent en partie les dernières observations de \textit{WISPR}. Toutefois, cela n'est probablement pas la seule source de striation fine observée par \textit{WISPR} dernièrement, où la contribution du champ magnetique reste encore à être évaluée précisement. \\

En réduisant considérablement l'écart entre la source et les points de mesures, la mission \textit{PSP} a déjà fourni beaucoup de données précieuses qui incitent à repenser les théories actuelles de formation du vent solaire lent. Les régions où les facteurs d'expansion sont élevés et favorables à la formation d'un vent solaire lent quasi-stationnaire sont également des environnements propices à la reconnexion magnétique, ces régions étant plus généralement appelées les couches (quasi-)séparatrices ou la S-web. La grande affluence d'événements de reconnexion magnétique dans la couronne a longtemps été soupçonnée d'être une source majeure de la variabilité observée du vent solaire lent. Dans le chapitre \ref{cha:dynamics}, nous avons considéré des théories dynamiques qui peuvent expliquer une partie de la variabilité du vent lent observé à distance et in situ à \textit{PSP}. Pour aller plus loin que les premiers "blobs" observés par \textit{SoHO} il y a 25 ans, nous avons suggéré une image plus générale de "blobs" qui se forment en même temps que des structures magnétiques torsadées. La présence de telles structures a été suspectée par des mesures in situ prises à \textit{STEREO} et plus récemment à \textit{PSP} et \textit{SolO}, mais \textit{PSP-WISPR} a permis pour la première fois d'observer à distance ces structures magnétiques torsadées de plus près comme le montre la section \ref{sec:dynamics_WISPR}. Les simulations avancées de WindPredict-AW ont été exploitées dans la section \ref{sec:dynamics_tearing} pour étudier la génération de ces champs torsadés à partir de la reconnexion magnétique à l'extrémité des streamers. Des structures de densité à périodicités multiples ont alors été produites avec des échelles de temps cohérentes avec les observations. En simulant leur aspect dans les images de \textit{WISPR}, j'ai trouvé une grande similarité avec des structures torsadées réelles qui ont été observées par \textit{WISPR}. Cependant, les observations de \textit{WISPR} ont révélé la présence de fluctuations de densité à plus petites échelles, qui ont également été mises en évidence lors d'une récente campagne \textit{STEREO-COR2} en champ profond. L'origine de ces structures reste largement inconnue mais nous avons montré dans la section \ref{sec:dynamics_griton2020} qu'elles peuvent être en partie générées par du chauffage impulsif à la base de la couronne. Ceci constitue une image élégante de la formation du vent solaire lent où les théories dynamiques fournissent l'énergie nécessaire à ce chauffage impulsif, à travers la reconnexion magnétique qui est connue pour se produire abondamment dans la basse couronne, genérant par ailleurs des fluctuations de densité qui se propagent ensuite dans le vent solaire. \\

Alors que \textit{WISPR} a capturé de nombreuses structures transitoires, des vents solaires lents avec une faible variabilité ont également été observés. Cela a été noté dans une analyse approfondie des mesures in situ prises à \textit{PSP} combinée à une évaluation précise de la connectivité que j'ai fournie pour cette étude (section \ref{sec:dynamics_insitu}). Des vents lents avec des propriétés globales différentes ont été identifiés, ce qui suggère non pas une variabilité \emph{temporelle} mais \emph{spatiale}. Leur différence semble être propre à la topologie magnétique de leur région source et corrobore donc l'image dressée dans le chapitre \ref{cha:stationnary} d'un vent solaire lent qui est texturé par la structure complexe du champ magnétique coronal ouvert. \\

Bien que les nouvelles observations réalisées par \textit{PSP} nous aient permis de déterminer plus précisément les sources du vent solaire lent et les mécanismes de formation associés, les études ultérieures devraient bénéficier de la grande quantité de données collectées par \textit{PSP} et maintenant \textit{Solar Orbiter} (\textit{SolO}) pour étendre les observations que nous avons faites au cas par cas. Pour tirer le meilleur parti de ces nouvelles mesures, nous avons mis en évidence la nécessité de développer de nouvelles méthodes qui utilisent des techniques d'imagerie synthétique couplées à des modèles MHD avancés de l'atmosphère solaire, mais aussi de nouveaux outils qui permettent une évaluation plus systématique des performances des modèles. Cependant, les modèles employés dans cette thèse rencontrent encore des difficultés importantes pour expliquer les observations les plus récentes de \textit{PSP} dans toute leur complexité. Nous avons particulièrement noté dans le chapitre \ref{cha:stationnary} que \textit{WISPR} fournit maintenant des tests plus rigoureux aux modèles de vent solaire actuels. \\

Comme présenté dans les sections \ref{subsubsec:intro_SSW_composition_FR} et \ref{subsubsec:intro_SSW_variability_FR}, une variabilité du vent lent est également observée en terme de sa composition en ions lourds. Comme il a été observé que les boucles coronales ont des signatures de composition similaires à celles du vent solaire lent mesurées in situ, nous avons montré dans le chapitre \ref{cha:ISAM_results} qu'une grande variabilité de la distribution des ions lourds le long des boucles coronales est également à prévoir. J'ai contribué au développement d'un modèle multi-espèces de l'atmosphère solaire appelé ISAM, qui résout le transport de l'Hydrogène et des espèces mineures d'une manière auto-consistante depuis la chromosphère jusqu'à la couronne. ISAM comprend également un traitement complet des interactions entre espèces qui nous a permis d'étudier en détail le rôle des effets collisionnels sur l'effet FIP, où j'ai notamment montré que les protons jouent un rôle primordial dans la séparation de l'Oxygène par exemple. Le fort couplage collisionnel de l'Oxygène avec l'Hydrogène par effets de friction et de diffusion thermique empêche la stratification complète de l'Oxygène, qui serait autrement beaucoup plus marquée avec un fort appauvrissement de l'Oxygène étant 16 fois plus lourd que l'Hydrogène. Les interactions Coulombiennes avec les protons ont également tendance à favoriser un effet FIP inverse à travers la région de transition où seuls les éléments à FIP moyen ou élevé auraient accès à la couronne, en étant invisibles à la "barrière" formée par les protons qui descendent dans la région de transition. Cet effet est susceptible d'être régulé dans les boucles dynamiques qui subissent des phases successives d'évaporation et de refroidissement, et d'être amplifié lors d'événements de reconnexion magnétique qui déposent une grande quantité d'énergie à la base de la couronne. Par ailleurs, cet effet FIP inverse "local" pourrait être fortement compensé plus haut dans la couronne comme le suggère des tests préliminaires qui incluent le Magnesium (faible FIP) et l'Helium (haut FIP). Bien que notre travail ne se soit concentré que sur l'analyse des seuls effets de diffusion sur l'Oxygène, ISAM s'est déjà révélé puissant pour tester certaines théories de l'effet FIP. C'est pourquoi le chapitre \ref{cha:ISAM_results} servira de base pour des applications ultérieures d'ISAM qui étudieront plus en profondeur l'effet FIP dans les boucles coronales, en préparant notamment l'intégration des interactions ondes-particules.

%% file: appendices/ISAM.tex
\chapter{Technical specificities of ISAM}

\section{Scaling of the equations}
\label{sec:ISAM_normalization}

All physical moments of the VDF that are solved in ISAM are scaled as follows:
\begin{subequations}
\begin{align}
    \Bar{n}_s &= \frac{n_s}{n_0} \\
    \Bar{u}_s &= \frac{u_s}{c_{s,0}} \\
    \Bar{T}_s^\parallel &= \frac{T_s^\parallel}{T_0} \\
    \Bar{T}_s^\perp &= \frac{T_s^\perp}{T_0} \\
    \Bar{\gamma}_s^\parallel &= \frac{q_s^\parallel}{p_s^\parallel}\sqrt{\frac{m_s}{k_b T_s^\parallel}} \\
    \Bar{\gamma}_s^\perp &= \frac{q_s^\perp}{p_s^\perp}\sqrt{\frac{m_s}{k_b T_s^\parallel}}
\end{align}
\end{subequations}
where here $c_{s,0}=\sqrt{k_b T_0/m_s}$ and $p_s^{\parallel,\perp}=n_s k_b T_s^{\parallel,\perp}$. In addition to the temperature and density, we consider the solar gravitational acceleration constant as a reference. These scaling parameters equal:
\begin{subequations}
\begin{align}
    n_0 &= 1\times 10^{20}\ m^{-3} \\
    T_0 &= 1\times 10^5\ K \\
    G_0 &= G M_\odot/R_\odot^2\ m.s^{-2}\simeq 273.9638\ m.s^{-2}
\end{align}
\end{subequations}
From this basic set we define the spatial and temporal characteristic scales as:
\begin{subequations}
\begin{align}
    t_0 &= c_{0,0}/G_0 \\
    r_0 &= c_{0,0} t_0
\end{align}
\end{subequations}
where $c_{0,0}=\sqrt{k_b T_0/m_0}$ is the thermal speed of the specie of reference. The specie of reference is chosen as the ion with the lowest mass which is therefore the proton with mass $m_0=m_p\simeq 1,007uma\simeq  1.6726\times 10^{-27}\ kg$. More generally, all atomic masses are normalized in the dalton or unified atomic mass unit $\mu_s=m_s/uma$ where $uma\simeq 1.6605\times^{-27}\ kg$. \\

Since the Alfvén-wave energy density $\mathcal{E}=\epsilon/\rho$ scales as the square of the velocity, its associated normalized quantity is defined as:
\begin{equation}
    \Bar{\mathcal{E}} = \mathcal{E}/c_{p,0}^2
\end{equation}
where for recall $c_{p,0}=\sqrt{k_b T_0/m_p}$ is the reference thermal speed for protons.

\section{Mass ratios for collision terms}
\label{sec:mass_ratios}

\subsection{Neutral-neutral collisions}

\begin{subequations}
\begin{align}
    z_{st}   &= -\frac{1}{5}  \\
    A^{*0p}_{st} &= 3\left(1-z_{st}\frac{m_t}{m_s+m_t}\right)\\ 
    A^{*0t}_{st} &= \left(1-z_{st}\frac{m_t}{m_s+m_t}\right)\\ 
    D^{*1pp}_{st} &= -\frac{3}{5}\left(5\left(\frac{m_s}{m_s+m_t}\right)^2+2\left(\frac{m_t}{m_s+m_t}\right)^2+4\frac{m_s}{m_s+m_t}\frac{m_t}{m_s+m_t}\right)\\ 
    D^{*1pt}_{st} &= \frac{3}{5}\left(4\frac{m_s}{m_s+m_t}+\frac{m_t}{m_s+m_t}\right)\frac{m_t}{m_s+m_t}\\ 
    D^{*1tp}_{st} &= \frac{1}{10}\left(4\frac{m_s}{m_s+m_t}+\frac{m_t}{m_s+m_t}\right)\frac{m_t}{m_s+m_t}\\ 
    D^{*1tt}_{st} &= -\frac{1}{10}\left(30\left(\frac{m_s}{m_s+m_t}\right)^2+13\left(\frac{m_t}{m_s+m_t}\right)^2+28\frac{m_s}{m_s+m_t}\frac{m_t}{m_s+m_t}\right)\\ 
    D^{*4pp}_{st} &= \frac{9}{5}\left(\frac{m_t}{m_s+m_t}\right)^2 \\ 
    D^{*4pt}_{st} &= \frac{9}{5}\left(\frac{m_t}{m_s+m_t}\right)^2\\ 
    D^{*4tp}_{st} &= \frac{3}{10}\left(\frac{m_t}{m_s+m_t}\right)^2\\ 
    D^{*4tt}_{st} &= \frac{3}{2}\left(\frac{m_t}{m_s+m_t}\right)^2 
\end{align}
\end{subequations}

\subsection{Ion-neutral (Maxwell) collisions}
$s$ and $t$ stand for ion and neutral respectively.
The reverse interaction is obtained by flipping the two indices.
\begin{subequations}
\begin{align}
    z_{st}   &= 0  \\
    A^{*0p}_{st} &= 3\left(1-z_{st}\frac{m_t}{m_s+m_t}\right)\\ 
    A^{*0t}_{st} &= \left(1-z_{st}\frac{m_t}{m_s+m_t}\right)\\ 
    D^{*1pp}_{st} &= -\frac{3}{5}\left(5\left(\frac{m_s}{m_s+m_t}\right)^2+2\left(\frac{m_t}{m_s+m_t}\right)^2+4\frac{m_s}{m_s+m_t}\frac{m_t}{m_s+m_t}\right)\\ 
    D^{*1pt}_{st} &= \frac{3}{5}\left(4\frac{m_s}{m_s+m_t}+\frac{m_t}{m_s+m_t}\right)\frac{m_t}{m_s+m_t}\\ 
    D^{*1tp}_{st} &= \frac{1}{10}\left(4\frac{m_s}{m_s+m_t}+\frac{m_t}{m_s+m_t}\right)\frac{m_t}{m_s+m_t}\\ 
    D^{*1tt}_{st} &= -\frac{1}{10}\left(30\left(\frac{m_s}{m_s+m_t}\right)^2+13\left(\frac{m_t}{m_s+m_t}\right)^2+28\frac{m_s}{m_s+m_t}\frac{m_t}{m_s+m_t}\right)\\ 
    D^{*4pp}_{st} &= \frac{9}{5}\left(\frac{m_t}{m_s+m_t}\right)^2 \\ 
    D^{*4pt}_{st} &= \frac{9}{5}\left(\frac{m_t}{m_s+m_t}\right)^2\\ 
    D^{*4tp}_{st} &= \frac{3}{10}\left(\frac{m_t}{m_s+m_t}\right)^2\\ 
    D^{*4tt}_{st} &= \frac{3}{2}\left(\frac{m_t}{m_s+m_t}\right)^2 
\end{align}
\end{subequations}

\subsection{Coulomb collisions}
\begin{subequations}
\begin{align}
    z_{st}   &= \frac{3}{5}    \\
    A^{*0p}_{st} &= 3\left(1-z_{st}\frac{m_t}{m_s+m_t}\right)\\ 
    A^{*0t}_{st} &= \left(1-z_{st}\frac{m_t}{m_s+m_t}\right)\\ 
    D^{*1pp}_{st} &= -\frac{3}{70}\left(70\left(\frac{m_s}{m_s+m_t}\right)^2+2\left(\frac{m_t}{m_s+m_t}\right)^2+35\frac{m_s}{m_s+m_t}\frac{m_t}{m_s+m_t}\right)\\ 
    D^{*1pt}_{st} &= \frac{3}{35}\left(49\frac{m_s}{m_s+m_t}+10\frac{m_t}{m_s+m_t}\right)\frac{m_t}{m_s+m_t}\\ 
    D^{*1tp}_{st} &= \frac{1}{70}\left(49\frac{m_s}{m_s+m_t}+10\frac{m_t}{m_s+m_t}\right)\frac{m_t}{m_s+m_t}\\ 
    D^{*1tt}_{st} &= -\frac{1}{35}\left(105\left(\frac{m_s}{m_s+m_t}\right)^2+8\left(\frac{m_t}{m_s+m_t}\right)^2+77\frac{m_s}{m_s+m_t}\frac{m_t}{m_s+m_t}\right)\\ 
    D^{*4pp}_{st} &= \frac{3}{70}\left(-21\frac{m_s}{m_s+m_t}+16\frac{m_t}{m_s+m_t}\right)\frac{m_t}{m_s+m_t}\\ 
    D^{*4pt}_{st} &= \frac{9}{35}\left(-7\frac{m_s}{m_s+m_t}+6\frac{m_t}{m_s+m_t}\right)\frac{m_t}{m_s+m_t}\\ 
    D^{*4tp}_{st} &= \frac{3}{70}\left(-7\frac{m_s}{m_s+m_t}+6\frac{m_t}{m_s+m_t}\right)\frac{m_t}{m_s+m_t}\\ 
    D^{*4tt}_{st} &= \frac{3}{35}\left(-7\frac{m_s}{m_s+m_t}+5\frac{m_t}{m_s+m_t}\right)\frac{m_t}{m_s+m_t} 
\end{align}
\end{subequations}

\section{Implementation of the improved radiative cooling recipe in ISAM}
\label{sec:ISAM_radloss_detailed}

\subsection{Upper chromosphere: the Carlsson et al. (2012)'s recipe}
The optically thin radiative loss profiles $L_{X_m}(T)$ have been directly extracted from figures 2, 4 and 5 in \citet{Carlsson2012}. For CaII and MgII and temperatures below $6000\ K$, I replaced the $L_{X_m}(T)$ values by those plotted in figures 12 and 13 in \citet{Carlsson2012} that account for radiative heating in the same transition lines. The empirical escape probability $E_{X_m}(\tau)$ that represents the medium opacity to radiation is extracted from figures 6,7 and 8 in \citet{Carlsson2012}. I also fetched their ionization fractions $n_{X_m}/n_{X}$ from figures 9, 10 and 11. For the $A_X$ quantities I use the coronal abundances of element $X$ relative to Hydrogen taken from \citet{Schmelz2012}. These recipes have been tabulated for HI, CaII and MgII and for chromospheric temperatures between $~3000\ K$ and $30 000\ K$. The remaining input parameter to the recipes is the optical depth parameter $\tau$ which is calculated by this formula:
\begin{equation}
    \tau =\begin{dcases*}
        -4\times 10^{-14} \int_{z=+\infty}^{z=0}n_{HI}dz & for HI \\
        -\int_{z=+\infty}^{z=0}\rho dz\quad (g.cm^{-2}) & for CaII and MgII
    \end{dcases*}
\end{equation}
where we take $\rho$ as the total mass density.

\subsection{Transition region and corona: \textit{ChiantiPy}}
The \textit{RadLoss} module of \textit{ChiantiPy} is used to compute the radiative losses for a given transition line of an atom $X$, electronic density and temperature array. The user can also ask for integrated radiative losses for all transitions of an atom or directly the summed up contributions from all lines of all elements including H, He, C, O, Ne, N, Mg, Si, S, Fe, Na, Al, Ar, Ca and Ni (ordered by decreasing relative abundance). 

As a first approach I ran \textit{ChiantiPy} to get the total radiative losses for all the above mentioned elements which can take up to several tens of minutes to compute all transition lines. As in equation \ref{eq:Carlsson2012_recipe} this implies to provide also \textit{ChiantiPy} with ionization fractions and abundances.

I selected the default ionization equilibria provided in the \textit{Chianti} database \citep{Dere2009,Dere2007} which agrees very well with the one given by \citet{Mazzotta1998}. This is not surprising as some of the ionization and recombination rates collected in the \textit{Chianti} database have been fetched from \citet{Mazzotta1998}. We further notice that most of these rates are also being used in ISAM (see section \ref{sec:ISAM_ioniz}). 

For the relative abundances I use the coronal abundances from \citet{Schmelz2012} that are provided in the file "\textit{sun\_coronal\_2012\_schmelz\_ext.abund}" in the \textit{Chianti} database.

\section{Numerical treatment of collisions and heating/cooling sources}
\label{sec:ISAM_col_numerical}

The collision terms are purely local in the sense that they do not involve coupling between adjacent altitude points. As a consequence the problem can be formatted as a matrix system:
\begin{align}
\label{eq:ISAM_col_num}
    \frac{Z^n-Z^0}{\Delta t} &= \frac{\Tilde{Z}-Z^0}{\Delta t}+AZ^n + B \\
    \text{where  }Z &=\begin{vmatrix}n_1 \\ u_1 \\ T_1^\parallel \\ T_1^\perp \\ q_1^\parallel \\ q_1^\perp \\ \vdots \end{vmatrix} \notag
\end{align}
where $Z$ is the solution vector of size $6\times k$ for $k$ species resolved. In the above formula the superscript $^0$, $\Tilde{}$ and $^n$ stands for the initial solution, the solution calculated by LCPFCT, and the final solution respectively. The square matrix $A$ contains all collision terms where the VDF moments from the solution vector are introduced. Because electrons are solved separately than neutrals and ions, the moments for electrons are not included in the solution vector $Z$. Their contribution to the collision terms is then accounted for in vector $B$.

\section{Numerical resolution of the Alfvén-wave propagation and dissipation}
\label{sec:ISAM_AW_numerical}

The method is very similar with the one already described in Appendix \ref{sec:ISAM_col_numerical}:
\begin{equation}
    \frac{\mathcal{E}^n-\mathcal{E}^0}{\Delta t} = \underbrace{\frac{\Tilde{\mathcal{E}}-\mathcal{E}^0}{\Delta t}}_{C}-\frac{Q_h}{\rho}\Biggr|^n
\end{equation}
where $\mathcal{E}=\epsilon/\rho$ is the actual quantity that we solve in LCPFCT.

The dissipation term given in equation \ref{eq:WindPredict_Qw} can be approximated as:
\begin{equation}
    \frac{Q_h^\pm}{\rho}\Biggr|^n \simeq \frac{\sqrt{\mathcal{E}^{\pm 0}}}{\mathcal{L}}\mathcal{E}^{\pm n}
\end{equation}

Similarly for the dissipation term given in equation \ref{eq:WindPredict_Qw} and that includes some form of Alfvén-wave reflection:
\begin{equation}
    \frac{Q_h}{\rho}\Biggr|^n \simeq c_d\frac{u_p^0+v_A^0}{v_A^0}\left|\frac{\partial}{\partial s}v_A \right|^0\mathcal{E}^n
\end{equation}
In both cases the dissipation term can be formulated as follows:
\begin{equation}
    \frac{Q_h}{\rho}\Biggr|^n = B \mathcal{E}^n
\end{equation}
where we dropped the $\pm$ subscripts for simplicity.

Therefore the linear superposition stage is equivalent at solving a first order partial differential equation:
\begin{equation}
    \frac{\partial \mathcal{E}}{\partial t}\Biggr|^n + B \mathcal{E}^n = C
\end{equation}
with solution:
\begin{equation}
    \mathcal{E}^n = \exp (B \Delta t) + \frac{C}{B}
\end{equation}

\section{Hyperbolocity criteria}
\label{sec:ISAM_hyperbol}

For a given grid profile, the numerical time step is systematically adjusted to meet the following conditions for hyperbolicity, below for the transport of parallel energy:
\begin{equation}
    \left(|u_s|+\left|\frac{3}{2}\frac{q_s^\parallel}{p_s^\parallel}\right|\right)\frac{\Delta t}{\Delta r} \lesssim 1
\end{equation}
or similarly for the transport of the perpendicular energy:
\begin{equation}
    \left(|u_s|+\left|\frac{q_s^\perp}{p_s^\perp}\right|\right)\frac{\Delta t}{\Delta r} \lesssim 1
\end{equation}

%% file: appendices/List_papers.tex
\chapter{List of publications in high impact journals}

\section{Main publications}

\hspace{\parindent} \textbf{Poirier N.}, Rouillard A.P., Kouloumvakos A., Przybylak A., Fargette N., Pobeda R., Réville V., Pinto R.F., Indurain M., Alexandre M., 2021, \textit{Exploiting white-light observations to improve estimates of magnetic connectivity}, FrASS, Volume 8, id.84, \href{https://ui.adsabs.harvard.edu/abs/2021FrASS...8...84P}{ADS LINK} \\

\textbf{Poirier N.}, Kouloumvakos A., Rouillard A.P., Pinto R.F., Vourlidas A., Stenborg G., Valette E., Howard R.A., Hess P., Thernisien A., Rich N., Griton L., Indurain M., Raouafi N.-E., Lavarra M., Réville V., 2020, \textit{Detailed Imaging of Coronal Rays with the Parker Solar Probe}, ApJS, Volume 246, Issue 2, id.60, \href{https://ui.adsabs.harvard.edu/abs/2020ApJS..246...60P}{ADS LINK} \\

Rouillard A.P., \textbf{Poirier N.}, Lavarra M., Bourdelle A., Dalmasse K., Kouloumvakos A., Vourlidas A., Kunkel V., Hess P., Howard R.A., Stenborg G., Raouafi N.-E., 2020, \textit{Modeling the Early Evolution of a Slow Coronal Mass Ejection Imaged by the Parker Solar Probe}, ApJS, Volume 246, Issue 2, id.72, \href{https://ui.adsabs.harvard.edu/abs/2020ApJS..246...72R}{ADS LINK} \\

Badman S.T., Brooks D.H., \textbf{Poirier N.}, Warren H.P., Petrie G., Rouillard A.P., Nick Arge C., Bale S.D., de Pablos Ag{\"u}ero D., Harra L., Jones S.I., Kouloumvakos A., Riley P., Panasenco O., Velli M., Wallace S., 2022, \textit{Constraining Global Coronal Models with Multiple Independent Observables}, ApJ, Volume 932, Issue 2, id.135, \href{https://ui.adsabs.harvard.edu/abs/2022ApJ...932..135B}{ADS LINK} \\

Griton L., Pinto R.F., \textbf{Poirier N.}, Kouloumvakos A., Lavarra M., Rouillard A.P., 2020, \textit{Coronal Bright Points as Possible Sources of Density Variations in the Solar Corona}, ApJ, Volume 893, Issue 1, id.64, \href{https://ui.adsabs.harvard.edu/abs/2020ApJ...893...64G}{ADS LINK} \\

Griton L., Rouillard A.P., \textbf{Poirier N.}, Issautier K., Moncuquet M., Pinto R.F., 2021, \textit{Source-dependent Properties of Two Slow Solar Wind States}, ApJ, Volume 910, Issue 1, id.63, \href{https://ui.adsabs.harvard.edu/abs/2021ApJ...910...63G}{ADS LINK} \\

Pinto R.F., \textbf{Poirier N.}, Rouillard A.P., Kouloumvakos A., Griton L., Fargette N., Kieokaew R., Lavraud B., Brun A.S., 2021, \textit{Solar wind rotation rate and shear at coronal hole boundaries. Possible consequences for magnetic field inversions}, A\&A, Volume 653, id.A92, \href{https://ui.adsabs.harvard.edu/abs/2021A&A...653A..92P}{ADS LINK} \\

Lavarra M, Rouillard A.P., \textbf{Poirier N.}, Blelly P.-L., Pinto R.F., Réville V., 2022, \textit{Testing a multi-specie model of the fast and slow solar wind}, in review \\

\section{Other publications}

\hspace{\parindent} Howard R.A., Vourlidas A., Bothmer V., Colaninno R.C., DeForest C.E., Gallagher B., Hall J.R., Hess P., Higginson A.K., Korendyke C.M., Kouloumvakos A., Lamy P.L., Liewer P.C., Linker J., Linton M., Penteado P., Plunkett S.P., \textbf{Poirier N.}, Raouafi N.E., Rich N., Rochus P., Rouillard A.P., Socker D.G., Stenborg G., Thernisien A.F., Viall N.M., 2019, \textit{Near-Sun observations of an F-corona decrease and K-corona fine structure}, Nature, Volume 576, Issue 7786, p.232-236, \href{https://ui.adsabs.harvard.edu/abs/2019Natur.576..232H}{ADS LINK} \\

Korreck K.E., Szabo A., Nieves-Chinchilla T., Lavraud B., Luhmann J., Niembro T., Higginson A.K., Alzate N., Wallace S., Paulson K., Rouillard A.P., Kouloumvakos A., \textbf{Poirier N.}, Kasper J.C., Case A.W., Stevens M.L., Bale S.D., Pulupa M., Whittlesey P., Livi R., Goetz K., Larson D., Malaspina D.M., Morgan H., Narock A.A., Schwadron N.A., Bonnell J., Harvey P., Wygant J., 2020, \textit{Source and Propagation of a Streamer Blowout Coronal Mass Ejection Observed by the Parker Solar Probe}, ApJS, Volume 246, Issue 2, id.69, \href{https://ui.adsabs.harvard.edu/abs/2020ApJS..246...69K}{ADS LINK} \\

Réville V., Fargette N., Rouillard A.P., Lavraud B., Velli M., Strugarek A., Parenti S., Brun A.S., Shi C., Kouloumvakos A., \textbf{Poirier N.}, Pinto R.F., Louarn P., Fedorov A., Owen C.J., Génot V., Horbury T.S., Laker R., O'Brien H., Angelini V., Fauchon-Jones E., Kasper J.C., 2022, \textit{Flux rope and dynamics of the heliospheric current sheet. Study of the Parker Solar Probe and Solar Orbiter conjunction of June 2020}, A\&A, Volume 659, id.A110, \href{https://ui.adsabs.harvard.edu/abs/2022A&A...659A.110R}{ADS LINK} \\

Kouloumvakos A., Vourlidas A., Rouillard A.P., Roelof E.C., Leske R., Pinto R.F., \textbf{Poirier N.}, 2020, \textit{The Solar Origin of Particle Events Measured by Parker Solar Probe}, ApJ, Volume 899, Issue 2, id.107, \href{https://ui.adsabs.harvard.edu/abs/2020ApJ...899..107K}{ADS LINK} \\

Réville V., Rouillard A.P., Velli M., Verdini A., Buchlin {\'E}ric, Lavarra M., \textbf{Poirier N.}, 2021, \textit{Investigating the origin of the FIP effect with a shell turbulence model}, FrASS, Volume 8, id.2, \href{https://ui.adsabs.harvard.edu/abs/2021FrASS...8....2R}{ADS LINK} \\

Rouillard A.P., Kouloumvakos A., Vourlidas A., Kasper J., Bale S., Raouafi N.-E., Lavraud B., Howard R.A., Stenborg G., Stevens M., \textbf{Poirier N.}, Davies J.A., Hess P., Higginson A.K., Lavarra M., Viall N.M., Korreck K., Pinto R.F., Griton L., Réville V., Louarn P., Wu Y., Dalmasse K., Génot V., Case A.W., Whittlesey P., Larson D., Halekas J.S., Livi R., Goetz K., Harvey P.R., MacDowall R.J., Malaspina D., Pulupa M., Bonnell J., de Witt T.D., Penou E., 2020, \textit{Relating Streamer Flows to Density and Magnetic Structures at the Parker Solar Probe}, ApJS, Volume 246, Issue 2, id.37, \href{https://ui.adsabs.harvard.edu/abs/2020ApJS..246...37R}{ADS LINK} \\

Rouillard A.P., Pinto R.F., Vourlidas A., De Groof A., Thompson W.T., Bemporad A., Dolei S., Indurain M., Buchlin E., Sasso C., Spadaro D., Dalmasse K., Hirzberger J., Zouganelis I., Strugarek A., Brun A.S., Alexandre M., Berghmans D., Raouafi N.-E., Wiegelmann T., Pagano P., Arge C.N., Nieves-Chinchilla T., Lavarra M., \textbf{Poirier N.}, Amari T., Aran A., Andretta V., Antonucci E., Anastasiadis A., Auchère F., Bellot Rubio L., Nicula B., Bonnin X., Bouchemit M., Budnik E., Caminade S., Cecconi B., Carlyle J., Cernuda I., Davila J.M., Etesi L., Espinosa Lara F., Fedorov A., Fineschi S., Fludra A., Génot V., Georgoulis M.K., Gilbert H.R., Giunta A., Gomez-Herrero R., Guest S., Haberreiter M., Hassler D., Henney C.J., Howard R.A., Horbury T.S., Janvier M., Jones S.I., Kozarev K., Kraaikamp E., Kouloumvakos A., Krucker S., Lagg A., Linker J., Lavraud B., Louarn P., Maksimovic M., Maloney S., Mann G., Masson A., Müller D., Önel H., Osuna P., Orozco Suarez D., Owen C.J., Papaioannou A., Pérez-Suárez D., Rodriguez-Pacheco J., Parenti S., Pariat E., Peter H., Plunkett S., Pomoell J., Raines J.M., Riethmüller T.L., Rich N., Rodriguez L., Romoli M., Sanchez L., Solanki S.K., St Cyr O.C., Straus T., Susino R., Teriaca L., del Toro Iniesta J.C., Ventura R., Verbeeck C., Vilmer N., Warmuth A., Walsh A.P., Watson C., Williams D., Wu Y., Zhukov A.N., 2020, \textit{Models and data analysis tools for the Solar Orbiter mission}, A\&A, Volume 642, id.A2, \href{https://ui.adsabs.harvard.edu/abs/2020A&A...642A...2R}{ADS LINK} \\

Lavraud B., Fargette N., Réville V., Szabo A., Huang J., Rouillard A.P., Viall N., Phan T.D., Kasper J.C., Bale S.D., Berthomier M., Bonnell J.W., Case A.W., Dudok de Wit T., Eastwood J.P., Génot V., Goetz K., Griton L.S., Halekas J.S., Harvey P., Kieokaew R., Klein K.G., Korreck K.E., Kouloumvakos A., Larson D.E., Lavarra M., Livi R., Louarn P., MacDowall R.J., Maksimovic M., Malaspina D., Nieves-Chinchilla T., Pinto R.F., \textbf{Poirier N.}, Pulupa M., Raouafi N.-E., Stevens M.L., Toledo-Redondo S., Whittlesey P.L., 2020, \textit{The Heliospheric Current Sheet and Plasma Sheet during Parker Solar Probe's First Orbit}, ApJL, Volume 894, Issue 2, id.L19, \href{https://ui.adsabs.harvard.edu/abs/2020ApJ...894L..19L}{ADS LINK} \\

%% file: appendices/Communications.tex
\chapter{List of communications}

\section{Orals}

 \hspace{\parindent} \textit{Simulating the FIP effect in coronal loops: modeling with a high-order multi-specie approach}, July 16-24 2022, COSPAR 44th Scientific Assembly. \\
 
 \textit{Simulating the FIP effect in coronal loops: modeling with a high-order multi-specie approach}, June 28 - July 1 2022, The 10th Coronal Loops Workshop. \\
 
 \textit{Interface entre la chromosphère et la couronne solaire: modélisation avec une approche 16-moments multi-espèces}, May 16-20 2022, Colloque du PNST. \\

\textit{Exploiting white-light observations to improve estimates of magnetic connectivity}, May 27 2021, SolO ISWG: SW sources and connection science. \\

\textit{Exploiting white-light observations to improve estimates of magnetic connectivity}, April 2 2021, LESIA science seminar. \\

\textit{Using modeling tools to interpret the coronal rays observed by WISPR from E1 to E6}, April 14 2021, WISPR Consortium meeting - virtual. \\

\textit{The CONNECT-TOOL: update on the white-light based selection of the best coronal models}, October 26-27 2020, MADAWG meeting. \\

\textit{Using white-light observations to constrain coronal models: application to E6}, October 15 2020, PSP SWG meeting. \\

\textit{The CONNECT-TOOL: current state/limitations \& next developments}, July 22 2020, MADAWG meeting. \\

\textit{Parker Solar Probe - First Results} (co-presenting with Naïs Fargette), February 2020, Journées scientifiques et techniques de l'IRAP. \\

\textit{Techniques to exploit remote-sensing observations from a rapidly moving spacecraft: application to Parker Solar Probe and Solar Orbiter}, January 2020, ISSI meeting: Exploring the Solar Wind In Regions Closer Than Ever Observed Before. \\

\textit{The Forming Slow Solar Wind Imaged along Streamer Rays by the Wide-Angle Imager on Parker Solar Probe}, December 9-13 2019, AGU Fall Meeting, \href{https://ui.adsabs.harvard.edu/abs/2019AGUFMSH12A..08P}{ADS LINK}. \\

\textit{Interpreting the first WISPR observations}, June 19-20 2019, First WISPR science meeting. \\

\section{Hybrid}

\hspace{\parindent} \textit{Simulating the FIP effect in coronal loops using a multi-species kinetic-fluid model}, May 23-27 2022, EGU General Assembly Conference. \\

\textit{Simulating the FIP effect in coronal loops using a multi-species kinetic-fluid model}, September 9 2021, ESPM-16 virtual edition. \\

\textit{Simulating the FIP effect in coronal loops using a multi-species kinetic-fluid model}, April 19-30 2021, EGU General Assembly Conference - Online edition, \href{https://ui.adsabs.harvard.edu/abs/2021EGUGA..23.8369P}{ADS LINK}. \\

\textit{The forming slow solar wind imaged along streamer rays by the wide-angle imager on Parker Solar Probe}, May 4-8 2020, EGU General Assembly Conference - Online edition, \href{https://ui.adsabs.harvard.edu/abs/2020EGUGA..2211552P}{ADS LINK}. \\

\section{Posters}

\hspace{\parindent} \textit{The many faces of Nicolas's PhD thesis}, May 2021, IRAP PhD day. \\

\textit{Simulating the FIP Effect in Coronal Loops Using a Multi-Species Kinetic-Fluid Model}, December 1-17 2020, AGU Fall Meeting - Online edition, \href{https://ui.adsabs.harvard.edu/abs/2020AGUFMSH0290009P}{ADS LINK}. \\

\textit{Evolution of the internal structure of CME flux ropes from the Sun to 1AU}, April 7-12 2019, EGU General Assembly Conference, \href{https://ui.adsabs.harvard.edu/abs/2019EGUGA..21.7591P}{ADS LINK}. \\

%% file: chapters/TableofContents.tex
\renewcommand{\contentsname}{Table of contents (full)}
\tableofcontents

%% file: chapters/ListofAbreviations.tex
\chapter{List of initials and abbreviations}

\begin{itemize}
    \item \textit{ACE}: \textit{Advanced Composition Explorer}
    \item ADAPT: Air force Data Assimilative Flux Transport
    \item \textit{AIA}: \textit{Atmospheric Imaging Assembly}
    \item AU: astronomical unit
    \item CBP: coronal bright point
    \item CDPP: Centre de Données de la Physique des Plasmas
    \item CE: charge-exchange
    \item CIR: corotating interaction regions
    \item CME: coronal mass ejection
    \item DI: direct ionization
    \item DR: dielectronic recombination
    \item EA: excitation-autoionization
    \item \textit{EIS}: \textit{EUV imaging spectrometer}
    \item \textit{EIT}: \textit{Extreme ultraviolet Imager Telescope}
    \item ESA: European Space Agency
    \item EUV: extreme ultraviolet
    \item FIP: first ionisation potential
    \item FITS: Flexible Image Transport System
    \item FOV: field-of-view
    \item FSW: fast solar wind
    \item GONG: Global Oscillation Network Group
    \item HCS: heliospheric current sheet
    \item HPS: heliospheric plasma sheet
    \item IDL: Interactive Data Language
    \item \textit{IMP}: \textit{Interplanetary Monitoring Platform}
    \item IRAP: Institut de Recherche en Astrophysique et Planétologie
    \item ISAM: Irap Solar Atmospheric Model
    \item ISSI: International Space Science Institute
    \item \textit{LASCO}: \textit{Large-Angle and Spectrometric Coronagraph}
    \item LOS: line-of-sight
    \item LTE: local thermal equilibrium
    \item MADAWG: Modelling and Data Analysis Working Group
    \item MAS: Magnetohydrodynamics Around a Sphere
    \item MCT: Magnetic Connectivity Tool
    \item MHD: magneto-hydro-dynamics
    \item MULTI-VP: Multiple Flux tube Solar Wind Model
    \item NASA: National Aeronautics and Space Administration
    \item NSO: National Solar Observatory
    \item PI: photoionization
    \item \textit{PSP}: \textit{Parker Solar Probe}
    \item RR: radiative recombination
    \item SCS: Schatten Current Sheet
    \item s/c: spacecraft
    \item \textit{SDO}: \textit{Solar Dynamics Observatory}
    \item \textit{SoHO}: \textit{Solar and Heliospheric Observatory}
    \item \textit{SolO}: \textit{Solar Orbiter}
    \item SSW: slow solar wind
    \item \textit{STEREO}: \textit{Solar-TErrestrial RElations Observatory}
    \item \textit{SWICS}: \textit{Solar Wind Ion Composition Spectrometer}
    \item TNE: thermal non-equilibrium
    \item TR: transition region
    \item WCS: World Coordinate System
    \item \textit{WISPR}: \textit{Wide-Field Imager for Parker Solar Probe}
    \item WKB: Wentzel–Kramers–Brillouin
    \item WL: white-light
    \item WSA: Wang-Sheeley-Arge
    \item WSO: Wilcox Solar Observatory
    \item 3-D: three-dimensional
    \item 2-D: two-dimensional
\end{itemize}

%% file: chapters/ListofFigures.tex
\listoffigures

%% file: Journal_abreviations.tex
\let\jnl@style=\rmfamily 
\def\ref@jnl#1{{\jnl@style#1}}%
\newcommand\aj{\ref@jnl{AJ}}
\newcommand\psj{\ref@jnl{PSJ}}
\newcommand\araa{\ref@jnl{ARA\&A}}
\newcommand\apj{\ref@jnl{ApJ}}
\newcommand\apjl{\ref@jnl{ApJL}}     
\newcommand\apjs{\ref@jnl{ApJS}}
\newcommand\ao{\ref@jnl{ApOpt}}
\newcommand\apss{\ref@jnl{Ap\&SS}}
\newcommand\aap{\ref@jnl{A\&A}}
\newcommand\aapr{\ref@jnl{A\&A~Rv}}
\newcommand\aaps{\ref@jnl{A\&AS}}
\newcommand\azh{\ref@jnl{AZh}}
\newcommand\baas{\ref@jnl{BAAS}}
\newcommand\icarus{\ref@jnl{Icarus}}
\newcommand\jaavso{\ref@jnl{JAAVSO}}  
\newcommand\jrasc{\ref@jnl{JRASC}}
\newcommand\memras{\ref@jnl{MmRAS}}
\newcommand\mnras{\ref@jnl{MNRAS}}
\newcommand\pra{\ref@jnl{PhRvA}}
\newcommand\prb{\ref@jnl{PhRvB}}
\newcommand\prc{\ref@jnl{PhRvC}}
\newcommand\prd{\ref@jnl{PhRvD}}
\newcommand\pre{\ref@jnl{PhRvE}}
\newcommand\prl{\ref@jnl{PhRvL}}
\newcommand\pasp{\ref@jnl{PASP}}
\newcommand\pasj{\ref@jnl{PASJ}}
\newcommand\qjras{\ref@jnl{QJRAS}}
\newcommand\skytel{\ref@jnl{S\&T}}
\newcommand\solphys{\ref@jnl{SoPh}}
\newcommand\sovast{\ref@jnl{Soviet~Ast.}}
\newcommand\ssr{\ref@jnl{SSRv}}
\newcommand\zap{\ref@jnl{ZA}}
\newcommand\nat{\ref@jnl{Nature}}
\newcommand\iaucirc{\ref@jnl{IAUC}}
\newcommand\aplett{\ref@jnl{Astrophys.~Lett.}}
\newcommand\apspr{\ref@jnl{Astrophys.~Space~Phys.~Res.}}
\newcommand\bain{\ref@jnl{BAN}}
\newcommand\fcp{\ref@jnl{FCPh}}
\newcommand\gca{\ref@jnl{GeoCoA}}
\newcommand\grl{\ref@jnl{Geophys.~Res.~Lett.}}
\newcommand\jcp{\ref@jnl{JChPh}}
\newcommand\jgr{\ref@jnl{J.~Geophys.~Res.}}
\newcommand\jqsrt{\ref@jnl{JQSRT}}
\newcommand\memsai{\ref@jnl{MmSAI}}
\newcommand\nphysa{\ref@jnl{NuPhA}}
\newcommand\physrep{\ref@jnl{PhR}}
\newcommand\physscr{\ref@jnl{PhyS}}
\newcommand\planss{\ref@jnl{Planet.~Space~Sci.}}
\newcommand\procspie{\ref@jnl{Proc.~SPIE}}

\newcommand\actaa{\ref@jnl{AcA}}
\newcommand\caa{\ref@jnl{ChA\&A}}
\newcommand\cjaa{\ref@jnl{ChJA\&A}}
\newcommand\jcap{\ref@jnl{JCAP}}
\newcommand\na{\ref@jnl{NewA}}
\newcommand\nar{\ref@jnl{NewAR}}
\newcommand\pasa{\ref@jnl{PASA}}
\newcommand\rmxaa{\ref@jnl{RMxAA}}

\newcommand\maps{\ref@jnl{M\&PS}}
\newcommand\aas{\ref@jnl{AAS Meeting Abstracts}}
\newcommand\dps{\ref@jnl{AAS/DPS Meeting Abstracts}}

\let\astap=\aap 
\let\apjlett=\apjl 
\let\apjsupp=\apjs 
\let\applopt=\ao